\documentclass[onecolumn]{aa}
\usepackage{natbib}
\bibpunct{(}{)}{;}{a}{}{,}
\usepackage{graphicx}
\usepackage{txfonts}
\usepackage[hang,bf,nooneline]{subfigure}
\usepackage[hang,bf,nooneline]{caption}
\usepackage{rotating}

\begin{document}

\title{EGRET Excess of Diffuse Galactic Gamma Rays\\ as Tracer of Dark Matter}

\author{W. de Boer\inst{1}, C. Sander\inst{1}, V. Zhukov\inst{1},
        A.V. Gladyshev\inst{2,3}, D.I. Kazakov\inst{2,3} }

   \offprints{W. de Boer}

\institute{Institut f\"ur Experimentelle Kernphysik, Universit\"at
           Karlsruhe (TH), P.O. Box 6980, 76128 Karlsruhe, Germany\\
   \email{Wim.de.Boer@cern.ch, sander@ekp.uni-karlsruhe.de, zhukov@ekp.uni-karlsruhe.de}
   \and
           Bogoliubov Laboratory of Theoretical Physics, Joint Institute for Nuclear Research,
           141980 Dubna, Moscow Region, Russia\\
   \email{gladysh@theor.jinr.ru, kazakovd@theor.jinr.ru}
   \and Institute of Theoretical and Experimental Physics, 117218, 25 B.Cheremushkinskaya, Moscow, Russia}

   \date{Received June 29, 2005; accepted August 17, 2005}

\authorrunning{de Boer \textit{et al.}}
\titlerunning{Diffuse Galactic Gamma Rays as Tracer of Dark Matter}

\abstract{ The public data from the EGRET space telescope on diffuse
Galactic gamma rays in the energy range from 0.1 to 10 GeV are
reanalyzed with the purpose of searching for signals of Dark Matter
annihilation (DMA). The analysis confirms the previously observed
excess for energies above 1 GeV in comparison with the expectations
from conventional Galactic models. In addition, the excess was found
to show all the key features of a signal from Dark Matter
Annihilation (DMA): a) the excess is observable in all sky
directions and has the same shape everywhere, thus pointing to a
common source; b) the shape corresponds to the expected spectrum of
the annihilation of non-relativistic massive particles into - among
others - neutral $\pi^0$ mesons, which decay into photons. From the
energy spectrum of the excess we deduce a WIMP mass between 50 and
100 GeV, while from the intensity of the excess in all sky
directions the shape of the halo could be reconstructed. The DM halo
is  consistent with  an almost spherical isothermal profile with
substructure in the Galactic plane in the form of toroidal rings at
4 and 14 kpc from the center. These rings lead to a peculiar shape
of the rotation curve, in agreement with the data, which proves that
the EGRET excess traces the Dark Matter.

}

\maketitle


\section{Introduction}

Cold Dark Matter (CDM) makes up 23\% of the energy of the Universe,
as deduced from the WMAP measurements of the temperature
anisotropies in the Cosmic Microwave Background, in combination with
data on the Hubble expansion and the density fluctuations in the
universe \citep{wmap}. The nature of the CDM is unknown, but one of
the most popular explanation for it is the neutralino, a stable
neutral particle predicted by Supersymmetry \citep{lspdm,jungman}.
The neutralinos, usually denoted by $\chi$, are spin 1/2 Majorana
particles, which can annihilate into pairs of Standard Model
particles, but other candidates are possible as well. The only
assumptions needed for  this analysis is that  DM particles were
produced in thermal equilibrium with all other particles in the
early Universe and its number density $n_\chi$ decreases from the
high value in the early universe to the present low number density
by annihilation, as it happened with  the number density of baryons
as well \citep{kolb}. The relic density of CDM  is inversely
proportional to $\langle\sigma v\rangle$, the averaged annihilation
cross section $\sigma$ of DM particles multiplied with their
relative velocity $v$. This inverse proportionality is obvious, if
one considers that a higher annihilation rate, given by
$\langle\sigma v\rangle n_\chi$, would have reduced the relic
density before freeze-out, i.e. the time, when the expansion rate of
the Universe, given by the Hubble constant, became equal to or
larger than the annihilation rate. The relation can be written as
\begin{equation}
\Omega_\chi h^2=\frac{m_\chi n_\chi}{\rho_c}\approx
(\frac{2\cdot 10^{-27} cm^3 s^{-1}}{\langle\sigma v\rangle})\label{wmap}.
\end{equation}
The nominator in the last part of this equation was calculated with CalcHEP \citep{calchep}  and 
found to be 30\% smaller than the value calculated in \citet{jungman}. For the
present value of $\Omega h^2=0.113 \pm 0.009$, as measured by WMAP
\citep{wmap}, the thermally averaged total cross section at the
freeze-out temperature of $m_\chi/22$ must have been around $2\cdot
10^{-26} ~{\rm cm^3s^{-1}}$. This is a cross section typical for
weak interactions and explains why the DM does not cluster strongly
in the center of galaxies, like the baryons do: the cross sections
are simply too small to have large energy losses when falling
towards the center. Therefore the DM particles are generically
called WIMPs, Weakly Interacting Massive Particles. All possible
enhancements of the annihilation rate from the clustering of DM
(usually called boost factor) will be calculated with respect to
this generic cross section, which basically only depends on the
value of the Hubble constant. Note that $\langle\sigma v\rangle$ as
calculated from Eq.~\ref{wmap}, is independent of the WIMP mass
(except for logarithmic corrections), as can be shown by a detailed
calculation \citep{kolb}.

The stable decay and fragmentation products from  Dark Matter
Annihilation (DMA) are neutrinos, photons, protons, antiprotons,
electrons and positrons. From these, the protons and electrons
disappear in the sea of many matter particles in the universe, but
the photons and antimatter particles may be detectable above the
background, generated by  particle interactions. Such  searches for
indirect Dark Matter detection have been actively pursued, see e.g.
the reviews by \citet{bergstrom} and \citet{sumner} or more recently
 by
\citet{Bertone:2004pz}. References to earlier work can be found in these reviews.

Gamma rays have the advantage that they
point back to the source and do not suffer energy losses, so they
are the ideal candidates to trace the dark matter density. The
charged components interact with Galactic matter and are deflected
by the Galactic magnetic field, so they do not point back to the
source.

A  detailed  distribution of gamma ray fluxes for energies between
0.03 and 10 GeV was obtained by the Energetic Gamma Ray Emission
Telescope EGRET, one of the four instruments on the Compton Gamma
Ray Observatory CGRO, which collected data during nine years, from
1991 to 2000. The diffuse component shows a clear excess of about a
factor two over the expected background from known nuclear
interactions, inverse Compton scattering and bremsstrahlung. The
excess was observed first by \citet{hunter}, while the most complete
calculation of the background is encoded in the GALPROP program
\citep{galprop1,galprop2}.
 This excess  was shown to possess all the key
features from dark matter annihilation (DMA), as discussed at
various conferences and workshops\citep{deboer1,deboer2,deboer3,deboer4}.
By fitting only
the {\it shapes} of the background and DMA signal the analysis
becomes independent of the absolute normalization of the background.
Therefore uncertainties in the background fluxes from e.g. the gas
densities and cosmic ray fluxes are eliminated.

Apart from fitting the shapes of DMA signal and background the
present analysis differs also from previous ones 
\citep[see e.g.][]{previous1,previous2,previous3,previous4,
previous5,previous6,previous7,
previous8,previous9,previous10,previous11,previous12,previous13,previous14,previous15,previous16}
by the fact that the fluxes {\it and} the energy spectrum in {\it
all} sky directions were considered simultaneously. This requires a
complete reanalysis of the publicly available EGRET data, since the
diffuse gamma ray data have been published only in a limited number
of sky directions.

Considering the excess in all sky directions with a sufficiently large
resolution allows to reconstruct the DM halo profile,
which  in turn can be used - in combination with the distribution of
visible matter - to reconstruct the shape of the rotation curve. The
absolute normalization of the DM density distribution or halo
profile can be obtained by requiring that the local rotation
velocity of our solar system at 8.3 kpc is 220 km/s. It will be
shown that the rotation curve, as obtained from the EGRET excess of
gamma rays, provides the first explanation for the peculiar change
of slope in the rotation curve at around 11 kpc, indicating that the
excess traces DM.

The paper has been organized as follows: Section \ref{fit} explains the fitting method,
Section \ref{analysis} describes the analysis of the EGRET data, Section \ref{halo}
describes the determination of the DM halo profile and comparison with the Galactic
rotation curve, Section \ref{objections} discusses  possible objections to
the DMA interpretation of the EGRET excess and Section \ref{summary} summarizes the results.

\section{Fitting Method for Indirect Dark Matter Annihilation}\label{fit}

As mentioned in the introduction, neutral particles play  a very special
role for indirect DM searches, since they point back to the source.
Therefore the gamma rays provide a perfect means to reconstruct the
halo profile of the DM by observing the intensity of the
gamma ray emissions in the various sky directions.

Of course there are different sources of diffuse gamma rays in the
Galaxy, so disentangling the annihilation signal is at first glance
not easy. However, the spectral {\it shapes} of the diffuse gamma ray
backgrounds and DMA signal are well known from accelerator experiments
and it is precisely the shape which was well measured by the EGRET telescope.
Furthermore, the shapes of the background and DMA signal are sufficiently different
to disentangle their contributions to the data by leaving the absolute
normalizations for background and DMA signal as free parameters in the fit.
We discuss first the galactic backgrounds, then the DMA signal and finally
the extragalactic background.

The galactic backgrounds are: decays of $\pi^0$ mesons
produced in nuclear interactions,  contributions from inverse
Compton scattering of electrons on photons and Bremsstrahlung from
electrons in the Coulomb field of nuclei. The best estimate of the
relative contributions is given by the GALPROP code
\citep{galprop,galprop1,galprop2}, which parametrizes  the gas
densities, cross sections and energy spectra for all processes of
interest and solves numerically the diffusion equation to obtain a
complete solution for the density map of all primary and secondary
nuclei. For this the so-called ``conventional'' GALPROP model was
used, which assumes the spectra of protons and electrons, as
measured locally in the solar system, to be representative for the
whole Galaxy. For protons, which have negligible energy losses, this
is a reasonable assumption; for electrons, which have larger energy
losses from ionization and Bremsstrahlung, this assumption may be
questioned. Therefore we have restricted the analysis to gamma ray
energies above 0.07 GeV, in which case the $\pi^0$ component starts
to become dominant: electron-induced gamma ray production is of the
same order of magnitude as the nuclei-induced gamma ray production
at this energy, but at 0.5 GeV the electron-induced component is
already below 10\% for the inner Galaxy and  below 25\% for other
regions. Therefore one is not too sensitive to  electron-induced
gamma rays in the region of the EGRET excess, which is maximal at
energies around 2-4 GeV. The ``conventional'' model is to be
contrasted with the ``optimized'' model \citep{optimized}, in which
case the electron and proton spectra are ``optimized'' to
explain the EGRET excess without DM by allowing them to deviate from the
locally observed spectra. Even the freedom for both the proton and electron
 spectra does not lead to
particular good fits of the background to the EGRET data, if all sky directions
are considered,
as will be discussed in the next section.

As mentioned  before, only the shape of the background
is needed for the fit, not the absolute normalization. Leaving the
normalization free in the fit for a given sky direction is
important, simply because one does not know the cosmic ray and gas
densities to better than about 20\% for a given sky direction. On
the other hand, one knows for  given cosmic ray spectra the shape of
the expected gamma rays quite well: for electron-induced gamma
production these processes can be easily calculated, while for the
nuclei-induced processes the gamma ray spectra are
known from the scattering of beams of nuclei on fixed targets. In the galaxy the
``beams'' are the cosmic rays, while the  gas in the disk is
the fixed target. The wealth of data on hadronic interactions have
resulted in the so called string fragmentation model, encoded e.g.
in the PYTHIA program from \citet{pythia}, which describes gamma ray
production from nuclear interactions
well.

The small contribution from the electron-induced gamma rays was
added to the dominant contribution from $\pi^0$ decays and the shape
of the total background thus obtained was used for a given sky
direction. The relative fraction from electron-induced and
nuclei-induced gamma rays  varies with sky direction and this
fraction was taken from GALPROP, but a fit with a constant fraction
yielded similar results for the DM profile. Given that we do not
attempt to determine the absolute normalization, the analysis is not
sensitive to the many GALPROP propagation parameters determining
absolute density profiles of the Galactic components. Furthermore
the propagation of the gamma rays is straightforward. The fitted
normalization factor of the background  in each direction was found
to agree within errors with the values determined from the GALPROP
code, as will be discussed later.

\begin{figure}
\begin{center}
 \includegraphics [width=0.5\textwidth,clip]{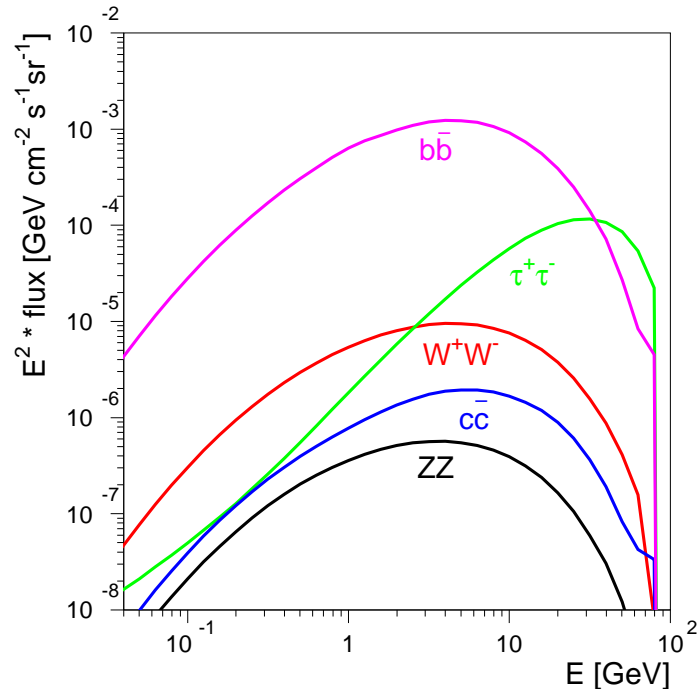}
 \caption[]{
 The expected shape of gamma ray spectra for the
 fragmentation of WIMPs into different final states (arbitrary
 normalization): heavy quark pairs (b,c), heavy gauge bosons (W,Z) and $\tau$ leptons.
 The annihilation of neutralinos, the preferred WIMP candidates,
 into  light final states is helicity suppressed \citep{jungman}.
 \label{gamma_spectra}}
\end{center}
\end{figure}

WIMPs are expected to annihilate into fermion-antifermion or gauge
boson pairs, so a large fraction will result in quark-antiquark
pairs, which produce typically 30-40 photons per annihilation in the
fragmentation process (mainly from $\pi^0$ decays). However, the
photons from DMA  are expected to have a
 spectrum significantly different from the ones from nuclear
interactions. This can be understood qualitatively as follows: the
WIMPs are strongly non-relativistic, so they annihilate almost at
rest. Therefore  DM annihilates into almost mono-energetic pairs of
particles with  an energy approximately equal to the WIMP mass. This
results in a rather energetic gamma ray spectrum with a sharp cut-off
 at the WIMP mass. Such gamma ray spectra from the
fragmentation of mono-energetic quark pairs have been measured
precisely at electron-positron colliders and the data are well
described by the  string fragmentation model mentioned above. The
expected gamma ray spectra  from this program are shown in Fig.
\ref{gamma_spectra} for a WIMP mass of 100 GeV and several
annihilation channels.  The difference  between various channels is
small, except for the $\tau$ final state, which has only a small
 $\pi^0$ multiplicity. The corresponding hard gamma ray spectrum 
from $\tau$ decays does not fit the data and excludes
therefore a large annihilation into   $\tau$-pairs. 


In addition to the Galactic background (GB) one expects a
contribution from the extragalactic background (EGBG). The origin of
these gamma rays can be other galaxies which may yield similar
contributions as our Galaxy, or quite different sources like Active
Galactic Nuclei (AGN), quasars or blazars. Since each extragalactic object
has individual properties, it is difficult to  predict  the
shape or the absolute value of this background component.
Experimentally the EGBG can be obtained by subtracting from the
EGRET data the Galactic contribution using the extrapolation method
pioneered by \citet{sreekumar}. Of course, the Galactic contribution
includes a contribution from Galactic DM, which is
dependent on the EGBG, so the EGBG can only be obtained in an
iterative procedure, as was done by \citet{sander}. This
contribution is taken to be of the same shape and same magnitude in
all sky directions. It starts to become important towards the galactic poles,
 where both the galactic
background and the DMA become small.
\begin{table}
\begin{center}
 \begin{tabular}{cccc} \hline
 Region & Longitude $l$ & Latitude $|b|$ & Description\\\hline
 A & 330-30 & 0-5 & Inner Galaxy\\
 B & 30-330 & 0-5 & Disk without inner Galaxy\\
 C & 90-270 & 0-10 & Outer Galaxy\\
 D & 0-360 & 10-20 & Low longitude \\
 E & 0-360 & 20-60 & High longitude\\
 F & 0-360 & 60-90 & Galactic Poles\\
 \hline
 \end{tabular}
 \end{center}
 \caption[skyregions]{The six sky regions mentioned in the text; the detailed halo profile
 was obtained from 180 independent sky directions, as described in the Appendix.}
 \label{t1}
 \end{table}

\begin{figure}
  \begin{center}
    \includegraphics[width=0.32\textwidth]{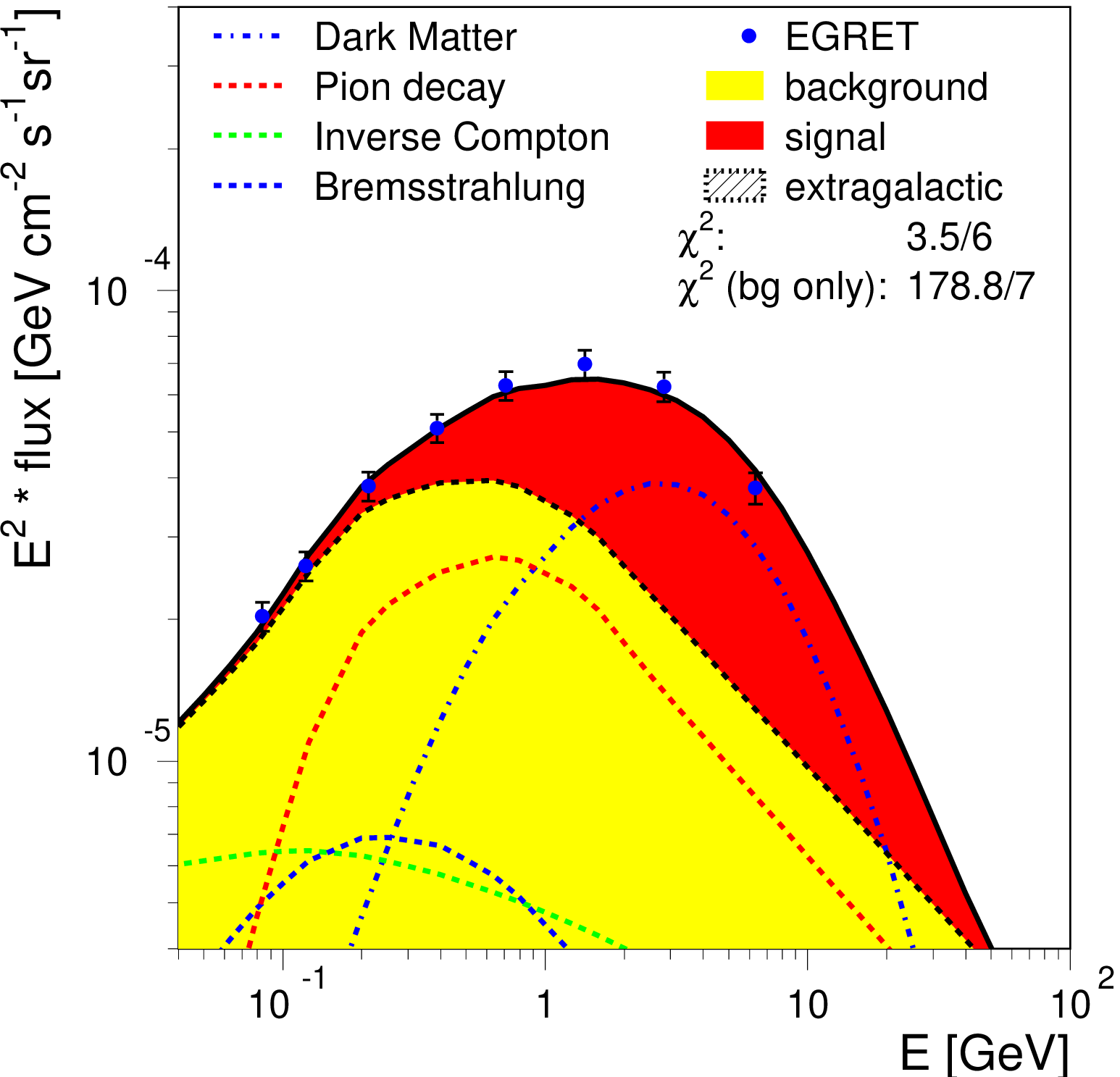}
    \includegraphics[width=0.32\textwidth]{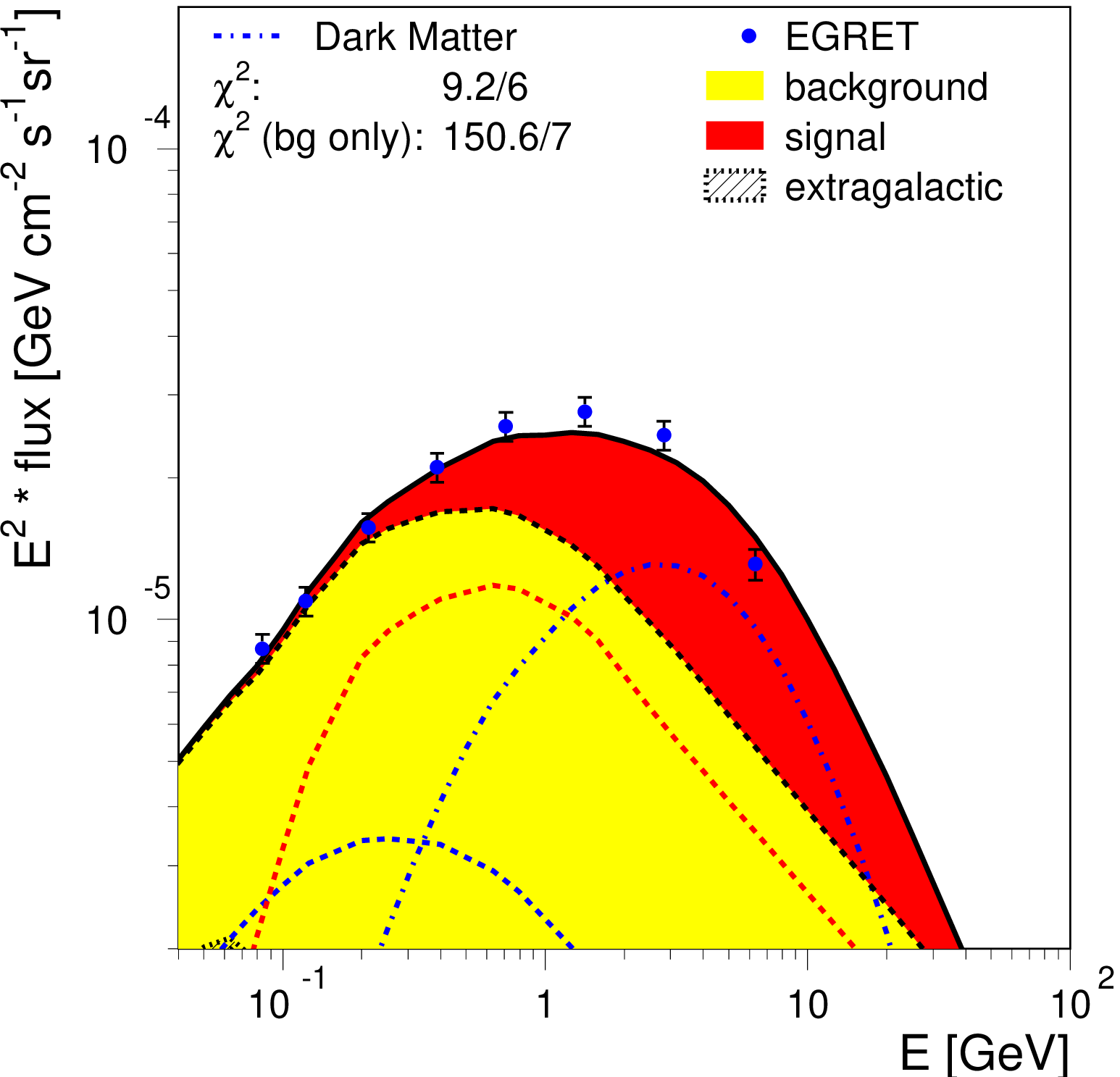}
    \includegraphics[width=0.32\textwidth]{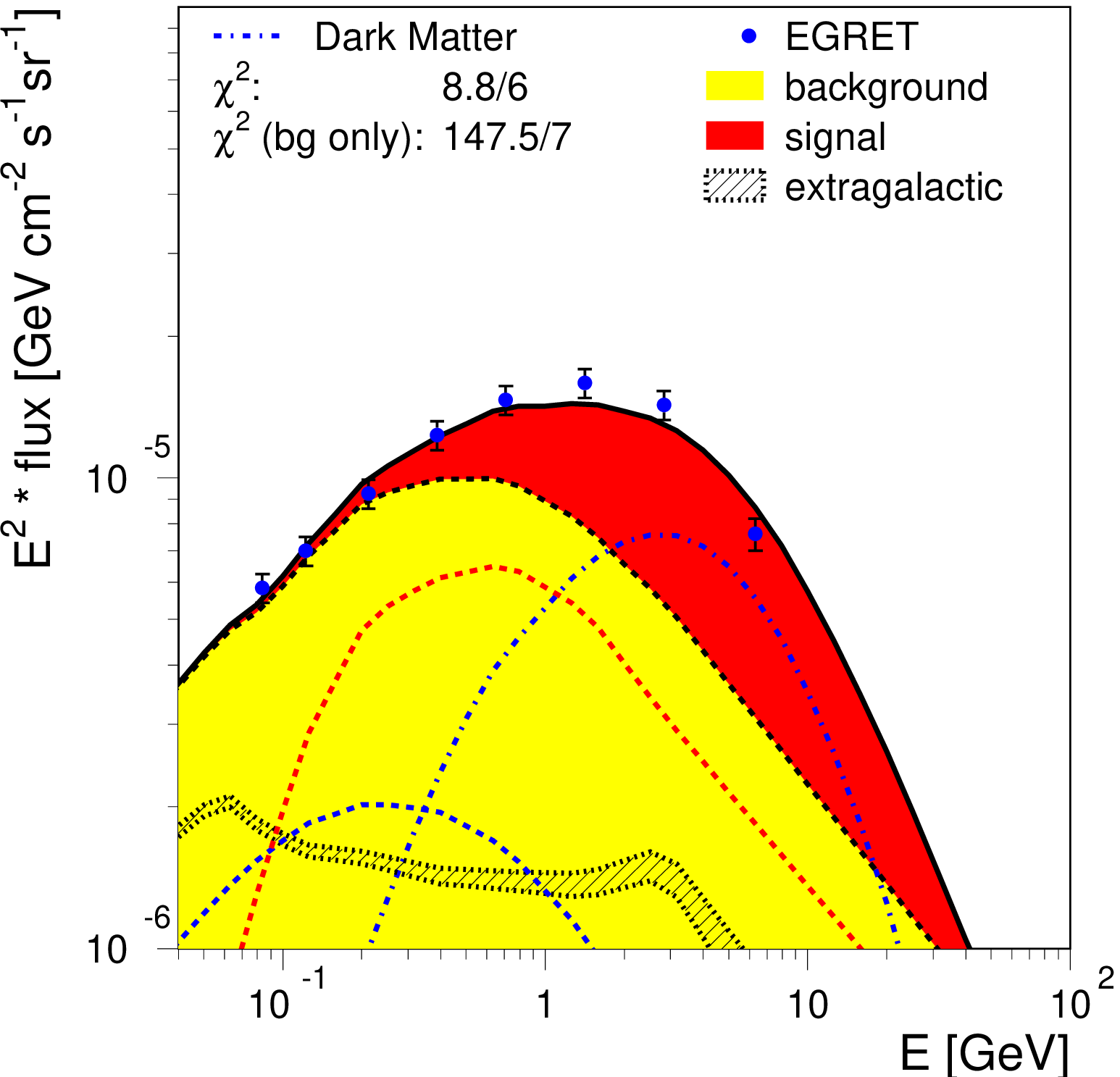}
    \includegraphics[width=0.32\textwidth]{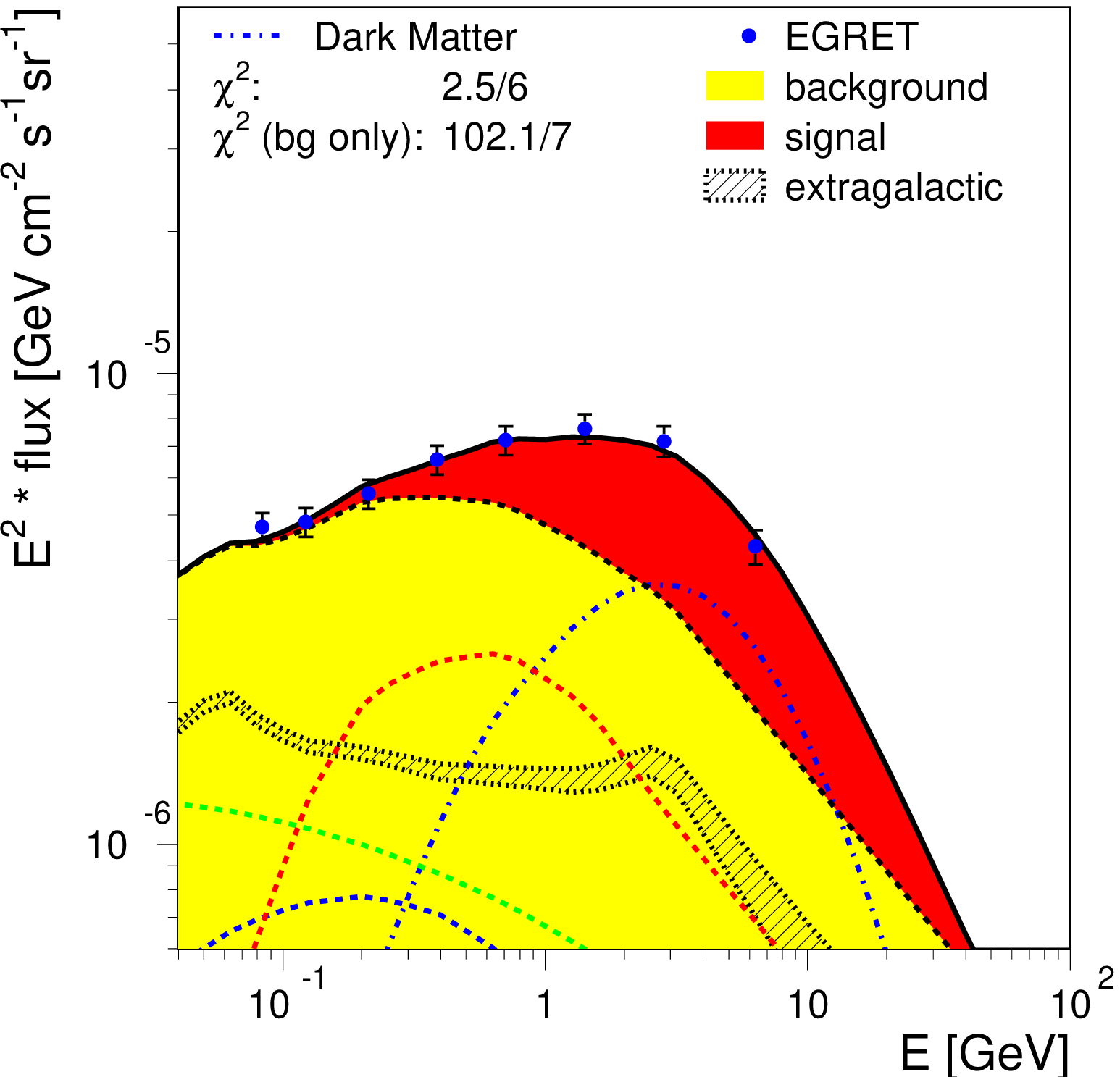}
    \includegraphics[width=0.32\textwidth]{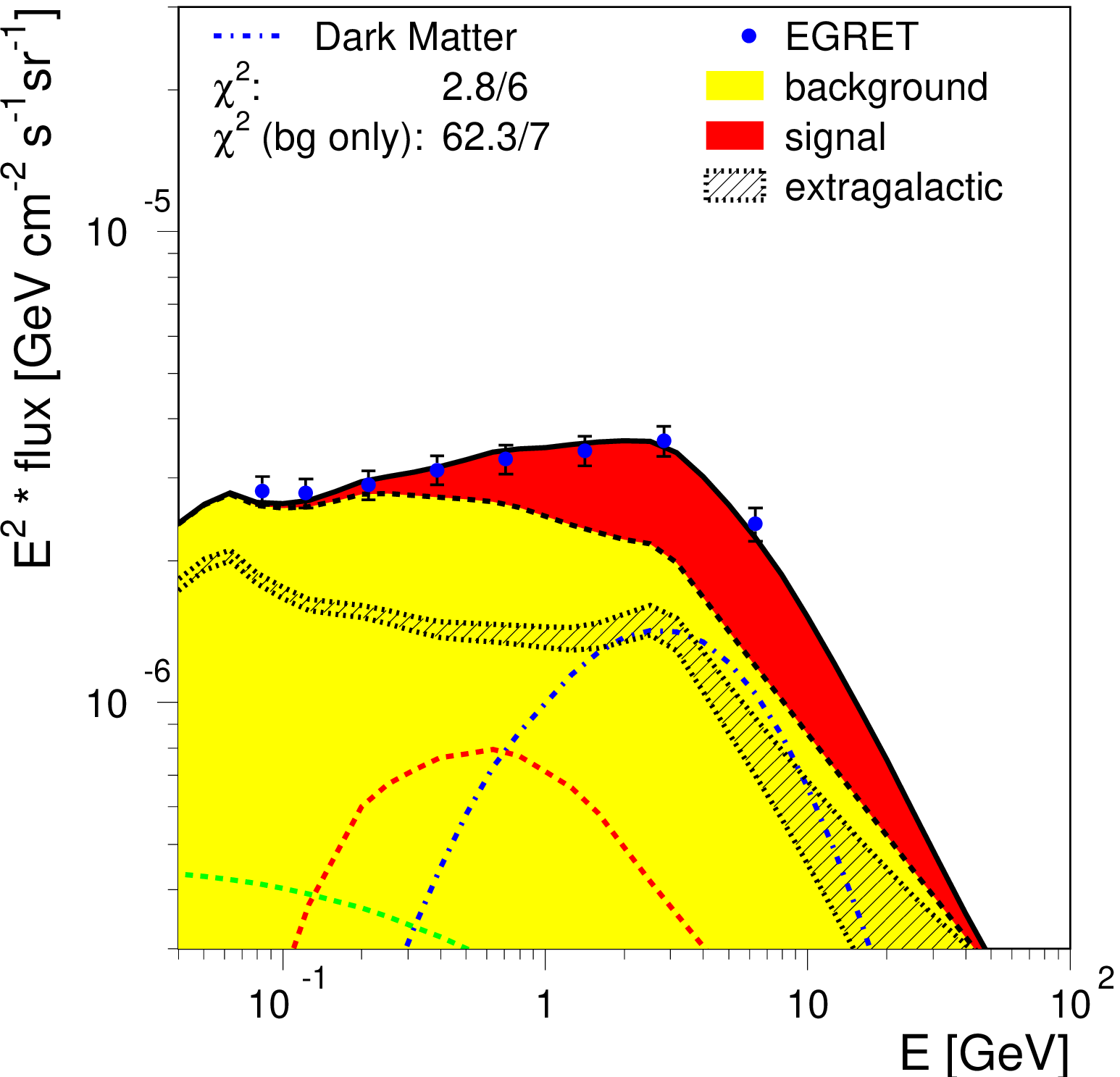}
    \includegraphics[width=0.32\textwidth]{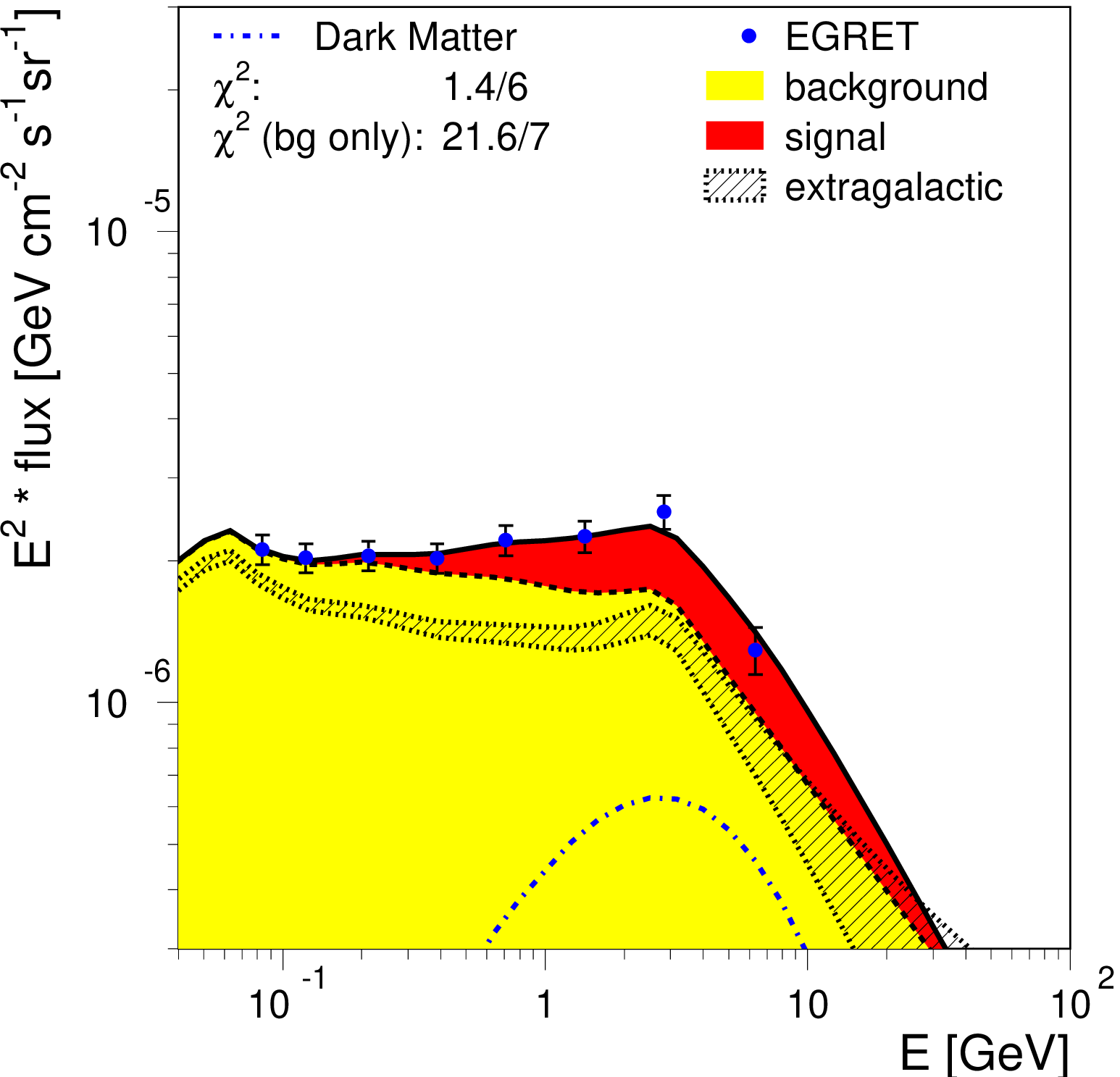}
    \caption[Spectrum of Conventional Model Including Dark
    Matter Annihilation]{Fit of the shapes of background and
    DMA signal to the EGRET data in the Galactic disk (top row,
    regions A, B,C from Table~\ref{t1})and outside the Galactic
    disk (bottom row, regions D,E,F). The light shaded (yellow)
    areas indicate the background using the shape of the
    conventional GALPROP model, while the dark shaded (red)
    areas are the signal contribution from DMA for a 60 GeV WIMP mass.
    The individual shapes of background and DMA have been indicated by dashed
    lines, while the extragalactic background is given by the
    hatched area. The $\chi^2$ of the background is determined
    with an independent background only fit, which yields
    a probability practically zero, as can be estimated from
    the indicated $\chi^2$ values for the statistically
    independent regions. The fit including DM has a total
    probability around 0.8.} \label{excess}
  \end{center}
\end{figure}

\section{Analysis of EGRET data}\label{analysis}
  The EGRET data is  publicly available as
high resolution (0.5x0.5 degree) sky maps from the NASA
archive\footnote{NASA archive:
http://cossc.gsfc.nasa.gov/archive/.}, which allows an analysis in
many independent sky directions after convolution with the
point-spread function, which is a function of energy and becomes
important  for gamma rays below 1 GeV. The data set with the
known point sources subtracted have been used. The point sources are
defined by a 5$\sigma$ enhancement above the diffuse background. In
general these point sources are only a small fraction of the total
gamma ray flux, so the analysis is not sensitive to the subtraction
of point sources. There are only  a few nearby strong sources,
which dominate the flux in these directions
and the subtraction causes an additional systematic uncertainty.
Therefore these  directions have been excluded in the determination
of the halo profile, which requires a fine scanning of all sky directions,
 as will be discussed in the next section. The
contribution of the subtracted point sources have been indicated in
the spectra for 180 independent sky directions, shown in the
Appendix.

The EGRET telescope was carefully calibrated at SLAC with a quasi
mono-energetic photon beam in the energy range of 0.02 to 10 GeV
\citep{egret_cal}. The efficiency and calibration during flight was
also carefully monitored \citep{egret_cal1}. Using Monte Carlo
simulations the energy range was recently extended up to higher
energies with a correspondingly larger uncertainty, mainly from the
self-vetoing of the detector by the back-scattering from the
electromagnetic calorimeter into the veto counters for high
energetic showers \citep{optimized,egret1}. Due to this uncertainty
only data below 10 GeV were used in the fits discussed below.
In total 8 energy ranges were used: 0.07-0.10 GeV; 0.10-0.15 GeV; 0.15-0.30 GeV;
0.30-0.50 GeV, 0.5-1.0 GeV; 1.0-2.0 GeV; 2.0-4.0 GeV; 4.0-10.0 GeV.
The data points have been plotted at the arithmetic mean of the low
and high endpoints of the bin, i.e. at $\sqrt{E_{low}E_{high}}$.

With the 9 years of data taking with the EGRET telescope
 180 independent sky directions can be
studied without statistical problems. However, the data is limited
by systematic uncertainties, which have to be taken into account
carefully. The overall normalization error is usually quoted as
15\%, but the relative point-to-point error is much smaller. The
latter can be determined by fitting the energy spectrum with a
polynomial and if a certain energy bin is left out of the fit, then
its variance with respect to the other energy points can be
determined to be between 3 and 7\%. Therefore,  if one fits only the
shape of the spectrum with a free normalization parameter, only
these relative errors between the energy points are the relevant
ones, which were taken to be 7\%.

\begin{figure}
  \begin{center}
    \includegraphics[width=0.8\textwidth]{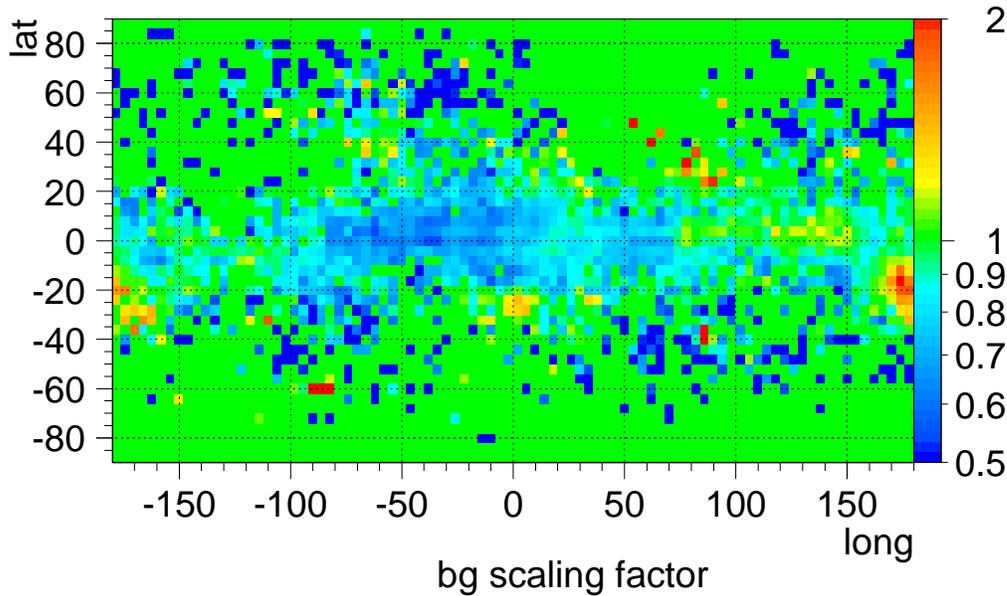}
    \caption[Sky Map of Background Scaling]{The ratio of the
    fitted background normalization and the absolute prediction
    of the conventional GALPROP model \citep{optimized} as function
    of latitude and longitude. For this plot a fine binning was
    used ($90\times 45$); for the whole sky the scaling is around 1,
    i.e. the background determination with our method is in good
    agreement with GALPROP, except for the disk region with
    latitudes below 50, where a systematic deviation of 20-30\% is seen. Since
    the density in the disk is known to be asymmetric, but GALPROP
    uses a symmetric parametrization, this discrepancy is
    understandable.} \label{bgscaling}
  \end{center}
\end{figure}

\begin{figure}
\begin{center}
 \includegraphics [width=0.45\textwidth,clip]{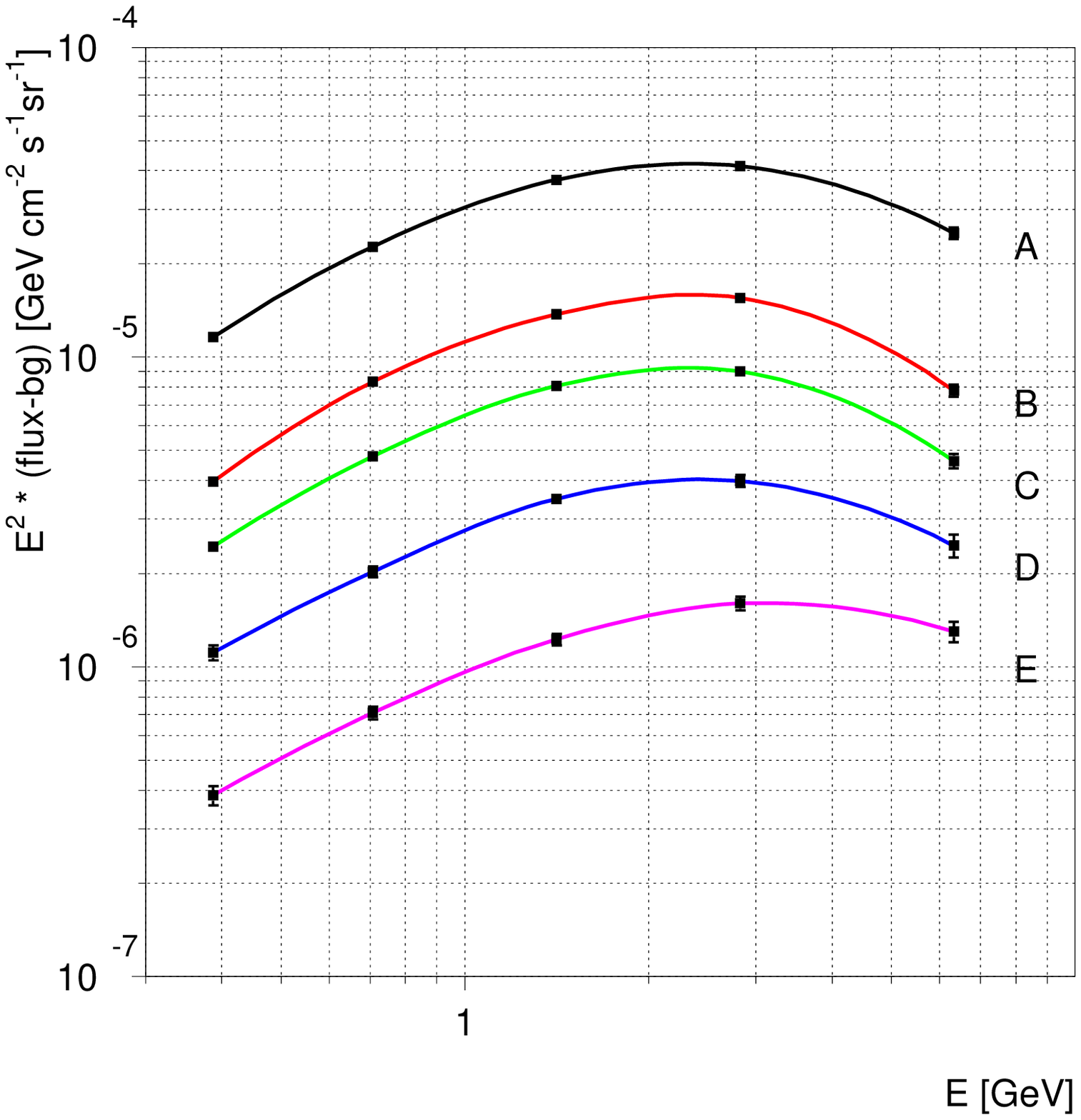}
 \includegraphics [width=0.45\textwidth,clip]{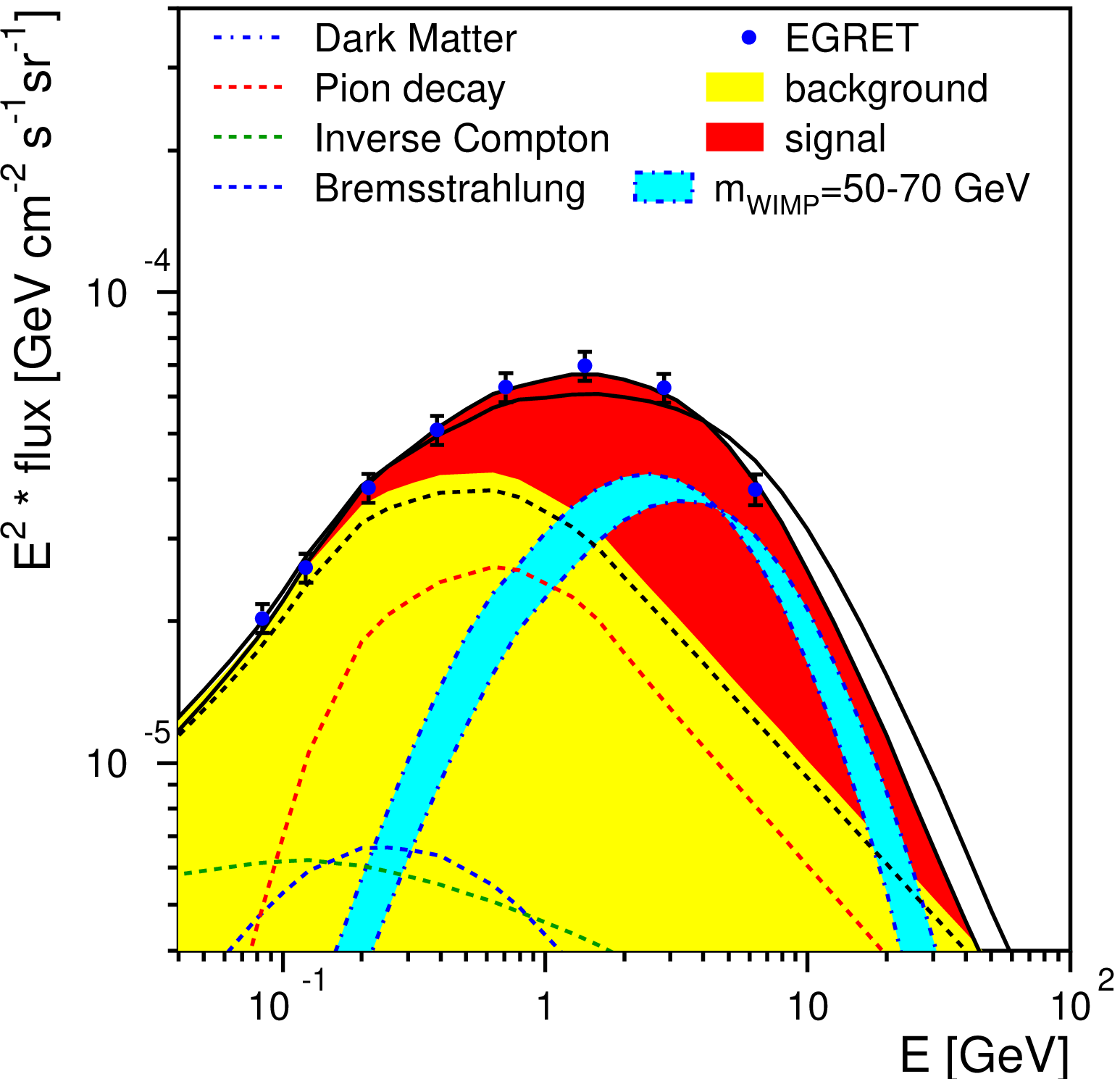}
 \caption[]{
 Left: The difference of the observed EGRET flux and fitted
 background for the various regions of Table \ref{t1}, i.e.
 the red area in Fig. \ref{excess}. One observes the same spectral
 shape for all regions, indicating a common source for the excess.
 Only the statistical errors have been plotted and the curves are
 fitted spline functions to guide the eye.
 Right: The influence of the WIMP mass on
 the spectrum:  the light shaded (blue)
 curve shows the influence of a WIMP mass variation between 50 and 70 GeV.
 The lower (upper)  solid line at the highest energies corresponds to
 the total flux for a 50 (70) GeV WIMP mass.
 \label{diff}}
\end{center}
\end{figure}

Fitting the known shapes of the three contributions (GB, EGBG, DMA)
 to the EGRET data, as discussed before in section \ref{fit}, yielded
astonishingly good fits, as shown in Fig. \ref{excess} for the 6
different sky directions given in Table \ref{t1}. For energies
between 0.07 and 0.5 GeV the flux is dominated by the background,
while above these energies a clear contribution from Dark Matter
annihilation is needed. The excess in different sky directions can
be explained by a single WIMP mass around 60 GeV and a single boost
factor of about 100. The free normalization of the background comes
out to be in reasonable agreement with the absolute predictions from
the GALPROP propagation model of our Galaxy
\citep{galprop,galprop1,galprop2}, as shown in Fig. \ref{bgscaling}
for a fine binning of the skymaps. Thus the fitting method yields
the correct normalization for the BG and the high energy excess for
a given background shape determines the  normalization for the DMA.
The excess in all sky directions has a similar shape, as
demonstrated in Fig. \ref{diff} for the first five sky regions of
Table \ref{t1}. The quality of the EGRET data can be appreciated
from Fig. \ref{diff}, where only the statistical errors are plotted.
They are only visible at large latitudes. The right hand side of
Fig. \ref{diff} shows the plots for WIMP masses of 50 and 70 GeV;
the 70 GeV WIMP mass clearly fits worse.
 WIMP masses below 50 GeV
lead to a too low relic density, since in that case one hits the $Z^0$-resonance
 and for WIMP masses below 40 GeV, i.e. on the other side
of the resonance,  the fit to the EGRET becomes worse.
Therefore a lower limit of 50 GeV is taken and the 95\% C.L. upper limit depends
somewhat on the background model: 70 GeV for the shape of the conventional
model and more like 100 GeV for the shape of the optimized model.
Therefore  the WIMP mass is estimated to be
 between 50 and 100 GeV.

Alternative explanations for the excess have been plentiful. Among
them: locally soft electron and proton spectra, implying that in
other regions of the Galaxy the spectra are harder, thus producing
harder gamma ray spectra. A summary of these discussions has been
given by \citet{optimized}, who find that hard proton spectra are
incompatible with the antiproton yield and hard electron spectra are
difficult to reconcile with the EGRET data up to 120 GeV.
However, they find that by modifying the electron and proton
injection spectra {\it simultaneously}, the
description of the data can be improved by increasing the fluxes
at high energies.
The energy dependence at high energies is kept with the same slope
as the locally measured spectra, which is required at least for
protons, since the energy loss time of protons above a few GeV is of
the order of the lifetime of the universe. Therefore it is hard to
have strong inhomogeneities in the proton spectra at high energies.
At low energies the fluxes are reduced by solar modulations, so
here the spectral shapes of protons and electrons have large
uncertainties and the shape can be optimized to fit the EGRET data.
\begin{figure}
  \begin{center}
    \includegraphics[width=0.32\textwidth]{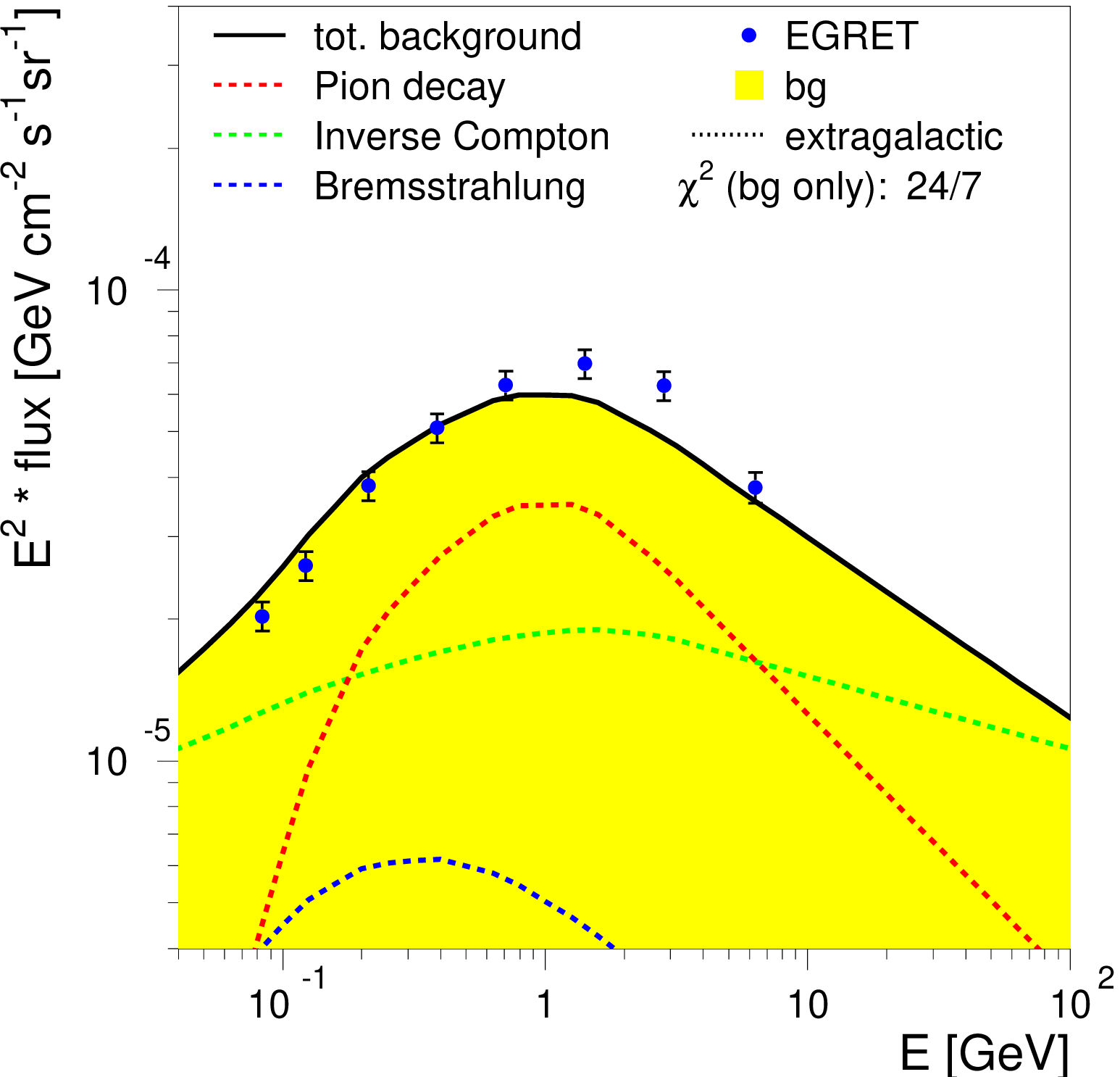}
    \includegraphics[width=0.32\textwidth]{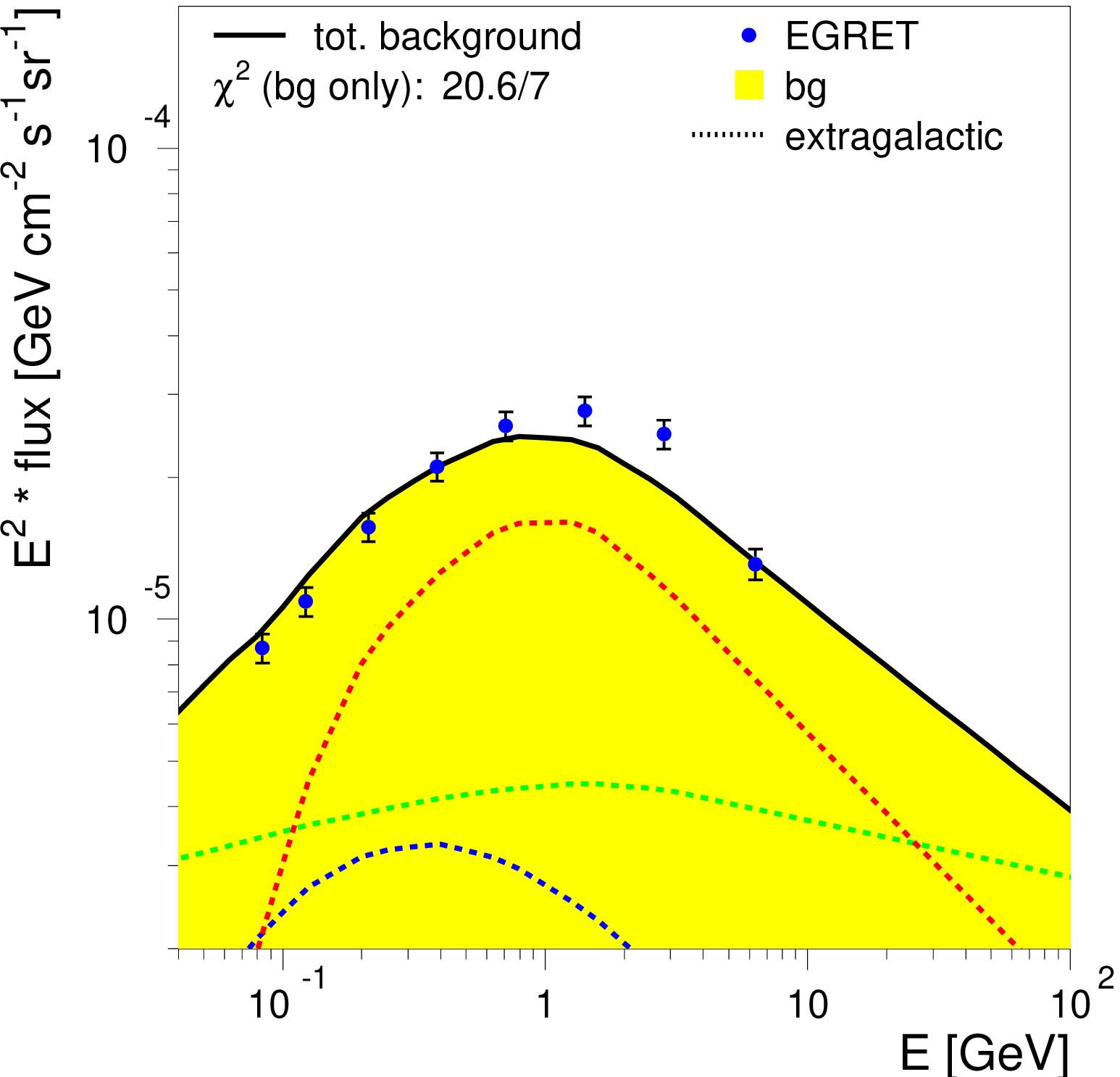}
    \includegraphics[width=0.32\textwidth]{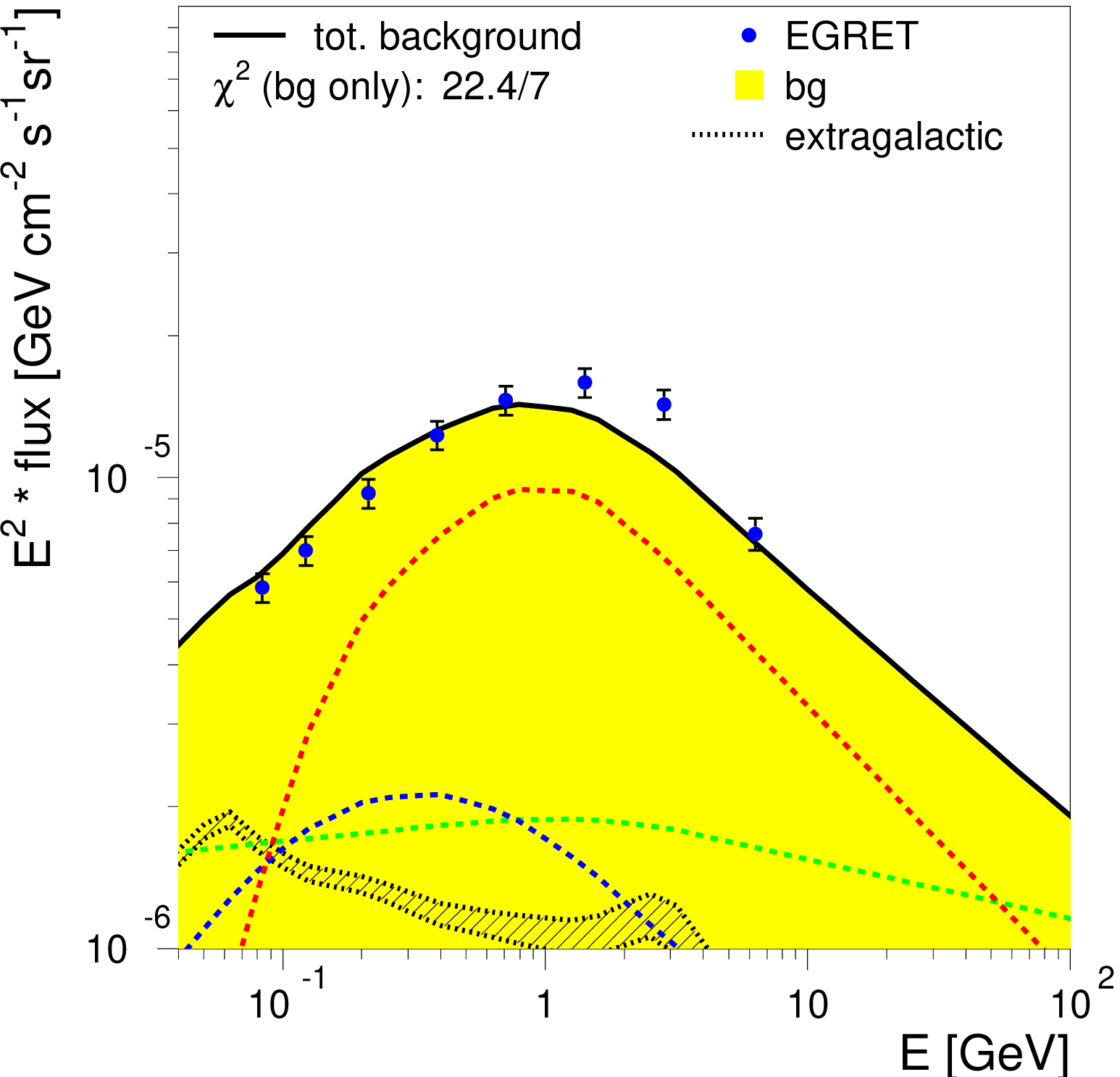}
    \includegraphics[width=0.32\textwidth]{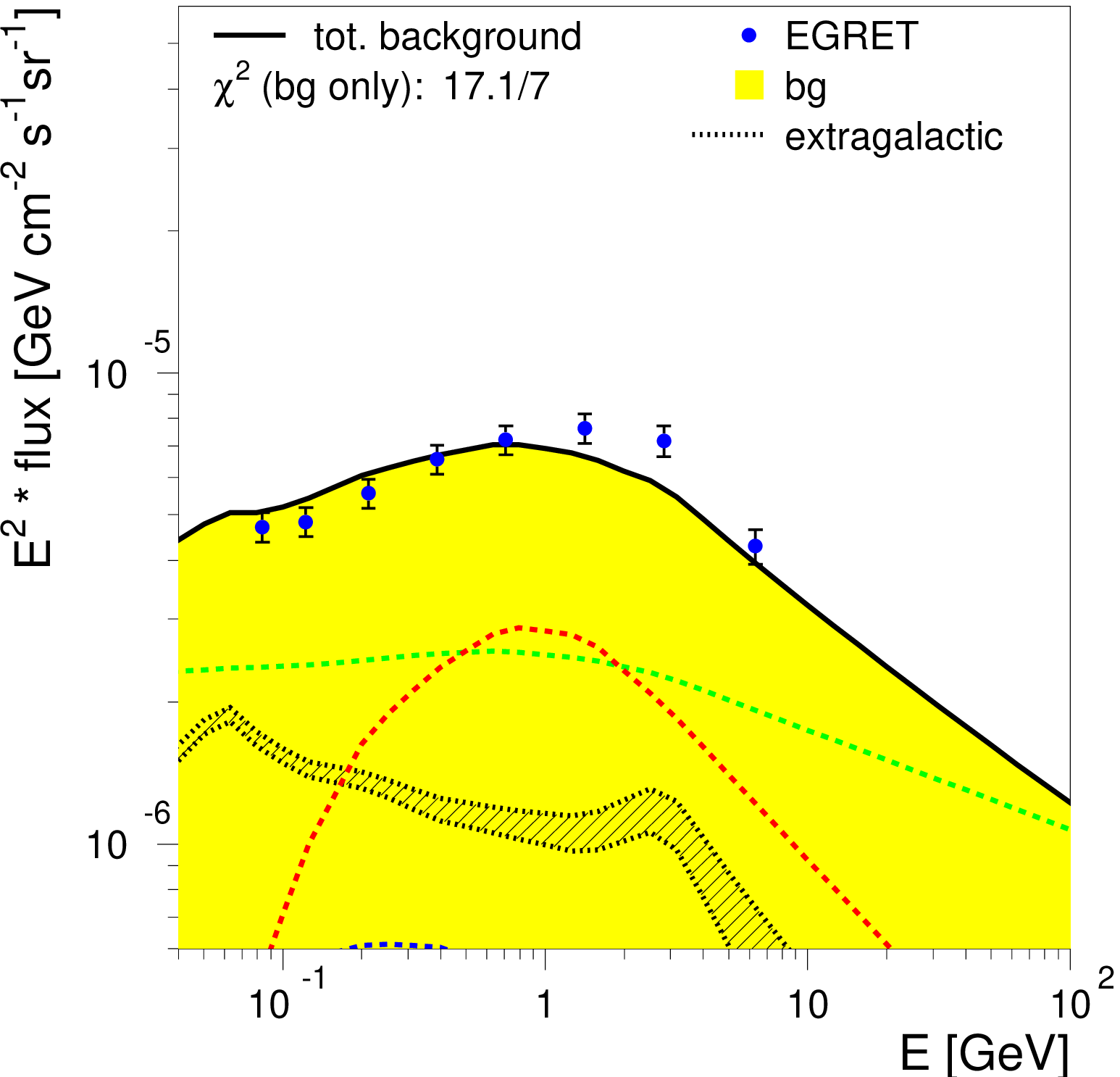}
    \includegraphics[width=0.32\textwidth]{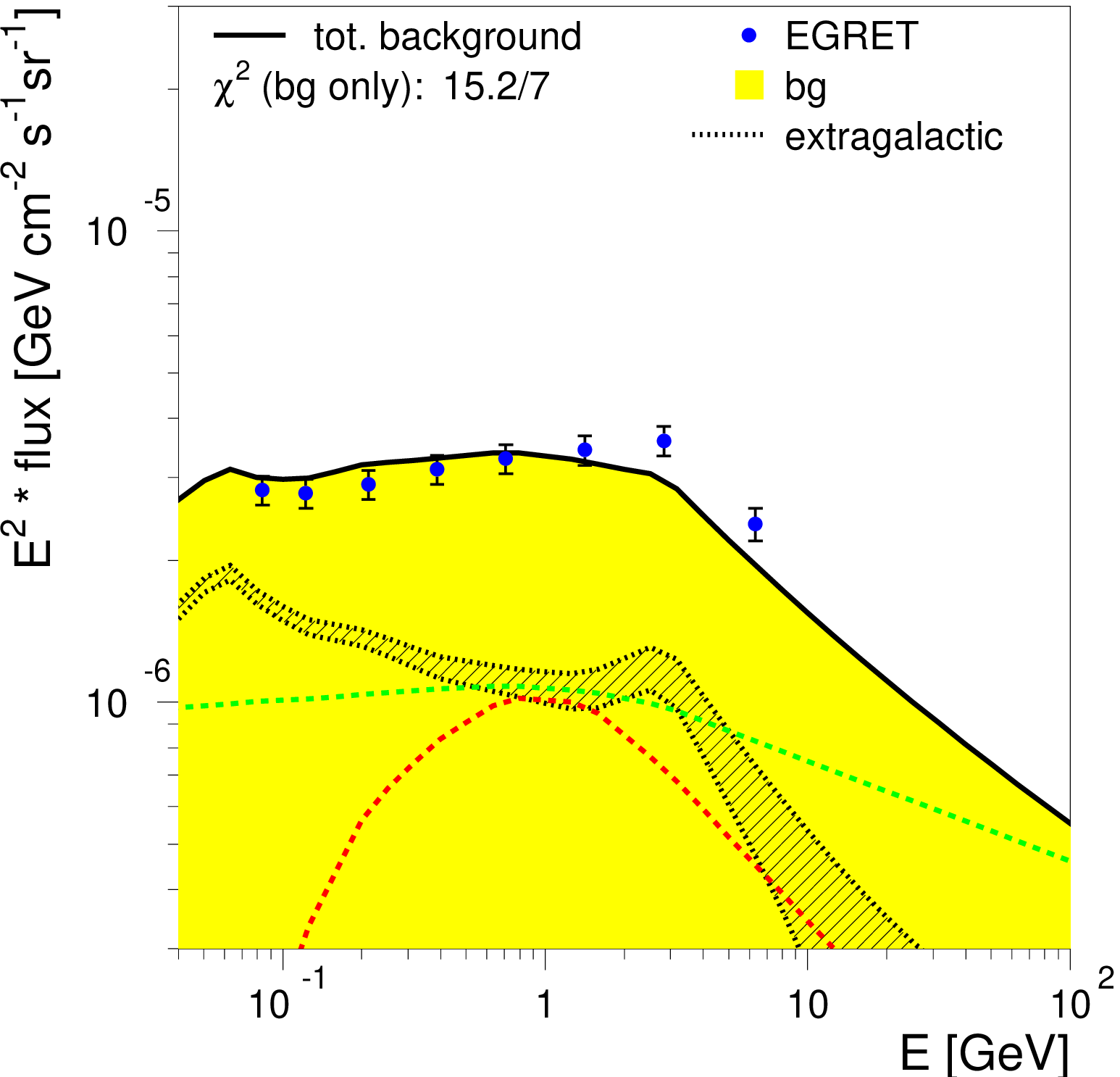}
    \includegraphics[width=0.32\textwidth]{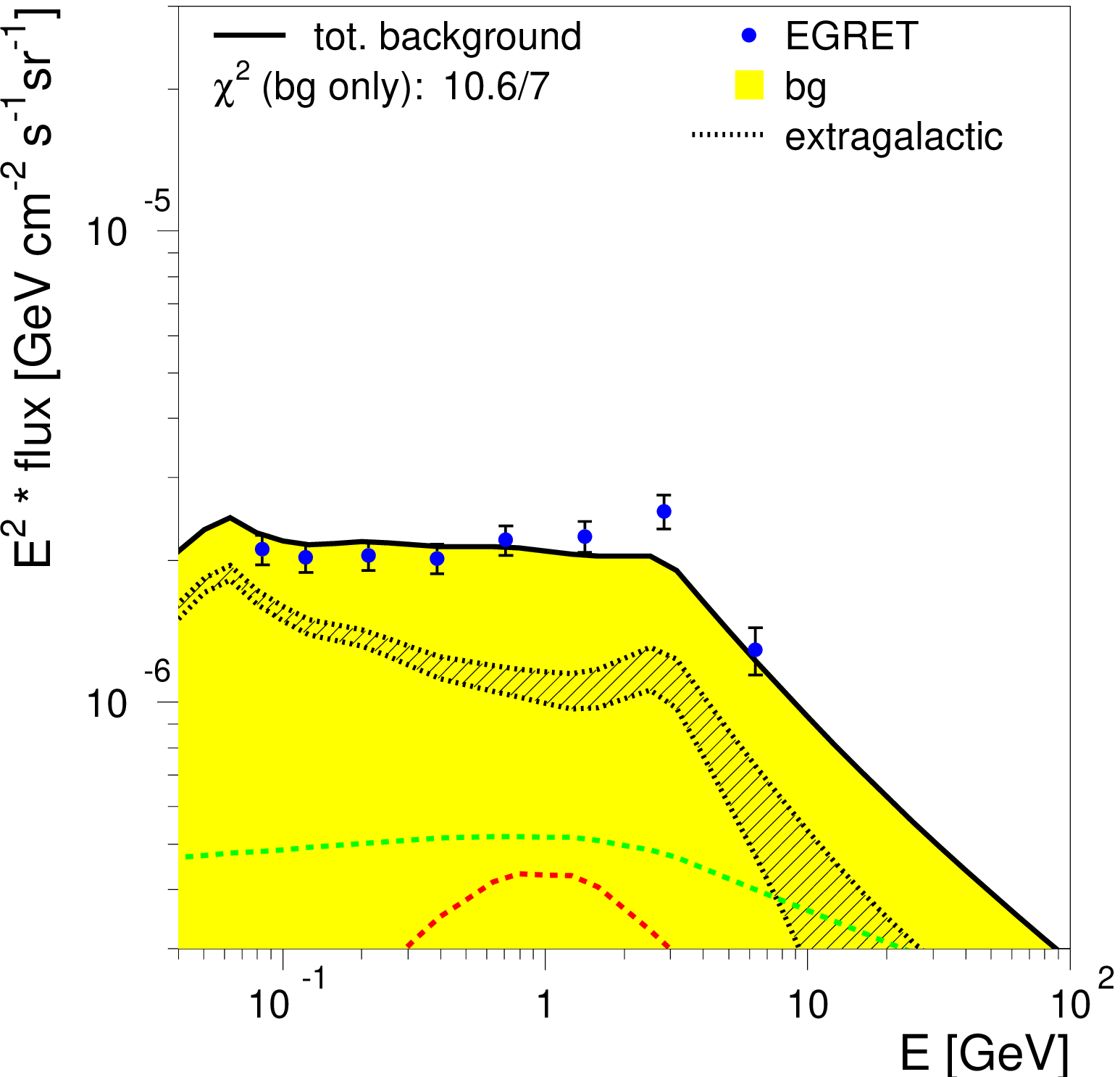}
    \caption[Spectrum of Optimized Model]{Fit results of the
    shape of the optimized model to the EGRET data; regions and
    coding as for Fig. \ref{excess}, but without DMA contribution.
    The excess  above 1 GeV
    can be improved by adding a DM contribution, in which
    case the normalization of the background  (solid line)
    will become lower again and fit the low energy data points.}
    \label{excess1}
  \end{center}
\end{figure}

\begin{figure}
  \begin{center}
    \includegraphics[width=0.32\textwidth]{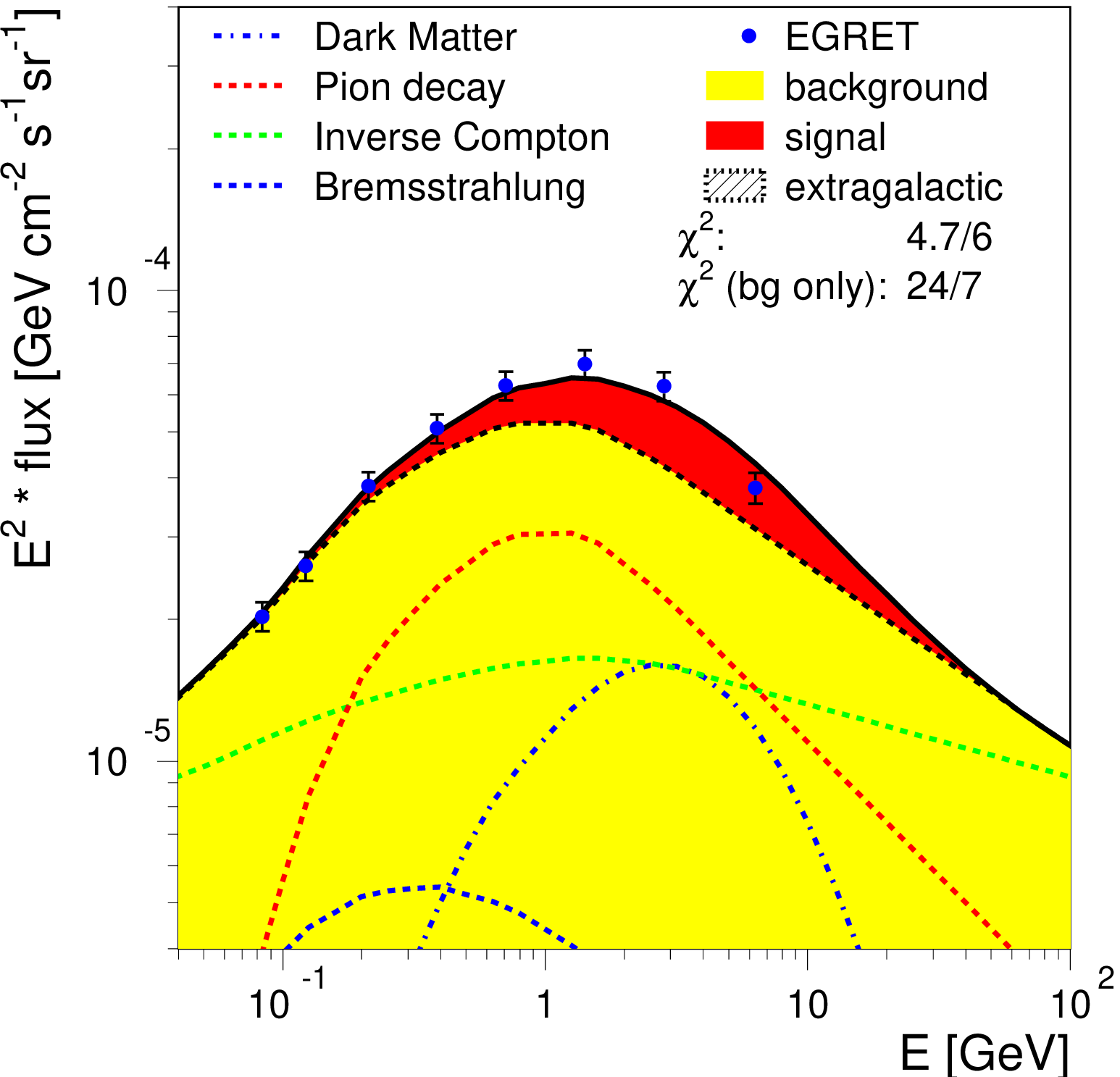}
    \includegraphics[width=0.32\textwidth]{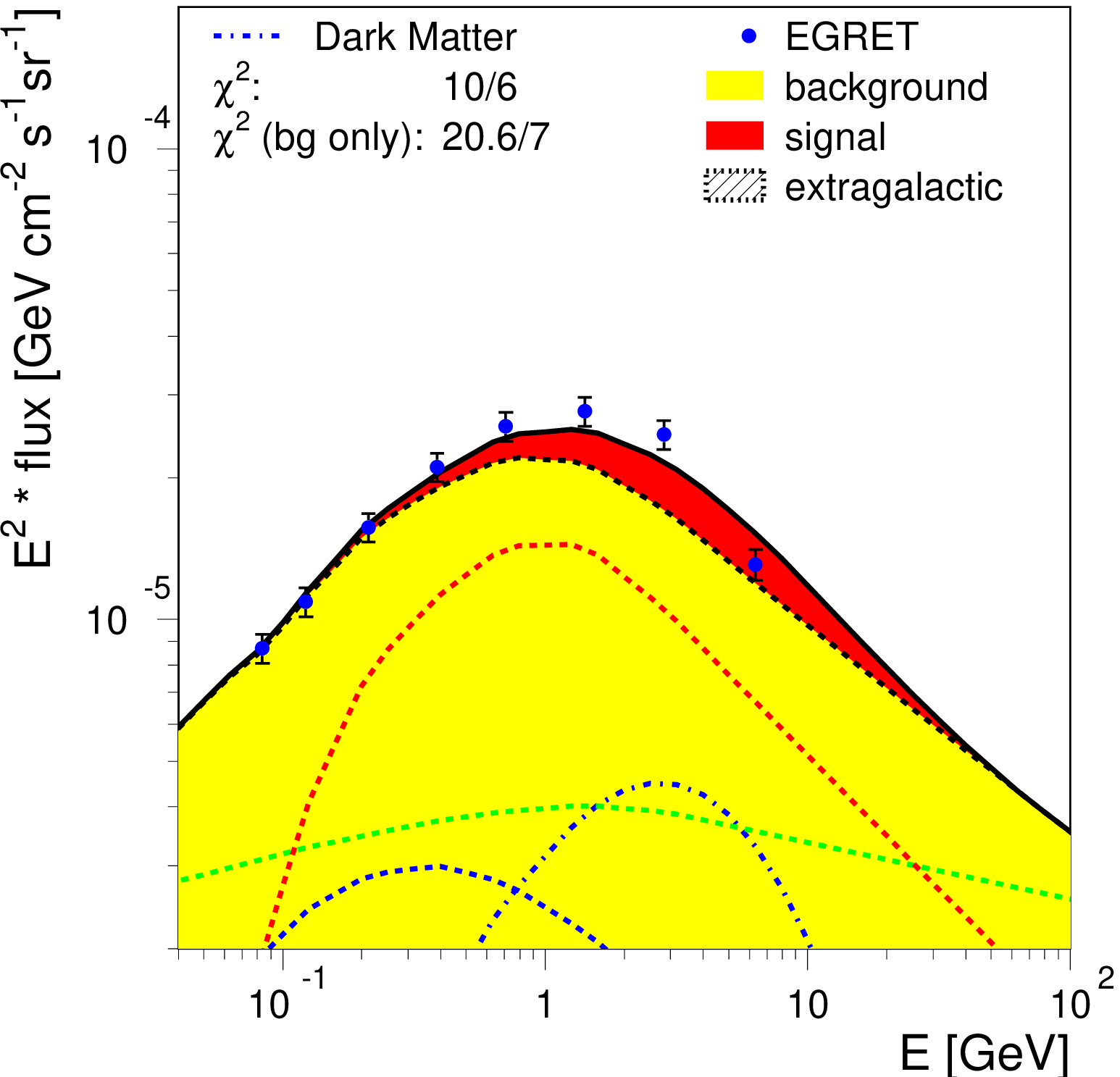}
    \includegraphics[width=0.32\textwidth]{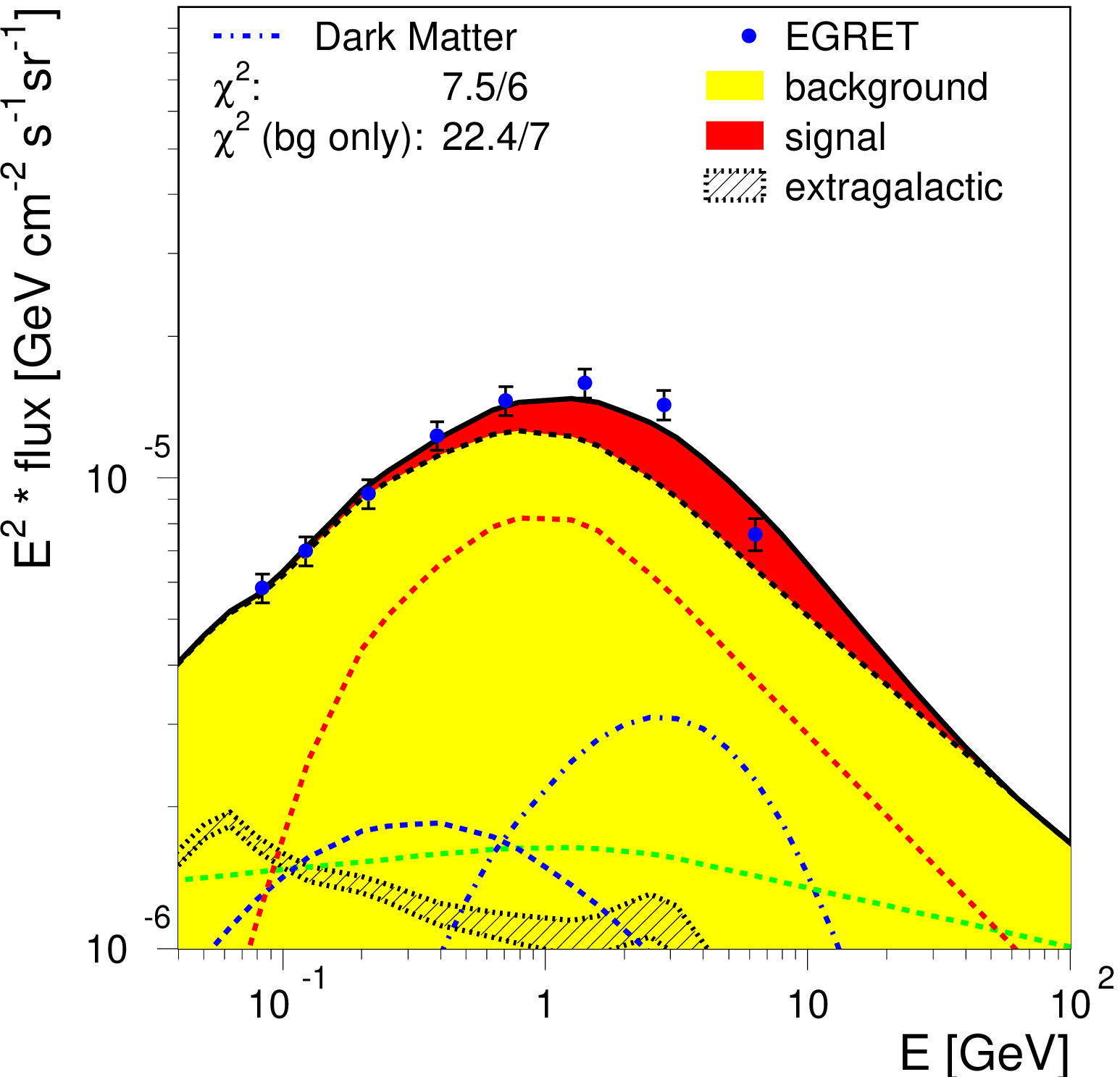}
    \includegraphics[width=0.32\textwidth]{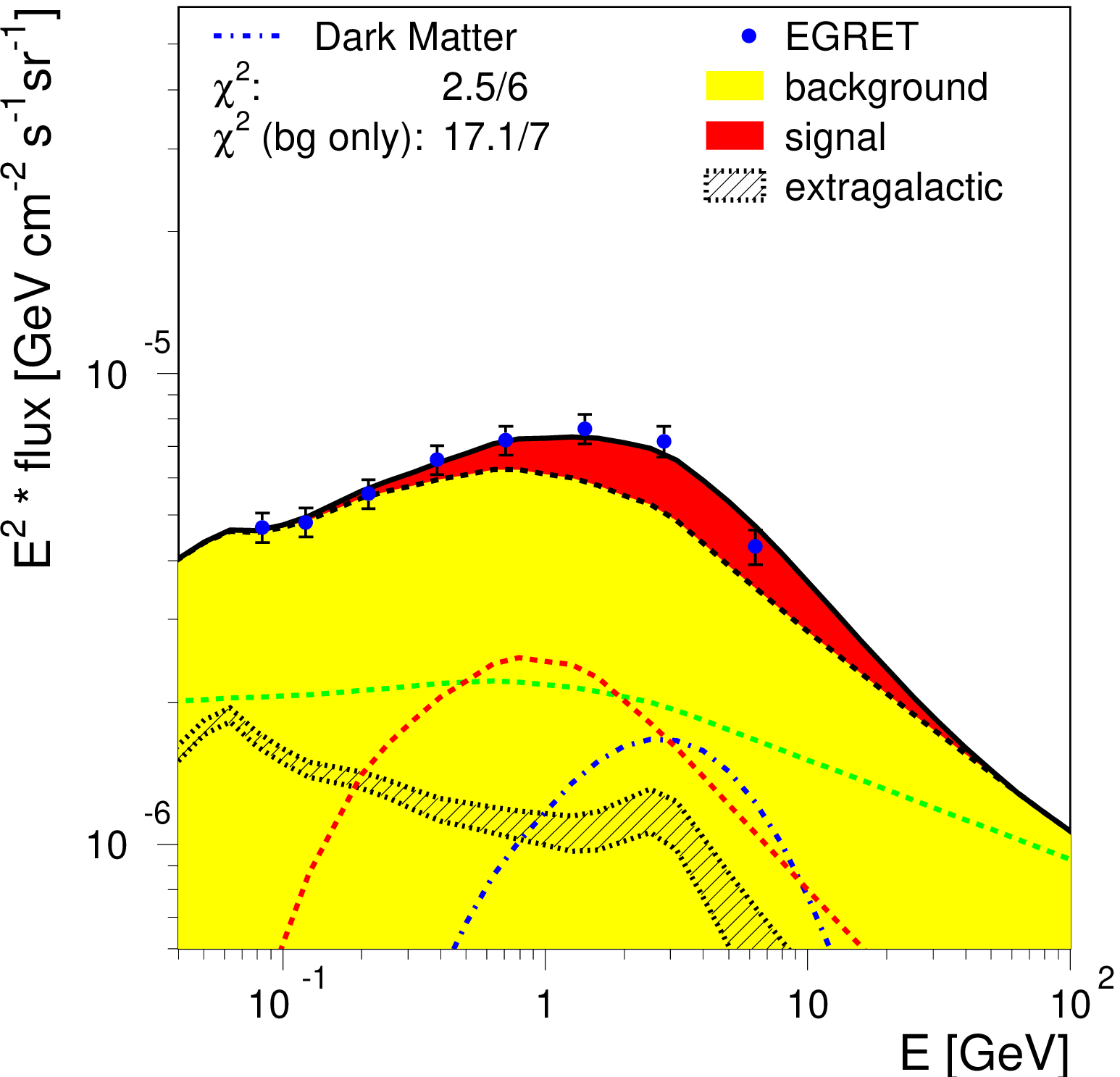}
    \includegraphics[width=0.32\textwidth]{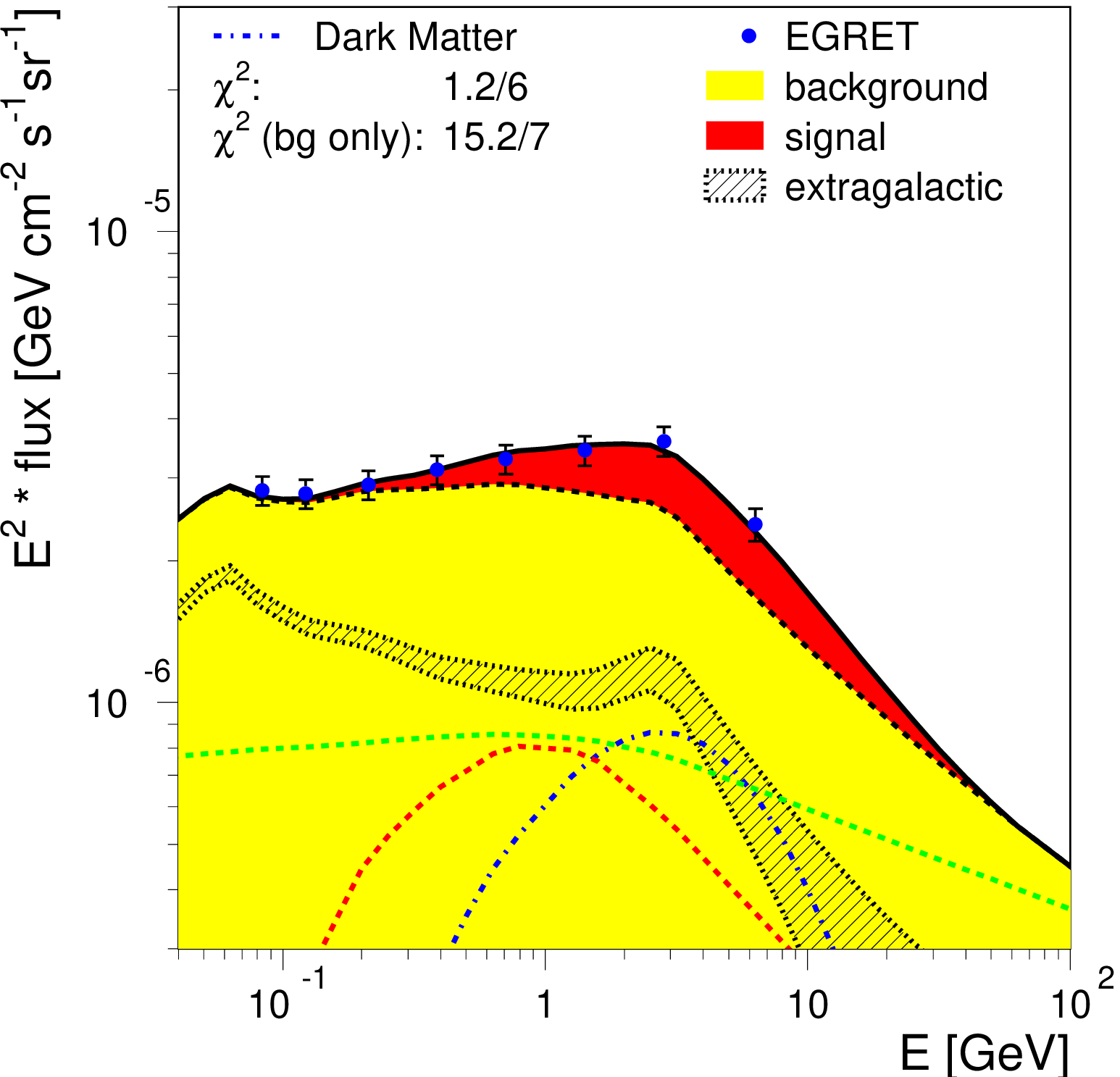}
    \includegraphics[width=0.32\textwidth]{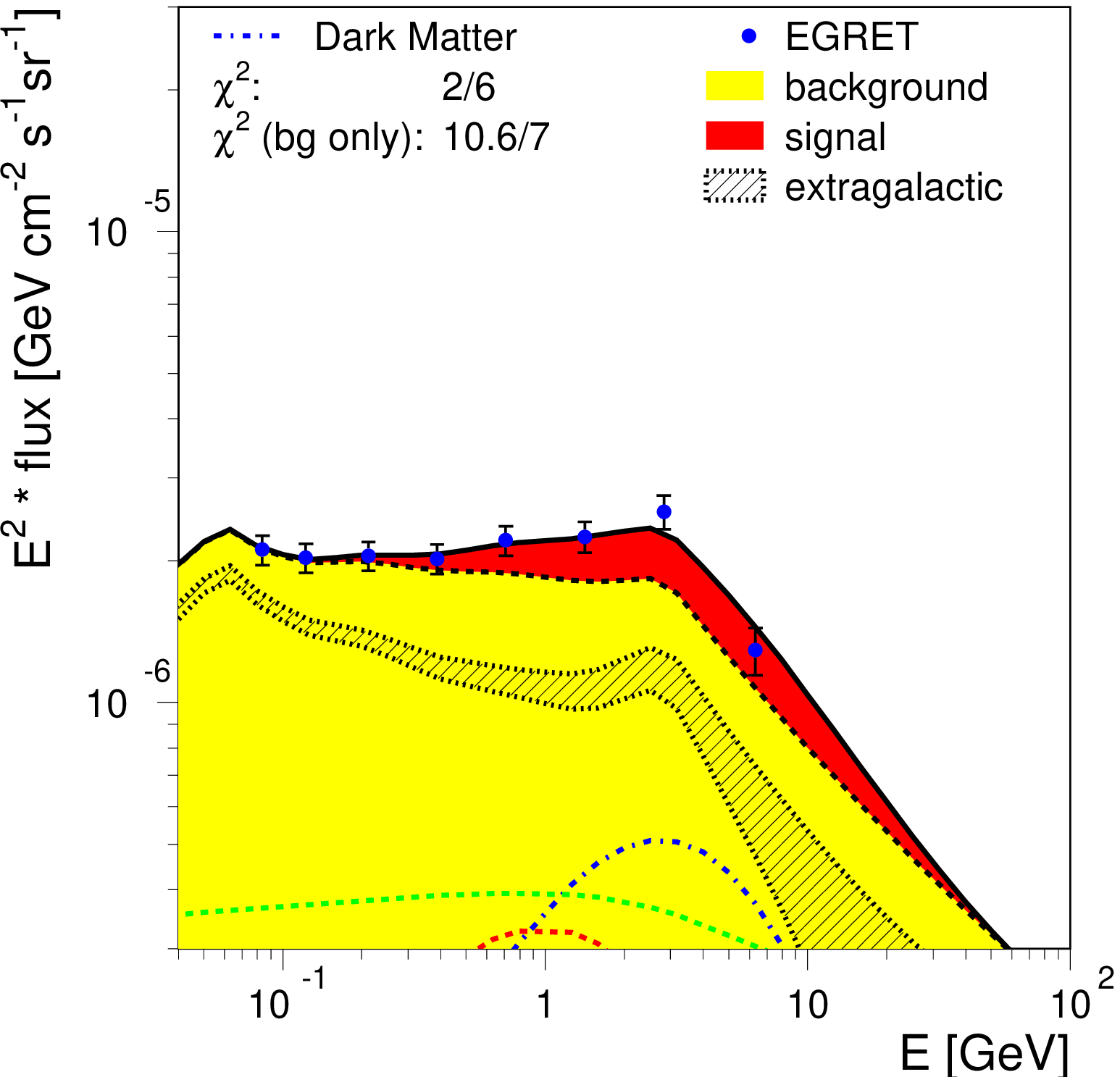}
    \caption[Spectrum of Optimized Model]{Fit results of the
    shape of the optimized model plus DMA to the EGRET data;
    region and coding as for Fig. \ref{excess}.
    The fit is now equally good
    as with the conventional background shape in Fig. \ref{excess},
     but the boost factor is roughly a factor three lower.}
    \label{excess2}
  \end{center}
\end{figure}

The problem with this "optimized solution" is however that the shape
of the gamma spectra is improved but still not   reproduced well, as
shown in Fig. \ref{excess1} for the regions of Table \ref{t1}. But
it is exactly the shape, which was well measured by EGRET, because
the relative errors between neighbouring energies are roughly half
of the normalization error of 15\%.  The probability, as calculated
from the total $\chi^2/d.o.f=110/42$, as indicated for each region
in Fig. \ref{excess1}, is below 10$^{-7}$. The fact that the shape
is not well fitted in the optimized model can also be seen from Fig.
9 in the original publication \citep{optimized}, which shows the
longitudinal profile for various energy bins: above 1 GeV the
prediction of the model is clearly too low, especially if one takes
into account that the statistical errors in the Galactic plane are
negligible, so the plotted errors of 15\% are correlated. And this
discrepancy above 1 GeV is observed in all directions, but this
energy range is exactly where DMA contributes. Adding DM to the
optimized model improves the fit probability from below $10^{-7}$ to
0.8, as shown in Fig. \ref{excess2}. Of course, the DM contribution
is smaller than in case of the conventional background in Fig.
\ref{excess}, which results in a reduction of the boost factor by
roughly a factor three. Similar results are obtained for the shape
proposed by \citet{kamae}. As with the optimized model, the absolute
prediction of \citet{kamae} overshoots the low energy data and
undershoots the high energy data, so if only the shape is fitted
with a free normalization factor, the excess is clearly present.
  Their proposed contribution
of diffractive pp-scattering only reduces the excess by 10-20\%
and if in addition the proposed harder
proton spectrum is used (spectral index -2.5 instead of -2.7 measured locally),
 the excess can be reduced by 30\%
 In all cases  the proposed background shape from \citet{kamae}  fits the
data in the different regions considerably worse than the optimized
model from \citet{optimized}, mainly because \citet{kamae}
 try to improve the fit by changing the proton spectrum only,
while in the optimized model both the electron spectrum {\it and} proton
spectrum are modified.

\begin{figure}
  \begin{center}
    \includegraphics[width=0.4\textwidth]{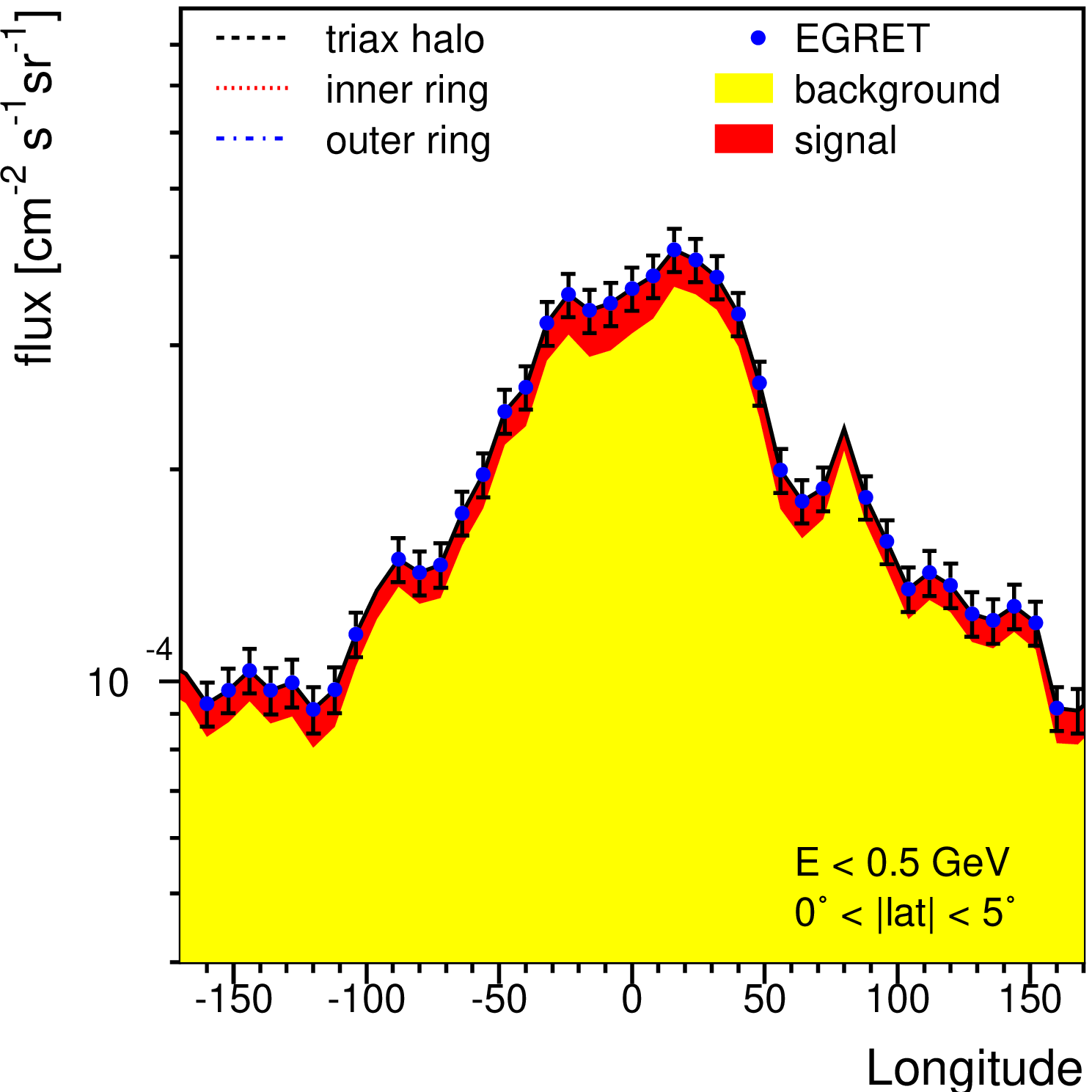}
    \includegraphics[width=0.4\textwidth]{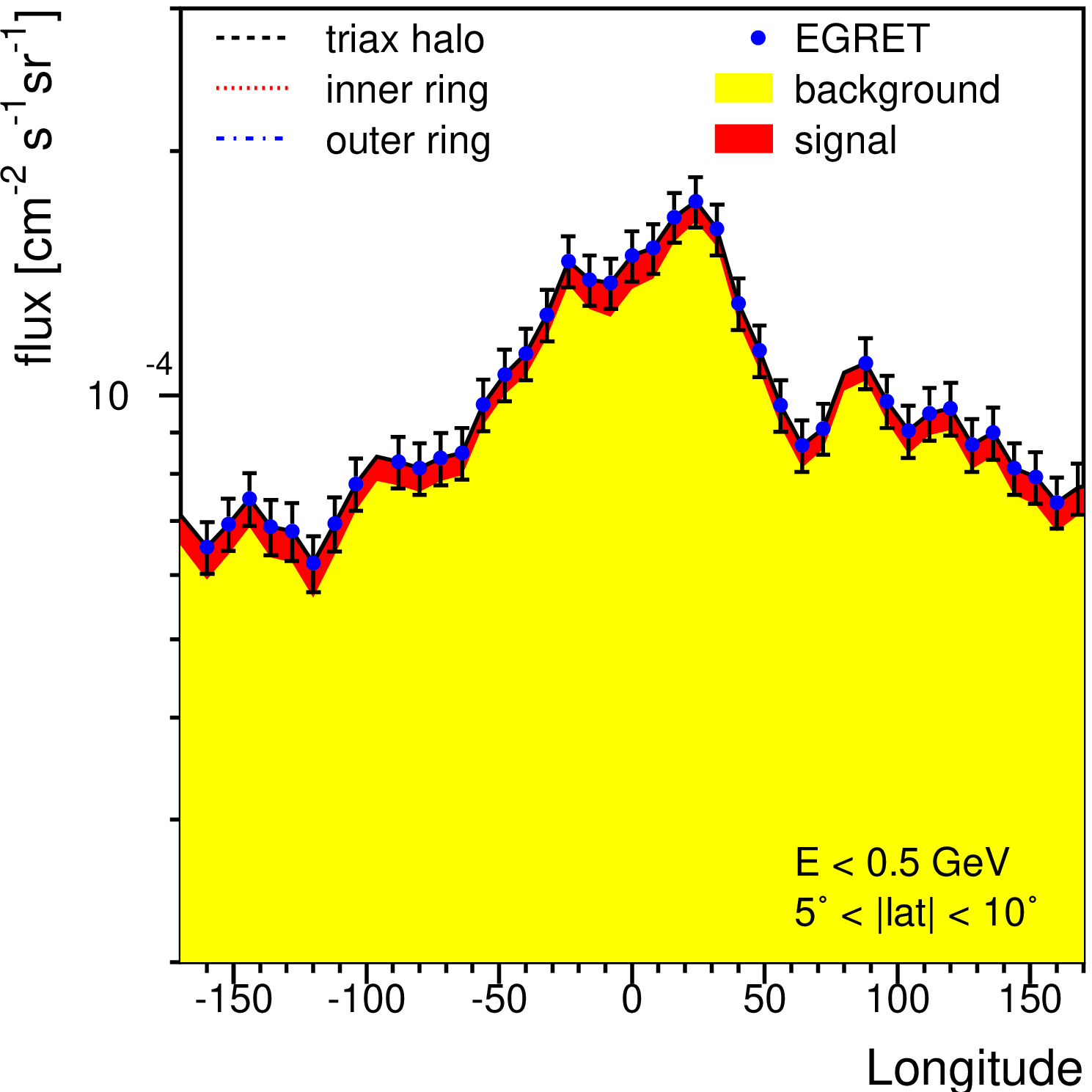}
    \includegraphics[width=0.4\textwidth]{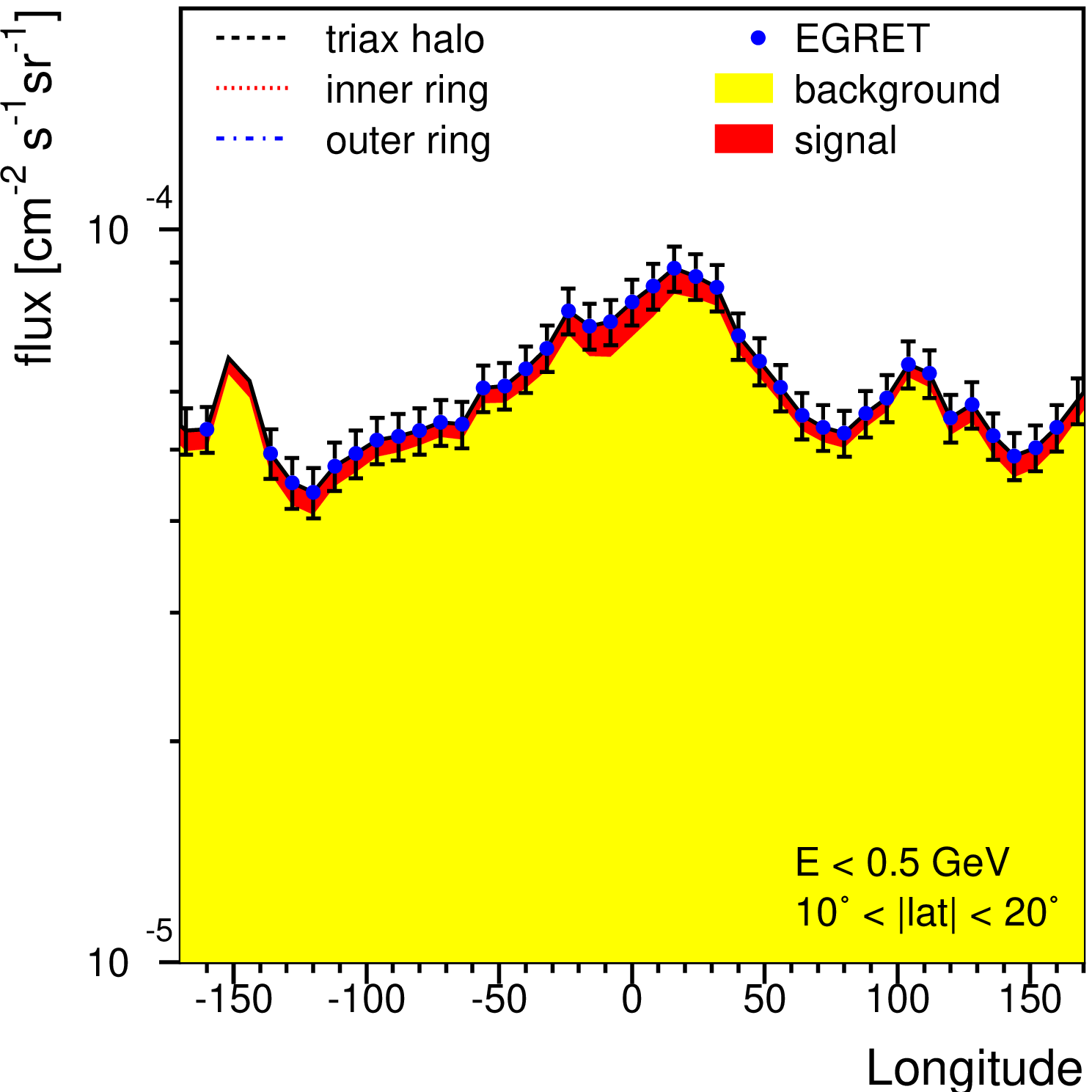}
    \includegraphics[width=0.4\textwidth]{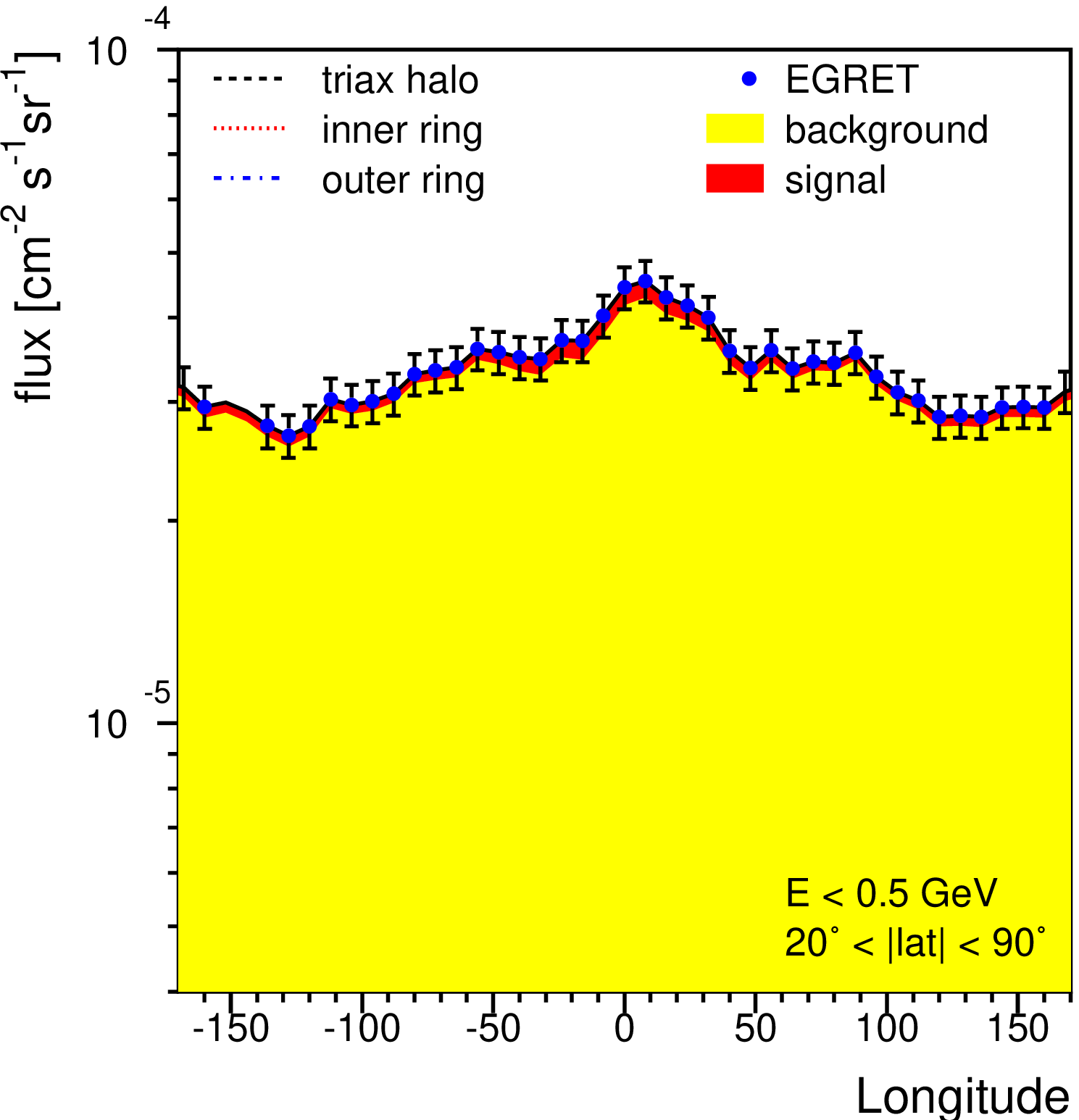}
    \caption[Fits for Pseudo-Isothermal Profile]{
The longitude distribution of diffuse Galactic gamma rays with energies
below 0.5 GeV for different latitudes
 The points represent the EGRET data.
 The contributions from the background and the almost negligible DMA for  energies
 below 0.5 GeV have been indicated by the light (yellow) and dark (red) shaded
 areas, respectively. The free normalization of the background for each bin
 provides a perfect description of the low energy data.} \label{long_low}
  \end{center}
\end{figure}

\begin{figure}
  \begin{center}
    \includegraphics[width=0.4\textwidth]{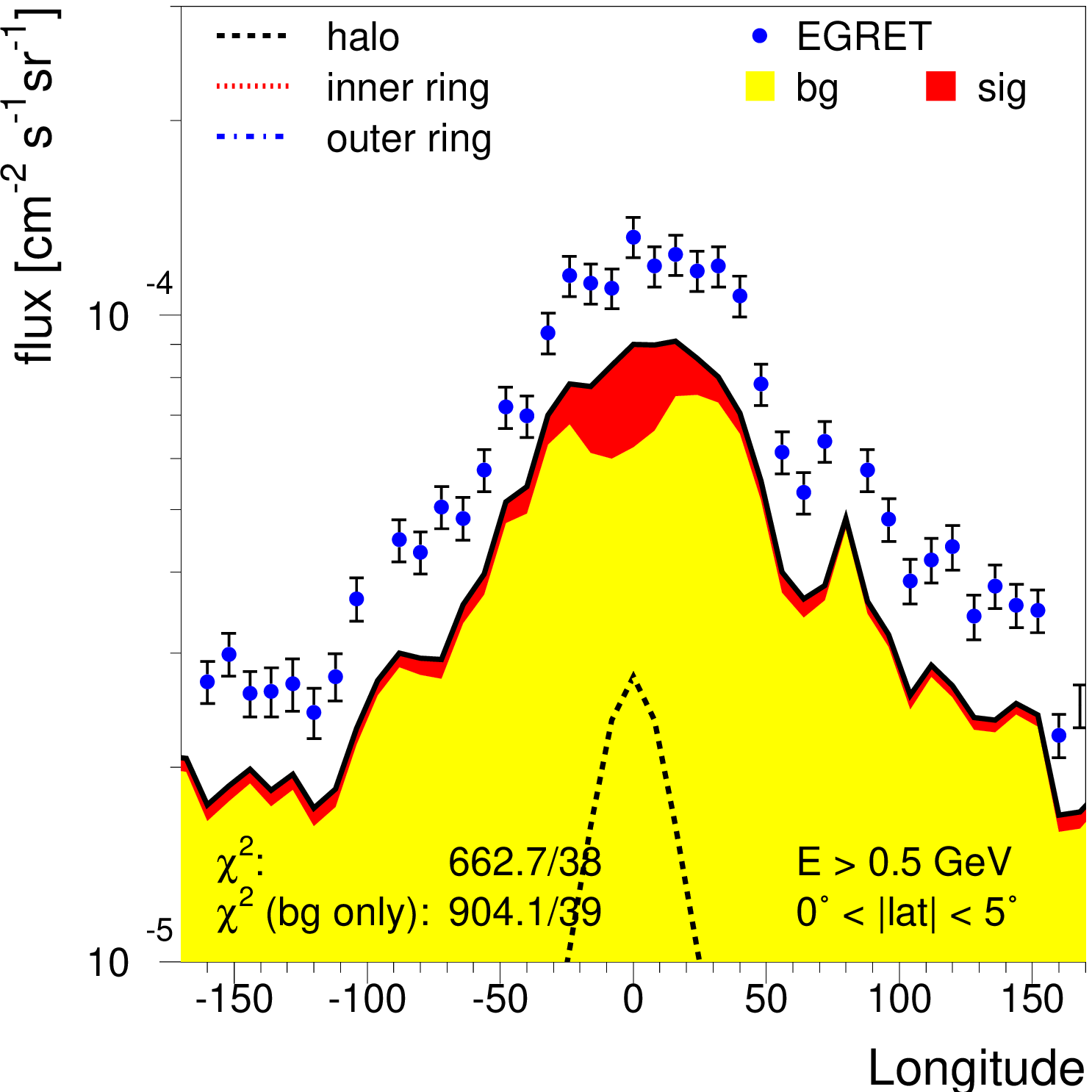}
    \includegraphics[width=0.4\textwidth]{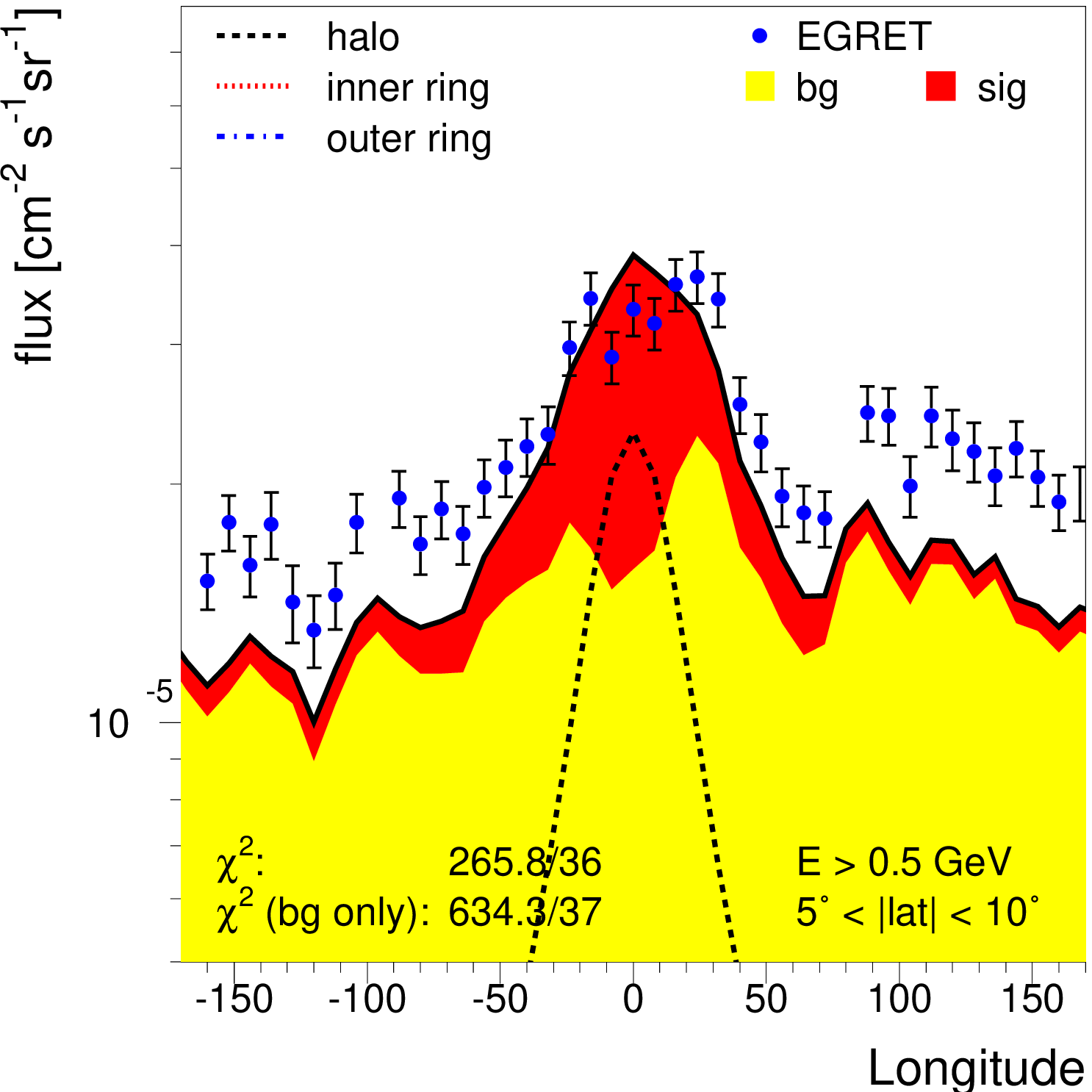}
    \includegraphics[width=0.4\textwidth]{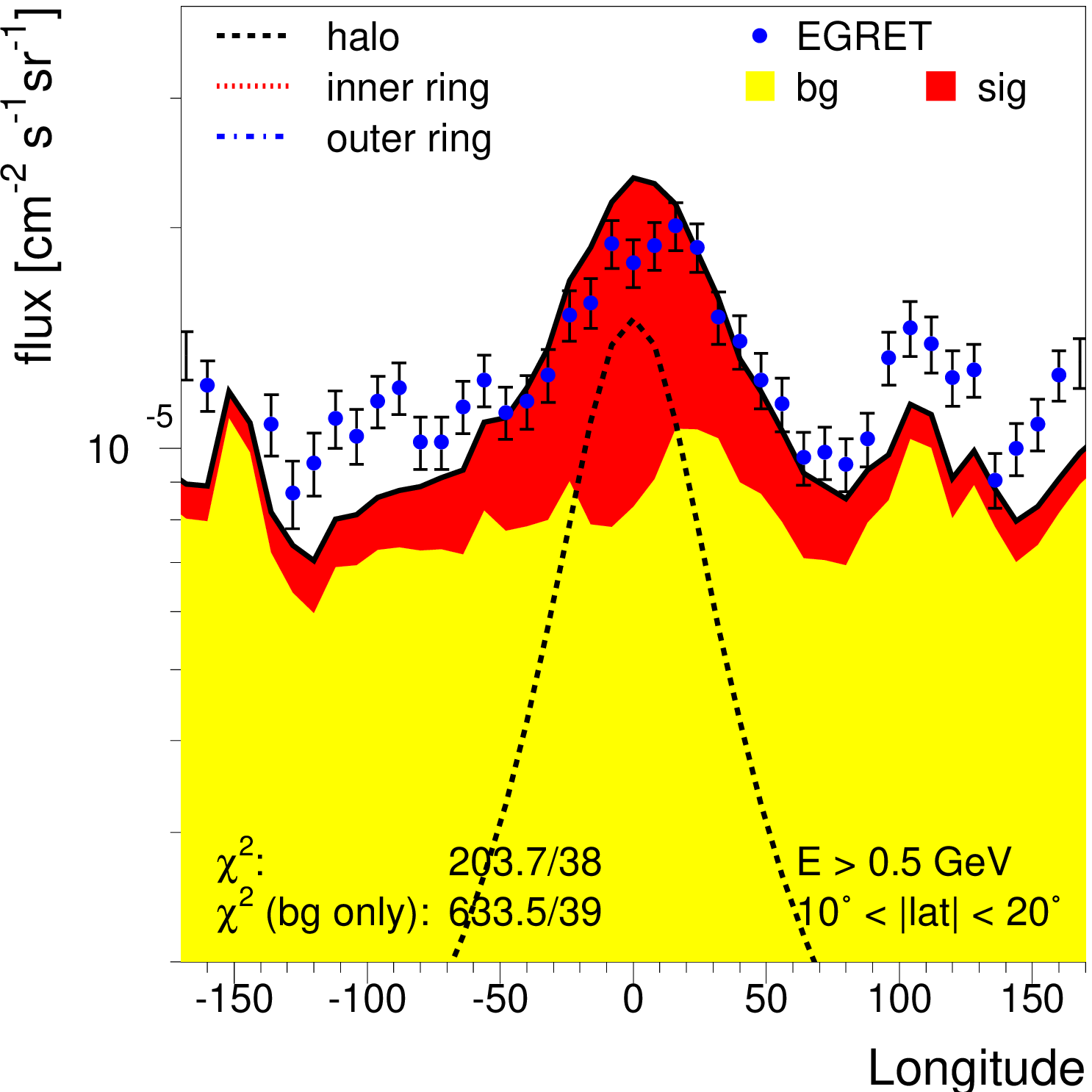}
    \includegraphics[width=0.4\textwidth]{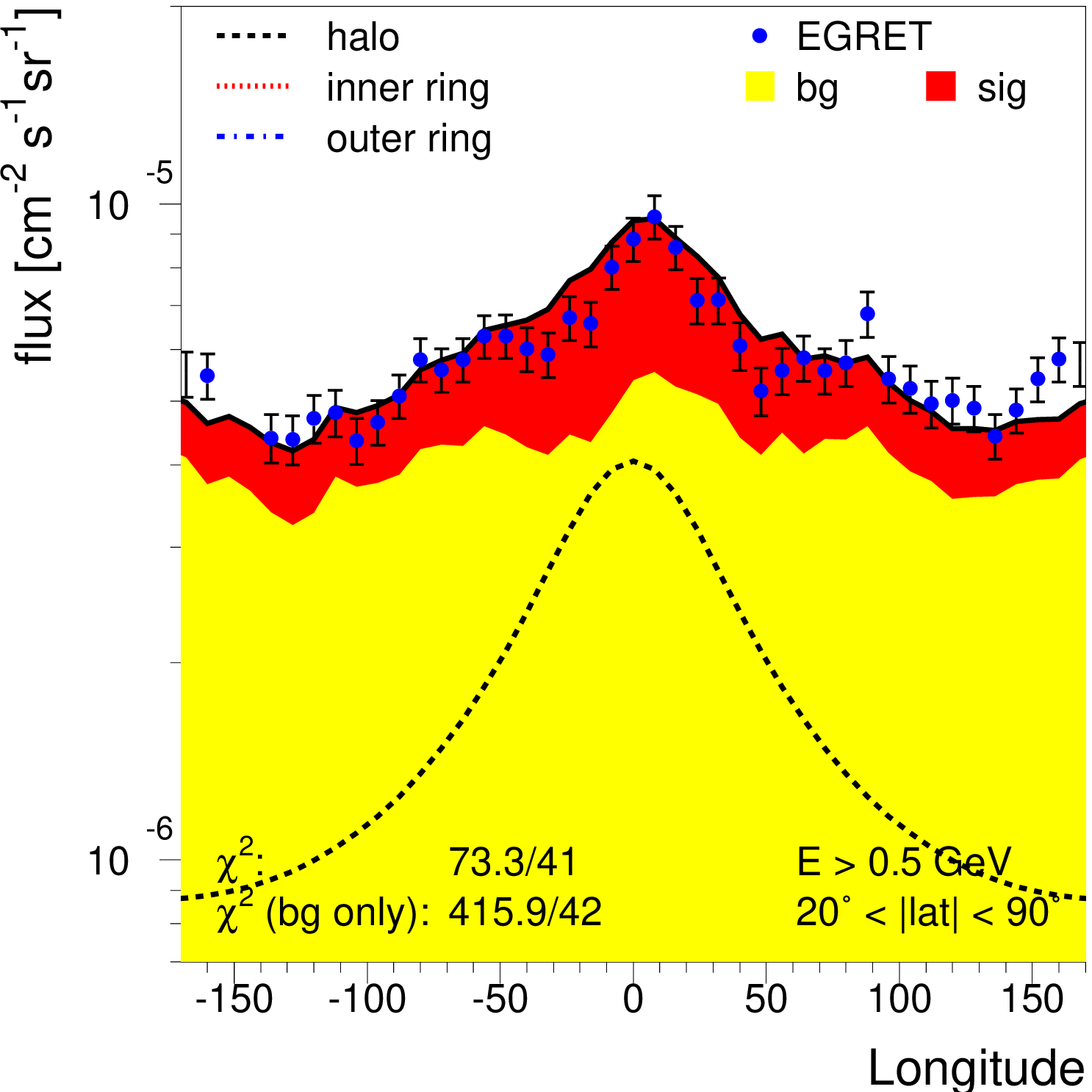}
    \caption[Fits for Pseudo-Isothermal Profile]{As in Fig. \ref{long_low}, but for
     for EGRET data above 0.5 GeV and DMA for an isothermal DM profile
     without ringlike substructures.} \label{long_high}
  \end{center}
\end{figure}

\begin{figure}
  \begin{center}
    \includegraphics[width=0.4\textwidth]{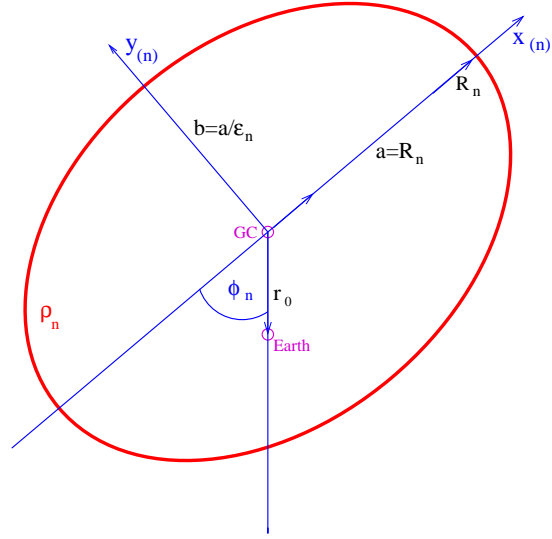}
    \caption[Schematic Picture of a DM Ring]{Schematic picture of
    a DM ring with  an elliptical shape and constant  DM density $\rho_n$ along the ring.
     The definitions are also valid
    for the triaxial halo component, if the coordinate system is rotated with
    the appropriate angle $\phi_{n}$ towards the Galactic center and the short axis $c$
     is modified by the ellipticity in the z-direction: $c=a/\epsilon_z$.
    } \label{haloschema}
  \end{center}
\end{figure}

\begin{figure}
  \begin{center}
    \includegraphics[width=0.4\textwidth]{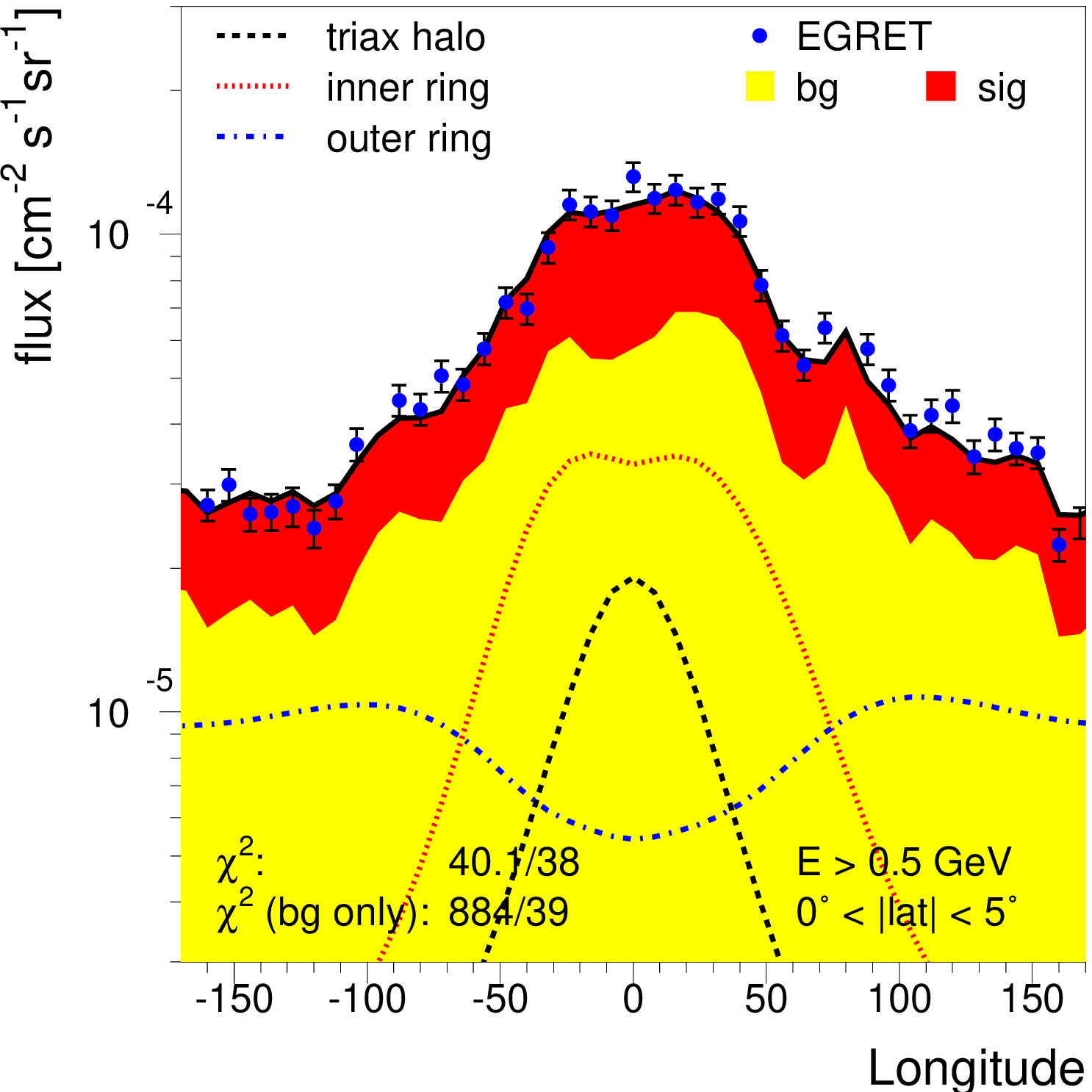}
    \includegraphics[width=0.4\textwidth]{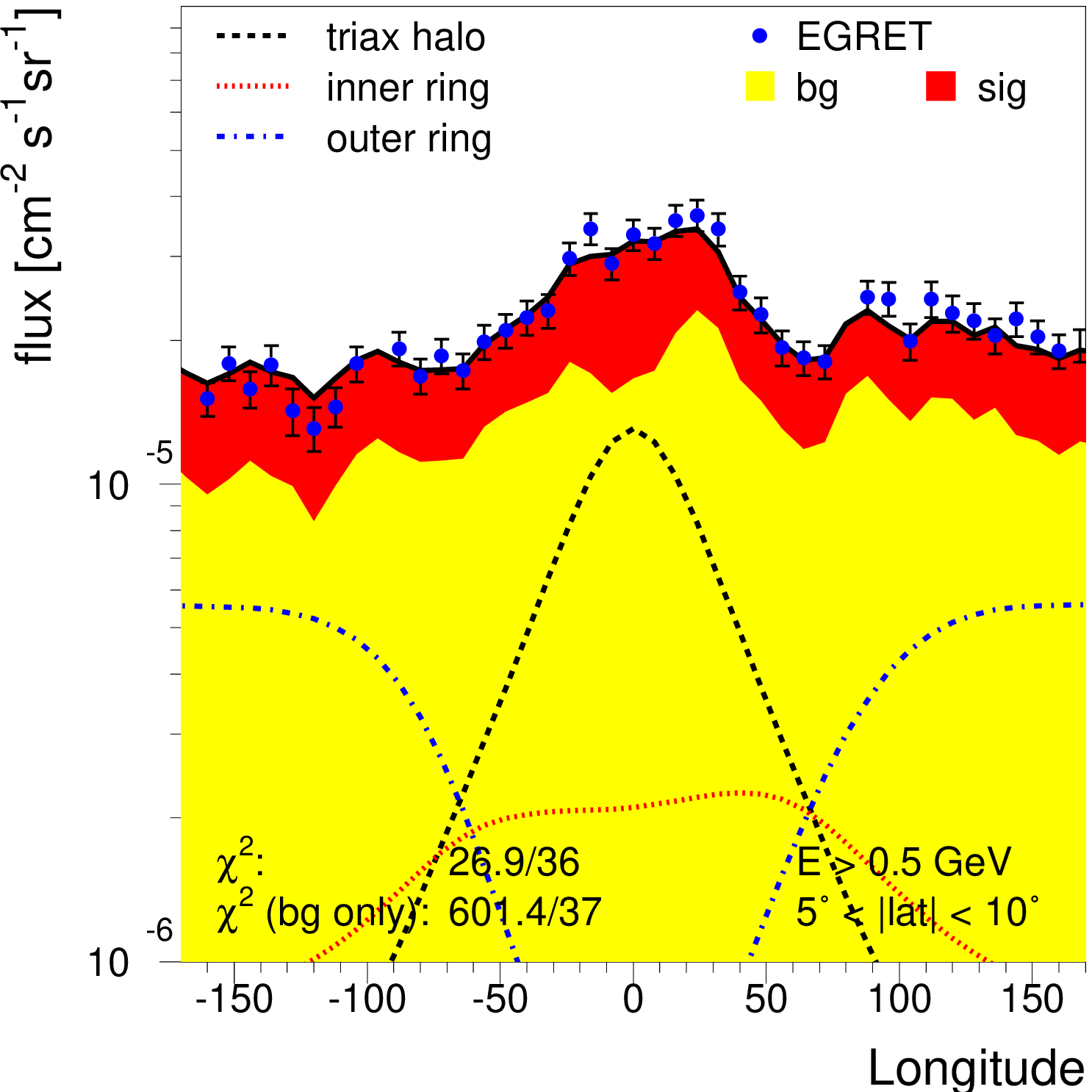}
    \includegraphics[width=0.4\textwidth]{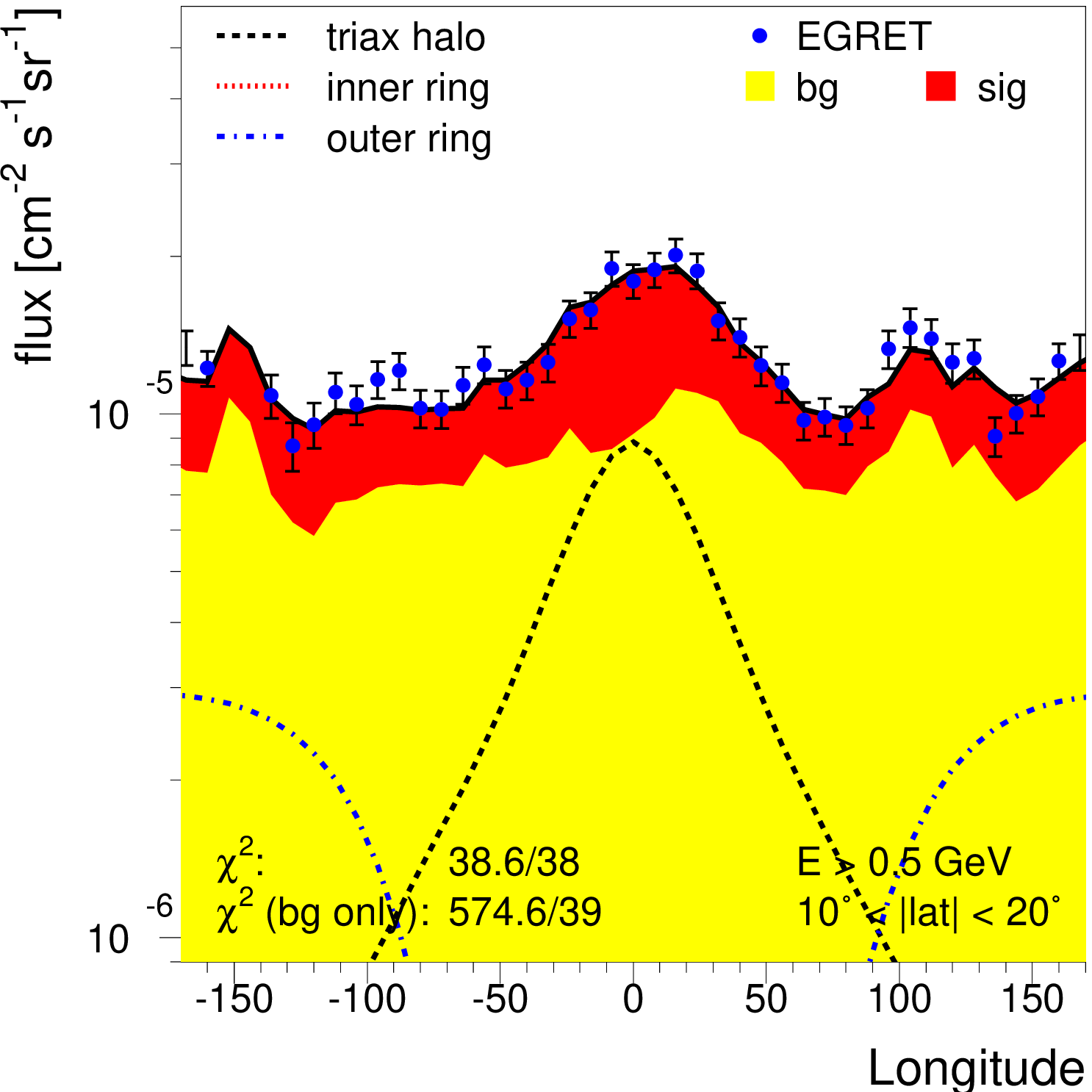}
    \includegraphics[width=0.4\textwidth]{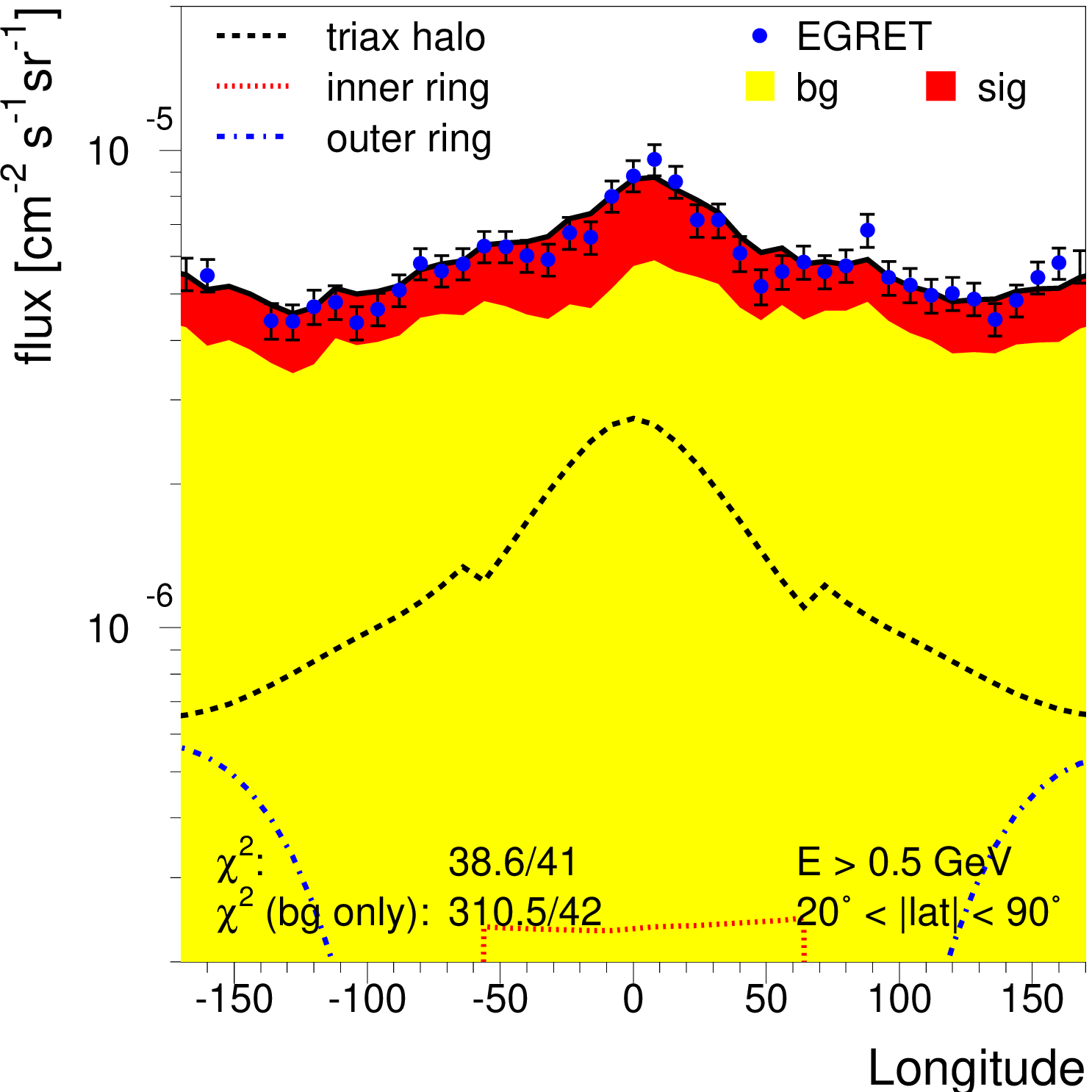}
    \caption[Fits for Pseudo-Isothermal Profile]{As in Fig. \ref{long_high}, but
     including DMA  for an isothermal
    DM profile with ringlike substructures at 4 and 14 kpc.} \label{long_rings}
  \end{center}
\end{figure}

\begin{figure}
\begin{center}
 \includegraphics [width=0.4\textwidth,clip]{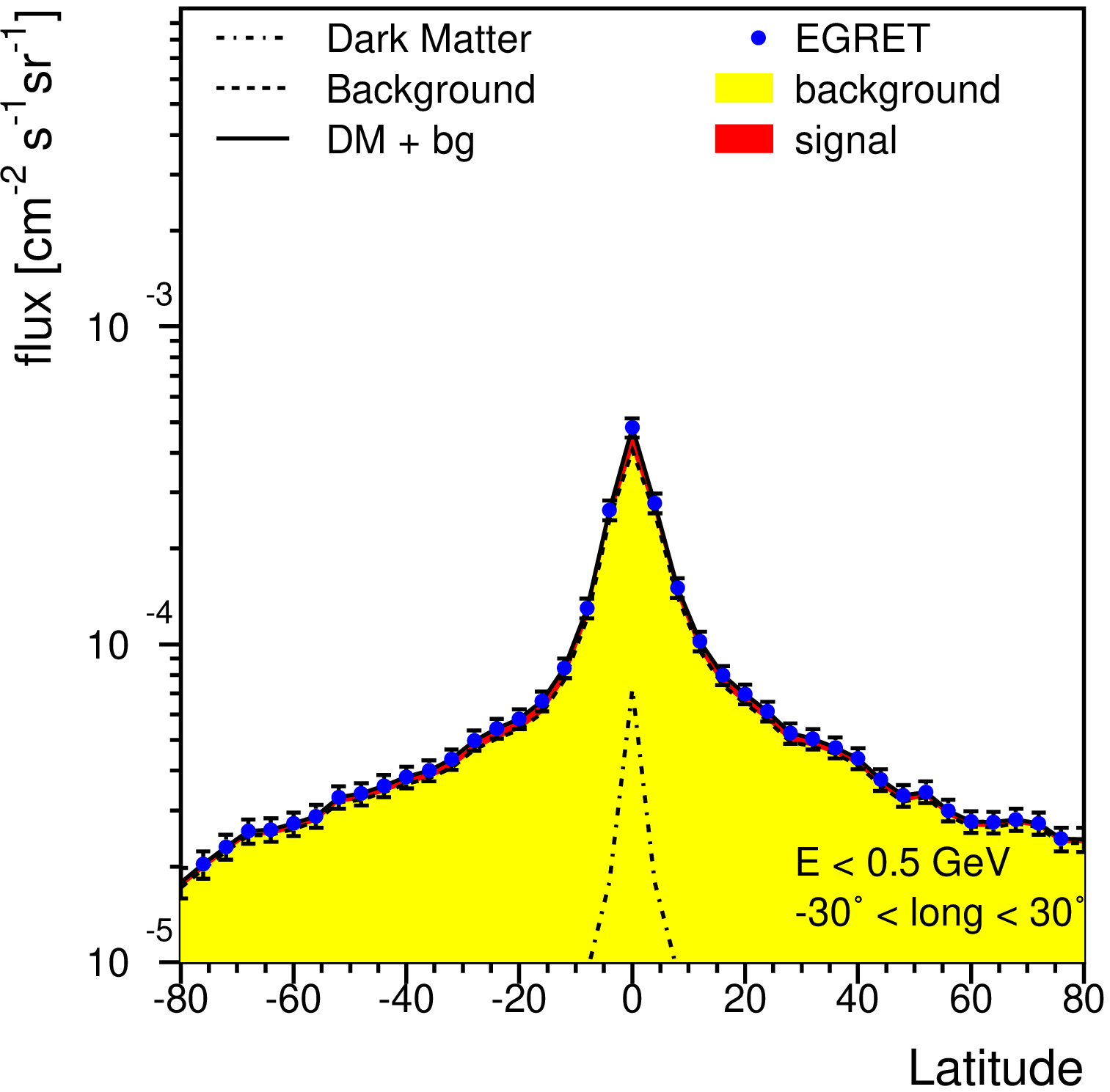}
 \includegraphics [width=0.4\textwidth,clip]{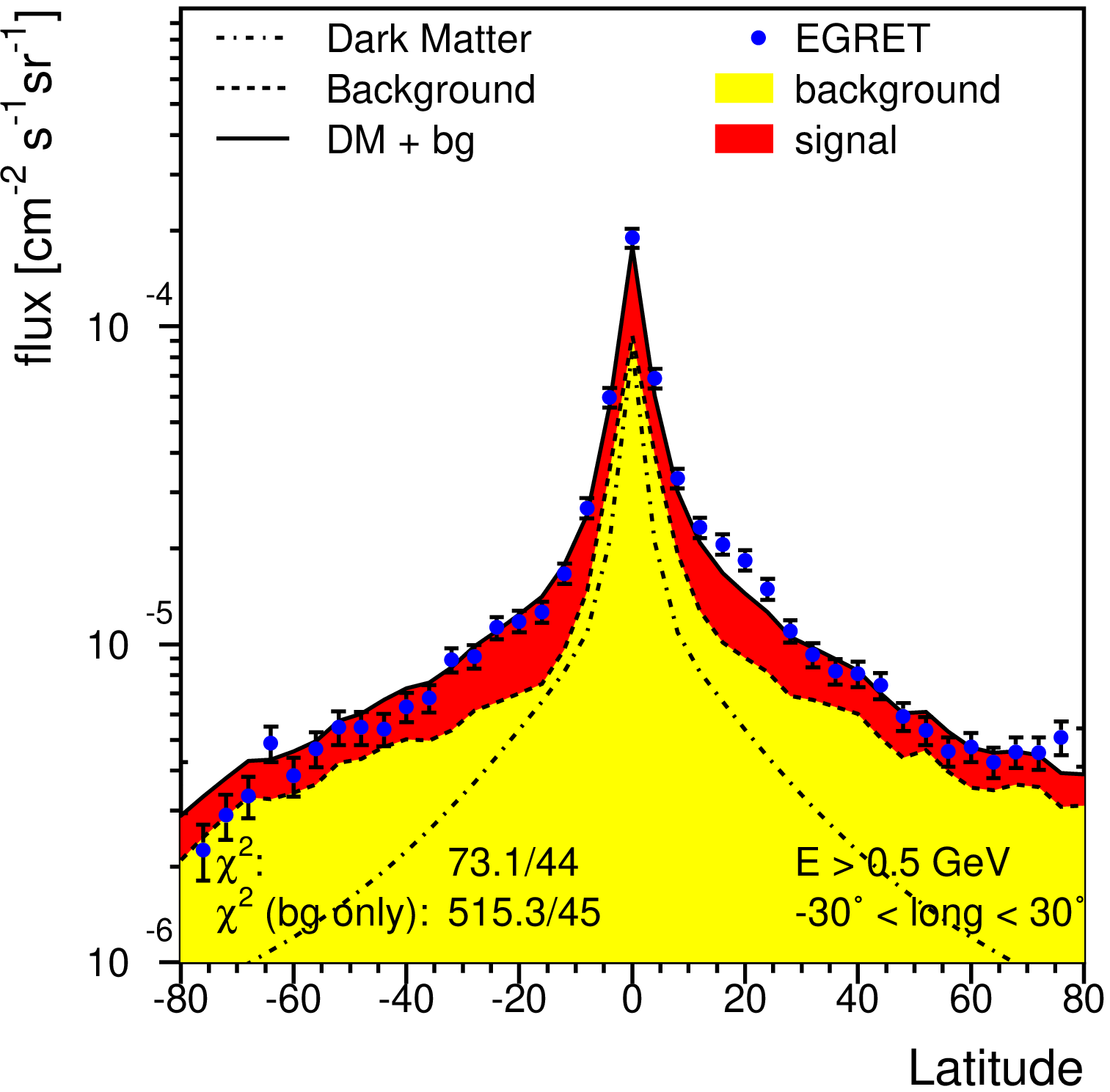}
 \caption[]{
 The latitude distribution of diffuse gamma-rays for longitudes
 $-30^\circ<l<30^\circ$ and two energy bins: $E_\gamma<0.5$ GeV (left)
 and $E_\gamma>0.5$ GeV (right). The points represent the EGRET data.
 The contributions from the background and the neutralino annihilation
 signal have been indicated by the light (yellow) and dark (red) shaded
 area, respectively. The fitted normalization factor of the background
 comes out in reasonable agreement with the GALPROP prediction, while
 the fitted normalization of the DM contribution corresponds to a WIMP mass
 around 60 GeV with a boost
 factor of about 100.}
 \label{lat}
\end{center}
\end{figure}

\section{Determination of the halo profile}\label{halo}

From the excess in the various sky directions one can obtain the
halo profile. However, this requires a finer sampling of the various
sky directions  than discussed above. Therefore  the fits to the 6
regions  were repeated for the following regions: the longitude
distributions are split into bins of 8$^\circ$ for four different
latitude ranges (absolute values of latitude: 0-5$^\circ$,
5-10$^\circ$, 10-20$^\circ$, 20-90$^\circ$), so one has 4x45=180
independent sky regions. The fit result for each of these regions has been
displayed in the Appendix. Also the fluxes from the point sources
are displayed, which have been subtracted from the data. In most
regions the point sources have a negligible contribution  except for
some disk  regions with strong pulsars (CRAB at $l\approx
-175^\circ$, GEMINGA at $l\approx -165^\circ$, VELA at $l\approx
-98^\circ$ and CYGNUS at $l\sim +80^\circ$). These regions have been
left out of the fit for the halo parameters.

 The differential gamma flux in a direction forming an angle $\psi$ with
the direction of the Galactic center is given by:
\begin{equation}
  \phi_\chi(E,\psi)=\frac{\langle \sigma v\rangle}{4\pi} \sum_f \frac{dN_f}{dE} b_f
  \int_{line~ of~ sight}B_l \frac{1}{2}\frac{\langle \rho_\chi\rangle^2}{M_\chi^2} dl_\psi
\label{gammafluxcont}
\end{equation}
where $b_f$ is the branching ratio into the tree-level annihilation
final state, while $dN_f/d E$ is the differential photon yield for
the final state $f$. The spectrum $dN_f/d E$ is similar for all hadronic
final states, as discussed before in section \ref{fit}. The cross
section $\langle\sigma v\rangle$ is taken from Eq. \ref{wmap} with
the WMAP value for $\Omega h^2$ \citep{wmap}.  So one observes from
this equation, that for a given excess and a known WIMP mass (around
60 GeV from the spectrum of the excess as discussed above), the only
unknown quantity in the direction $\psi$ is the  square of the
averaged WIMP mass density  $\rho_\chi$ multiplied by the boost
factor, which represents the enhancement of the annihilation
rate by the  clustering of DM. 
The hierarchical formation of galaxies from small clumps leads to a
spectrum of clump masses starting may be as low as $10^{-6}$ solar masses
\citep{dokuchaev,moore}. 
The DMA is  higher in the clumps than outside, since the local density
is there  higher and DMA is proportional to the local density
squared, i.e. $<\rho_{DM}^2>$. This can be considerably larger than
$<\rho_{DM}>^2$ and the ratio is the enhancement or boost factor, i.e.
B=$<\rho_{DM}^2>$/$<\rho_{DM}>^2$.
The rotation curve is only sensitive to  the total mass, i.e. $<\rho_{DM}>$.
The boost factor can be obtained from the fitted normalization and
can be large, from a few tens to a few thousands, depending on the
unknown details of the DM clustering.
In general, the boost factor towards the Galactic center may be
lower than in other directions because of the likely tidal
disruption of small DM clusters by the fly-by from stars. In this
case the flux is proportional to $B(r)<\rho(r)^n>$, where $n$ can
vary between 1 and 2 depending on the DM clustering: $n=2$ if no
clustering and $n=1$ if DMA is predominantly in non-overlapping
clusters. Consequently one has many alternatives to fit, which are
outside the scope of the present paper. Therefore we concentrate on
a boost factor independent of $r$ and $n=2$, which turns out to
yield a good fit.

 If one assumes that the clustering
is similar in all directions, i.e. the same boost factors in all
directions, the DM density profile  $\rho_\chi (r)$ can be determined from the
excess in the various directions. The most transparent way to do this is the
following: search for a  functional form of
the profile as function of the distance from the Galactic center and see which form
yields the best fit.
A survey of the optical rotation curves of 400 galaxies
shows that the halo distributions of most of them can be fitted
either with the Navarro-Frenk-White (NFW) or the pseudo-isothermal
profile or both \citep{Jimenez:2002vy}. These halo profiles can be
parametrized as follows:
\begin{equation}
 \rho (r)=\rho_0\cdot (\frac{r}{a})^{-\gamma}\left[1+
 (\frac{r}{a})^\alpha\right] ^{\frac{\gamma-\beta}{\alpha}},
\label{profile0}
 \end{equation}
 where $a$ is a scale radius and the slopes $\alpha$, $\beta$ and
 $\gamma$ can be roughly thought of as the radial dependence at $r\approx a$,
 $r>>a$ and $r<<a$, respectively.
The cuspy NFW profile \citep{nfw} is defined by  $(\alpha, \beta,
\gamma)$ =(1,3,1) for a scale $a=10$ kpc, while the Moore profile
with $\gamma=1.2$ is even more cuspy \citep{moore1}. The isothermal
profile with  $(\alpha, \beta, \gamma)$ =(2,2,0) has no cusp
($\gamma=0$), but a core which is taken to be the size of the inner
Galaxy, i.e. $a=5$ kpc and  $\beta=2$ implies a flat rotation curve.
The EGRET excess towards the Galactic center does not show a cusp, but is perfectly
flat near the center as expected for a cored profile, so
  a cored isothermal profile was fitted to the excess in 180 sky directions.

 The fit results are shown in Fig.
\ref{long_low} for gamma energies below 0.5 GeV and in Fig.
\ref{long_high} for gamma energies above 0.5 GeV as function of
longitude for various latitudes, i.e. one determines the flux
towards the earth by looking around in a full circle either in the
 Galactic plane or at various angles above and below the
disk. Clearly, the data are well described in all directions for
data below 0.5 GeV with hardly any contribution from DMA, while
for data above 0.5 GeV only the data towards the Galactic poles  are
reasonably well described by the background plus DM component (dark
(red) contribution). Towards the Galactic center and Galactic
anticenter the isothermal profile does not provide enough DM, as is
obvious in the upper panels of Fig. \ref{long_high}. There are
several reasons why this might be so. For example the infall of a
satellite galaxy into the gravitational potential of larger galaxy can
lead to toroidal ringlike DM overdensities \citep{hayashi}.
When ringlike structures were added with the radius and
widths of the ring in and out of the plane as free
parameters, the fit quickly converged for only two toroidal ringlike
structures, namely at radii of 4 and 14 kpc. The enhanced gamma
radiation at 14 kpc was already observed in the original
discovery paper of the excess \citep{hunter} and called
``cosmic enhancement factor''.
Note that the radius
of the ring can be determined from the longitude profile in the
plane of the Galaxy, i.e. at low latitudes, because the solar system
is not at the center, so if the density is constant along the ring,
different segments of the ring yield different fluxes which depend
on the radius, orientation and ellipticity of the ring. The extent
of the ring above the plane can be obtained from the longitude
distribution for higher latitudes. It should be noted that the assumption of
a constant density along the ring is not necessarily true. However, the fit is not very sensitive
towards the ring density on the opposite side of the galactic center,
so only the ring segment nearest to the solar system is assumed to have a
constant density; this assumption yields a reasonable fit.
Although the overdensities from the infall of a satellite galaxy do not produce
rings, but at most ringlike segments during each passage near the pericenter, the precession
of the orbit can lead to various segments along the pericenter.
Since the analysis prefers
 a non-zero density over an angular range  close to 360 degrees
  we continue to speak of ``rings'' of DM, although this does not mean at all perfect
   circularity.

With the rings the total DM halo profile can be parametrized as:
\begin{equation}
  \rho_\chi (\vec r) = \rho_0\cdot \left(\frac{\tilde r}{r_0}\right)^{-\gamma}\left[\frac{1+
  \left(\frac{\tilde r}{a}\right)^\alpha}{1+\left(\frac{r_0}{a}\right)^\alpha}\right]^{\frac{\gamma-\beta}
  {\alpha}} + \sum_{n=1}^2 \rho_n \exp\left(-\frac{\left(\tilde{r}_{gc,n}-{R_n}\right)^2}
  {2\cdot\sigma_{R,n}^2}-\left\vert\frac{z}{\sigma_{z,n}}\right\vert \right)\label{profile1}
\end{equation}
with
\begin{equation}
  \tilde{r} = \sqrt{x^2+\frac{y^2}{\varepsilon_{xy}^2}+\frac{z^2}{\varepsilon_z^2}}, \hspace*{1cm}
  \tilde{r}_{gc,n} = \sqrt{x_{(n)}^2+\frac{y_{(n)}^2}{\varepsilon_{xy,n}^2}};
\end{equation}
 $\varepsilon_{xy}$ and $\varepsilon_z$ ($\varepsilon_{xy,n}$)
are the ellipticities of the triaxial halo profile and rings,
respectively. The first term of Eq. \ref{profile1} has been modified with respect to
Eq. \ref{profile0} in order  that $\rho_0$
represents the DM density in the solar system, i.e. at $r=r_0$ $\rho=\rho_0$.
Other degrees of freedom are the angles with respect
to the axis earth - Galactic center of the halo $\phi_{gc}$ and of
the rings $\phi_n$, i.e. each component has its own coordinate
system which is rotated around the $z$-axis. The maximum WIMP
density of a ring $\rho_{n}$ is reached in the Galactic plane ($z =
0$) at a distance from the Galactic center $\tilde{r}_{gc,n}=R_n$.
 Figure \ref{haloschema} shows a schematic picture of a ring with the different
definitions.

The radial width of the outer ring is taken to be different for the
inner and outer side as can happen for the infall of a satellite galaxy,
which is disrupted most strongly near the pericenter  and the matter
will be distributed towards larger distances, so the shape was modified
to fall off to zero at radii smaller than the pericenter within a
distance $d_{r}$ using two quadratic functions: $ \rho(r) =  a\cdot
(r-(R_n-d_r))^2~    \mbox{\rm for }~ (R_n-d_r)<r<(R_n-d_r/2)$ and $
\rho(r) =  \rho_n-a\cdot (r-R_n)^2  ~ \mbox{\rm for }
~(R_n-d_r/2)<r<R_n.$

The parameters of the halo model and the boost-factor are varied to
minimize the following $\chi^2$ function:
\begin{equation}
  \chi^2=\sum\limits_{i,j}\left({f^{i,j}\cdot\phi_{\mbox{\scriptsize{bg}}}^{i,j}+
  B\cdot\phi_{\mbox{\scriptsize{dm}}}^{i,j}+\phi_{\mbox{\scriptsize{eg}}}-
  \phi_{\mbox{\scriptsize{exp}}}^{i,j}\over\sigma_{\mbox{\scriptsize{}}}^{i,j}}\right)^2,
\end{equation}
where $i$ and $j$ denote the different bins in longitude and
latitude and $f^{i,j}$ and $B$ are the normalization factors of
the background and DMA. Note that the boost factor $B$ and the
extragalactic flux $\phi_{eg}$ were taken to be
the same in all directions, i.e. independent of $i$ and $j$. The
scaling factor $\rho_0$ of the ``spherical'' component (=isothermal profile) is not a
free parameter, since it is scaled to get the rotation velocity at
$r_0$ correct. This means that if the fit requires a more massive
inner ring, the  density of the spherical  component is adjusted
automatically. Note that in total one fits 180 independent sky
directions, each with 8 data points above 0.07 GeV, so one has a
total of more than 1400 data points, which is enough to determine
the halo parameters. In addition, the parameters of the different
contributions are determined by completely independent data: the
outer (inner) ring is determined by the flux {\it in} the plane of
the disk away (towards)  the Galactic center, while the isothermal
profile is mainly determined by the fluxes {\it outside} the
Galactic disk. As a test of the convergence and angular resolution
a fit with 360 sky directions was performed as well, which yielded
practically identical results.

All sky directions are now well described by the basic isothermal
profile plus the substructure in the form of two toroidal rings,
as shown in Fig. \ref{long_rings} with the contributions of the
two rings  indicated separately. They clearly dominate for low
latitudes, but are small for latitudes above 10 degrees.
The latitude distributions are also well described, as shown
in Fig. \ref{lat} for the
 direction towards the Galactic center. The fit results of the parameters are
 summarized in Table \ref{t2}. The errors in the parameters are
 mainly systematic, e.g. depending on the fact that we took the
 boost factor to be the same in all directions etc. Determining
 these systematic errors is outside the scope of the present paper,
 but the qualitative picture  of two ringlike substructures is
 independent of such details.
\begin{table}
  \begin{center}
    \begin{tabular}{|c|c|c|}
      \hline
      parameter & halo without rings & halo with rings  \\
      \hline
      $\rho_0$ [GeV cm$^{-3}$] & 0.57  & 0.5   \\
      $r_0$ [kpc]              & 8.3   & 8.3   \\
      $\alpha$                 & 2     & 2     \\
      $\beta$                  & 2     & 2     \\
      $\gamma$                 & 0     & 0     \\
      $a$ [kpc]                & 5     & 5     \\
      $\varepsilon_{xy}$       & 0.7   & 0.8   \\
      $\varepsilon_{z}$        & 0.6   & 0.75  \\
      $\phi_{gc}$ [$^\circ$]   & 0     & 90    \\
      $M_{200}$ [M$_\odot$]    & $2.8\cdot 10^{12}$ & $3.4\cdot 10^{12}$  \\
      $R_{200}$ [kpc]          & 290   & 310   \\
      \hline
      $\rho_1$ [GeV cm$^{-3}$] & - & 4.5 \\
      $R_1$ [kpc]              & - & 4.15  \\
      $\sigma_{r,1}$ [kpc]     & - & 4.15  \\
      $\sigma_{z,1}$ [kpc]     & - & 0.17  \\
      $\varepsilon_{xy,1}$     & - & 0.8 \\
      $\phi_1$ [$^\circ$]      & - & -70  \\
      $M_1$ [M$_\odot$]        & - &  $9.3\cdot 10^{9}$ \\
      \hline
      $\rho_2$ [GeV cm$^{-3}$] & - & 1.85\\
      $R_2$ [kpc]              & - & 12.9 \\
      $\sigma_{r,2}$ [kpc]     & - & 3.3  \\
      $d_{n}$ [kpc]            & - & 4  \\
      $\sigma_{z,2}$ [kpc]     & - & 1.7  \\
      $\varepsilon_{xy,2}$     & - & 0.95  \\
      $\phi_2$ [$^\circ$]      & - & -20   \\
      $M_2$ [M$_\odot$]        & - & $\cdot 10^{10}$ \\
      \hline
      \hline
      $\chi^2$/d.o.f.      & 1206/157 & 144.2/157  \\
      probability          & 0 & $0.74$   \\
      \hline
    \end{tabular}
  \end{center}
  \caption[Fit Results for Different Halo Models]{Fit results
   with and without the ringlike substructures. The
  triaxial halo  and the two ring parameters with subscript
  1 and 2 are given. The parameters $r_0, \alpha ,\beta ,\gamma , a_0$ are
  fixed by the definition of the pseudo-isothermal profile
  and  $M_{200} , R_{200},  M_1, M_2$ are derived values, so  the remaining
  14 values have been optimized.} \label{t2}
\end{table}

The boost factor for the profile with rings is around 100, if one
assumes the DM annihilation cross section at the WIMP decoupling temperature
 $m_\chi/22\approx 3$ GeV 
in the early universe  to be still valid for the low
kinetic energies in the present universe. The independence of the center-of-mass energy
of the annihilation cross section is true, if the
annihilation proceeds via the exchange of spin-less particles, like
Higgs bosons. But the annihilation depends strongly on momenta for
DMA via the exchange of a particle with spin, like the $Z$-boson.
The  resonant $Z$-exchange becomes dominant if the WIMP mass is
close to half the Z-boson (around 45 GeV) and then the annihilation
cross section at  the temperatures of the present universe is much smaller, thus
requiring boost factors of $10^3$ or more for WIMP masses below 50
GeV. Unfortunately, such large boost factors are not necessarily
excluded, so one cannot determine a lower limit on the WIMP mass
from the boost factors alone.
\subsection{Ring structure}\label{ring}

\begin{figure}
\begin{center}
 \includegraphics [width=0.4\textwidth,clip]{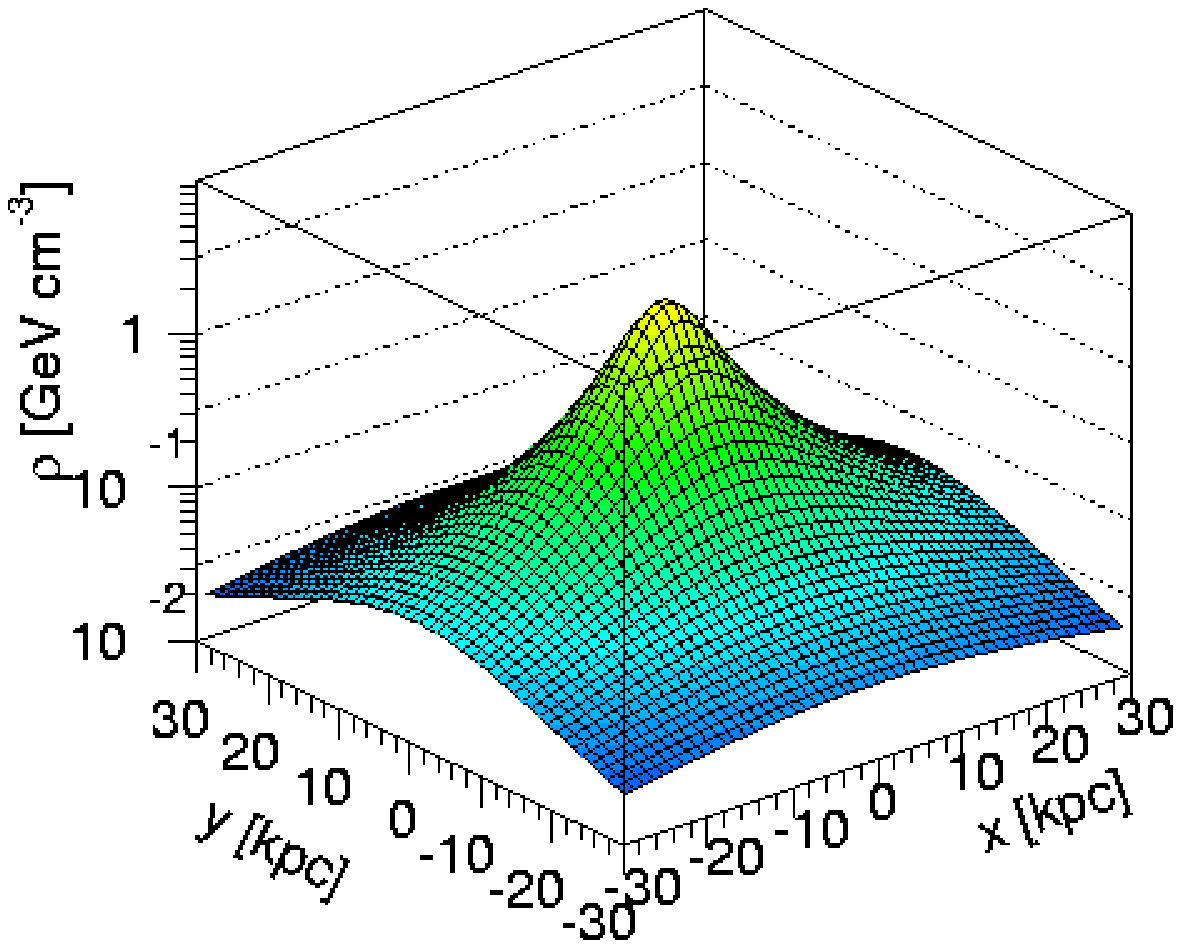}
 \includegraphics [width=0.4\textwidth,clip]{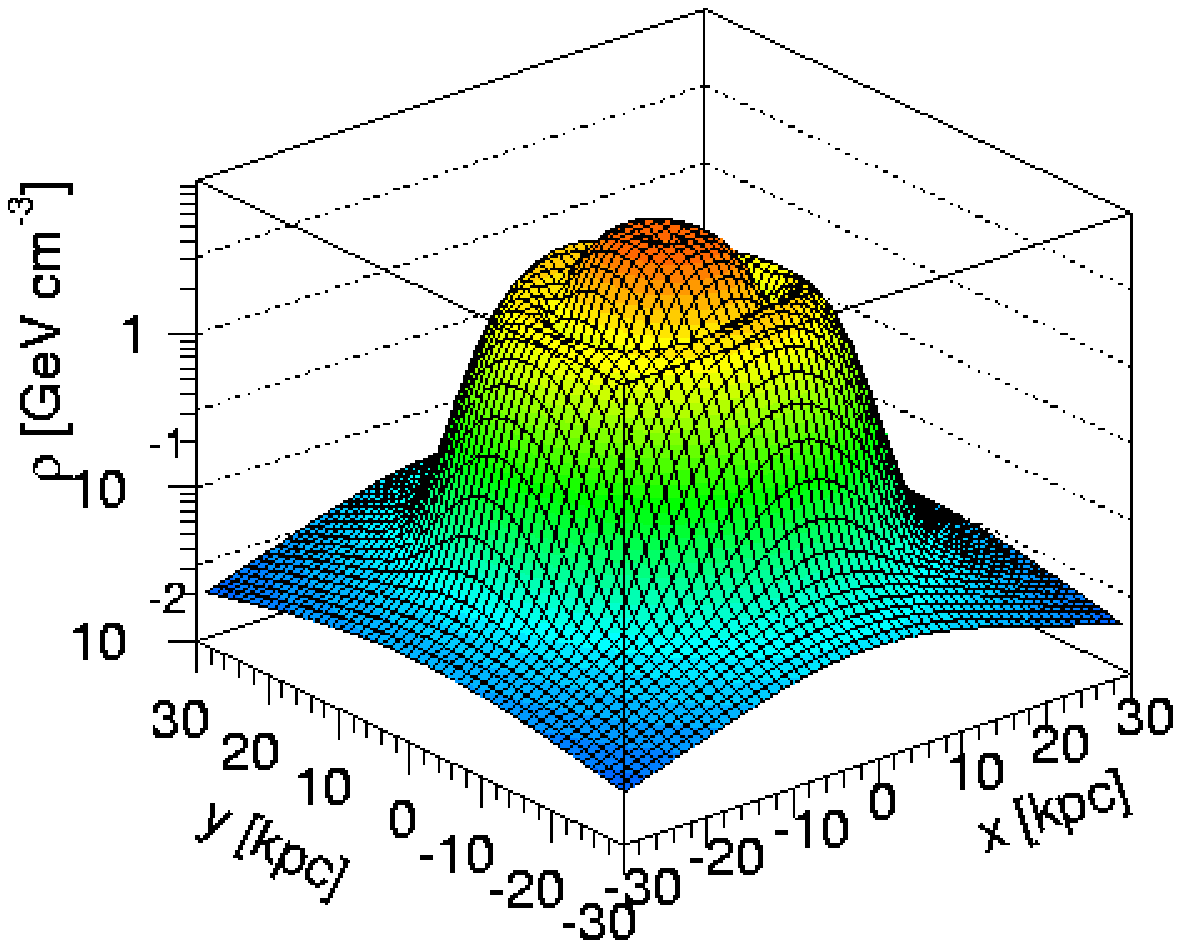}
 \includegraphics [width=0.4\textwidth,clip]{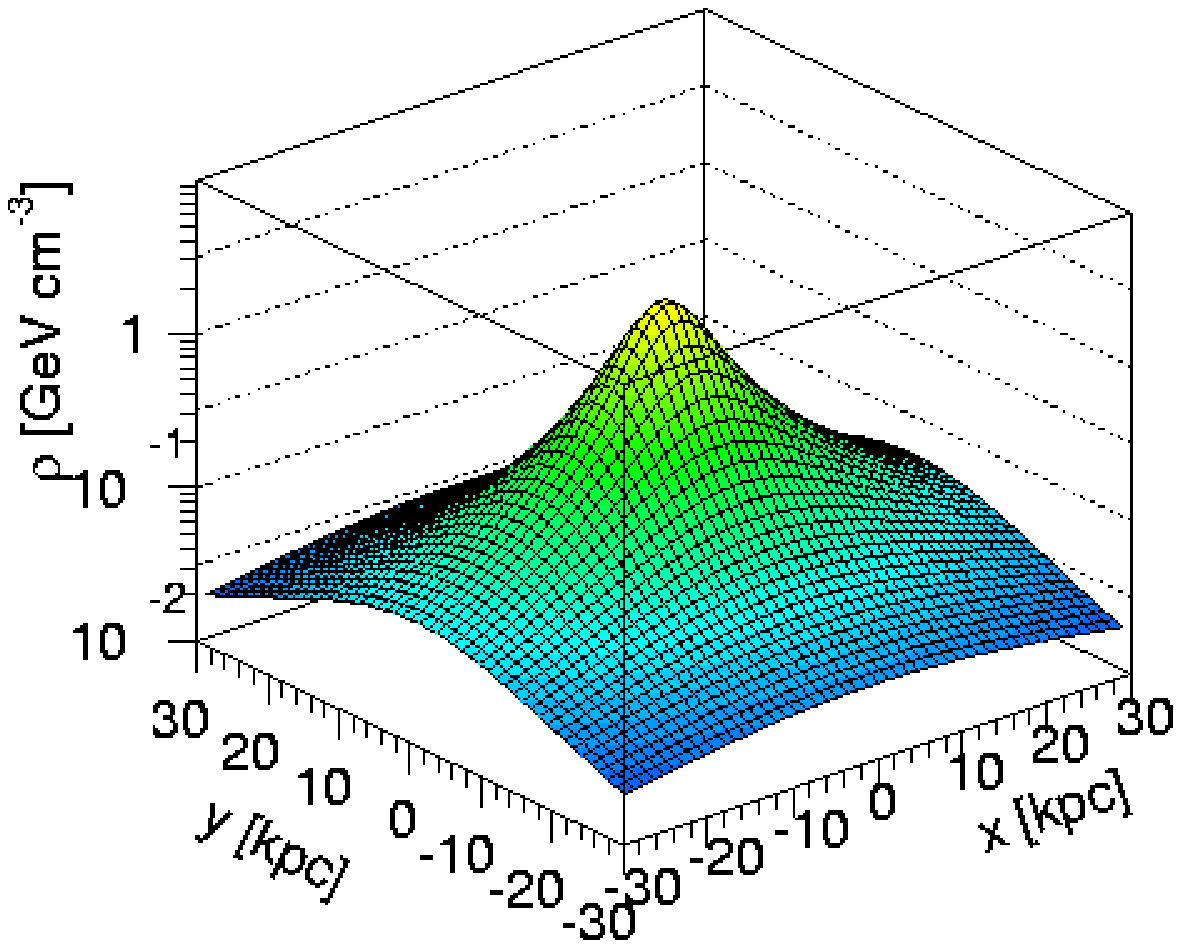}
 \includegraphics [width=0.4\textwidth,clip]{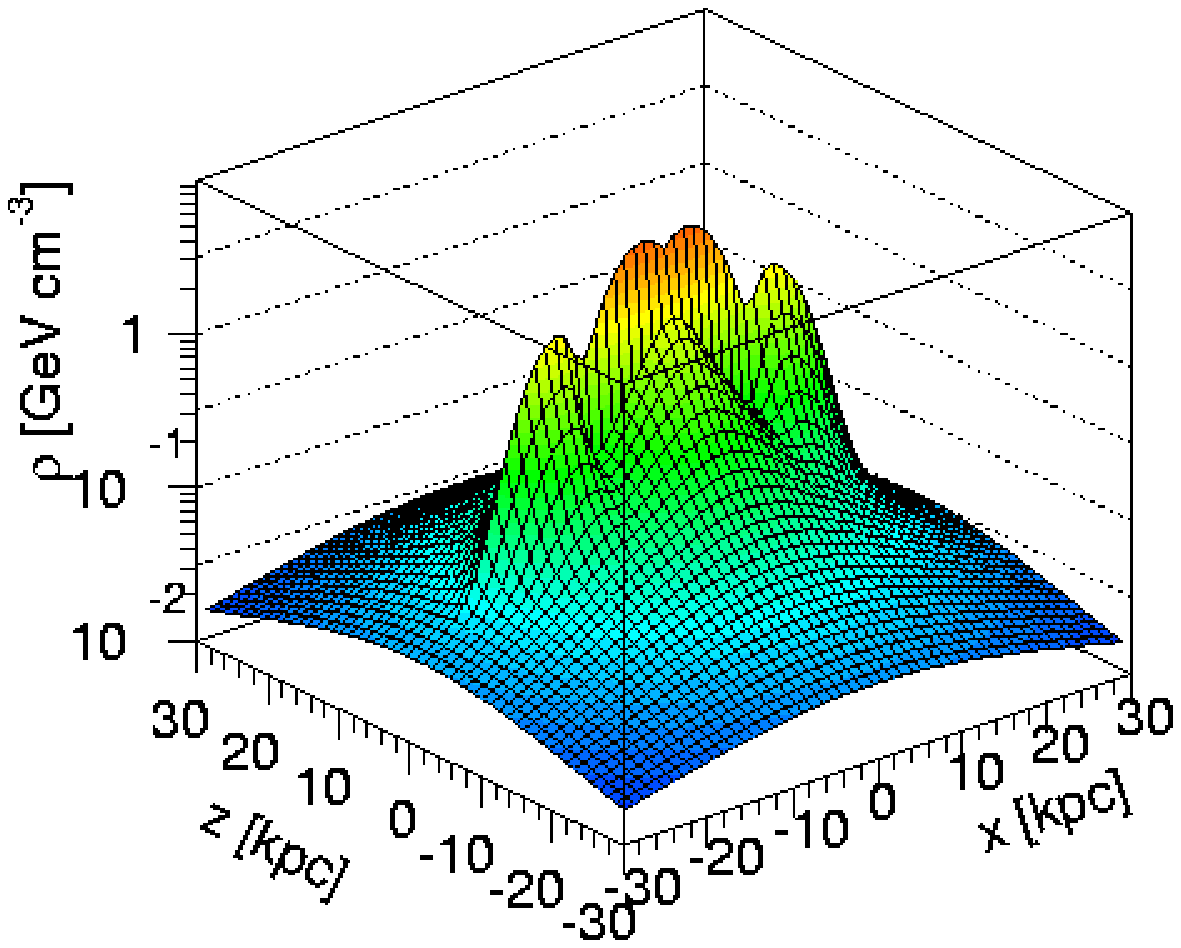}
 \caption[]{
  3D-presentations of the isothermal haloprofile in the Galactic
  plane (xy-projection) (top row) and
  perpendicular to the disk (xz-plane) (bottom row)
  without (left) and with (right)  toroidal ringlike structures at 4 and 14 kpc. }
 \label{profile}
\end{center}
\end{figure}

\begin{figure}[]
\begin{center}
 \includegraphics [width=0.7\textwidth,clip]{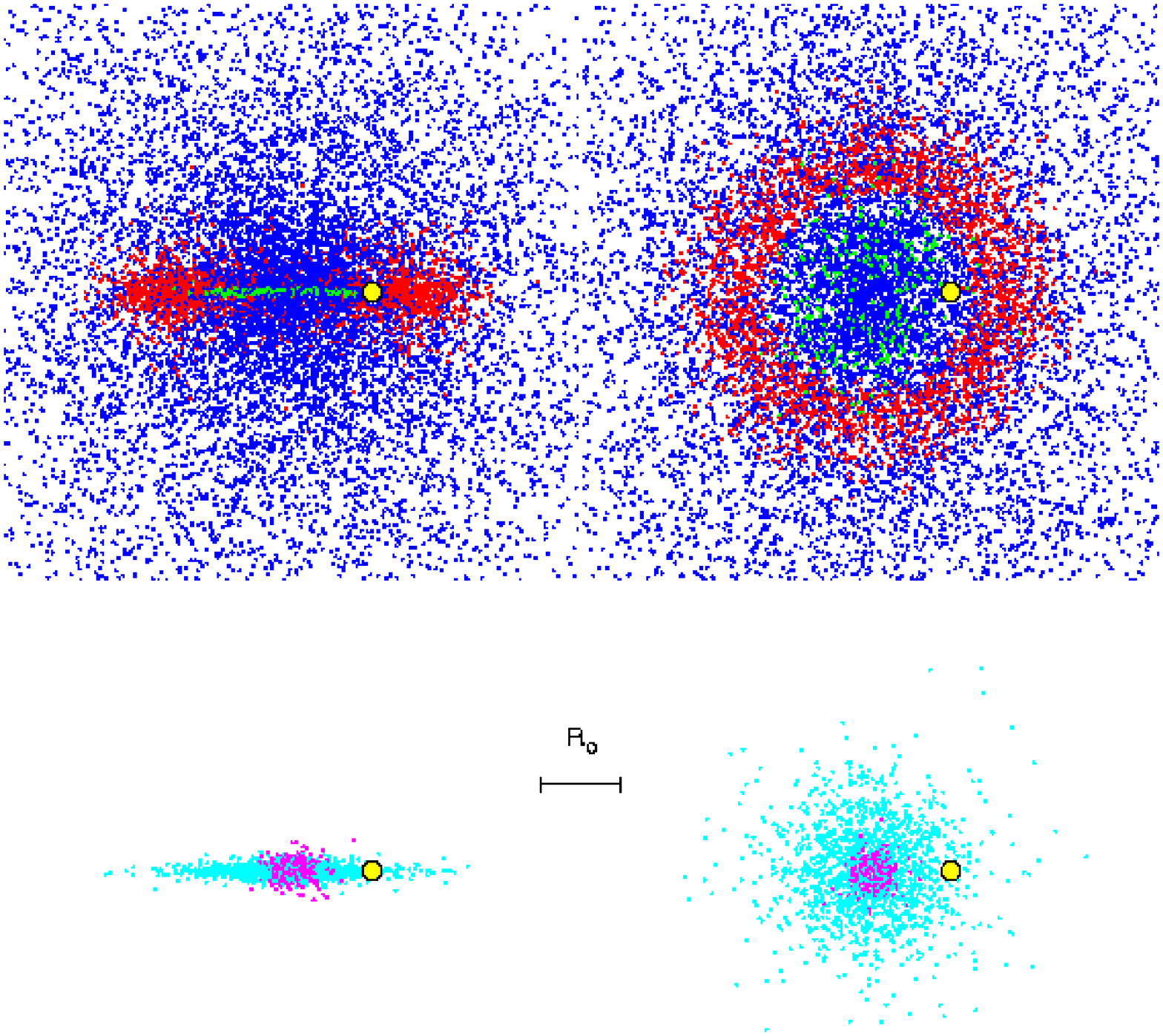}
 \caption[]{Visualization of the DM (top) and baryonic matter
 (bottom) distribution of the Milky Way in an edge-on (left)
 and top (right) view of the disk. The blue dots mark the
 modified isothermal profile; the red ones mark the outer
 and the green ones the inner ring; cyan marks the
 exponential disk and pink the stellar bulge; the large yellow
 circle marks the position of the sun; the density of points of
 each component is proportional to their mass fraction.}
 \label{scatter}
\end{center}
\end{figure}

Fig. \ref{profile}  displays the halo distribution in the disk
(xy-plane) and perpendicular to the plane (xz-plane) in a 3D plot,
while Fig. \ref{scatter} shows it in the projections in comparison
with the distribution of baryonic matter. The contributions from the
inner and outer rings at radii of 4.2 and 14 kpc, respectively can
be clearly seen. The maximum ring densities   of  the inner (outer)
 ring are about a factor of 6 (7)  higher than the isothermal
 profile at these maxima.

 The maximum density of the outer ring is at a radius of
14 kpc with a width   of about 3.3 kpc in radius and 1.7 kpc
perpendicular to the disk. These coordinates coincide with the ring
of stars observed in the plane of the Galaxy at a distance of 14-18
kpc from the Galactic center
\citep{newberg,ibata,yanny,rocha-pinto1}. These stars show a much
smaller velocity dispersion (10-30 km/s) and a larger z-distribution
than the thick disk, so the ring cannot be considered an extension
of the disk. A viable alternative is the infall of a satellite
galaxy \citep{yanny,helmi,rocha-pinto,penarrubia,penarrubia1}, for
which one expects in addition to the visible stars a DM component.
From the size of the ring and its peak density one can estimate the
amount of DM in the outer ring to be around 9.10$^{10}$ solar
masses. Since the gamma ray excess is best fitted with a full
360$^\circ$ of the sky, one can extrapolate the observed $100^\circ$
of visible stars to obtain a total visible mass of  $\approx
10^8-10^9$ solar masses \citep{yanny,ibata}, so the baryonic matter
in the outer ring is only a small fraction of its total mass.

The inner ring at 4.2 kpc with a width of 4.2 kpc in radius and 0.2
kpc in $z$ is more difficult to interpret, since the density of the
inner region is modified by adiabatic compression
\citep{adiabatic,adiabatic1,adiabatic2} and interactions between the
bar and the halo \citep{weinberg,athanassoula}. However, it is
interesting to note that its radius coincide with the ring of cold
dense molecular hydrogen gas, which reaches a maximum density at 4.5
kpc and has a width around 2 kpc  \cite{gordon,hunter}.
At the same radius a toroidal structure of dust has been observed,
which provides shelter against dissociating UV radiation and
allows atomic hydrogen
to coalesce at the  surfaces into molecular hydrogen.
Therefore  a ring of molecular hydrogen
suggests a gravitational potential well
in agreement with the EGRET excess in this region.

\subsection{Comparison with rotation curve}\label{rotation}

\begin{figure}
\begin{center}
 \includegraphics [width=0.7\textwidth,clip]{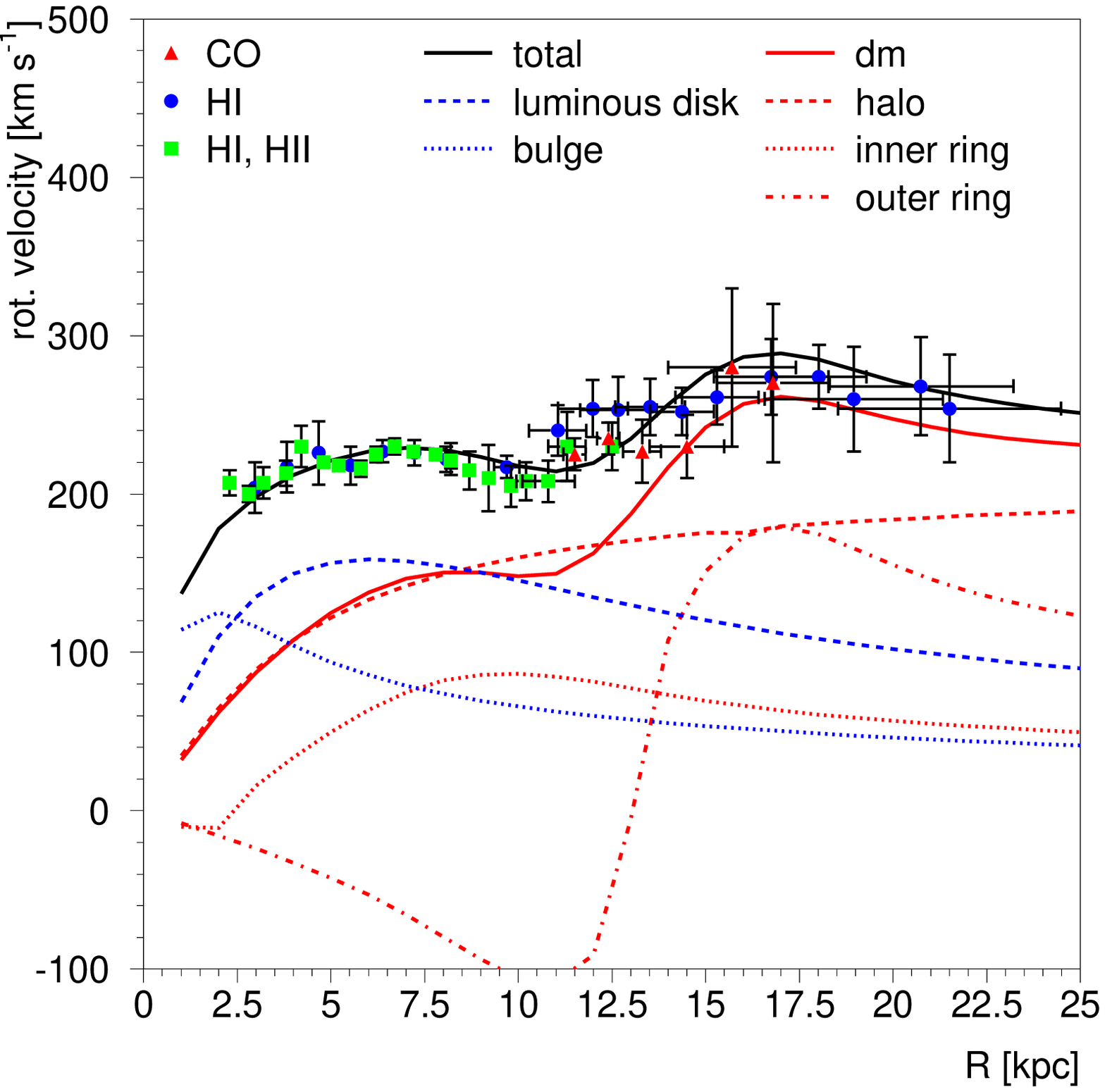}
 \caption[]{
 The rotation curve of the Galaxy for the DM halo profile of
 Fig. \ref{profile}. The data are from
 \citet{honma,brand,brand1,brand2,schneider}. The contributions
 from the individual mass components have been indicated. Note the negative
 contribution of the massive  ring of DM at 14 kpc, which
 exerts an outward and hence negative force on a tracer
 inside that ring.}
 \label{rot}
\end{center}
\end{figure}

\begin{figure}[tbp]
  \begin{center}
    \includegraphics[width=0.45\textwidth]{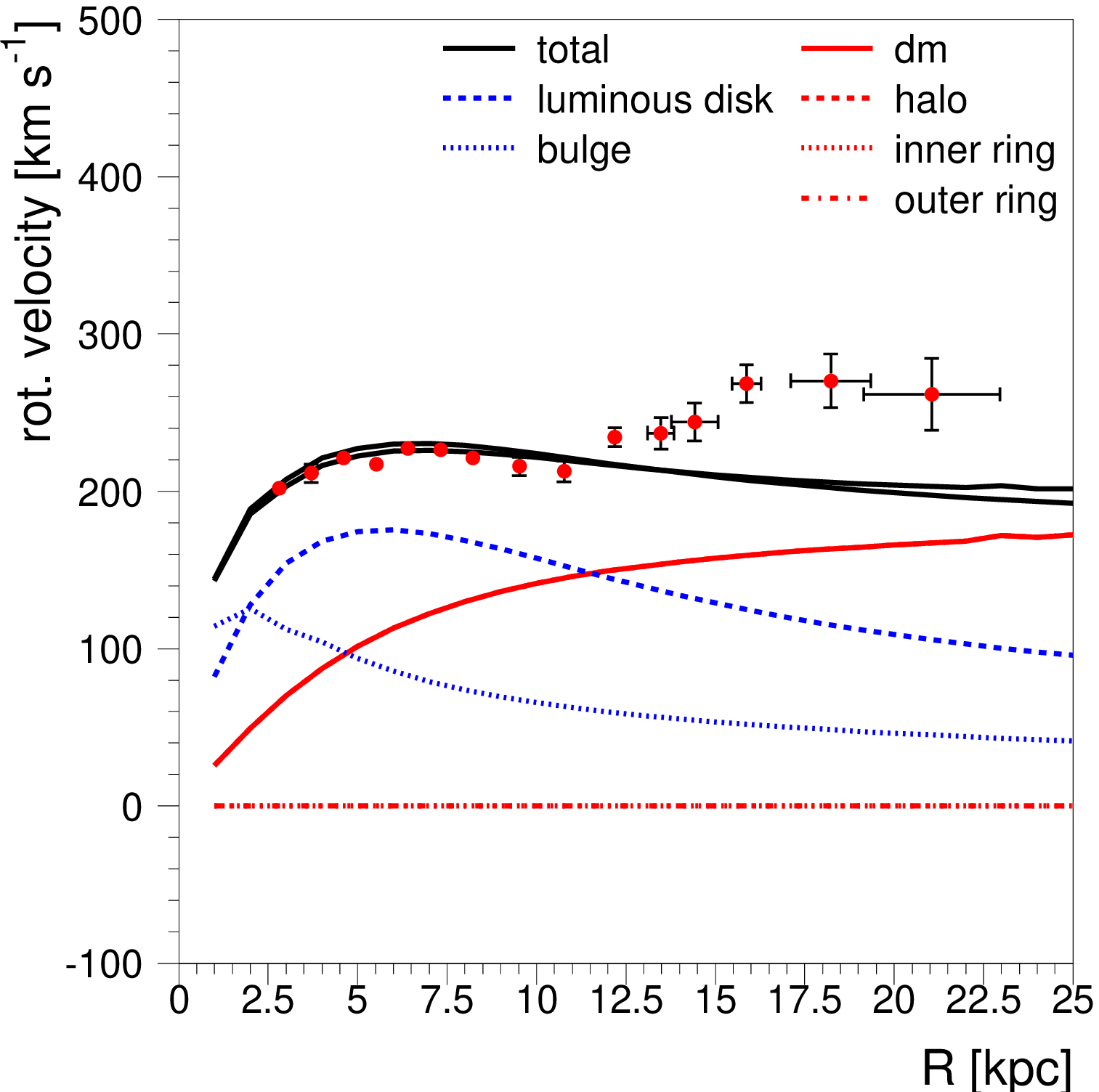}
    \includegraphics[width=0.45\textwidth]{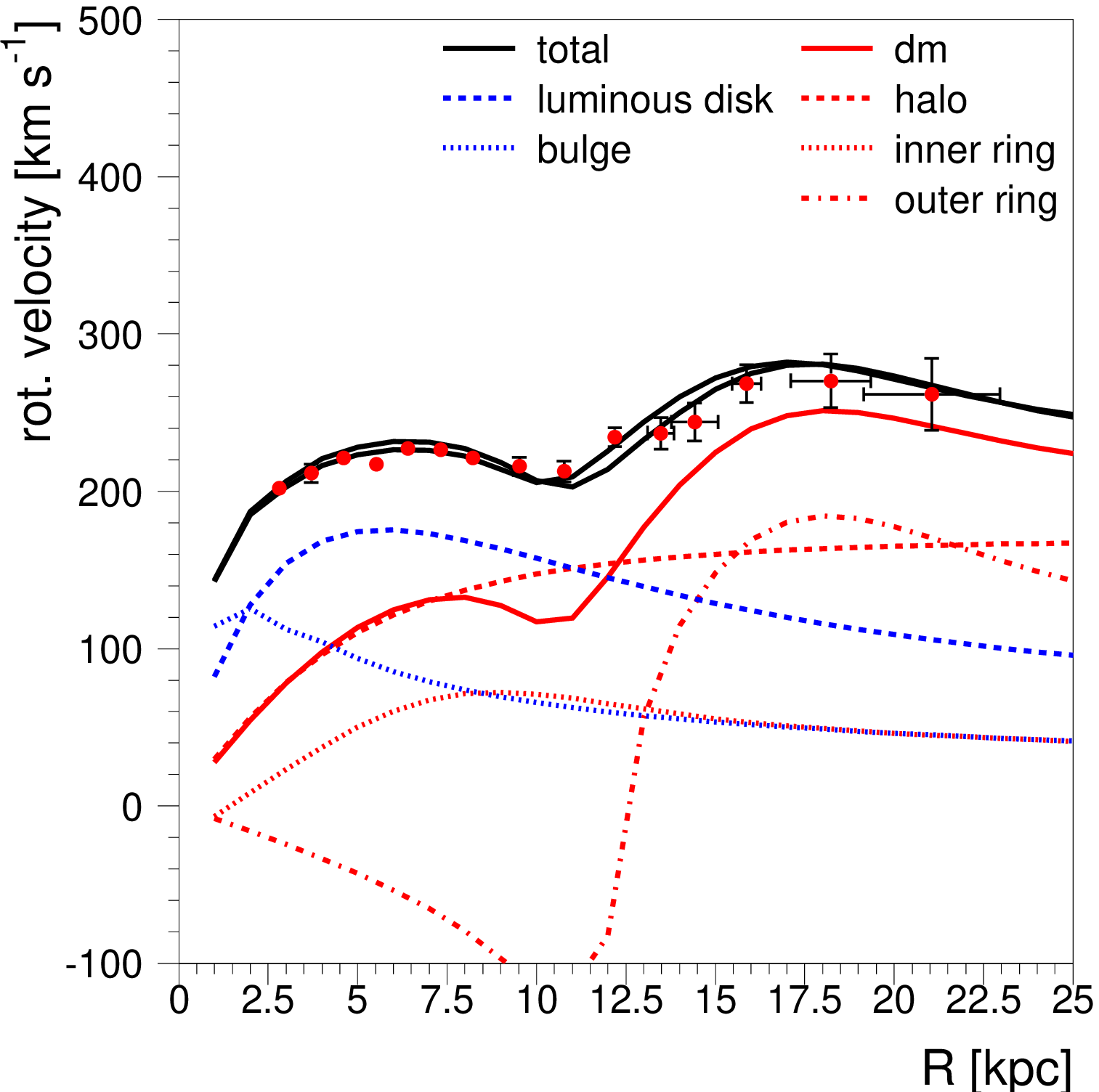}
    \caption[Rotation Curves for Pseudo-Isothermal Profiles]{Rotation
    curve for the pseudo-isothermal profile without (left) and with (
    right) rings in the Galactic plane; here the different data sets
    on rotation velocities are the same as in Fig. \ref{rot}, but for
    clarity a weighted average has been taken. One can see that the
    pseudo-isothermal profile without rings does not provide a good fit to the data,
    while the  inclusion of the toroidal substructures provides a
    perfect fit. The rotation curves (solid black lines) are
    calculated in two perpendicular directions: along the long and
    medium axis of the triaxial halo, which both lie in the plane
    of the disk.}
    \label{rot1}
  \end{center}
\end{figure}

It is interesting to note that the present analysis is able to trace
the mass distribution of DM in our Galaxy, since the mass is given
by the WIMP number density $n_\chi$ times the WIMP mass $m_\chi$.
The first one is obtained from the gamma ray flux, 
the second from the gamma ray spectrum.
The relative contributions of the rings and pseudo-isothermal
profile are obtained from the intensity of the EGRET excess and the
absolute normalization of the total mass  is obtained by requiring
that the local rotation velocity of our solar system at 8.3 kpc is
220 km/s. For the WIMP mass of 60 GeV from the spectrum and the halo
parameters from Table \ref{t2} we can immediately calculate the mass
in the rings and the mass in the pseudo-isothermal profile. For the
mass in the outer (inner) ring one finds a value around $9\cdot10^{10}$
 ($9\cdot10^9$)  solar masses, which is only a
small fraction of the total mass of $3\cdot 10^{12}$ solar masses
inside a radius $R_{200}$ of 310 kpc. The latter radius represents
the volume with an averaged overdensity of 200 times the critical
density of the universe. However, the  mass in the outer ring  is
about 50\% of the mass of the Galaxy inside its radius. Therefore
one expects a significant influence on the rotation curve, which
then should show a minimum below and maximum above this radius,
since the rotation speed squared is proportional to the derivative
of the potential. Note that the absolute value of the masses in the rings
is not sensitive to the background model, since the absolute mass scale
is set by the normalization to the rotation curve, so  a different
background model, like the shape from the optimized model will
change the boost factor, but not the absolute masses.

A minimum in the rotation curve inside the outer ring radius and maximum outside this radius 
is indeed observed,
as shown in Fig. \ref{rot}. The data were
taken from \citet{honma} using HI gas, from \citet{brand, brand1,brand2}  using
HII gas  and \citet{schneider}, who use CO clouds. The
contributions from each of the mass terms have been shown
separately. The  baryonic matter distribution was taken
from \citet{olling}. The basic explanation for the negative
contribution from the outer ring  is that a tracer star at the
inside of the ring at 14 kpc feels an outward force from the ring,
thus a negative  contribution to the rotation velocity. If one just
takes the contributions from the visible matter and the isothermal
profile without rings, the data cannot be described, as shown on the
left hand side of Fig. \ref{rot1}. Here the data points were
obtained from the ones in Fig. \ref{rot} by taking a weighted
average. With the rings a perfect description is obtained, as shown
on the right hand side of Fig. \ref{rot1}. Here two rotation curves
were calculated: one along the long axis of the triaxial halo
profile, which was found to be in the Galactic plane and one
perpendicular to this axis in the plane. Since the ellipticity is small
($\epsilon_{xy}\approx\epsilon_z\approx 0.8$, see Table \ref{t2}), the
difference is small.

Usually the rotation curves with inhomogeneous mass distributions
are calculated by solving the Poisson equation, which yields
the gravitational potential at a given position: $\Phi(r,\theta,\phi)$.
The rotation velocity for
a circular orbit at a radius $r$ can then be calculated by requiring
that  the resulting gravitational force on a tracer star  equals the
centrifugal force, i.e.
\begin{equation}
v^2/r=F_G/m\equiv d\Phi(r)/dr.
\label{rc}
\end{equation}
However, the contribution from  ringlike structures is not easily obtained
from a standard Poisson solver, since these are usually optimized for
spherical symmetries. They did not converge for the ringlike
structures in our case. Therefore the derivative of the potential
instead of the potential was calculated numerically, which is what is needed for
the rotation curve and avoids  an additional numerical derivation.

To be more precise, the following was done. The gravitational potential $\Phi$ of the
Poisson equation can be written in spherical coordinates
($x=r\cos\phi \sin\theta,\ y=r\sin\phi \sin\theta,\ z=r\cos\theta $) as:
\begin{equation}
\Phi(r,\theta,\phi)=-\int_0^\infty r'^2dr'\int_{-1}^{1}d\cos\theta'
\int_0^{2\pi}d\phi'\frac{\rho(r',\theta',\phi')}
{\sqrt{r^2+r'^2-2rr'\sin\theta\sin\theta'\cos(\phi-\phi')
-2rr'\cos\theta\cos\theta'}}
\label{phi1}
\end{equation}
or in the plane in the direction $\phi=0, \theta=\pi/2$:
\begin{equation}
\Phi(r,\pi/2,0)=-\int_0^\infty r'^2dr'\int_{-1}^{1}d\cos\theta'
\int_0^{2\pi}d\phi'\frac{\rho(r',\theta',\phi')}{\sqrt{r^2+r'^2
-2rr'\sin\theta'\cos(\phi-\phi')}}
\label{phi2}
\end{equation}
Note that $\rho$ includes all masses. The rotation velocity for
a circular orbit at a radius $r$ can then be calculated
in the direction $\phi=0, \theta=\pi/2$ (using Eq. \ref{rc}):
\begin{equation}
{v^2(r)\over r}=\frac{d\Phi(r)}{dr}=\int_0^\infty
r'^2dr'\int_{-1}^{1}d\cos\theta'
\int_0^{2\pi}d\phi'\frac{\rho(r',\theta',\phi')
(r-r'\sin\theta'\cos(\phi-\phi'))}
{(r^2+r'^2-2rr'\sin\theta'\cos(\phi-\phi'))^{3/2}}. \label{v2}
\end{equation}
This threefold integral was integrated numerically to obtain the
contribution from all mass elements in the halo.  The contributions
of the bulge, disk and DM contributions from the isothermal halo plus rings
 have been indicated separately in Fig. \ref{rot}.
The negative contribution from a ring is expressed by the fact
that the derivative of the gravitational potential $\Phi$ changes
its sign, when crossing the maximum of the ring and so does the
contribution to $v^2$ (see term $r-r'$ in numerator of Eq.
\ref{v2}). This implies an outward gravitational force exerted by
the ring for a tracer inside the ring and an inward force for a
tracer outside the ring. The  hitherto mysterious
change of sign of the slope near $r=1.3r_0$ finds then its natural
explanation in the large ring of DM at $r=14$ kpc, whose mass is
determined by the excess of energetic gamma rays.

\subsection{Galactic parameters}

From the baryonic density profile and
the DM profile determined in this paper, one can determine the
following basic properties of our Galaxy:
\begin{itemize}
\item The radius containing an average density 200 times the critical
density equals $R_{200}=310 ~kpc$
\item The total DM mass inside this
radius  is $M_{200}=3.0\cdot10^{12}M_\odot$  to be compared with a
visible mass of 5.5$\cdot10^{10}$ $M_\odot$
\item The inner (outer)
ring contribute 0.3 (3) \% to the total DM mass
\item  The fraction
$f_d=M_{baryonic}/M_{200}=0.02$
\item The concentration parameter $c=R_{200}/a=310/5=63$.
\end{itemize}
It should be remembered that these parameters were obtained assuming
an isothermal profile for the DM with a constant boost factor.
Assuming a smaller  boost factor in the bulge because of tidal
disruptions there, will increase the mass in the center, since the
flux is proportional to $B(r)<\rho^n>$, as discussed before. This
reduces the DM mass in the outer regions, so the numbers given above
should be considered an upper limit, but these parameters are well
inside the range found for other galaxies with  an isothermal
profile \citep{Jimenez:2002vy} and earlier mass estimates of the
Galaxy \citep{wilkinson}.

\section{Possible objections to the DMA interpretation}\label{objections}

The DMA interpretation of the EGRET excess would mean that DM is
not so dark anymore, but DM is visible from the 30-40 flashes of
energetic gamma rays for each annihilation. This would be great,
but are there more mundane explanations? Attempts to modify the
electron and proton spectra from the locally measured spectra do
not describe the shape of the EGRET data in all sky directions, as
discussed in detail before by comparing the EGRET data with the
``optimized model''. Here we summarize some  other  possible
objections.

\begin{enumerate}
\item
Are the EGRET data reliable enough to make such strong
conclusions? The EGRET detector was calibrated in a quasi
mono-energetic gamma ray beam at SLAC, so its response
is well known \citep{egret_cal}. Also the monitoring during the flight was done
carefully \citep{egret_cal1}. We have only used data in the energy
range between 0.07 and 10 GeV, where the efficiency is more or
less flat. So  the 9 years flight provided accurate and reliable
data, especially it would be hard to believe in an undetected
calibration problem, which would only effect the data above 0.5
GeV and fake the gamma ray spectrum from the fragmentation of
mono-energetic quarks.
\item
The gamma ray spectrum above 0.07 GeV starts to be dominated by
pp-interactions and is therefore strongly dependent on the proton
energy spectrum. This cosmic ray spectrum
 was measured only locally in the solar neighborhood. Could a harder spectrum
 near the Galactic center, where protons can be accelerated by the many supernovae
 there, explain the EGRET excess? No, first of all the diffusion times are much
 shorter than the energy loss times of
 protons with energies above a few GeV,
 so  one expects
 everywhere the same energy spectrum. This is proven by the fact, that
 the gamma ray spectrum for the Galactic center and the Galactic
 anti-center can be described by the {\it same} background shape.
\item Is the background well enough known to provide evidence for DMA?
The background is dominated by pp-collisions with a reasonably well known shape and fitting
the normalization yields a ``self-calibrating'' background.
Trying to obtain a harder gamma ray spectrum by hardening the proton spectrum increases
usually not only the high energy gamma rays, but also contributes to
the low energy part of the spectrum.
Fitting this ``wrong'' shape with a free
normalization reduces then the low energy
excess again and recovers the high energy excess.
Note that this ``self-calibration'' of the background also takes care of
gas clouds, ringlike or asymmetric structures in the background, uncertainties
in the absolute value of the total cross sections, etc.

An alternative way of formulating the problem of  models without
DMA: if the shape of the EGRET excess can be explained perfectly in
all sky directions by a gamma contribution originating from the
fragmentation of mono-energetic quarks, it is very difficult to
replace such a contribution by an excess from nuclei (quarks) or
electrons with a steeply falling energy spectrum, especially if one takes
into account that the
spatial distribution of gamma rays from DMA is different from the
gamma rays from the background.
\item
  Is it possible to explain
the excess in diffuse gamma rays with unresolved point sources? This
is unlikely, first of all since the known point sources
\citep{egret_cat} are only a small fraction of the diffuse gamma
rays and the majority of the resolved sources has a rather soft
spectrum, typically well below 1 GeV, as can be seen from the plots
in the Appendix. If this part of the spectrum would be dominated by
unresolved sources, then the {\it diffuse} component below 1 GeV would become
lower after subtracting the hypothetical unresolved sources,
which in turn would lead to a lower
normalization of the background and a correspondingly stronger
excess for a fixed background shape. So arguing against DMA by unresolved sources
goes in the wrong direction.
\item One observes a ring of molecular hydrogen near the inner ring
and a ring of atomic hydrogen near the outer ring. Could this excess
of hydrogen not be responsible for the excess of the gamma rays? No,
our method  of fitting only the shapes with a free normalization
implies that this analysis is insensitive to density fluctuations of
the background, which change the normalization, not the shape.
\item
Is one not over-interpreting the EGRET data by
fitting so many parameters for the different components: triaxial halo, inner ring and outer ring?
No, first of all the excess and enhancement in a ringlike structure at 14 kpc was
already discovered in the original paper by \citet{hunter}. What we did is just trying
to see if the excess fits: a)  a single WIMP mass in all directions;
b)  an isothermal DM profile
plus the substructure; c) the Galactic rotation curve.
The DM halo components are determined by {\it independent} sky directions:
the outer ring parameters are determined mainly by  30 sky
directions towards the Galactic anti-center, the inner ring
parameters by ca. 15 sky directions towards the Galactic center
and the triaxial halo parameters by ca. 130 sky directions out of
the Galactic plane. And the most remarkable thing is that all
these independent sky directions all show an excess, which can be
explained by a single WIMP mass around 60 GeV. This is like having
180 independent experiments at an accelerator all saying we see a
significant excess of gamma rays corresponding to $\pi^0$
production from mono-energetic quarks. Then asked what mass they
need to describe the excess, they all say 60 GeV!
\item
The  outer rotation curve of our
Galaxy  has  large uncertainties from the distance $r_0$
between the Sun and the Galactic center and is determined with a
different method than the inner rotation curve. Can this fake the
good agreement between the calculated rotation curve from the gamma ray excess and
the measured rotation curve? The outer rotation curve  indeed depends strongly on $r_0$,
as shown by \citet{honma}, who varied $r_0$ between 7 and 8.5 kpc.
At present one knows from the kinematics of the stars near the black
hole at the center of our Galaxy that $r_0=8\pm 0.4$ kpc
 \citep{rnull}, so the distance is already reasonably well known. But
whatever the value of $r_0$, the change in slope around 1.3$r_0$ is
always present, indicating a ringlike DM structure is always needed.
Furthermore the outer rotation curve shows first the same decrease
as the inner rotation curve and only then changes the slope, so the
different methods  agree between $r_0$ and 1.3$r_0$
\item
How can one be sure that the outer ring originated from the tidal
disruption of a rather massive satellite galaxy, so one can expect
an enhanced DM density in the ring? One finds three independent
ringlike structures: stars, atomic hydrogen gas and excess of gamma
radiation. The stars show a scale height of several kpc and a low
velocity dispersion, so they cannot be part of the Galactic disk.
Therefore the infall of a satellite galaxy is the natural
explanation. Since the tidal forces are proportional to $1/r^3$, the
satellite will be disrupted most strongly at its pericenter, which
can lead to  DM density enhancements  at the pericenter
after a few orbits \citep{hayashi}. Some of the stars and gas may be caught
in this potential well. All three are found at 14 kpc with the stars all
being old  and more than 90\%
of the mass being DM as deduced from the strong
EGRET excess at this radius.
\item
The outer ring at 14 kpc has a mass around  $9\cdot10^{10}$
solar masses. This is  around 50\% of the total mass inside the
ring and one may worry about the disk stability of the Milky Way by
the infall of such a heavy Galaxy. However, large spiral galaxies
show bumps of similar size \citep{sofue1}, so it seems not to be
uncommon to have masses of this size forming ringlike structures.
Furthermore, the stars near the 14 kpc ring are all very old,
so the infall of the satellite galaxy may have been very early,
in which case the disk might have grown after the infall.
\item
Is it not peculiar that if a ringlike structure originates from the
infall of satellite galaxy, that it lies in the plane of the Galaxy? No, in
principle the infall can happen in all directions with respect to
the plane, but  the angular momenta of the inner halo and a baryonic
disk tend to align after a certain time by tidal torques
 \citep{bailin}. This explains the enhanced DMA in the disk,
which is orthogonal to the prejudice that
DM should be distributed more or less spherically.
\end{enumerate}
\section{Summary}\label{summary}
If Dark Matter is a thermal relic from the early Universe,
then it is known to annihilate, since the  small amount of relic
density measured nowadays requires a large reduction in its
number density. The annihilation cross section can be obtained directly
 from its inverse proportionality to the relic density, the latter being
well known from  precision cosmology experiments \citep{wmap}.
The annihilation into quark pairs will produce $\pi^0$ mesons during the
fragmentation into hadrons, which in turn will decay into gamma rays.
Since  DM is cold, i.e. non-relativistic, the fragmenting quarks
have an initial energy equal to the WIMP mass. The gamma spectrum
from such mono-energetic quarks is well known from electron-positron
colliders, which produce exactly such states. For heavy WIMP masses
the gamma spectrum is considerably harder than the background
spectrum, mainly from $\pi^0$ mesons produced in pp-collisions from
cosmic rays with gas in the disk. Such an excess of hard gamma rays
has indeed been observed by the EGRET satellite and the relative contributions
from background and DM annihilation signal can be obtained by
fitting their different shapes with a free normalization factor
for background and signal.

The results of the analysis can be summarized as follows:
\begin{itemize}
\item By analyzing the EGRET data in 180 independent sky directions
 we find first of all an excess in each direction, as expected for DMA,
 and secondly, the spectral shape of the excess is the {\it same} in
{\it each} direction and corresponds to a  WIMP mass around 60 GeV.
\item
The {\it flux} of the excess determines the halo profile, i.e.  the
number density $n_\chi$ of the WIMPs. Together with the WIMP mass
$m_\chi$ from the {\it spectrum} of the excess  one can
reconstruct the DM mass distribution (=$n_\chi~\cdot~m_\chi$) in the
Galaxy, which in turn can be used - in combination with the visible matter -
 to calculate the rotation curve.
The result explains   the hitherto unexplained change of slope
in the outer rotation curve.
\end{itemize}

The results mentioned above make no assumption on the nature of the
Dark Matter, except that its annihilation produces hard gamma rays
from quark fragmentation. The fitted normalization of the background
flux comes out to be close to the absolute prediction of the GALPROP
conventional propagation model of our Galaxy \citep{galprop2}, while
the normalization of the DM signal corresponds to a boost factor
from 20 upwards. Such a boost factor from the clustering of DM was
calculated with respect to the annihilation cross section from Eq.
\ref{wmap}, which is the cross section at the freeze-out temperature
of a few GeV.  At the present time the temperature of the universe
is much lower, which could reduce the annihilation cross section,
thus increasing the boost factor.

A good WIMP candidate is the neutralino of Supersymmetry. For the
WIMP mass in the range of 50-100 GeV this has very much the
properties of a spin 1/2 photon, which would imply that  DM is the
supersymmetric partner of the cosmic microwave background (CMB).
The present analysis is perfectly consistent with such a scenario,
if the supersymmetric partners of the quarks and leptons are
around 1 TeV. Details about the connection with Supersymmetry have
been discussed elsewhwere\citep{deboer2,deboer3,deboer4,sander}.

It should be emphasized that  the excess of diffuse gamma rays has a
statistical significance of at least 10 $\sigma$ if compared with
the conventional shape of the background. This combined with all
features mentioned above provides an intriguing hint that this
excess is a) indeed  indirect evidence for Dark Matter Annihilation
and b) traces the DM in our Galaxy, as proven by  the fact that we
can reconstruct the rotation curve of our Galaxy from the gamma
rays.

\acknowledgements We thank I.V. Moskalenko, O. Reimer and A. Strong
for sha\-ring with us all their knowledge about our Galaxy and the
EGRET data and allowing us to use their implementation of the EGRET
analysis software in the GALPROP program.

We also like to thank the EGRET Science Team   for their hard work
for collecting and calibrating the data and the NASA for their
support in making satellite data publicly available.

 This work was supported by the BMBF (Bundesministerium f\"ur Bildung und Forschung) via the DLR
(Deutsches Zentrum f\"ur Luft- und Raumfahrt), a grant from the DFG
(Deutsche Forschungsgemeinschaft, Grant 436 RUS 113/626/0-1), RFBR
(Russian Foundation for Basic Research, Grant 05-02-17603), and the
Heisenberg-Landau Program.

\bibliographystyle{aa}

\appendix

\section{Fits to the EGRET data for 180 independent  sky directions.} \label{allregions}

The fits of the shapes of the gamma ray spectrum from the
conventional background model and DMA signal to the
EGRET data are shown for  180  regions of the sky:
the longitude
distributions are split into bins of 8$^\circ$ for four different
latitude ranges (absolute values of latitude: 0-5$^\circ$,
5-10$^\circ$, 10-20$^\circ$, 20-90$^\circ$), so one has 4x45=180
angular bins.
 The background scalings
and the boost factors are free parameters. The contribution from the extragalactic background, as
determined by \citet{sander}, and the spectra of the subtracted EGRET point sources
from \citet{egret_cat} are
shown as well.
\newpage
  \begin{center}
    \framebox[0.21\textwidth][c]{$\vert \mbox{lat}\vert<5^\circ$}
    \framebox[0.21\textwidth][c]{$5^\circ<\vert \mbox{lat}\vert<10^\circ$}
    \framebox[0.21\textwidth][c]{$10^\circ<\vert \mbox{lat}\vert<20^\circ$}
    \framebox[0.21\textwidth][c]{$20^\circ<\vert \mbox{lat}\vert<90^\circ$}\\
    \hspace{-1cm}
    \begin{turn}{90} \framebox[0.21\textwidth][c]{{\scriptsize $-180^\circ<\mbox{long}<-172^\circ$}} \end{turn}
    \includegraphics[width=0.21\textwidth]{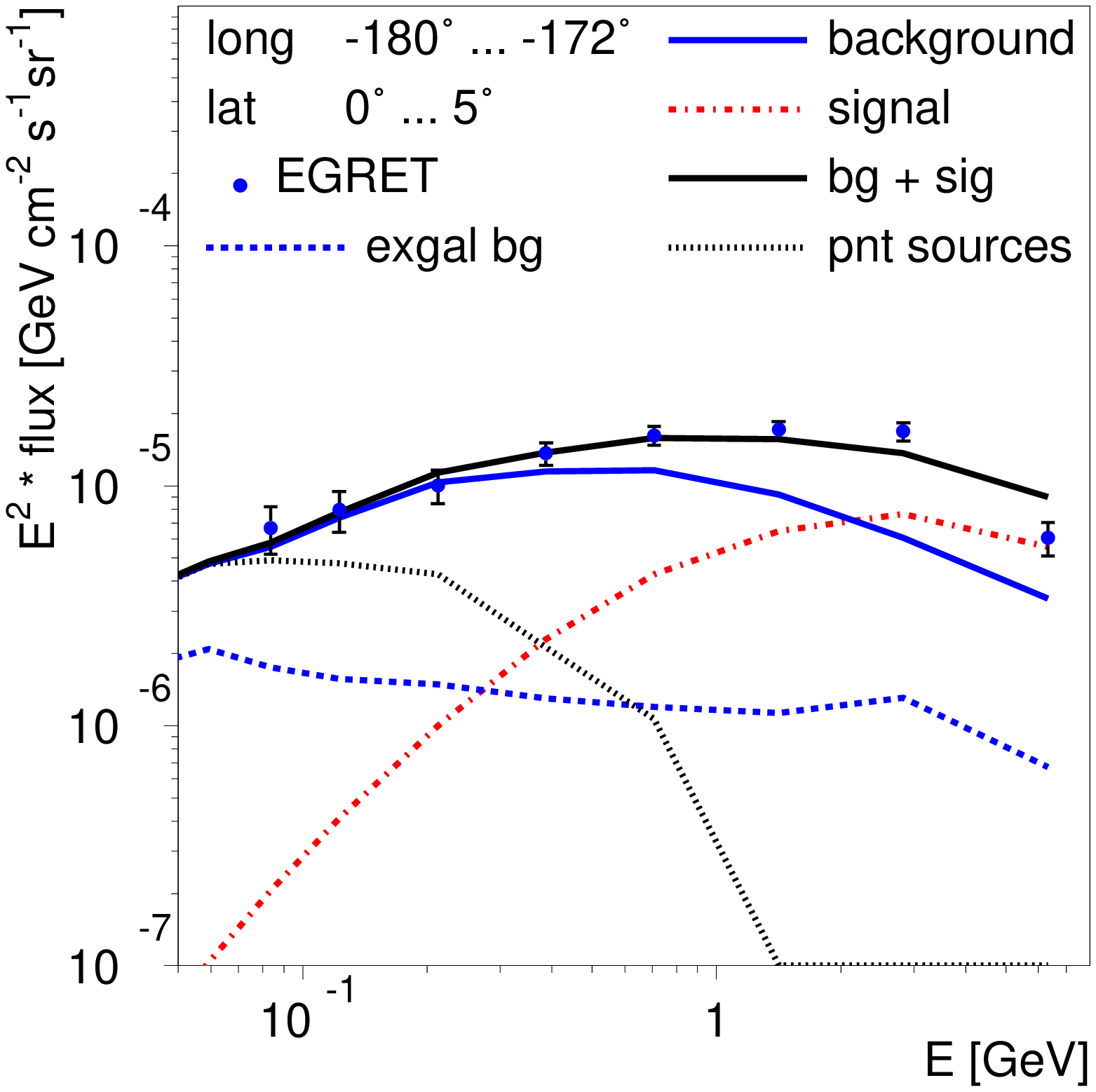}
    \includegraphics[width=0.21\textwidth]{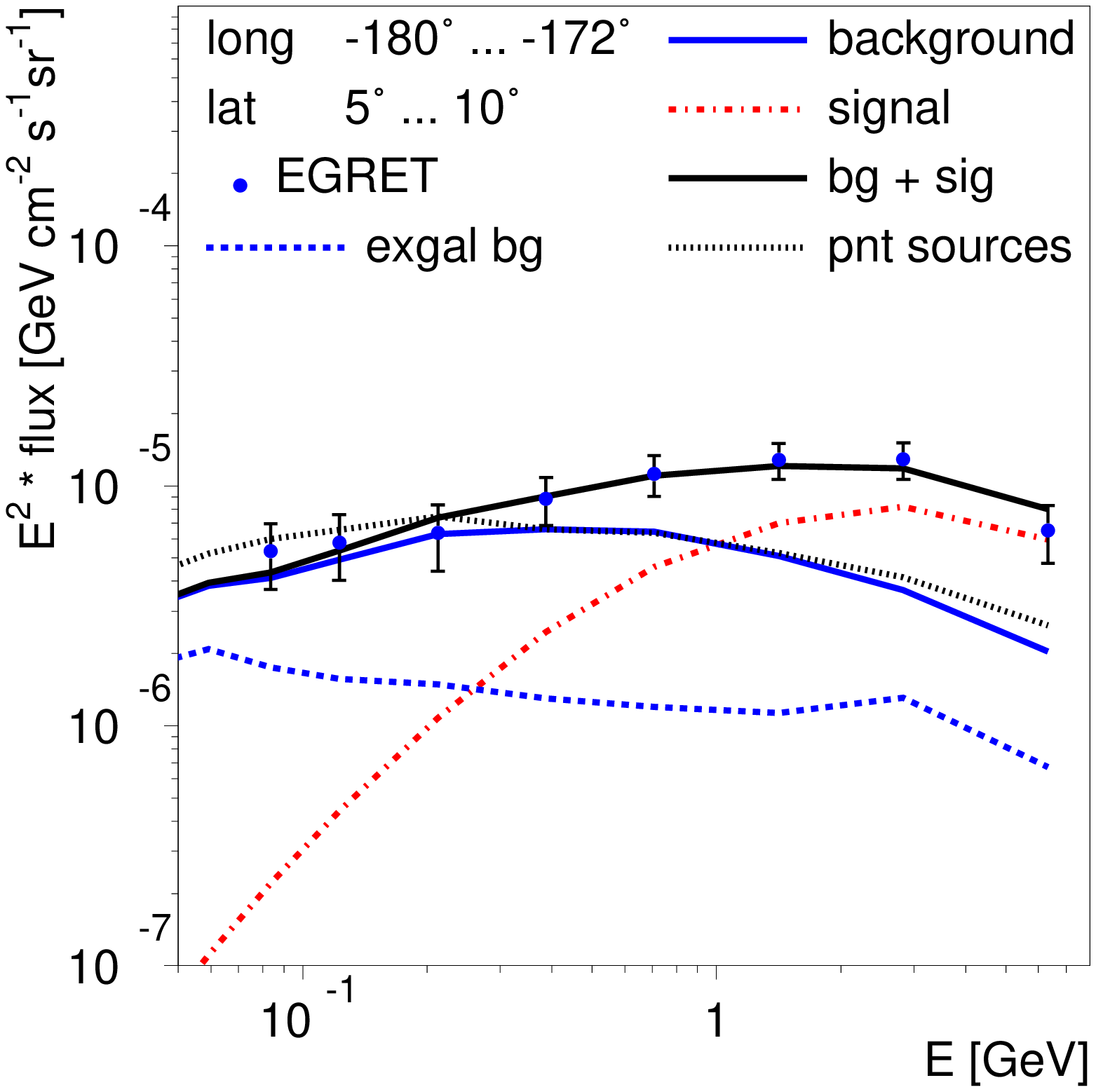}
    \includegraphics[width=0.21\textwidth]{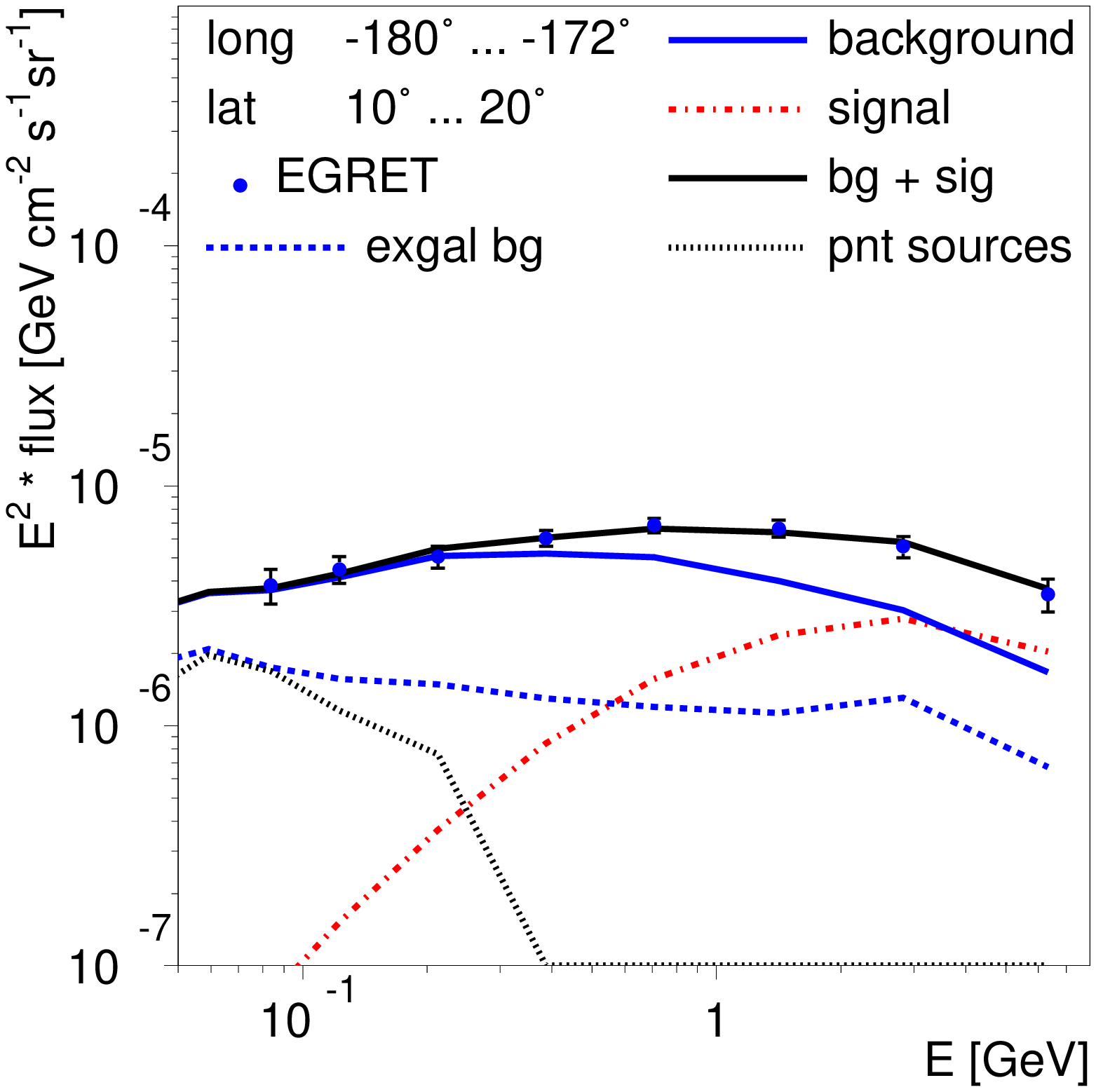}
    \includegraphics[width=0.21\textwidth]{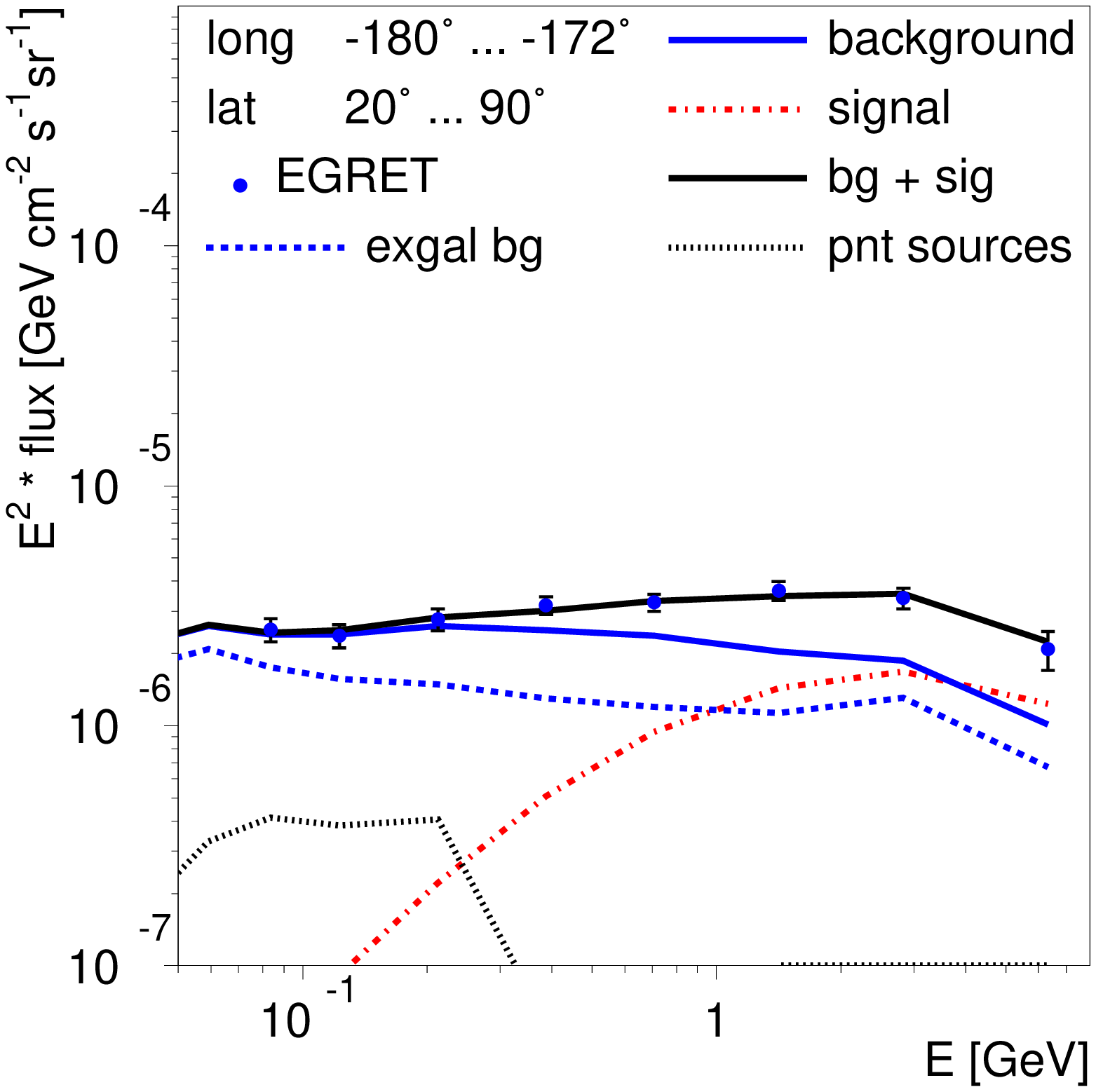}\\
    \hspace{-1cm}
    \begin{turn}{90} \framebox[0.21\textwidth][c]{{\scriptsize $-172^\circ<\mbox{long}<-164^\circ$}} \end{turn}
    \includegraphics[width=0.21\textwidth]{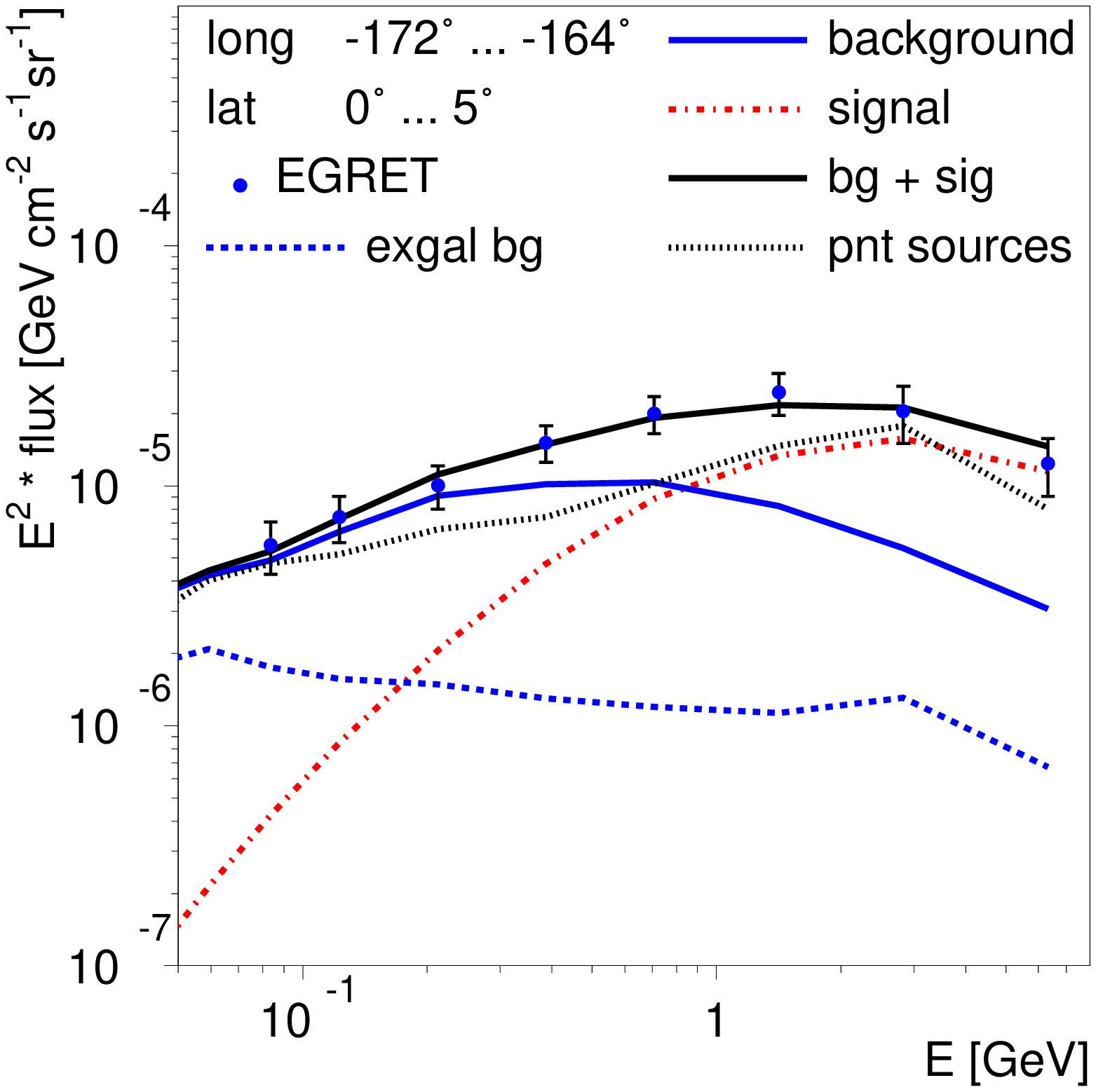}
    \includegraphics[width=0.21\textwidth]{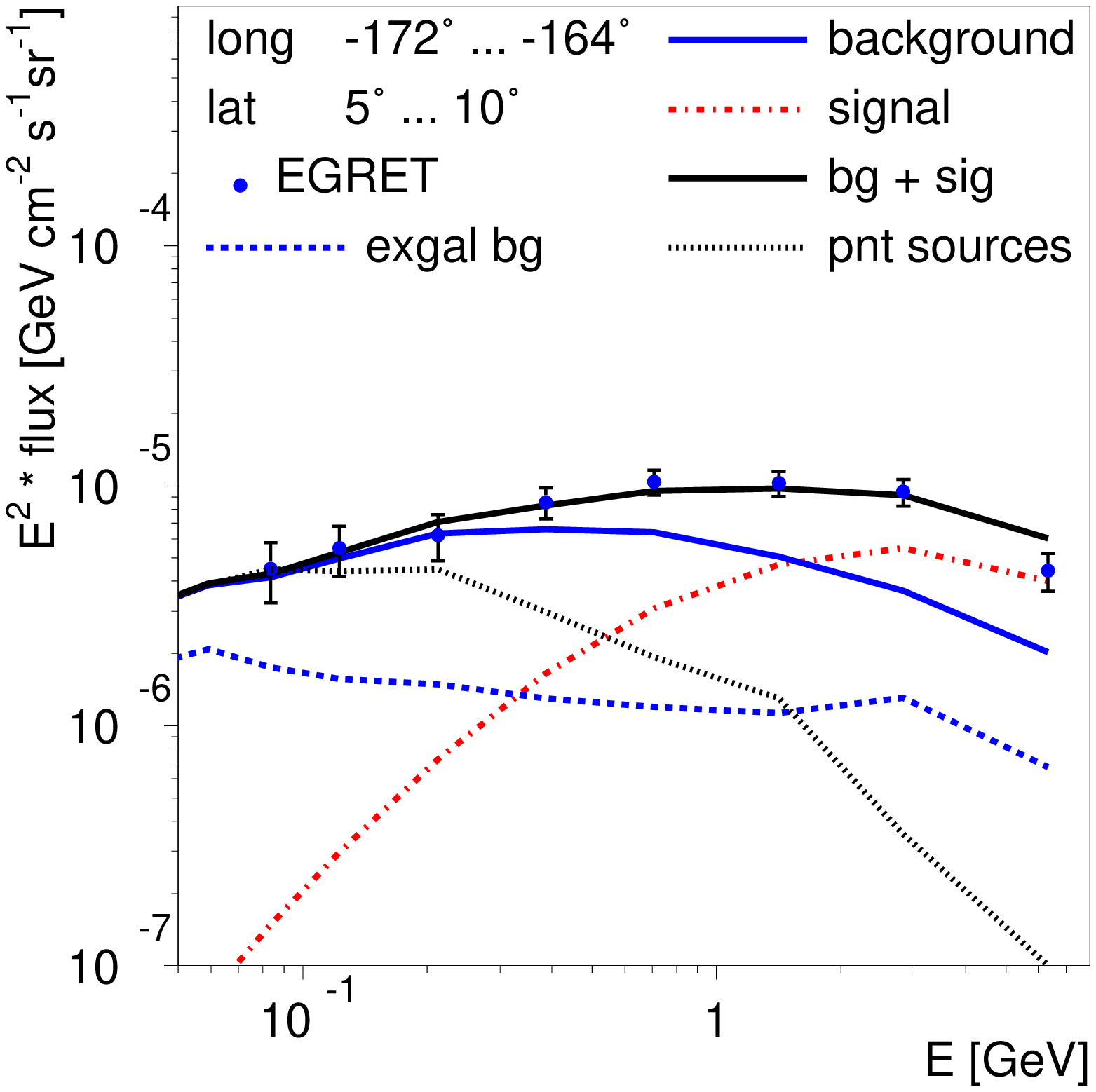}
    \includegraphics[width=0.21\textwidth]{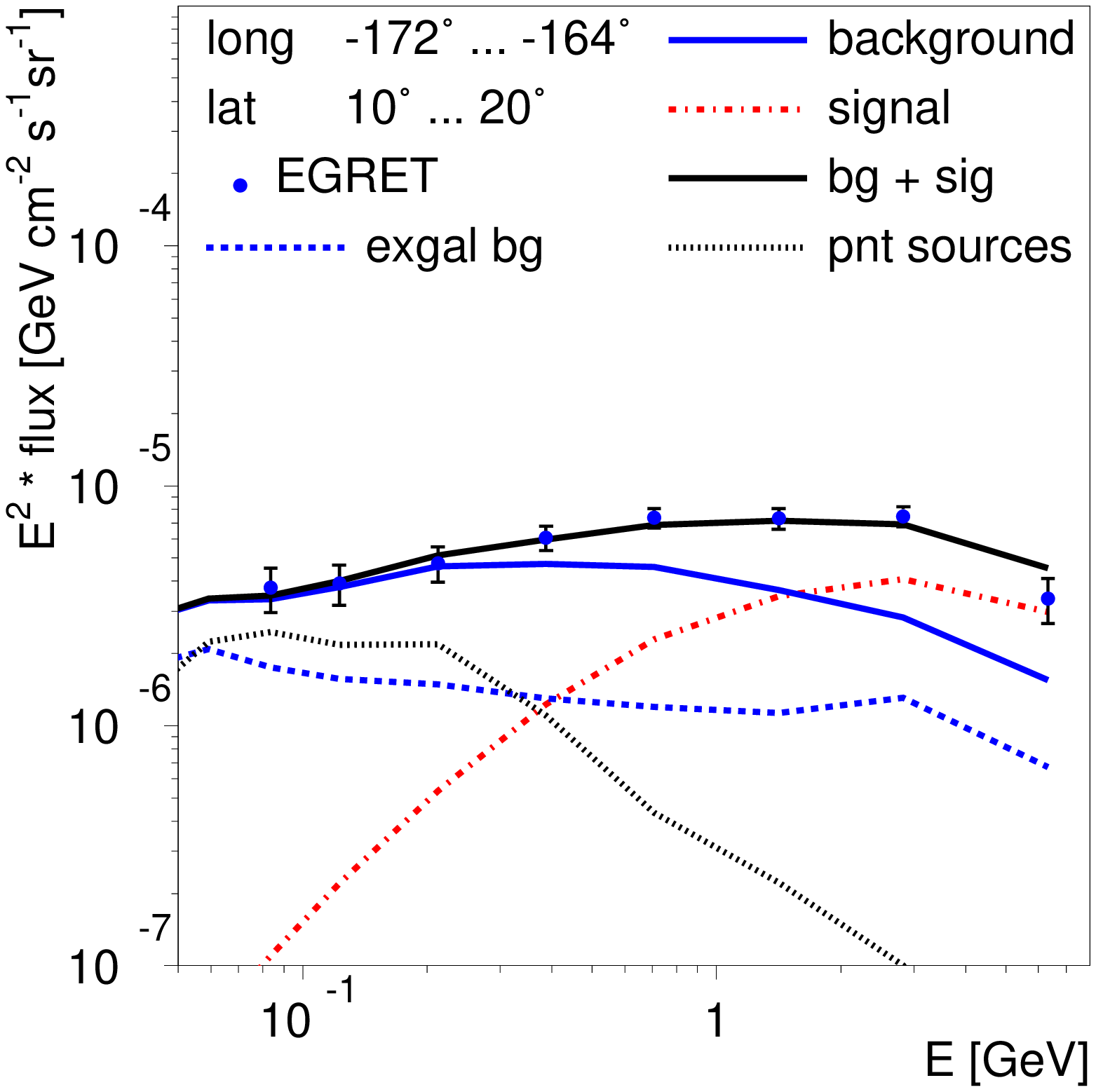}
    \includegraphics[width=0.21\textwidth]{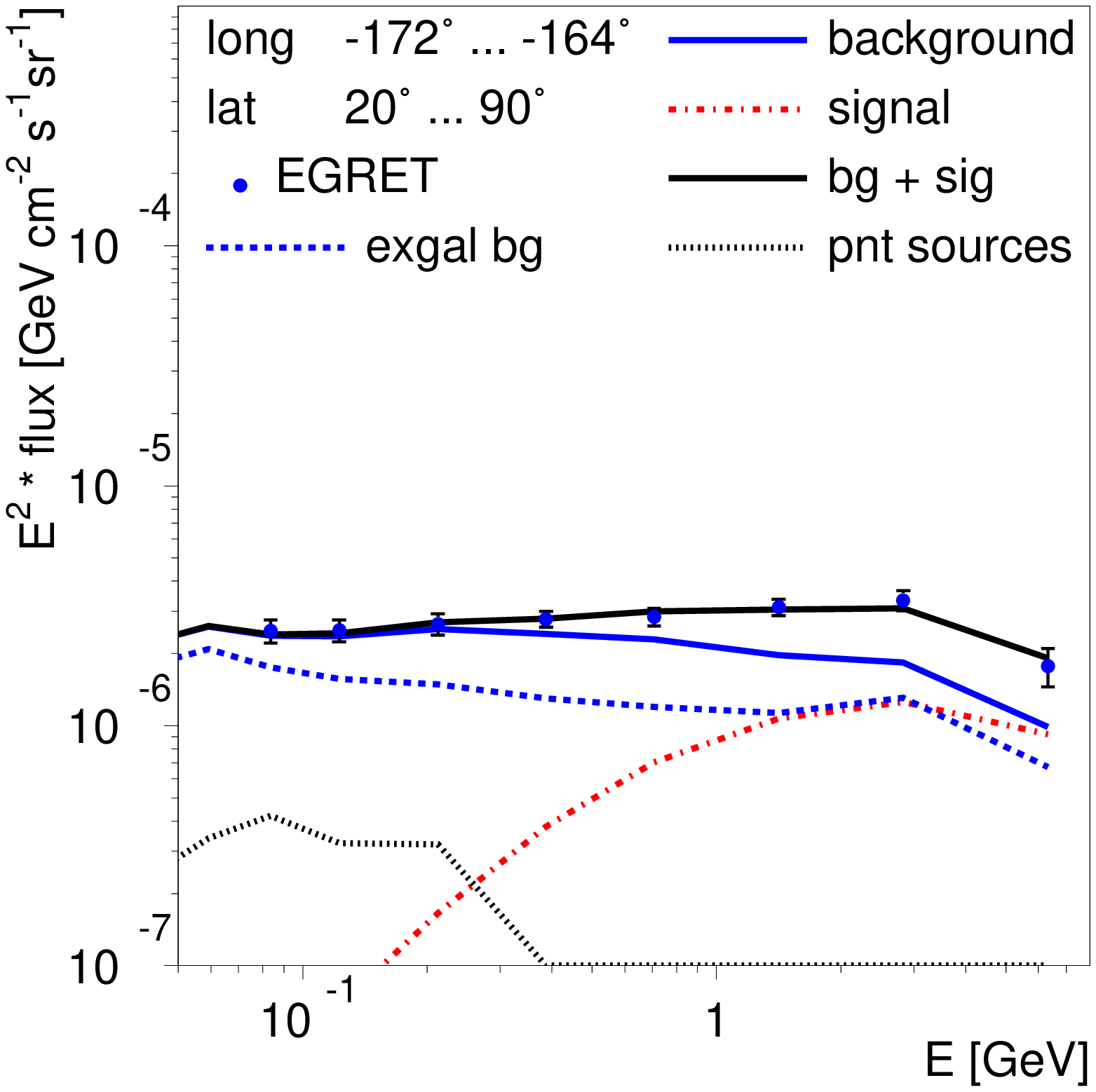}\\
    \hspace{-1cm}
    \begin{turn}{90} \framebox[0.21\textwidth][c]{{\scriptsize $-164^\circ<\mbox{long}<-156^\circ$}} \end{turn}
    \includegraphics[width=0.21\textwidth]{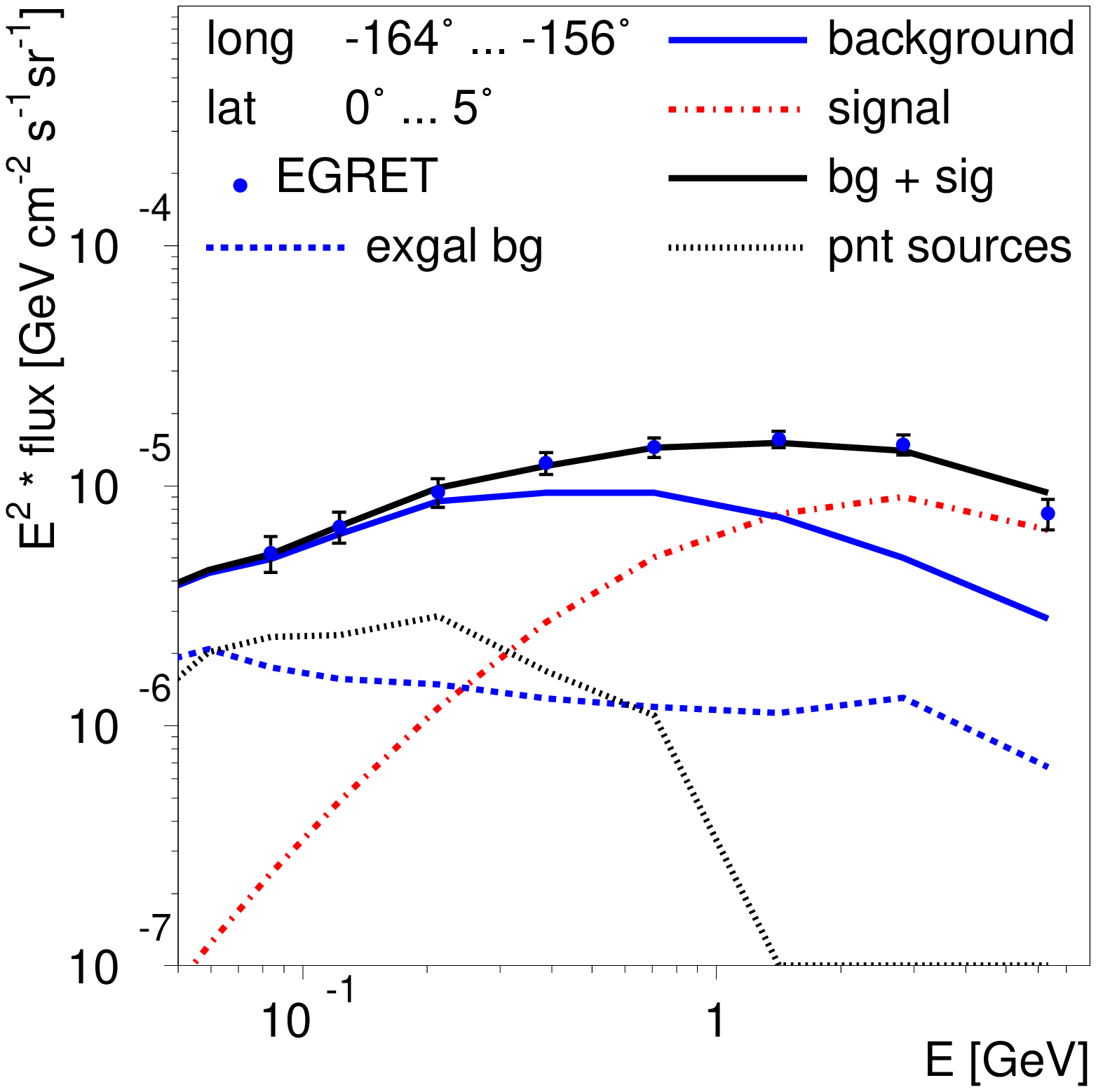}
    \includegraphics[width=0.21\textwidth]{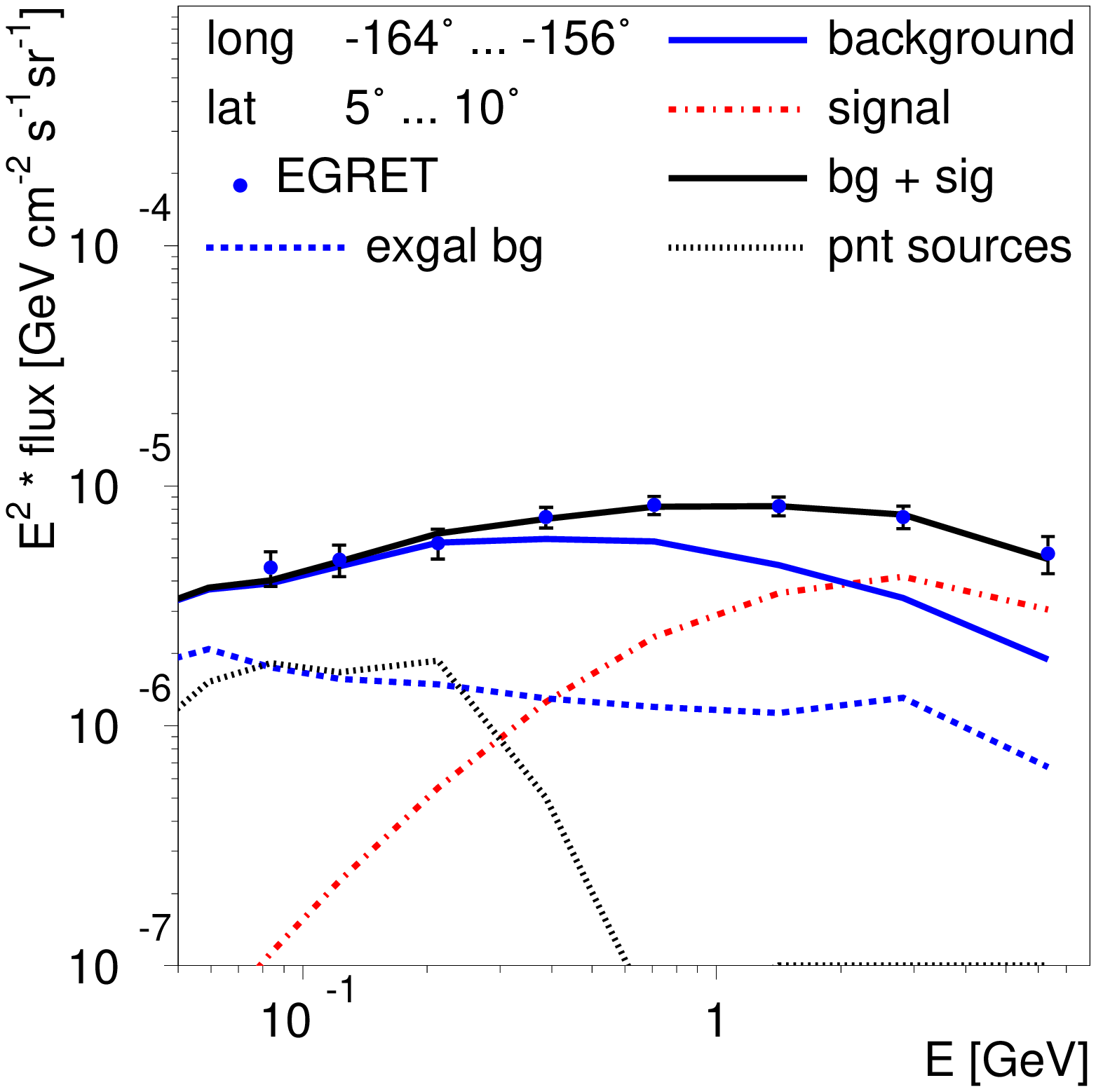}
    \includegraphics[width=0.21\textwidth]{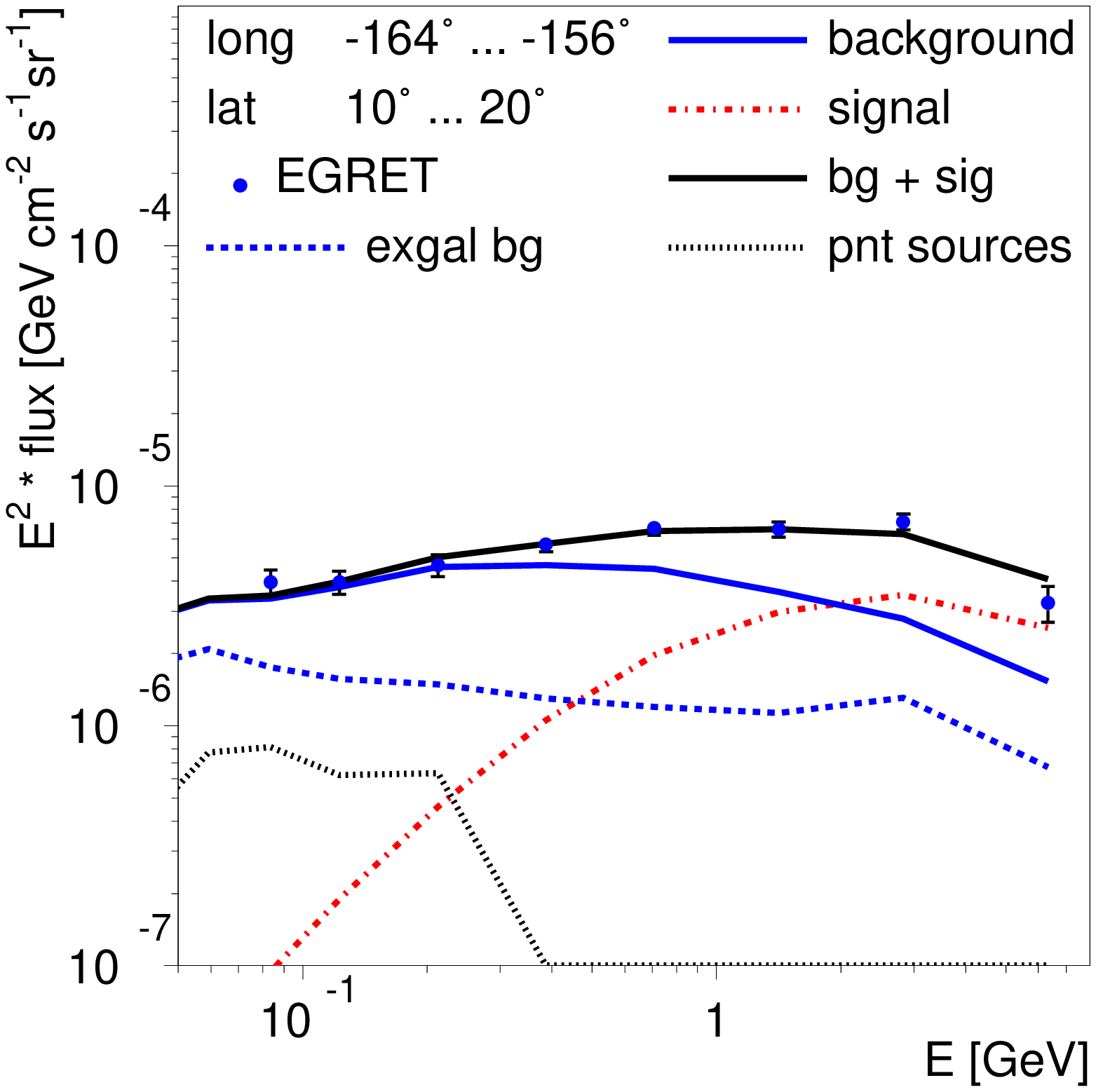}
    \includegraphics[width=0.21\textwidth]{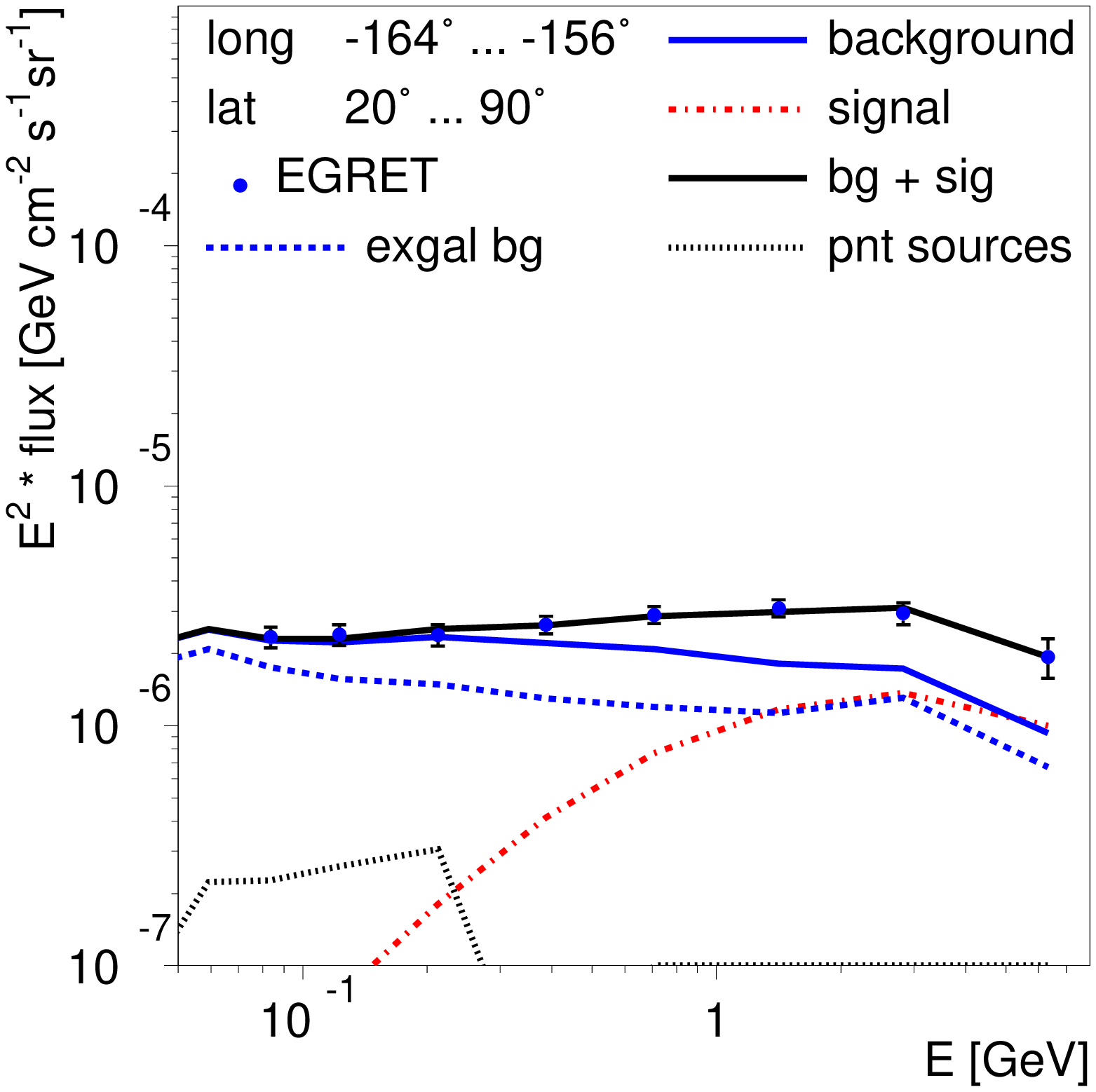}\\
    \hspace{-1cm}
    \begin{turn}{90} \framebox[0.21\textwidth][c]{{\scriptsize $-156^\circ<\mbox{long}<-148^\circ$}} \end{turn}
    \includegraphics[width=0.21\textwidth]{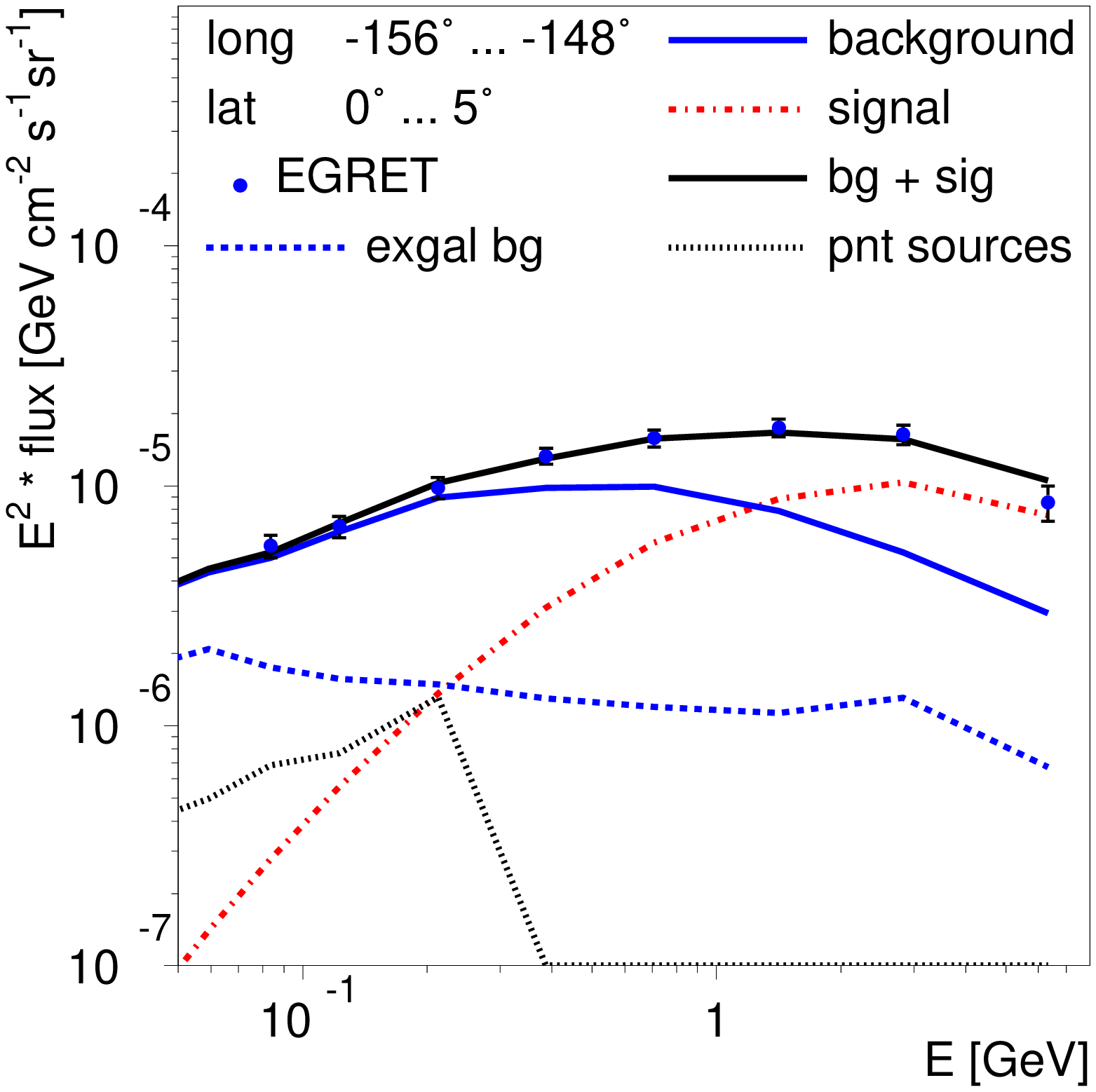}
    \includegraphics[width=0.21\textwidth]{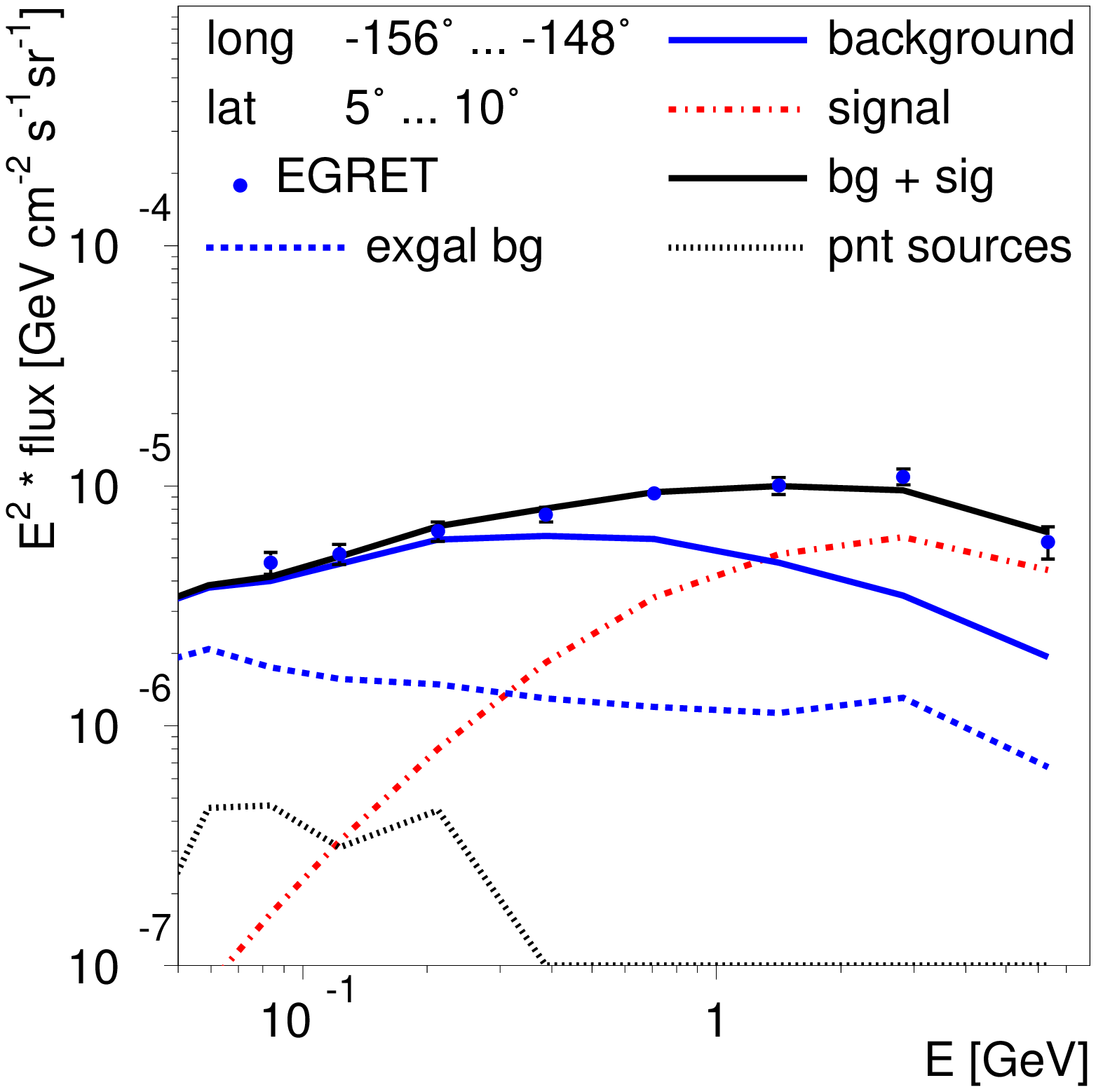}
    \includegraphics[width=0.21\textwidth]{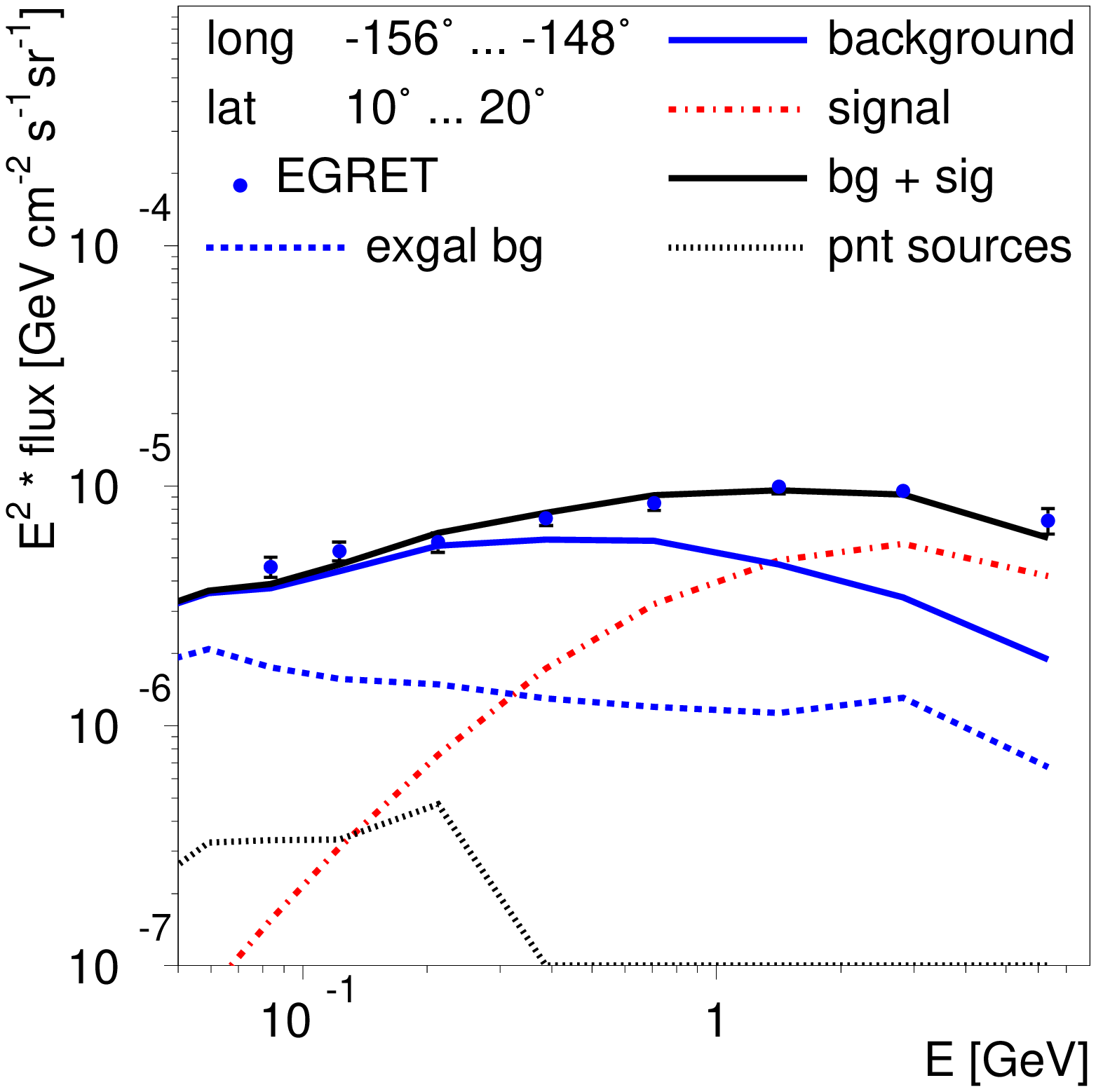}
    \includegraphics[width=0.21\textwidth]{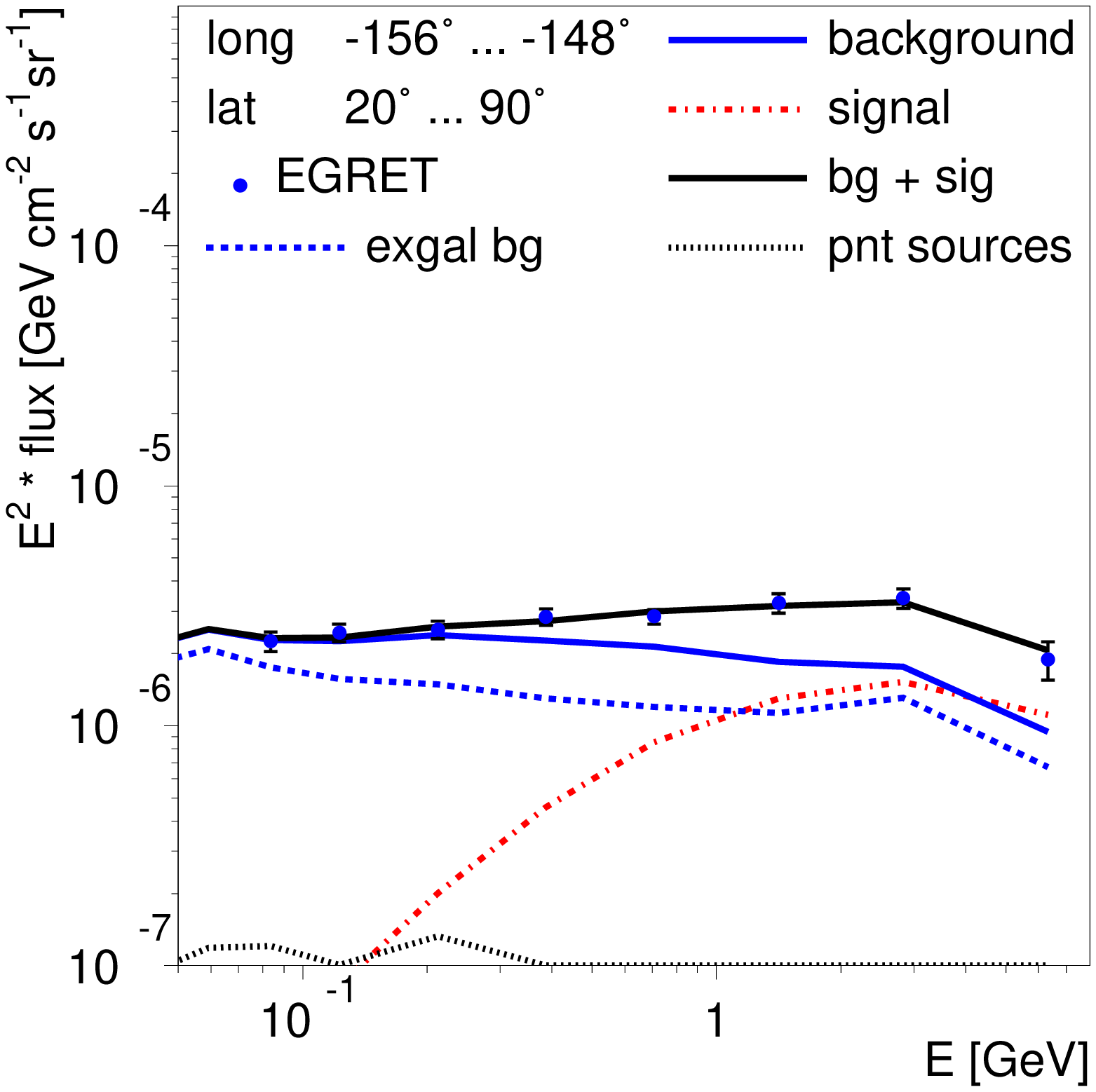}\\
    \hspace{-1cm}
    \begin{turn}{90} \framebox[0.21\textwidth][c]{{\scriptsize $-148^\circ<\mbox{long}<-140^\circ$}} \end{turn}
    \includegraphics[width=0.21\textwidth]{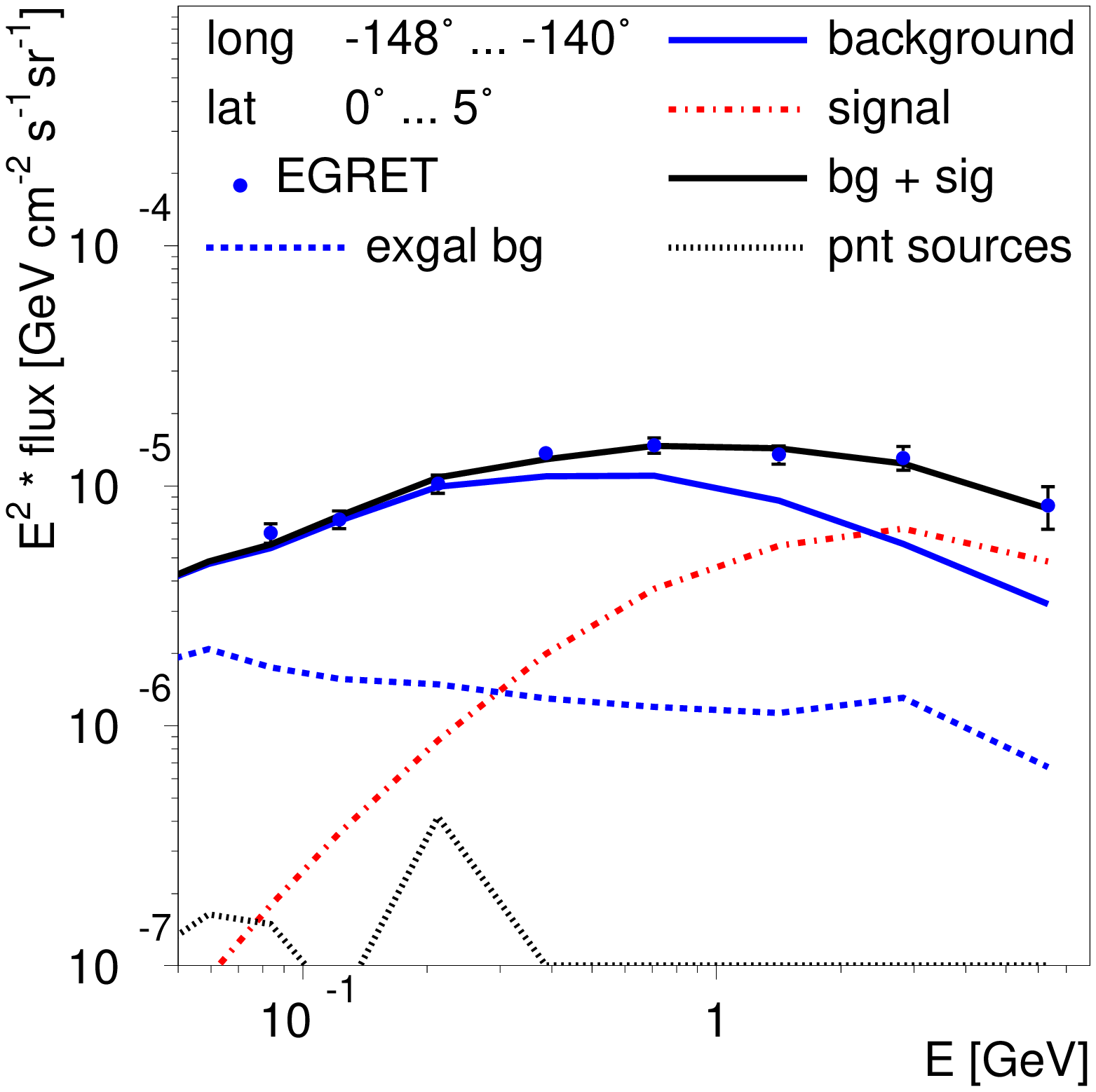}
    \includegraphics[width=0.21\textwidth]{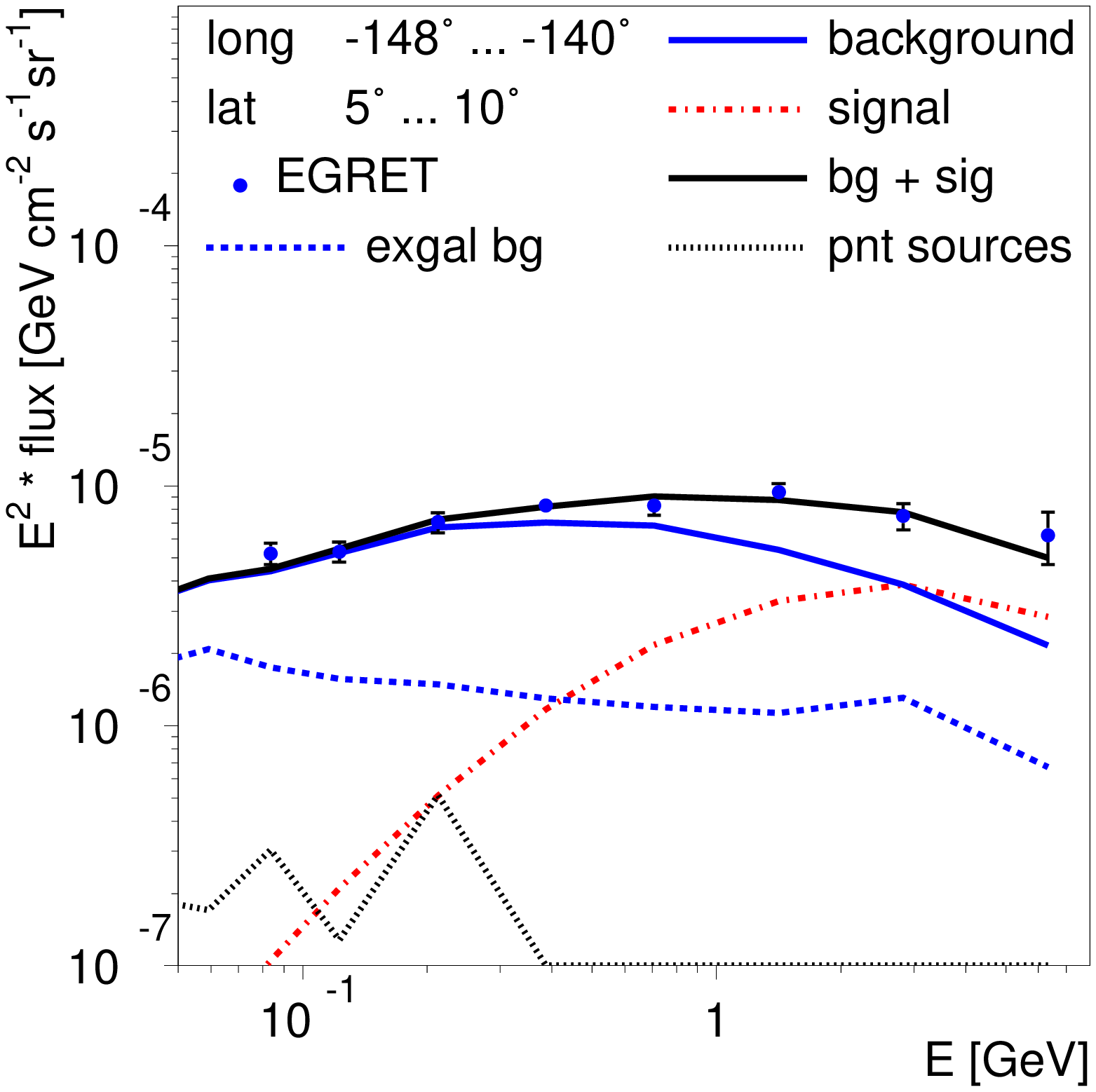}
    \includegraphics[width=0.21\textwidth]{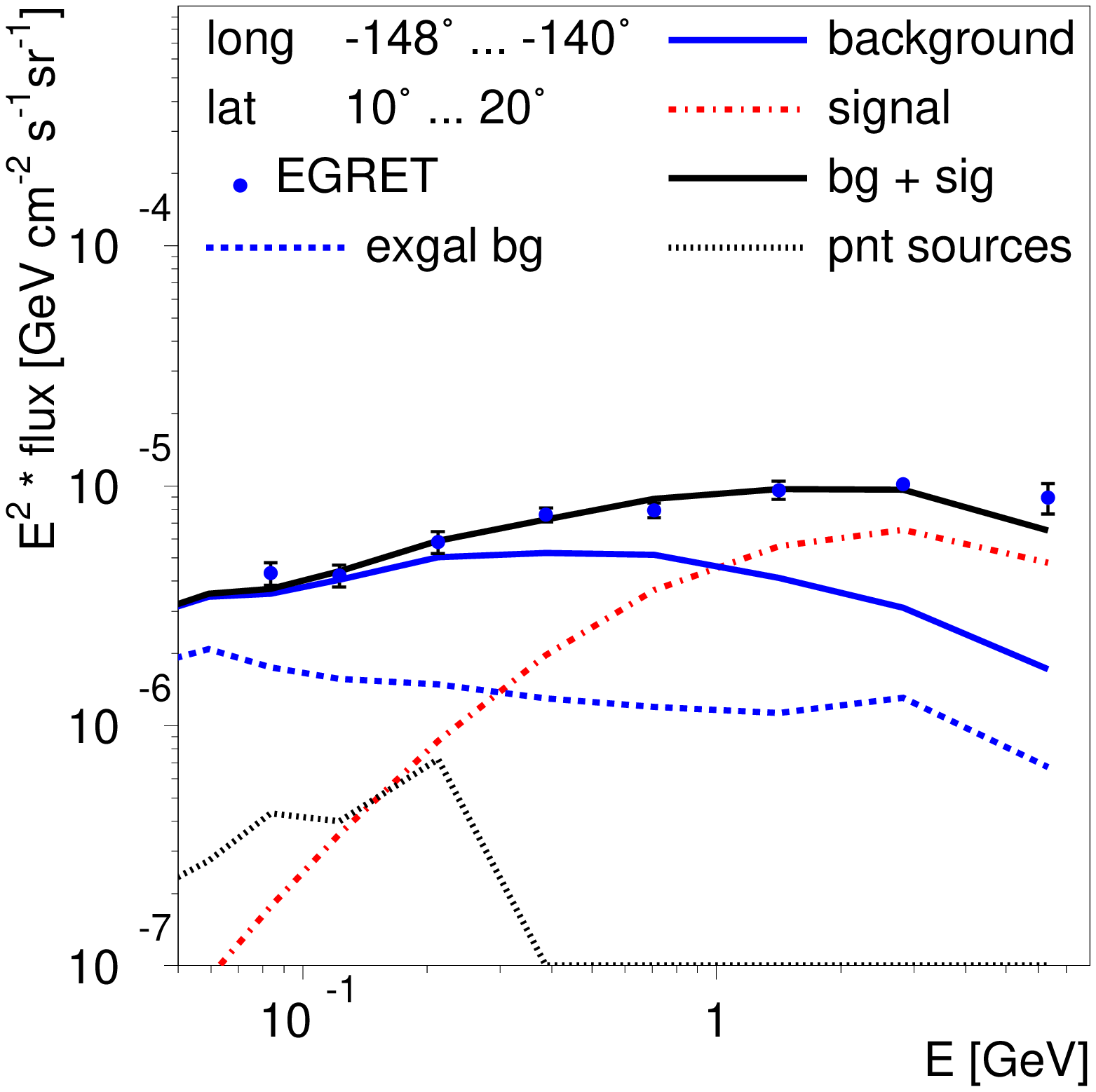}
    \includegraphics[width=0.21\textwidth]{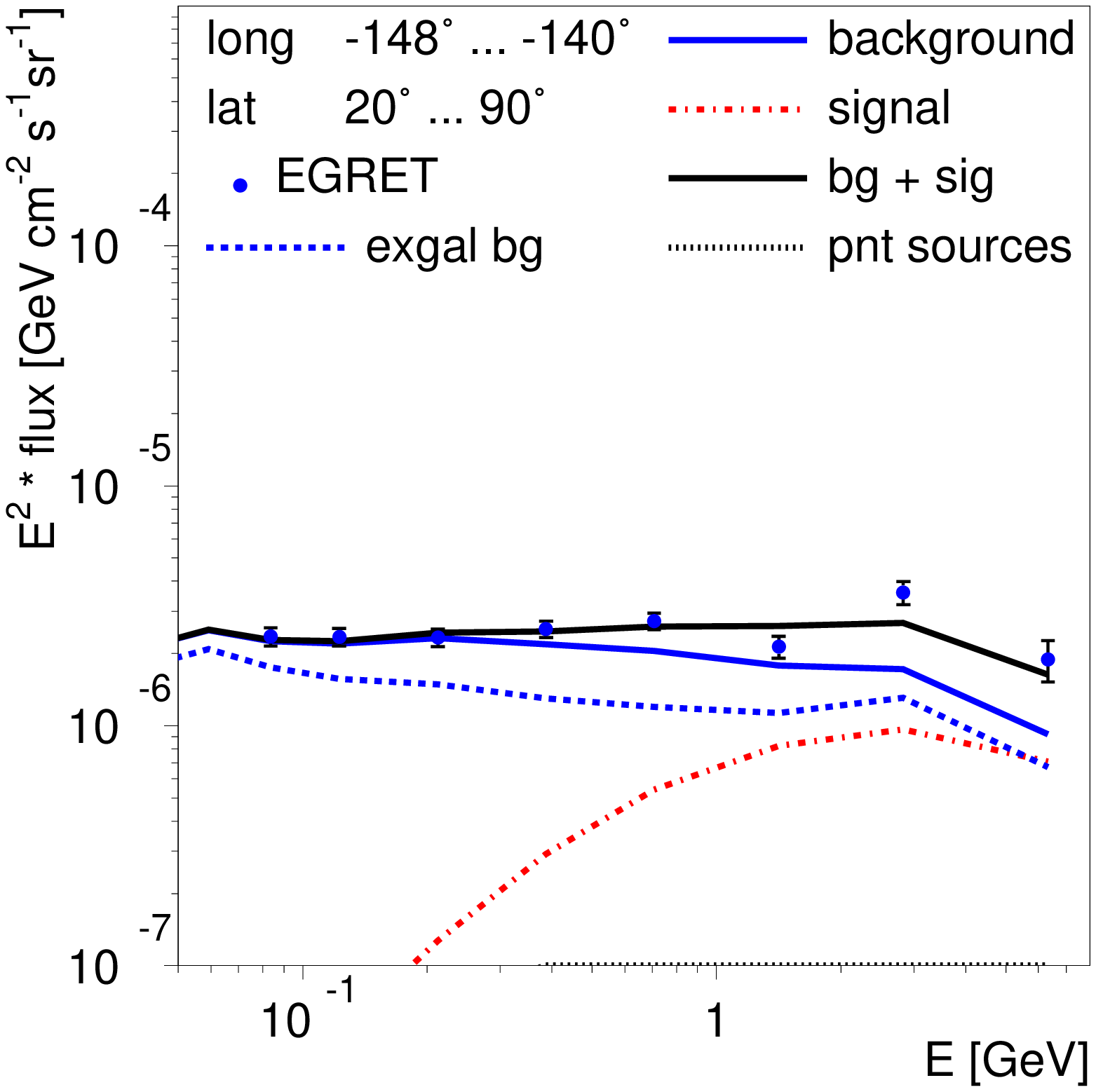}\\
    \hspace{-1cm}
    \begin{turn}{90} \framebox[0.21\textwidth][c]{{\scriptsize $-140^\circ<\mbox{long}<-132^\circ$}} \end{turn}
    \includegraphics[width=0.21\textwidth]{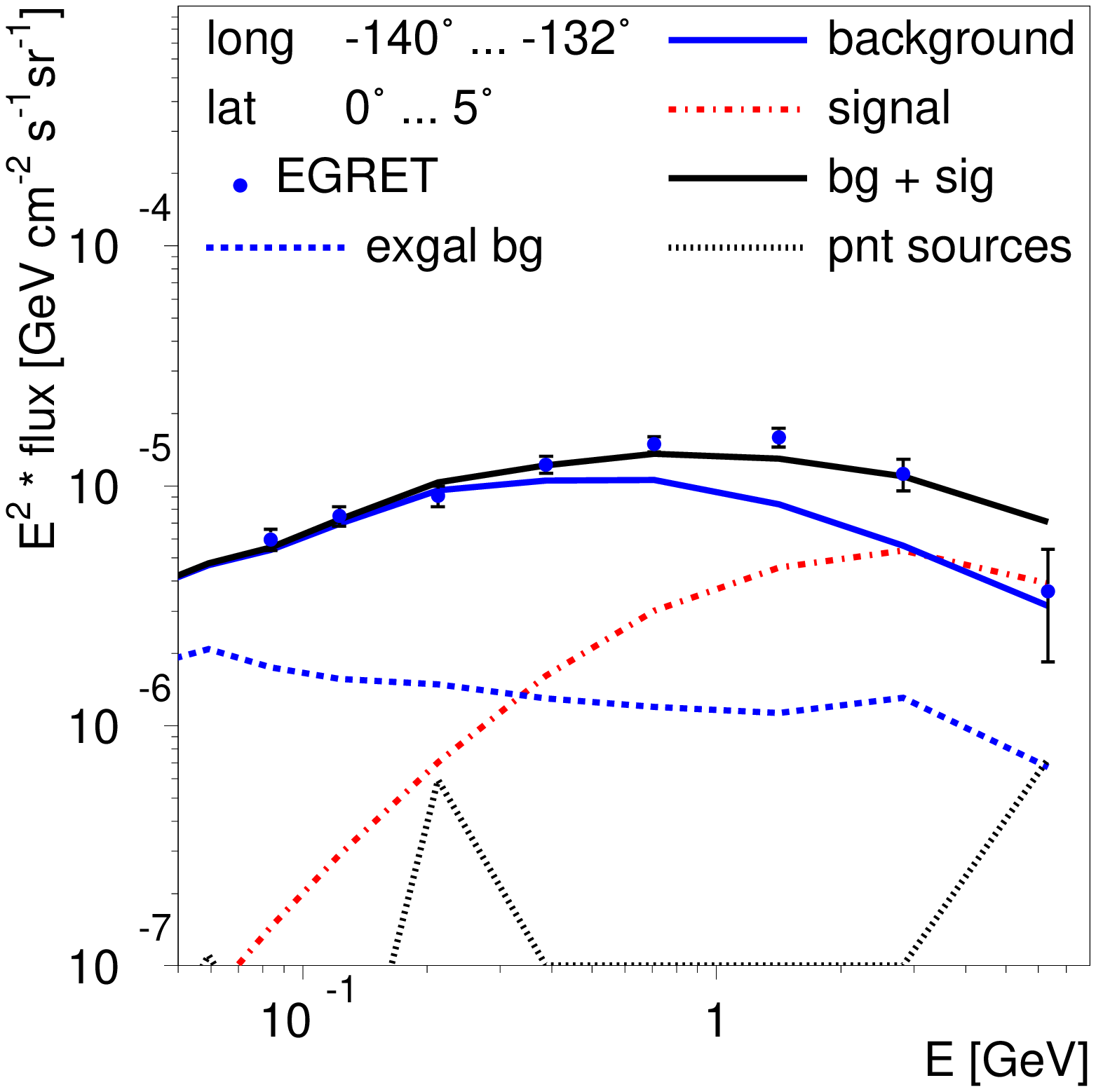}
    \includegraphics[width=0.21\textwidth]{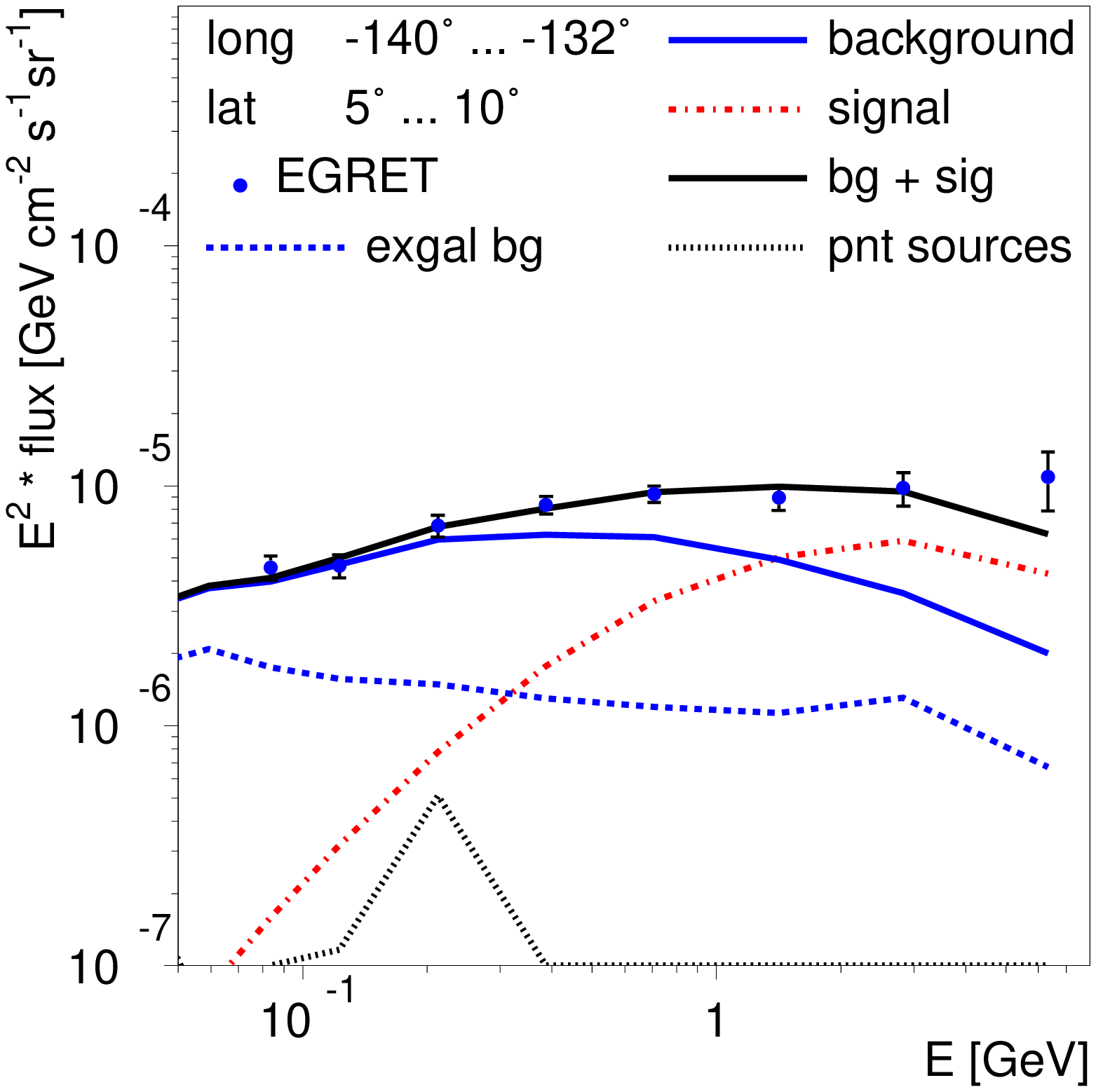}
    \includegraphics[width=0.21\textwidth]{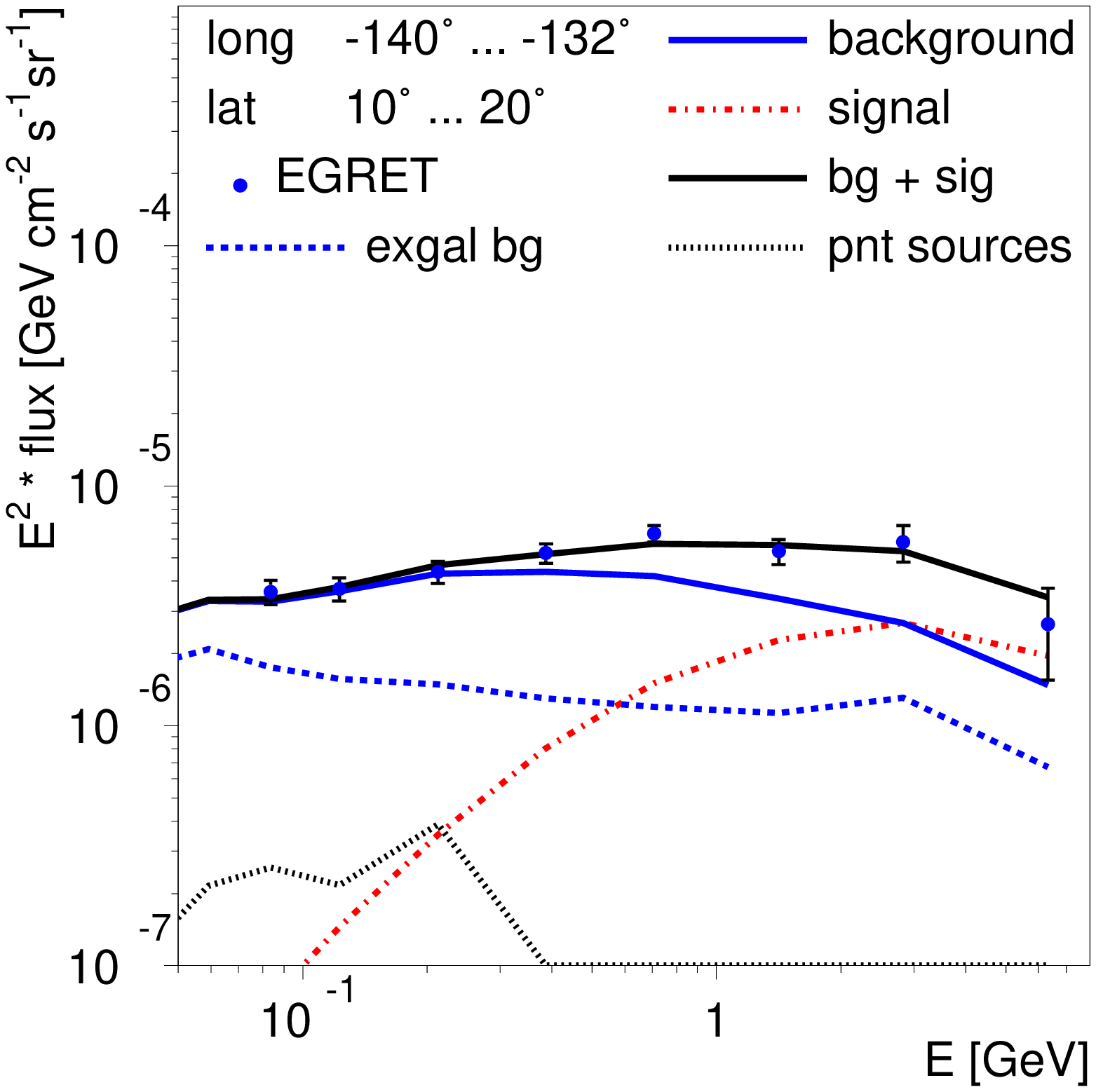}
    \includegraphics[width=0.21\textwidth]{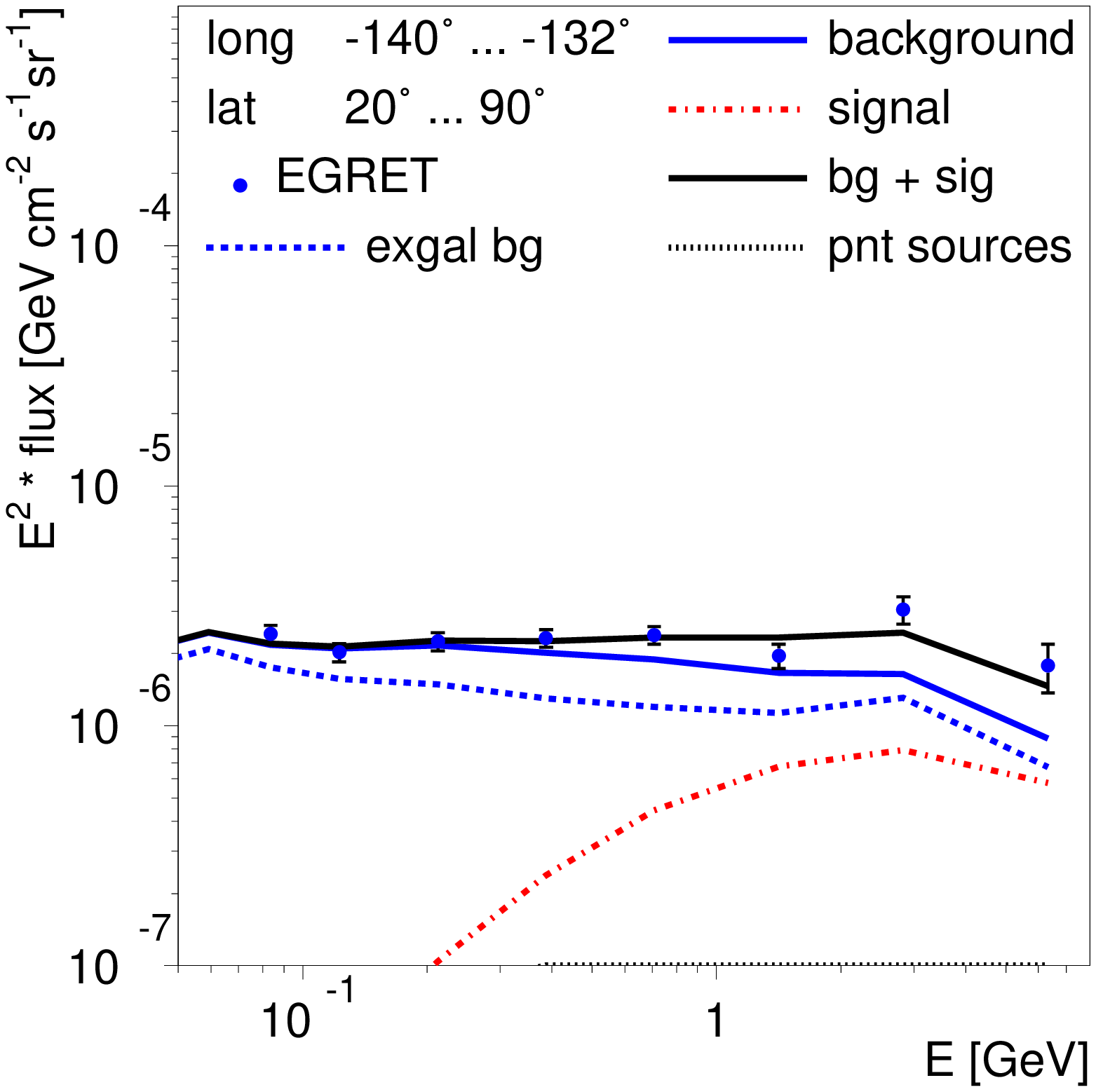}\\
  \end{center}
  \clearpage
  \begin{center}
    \framebox[0.21\textwidth][c]{$\vert \mbox{lat}\vert<5^\circ$}
    \framebox[0.21\textwidth][c]{$5^\circ<\vert \mbox{lat}\vert<10^\circ$}
    \framebox[0.21\textwidth][c]{$10^\circ<\vert \mbox{lat}\vert<20^\circ$}
    \framebox[0.21\textwidth][c]{$20^\circ<\vert \mbox{lat}\vert<90^\circ$}\\
    \hspace{-1cm}
    \begin{turn}{90} \framebox[0.21\textwidth][c]{{\scriptsize $-132^\circ<\mbox{long}<-124^\circ$}} \end{turn}
    \includegraphics[width=0.21\textwidth]{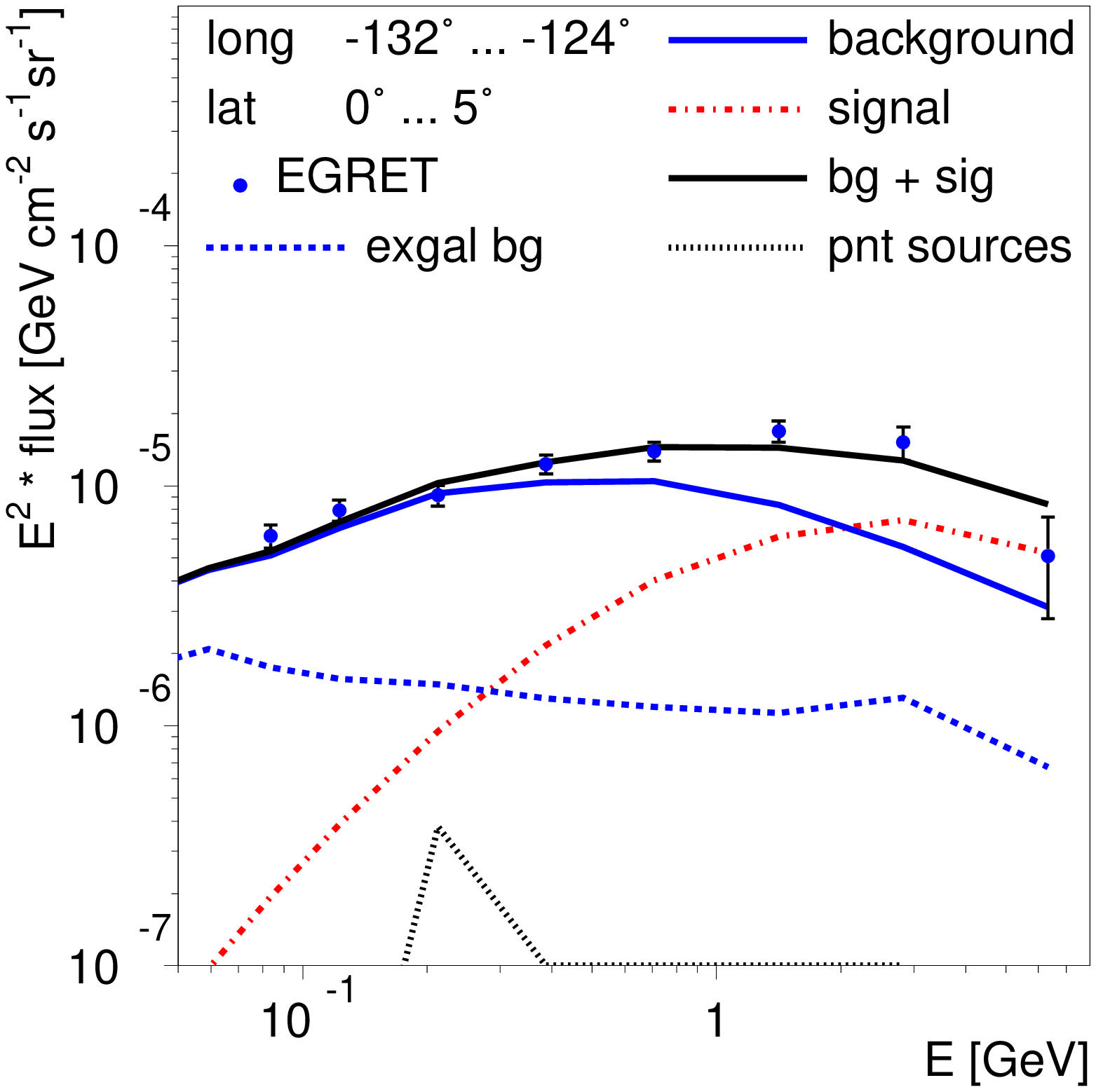}
    \includegraphics[width=0.21\textwidth]{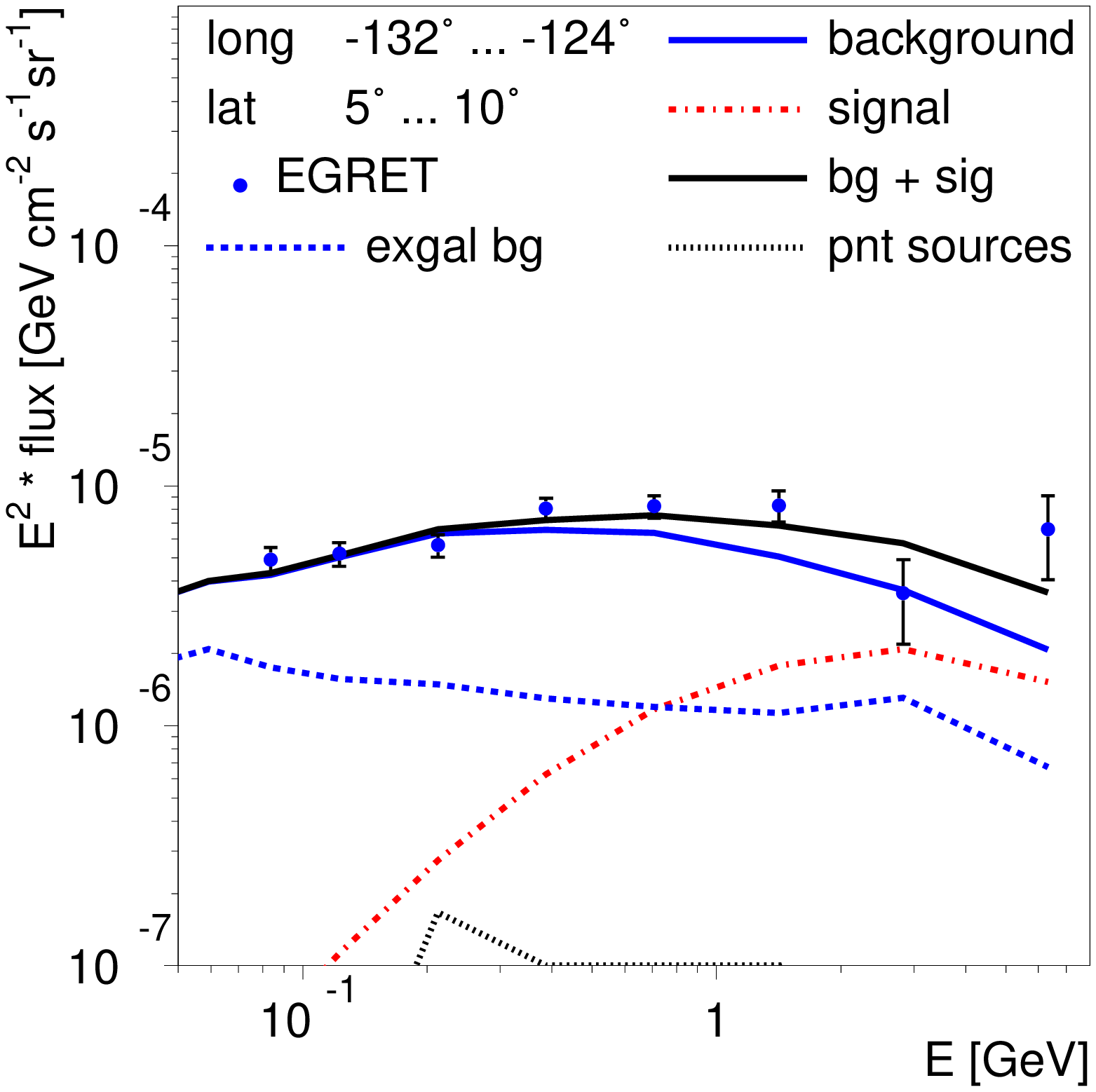}
    \includegraphics[width=0.21\textwidth]{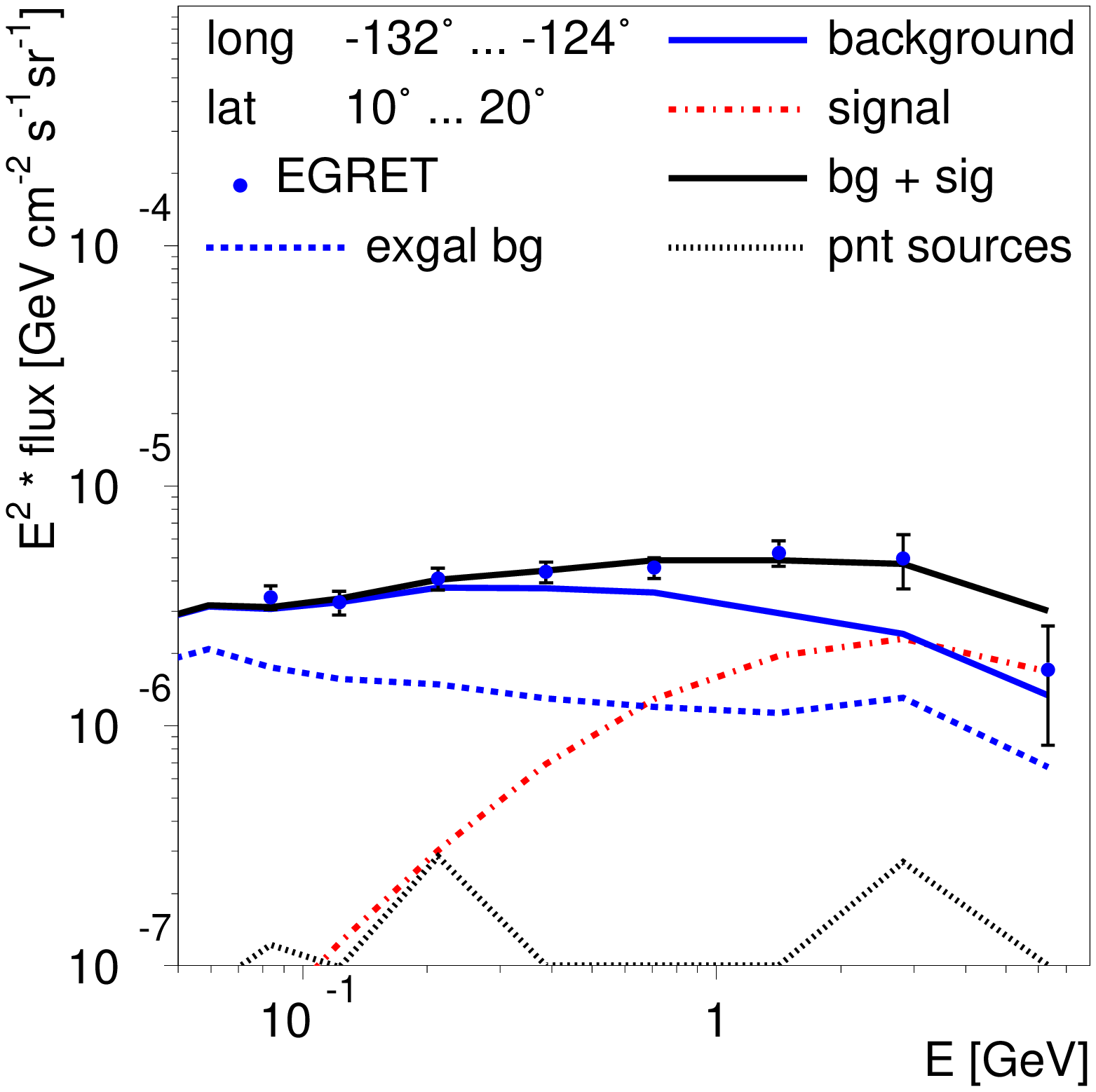}
    \includegraphics[width=0.21\textwidth]{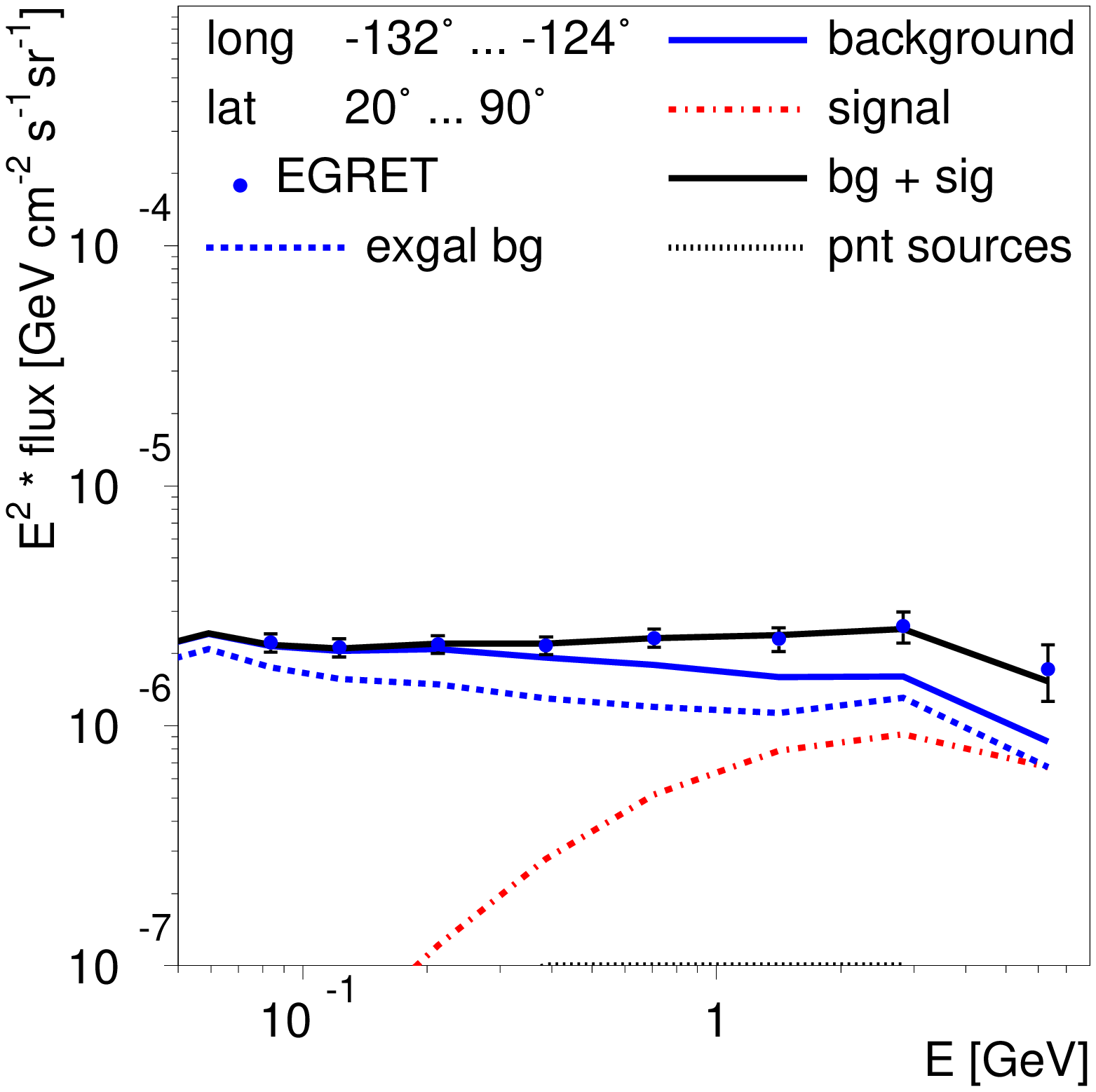}\\
    \hspace{-1cm}
    \begin{turn}{90} \framebox[0.21\textwidth][c]{{\scriptsize $-124^\circ<\mbox{long}<-116^\circ$}} \end{turn}
    \includegraphics[width=0.21\textwidth]{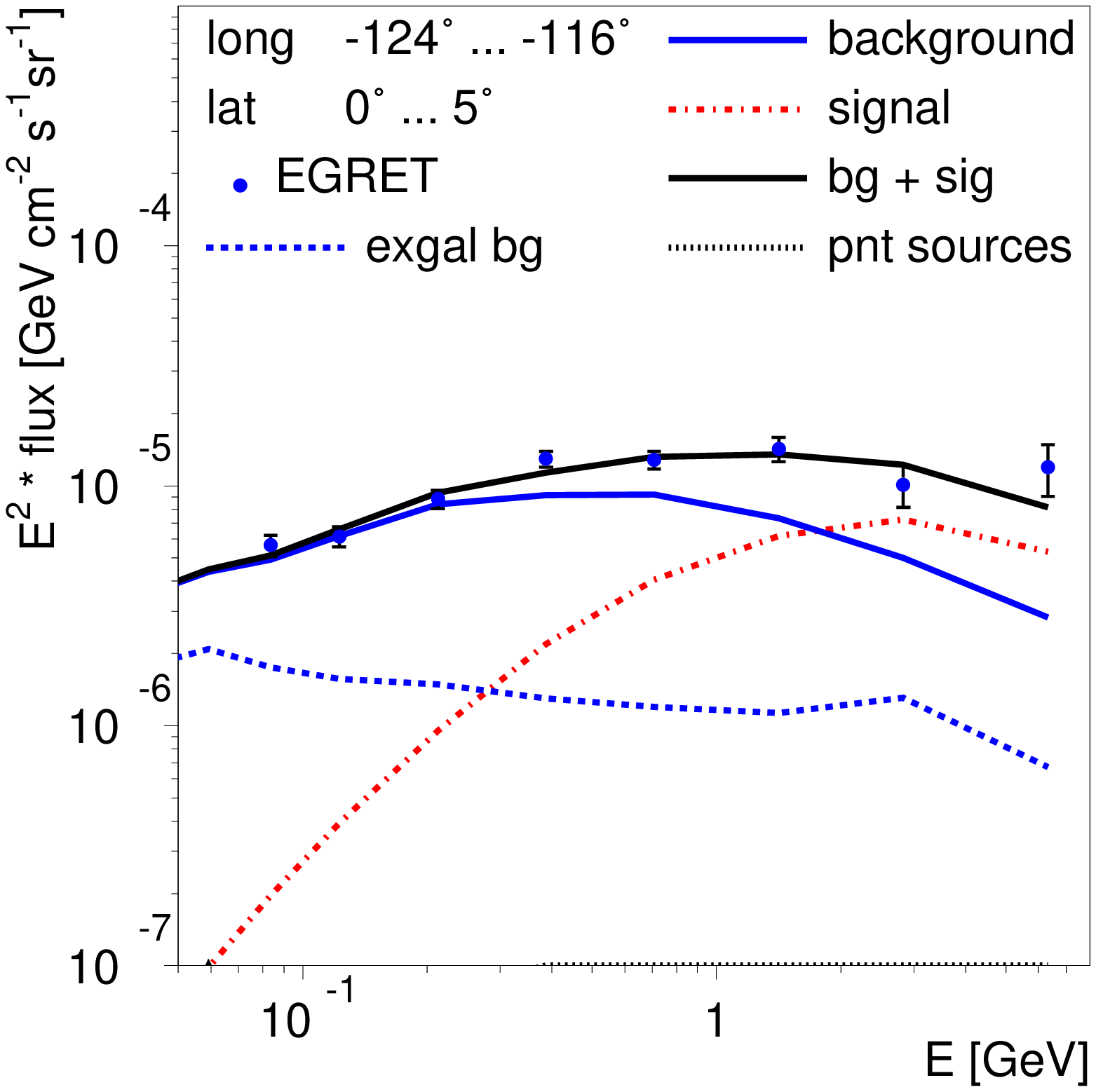}
    \includegraphics[width=0.21\textwidth]{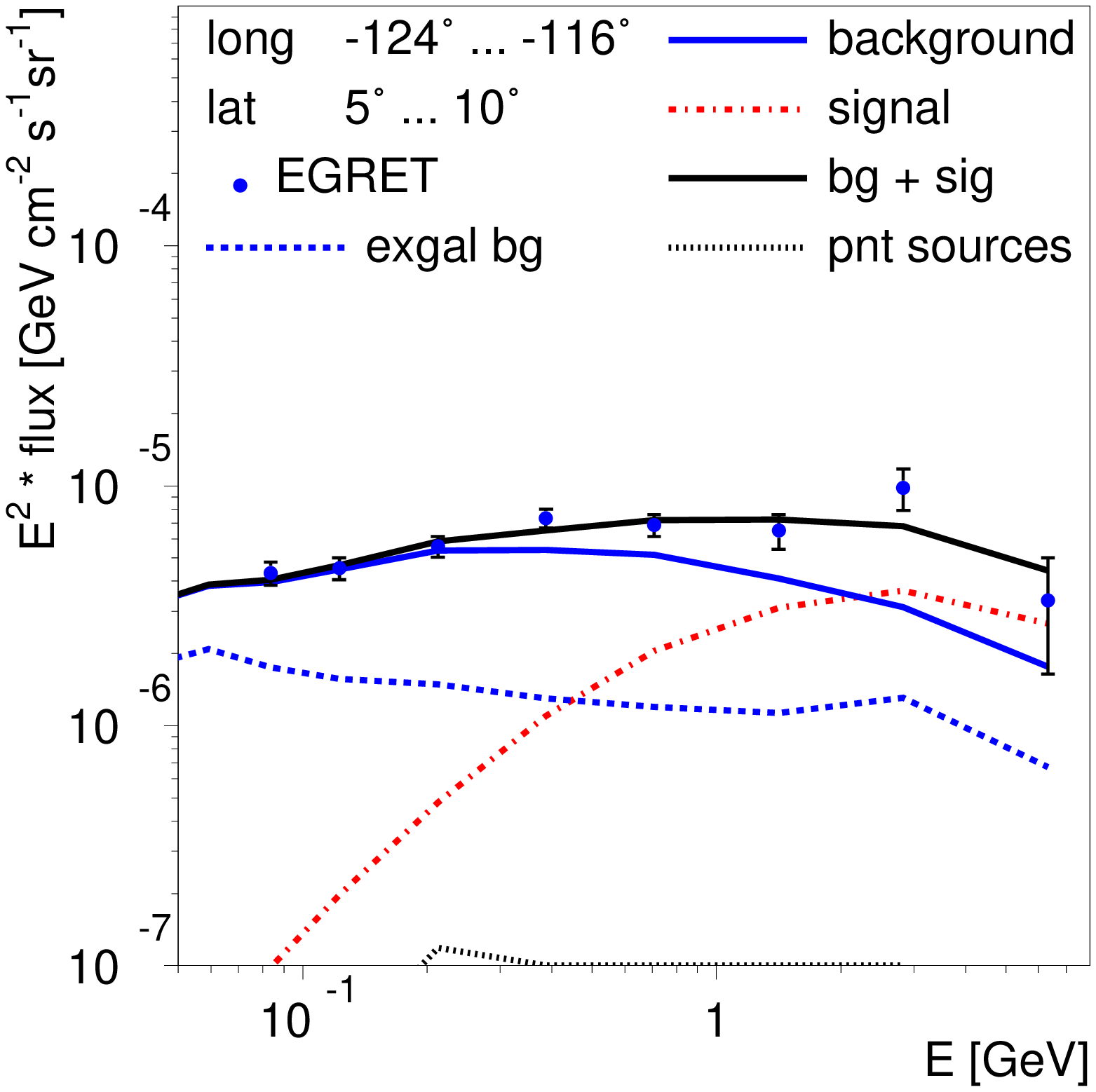}
    \includegraphics[width=0.21\textwidth]{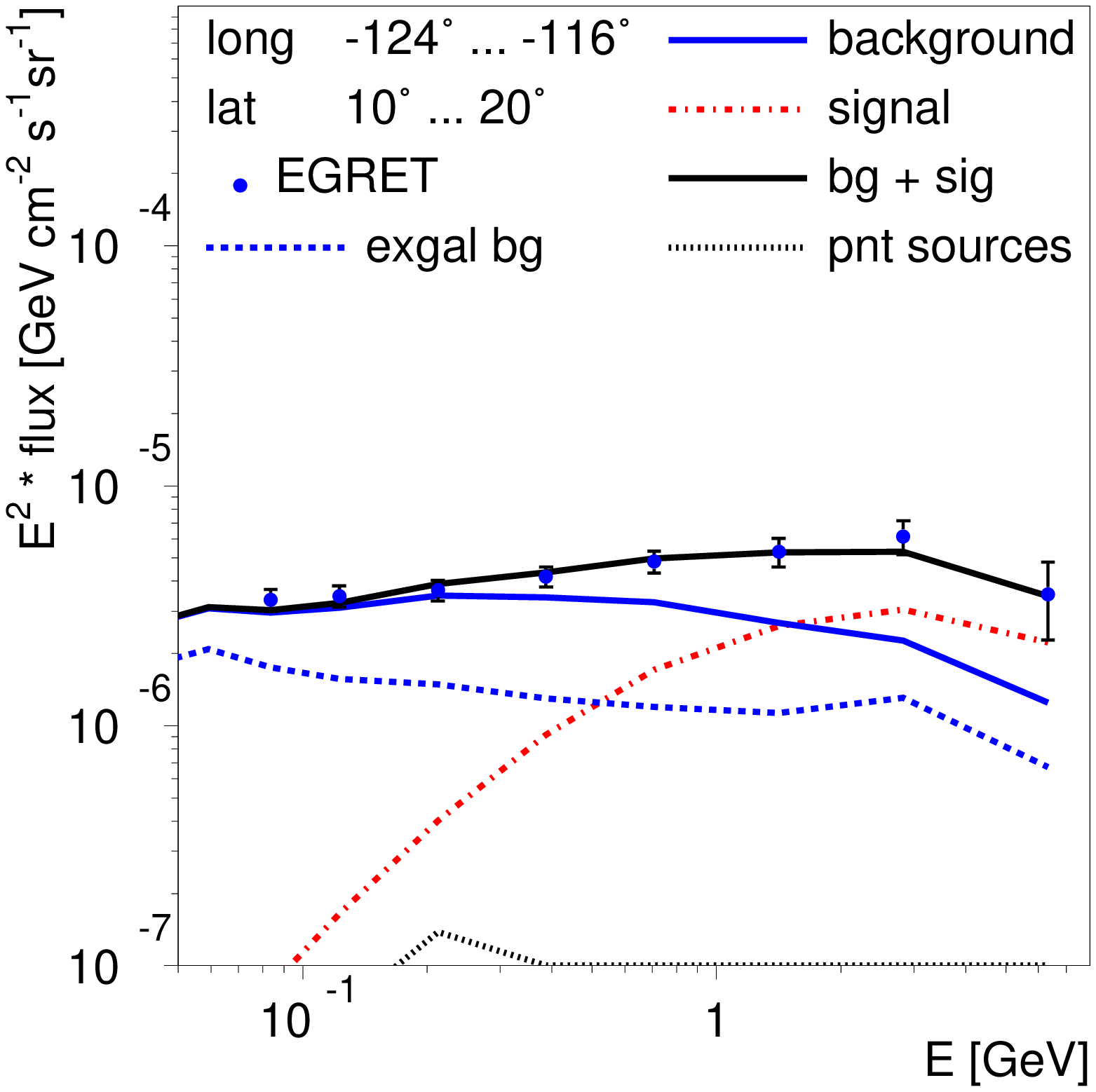}
    \includegraphics[width=0.21\textwidth]{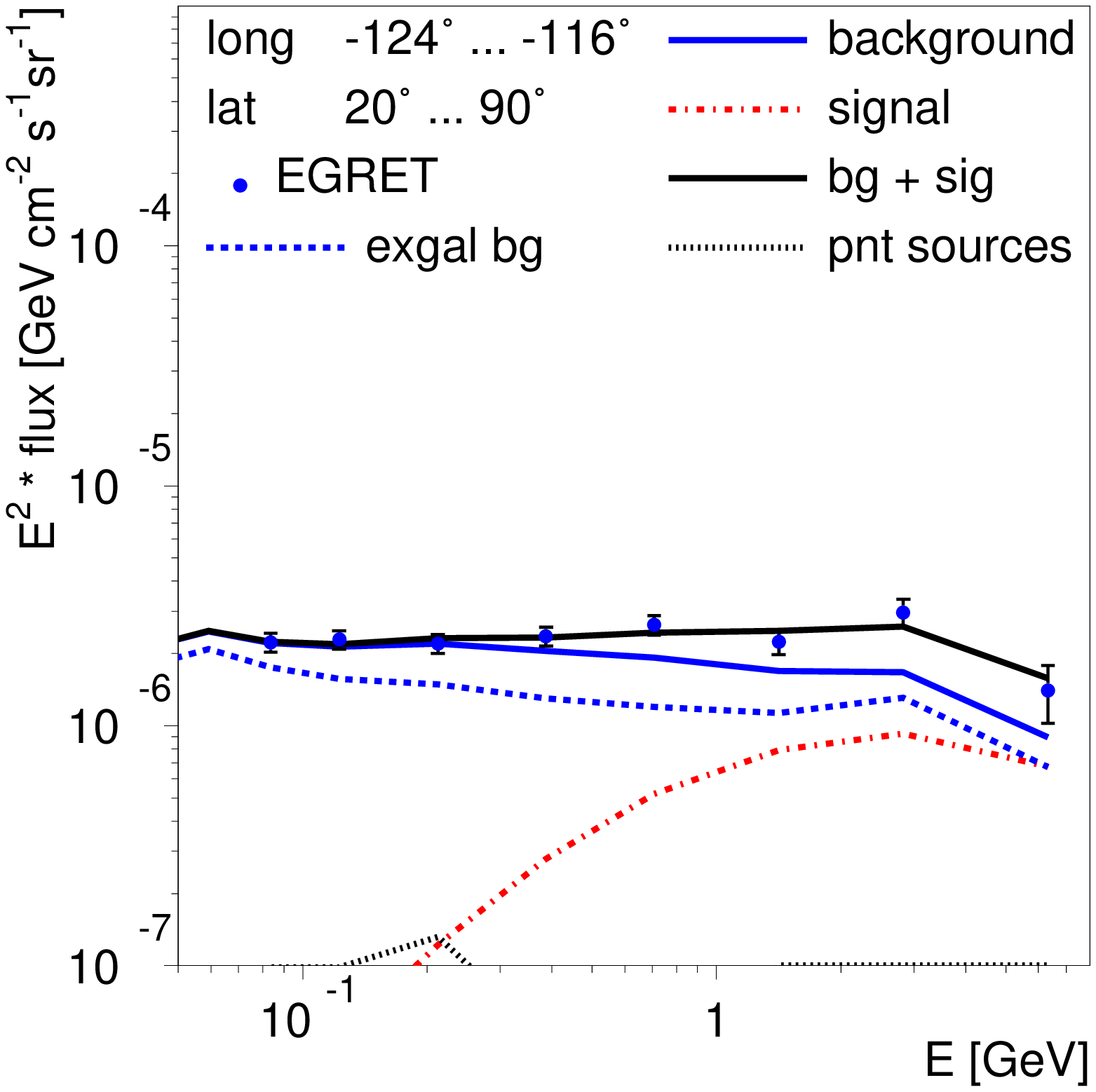}\\
    \hspace{-1cm}
    \begin{turn}{90} \framebox[0.21\textwidth][c]{{\scriptsize $-116^\circ<\mbox{long}<-108^\circ$}} \end{turn}
    \includegraphics[width=0.21\textwidth]{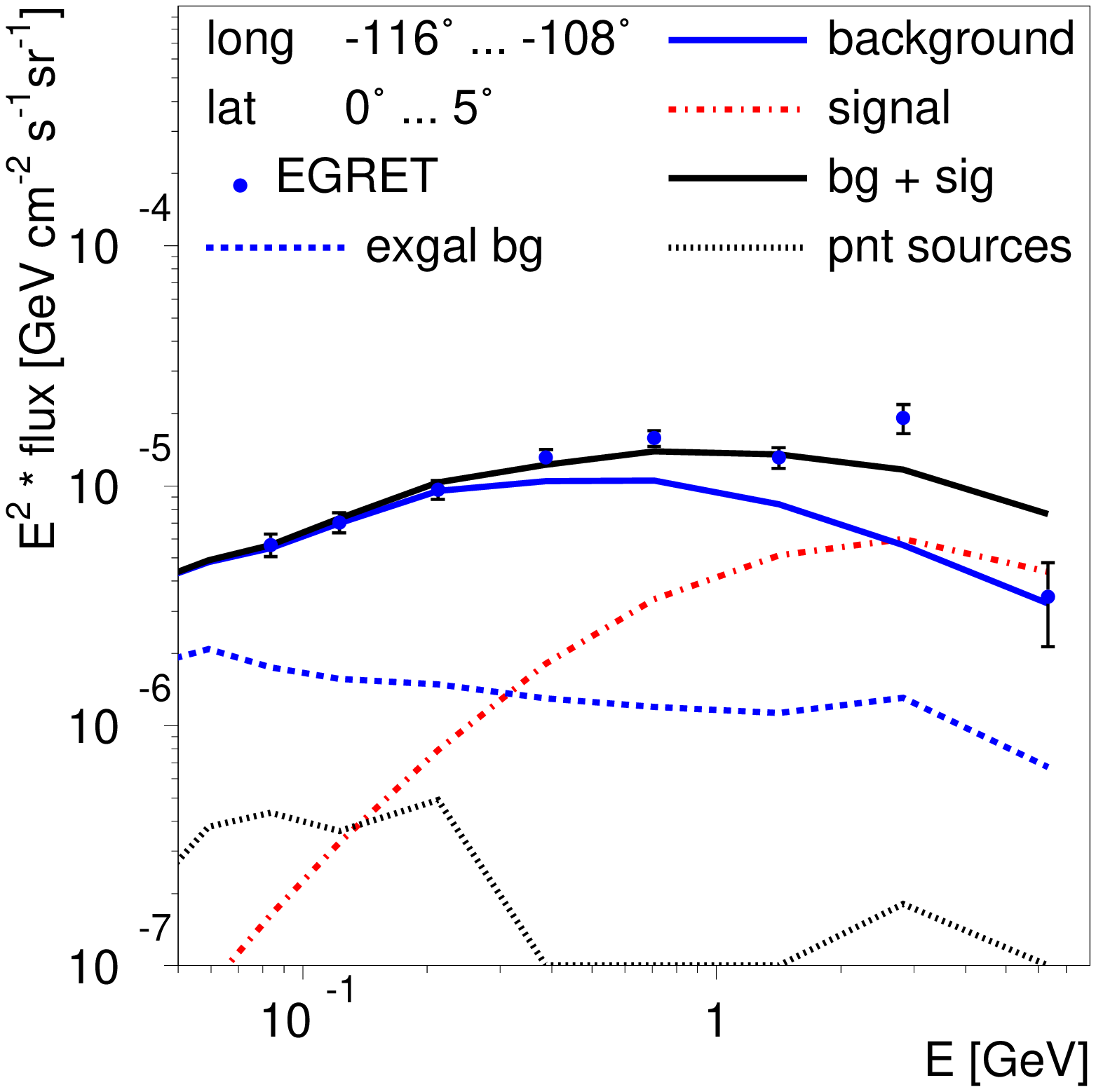}
    \includegraphics[width=0.21\textwidth]{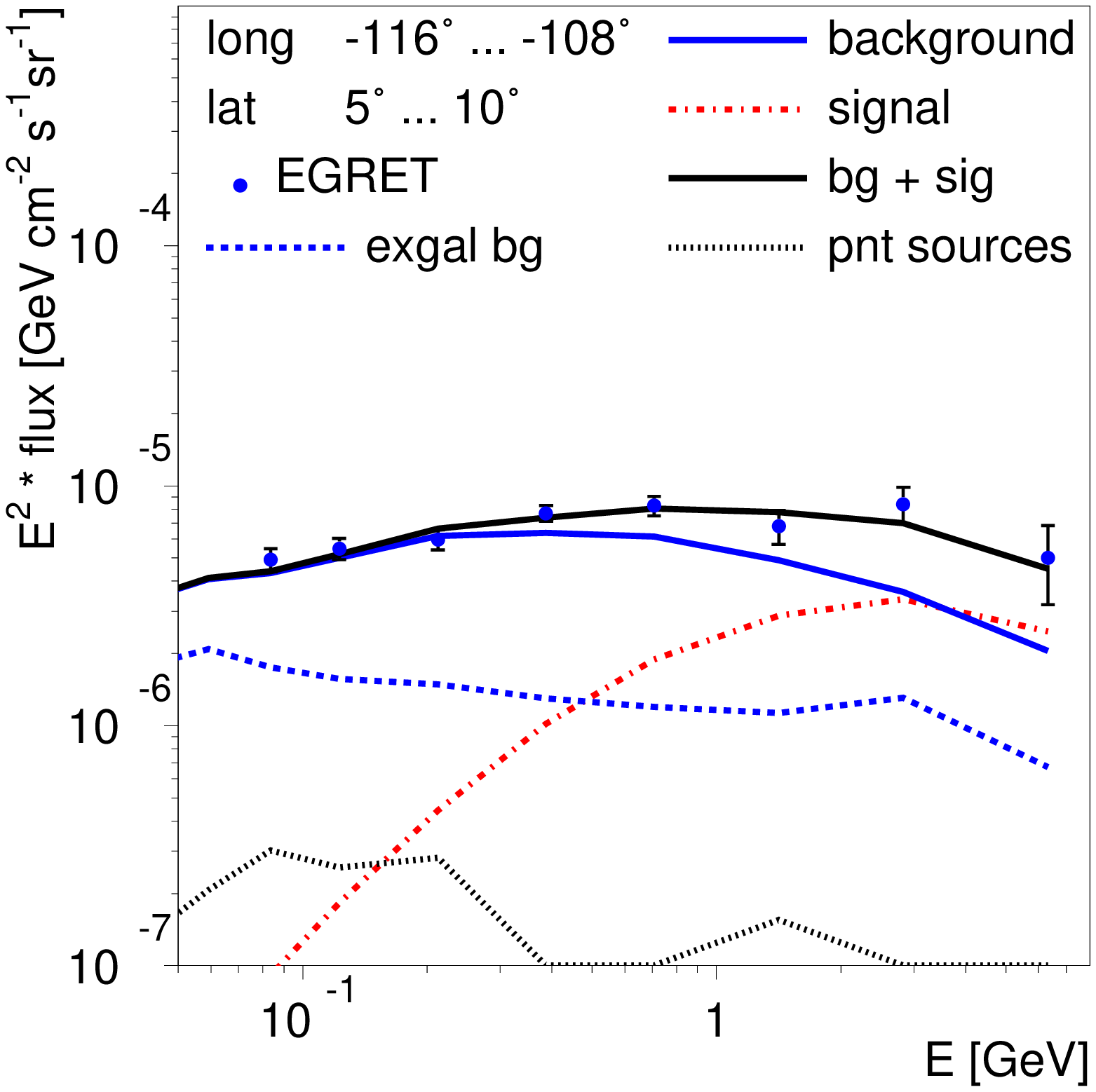}
    \includegraphics[width=0.21\textwidth]{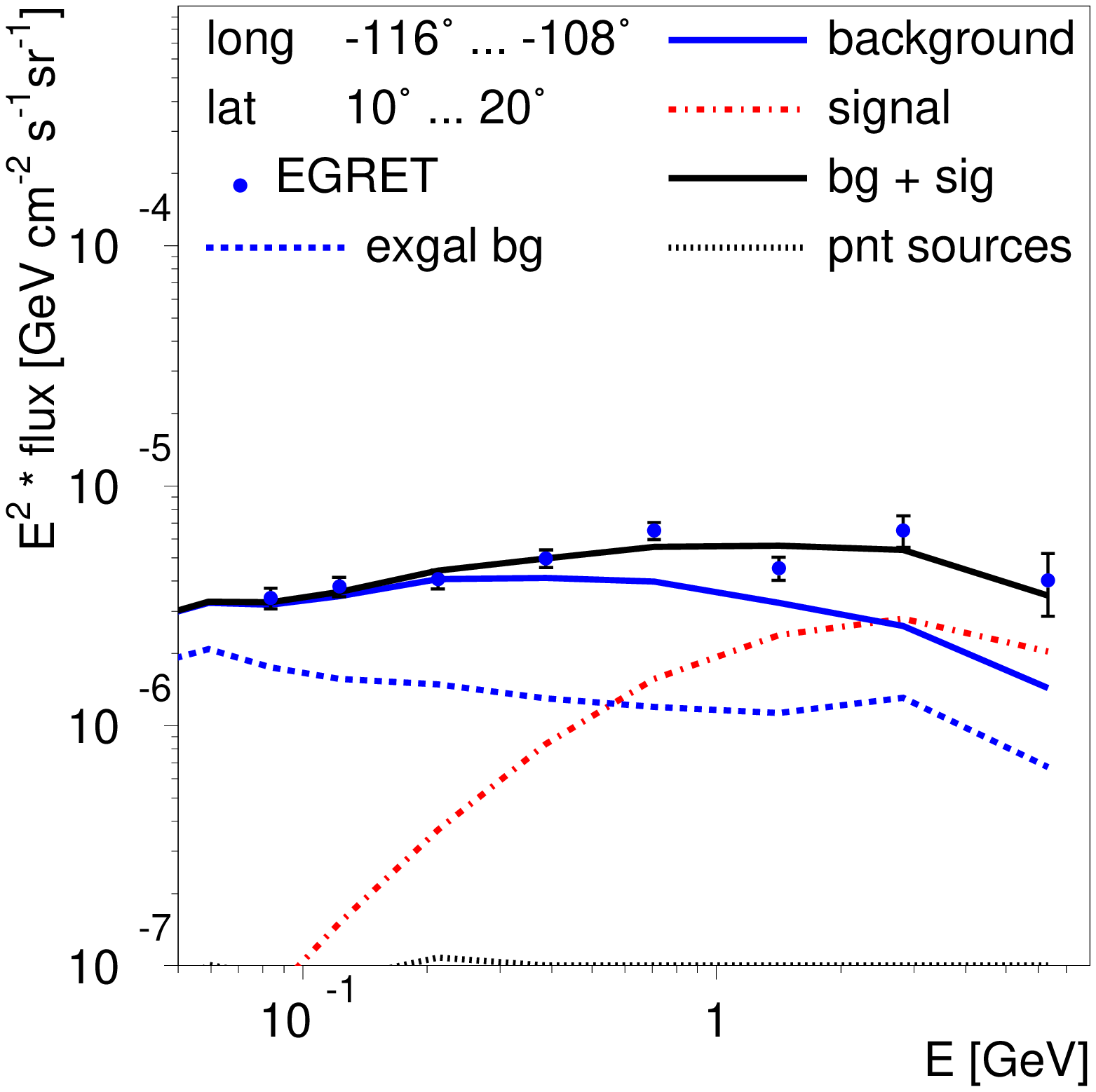}
    \includegraphics[width=0.21\textwidth]{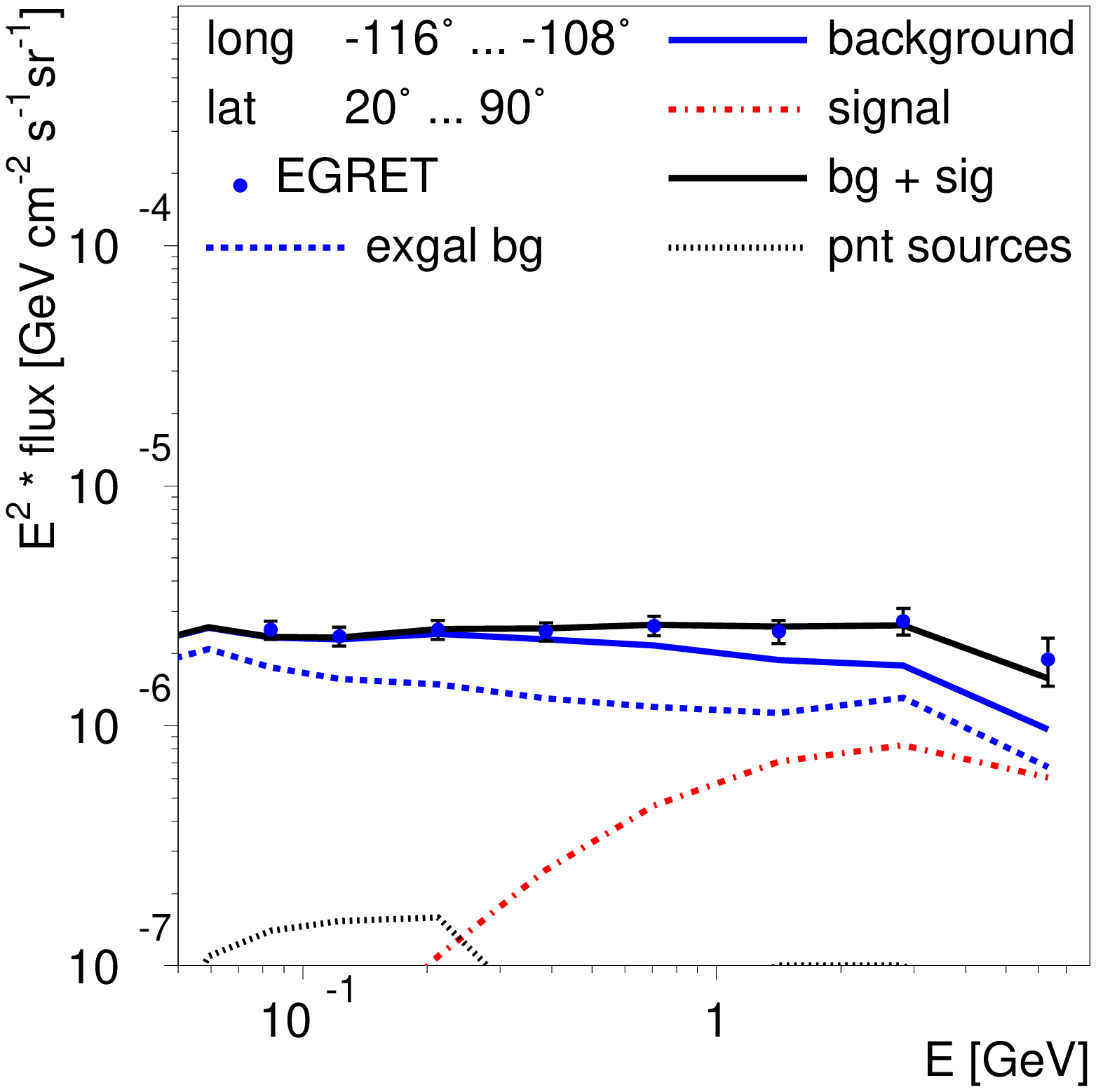}\\
    \hspace{-1cm}
    \begin{turn}{90} \framebox[0.21\textwidth][c]{{\scriptsize $-108^\circ<\mbox{long}<-100^\circ$}} \end{turn}
    \includegraphics[width=0.21\textwidth]{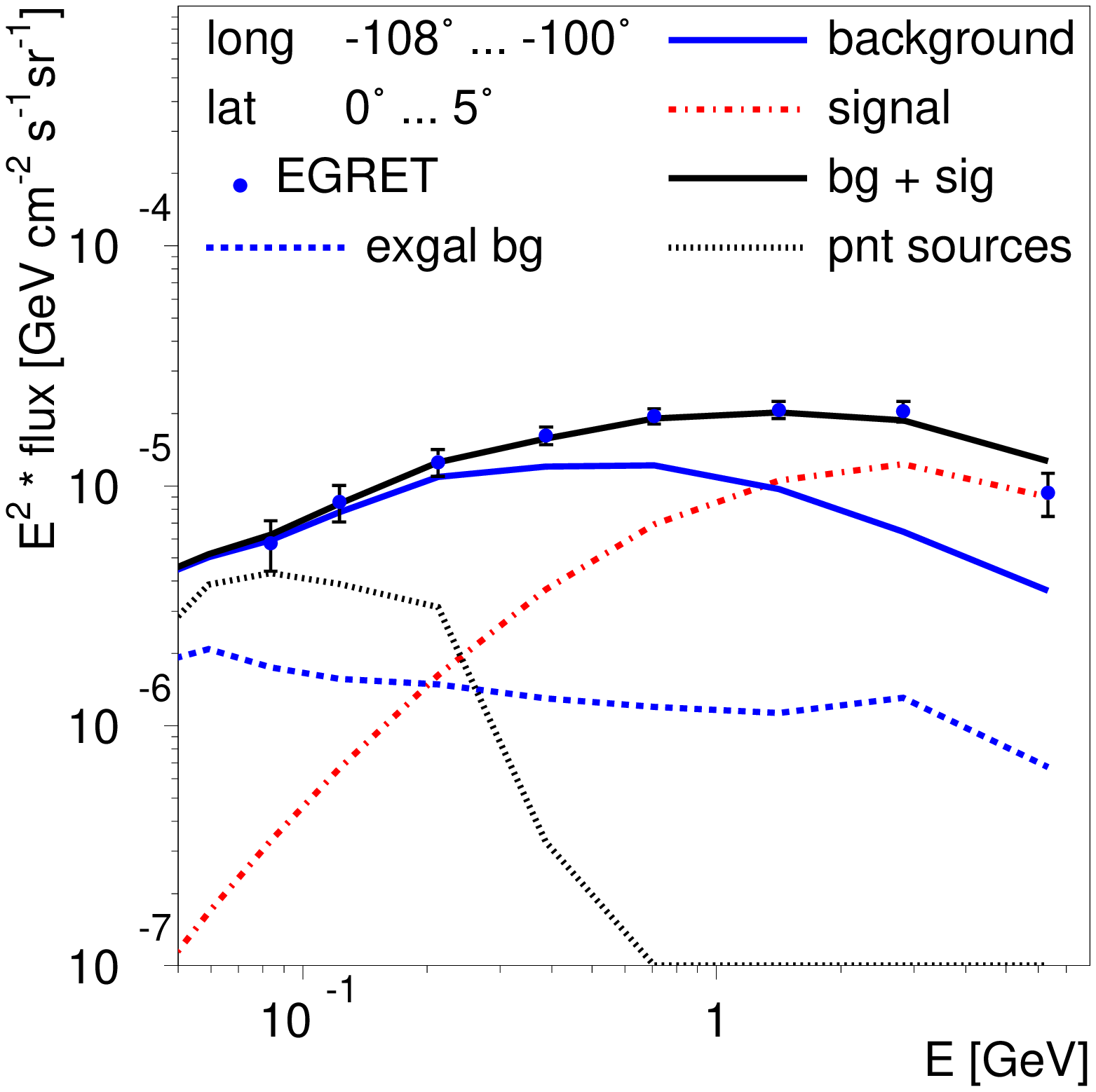}
    \includegraphics[width=0.21\textwidth]{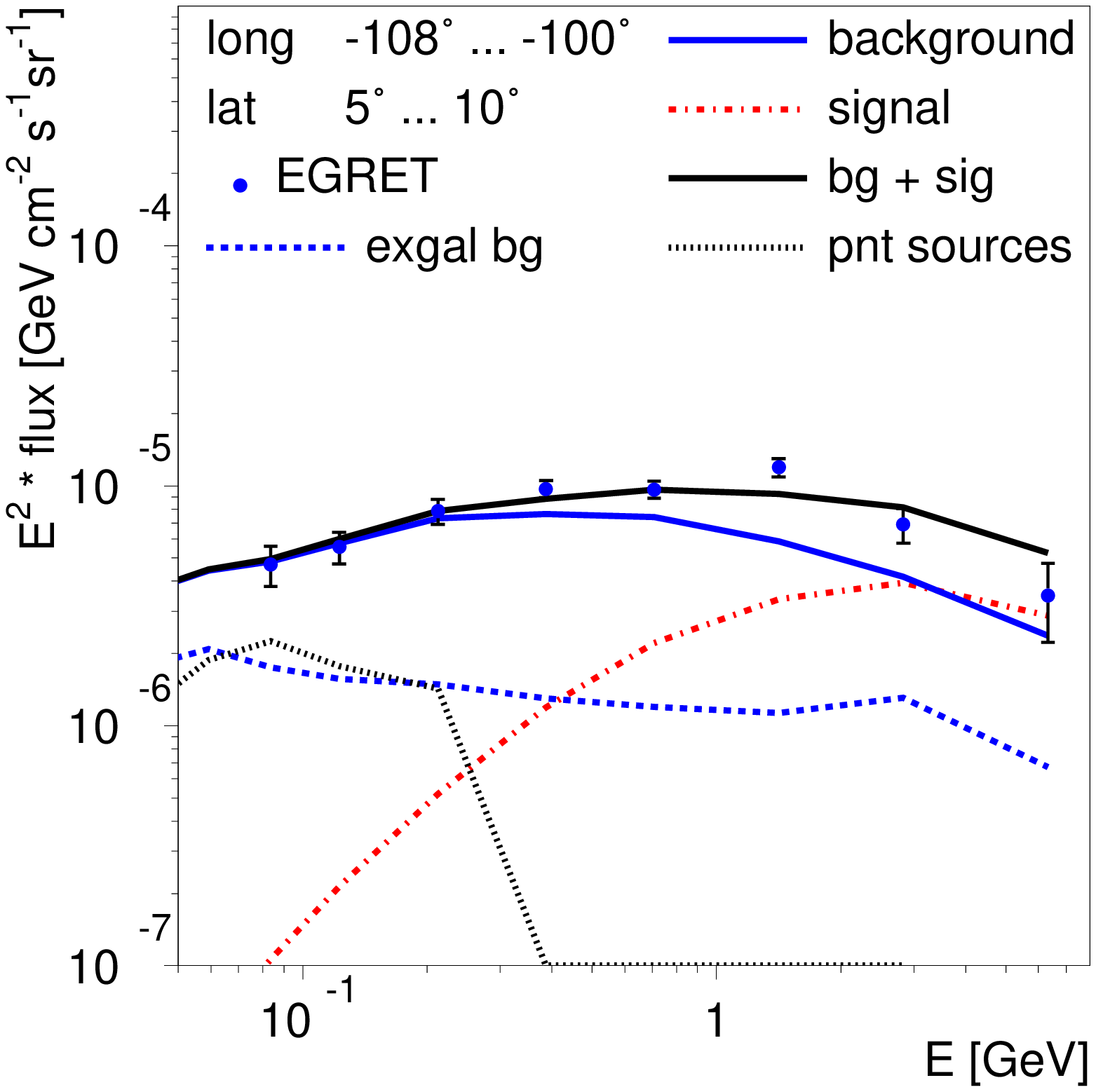}
    \includegraphics[width=0.21\textwidth]{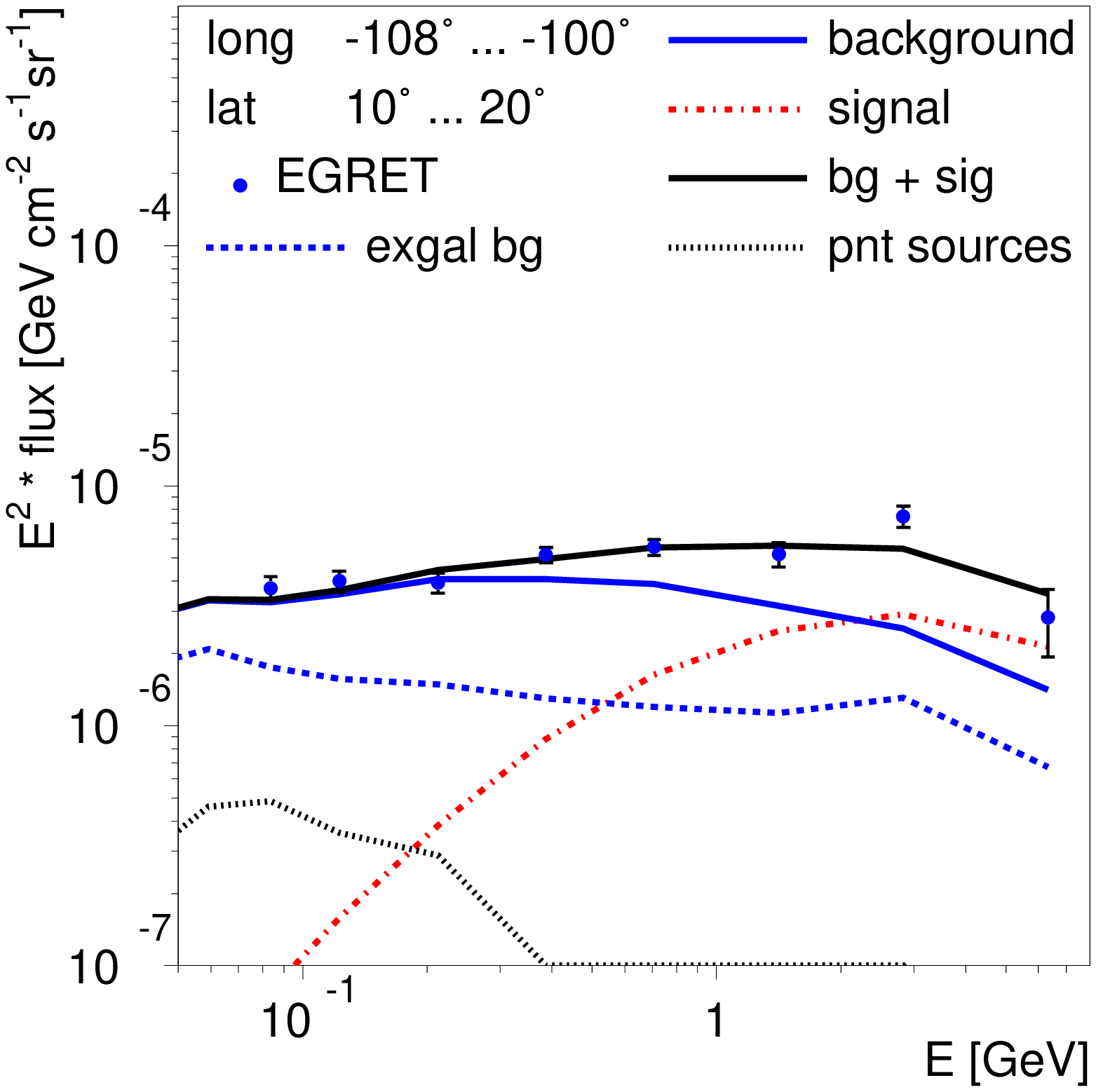}
    \includegraphics[width=0.21\textwidth]{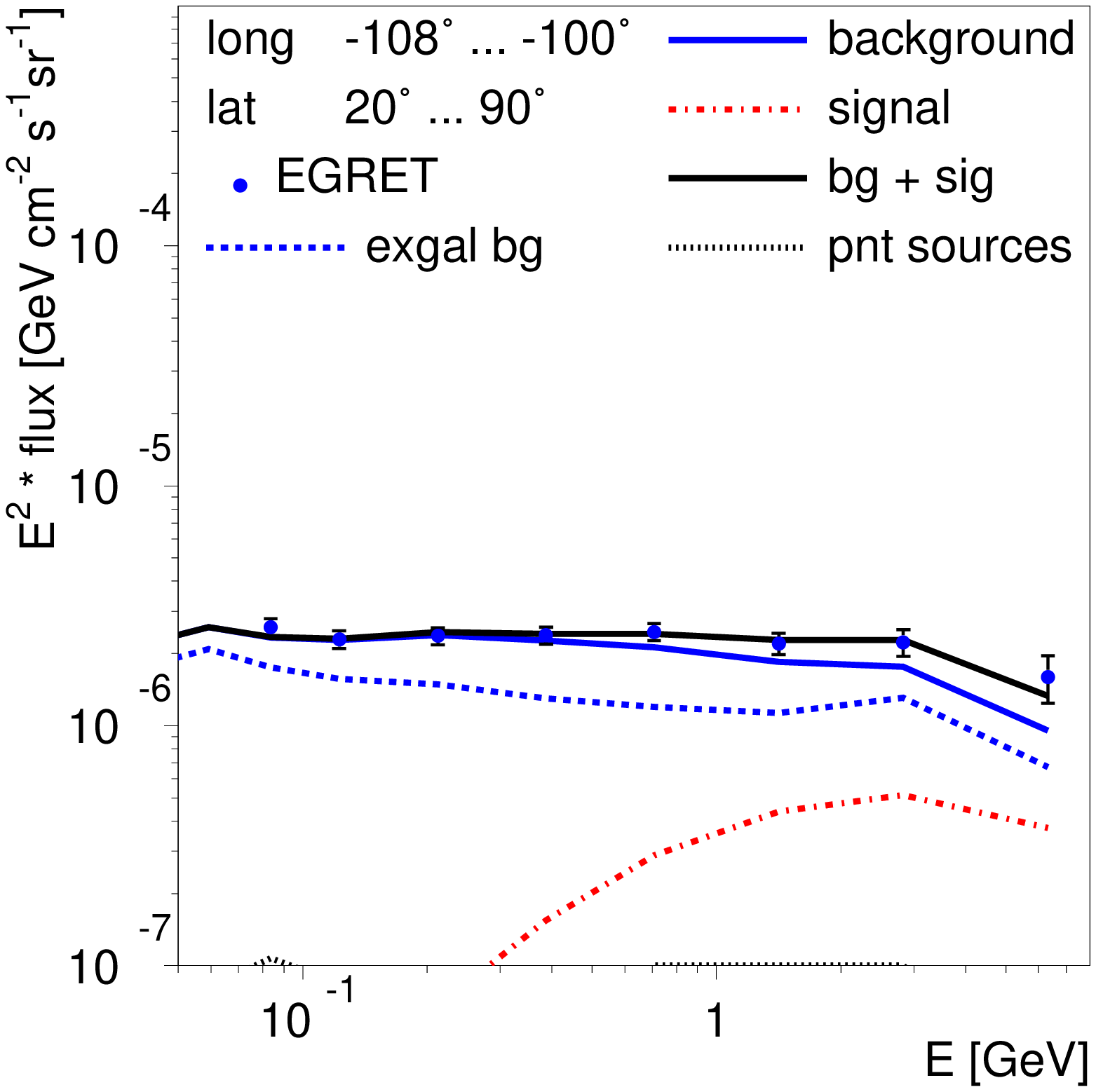}\\
    \hspace{-1cm}
    \begin{turn}{90} \framebox[0.21\textwidth][c]{{\scriptsize $-100^\circ<\mbox{long}<-92^\circ$}} \end{turn}
    \includegraphics[width=0.21\textwidth]{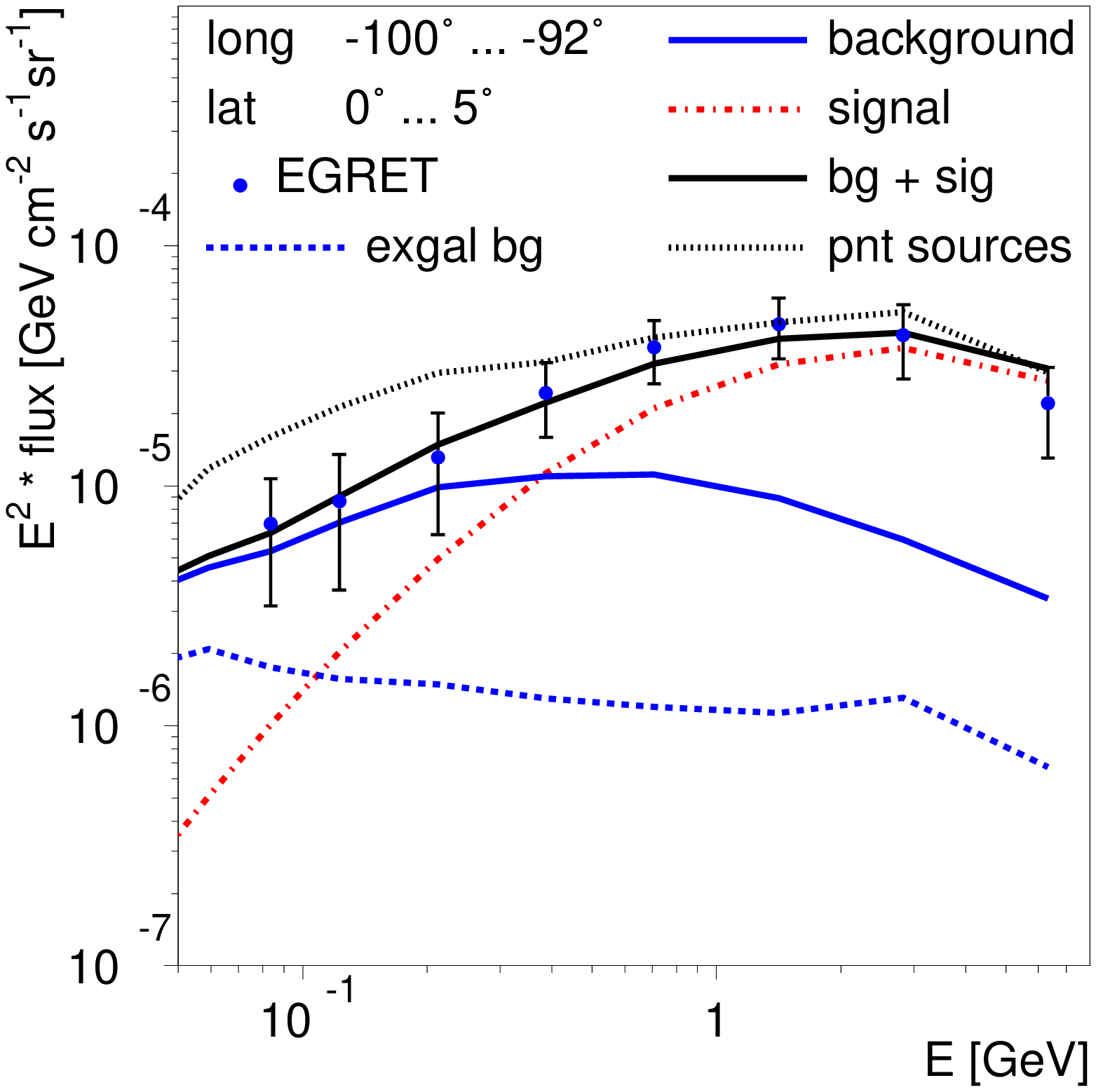}
    \includegraphics[width=0.21\textwidth]{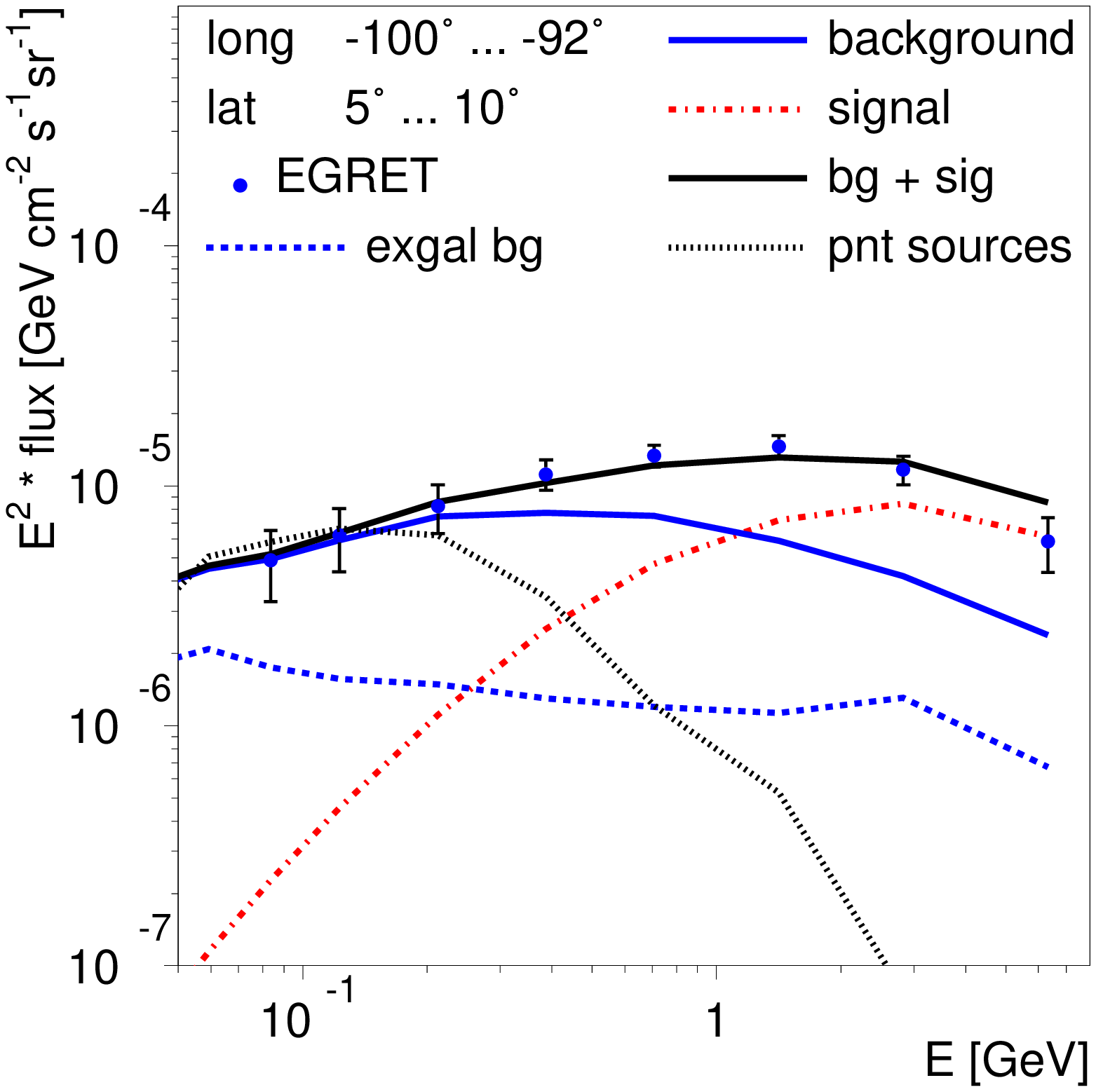}
    \includegraphics[width=0.21\textwidth]{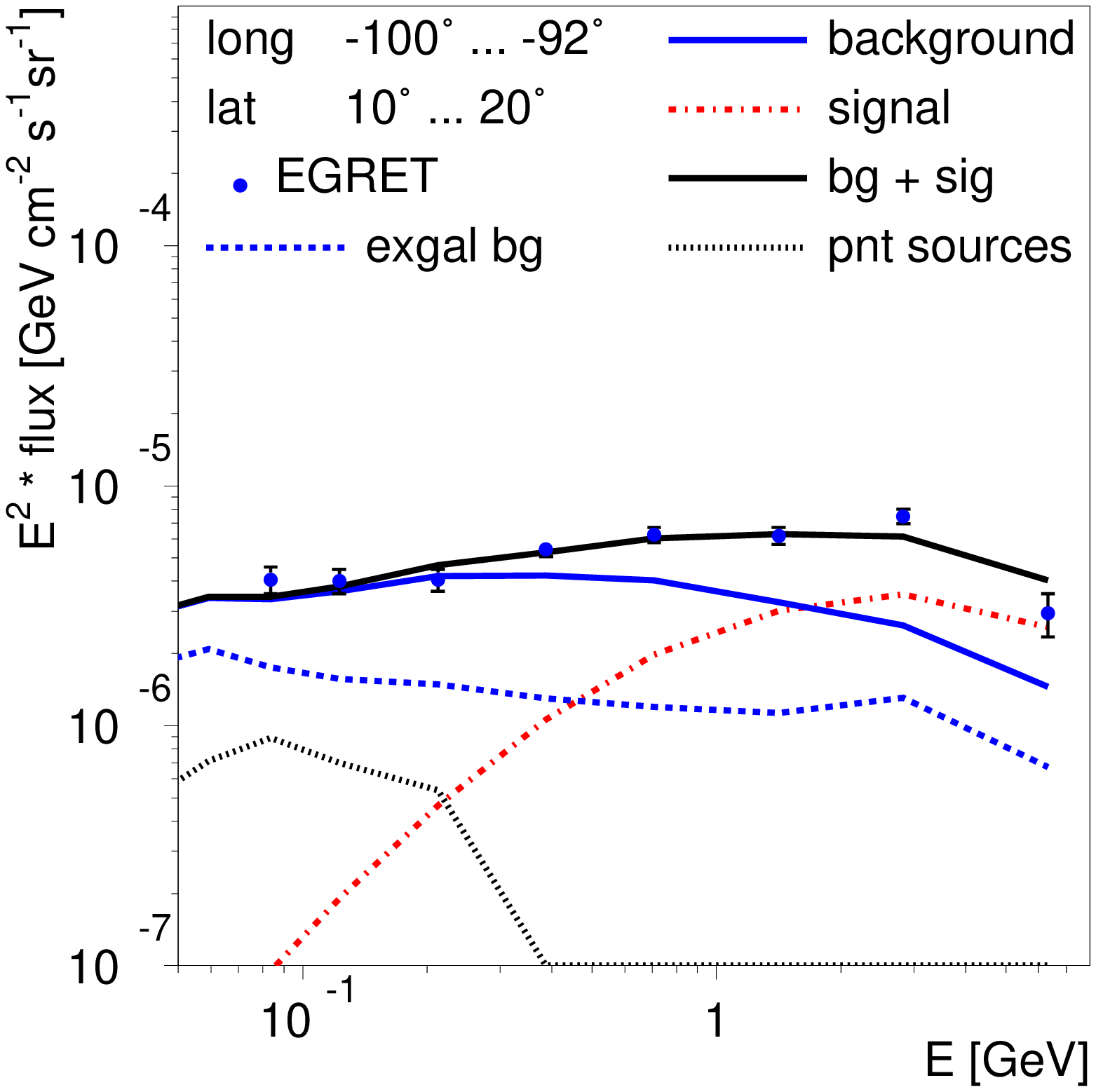}
    \includegraphics[width=0.21\textwidth]{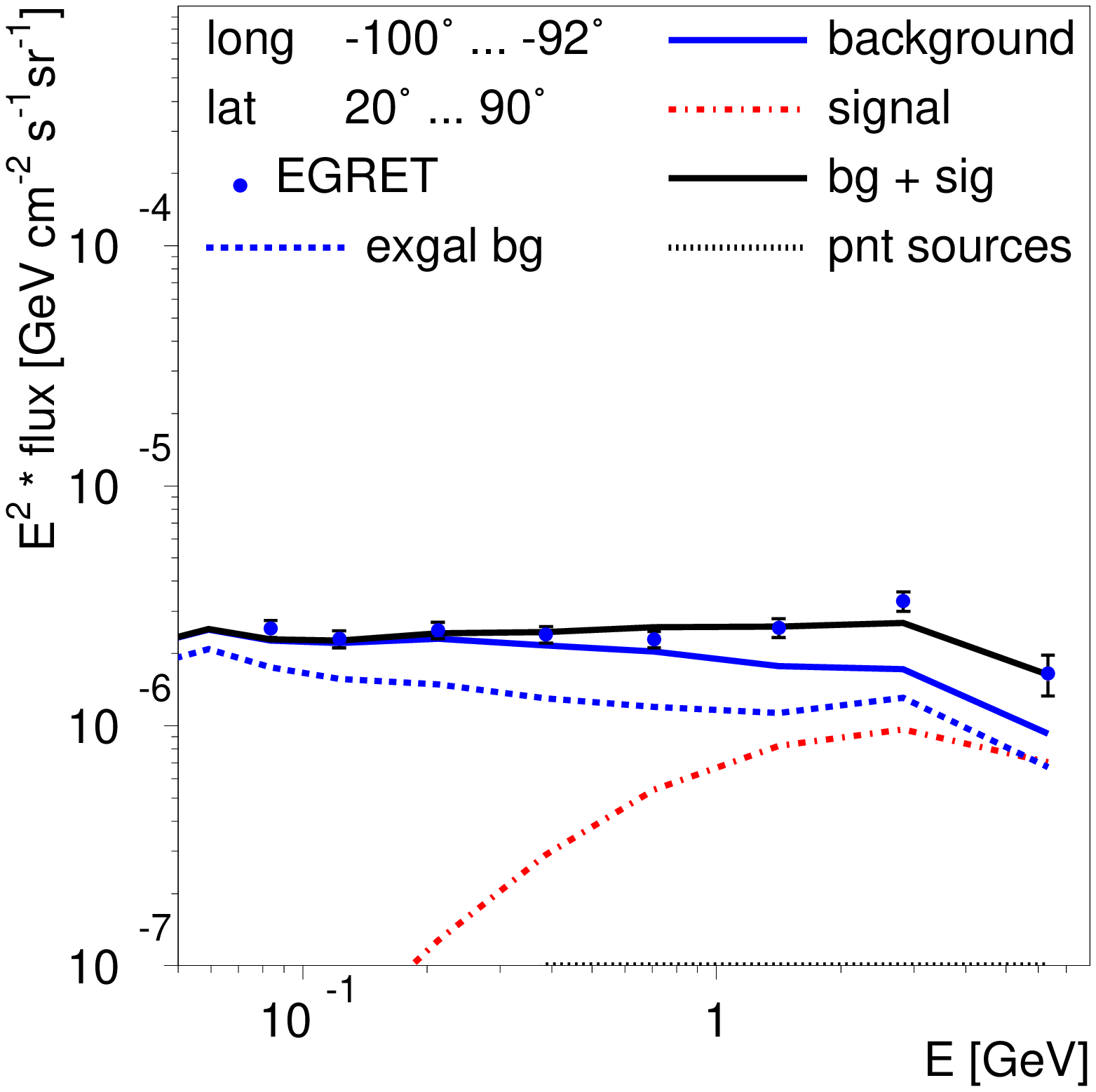}\\
    \hspace{-1cm}
    \begin{turn}{90} \framebox[0.21\textwidth][c]{{\scriptsize $-92^\circ<\mbox{long}<-84^\circ$}} \end{turn}
    \includegraphics[width=0.21\textwidth]{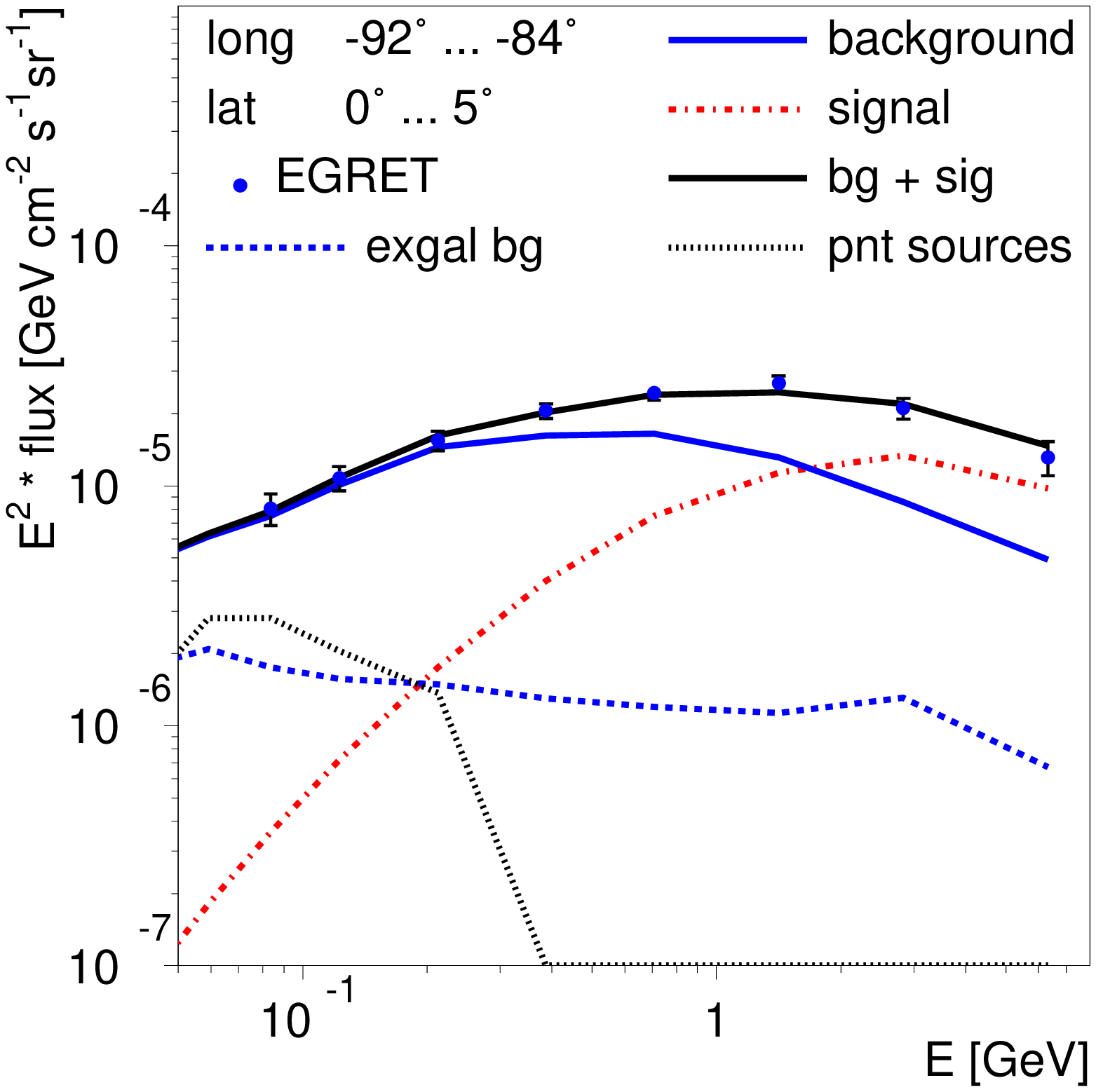}
    \includegraphics[width=0.21\textwidth]{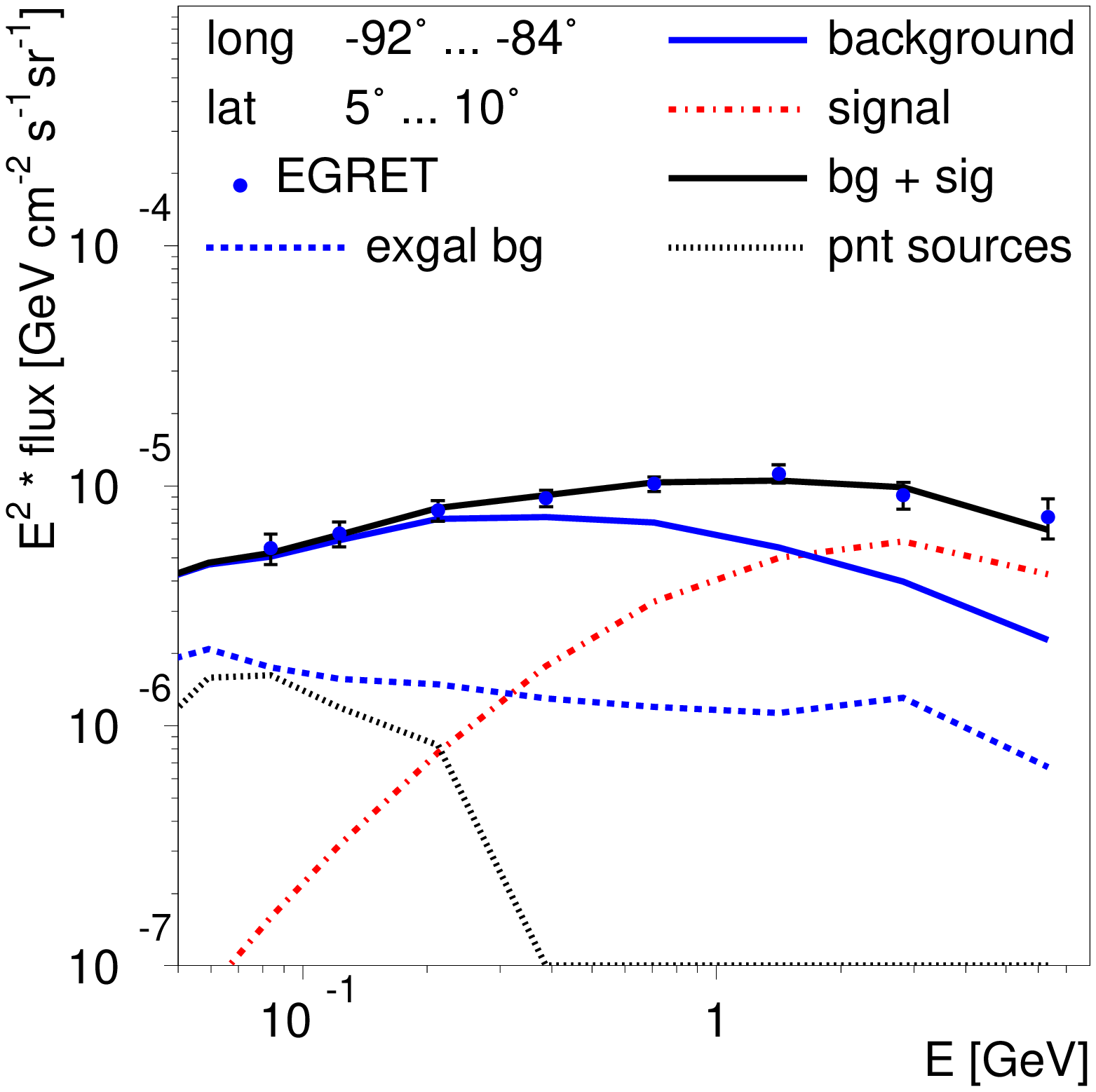}
    \includegraphics[width=0.21\textwidth]{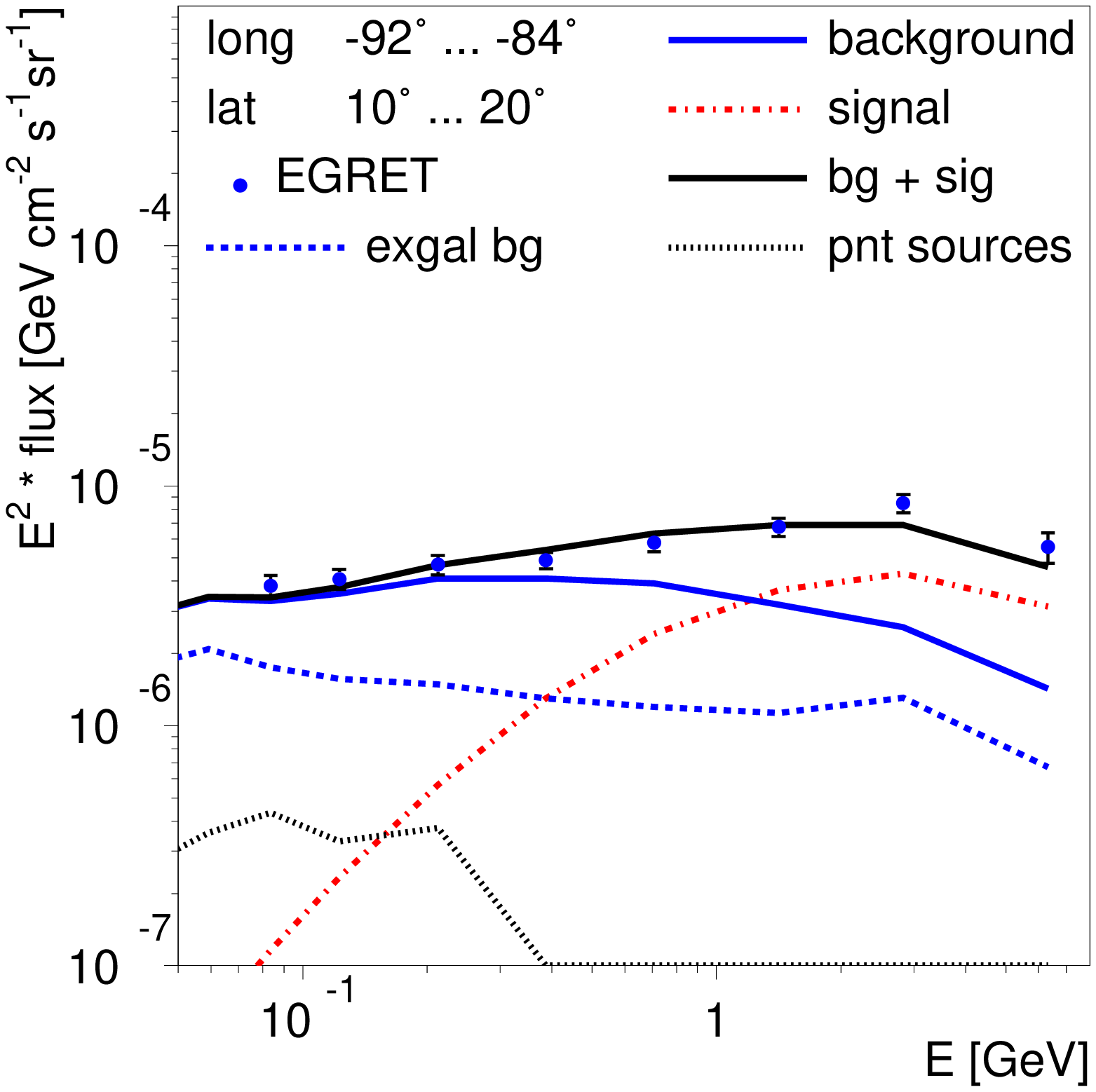}
    \includegraphics[width=0.21\textwidth]{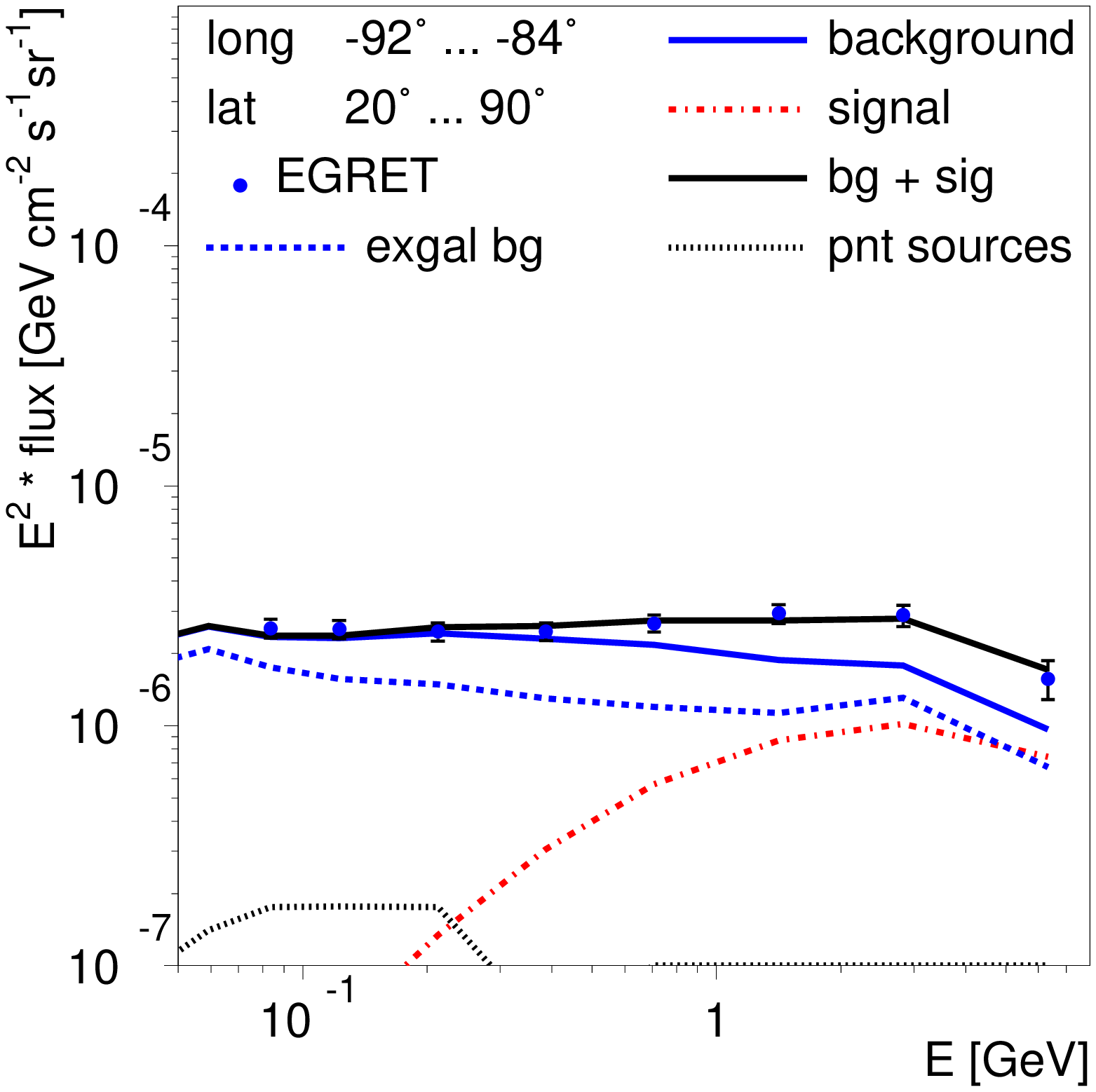}\\
  \end{center}
  \clearpage
  \begin{center}
    \framebox[0.21\textwidth][c]{$\vert \mbox{lat}\vert<5^\circ$}
    \framebox[0.21\textwidth][c]{$5^\circ<\vert \mbox{lat}\vert<10^\circ$}
    \framebox[0.21\textwidth][c]{$10^\circ<\vert \mbox{lat}\vert<20^\circ$}
    \framebox[0.21\textwidth][c]{$20^\circ<\vert \mbox{lat}\vert<90^\circ$}\\
    \hspace{-1cm}
    \begin{turn}{90} \framebox[0.21\textwidth][c]{{\scriptsize $-84^\circ<\mbox{long}<-76^\circ$}} \end{turn}
    \includegraphics[width=0.21\textwidth]{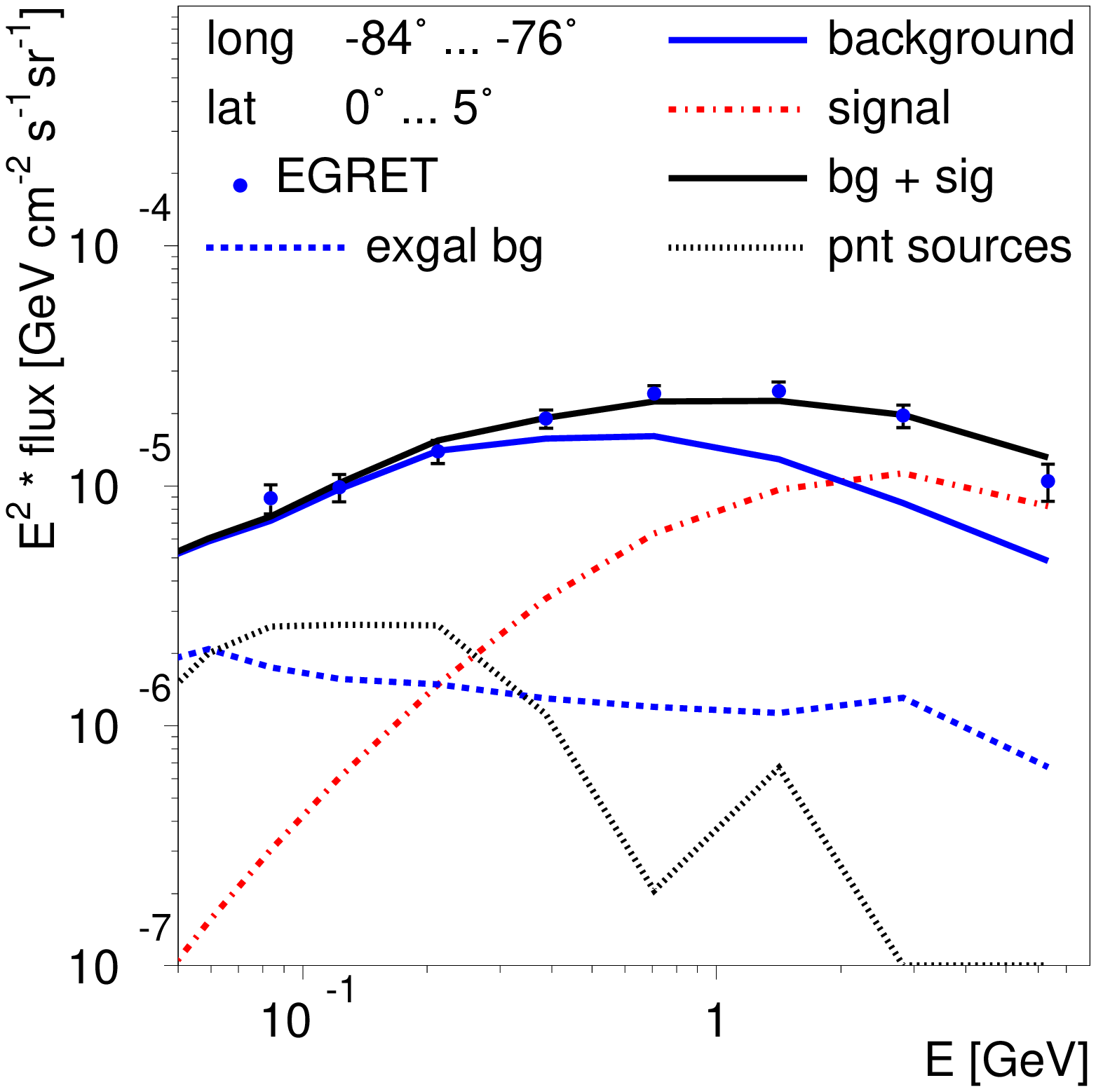}
    \includegraphics[width=0.21\textwidth]{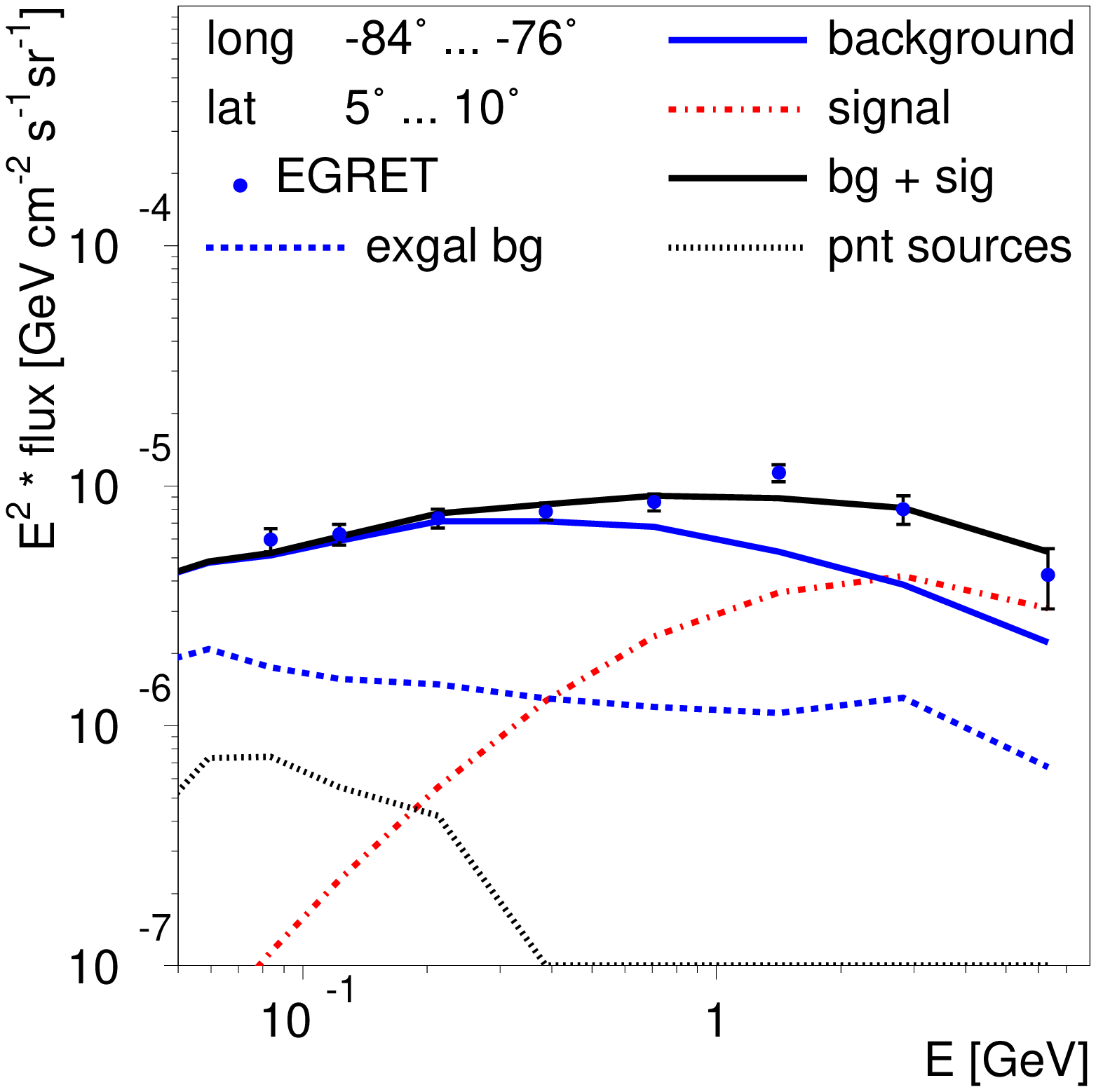}
    \includegraphics[width=0.21\textwidth]{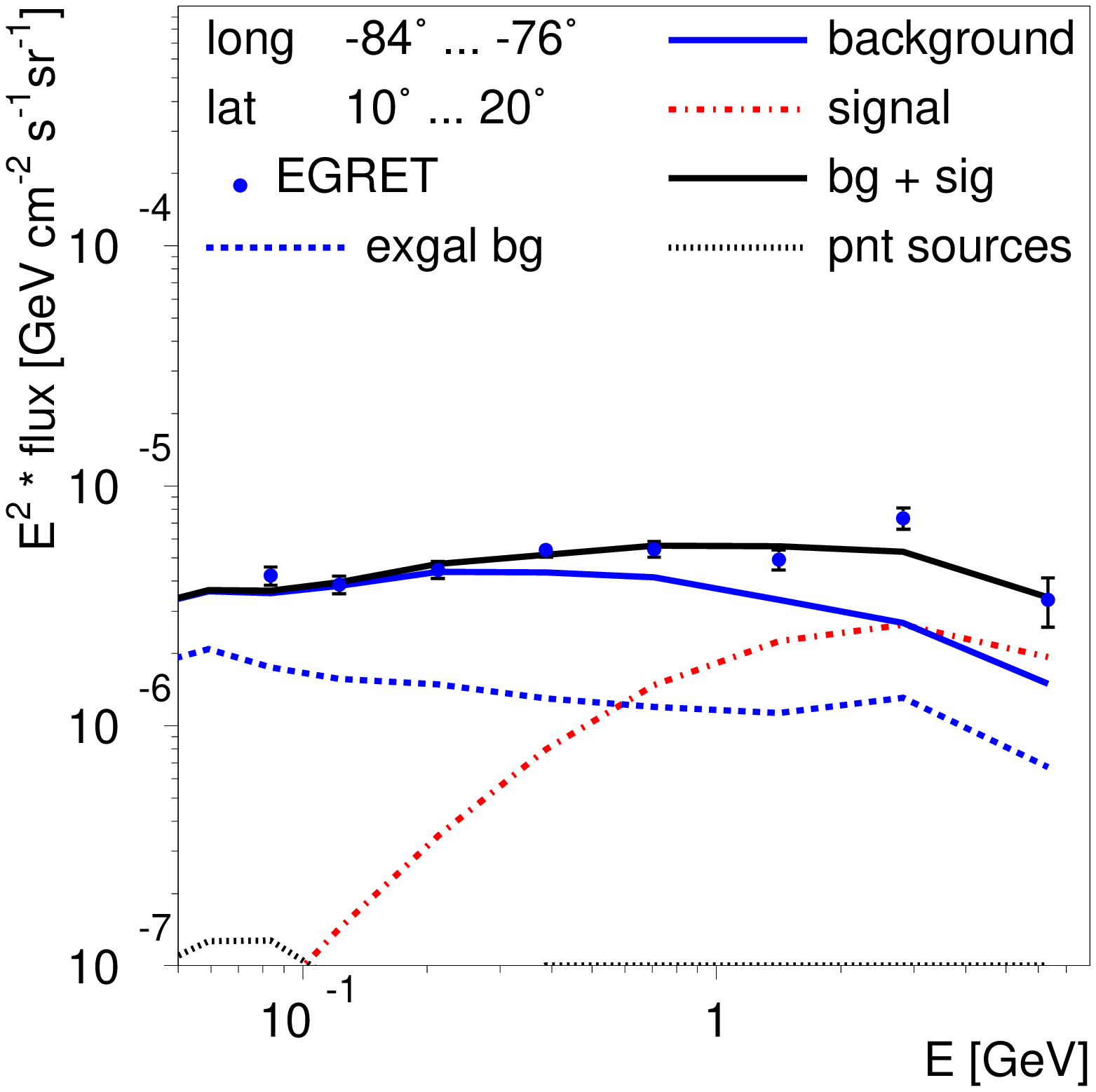}
    \includegraphics[width=0.21\textwidth]{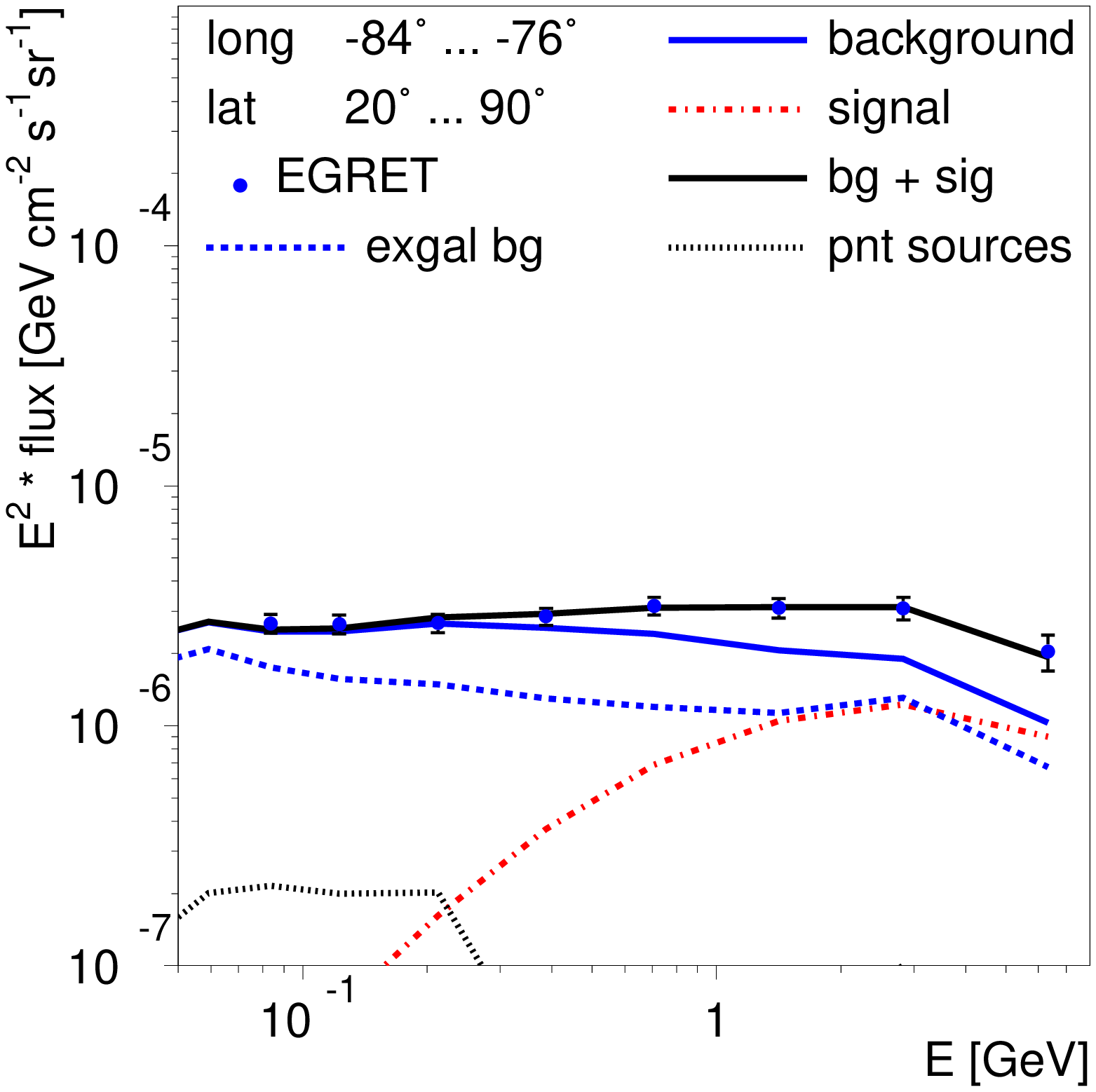}\\
    \hspace{-1cm}
    \begin{turn}{90} \framebox[0.21\textwidth][c]{{\scriptsize $-76^\circ<\mbox{long}<-68^\circ$}} \end{turn}
    \includegraphics[width=0.21\textwidth]{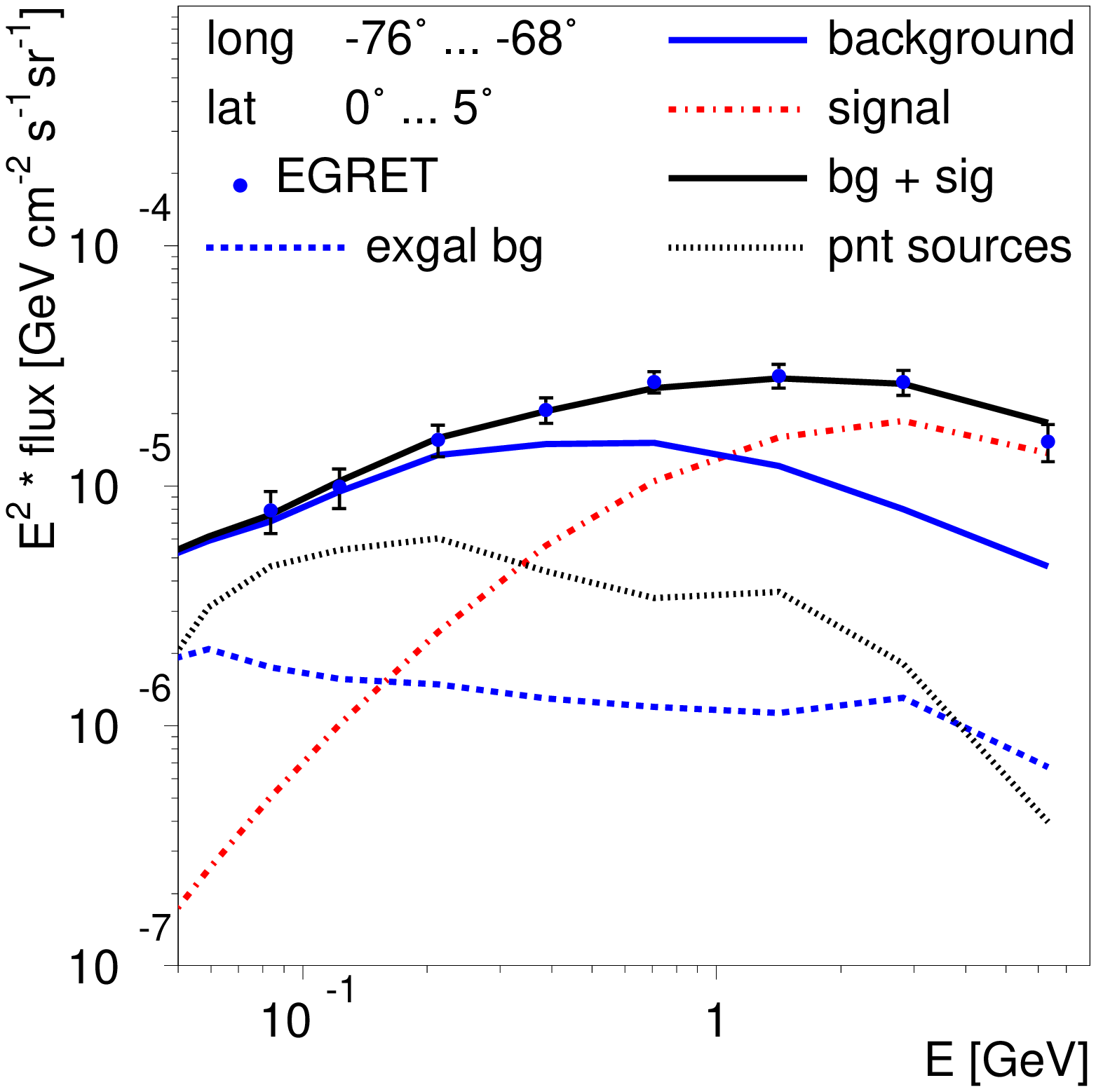}
    \includegraphics[width=0.21\textwidth]{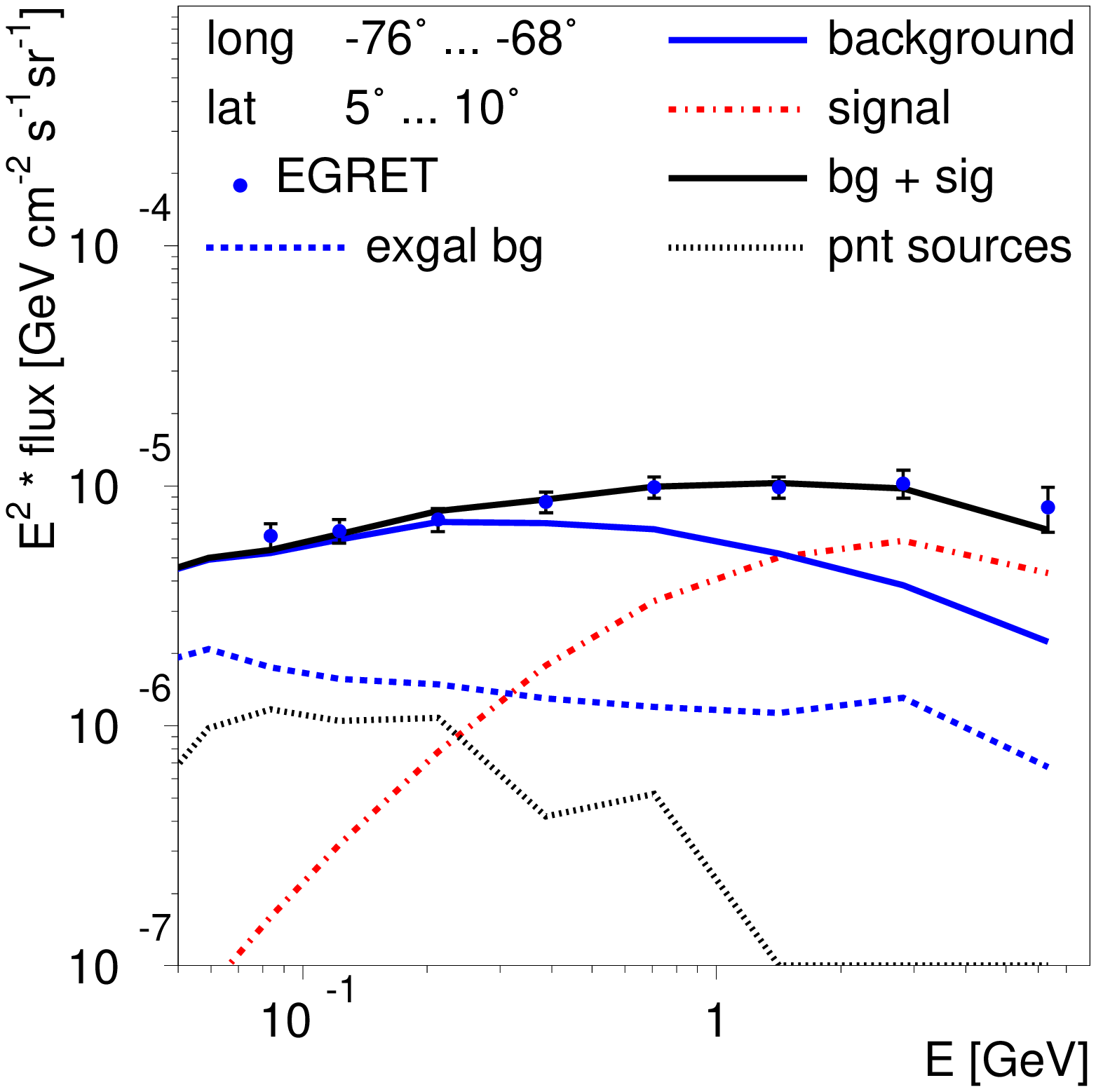}
    \includegraphics[width=0.21\textwidth]{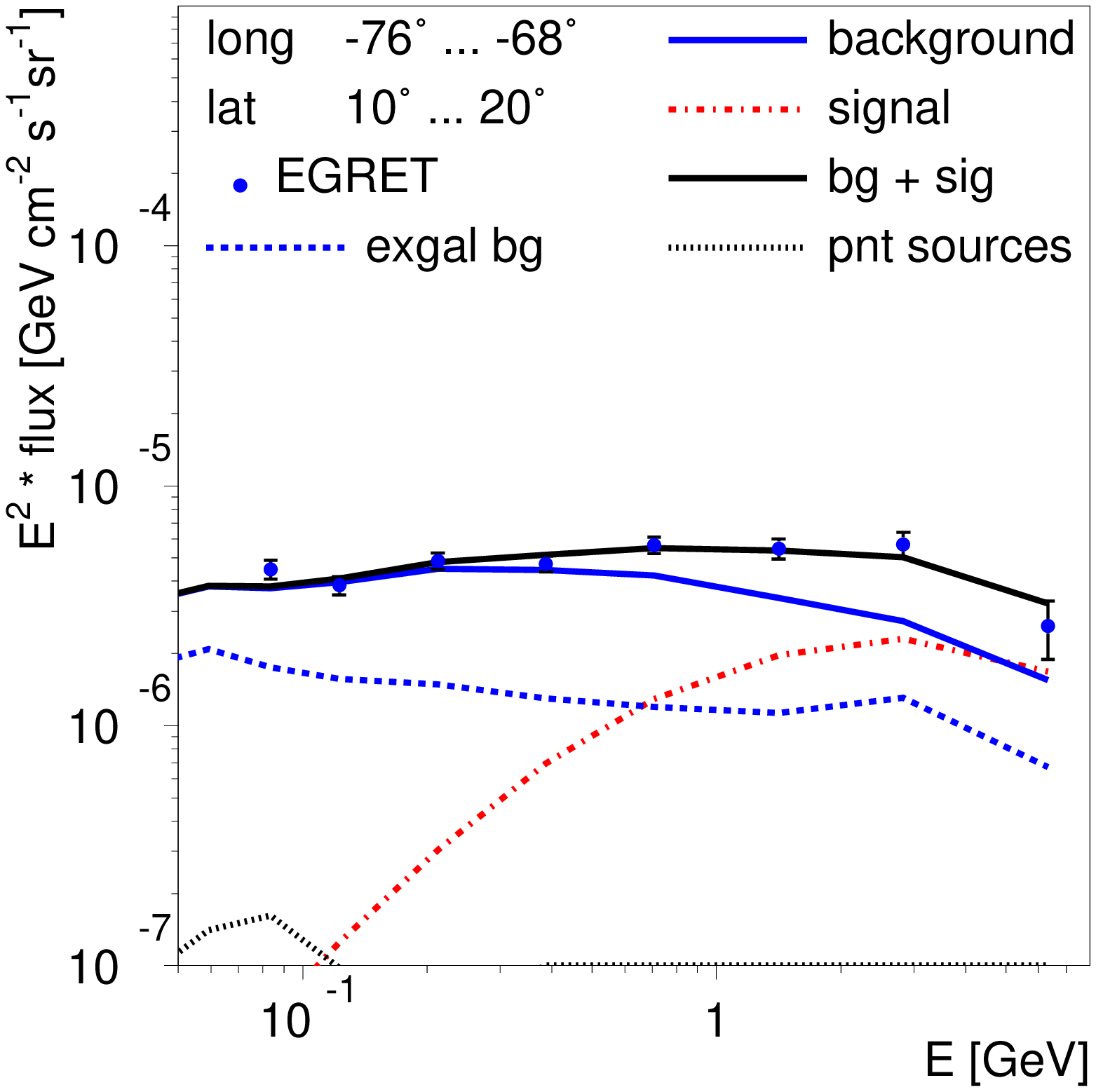}
    \includegraphics[width=0.21\textwidth]{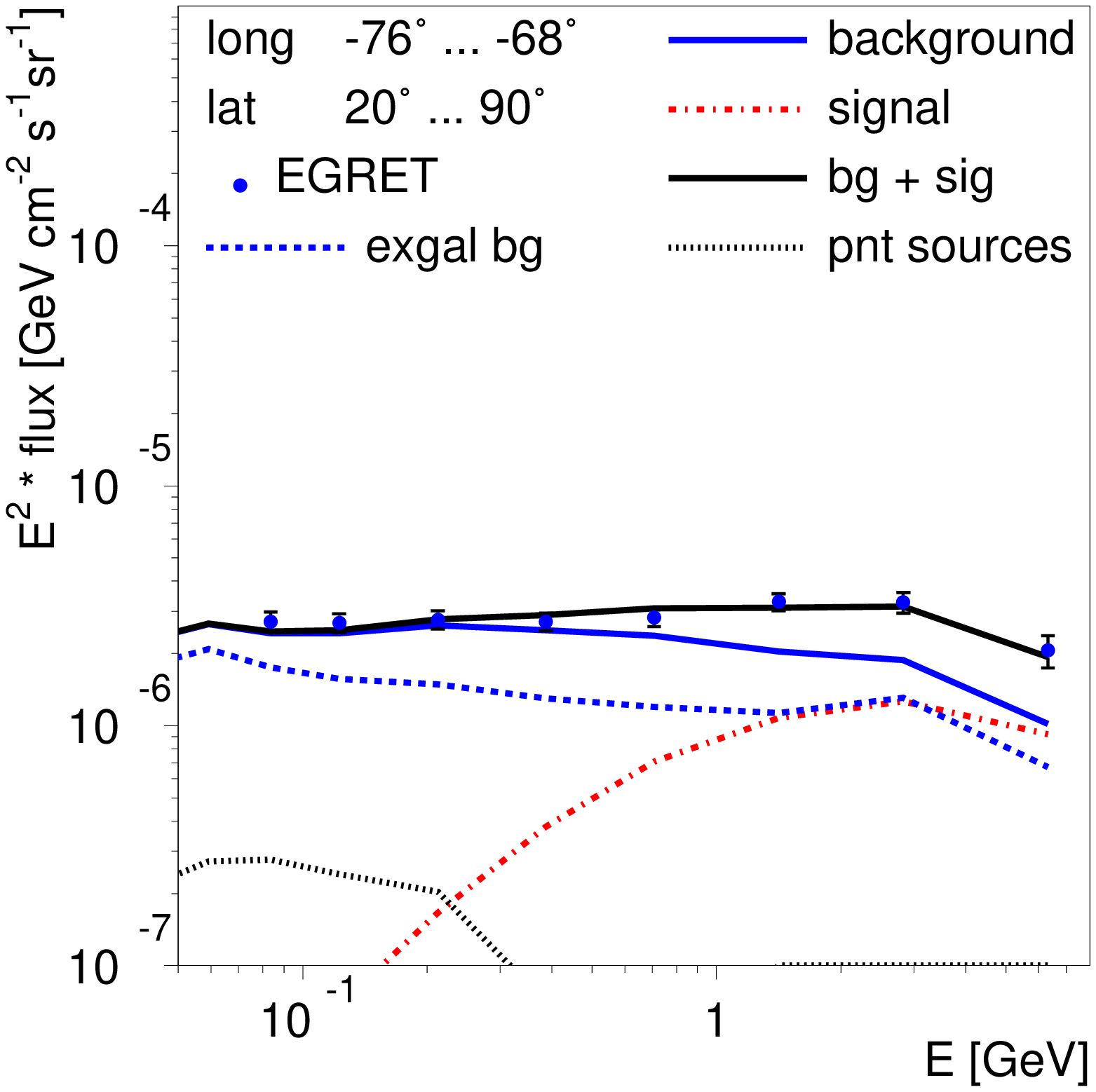}\\
    \hspace{-1cm}
    \begin{turn}{90} \framebox[0.21\textwidth][c]{{\scriptsize $-68^\circ<\mbox{long}<-60^\circ$}} \end{turn}
    \includegraphics[width=0.21\textwidth]{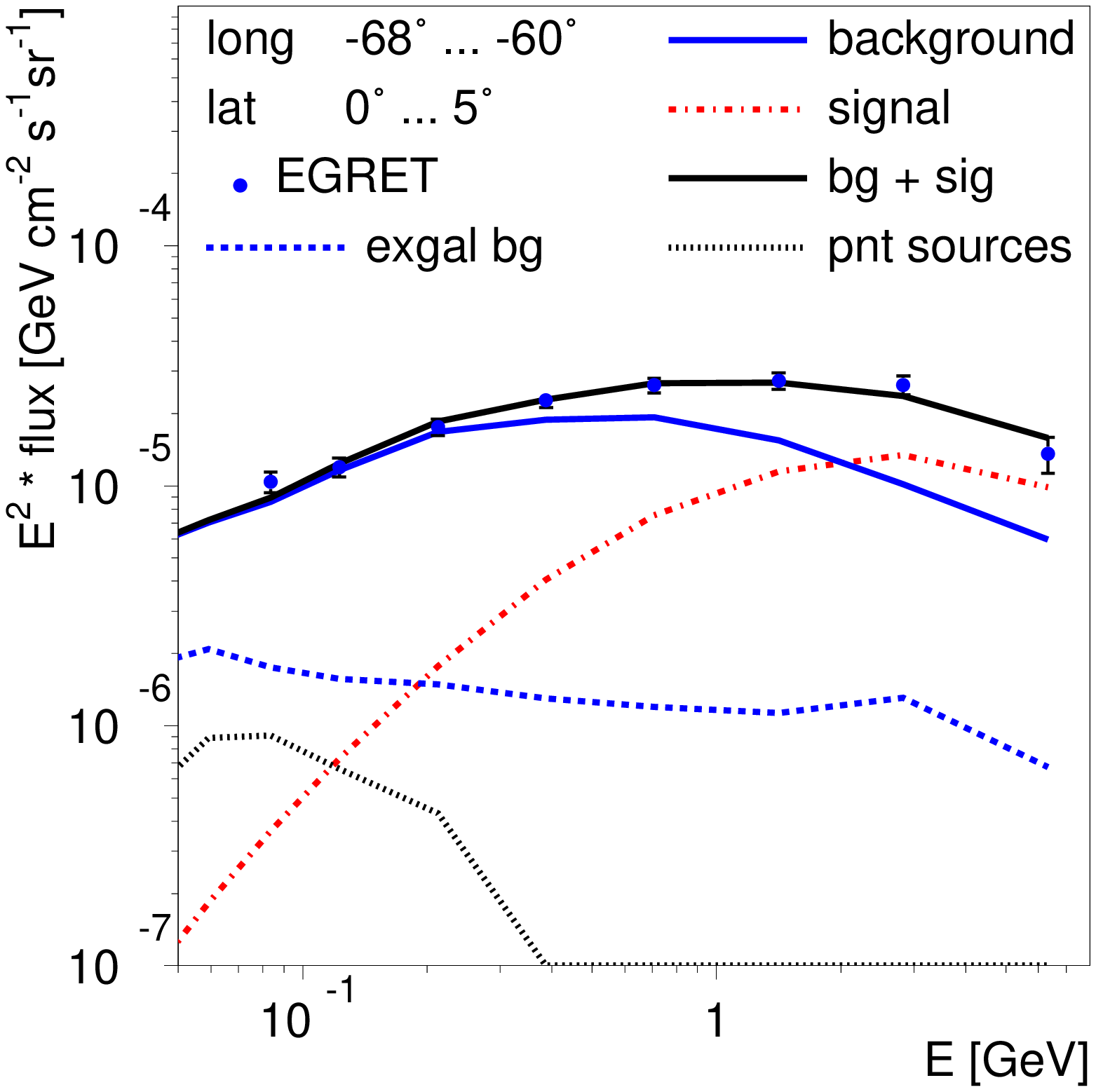}
    \includegraphics[width=0.21\textwidth]{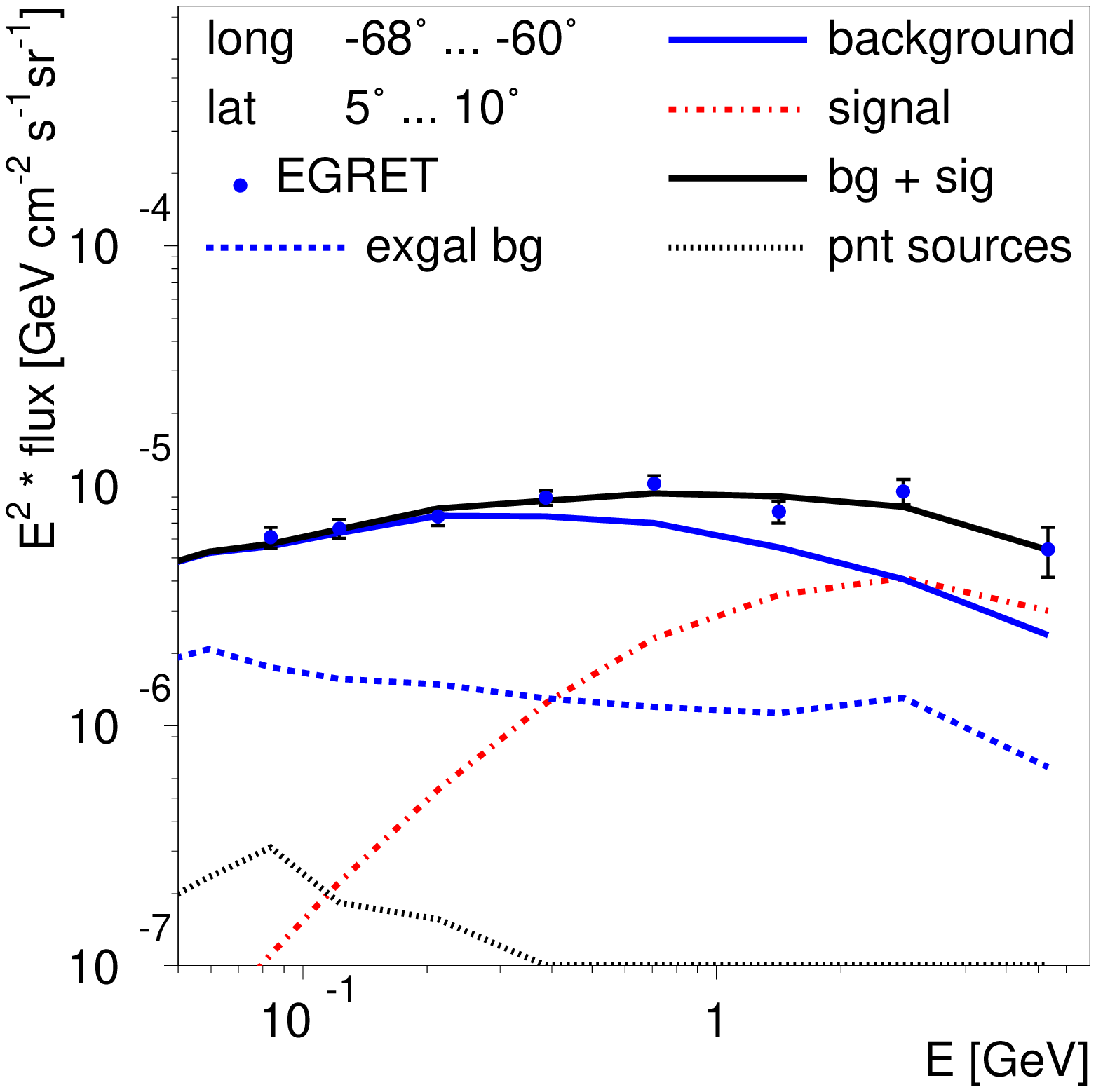}
    \includegraphics[width=0.21\textwidth]{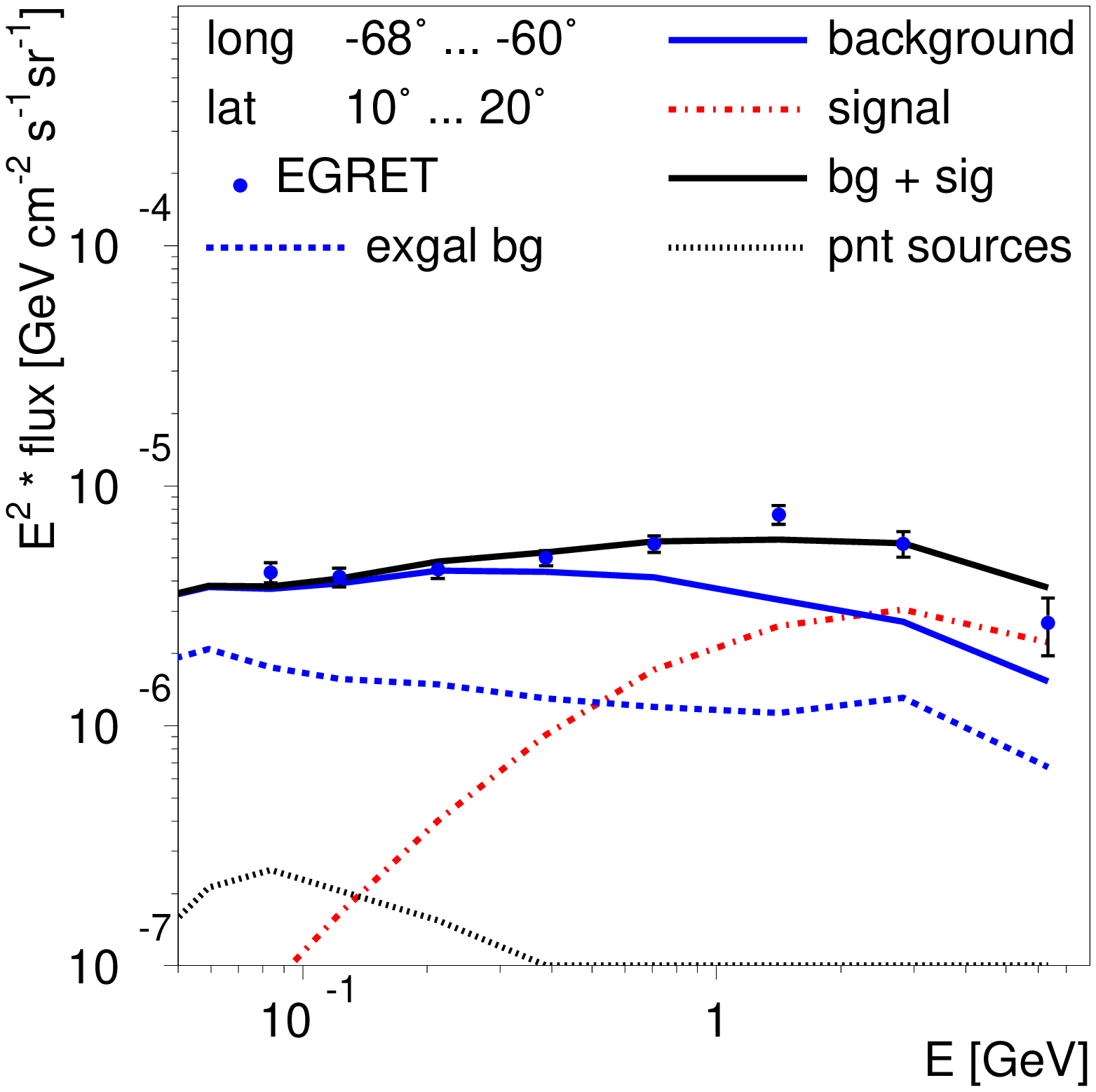}
    \includegraphics[width=0.21\textwidth]{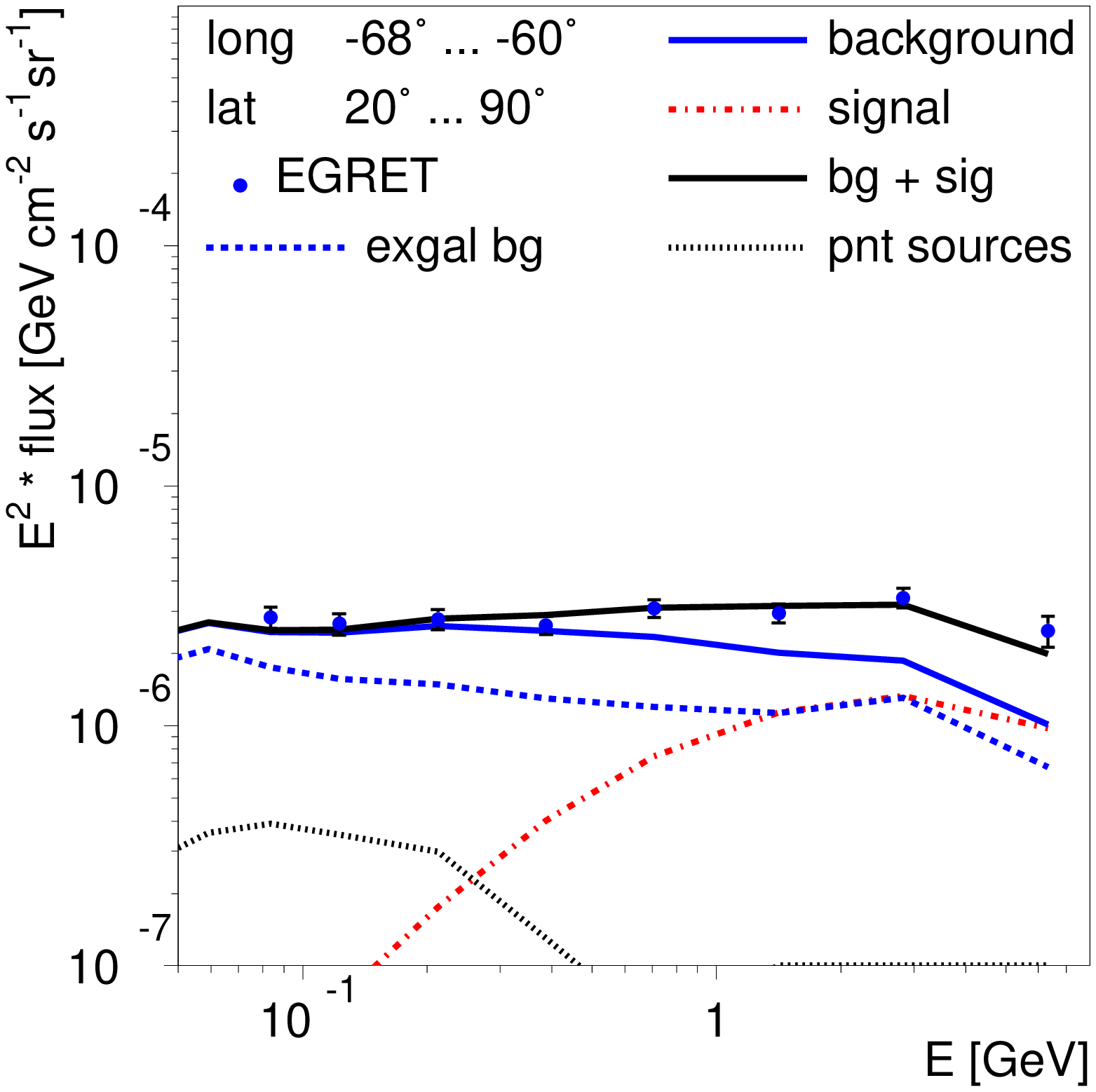}\\
    \hspace{-1cm}
    \begin{turn}{90} \framebox[0.21\textwidth][c]{{\scriptsize $-60^\circ<\mbox{long}<-52^\circ$}} \end{turn}
    \includegraphics[width=0.21\textwidth]{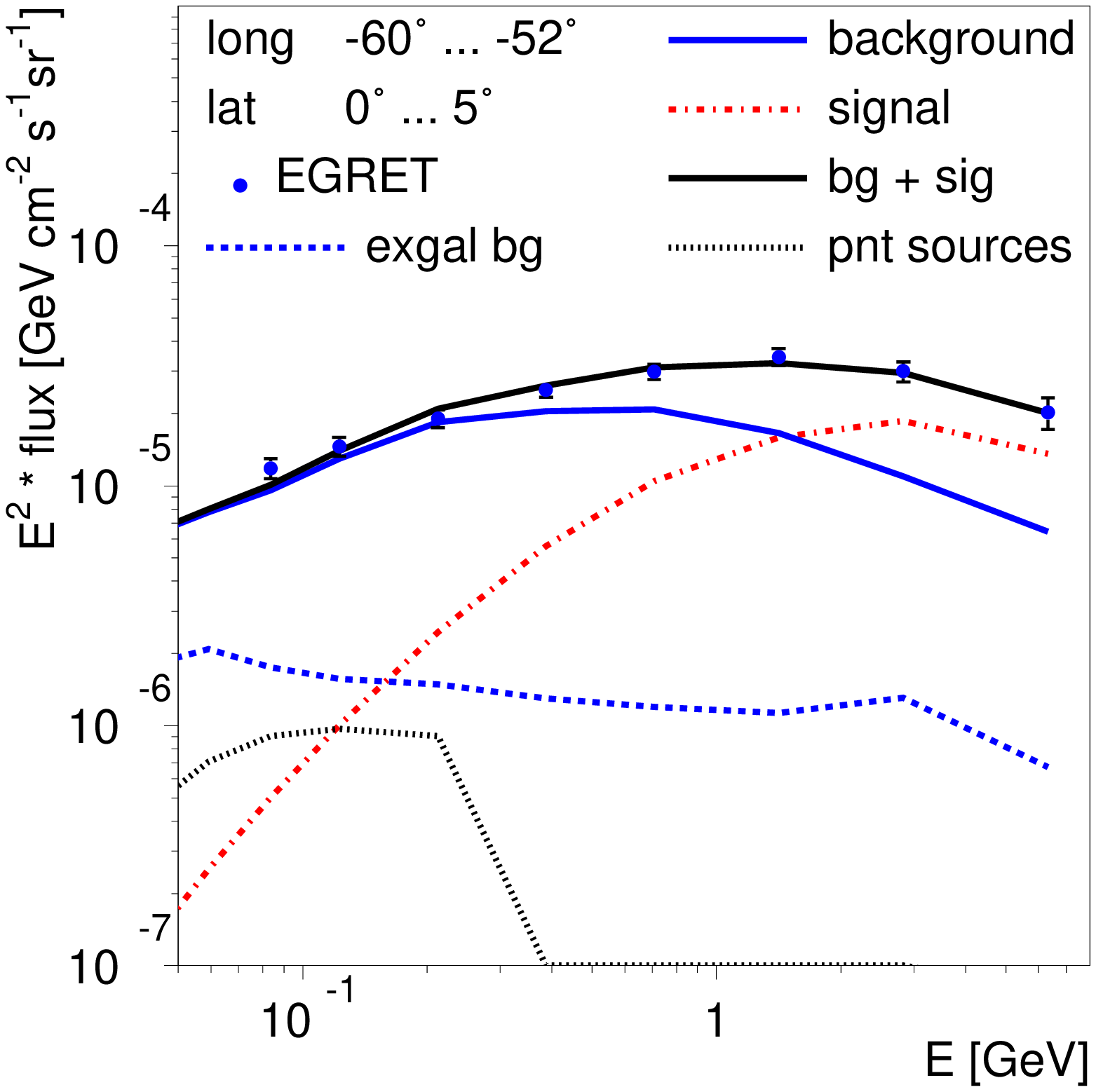}
    \includegraphics[width=0.21\textwidth]{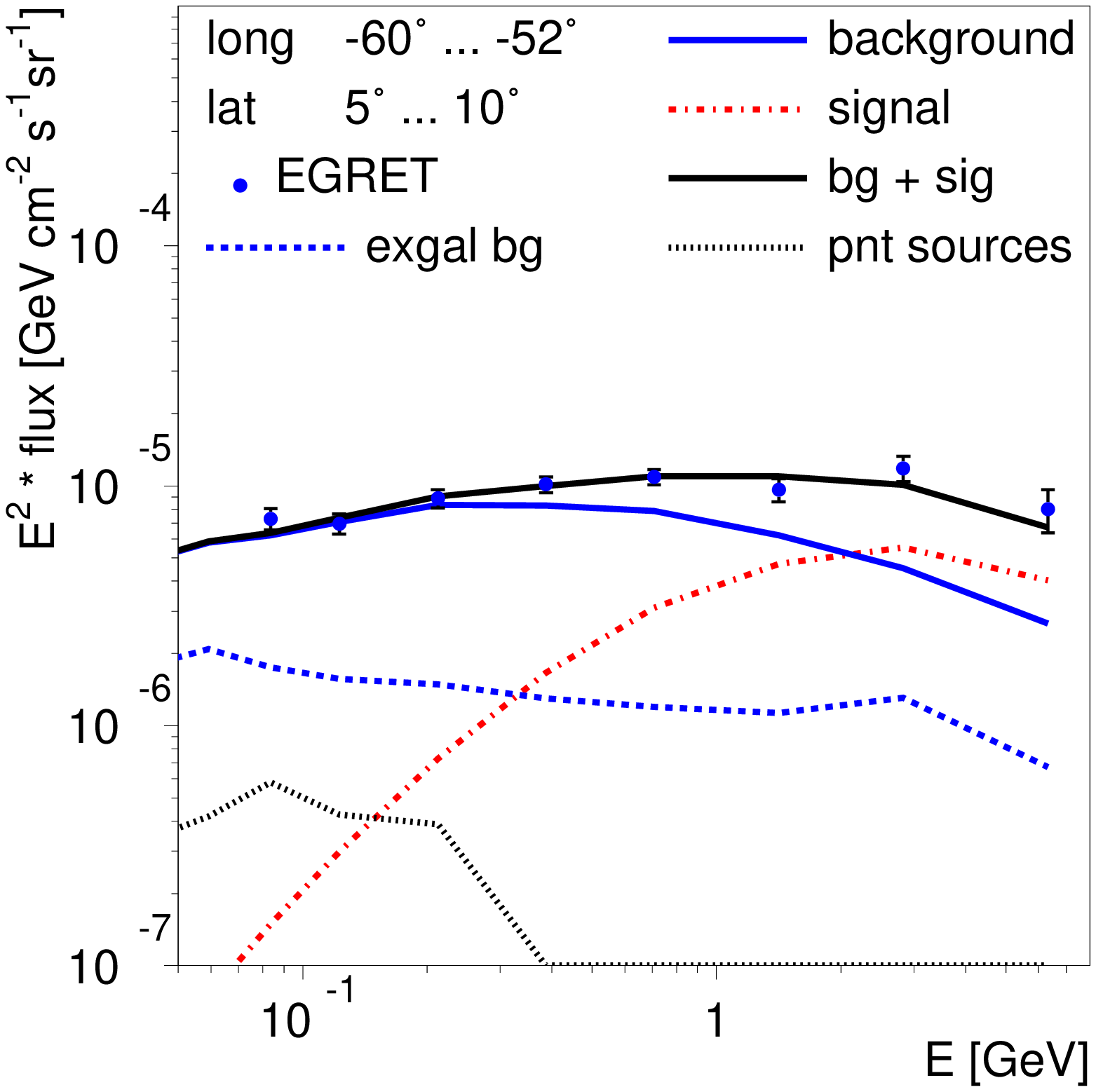}
    \includegraphics[width=0.21\textwidth]{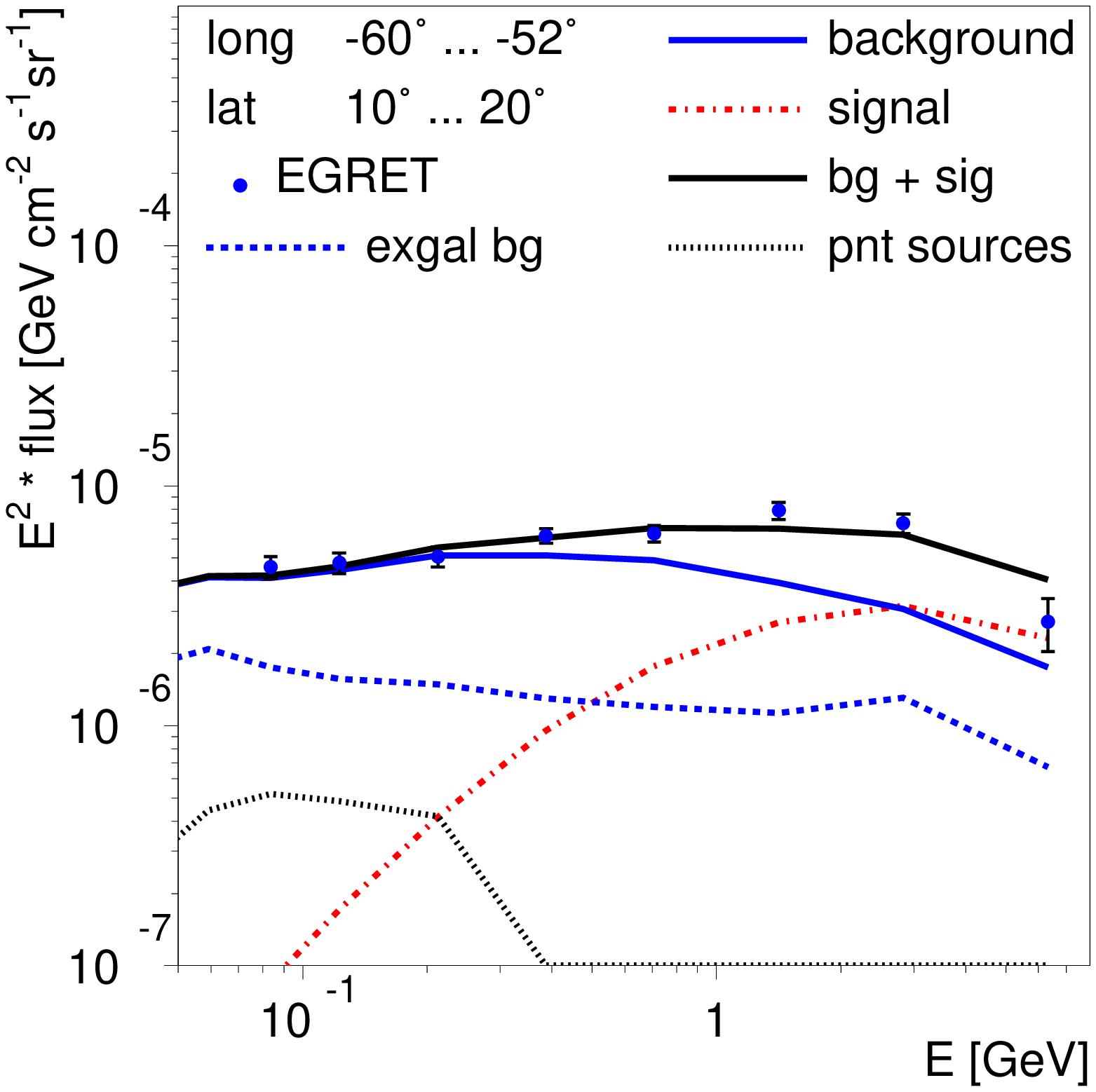}
    \includegraphics[width=0.21\textwidth]{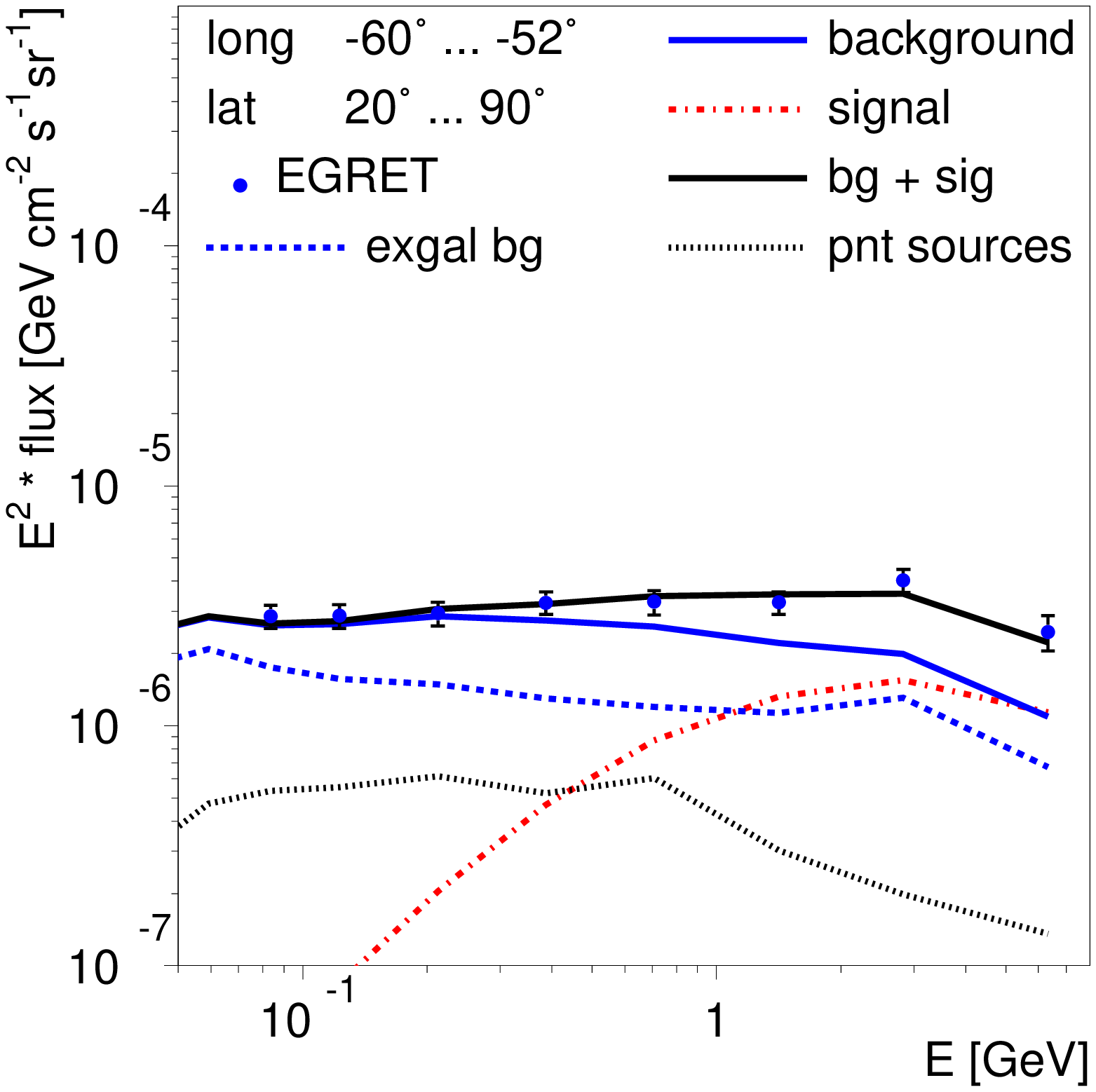}\\
    \hspace{-1cm}
    \begin{turn}{90} \framebox[0.21\textwidth][c]{{\scriptsize $-52^\circ<\mbox{long}<-44^\circ$}} \end{turn}
    \includegraphics[width=0.21\textwidth]{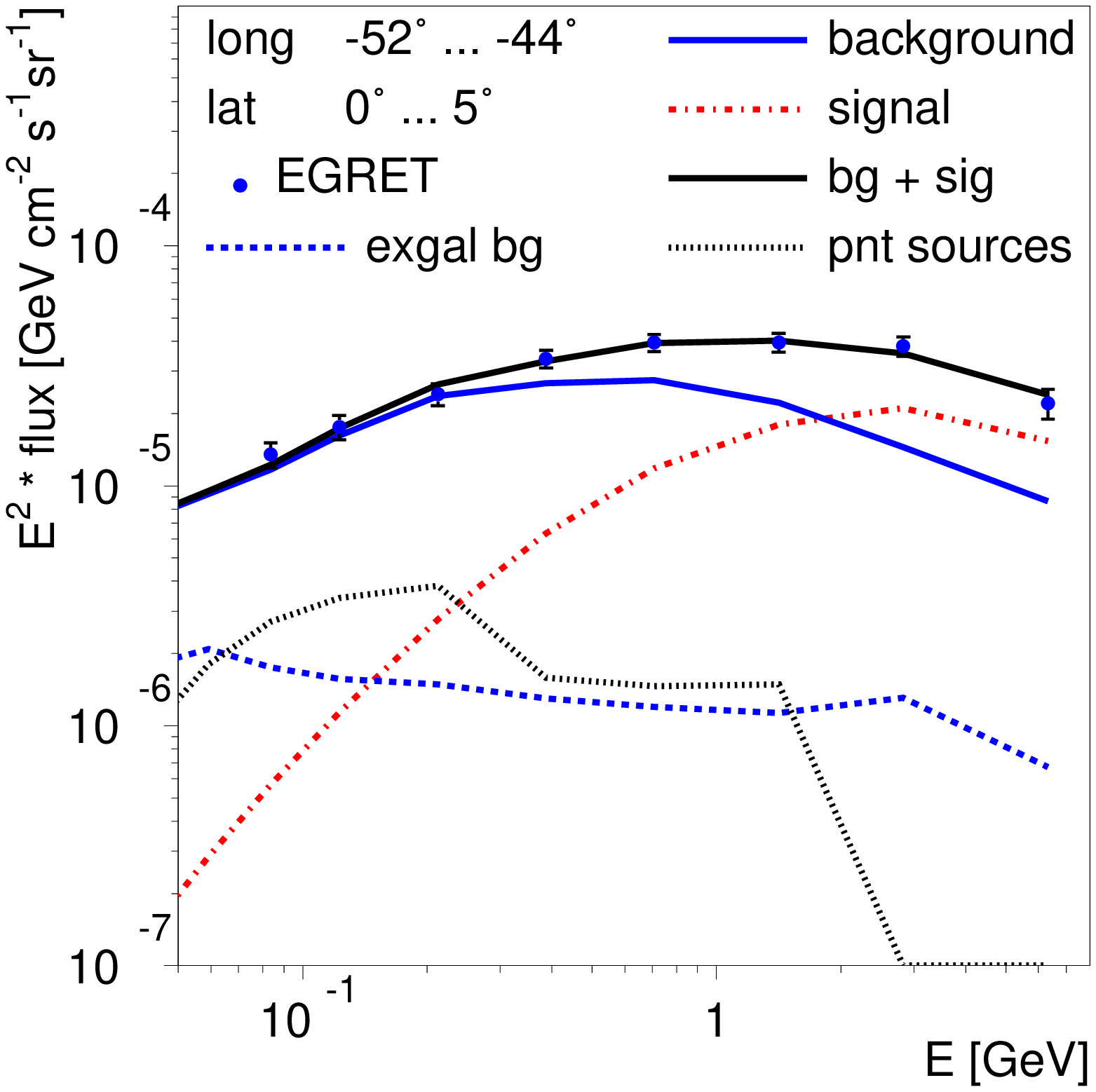}
    \includegraphics[width=0.21\textwidth]{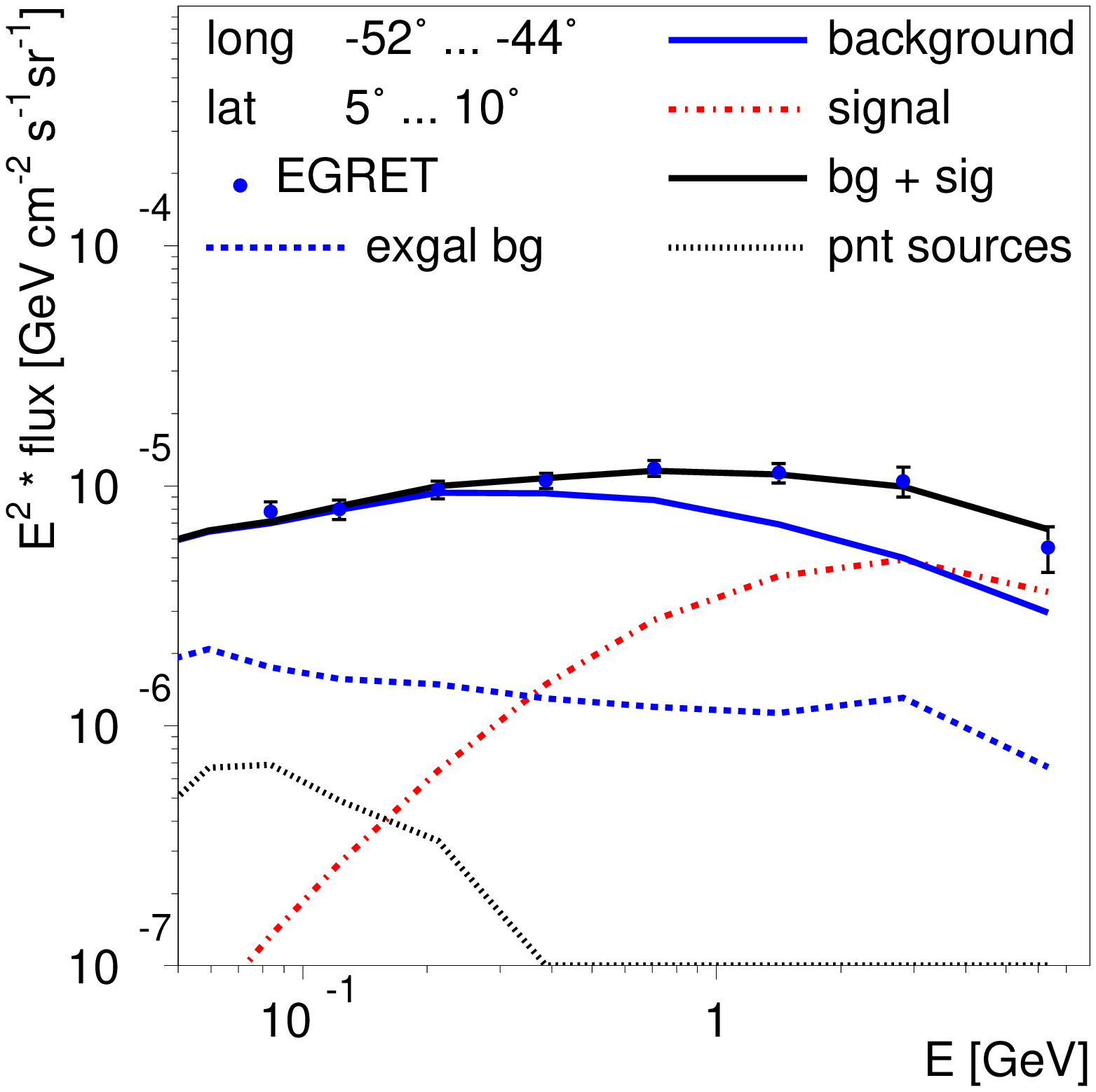}
    \includegraphics[width=0.21\textwidth]{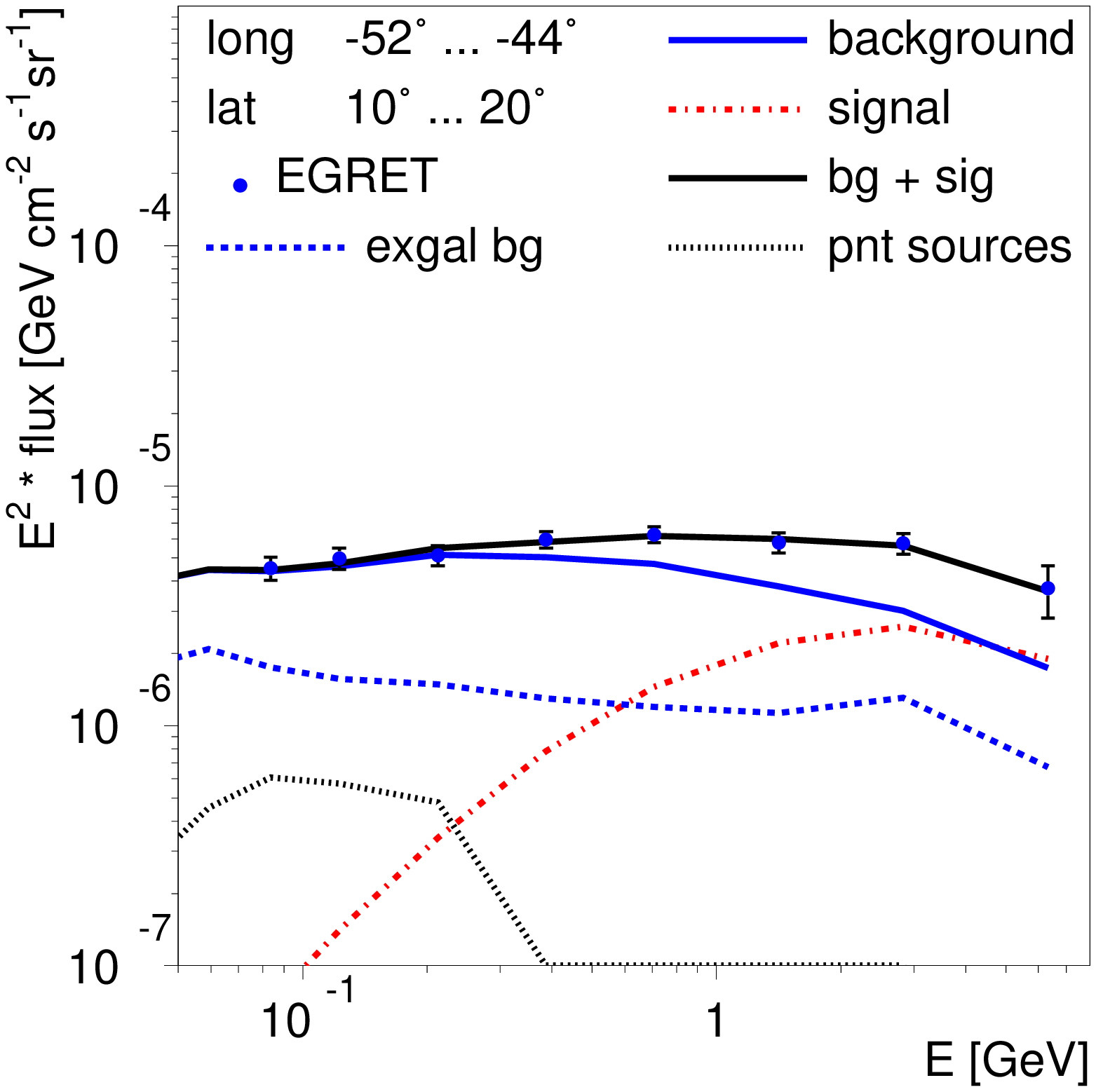}
    \includegraphics[width=0.21\textwidth]{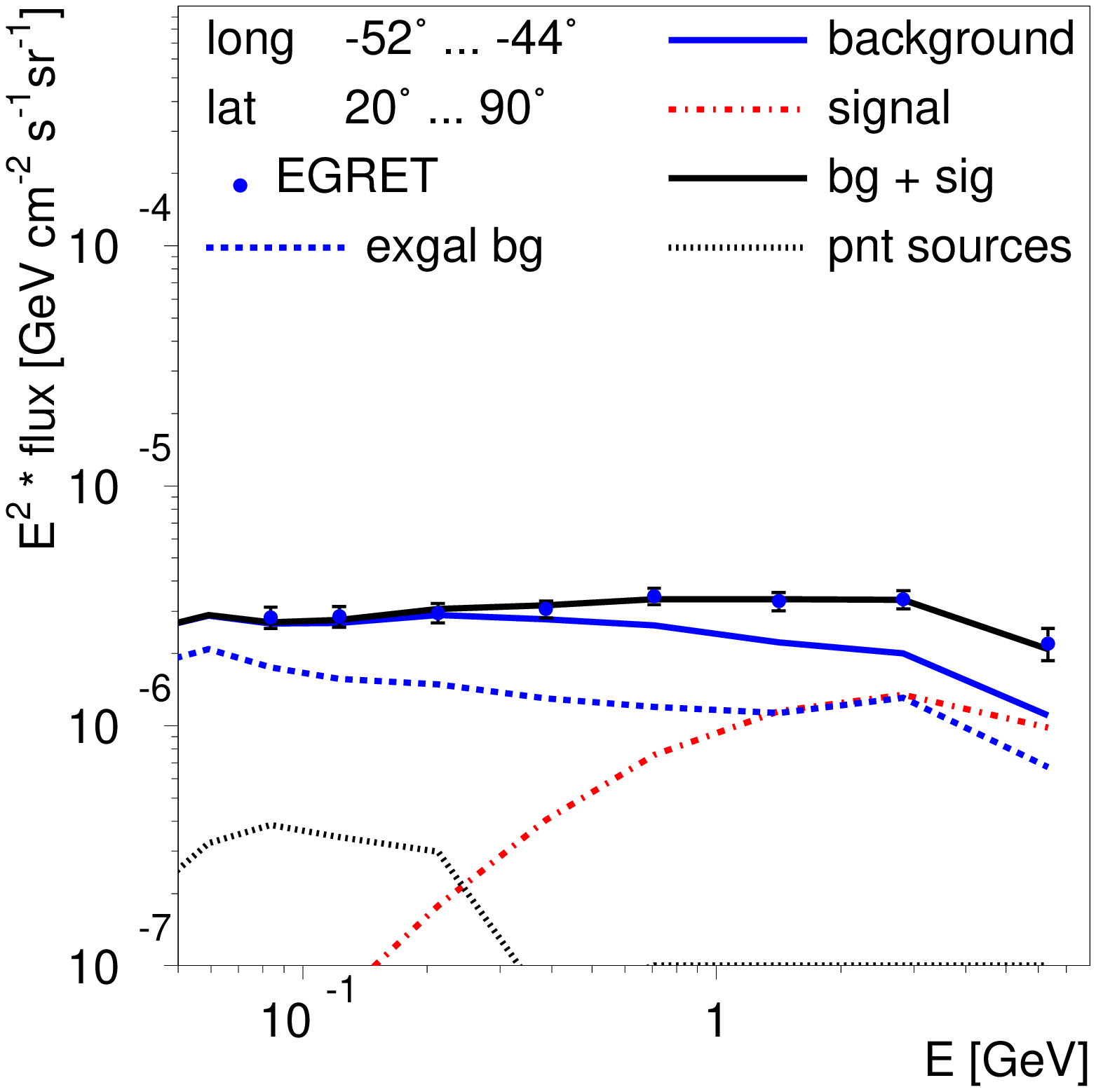}\\
    \hspace{-1cm}
    \begin{turn}{90} \framebox[0.21\textwidth][c]{{\scriptsize $-44^\circ<\mbox{long}<-36^\circ$}} \end{turn}
    \includegraphics[width=0.21\textwidth]{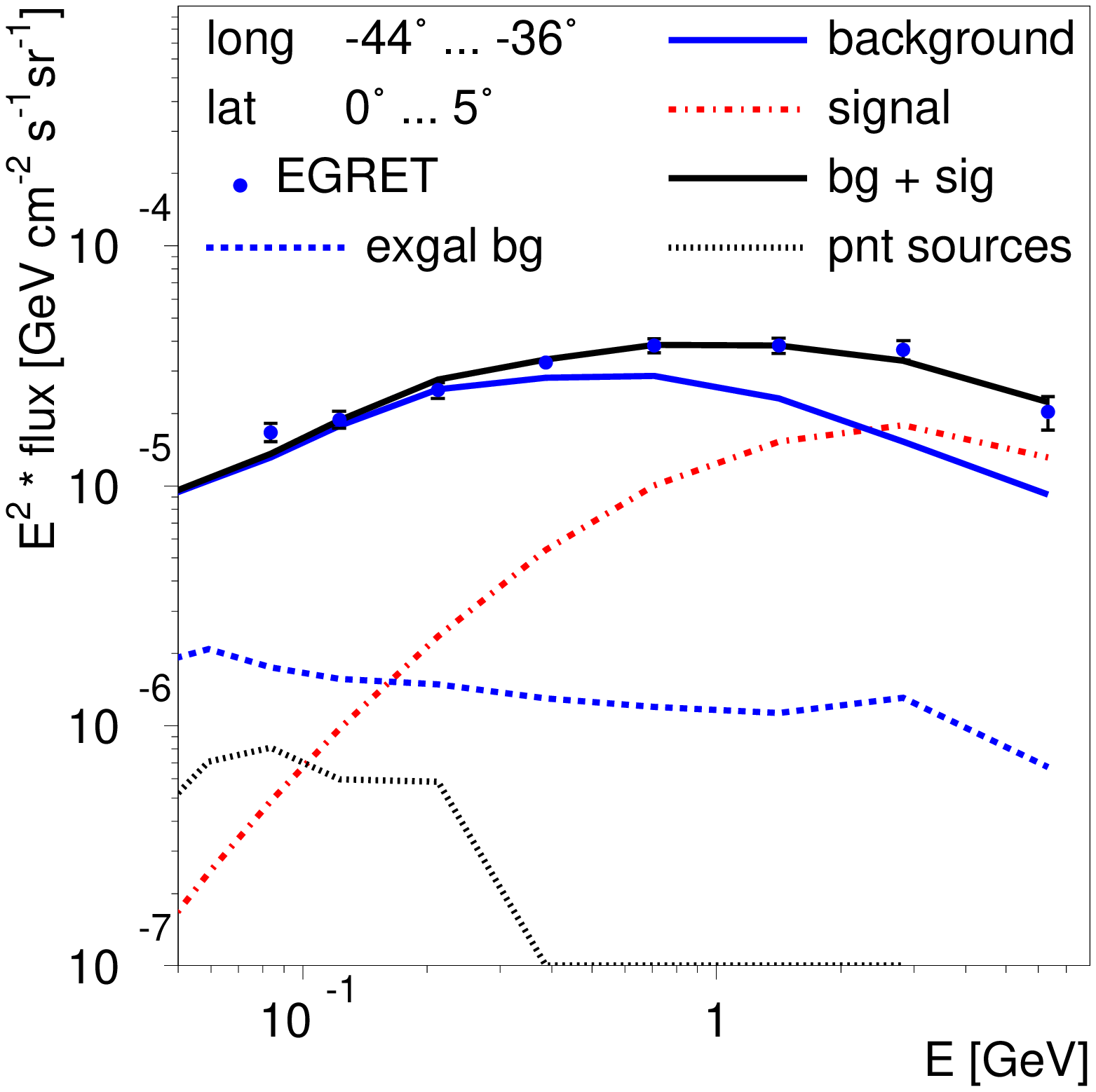}
    \includegraphics[width=0.21\textwidth]{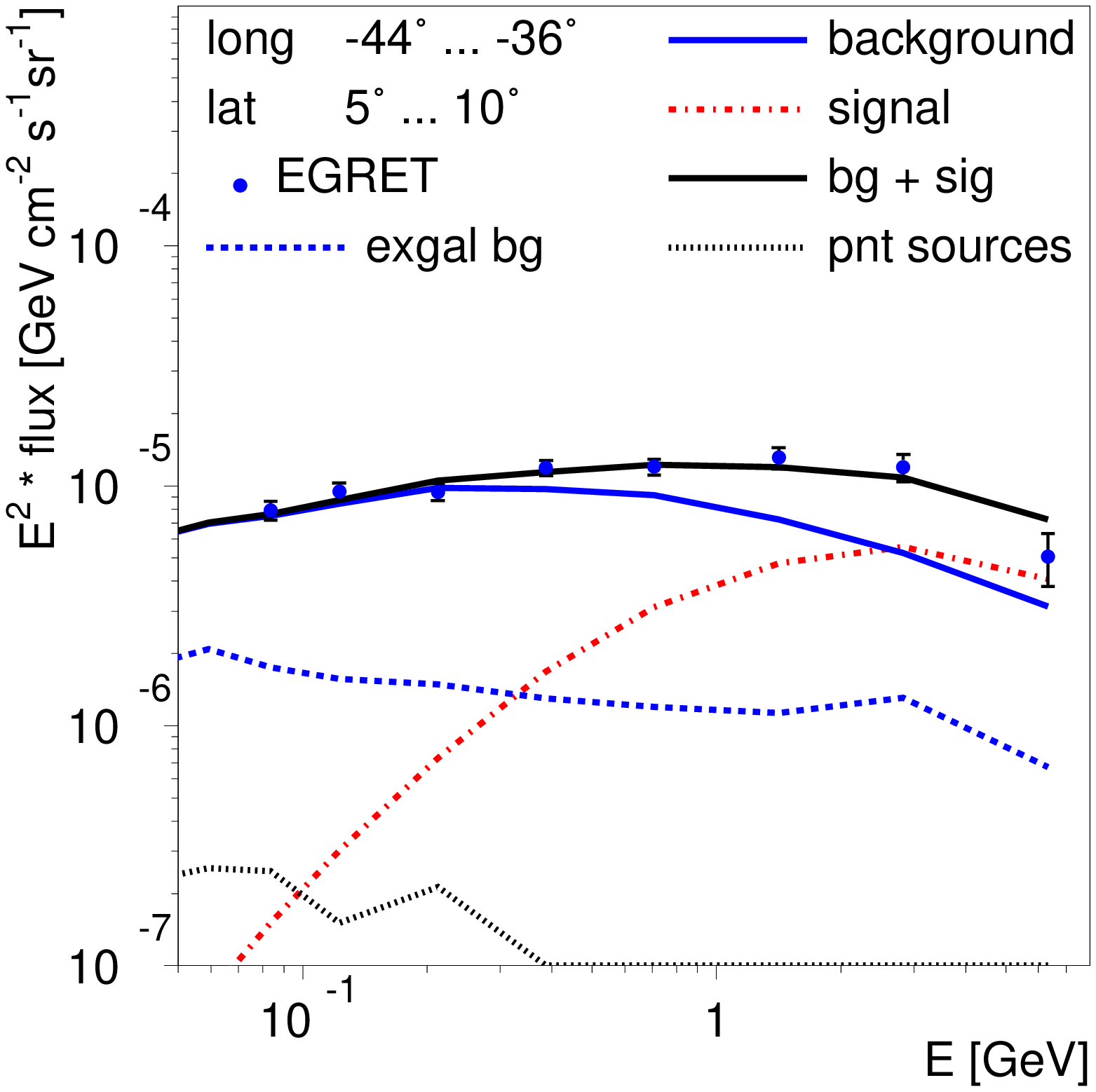}
    \includegraphics[width=0.21\textwidth]{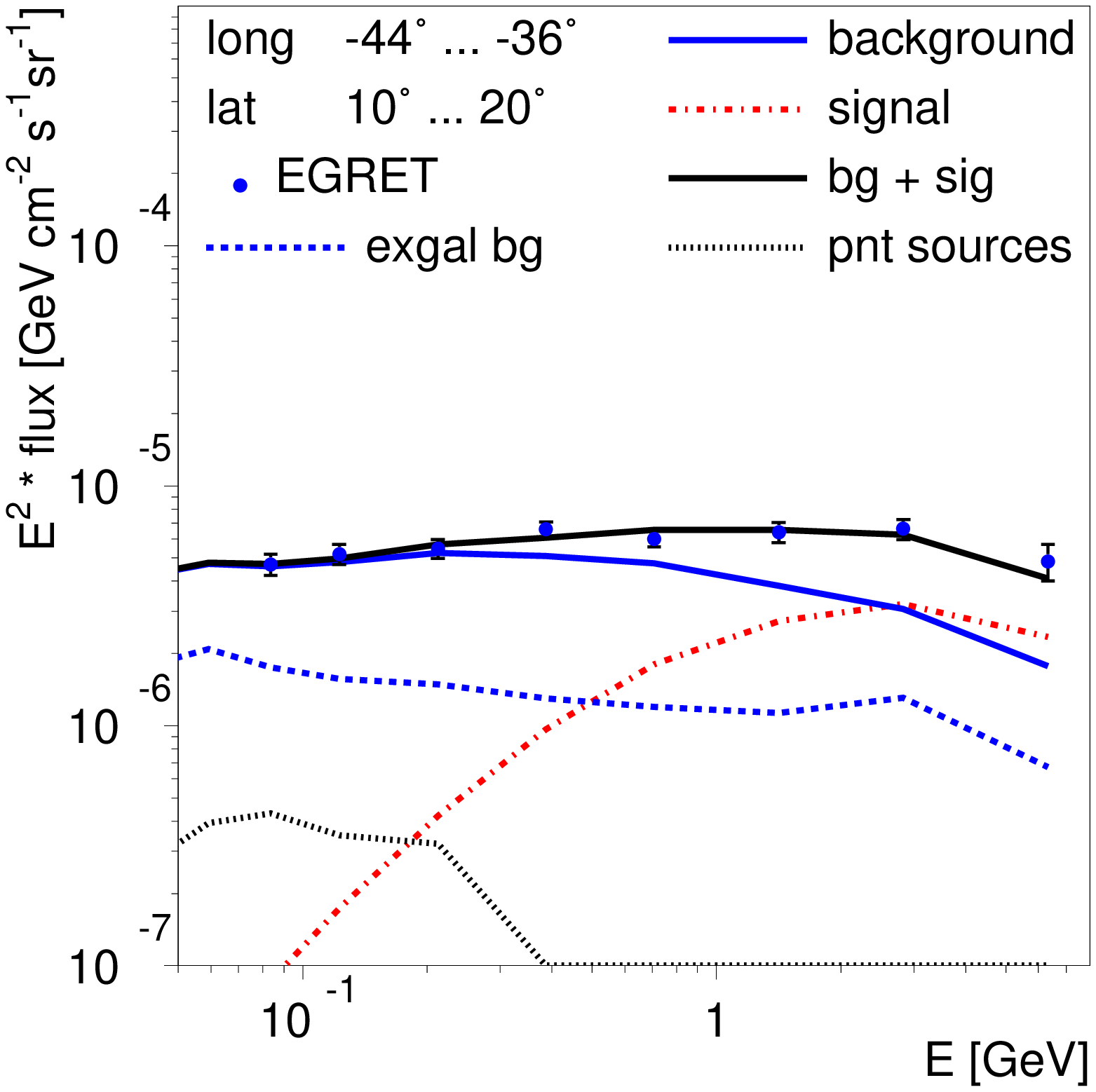}
    \includegraphics[width=0.21\textwidth]{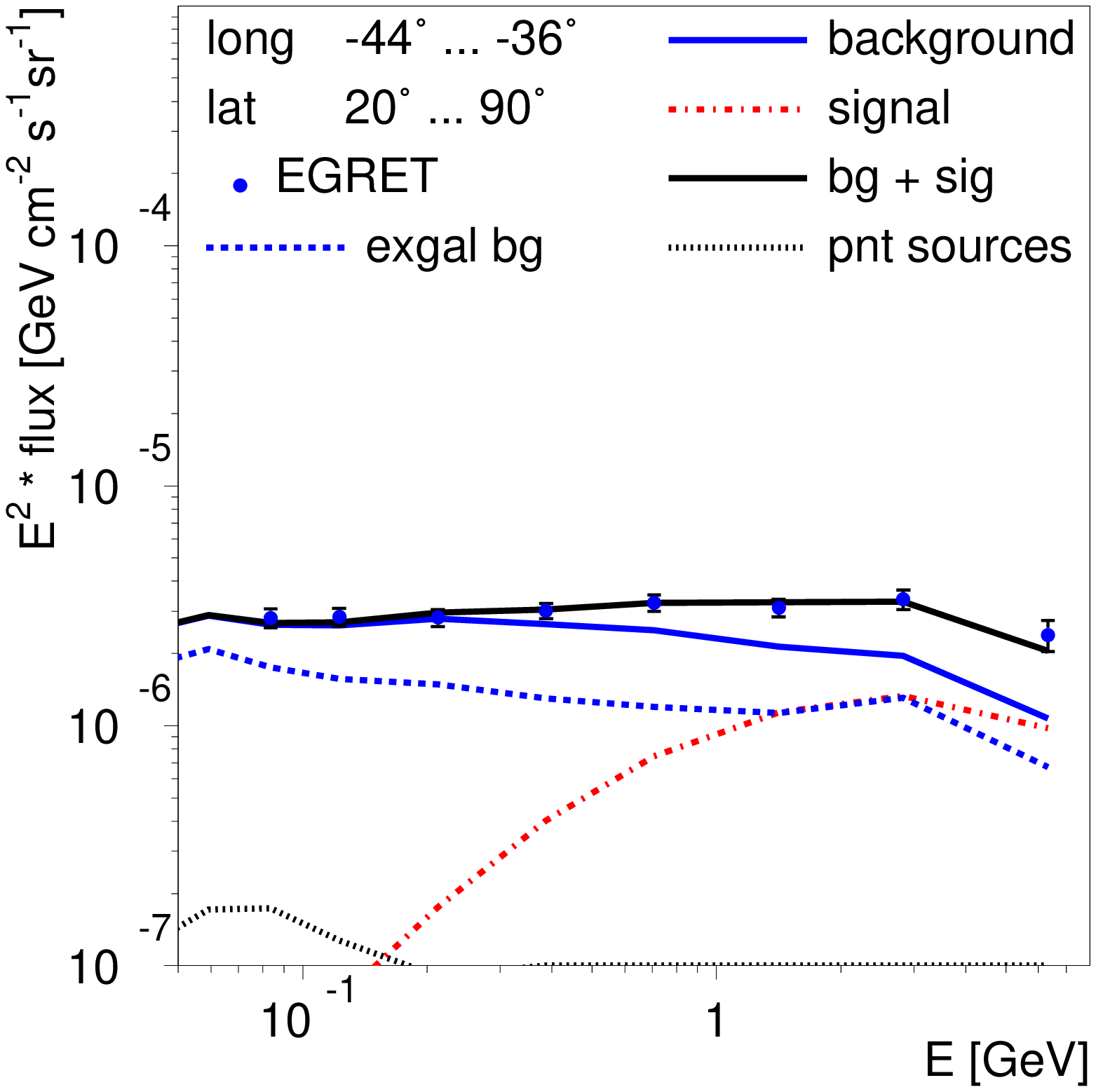}\\
  \end{center}
  \clearpage
  \begin{center}
    \framebox[0.21\textwidth][c]{$\vert \mbox{lat}\vert<5^\circ$}
    \framebox[0.21\textwidth][c]{$5^\circ<\vert \mbox{lat}\vert<10^\circ$}
    \framebox[0.21\textwidth][c]{$10^\circ<\vert \mbox{lat}\vert<20^\circ$}
    \framebox[0.21\textwidth][c]{$20^\circ<\vert \mbox{lat}\vert<90^\circ$}\\
    \hspace{-1cm}
    \begin{turn}{90} \framebox[0.21\textwidth][c]{{\scriptsize $-36^\circ<\mbox{long}<-28^\circ$}} \end{turn}
    \includegraphics[width=0.21\textwidth]{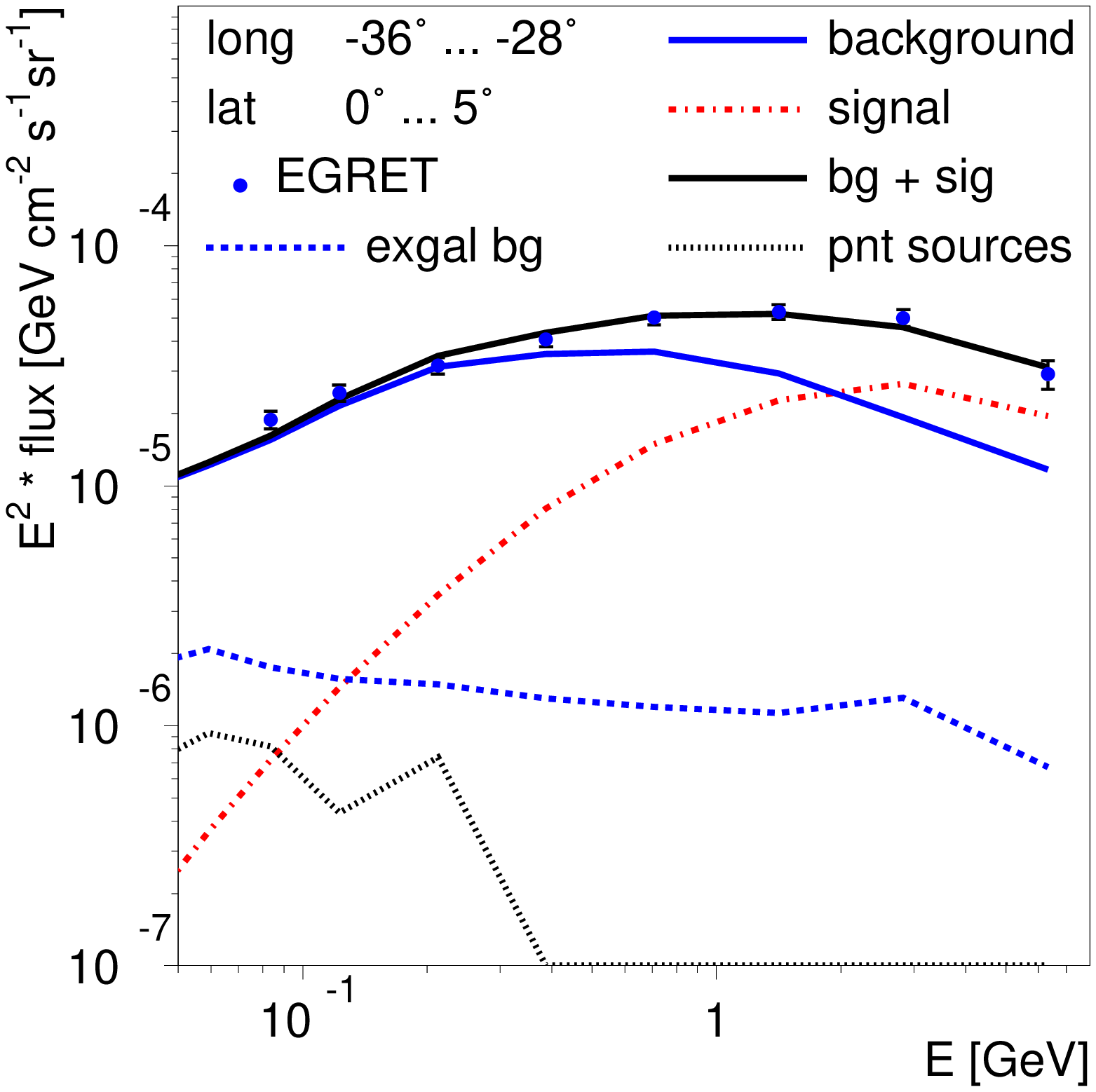}
    \includegraphics[width=0.21\textwidth]{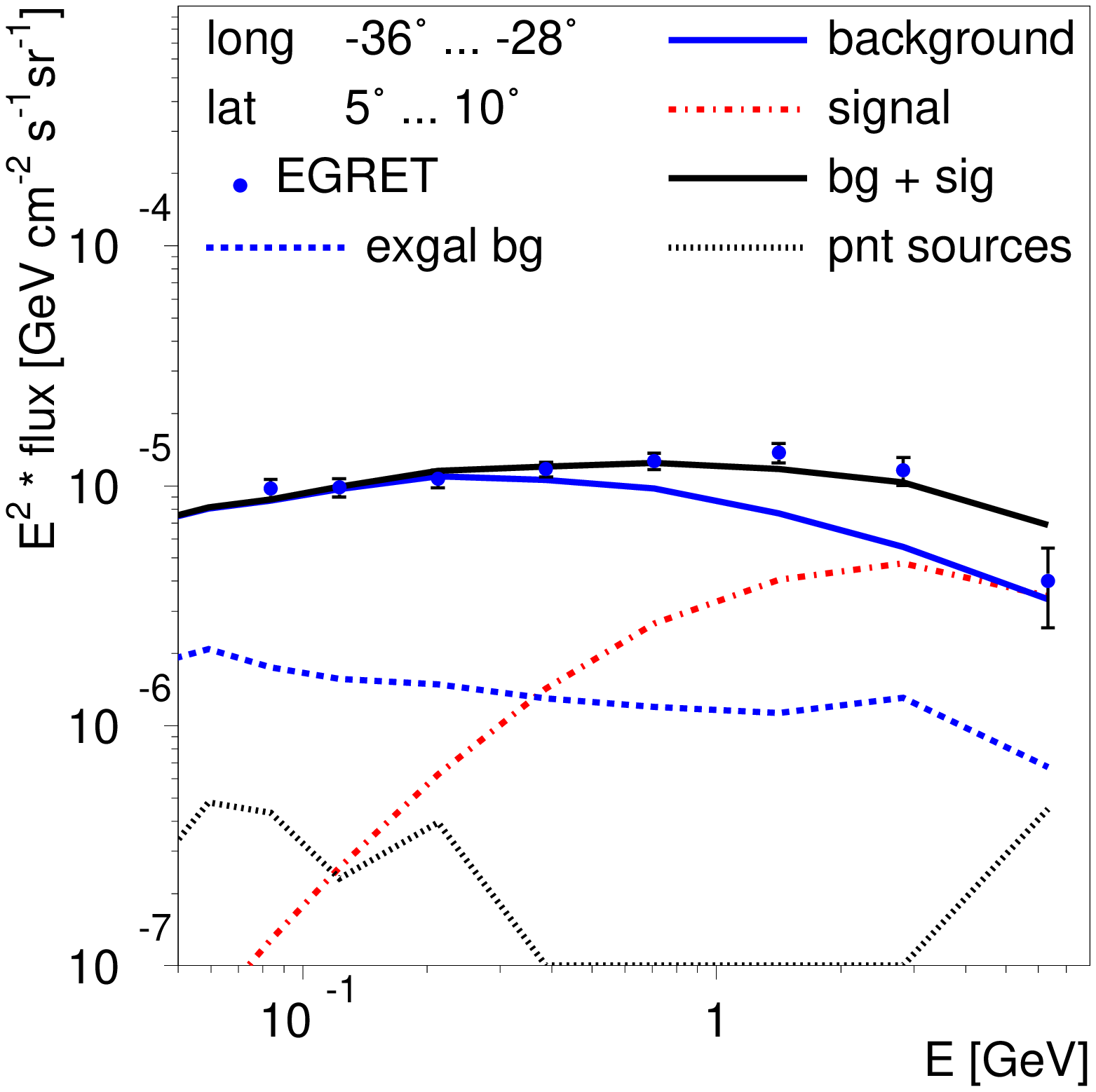}
    \includegraphics[width=0.21\textwidth]{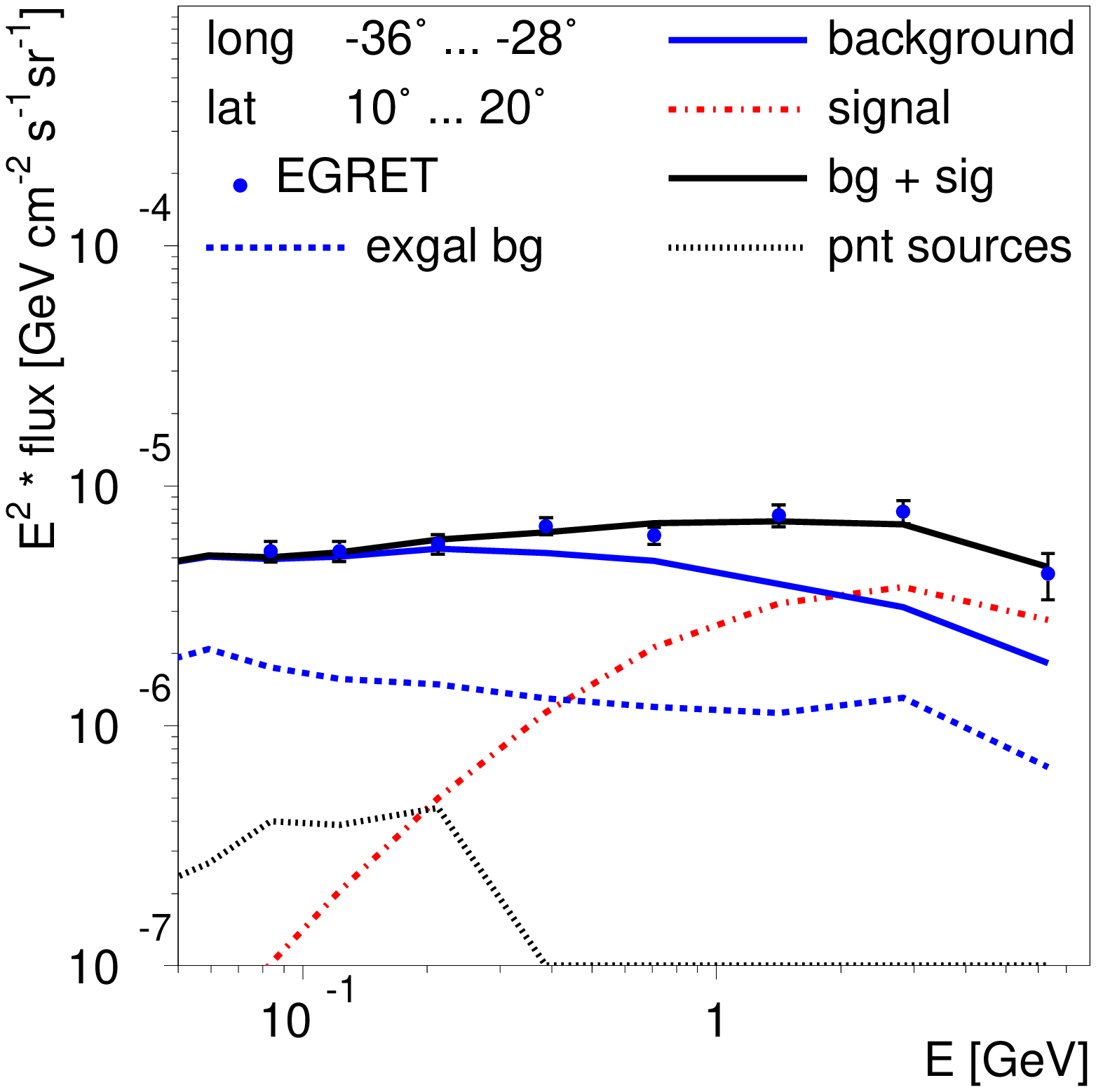}
    \includegraphics[width=0.21\textwidth]{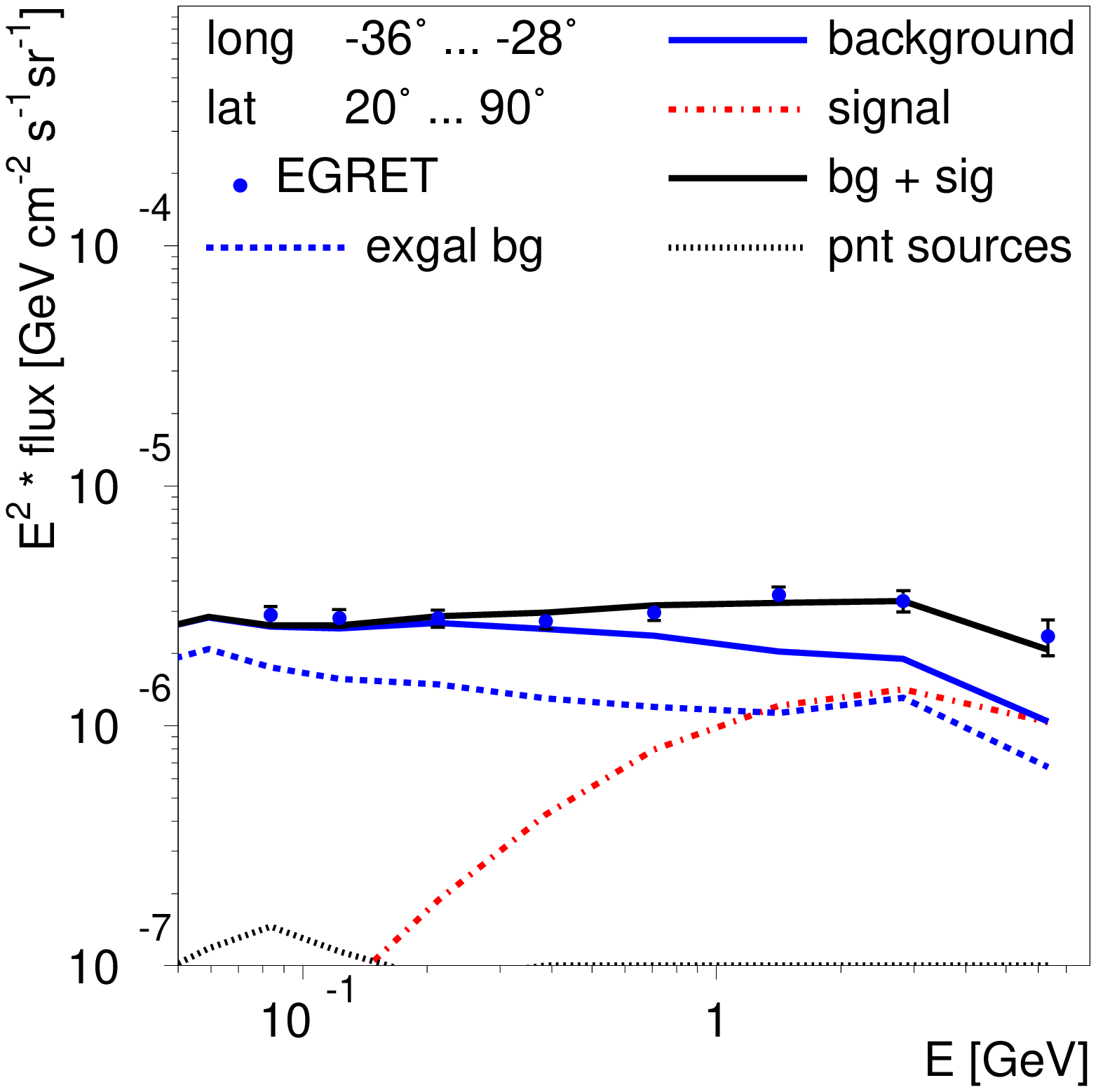}\\
    \hspace{-1cm}
    \begin{turn}{90} \framebox[0.21\textwidth][c]{{\scriptsize $-28^\circ<\mbox{long}<-20^\circ$}} \end{turn}
    \includegraphics[width=0.21\textwidth]{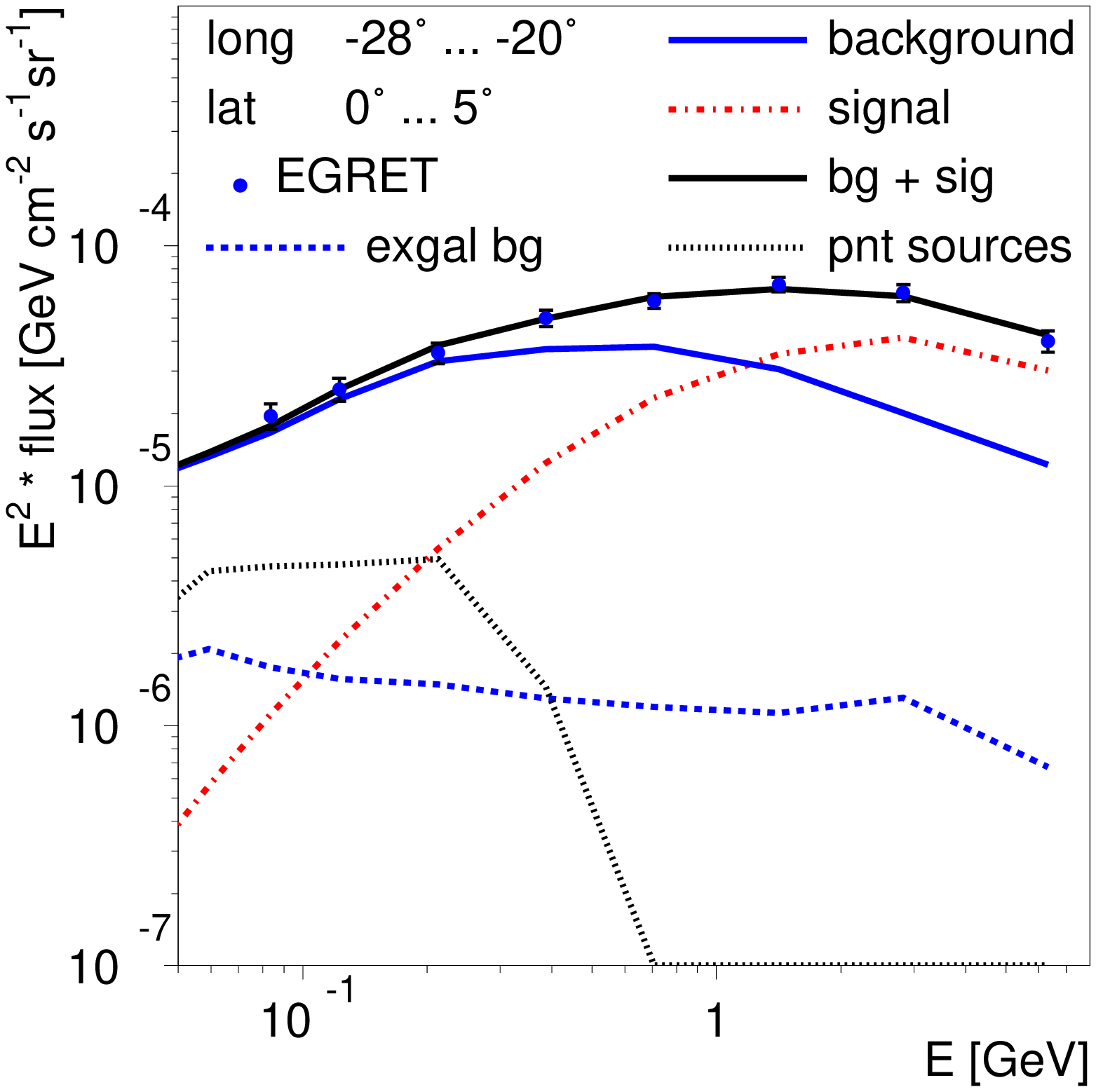}
    \includegraphics[width=0.21\textwidth]{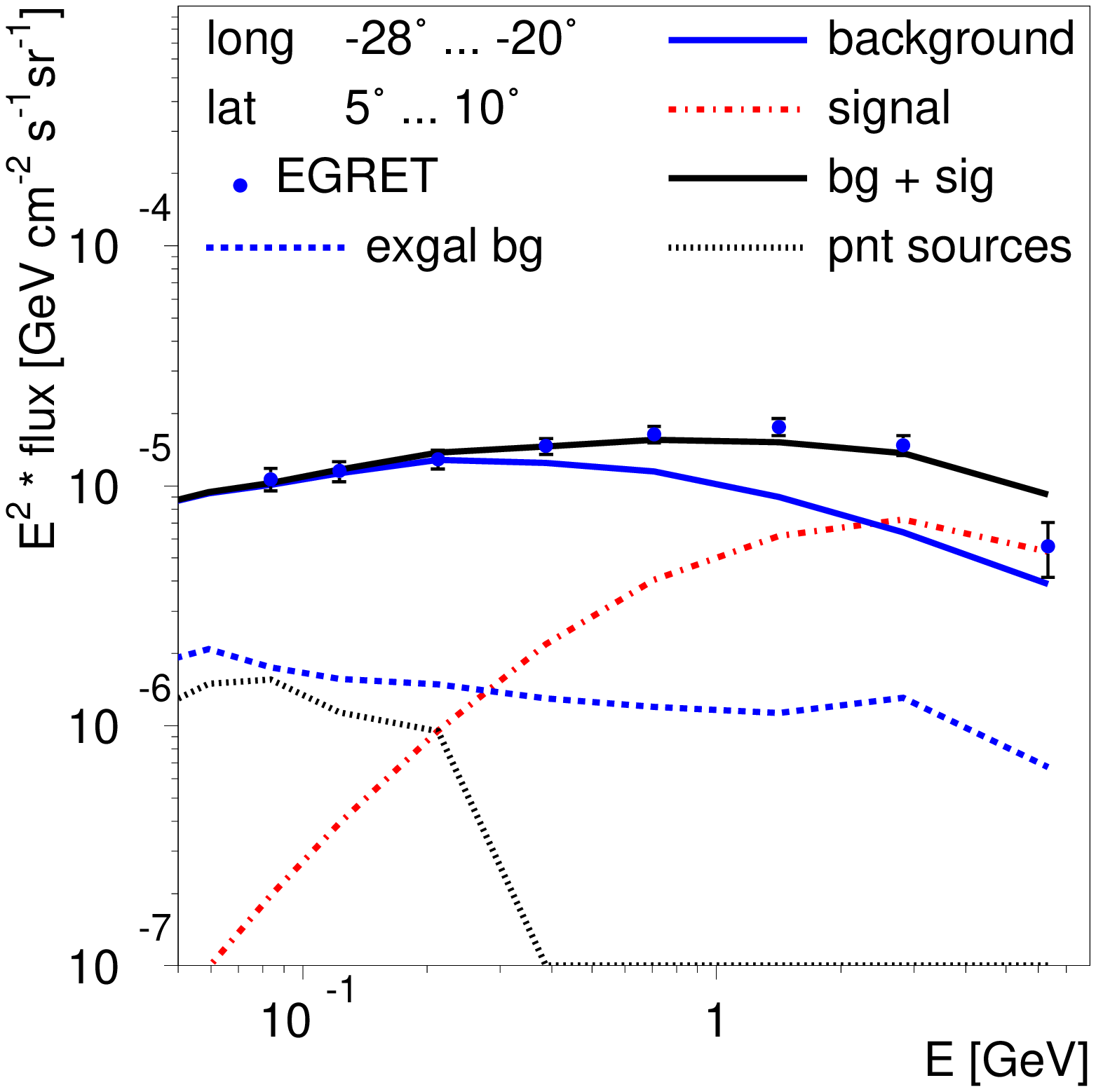}
    \includegraphics[width=0.21\textwidth]{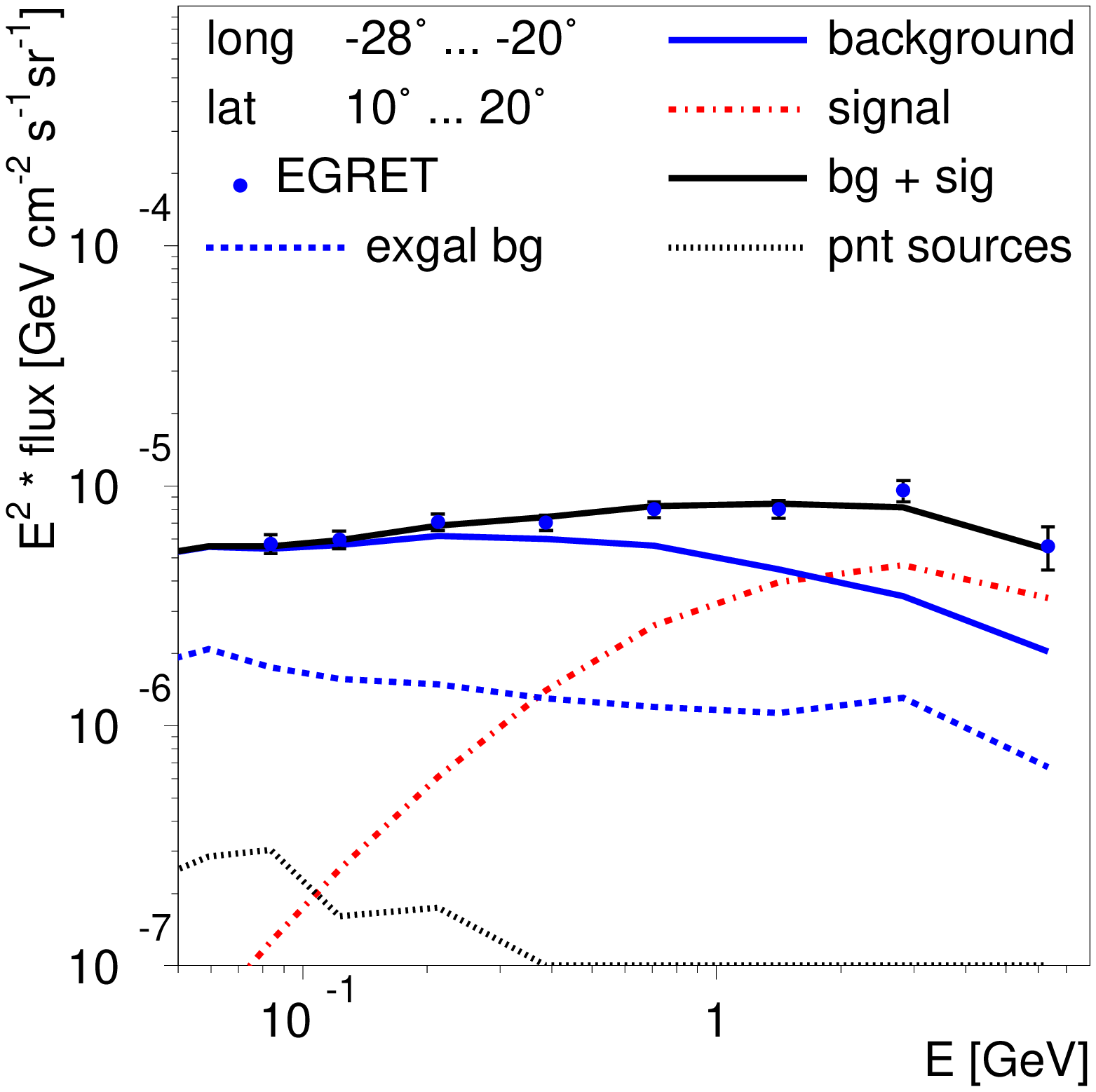}
    \includegraphics[width=0.21\textwidth]{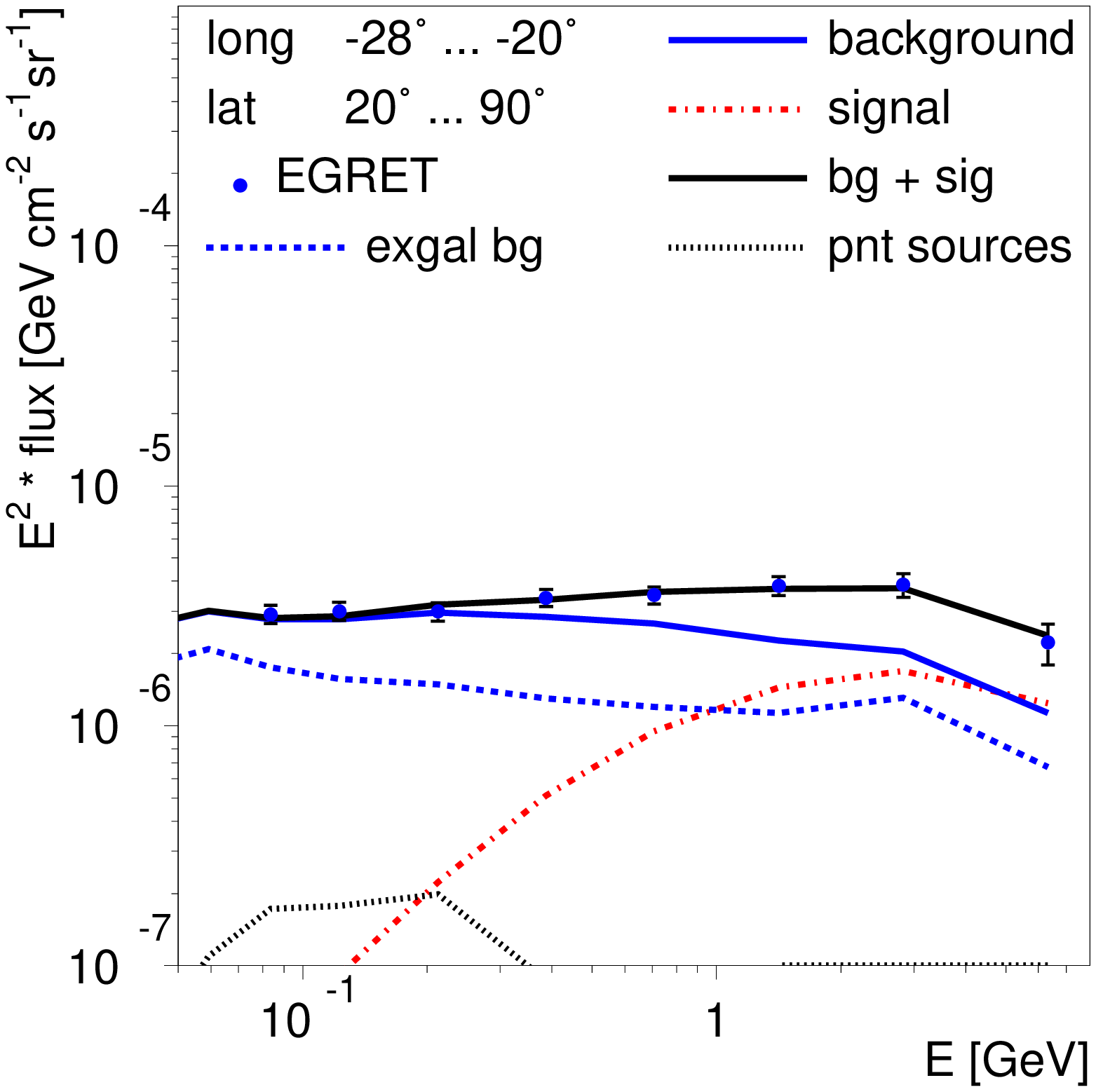}\\
    \hspace{-1cm}
    \begin{turn}{90} \framebox[0.21\textwidth][c]{{\scriptsize $-20^\circ<\mbox{long}<-12^\circ$}} \end{turn}
    \includegraphics[width=0.21\textwidth]{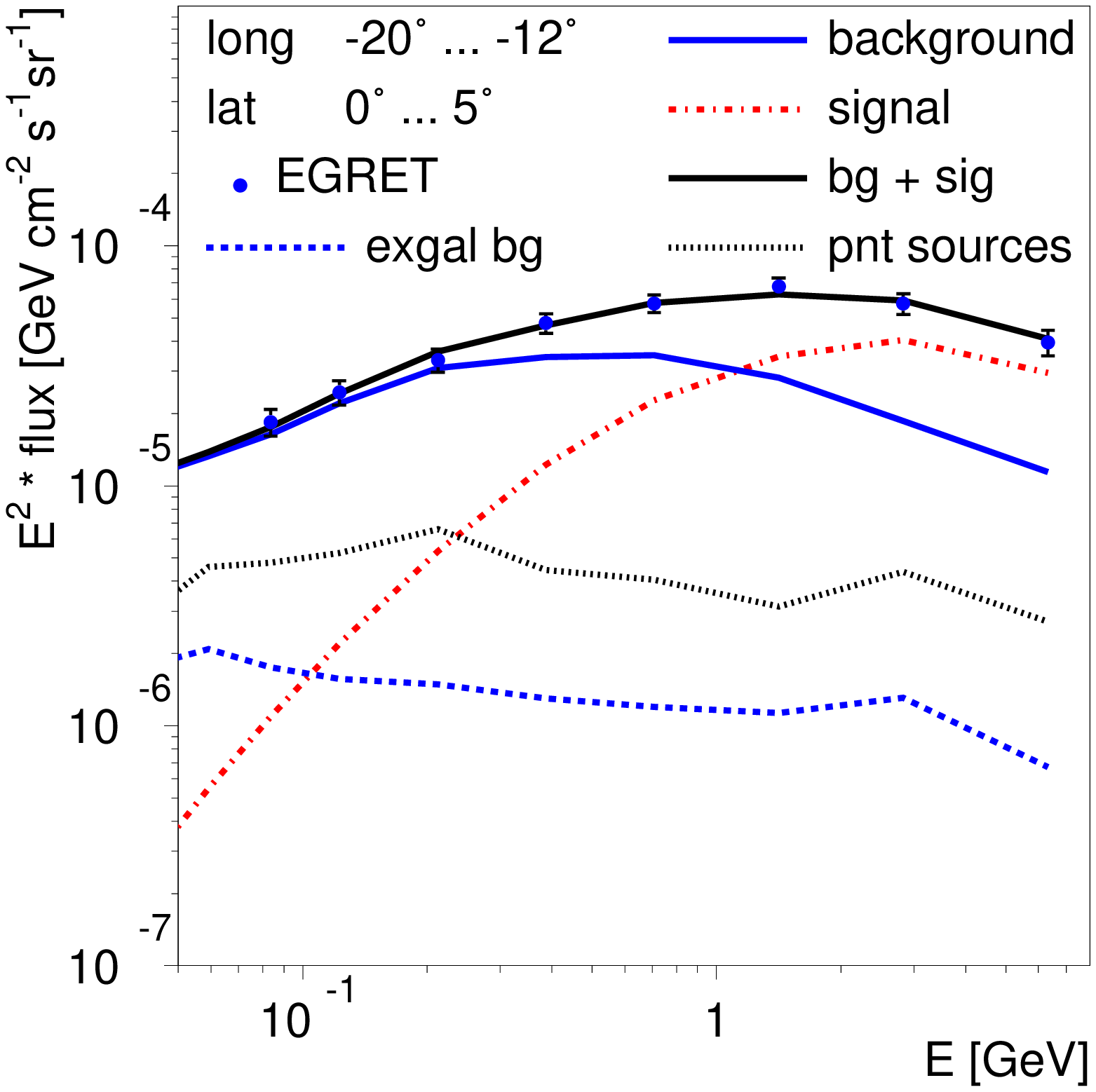}
    \includegraphics[width=0.21\textwidth]{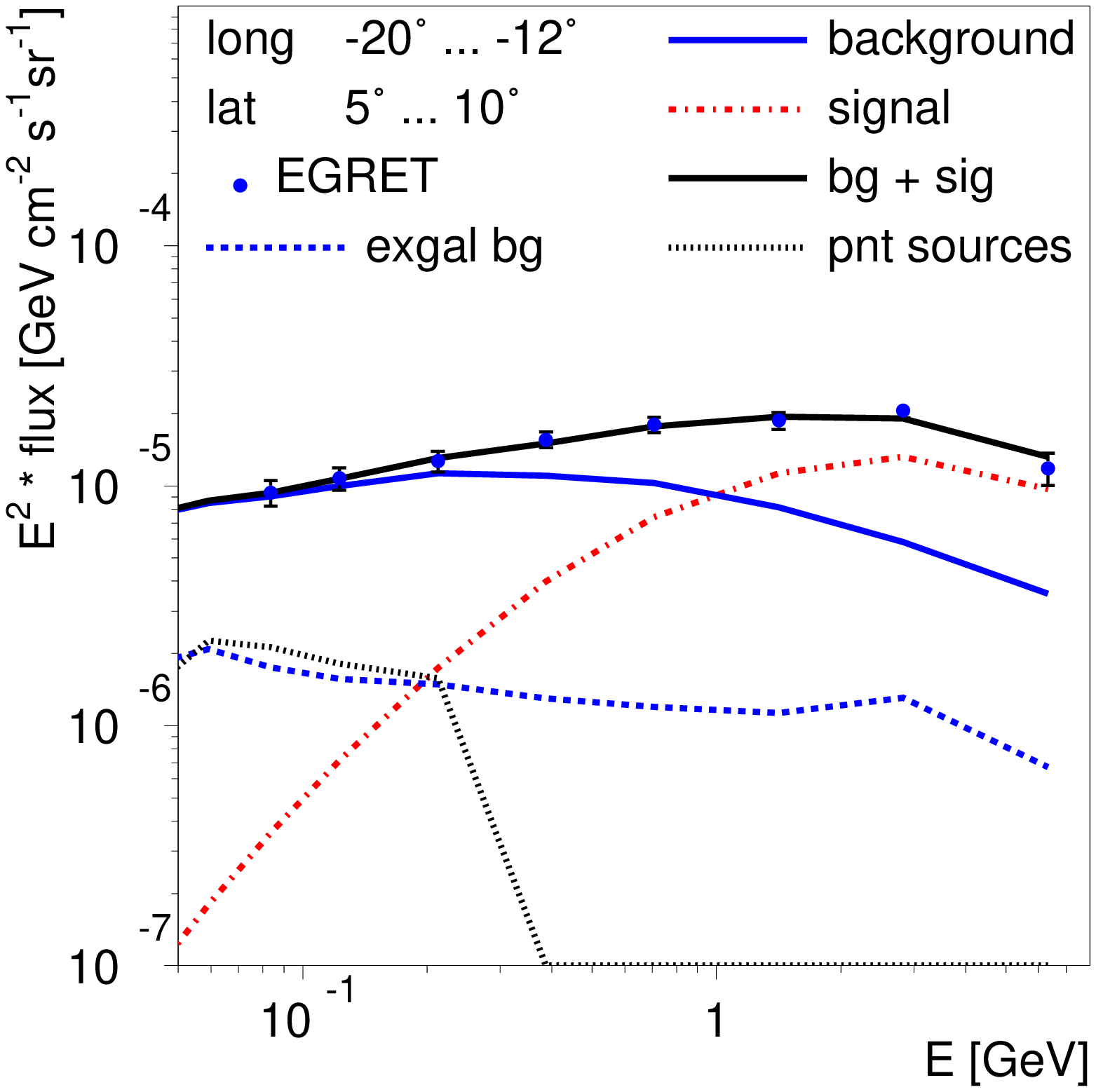}
    \includegraphics[width=0.21\textwidth]{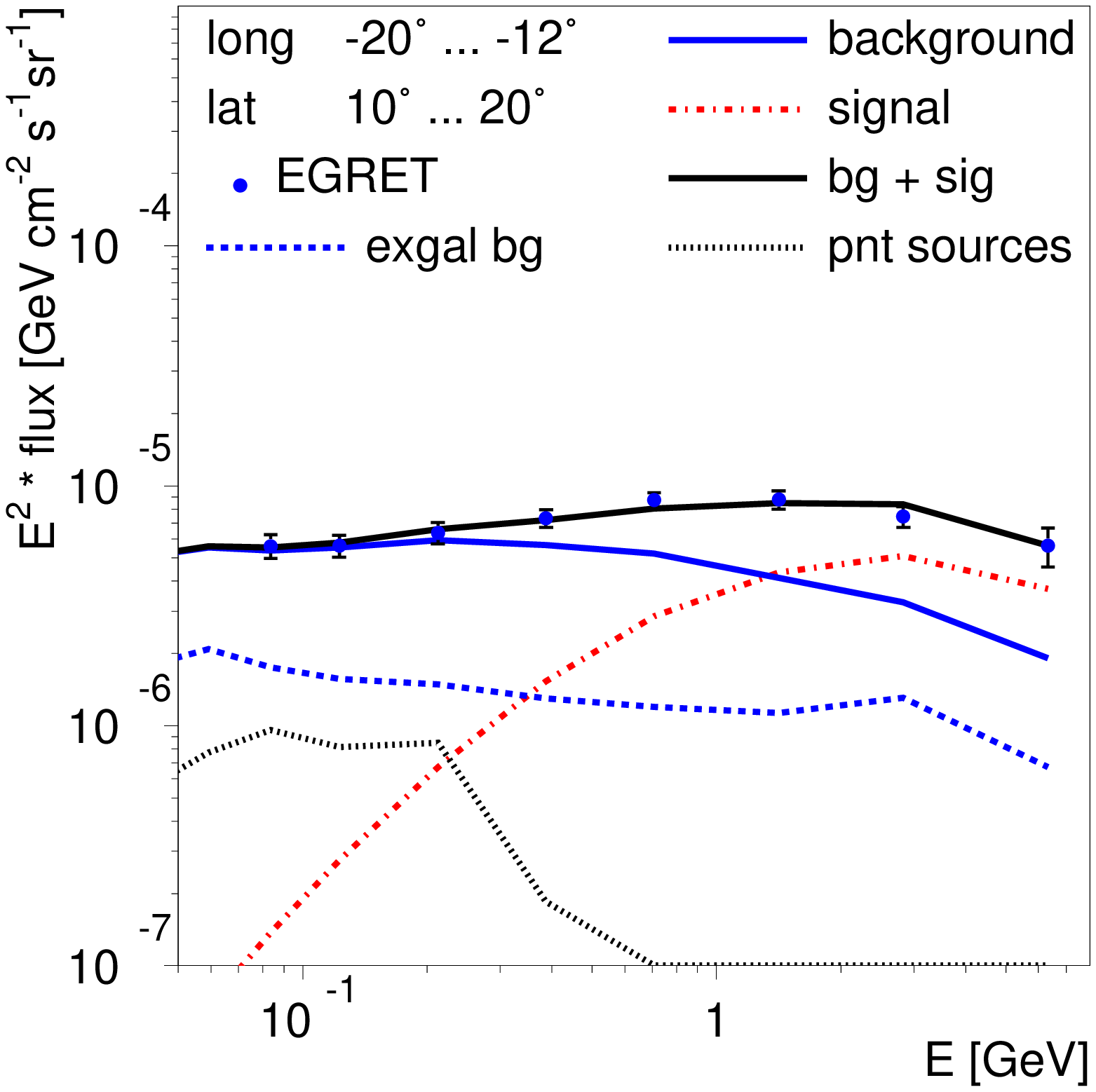}
    \includegraphics[width=0.21\textwidth]{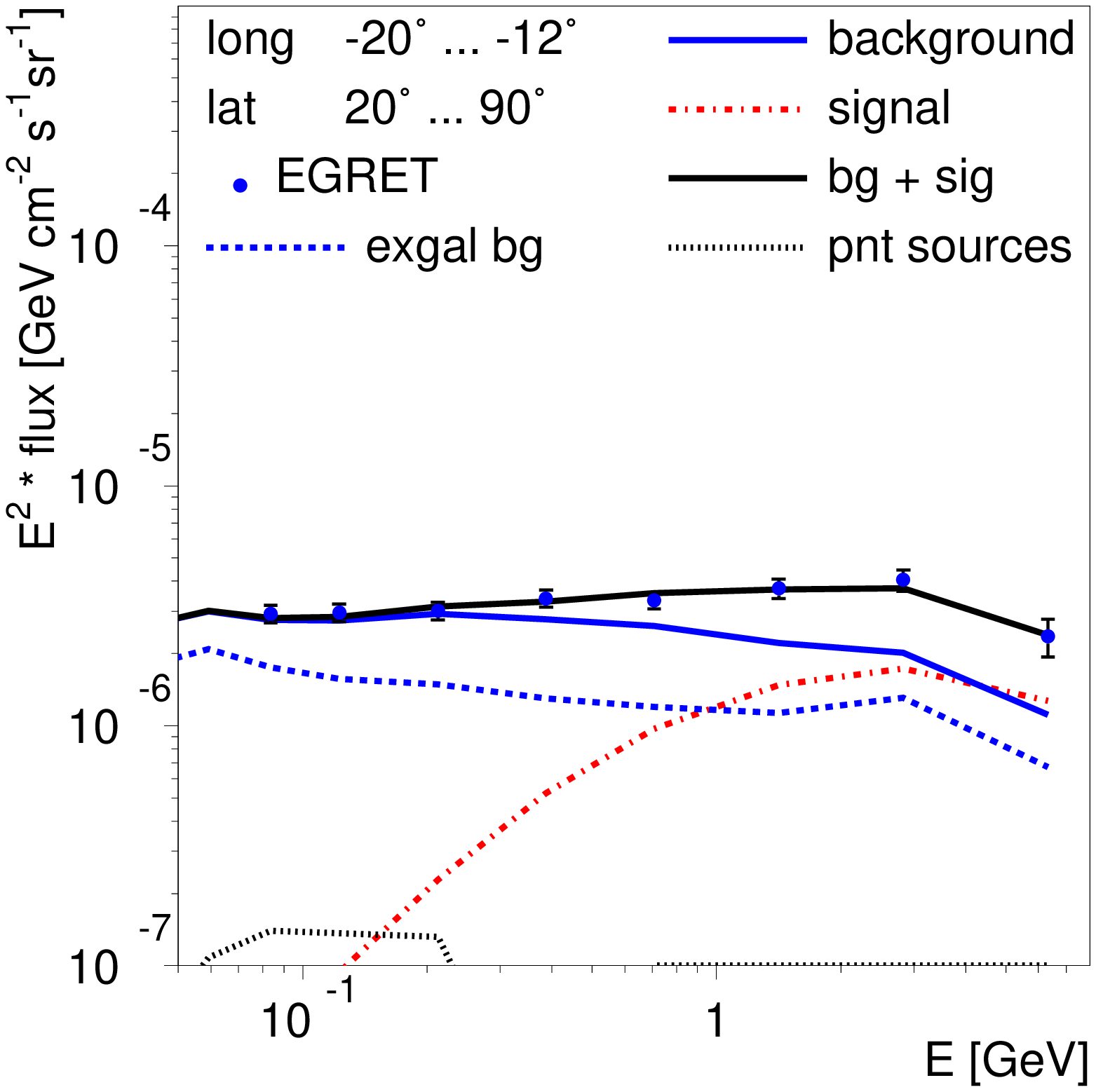}\\
    \hspace{-1cm}
    \begin{turn}{90} \framebox[0.21\textwidth][c]{{\scriptsize $-12^\circ<\mbox{long}<-4^\circ$}} \end{turn}
    \includegraphics[width=0.21\textwidth]{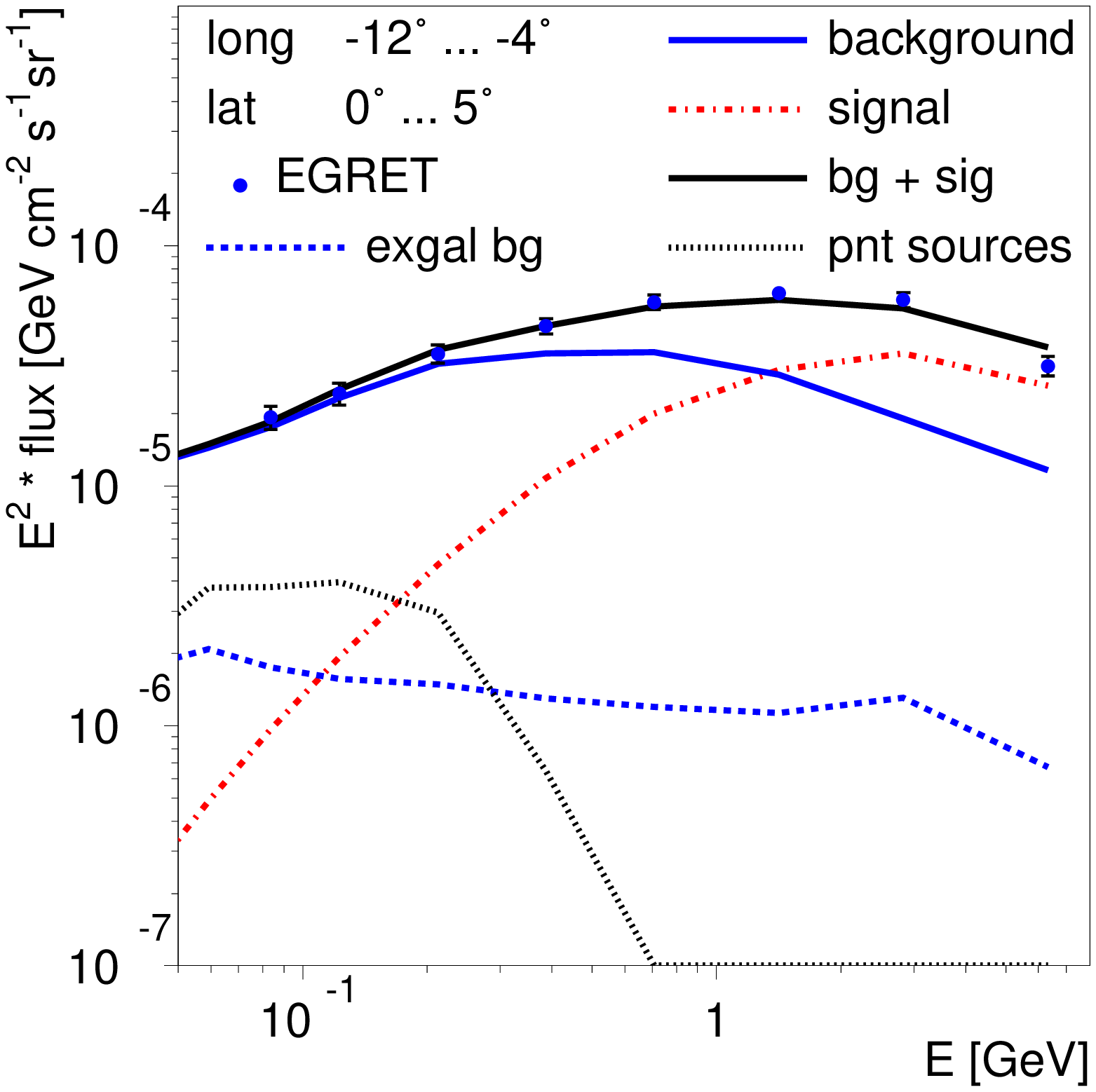}
    \includegraphics[width=0.21\textwidth]{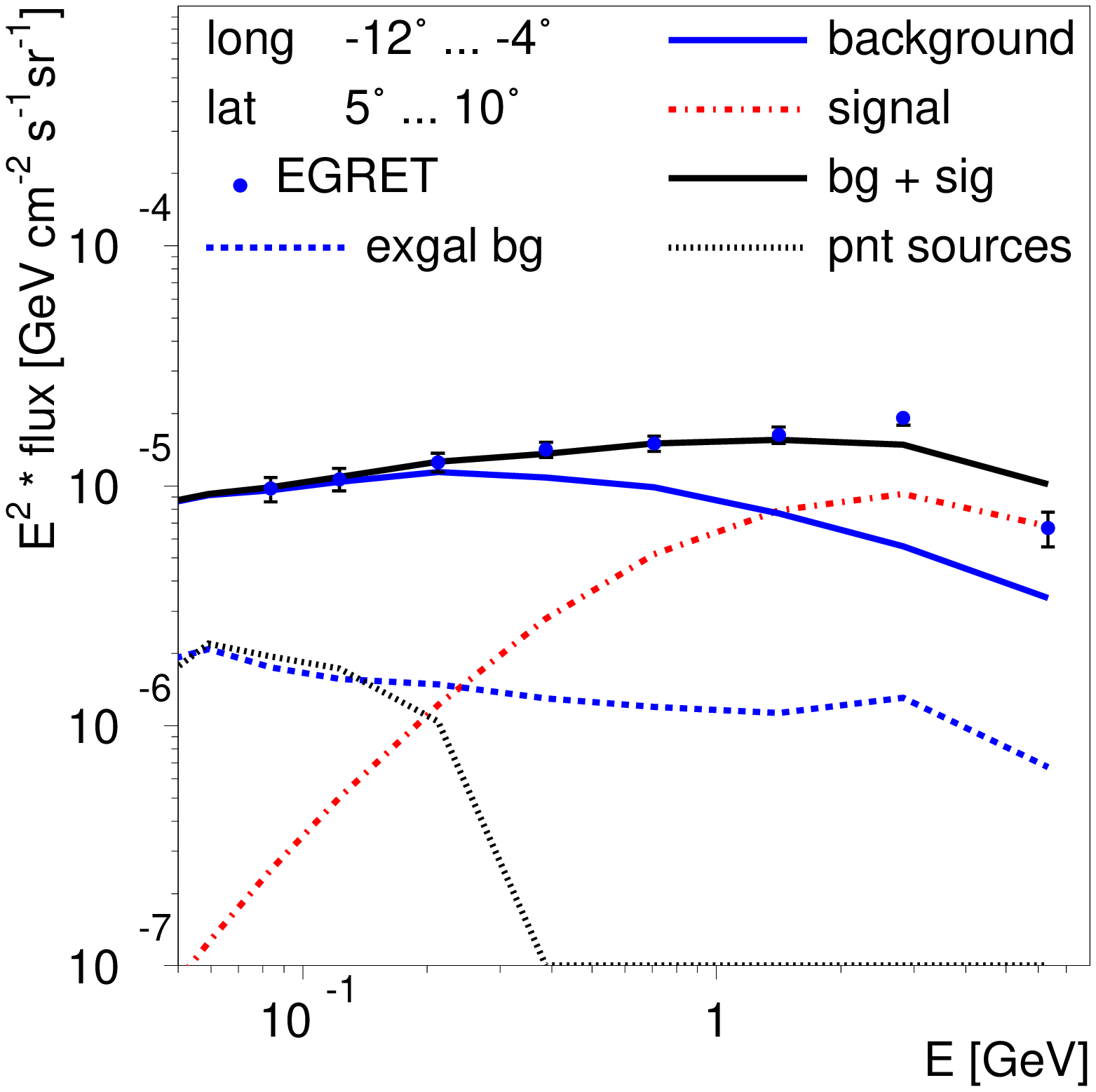}
    \includegraphics[width=0.21\textwidth]{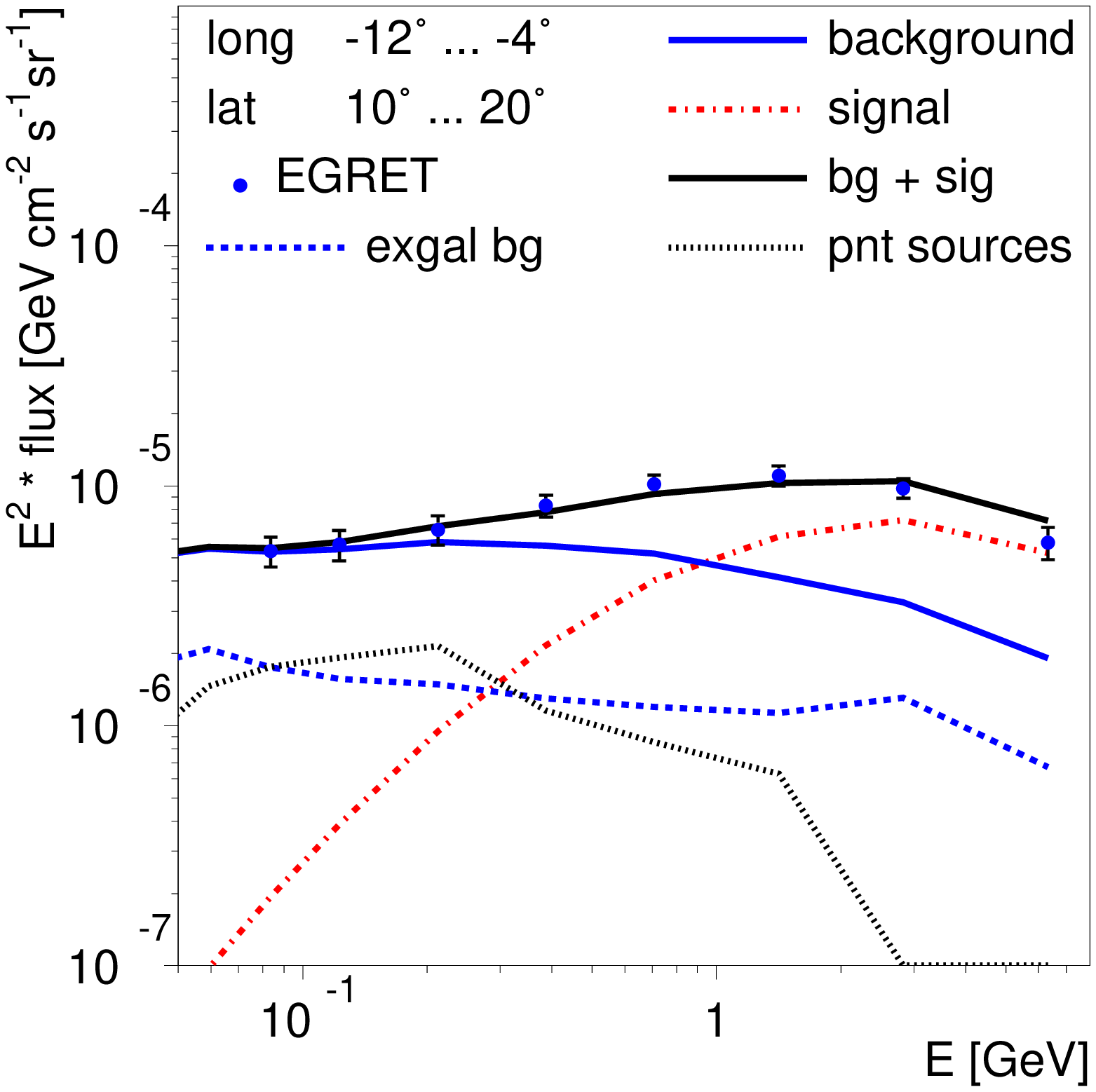}
    \includegraphics[width=0.21\textwidth]{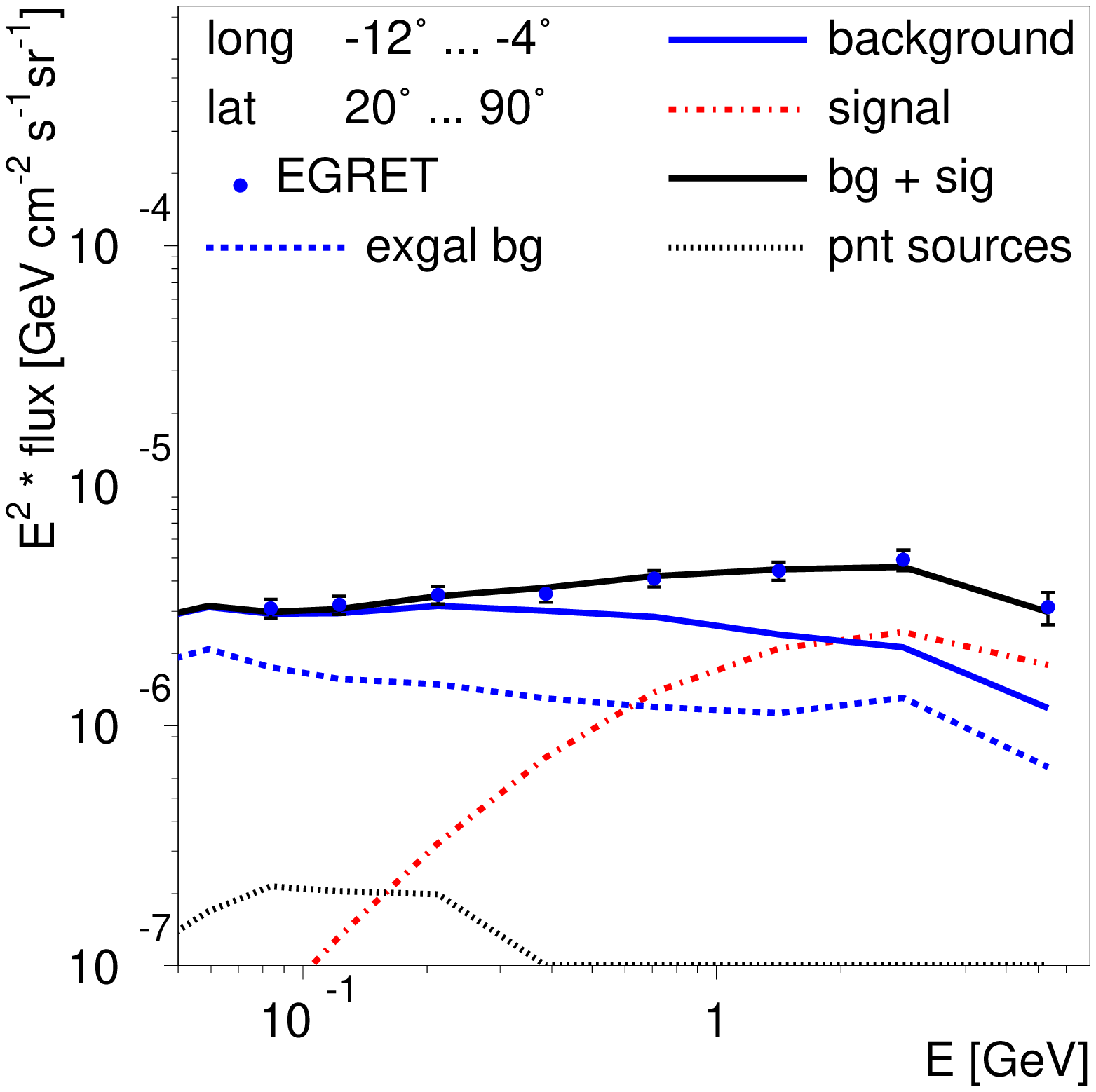}\\
    \hspace{-1cm}
    \begin{turn}{90} \framebox[0.21\textwidth][c]{{\scriptsize $-4^\circ<\mbox{long}<4^\circ$}} \end{turn}
    \includegraphics[width=0.21\textwidth]{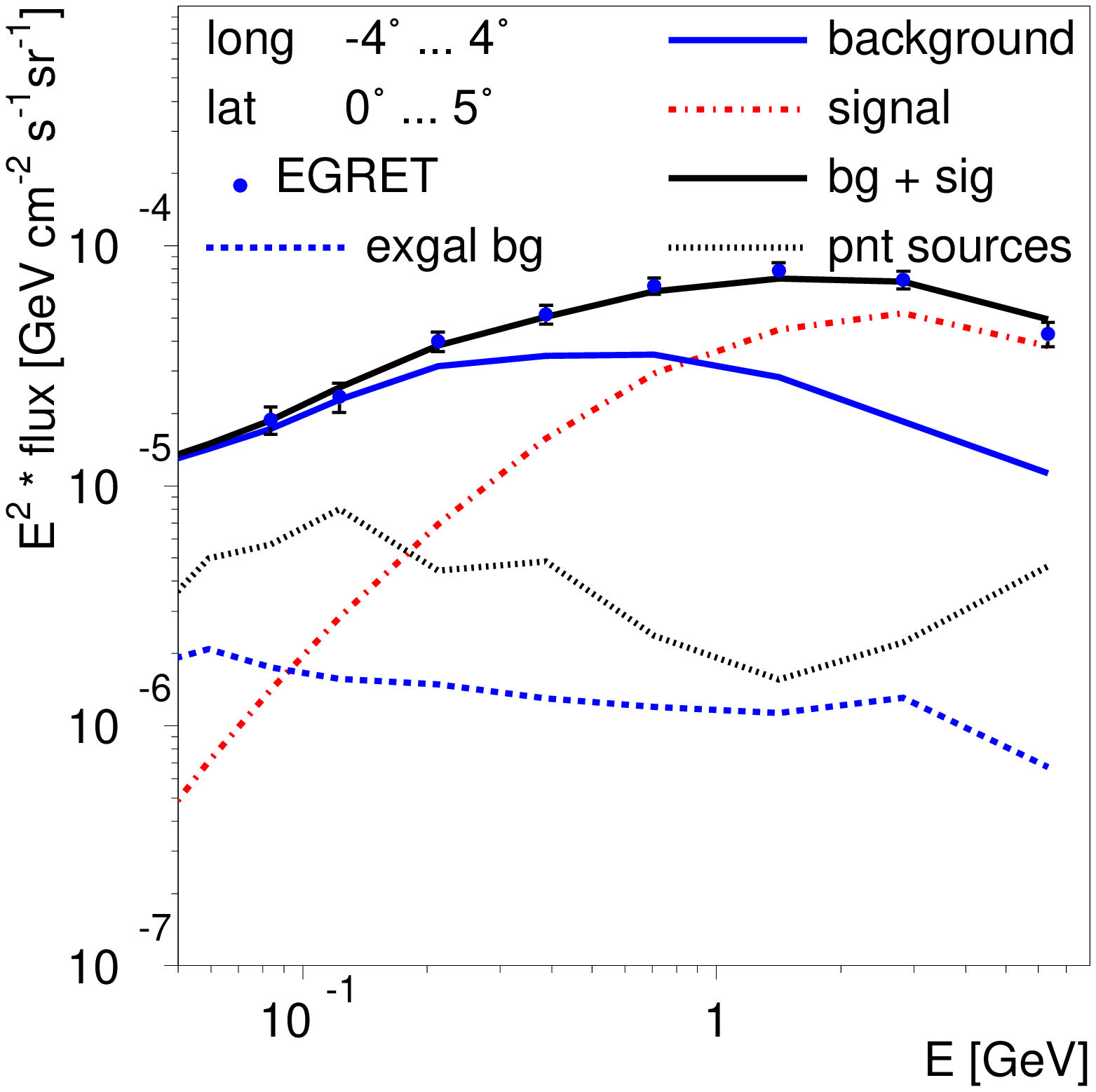}
    \includegraphics[width=0.21\textwidth]{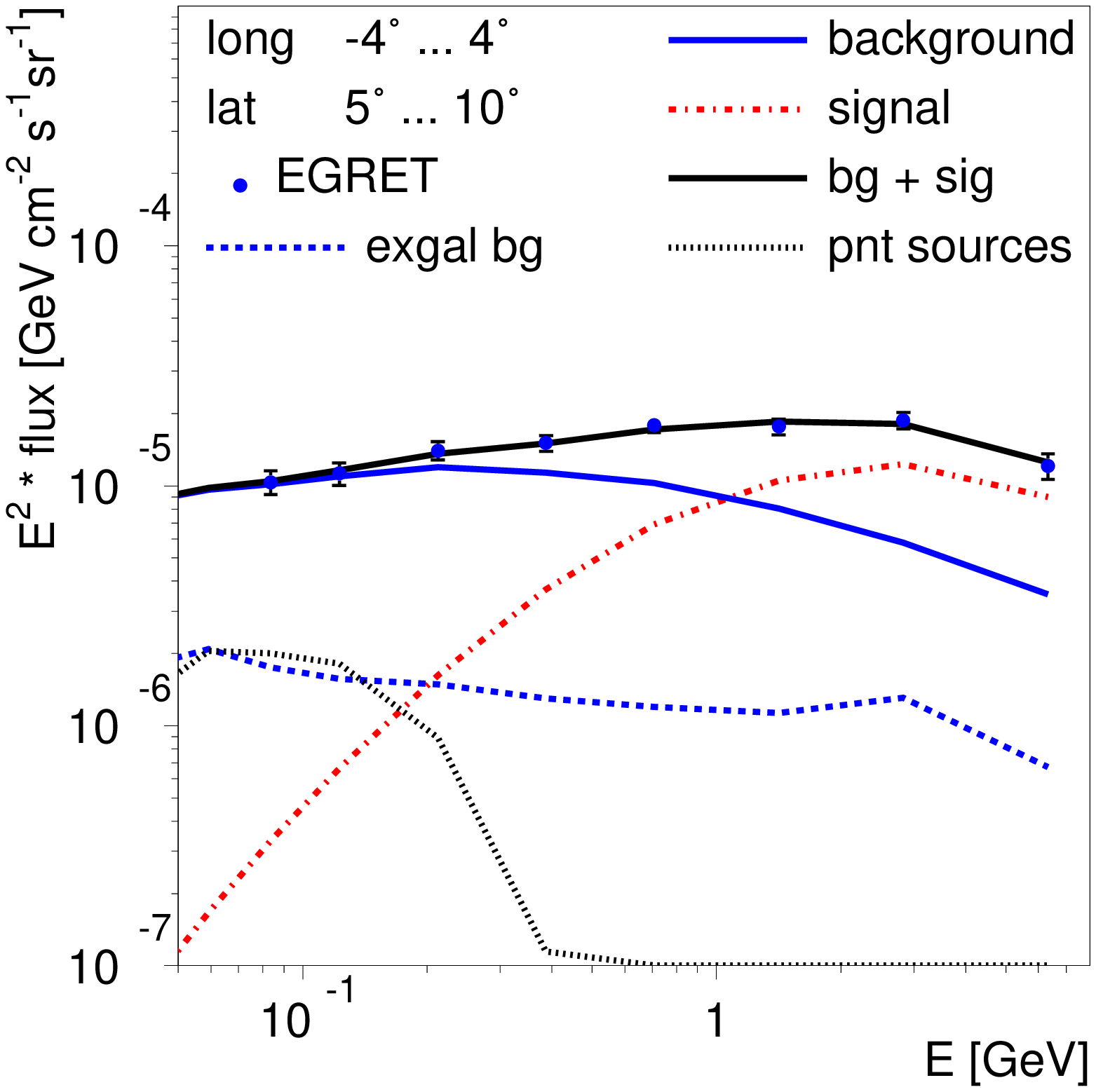}
    \includegraphics[width=0.21\textwidth]{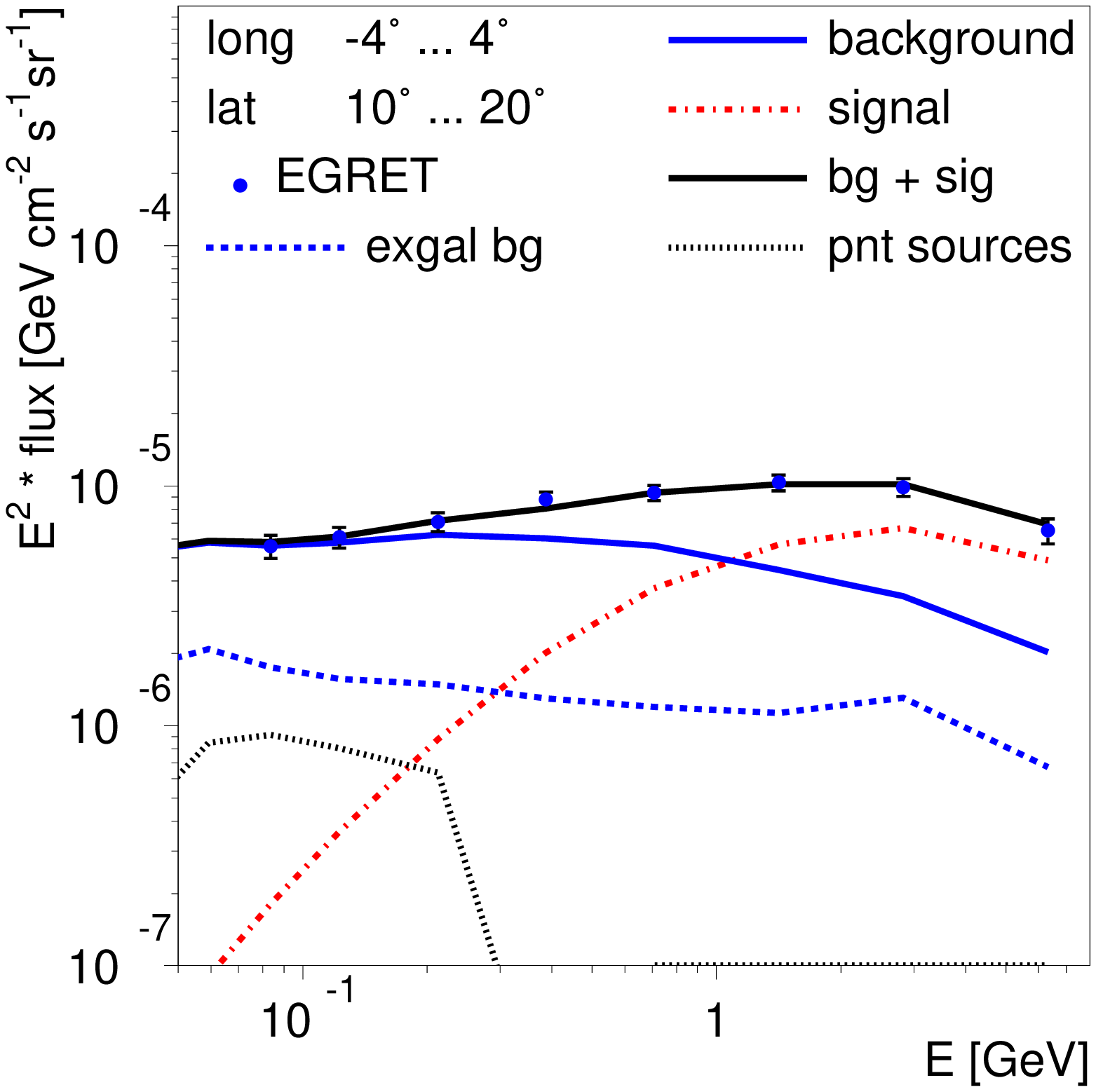}
    \includegraphics[width=0.21\textwidth]{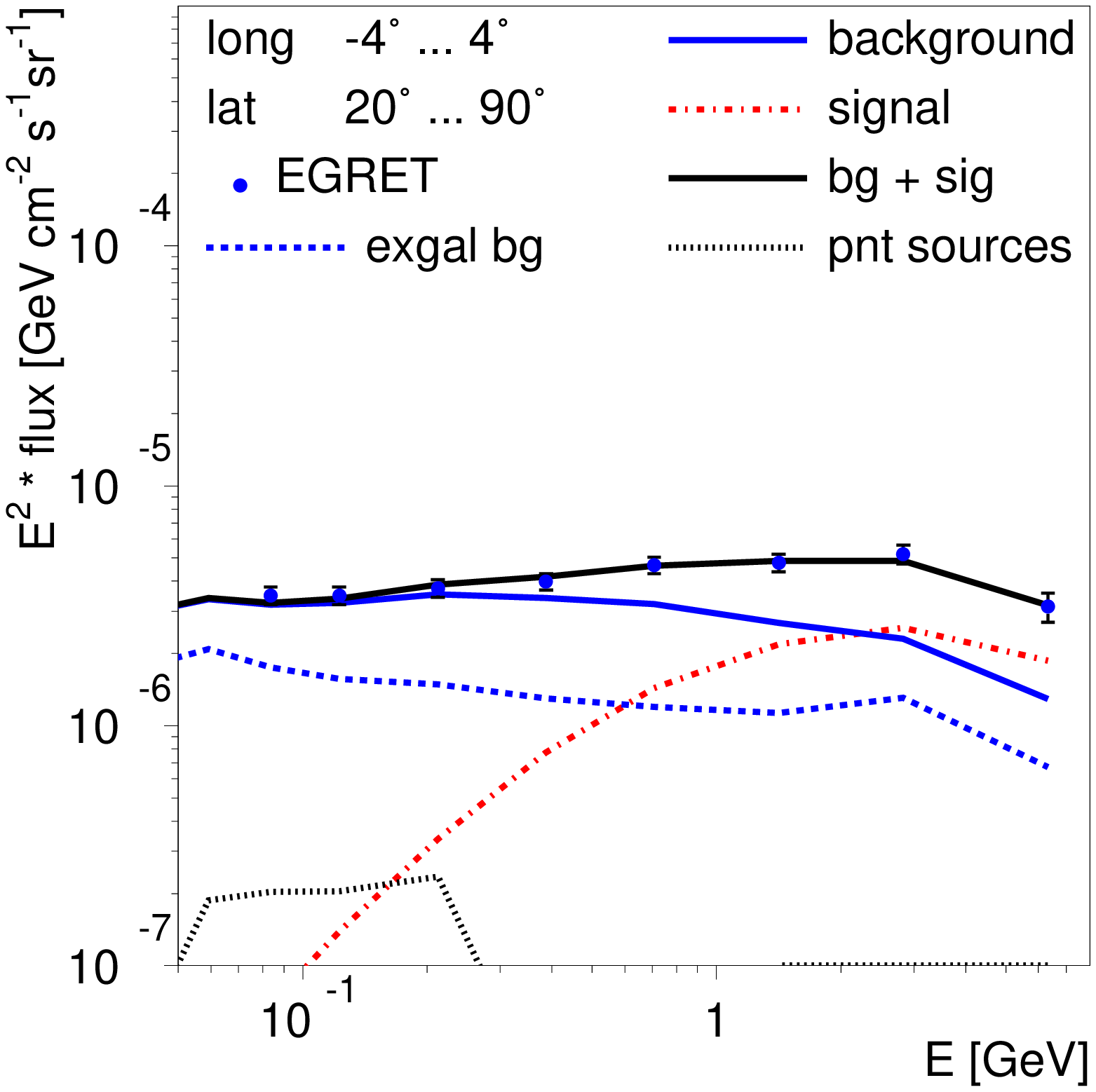}\\
    \hspace{-1cm}
    \begin{turn}{90} \framebox[0.21\textwidth][c]{{\scriptsize $4^\circ<\mbox{long}<12^\circ$}} \end{turn}
    \includegraphics[width=0.21\textwidth]{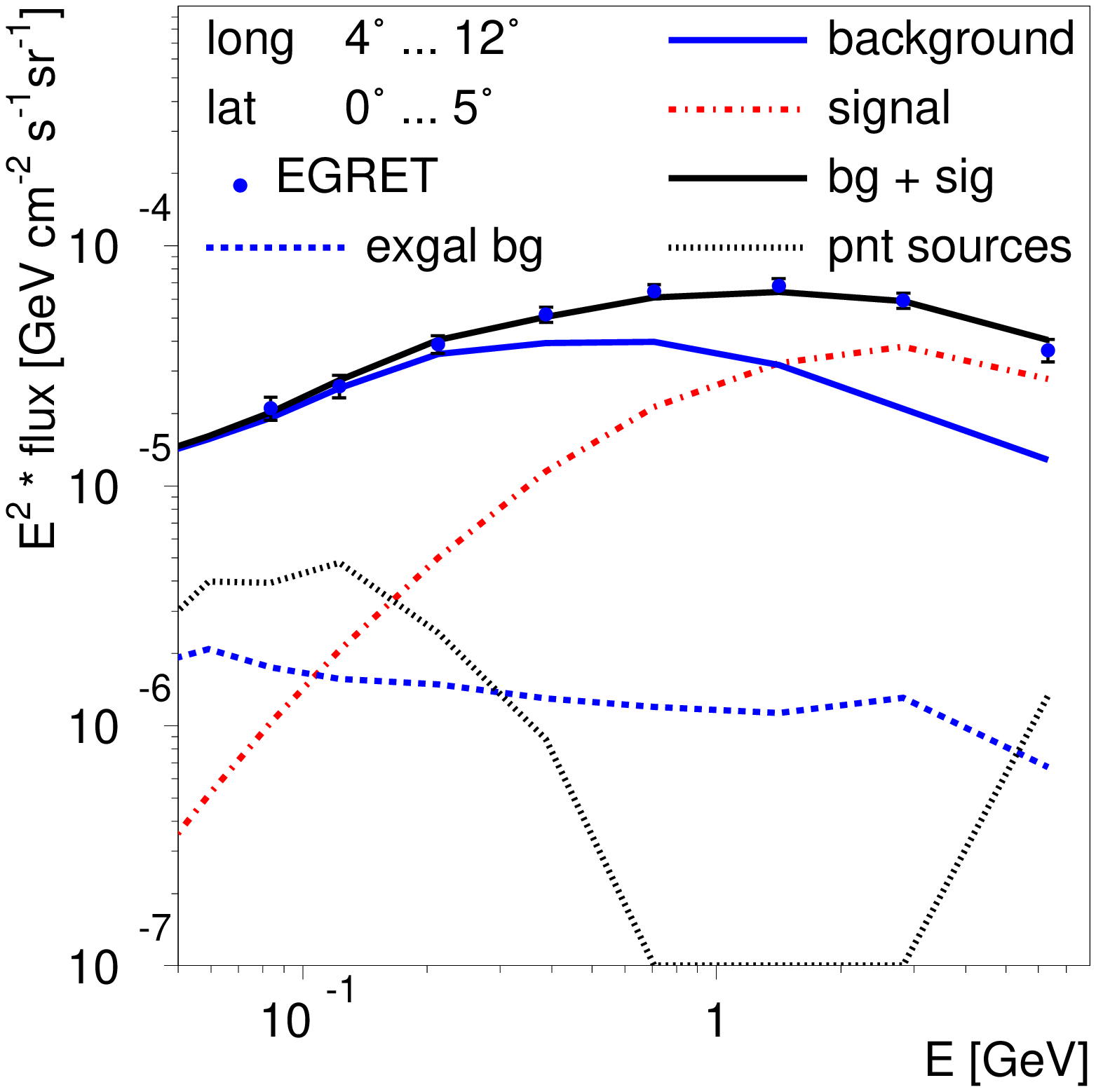}
    \includegraphics[width=0.21\textwidth]{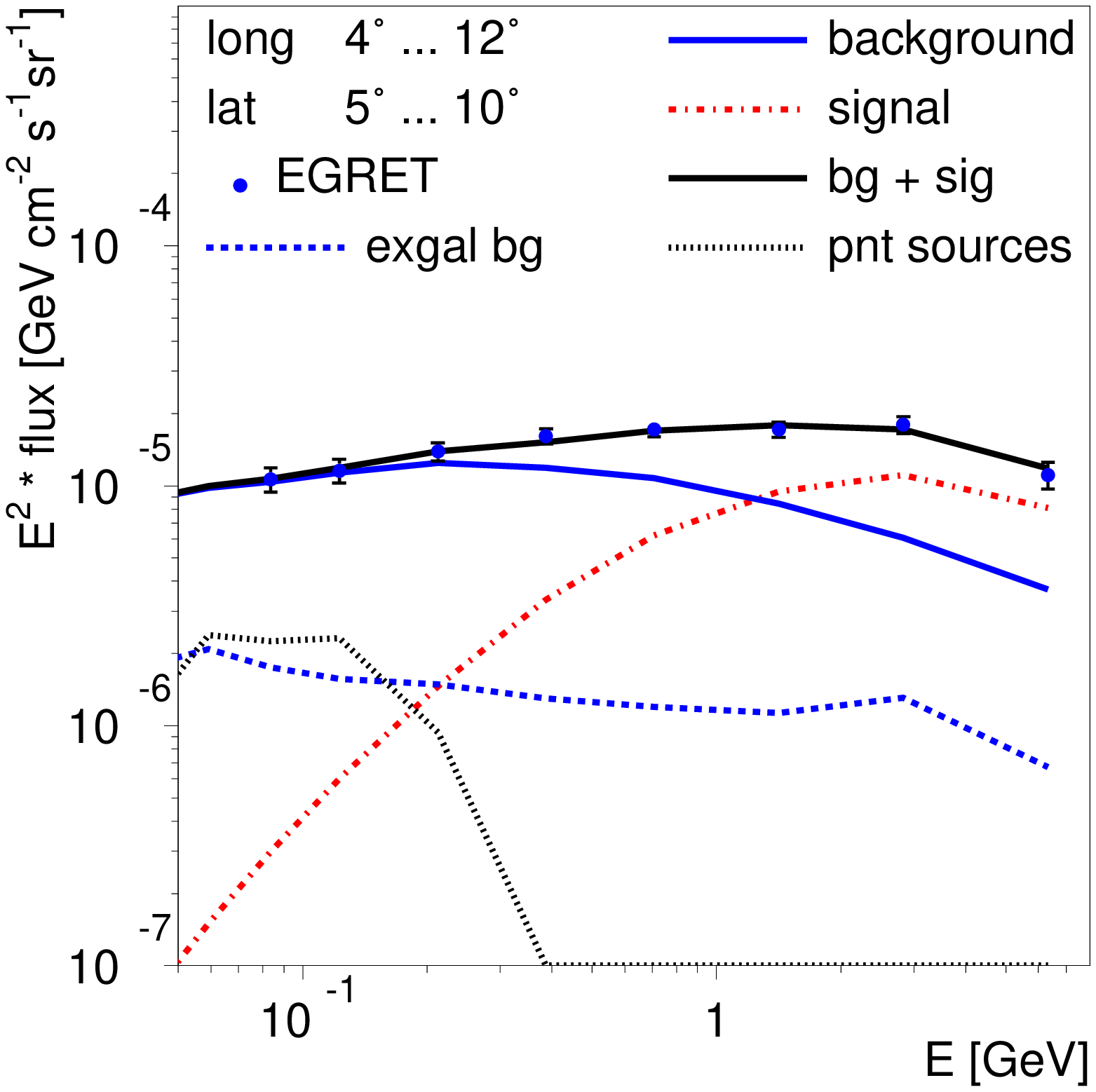}
    \includegraphics[width=0.21\textwidth]{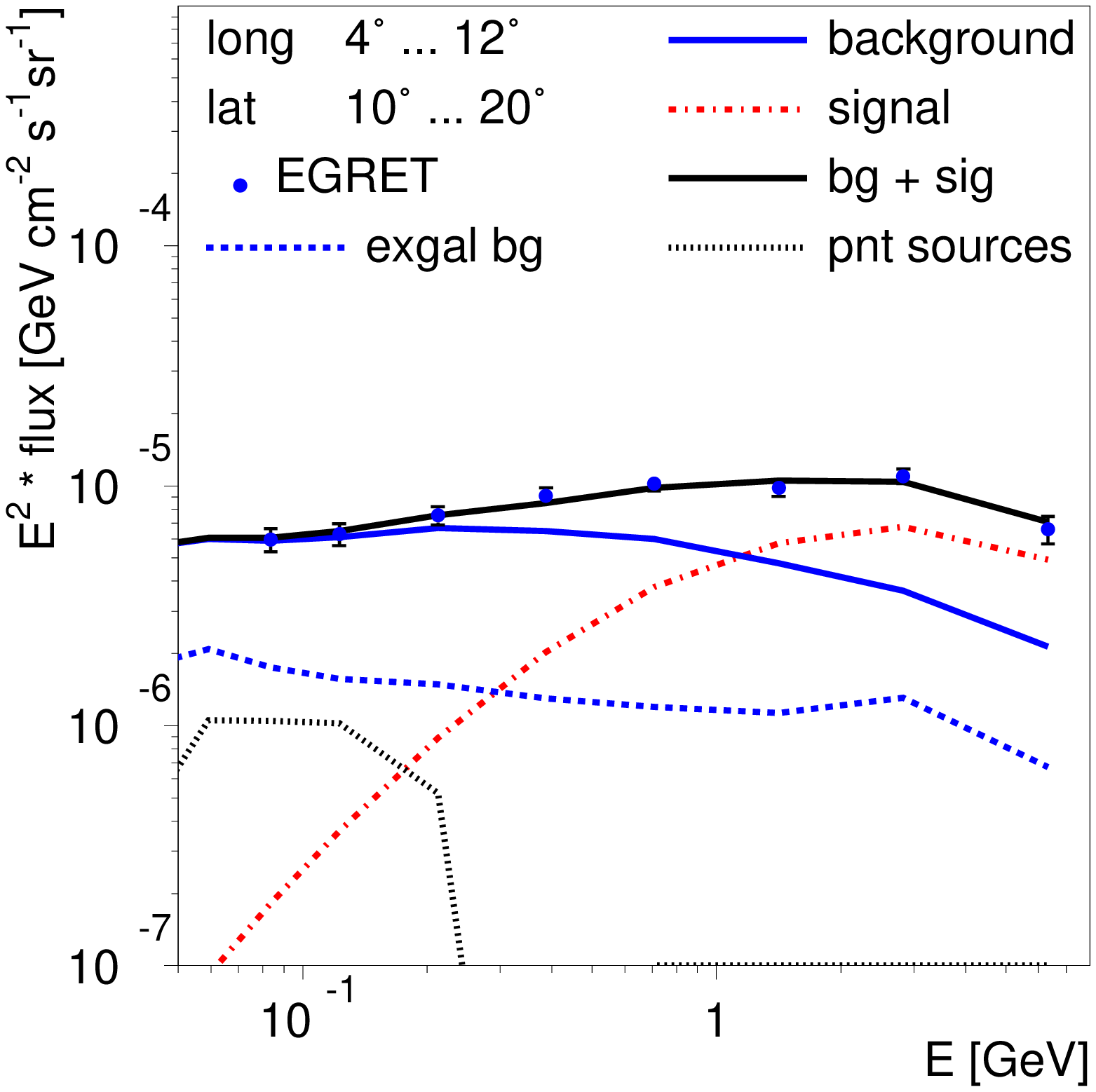}
    \includegraphics[width=0.21\textwidth]{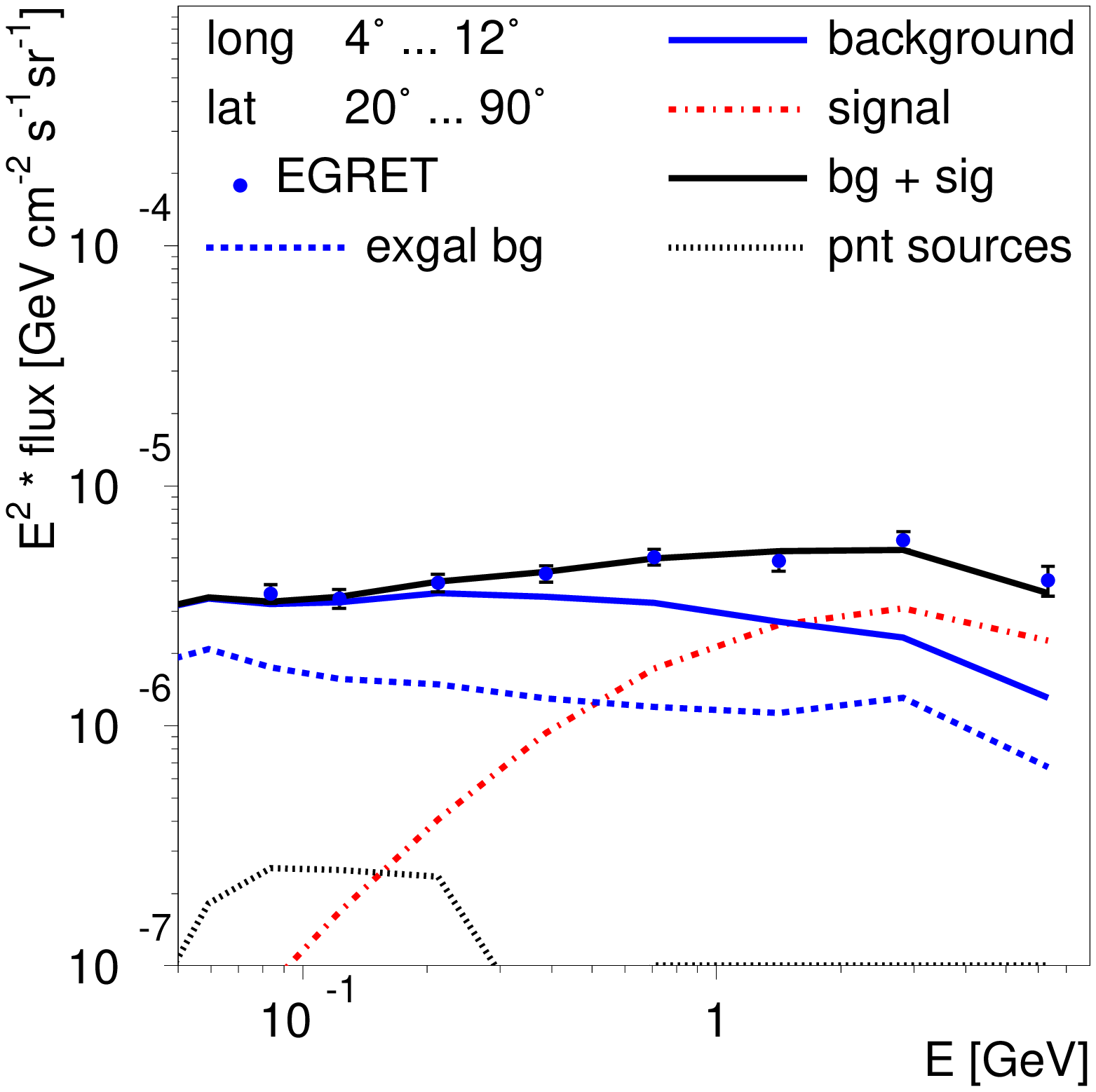}\\
  \end{center}
  \clearpage
  \begin{center}
    \framebox[0.21\textwidth][c]{$\vert \mbox{lat}\vert<5^\circ$}
    \framebox[0.21\textwidth][c]{$5^\circ<\vert \mbox{lat}\vert<10^\circ$}
    \framebox[0.21\textwidth][c]{$10^\circ<\vert \mbox{lat}\vert<20^\circ$}
    \framebox[0.21\textwidth][c]{$20^\circ<\vert \mbox{lat}\vert<90^\circ$}\\
    \hspace{-1cm}
    \begin{turn}{90} \framebox[0.21\textwidth][c]{{\scriptsize $12^\circ<\mbox{long}<20^\circ$}} \end{turn}
    \includegraphics[width=0.21\textwidth]{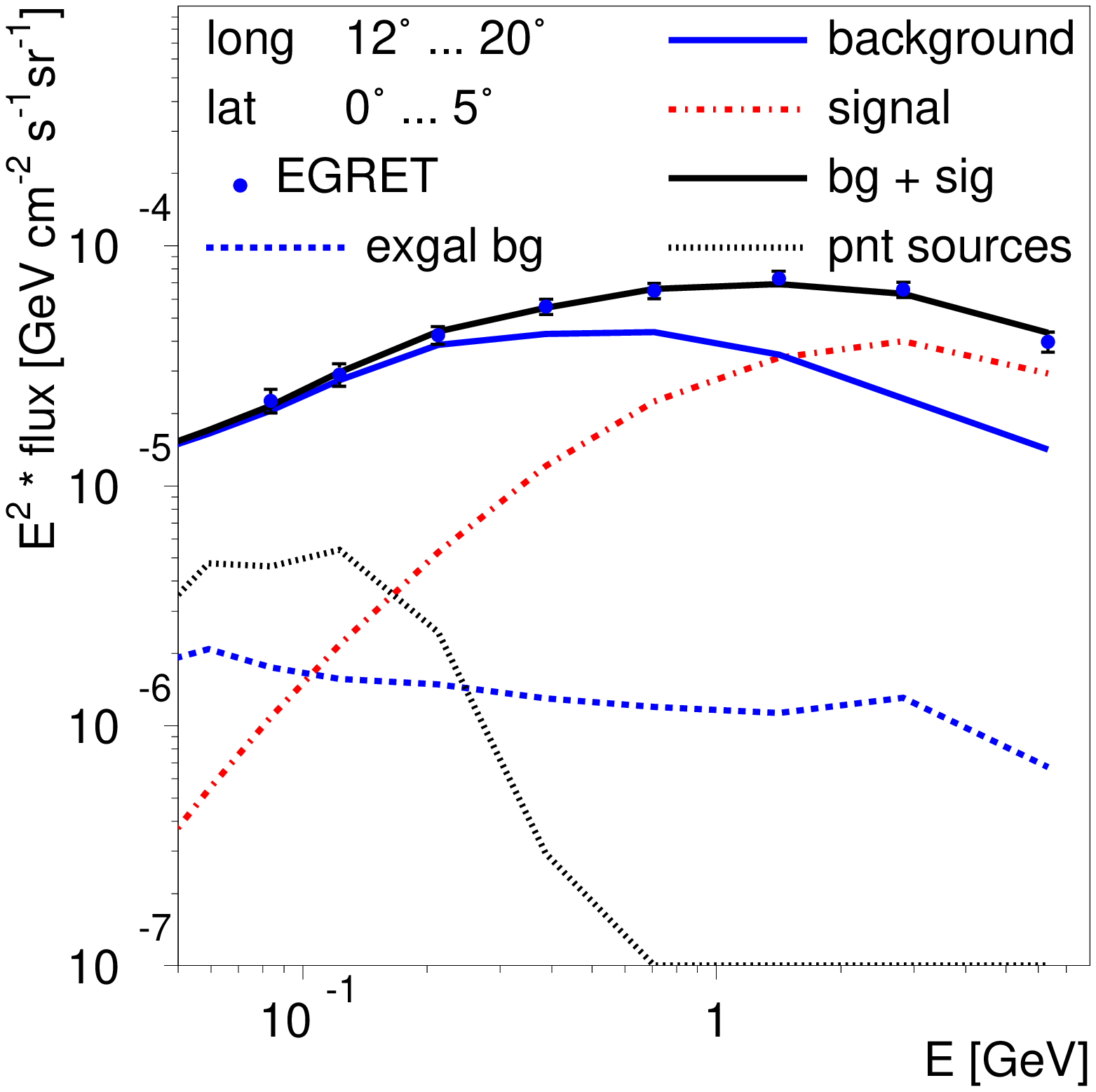}
    \includegraphics[width=0.21\textwidth]{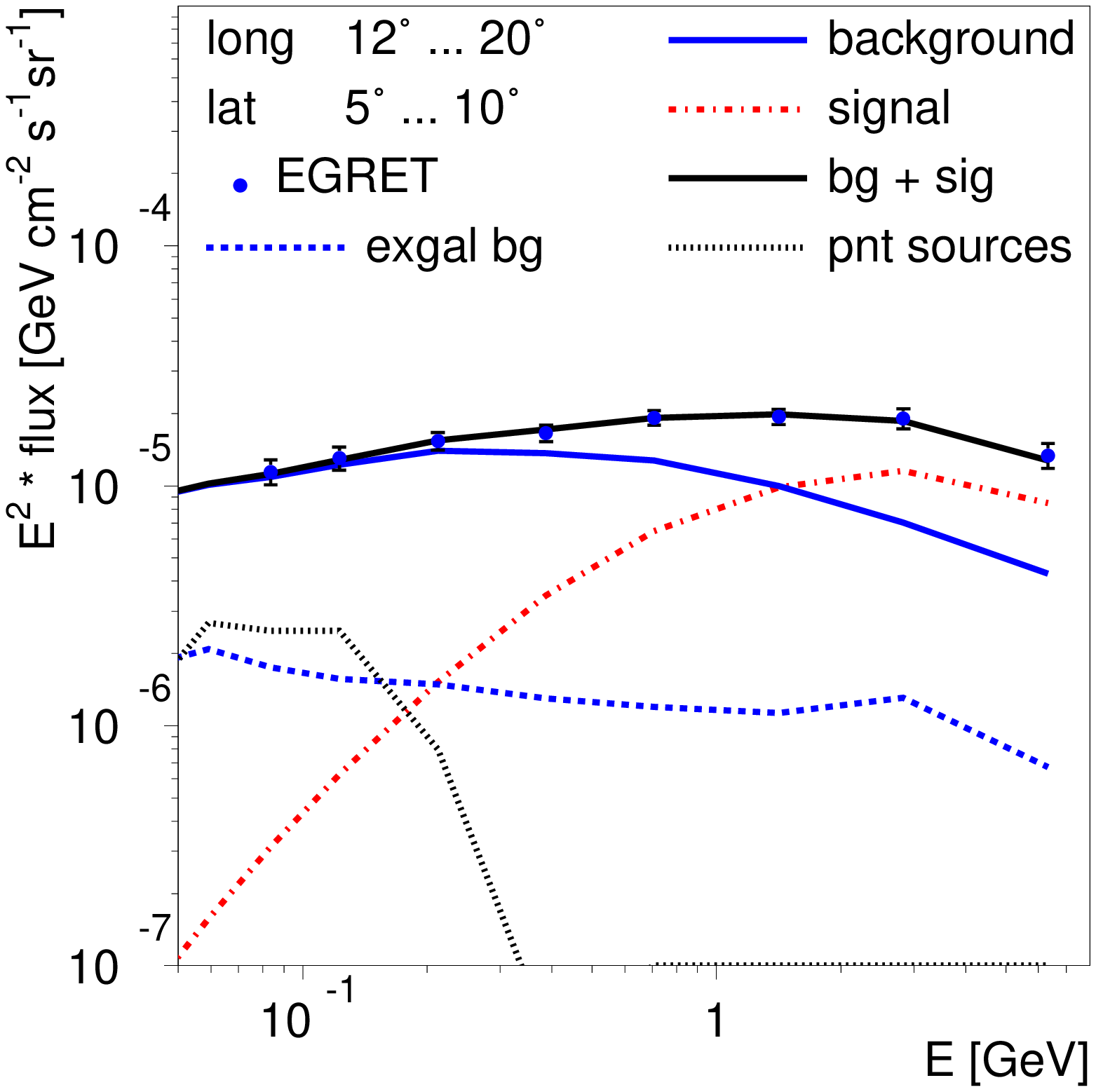}
    \includegraphics[width=0.21\textwidth]{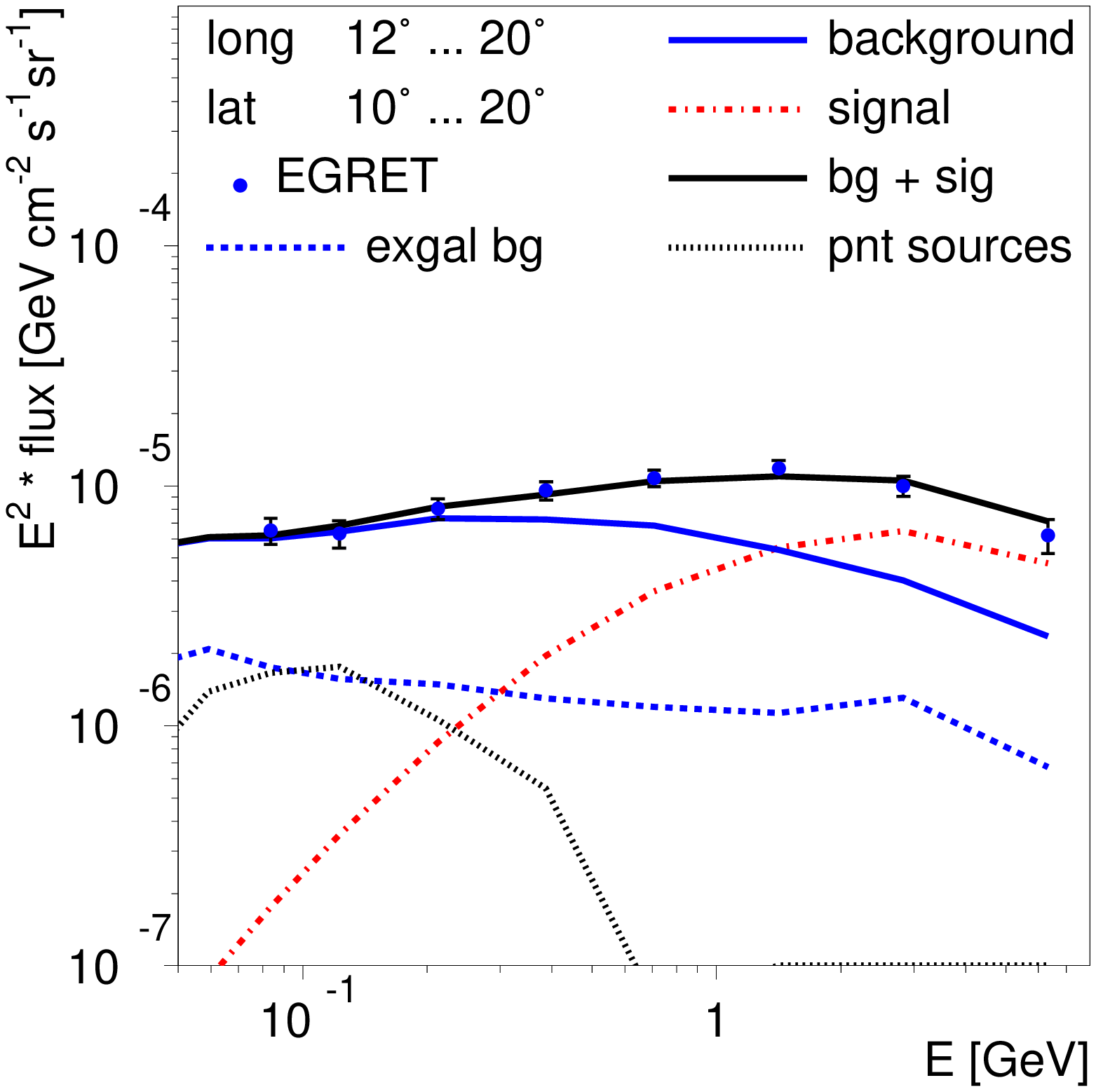}
    \includegraphics[width=0.21\textwidth]{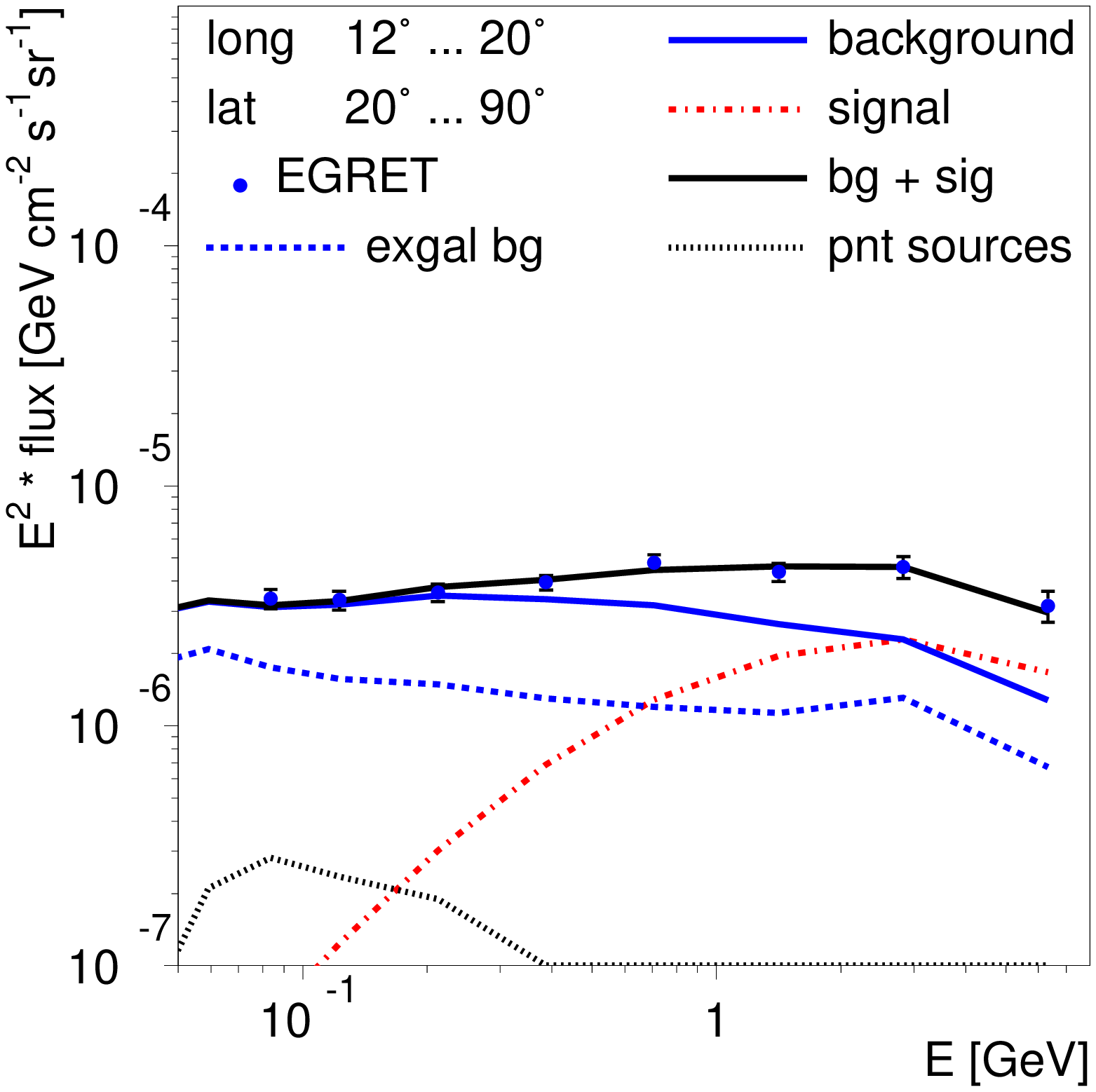}\\
    \hspace{-1cm}
    \begin{turn}{90} \framebox[0.21\textwidth][c]{{\scriptsize $20^\circ<\mbox{long}<28^\circ$}} \end{turn}
    \includegraphics[width=0.21\textwidth]{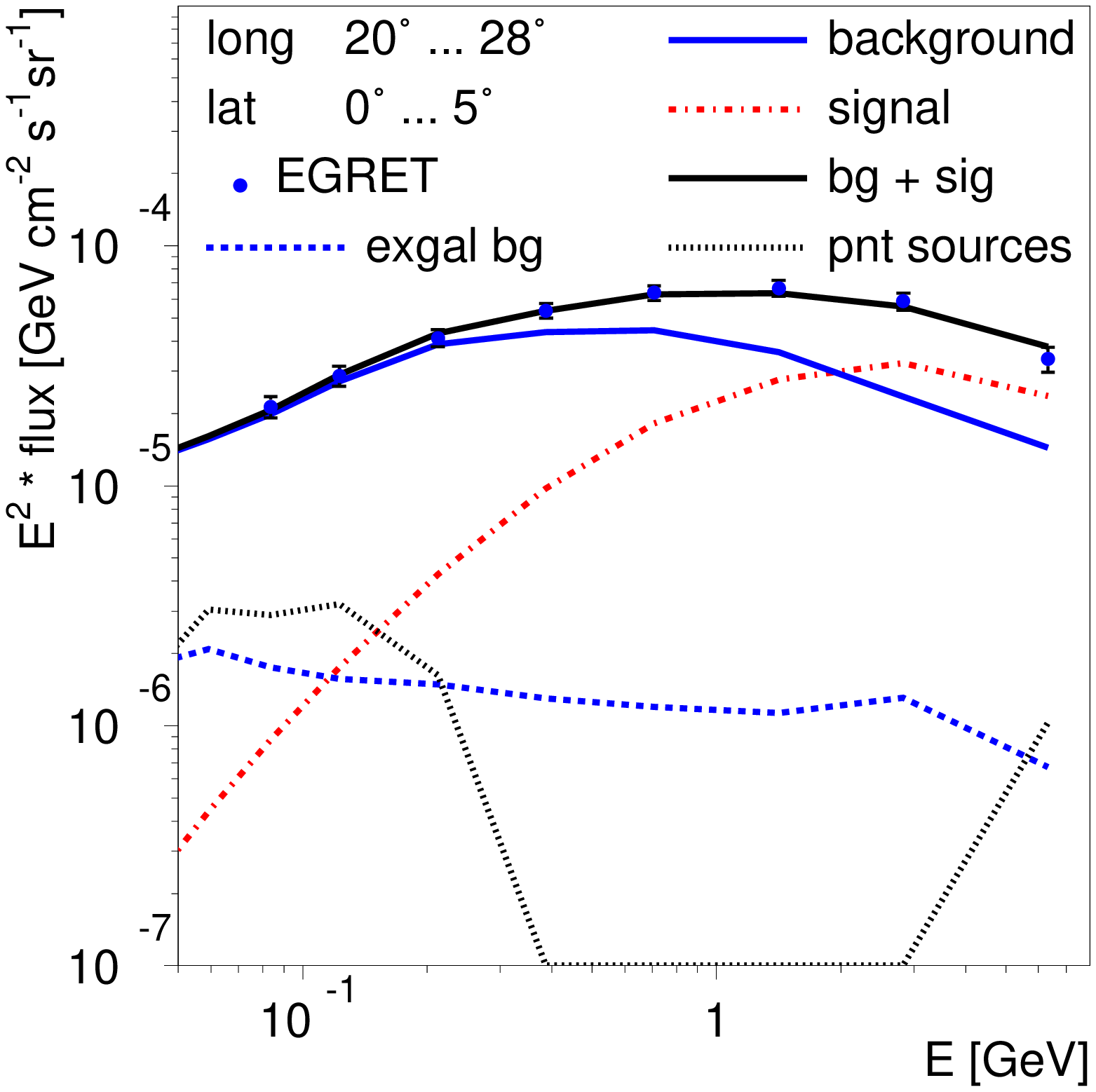}
    \includegraphics[width=0.21\textwidth]{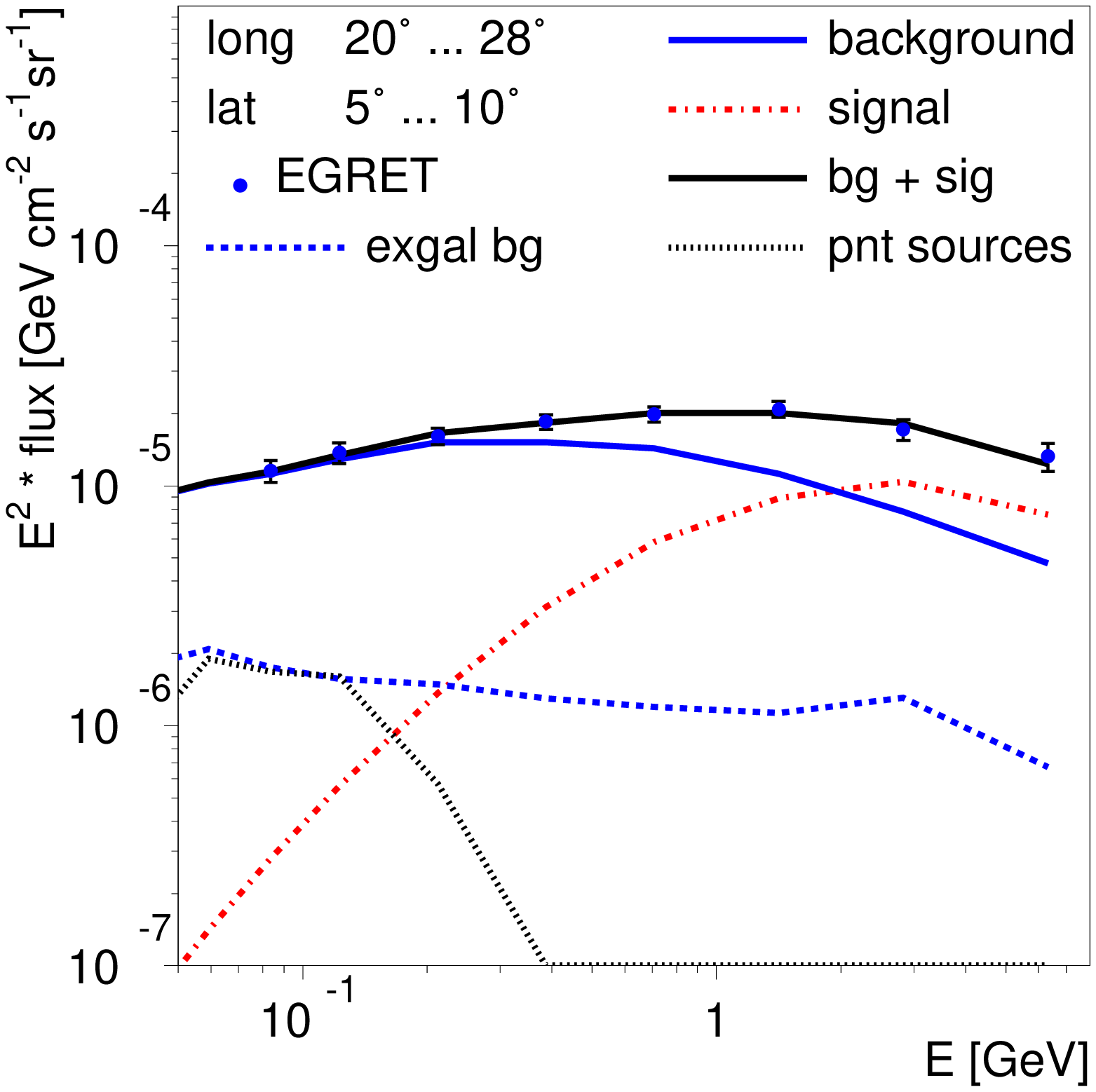}
    \includegraphics[width=0.21\textwidth]{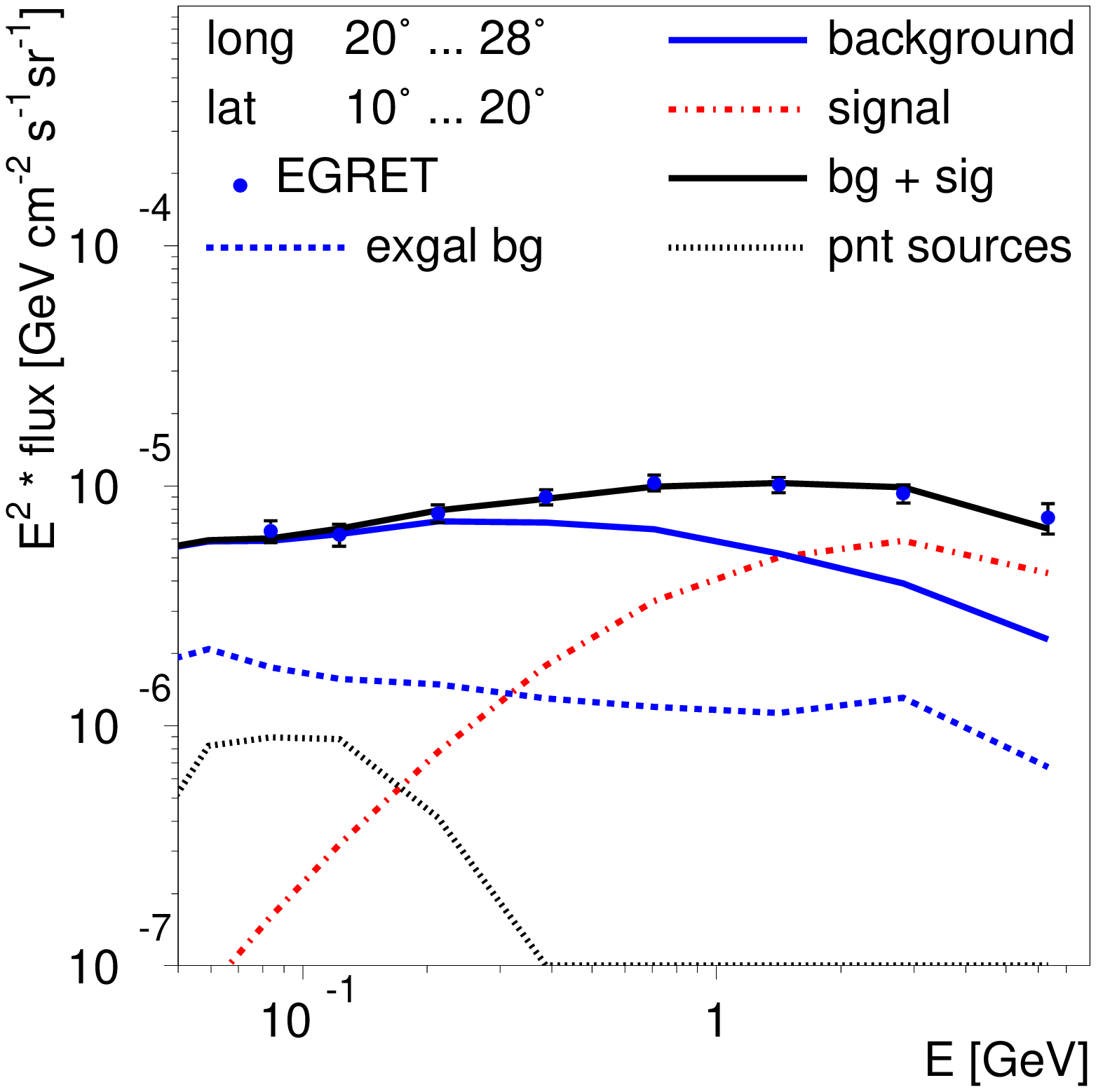}
    \includegraphics[width=0.21\textwidth]{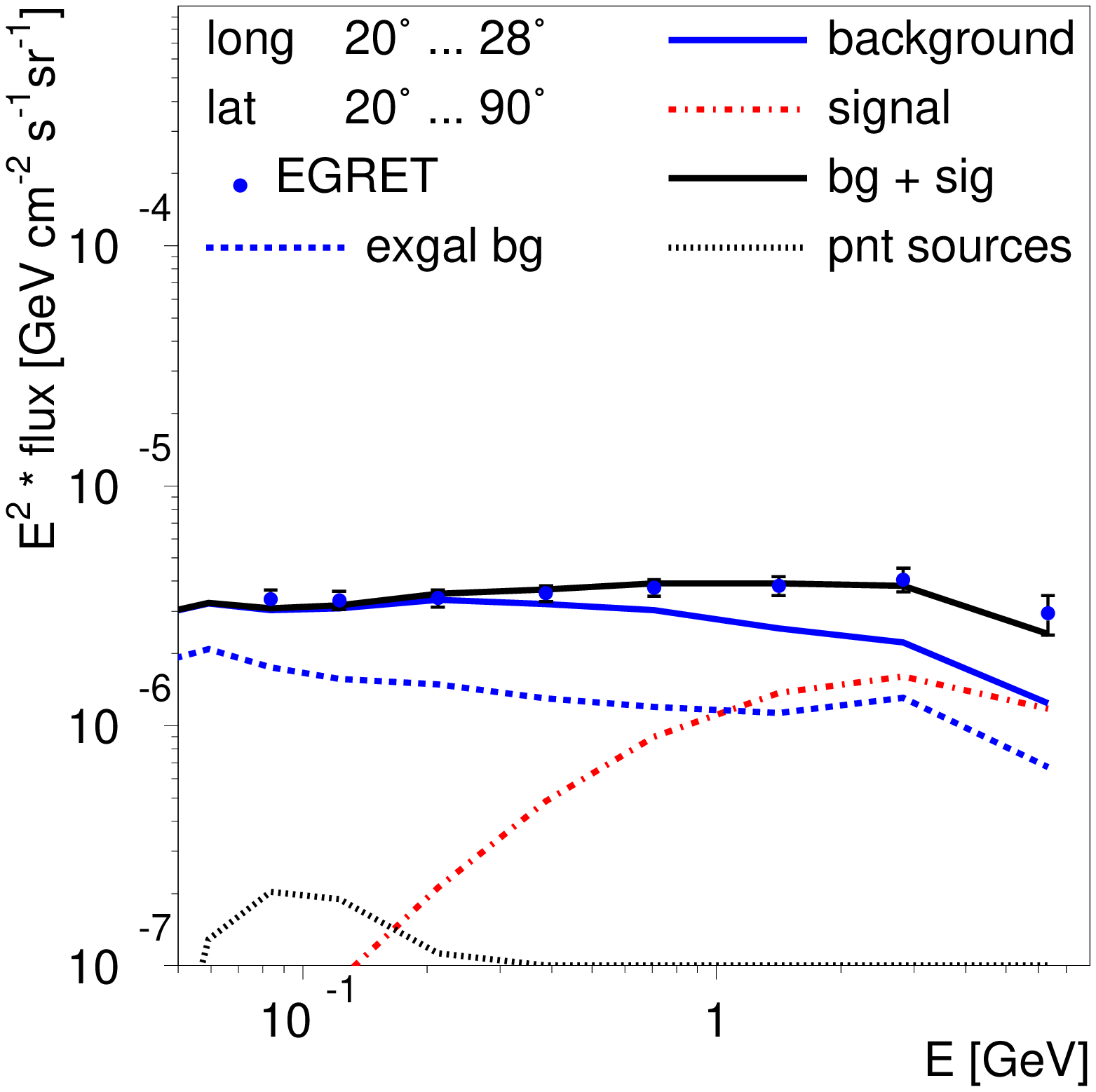}\\
    \hspace{-1cm}
    \begin{turn}{90} \framebox[0.21\textwidth][c]{{\scriptsize $28^\circ<\mbox{long}<36^\circ$}} \end{turn}
    \includegraphics[width=0.21\textwidth]{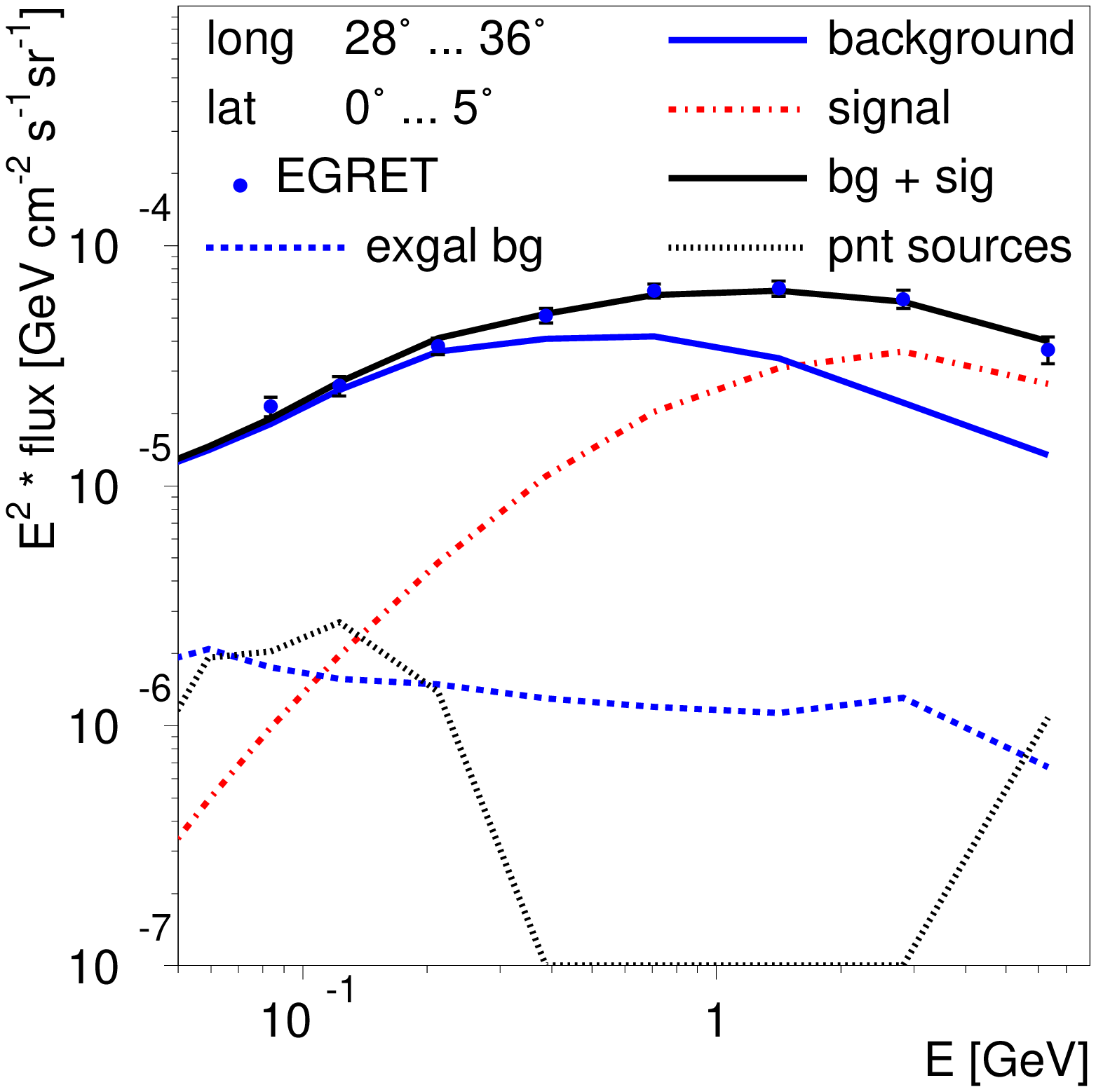}
    \includegraphics[width=0.21\textwidth]{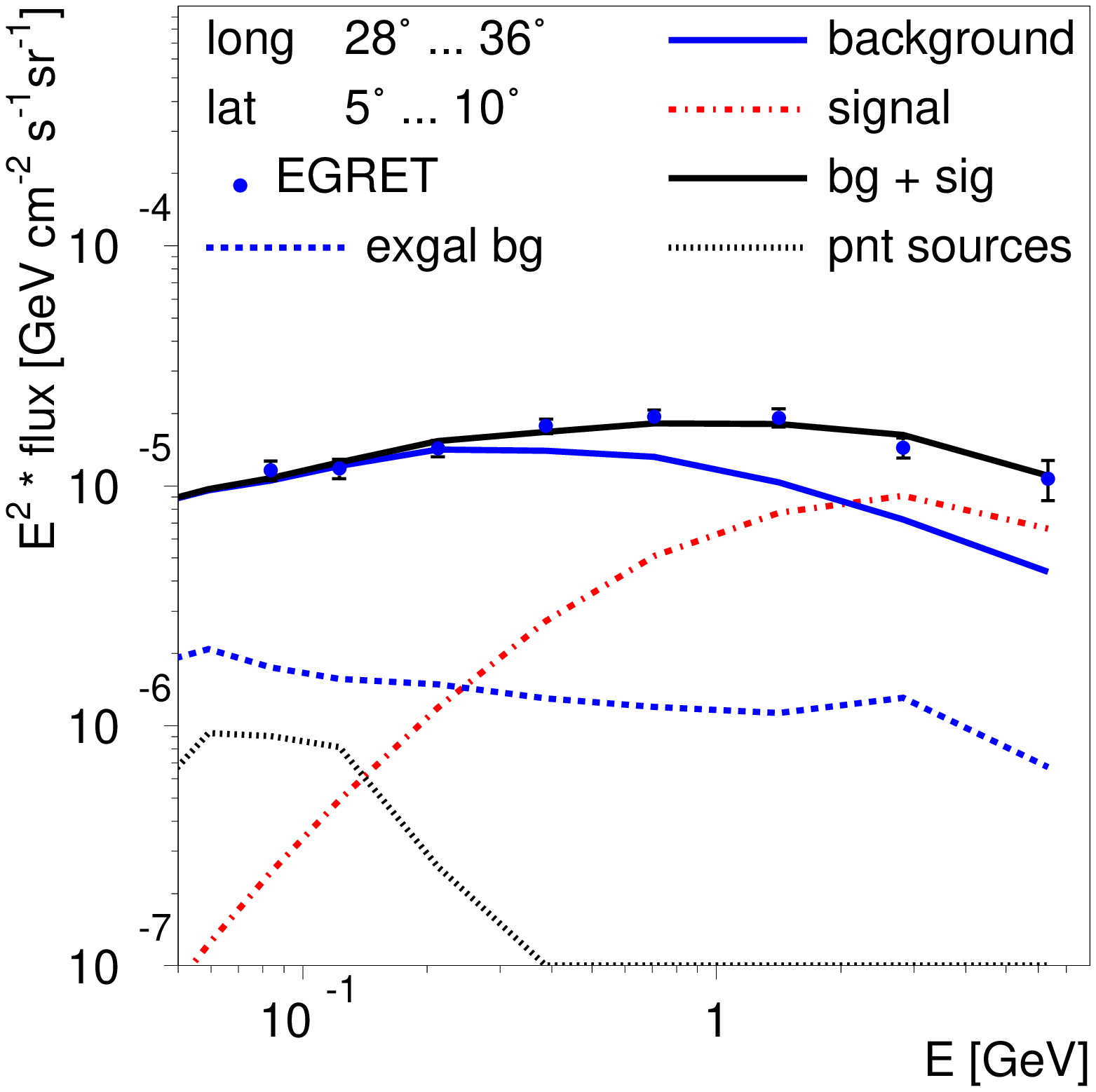}
    \includegraphics[width=0.21\textwidth]{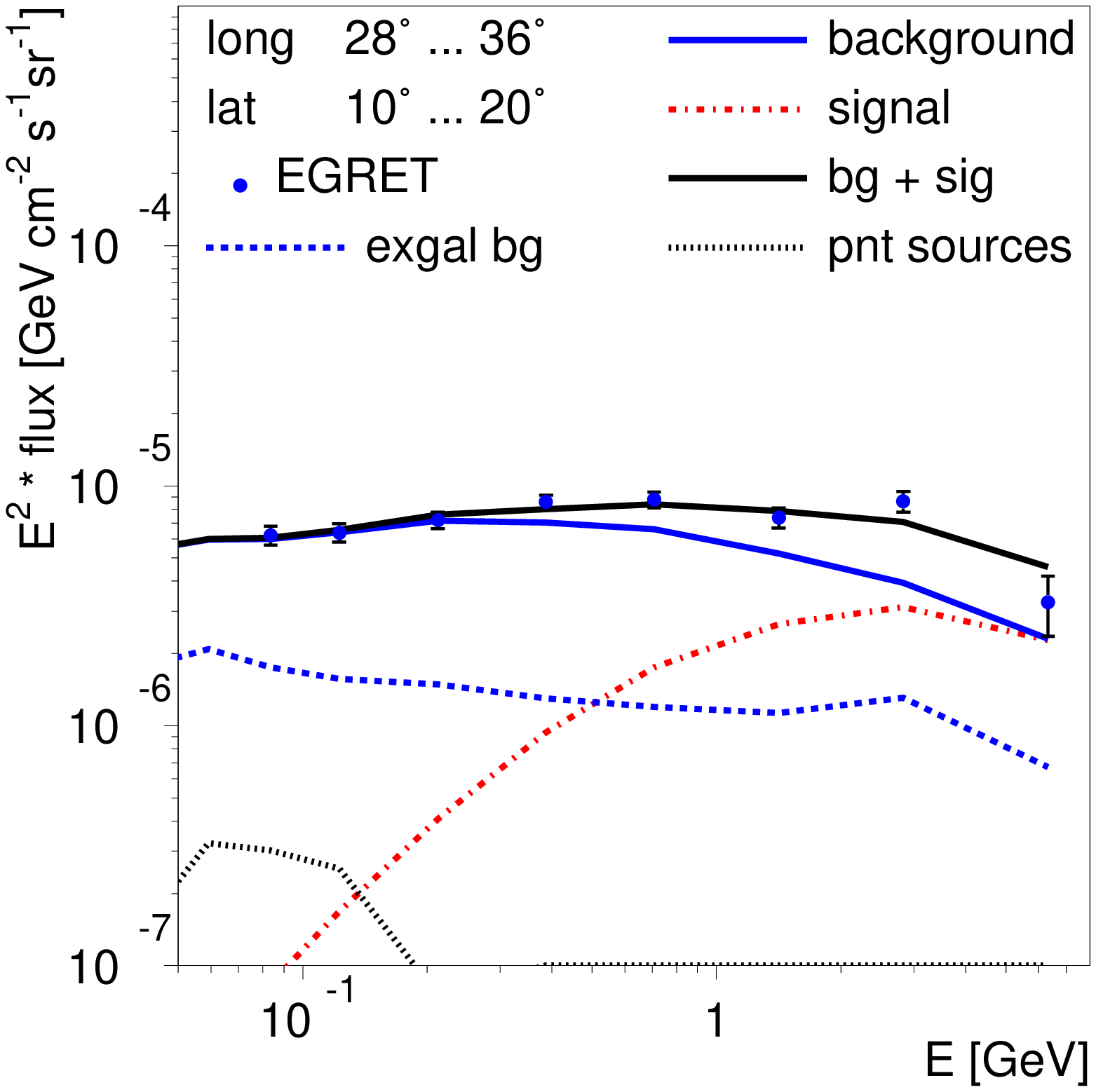}
    \includegraphics[width=0.21\textwidth]{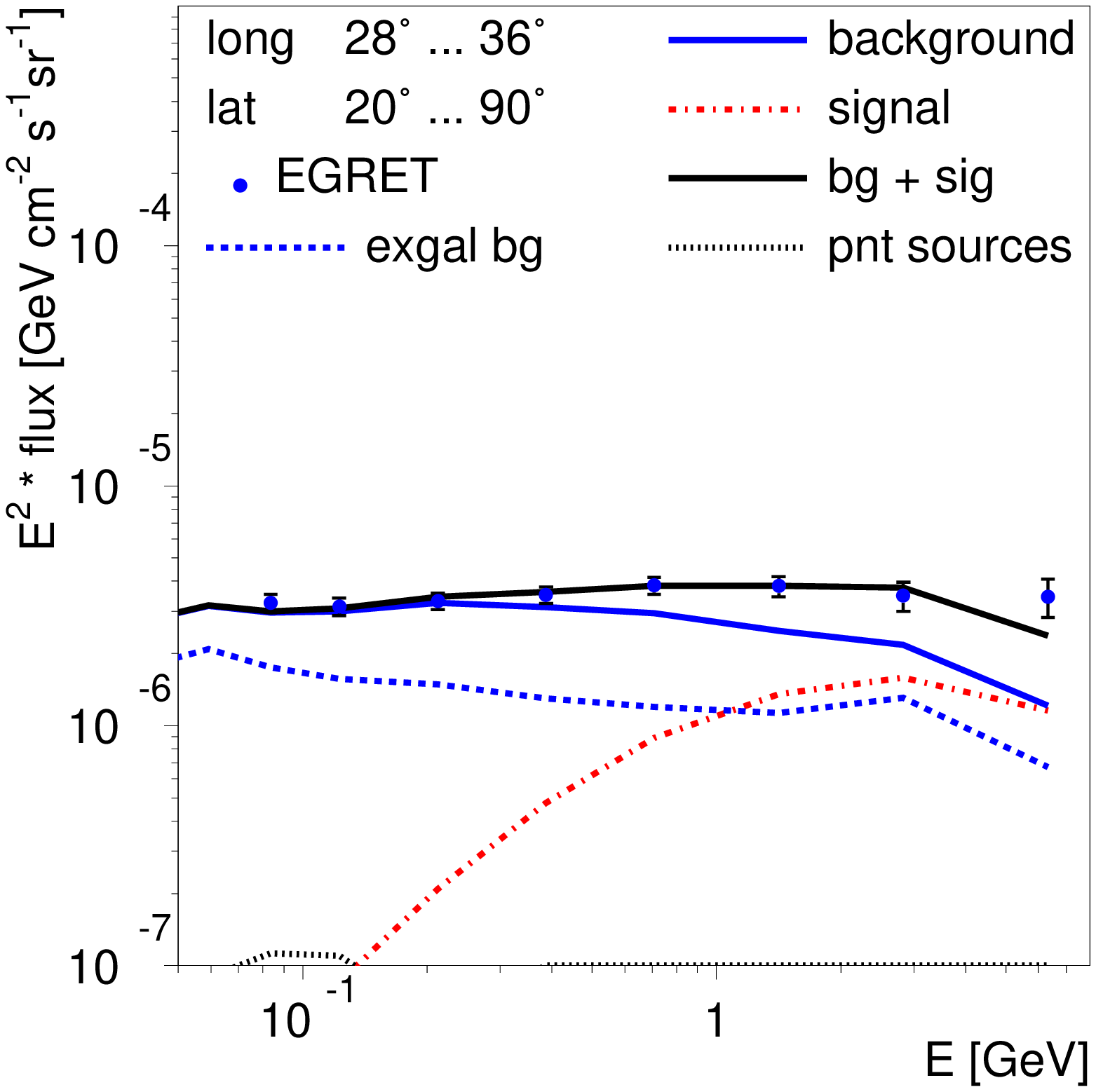}\\
    \hspace{-1cm}
    \begin{turn}{90} \framebox[0.21\textwidth][c]{{\scriptsize $36^\circ<\mbox{long}<44^\circ$}} \end{turn}
    \includegraphics[width=0.21\textwidth]{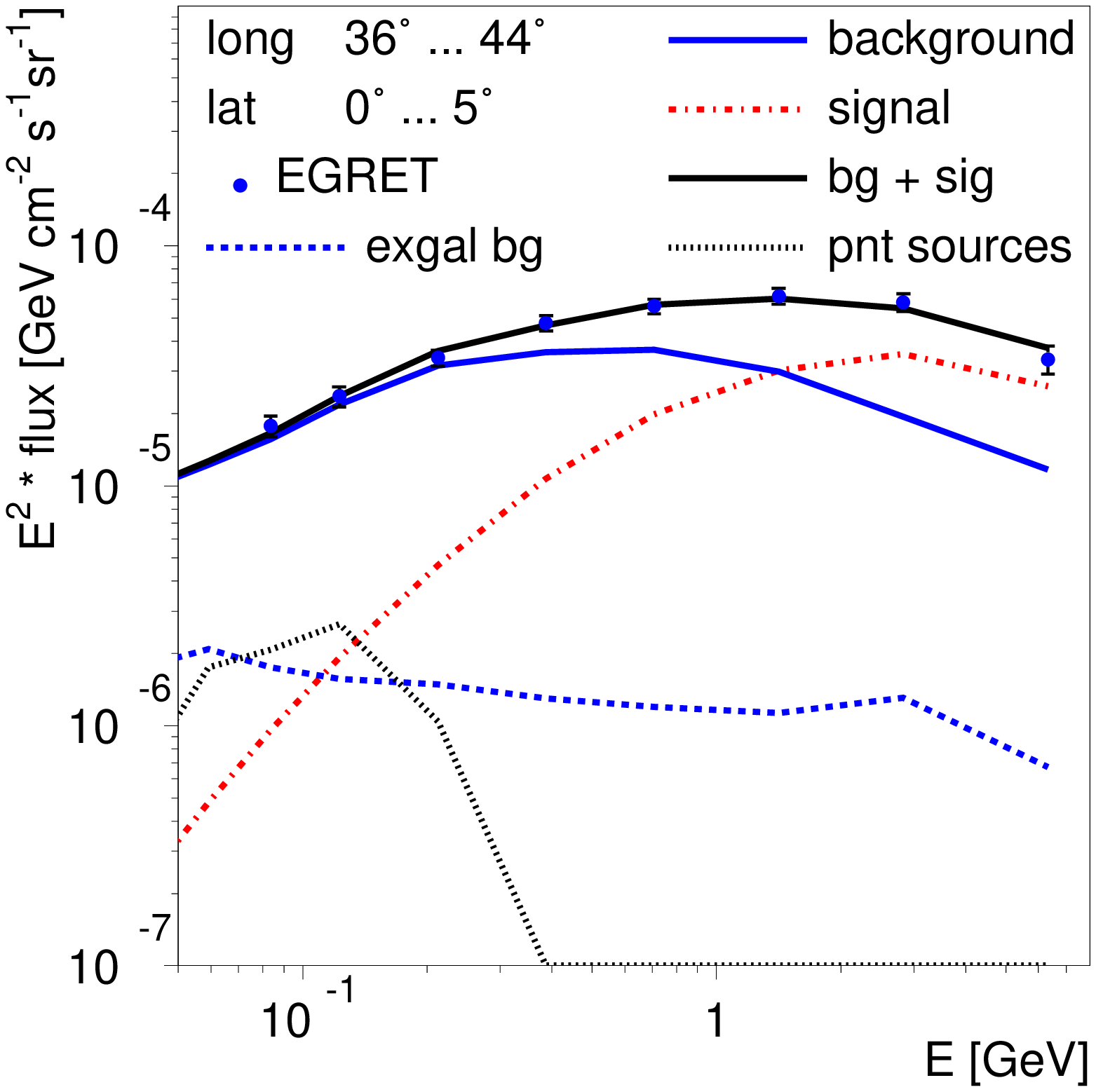}
    \includegraphics[width=0.21\textwidth]{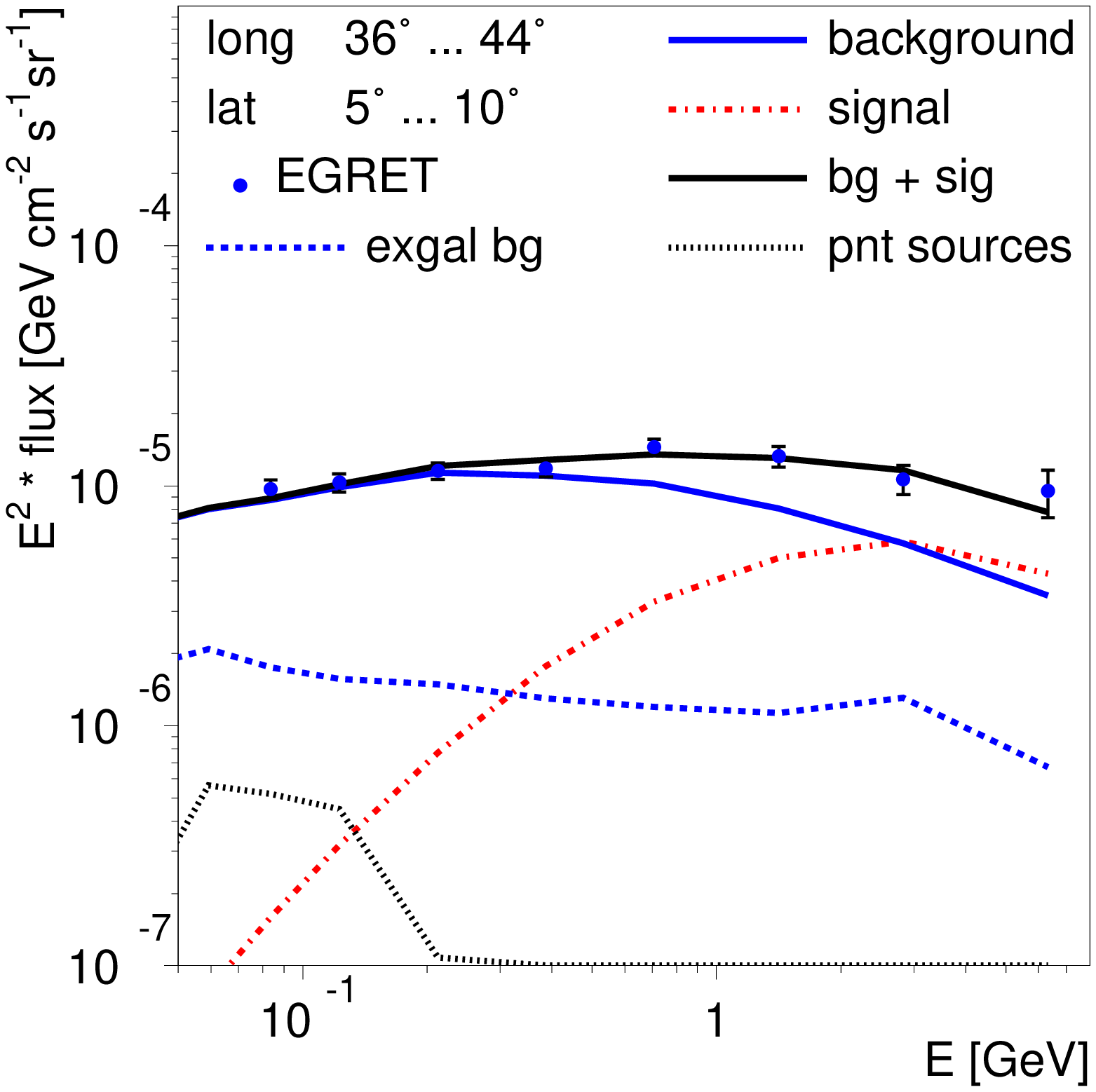}
    \includegraphics[width=0.21\textwidth]{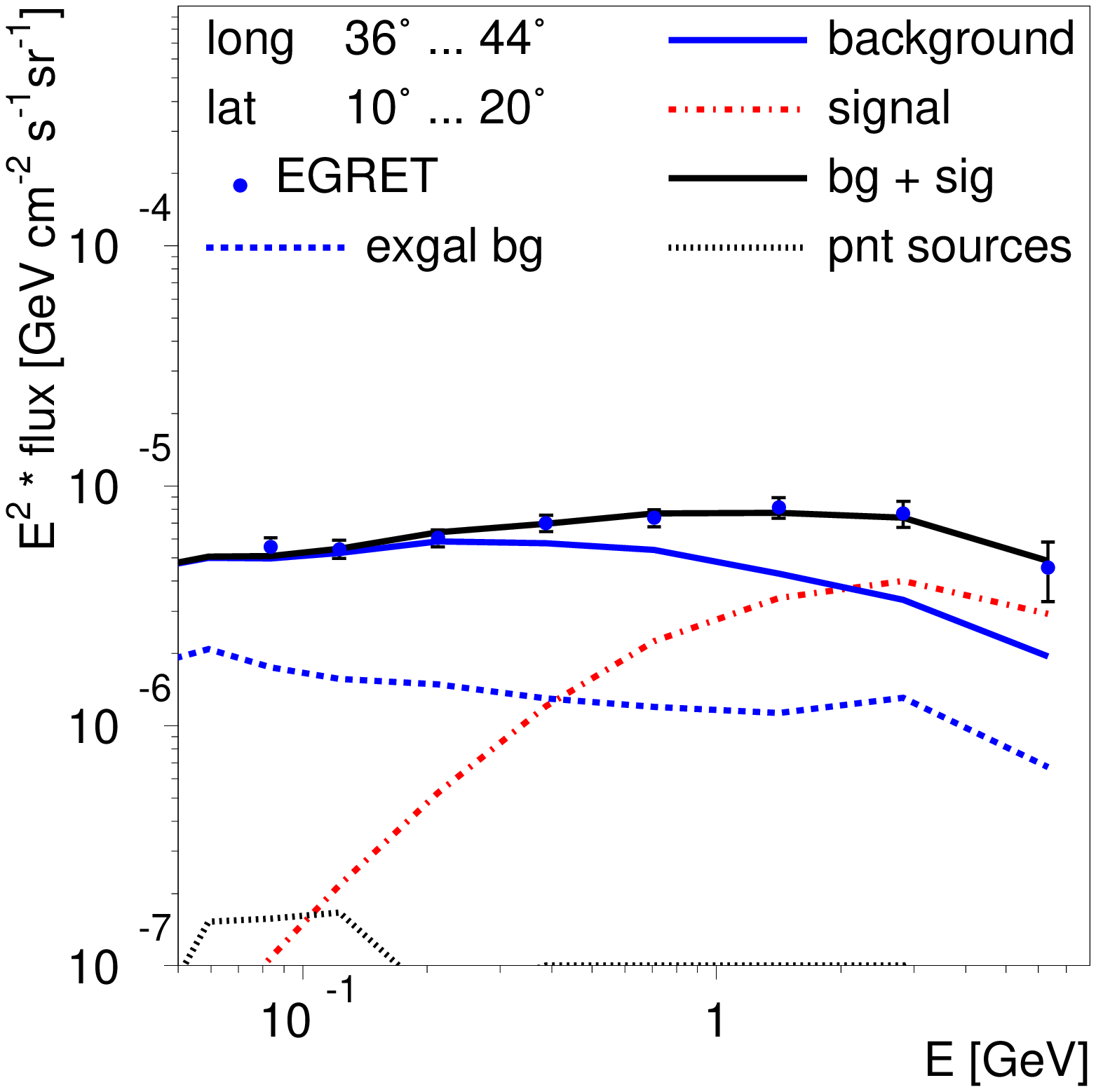}
    \includegraphics[width=0.21\textwidth]{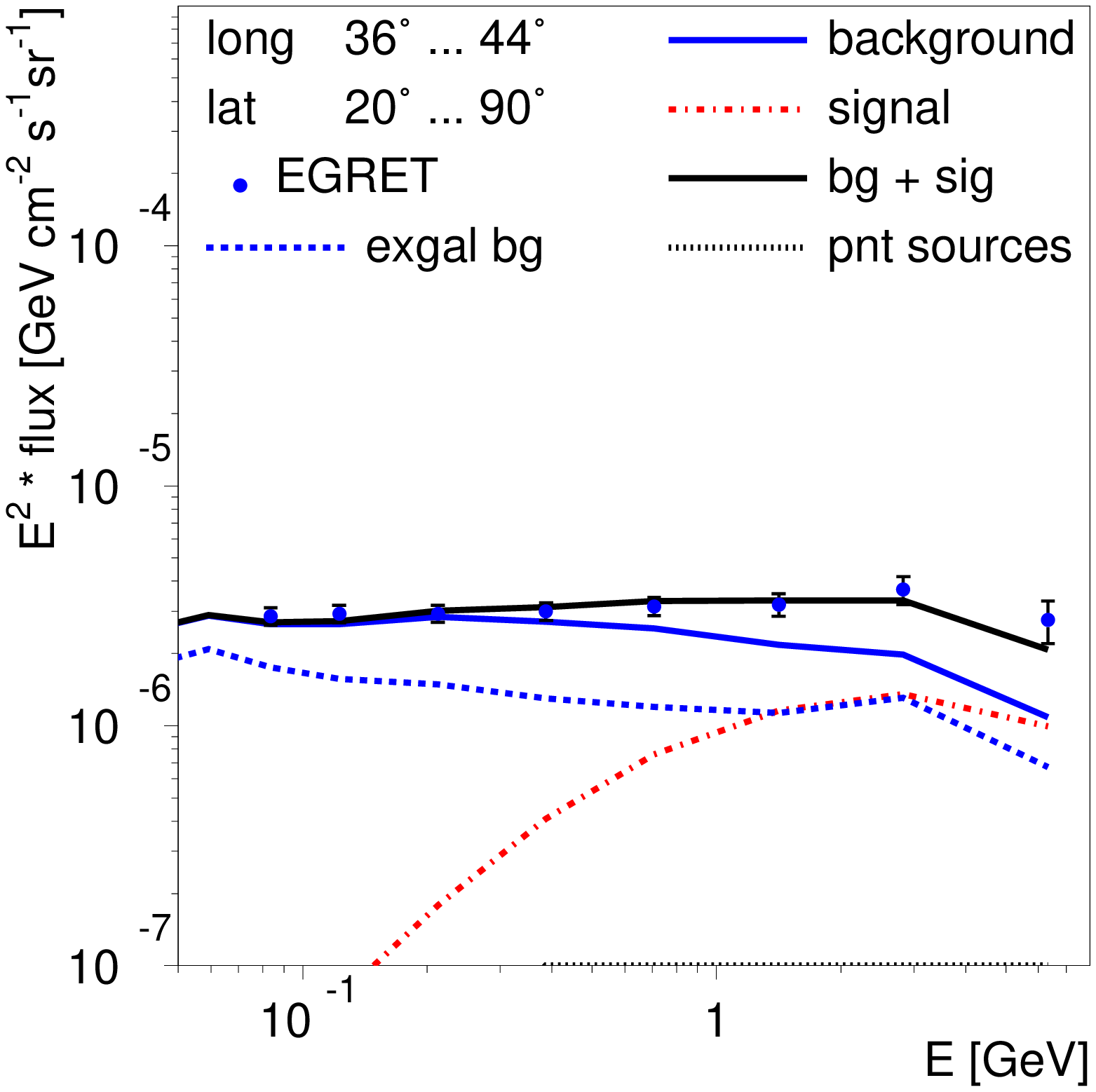}\\
    \hspace{-1cm}
    \begin{turn}{90} \framebox[0.21\textwidth][c]{{\scriptsize $44^\circ<\mbox{long}<52^\circ$}} \end{turn}
    \includegraphics[width=0.21\textwidth]{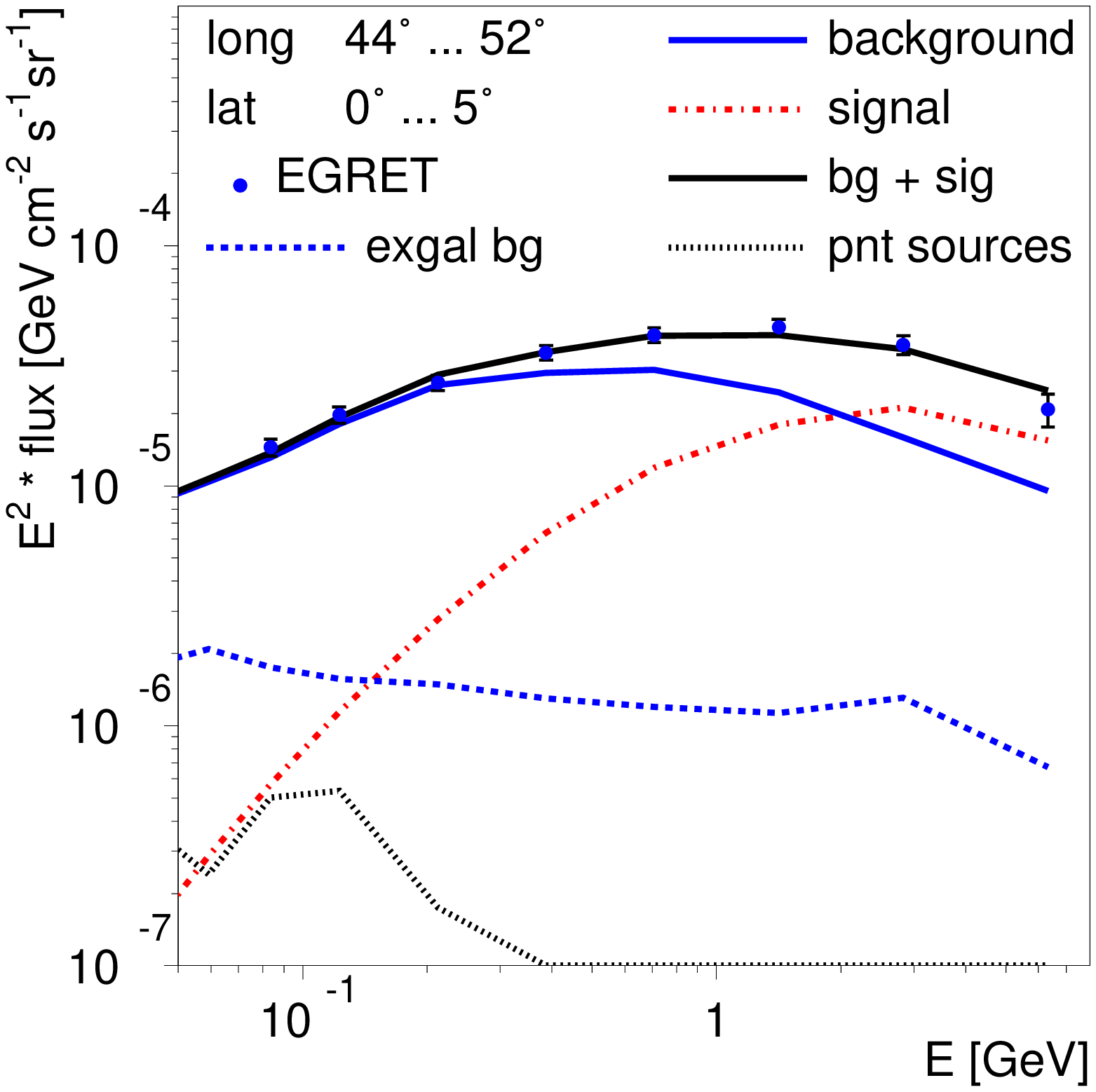}
    \includegraphics[width=0.21\textwidth]{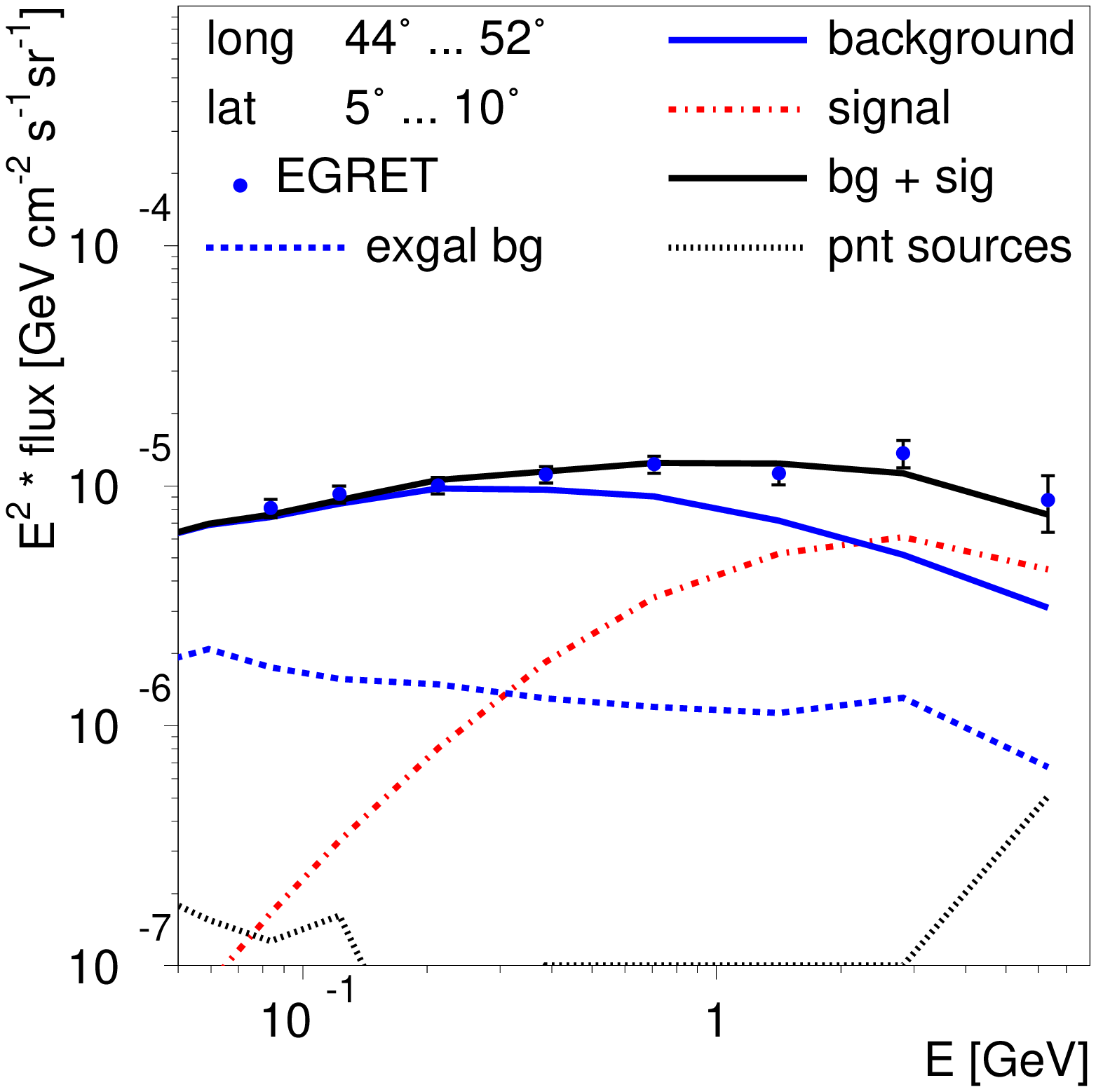}
    \includegraphics[width=0.21\textwidth]{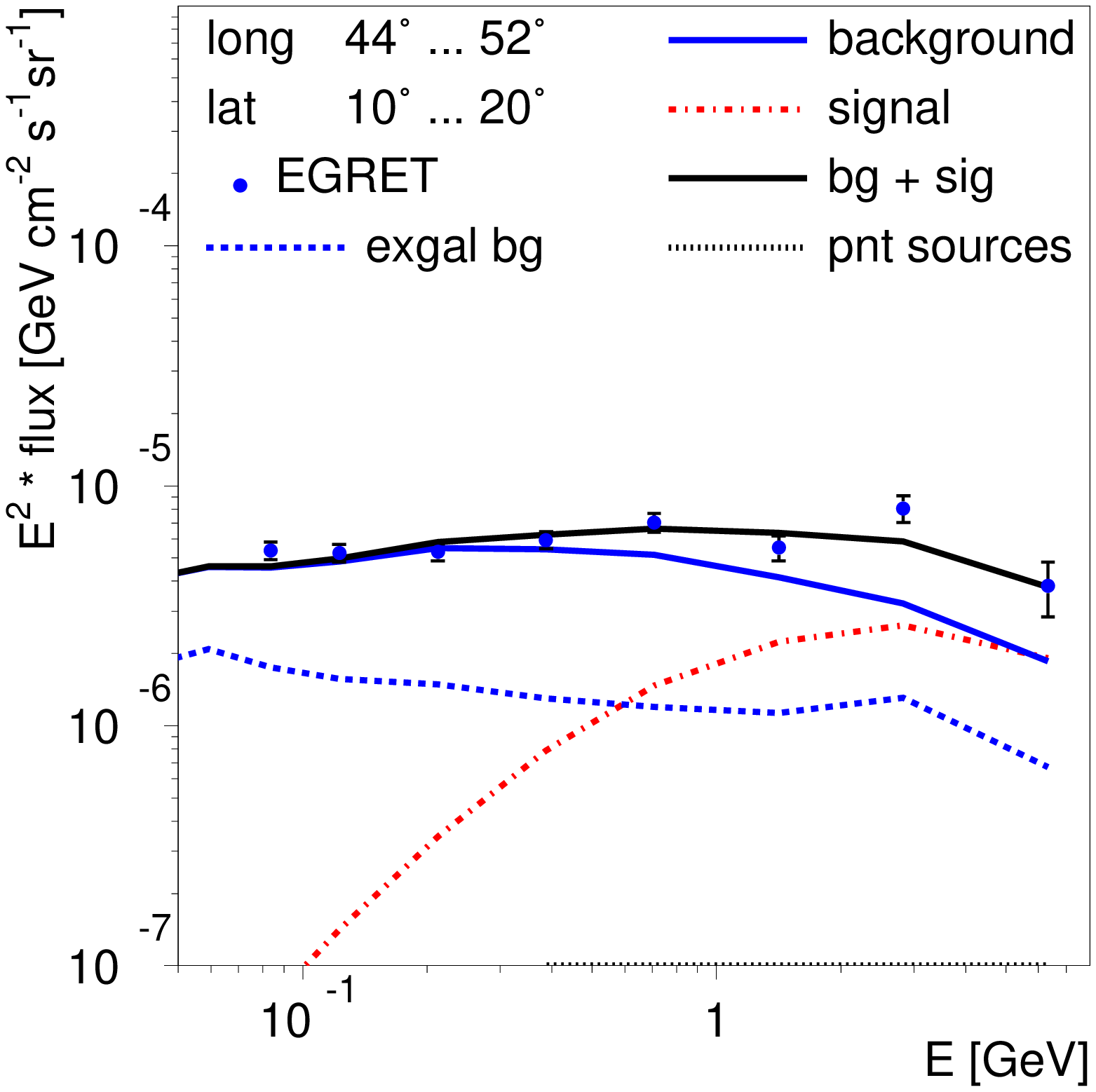}
    \includegraphics[width=0.21\textwidth]{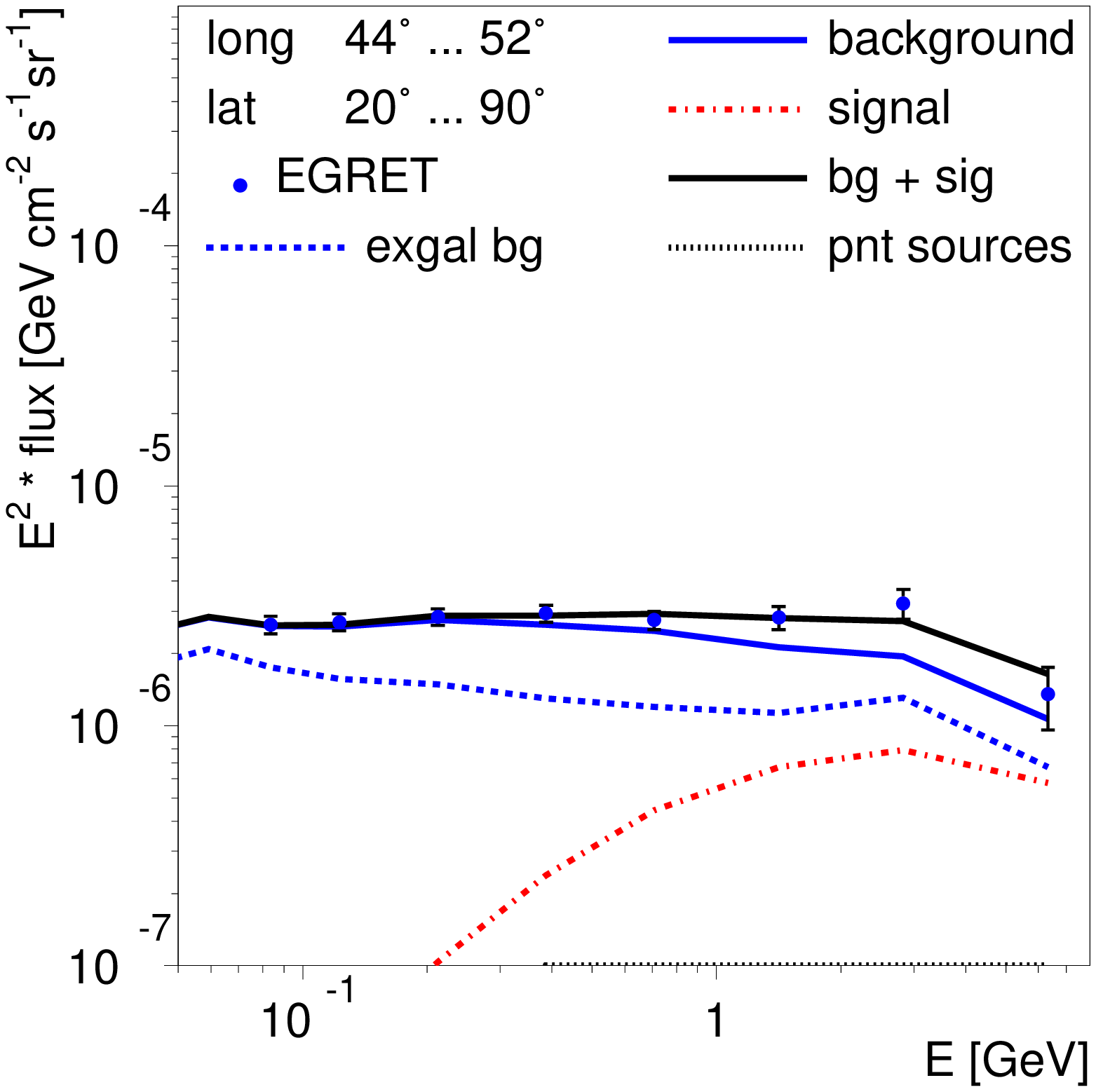}\\
    \hspace{-1cm}
    \begin{turn}{90} \framebox[0.21\textwidth][c]{{\scriptsize $52^\circ<\mbox{long}<60^\circ$}} \end{turn}
    \includegraphics[width=0.21\textwidth]{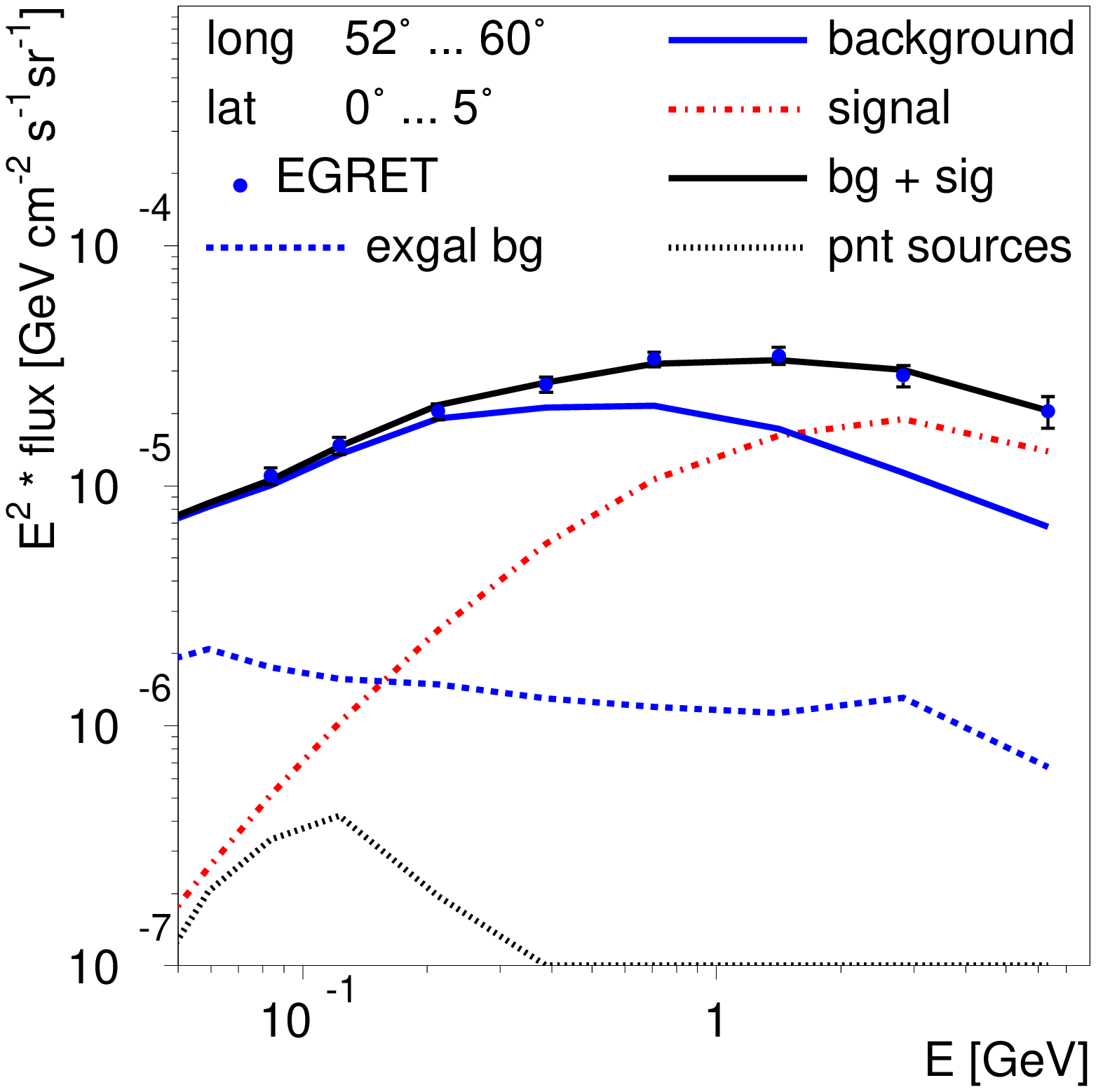}
    \includegraphics[width=0.21\textwidth]{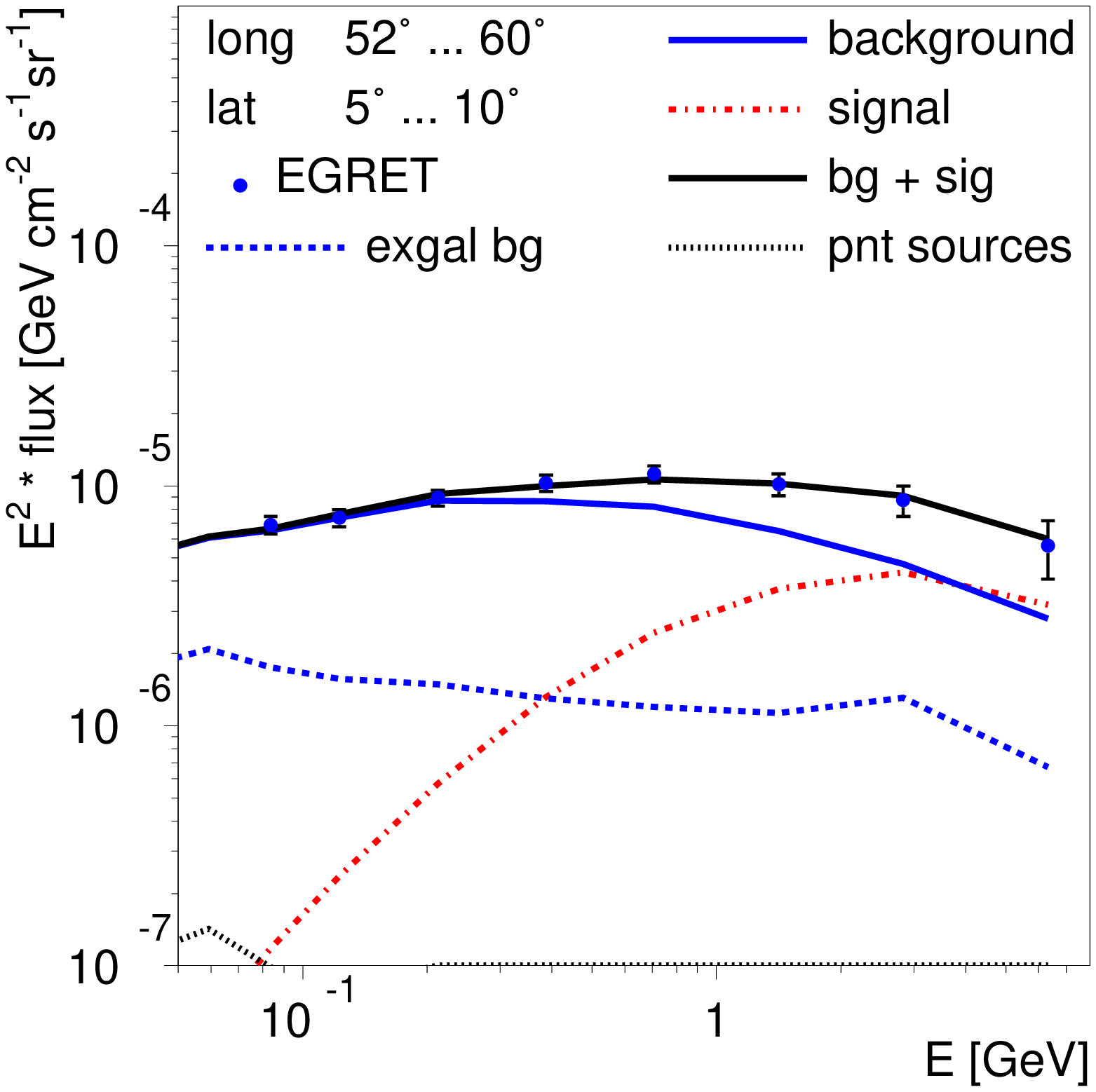}
    \includegraphics[width=0.21\textwidth]{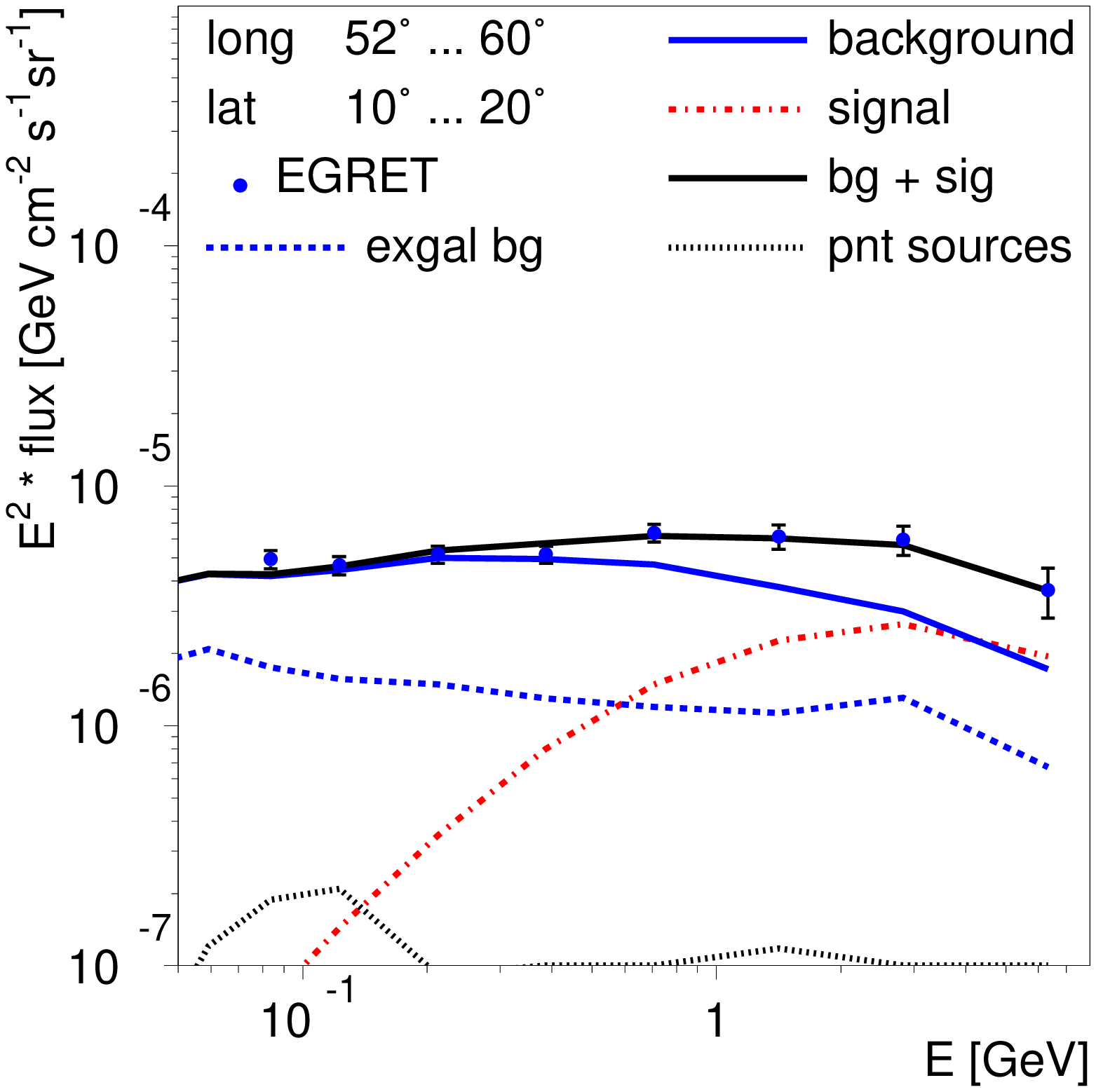}
    \includegraphics[width=0.21\textwidth]{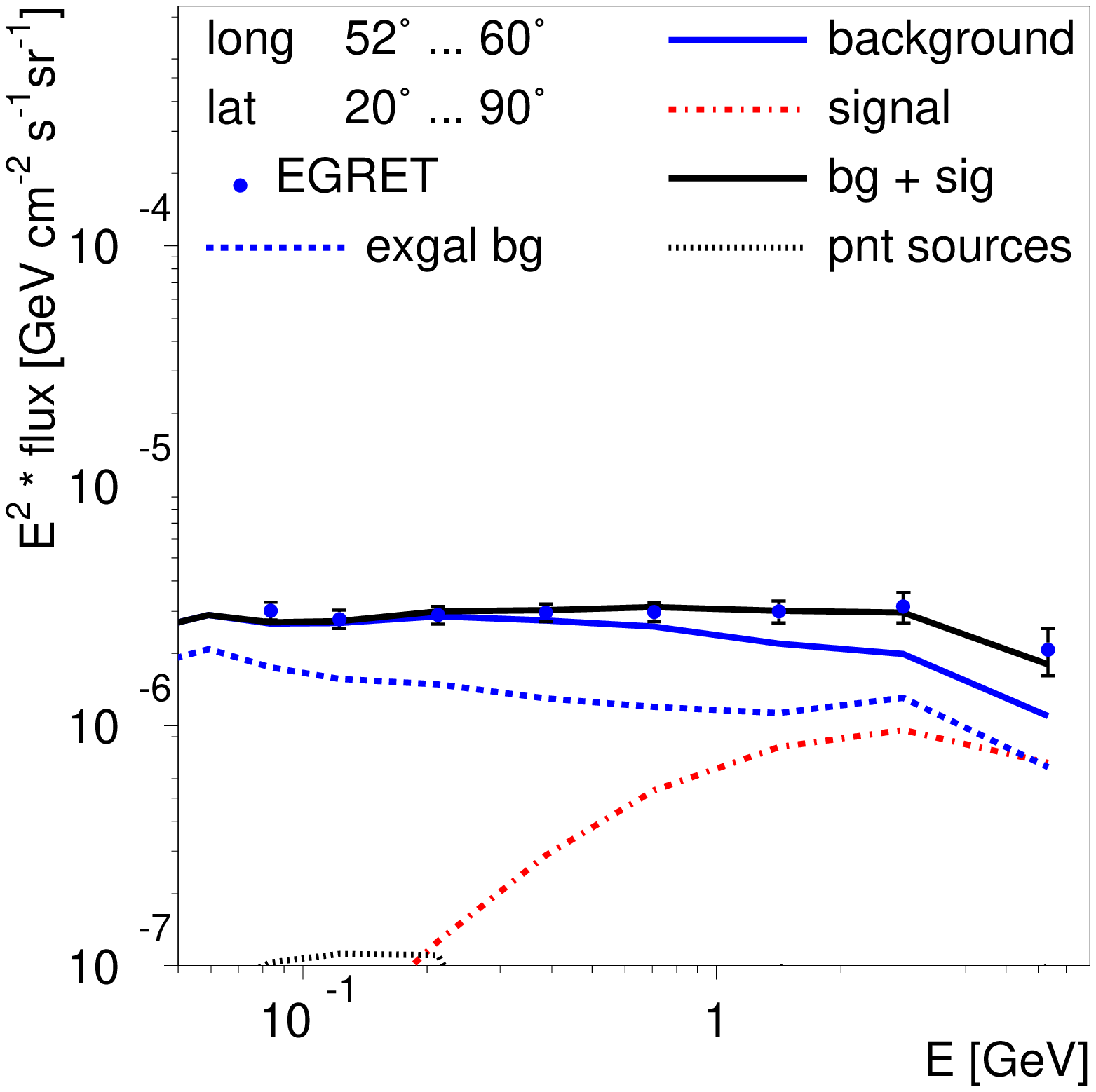}\\
  \end{center}
  \clearpage
  \begin{center}
    \framebox[0.21\textwidth][c]{$\vert \mbox{lat}\vert<5^\circ$}
    \framebox[0.21\textwidth][c]{$5^\circ<\vert \mbox{lat}\vert<10^\circ$}
    \framebox[0.21\textwidth][c]{$10^\circ<\vert \mbox{lat}\vert<20^\circ$}
    \framebox[0.21\textwidth][c]{$20^\circ<\vert \mbox{lat}\vert<90^\circ$}\\
    \hspace{-1cm}
    \begin{turn}{90} \framebox[0.21\textwidth][c]{{\scriptsize $60^\circ<\mbox{long}<68^\circ$}} \end{turn}
    \includegraphics[width=0.21\textwidth]{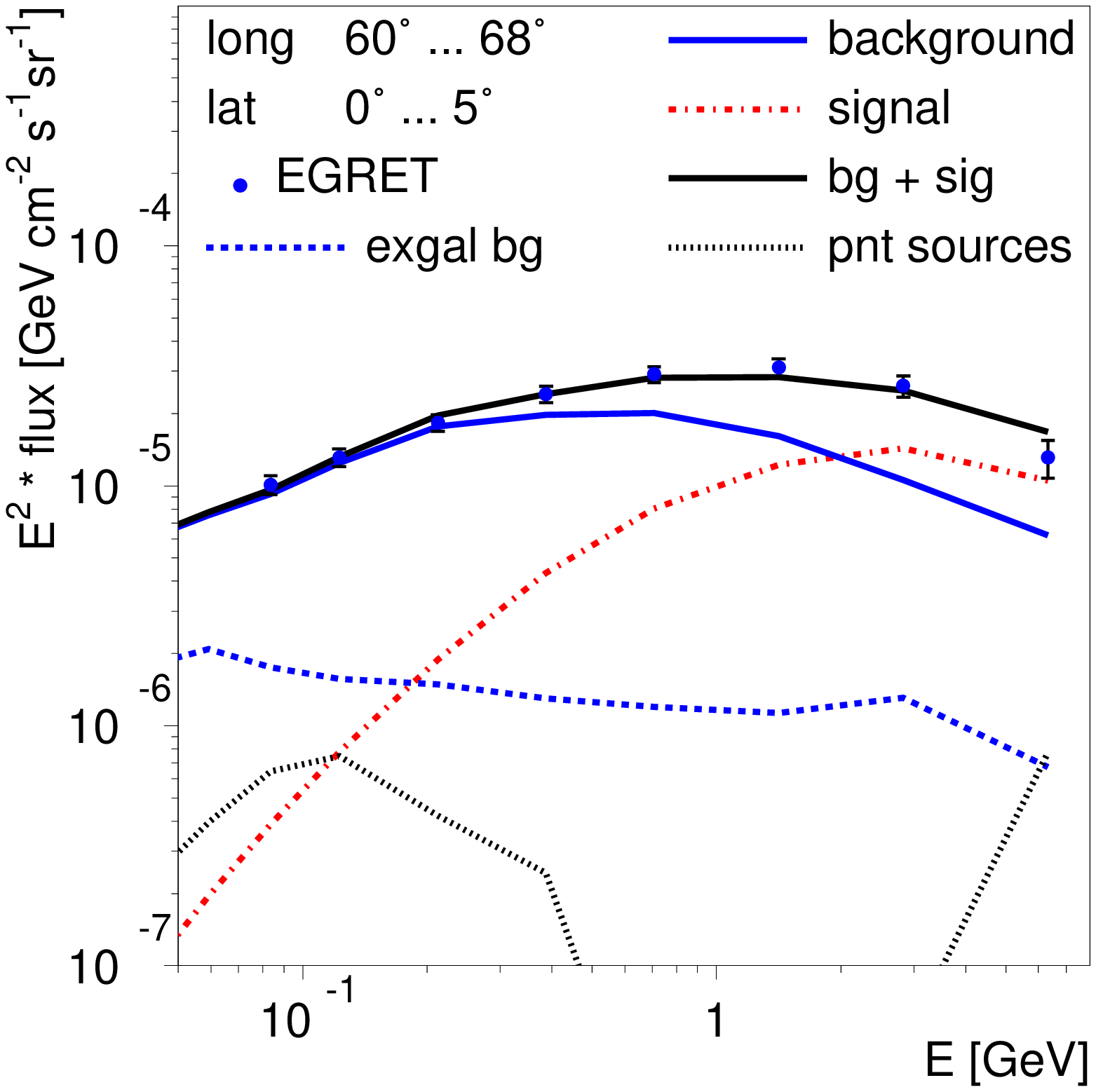}
    \includegraphics[width=0.21\textwidth]{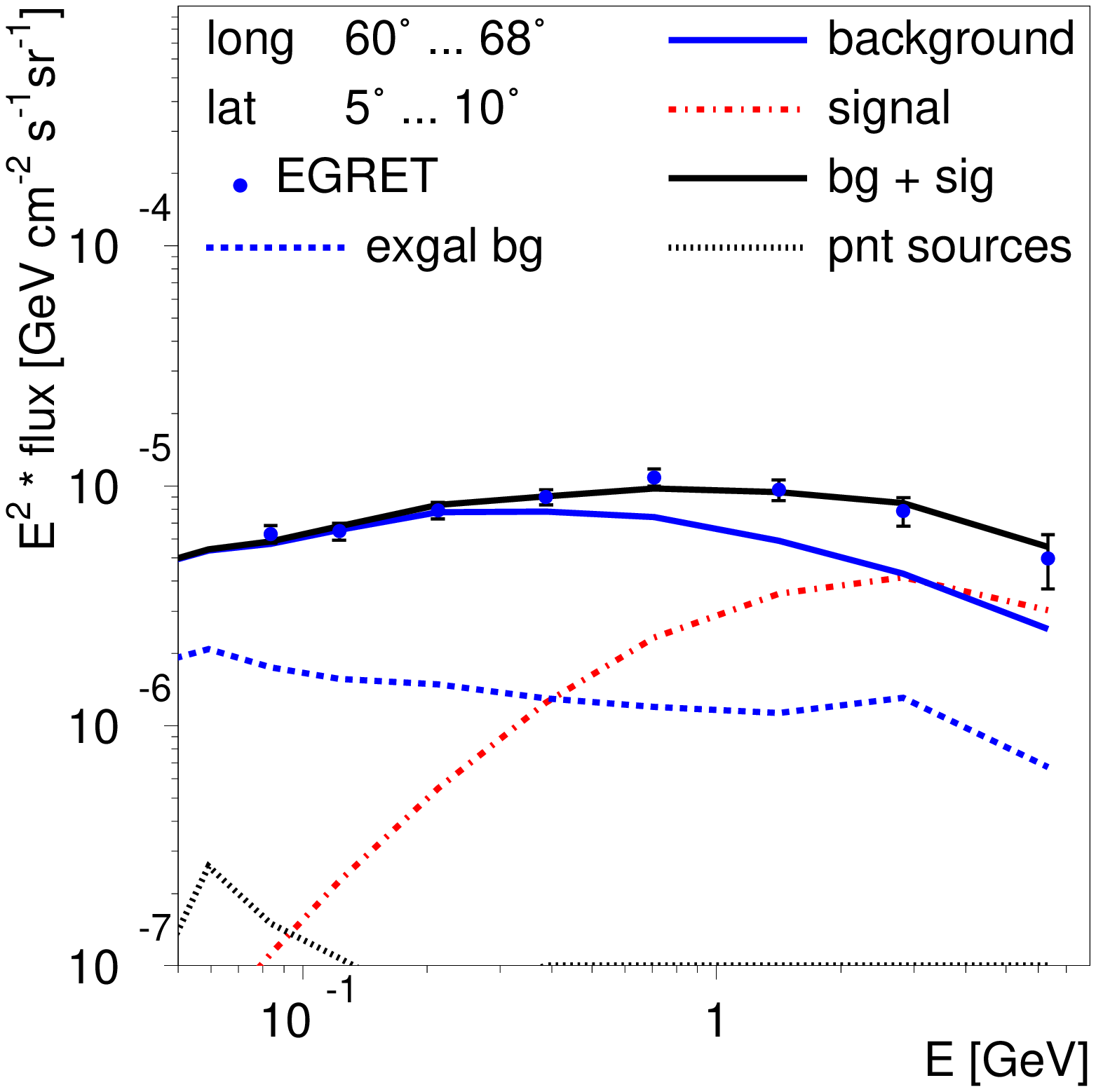}
    \includegraphics[width=0.21\textwidth]{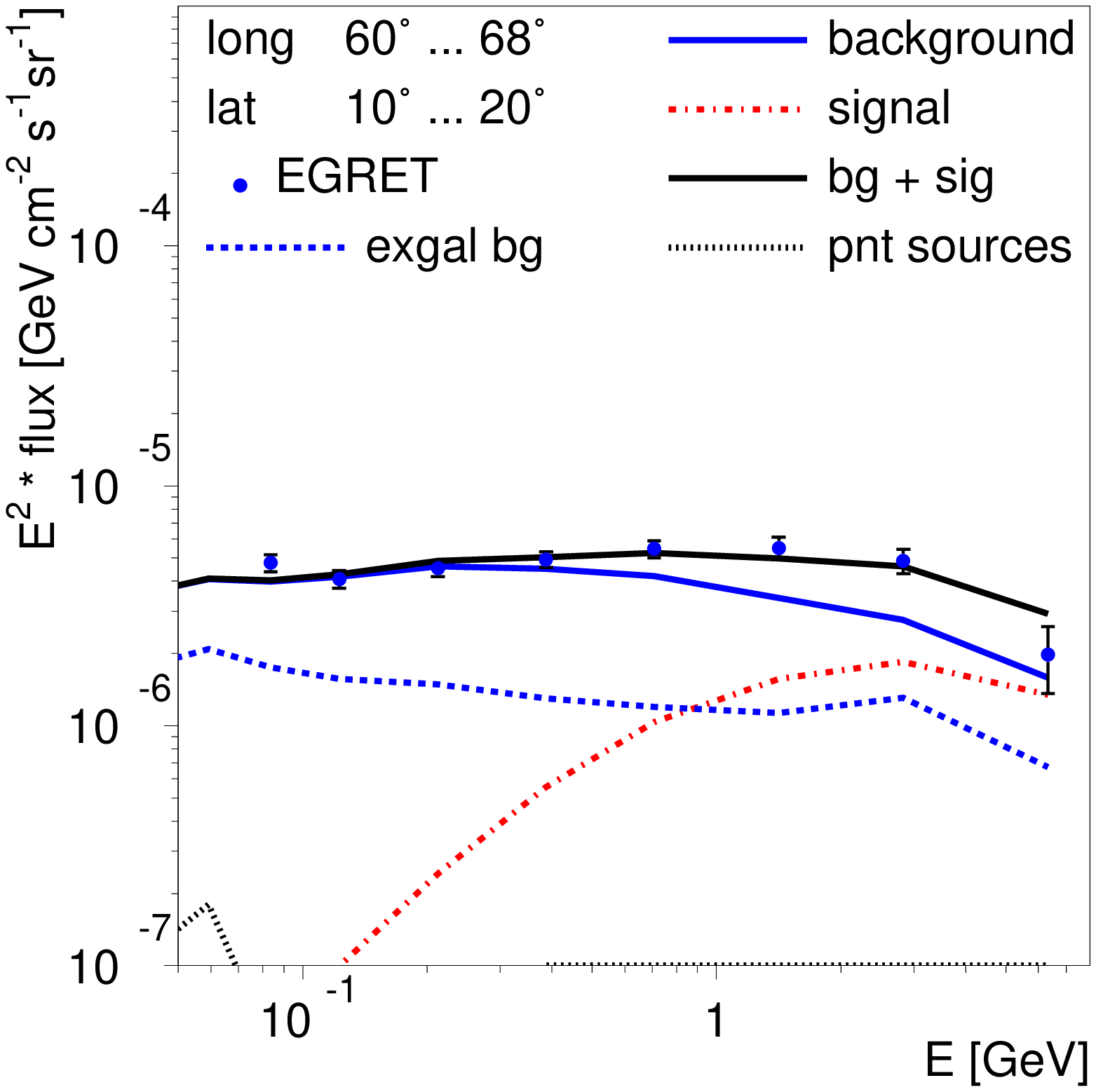}
    \includegraphics[width=0.21\textwidth]{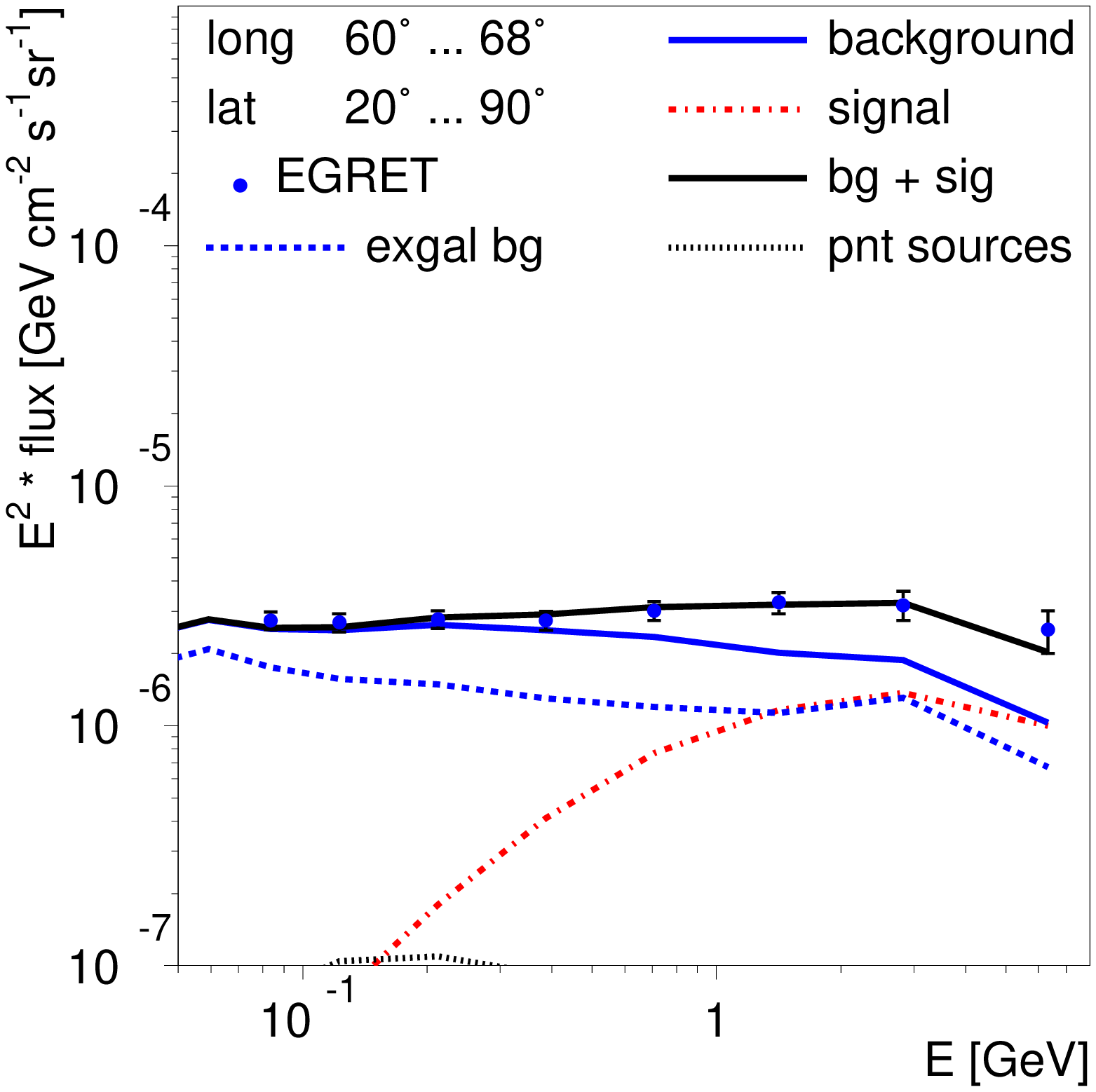}\\
    \hspace{-1cm}
    \begin{turn}{90} \framebox[0.21\textwidth][c]{{\scriptsize $68^\circ<\mbox{long}<76^\circ$}} \end{turn}
    \includegraphics[width=0.21\textwidth]{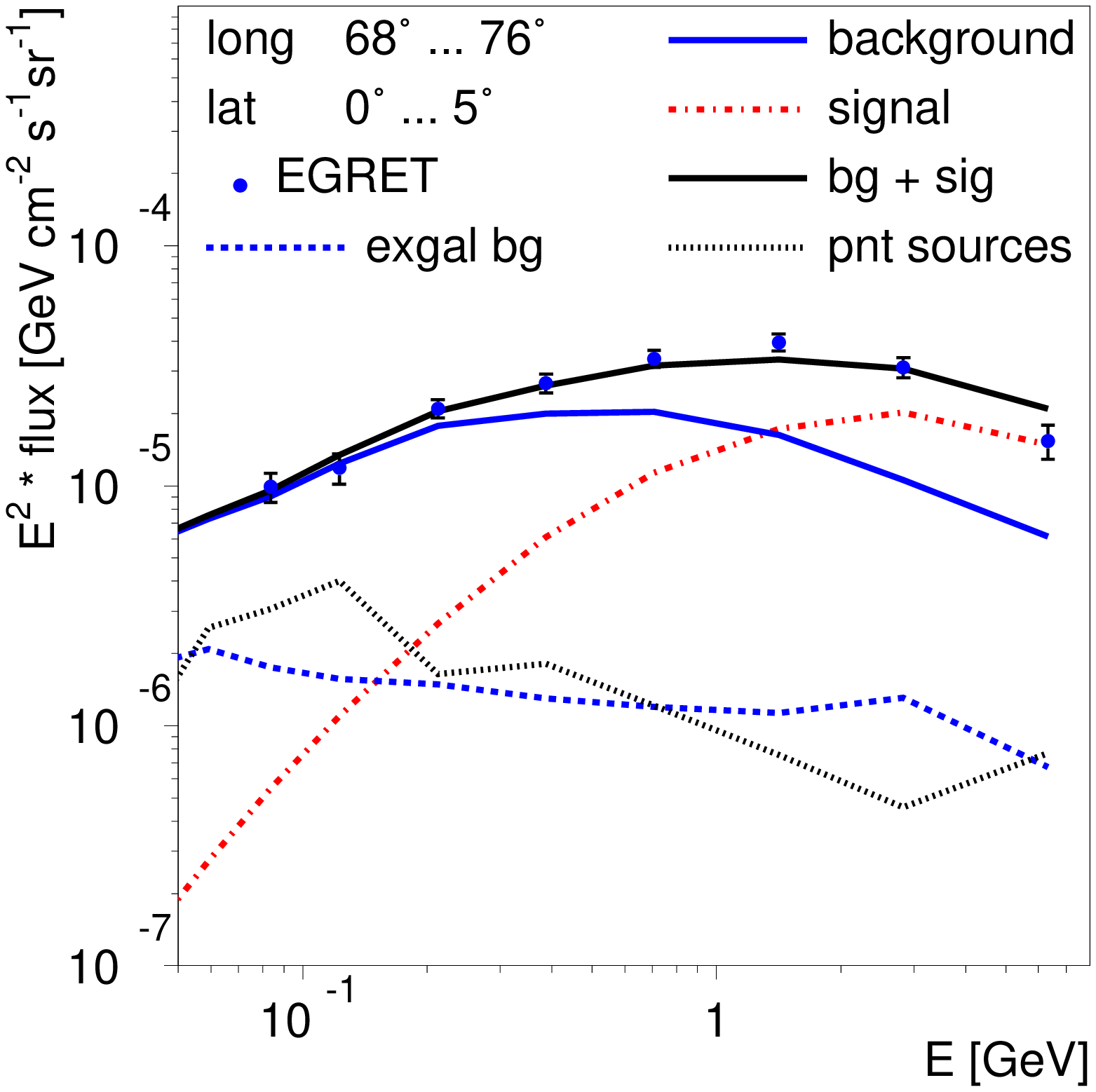}
    \includegraphics[width=0.21\textwidth]{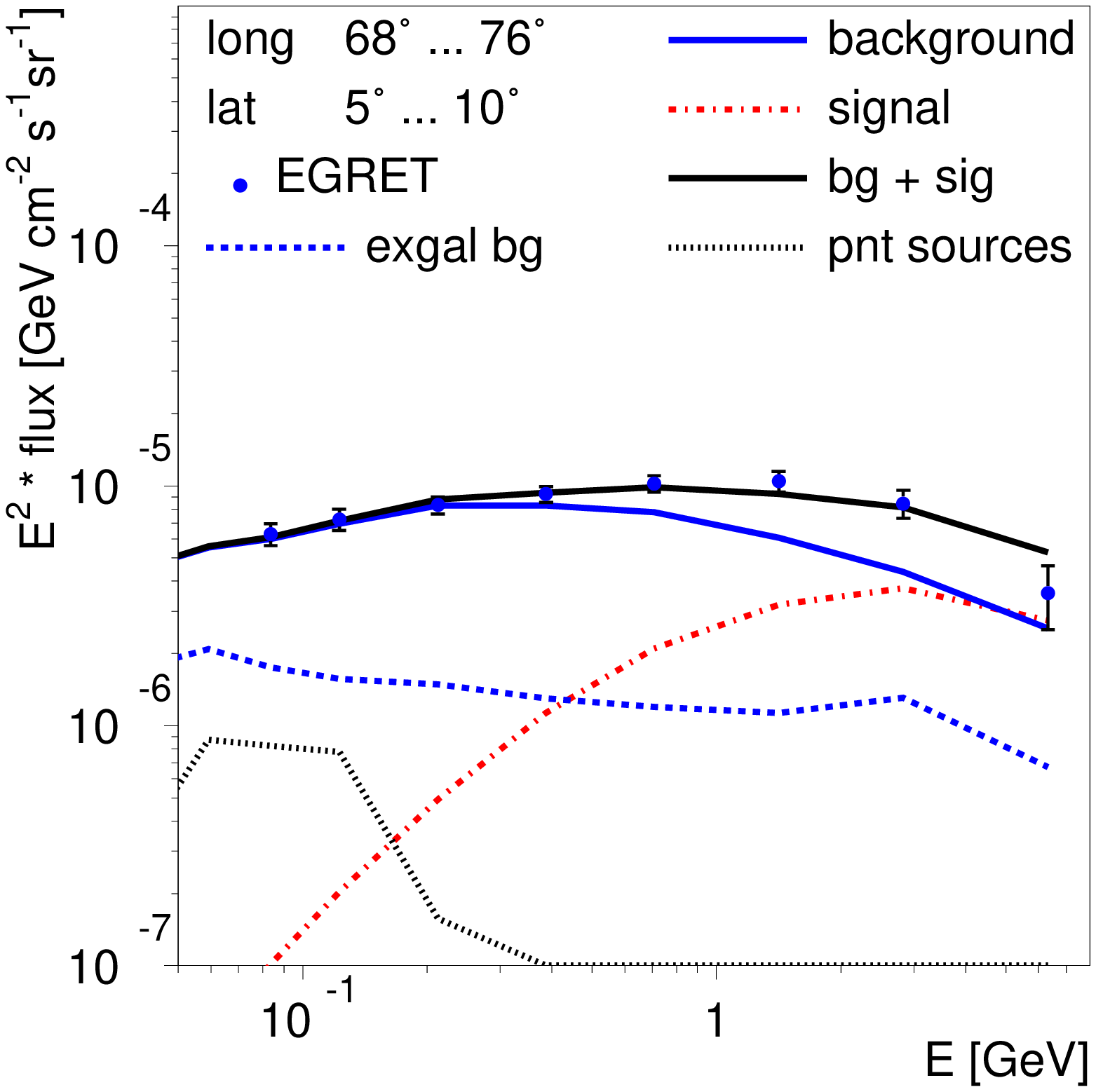}
    \includegraphics[width=0.21\textwidth]{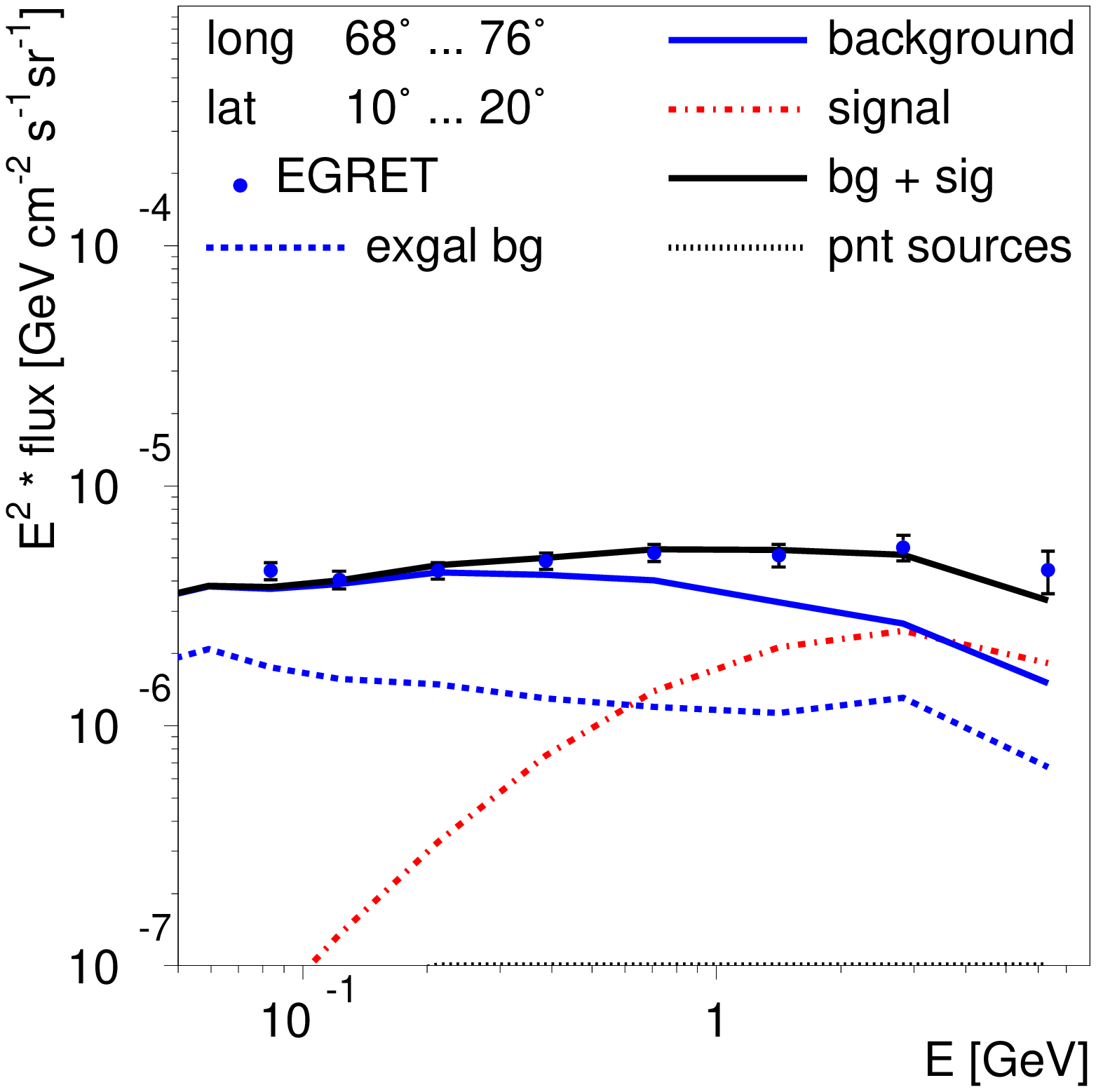}
    \includegraphics[width=0.21\textwidth]{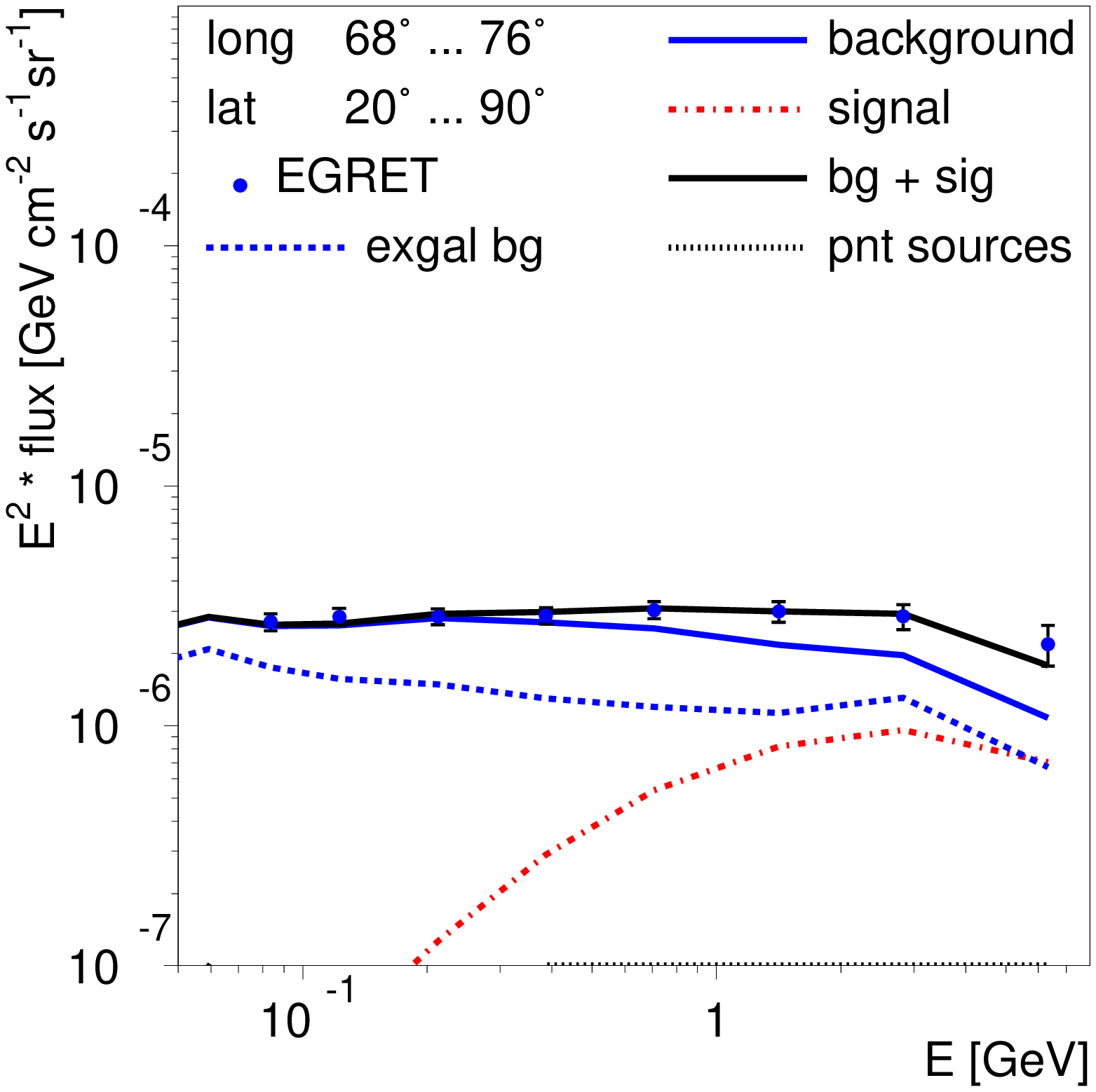}\\
    \hspace{-1cm}
    \begin{turn}{90} \framebox[0.21\textwidth][c]{{\scriptsize $76^\circ<\mbox{long}<84^\circ$}} \end{turn}
    \includegraphics[width=0.21\textwidth]{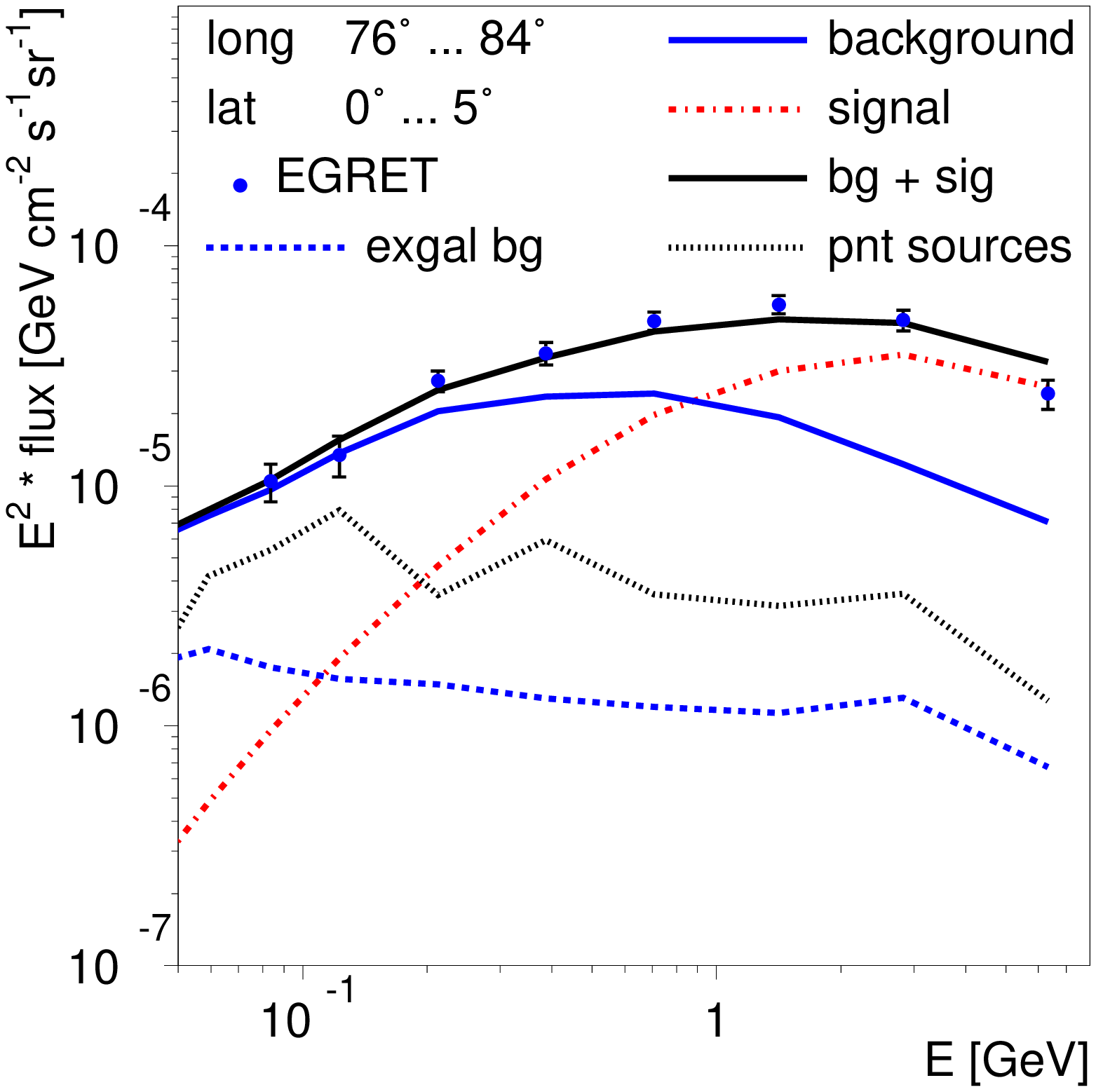}
    \includegraphics[width=0.21\textwidth]{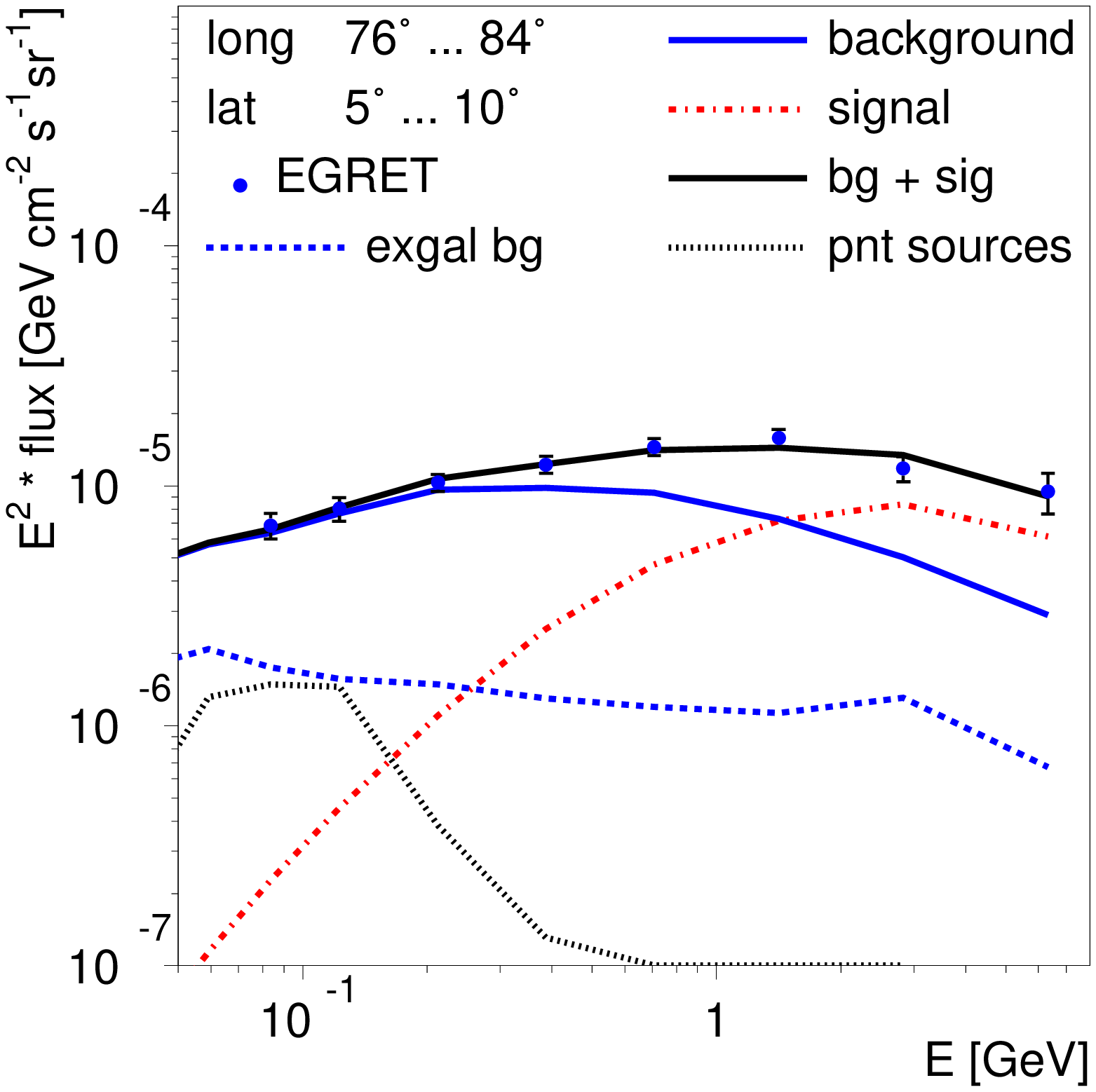}
    \includegraphics[width=0.21\textwidth]{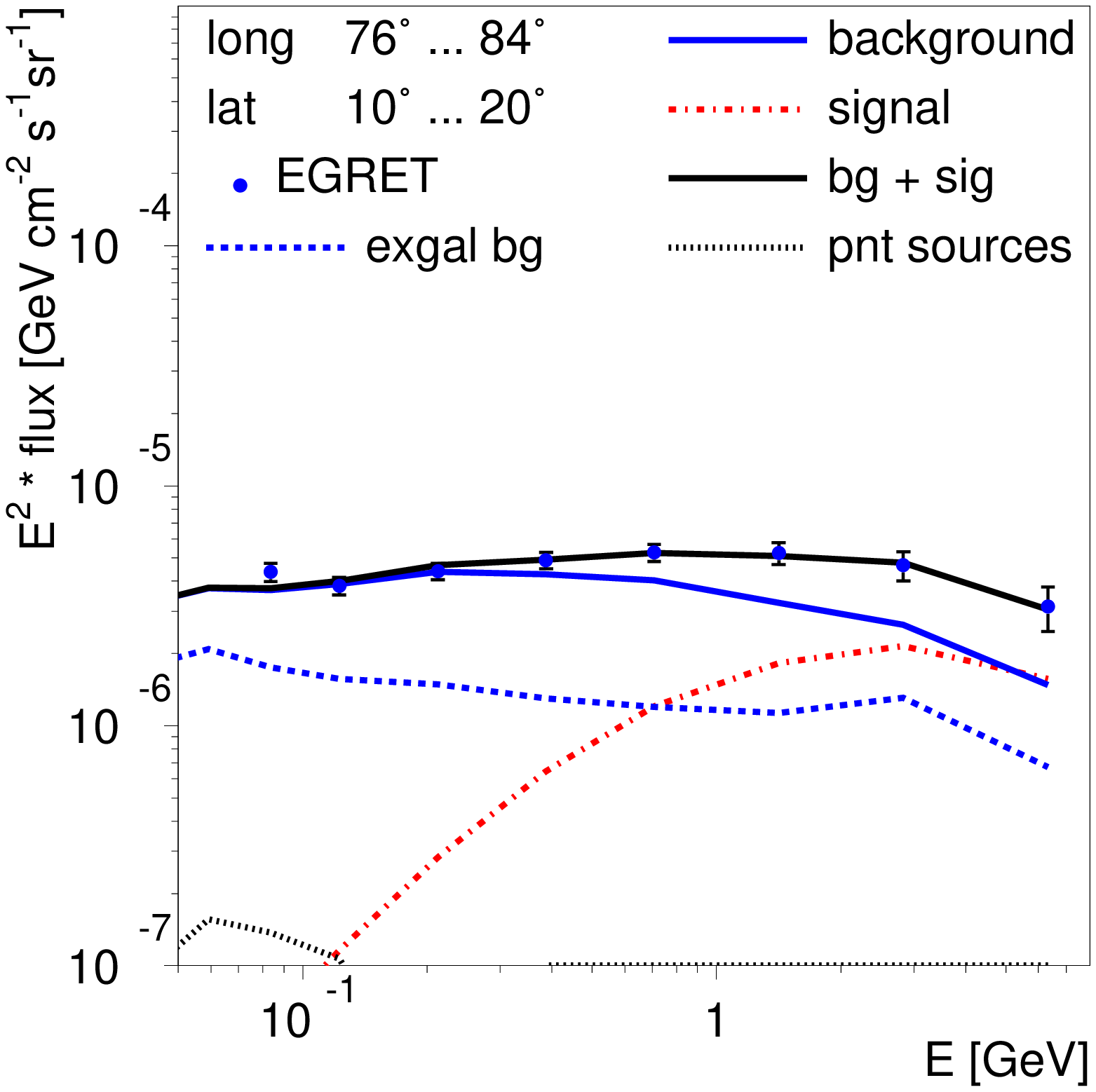}
    \includegraphics[width=0.21\textwidth]{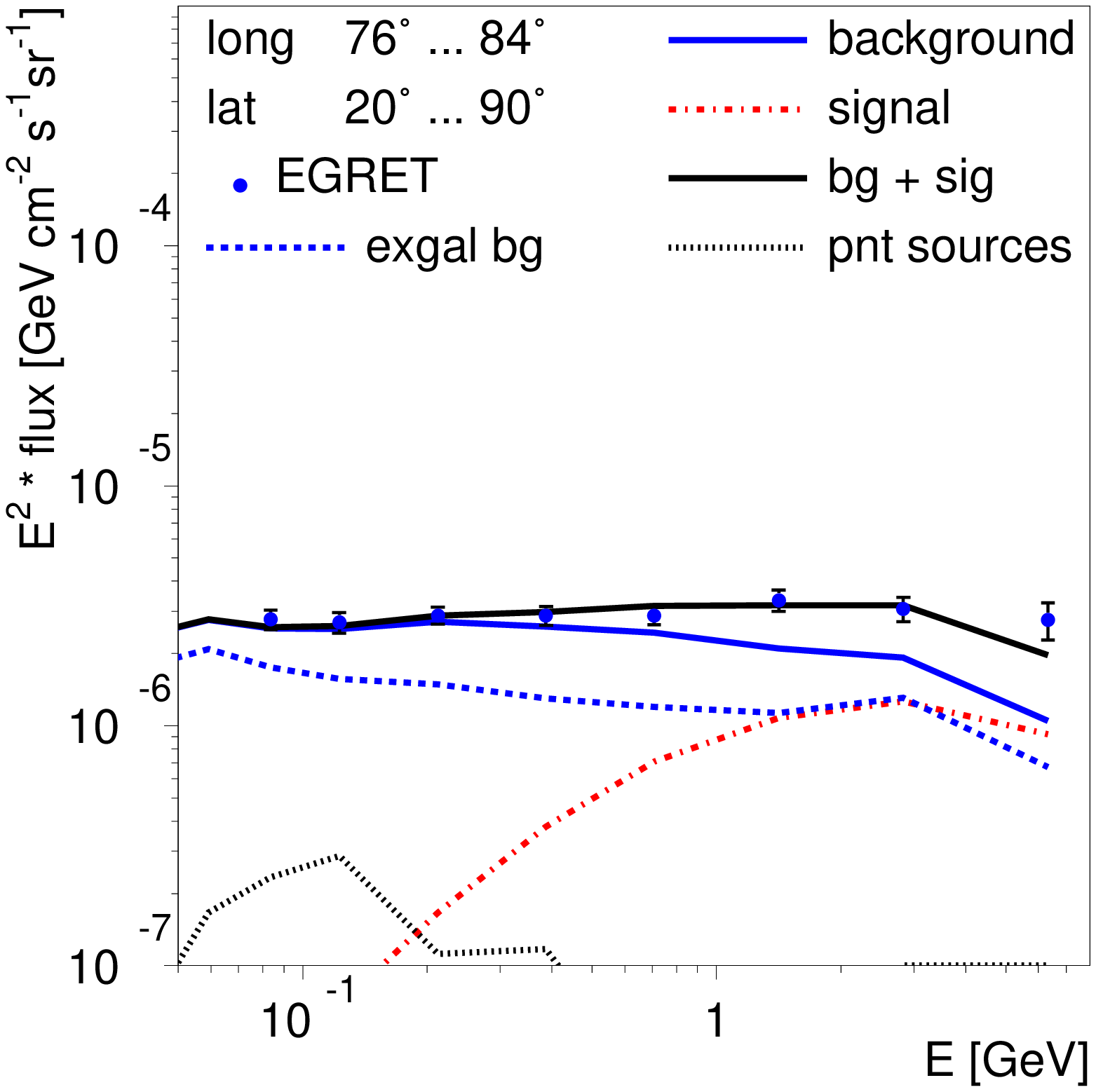}\\
    \hspace{-1cm}
    \begin{turn}{90} \framebox[0.21\textwidth][c]{{\scriptsize $84^\circ<\mbox{long}<92^\circ$}} \end{turn}
    \includegraphics[width=0.21\textwidth]{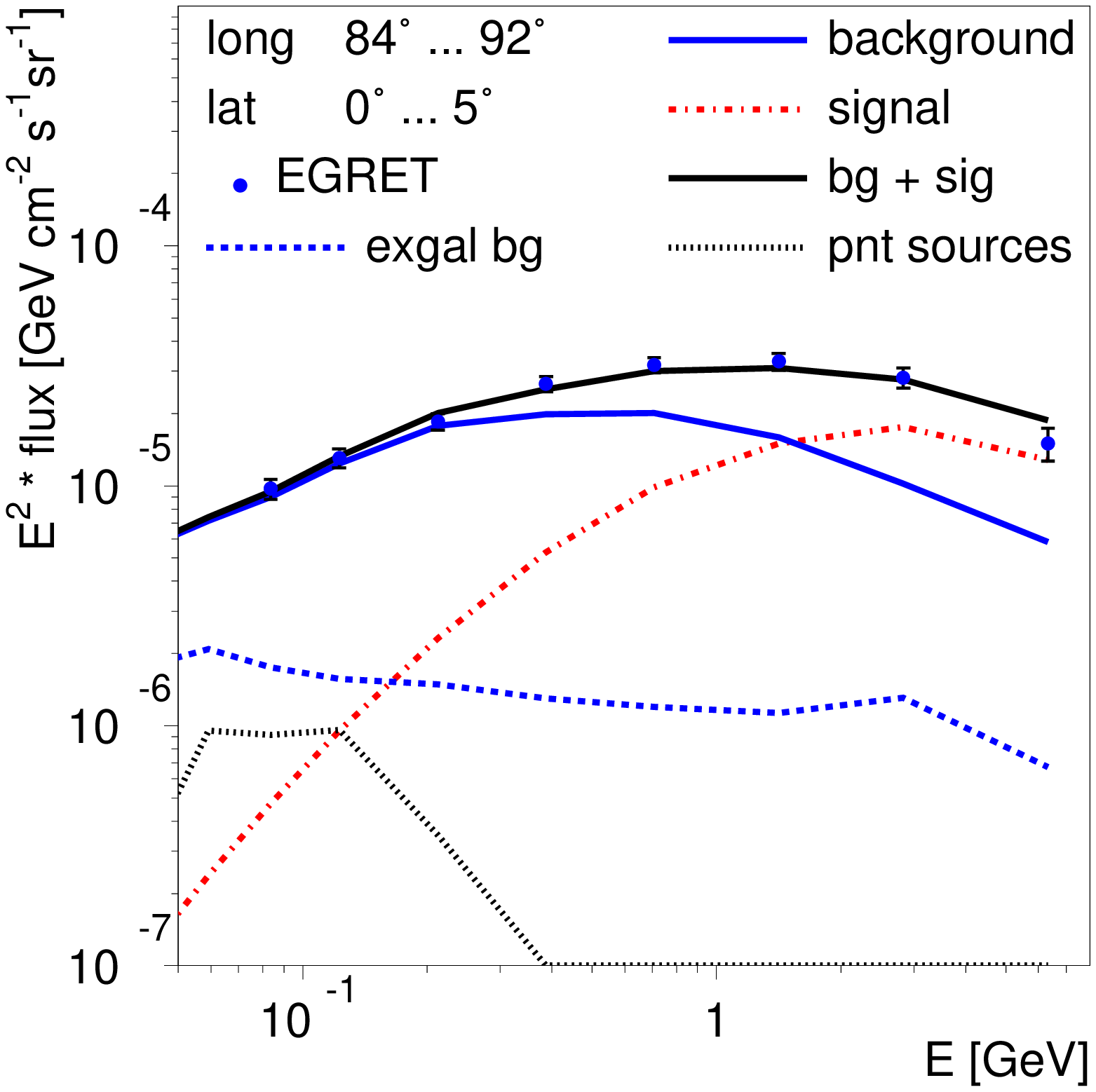}
    \includegraphics[width=0.21\textwidth]{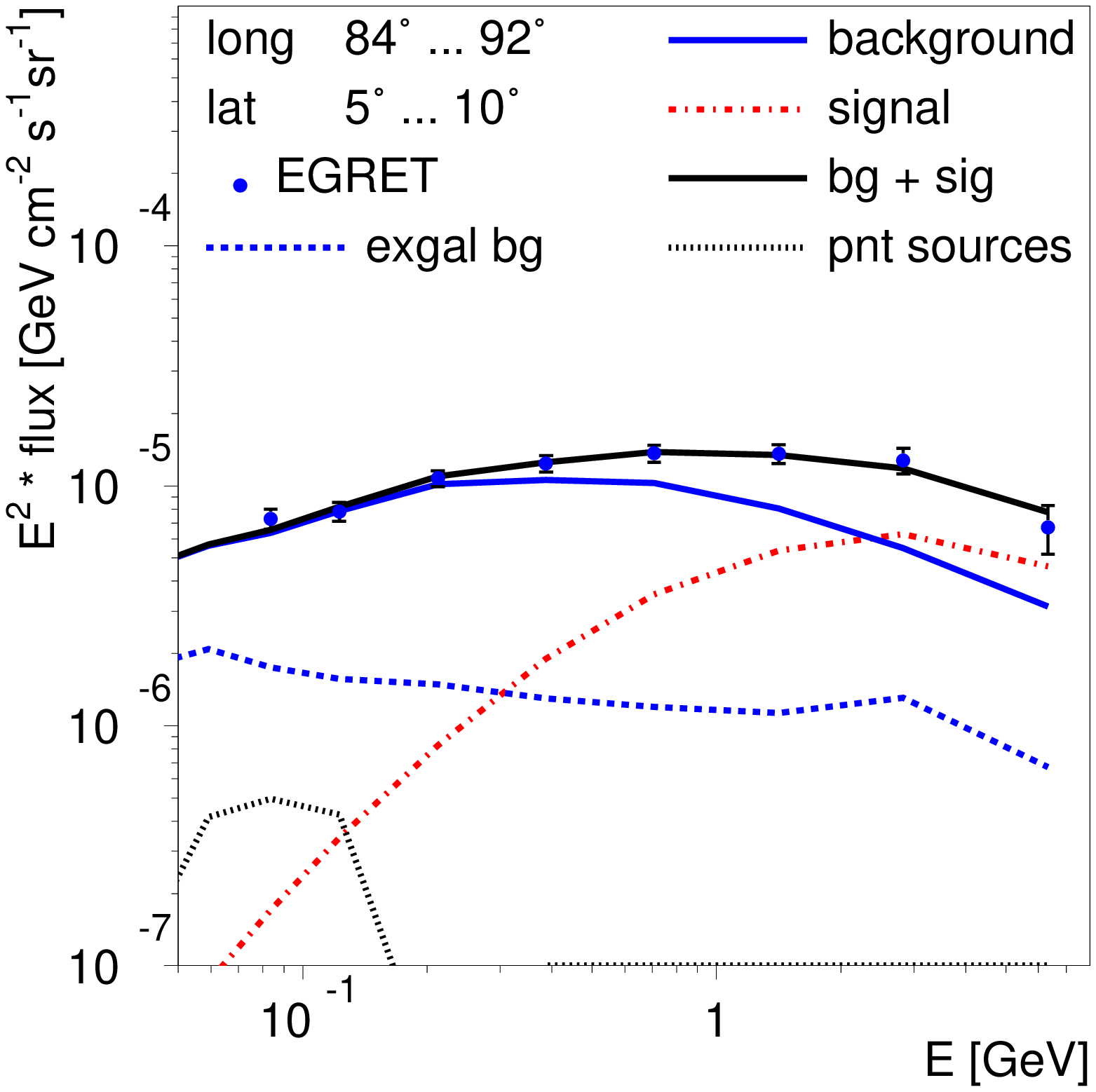}
    \includegraphics[width=0.21\textwidth]{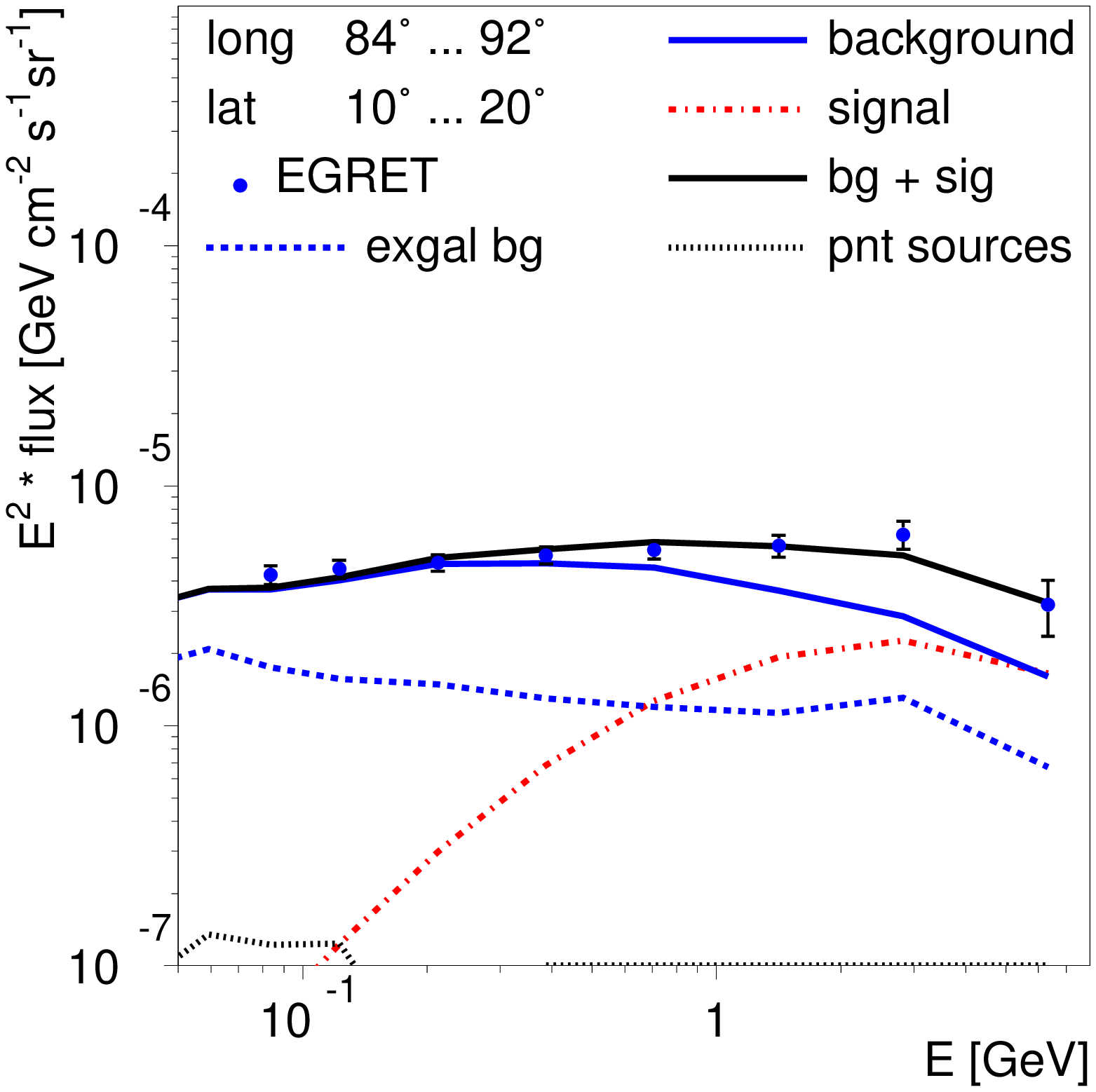}
    \includegraphics[width=0.21\textwidth]{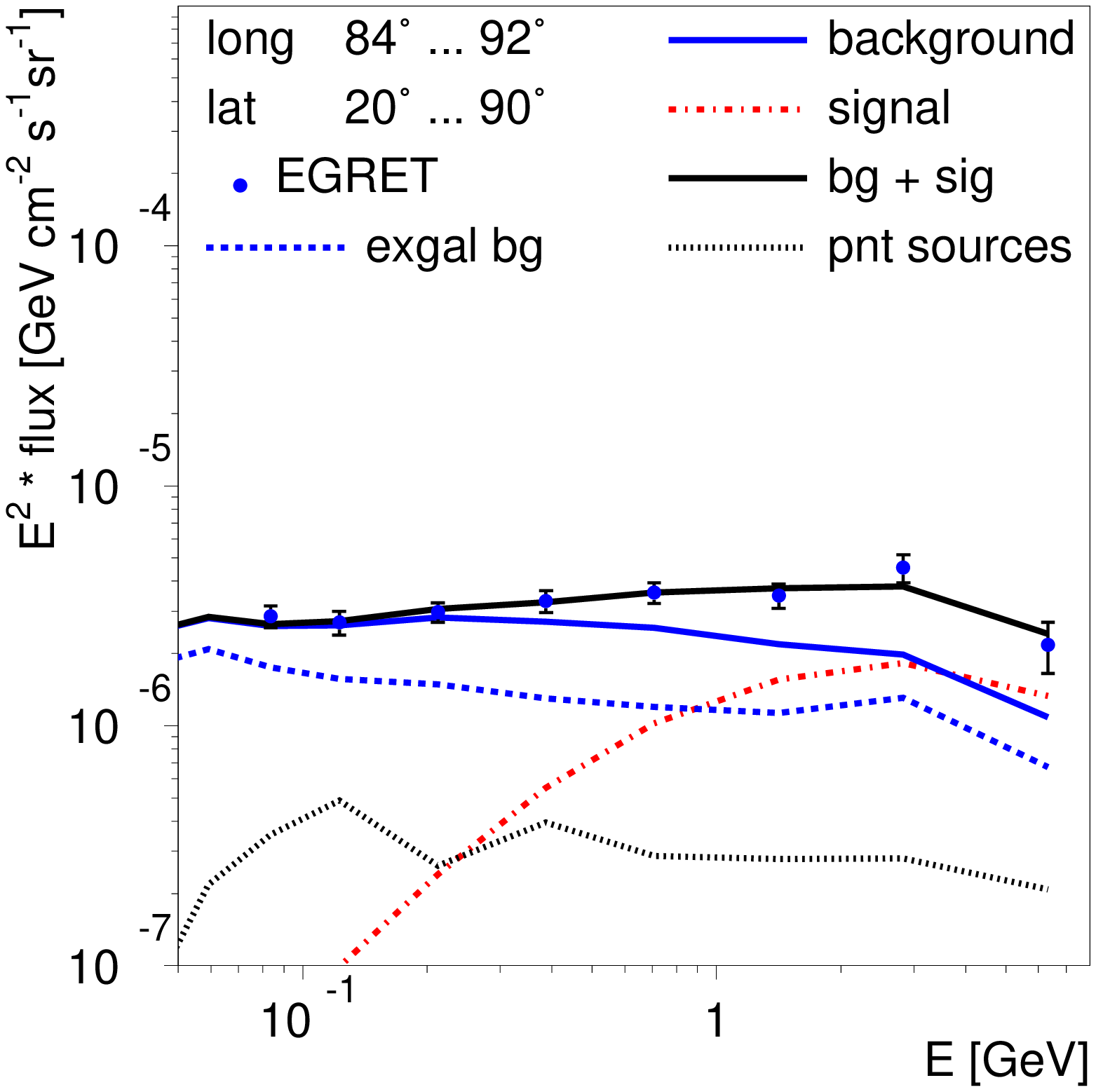}\\
    \hspace{-1cm}
    \begin{turn}{90} \framebox[0.21\textwidth][c]{{\scriptsize $92^\circ<\mbox{long}<100^\circ$}} \end{turn}
    \includegraphics[width=0.21\textwidth]{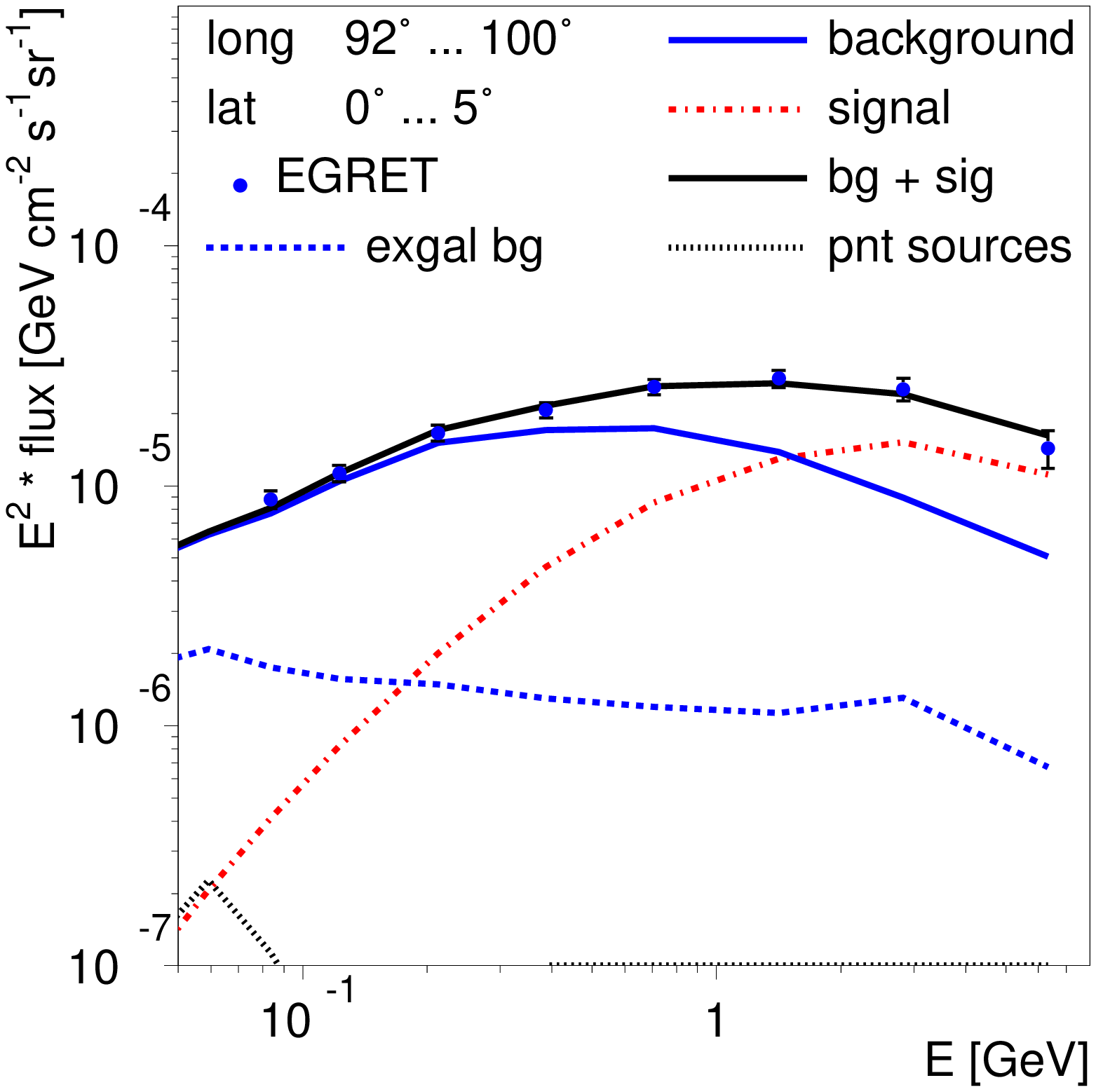}
    \includegraphics[width=0.21\textwidth]{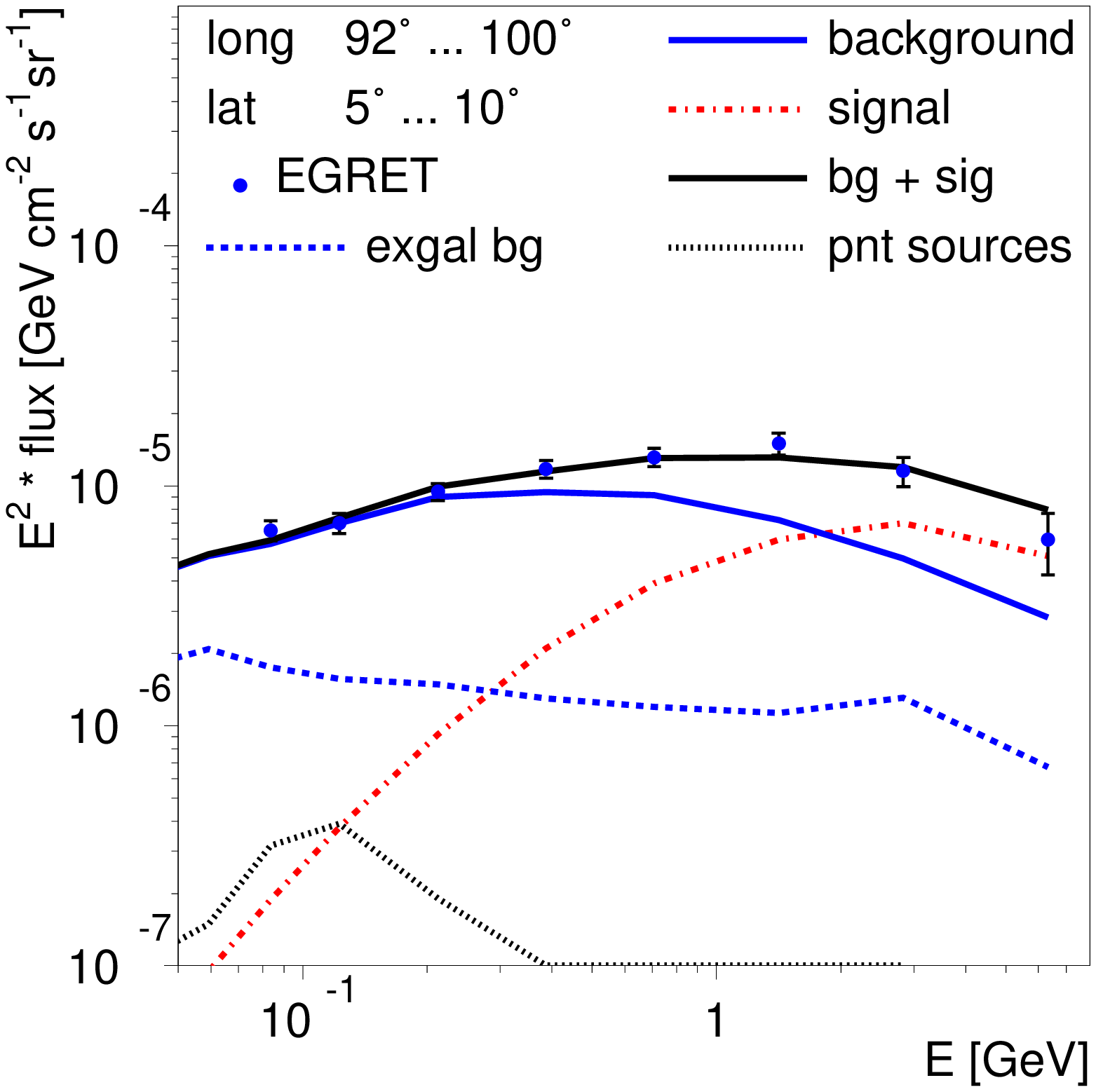}
    \includegraphics[width=0.21\textwidth]{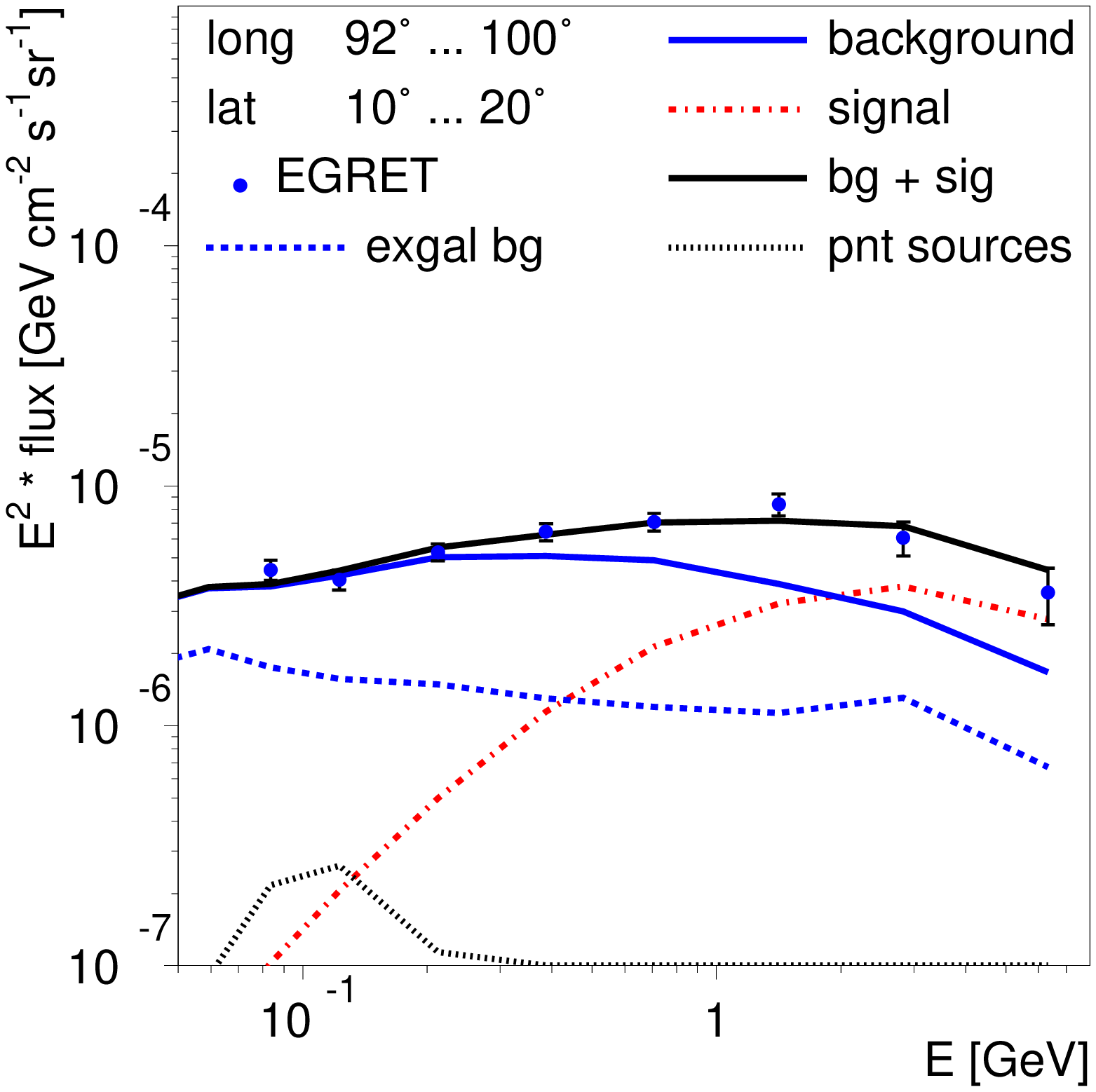}
    \includegraphics[width=0.21\textwidth]{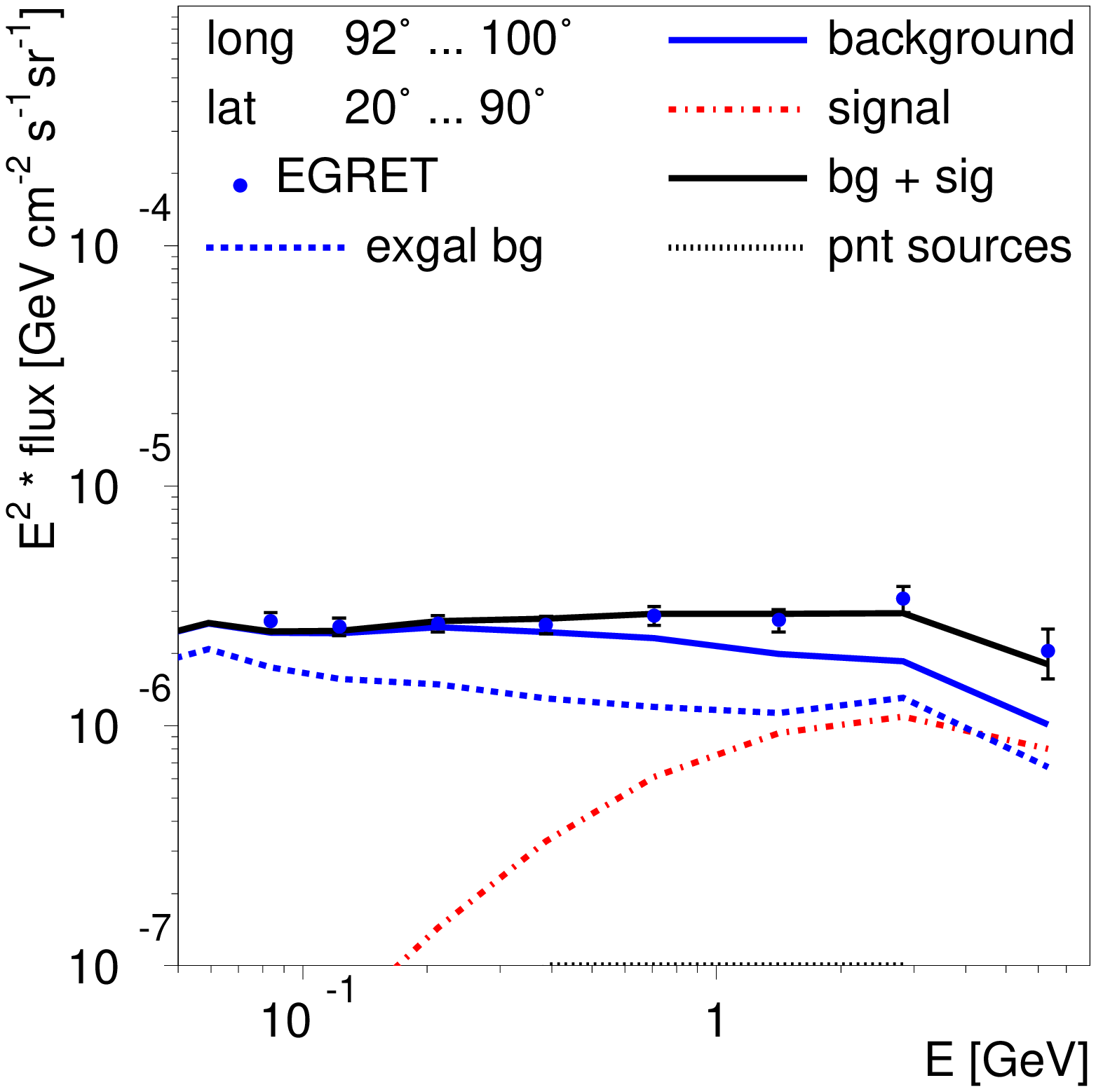}\\
    \hspace{-1cm}
    \begin{turn}{90} \framebox[0.21\textwidth][c]{{\scriptsize $100^\circ<\mbox{long}<108^\circ$}} \end{turn}
    \includegraphics[width=0.21\textwidth]{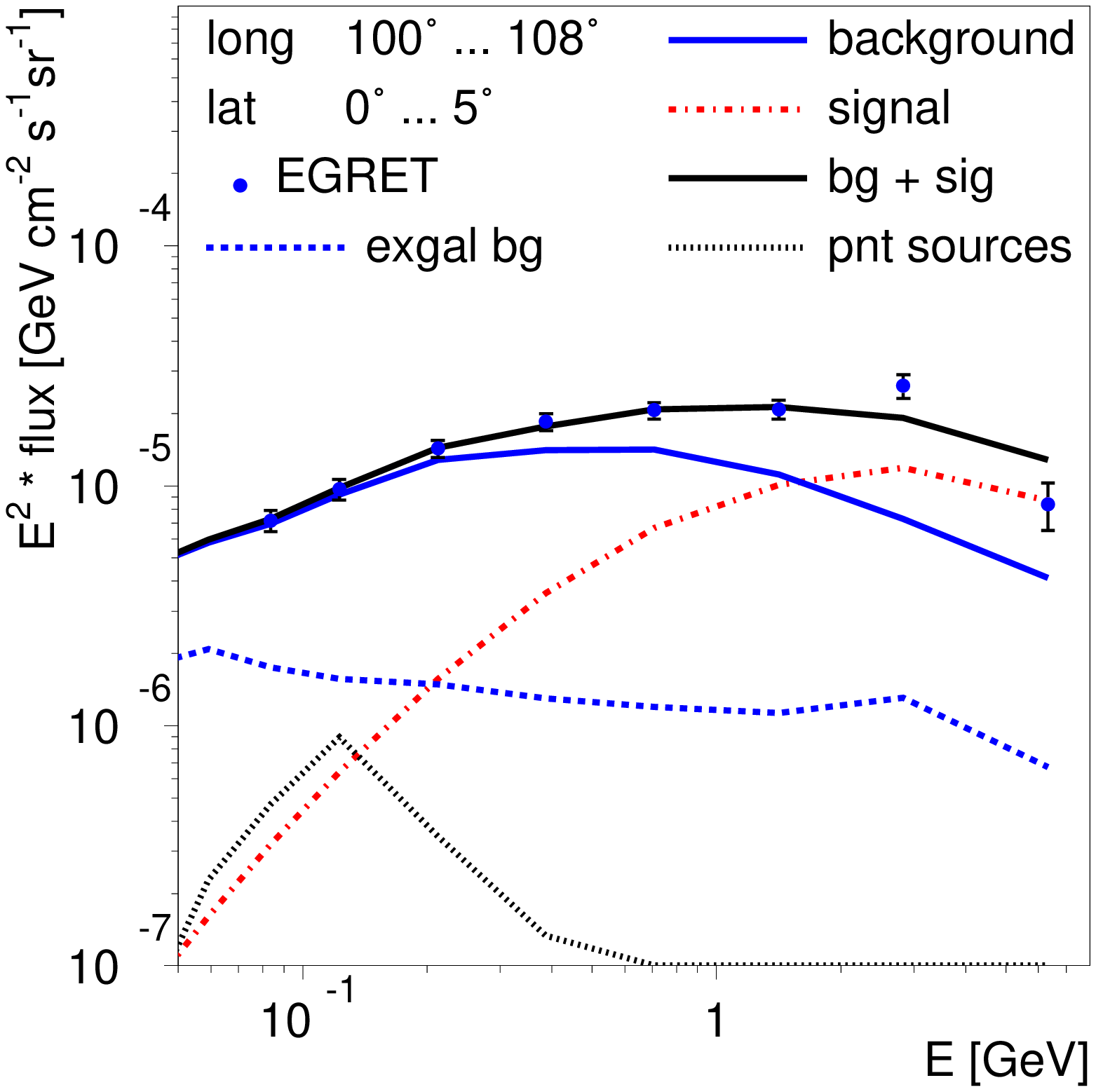}
    \includegraphics[width=0.21\textwidth]{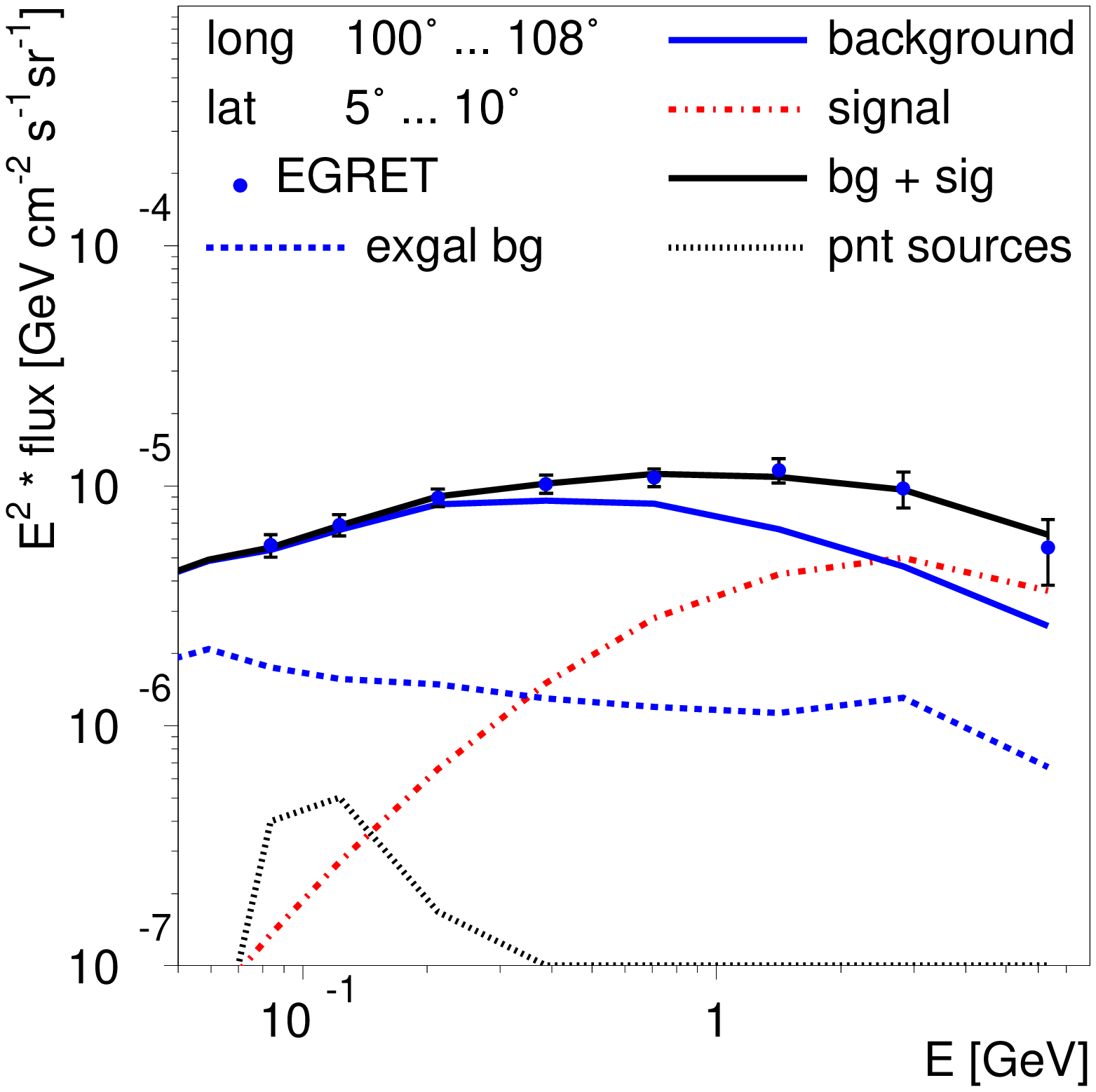}
    \includegraphics[width=0.21\textwidth]{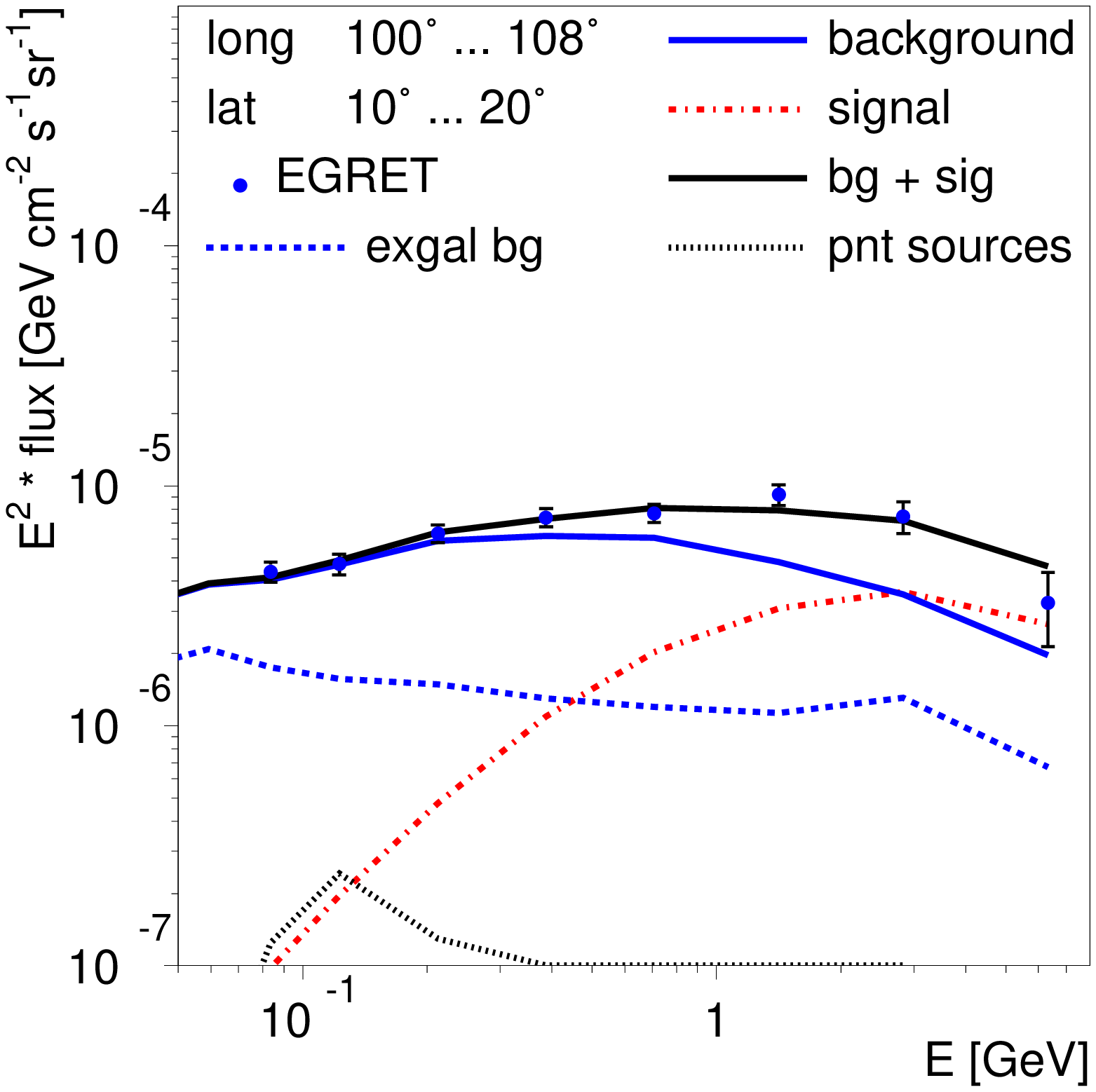}
    \includegraphics[width=0.21\textwidth]{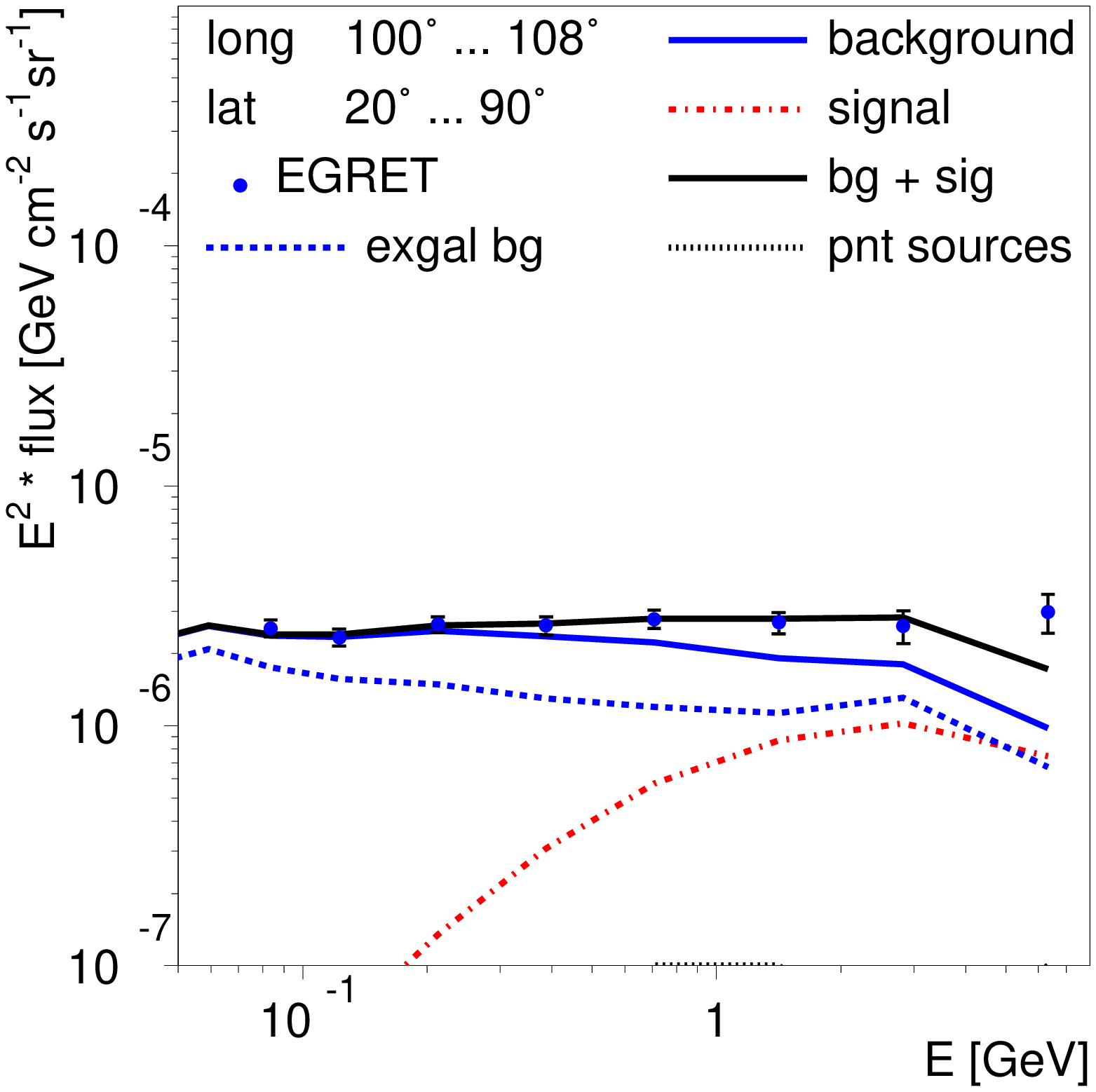}\\
  \end{center}
  \clearpage
  \begin{center}
    \framebox[0.21\textwidth][c]{$\vert \mbox{lat}\vert<5^\circ$}
    \framebox[0.21\textwidth][c]{$5^\circ<\vert \mbox{lat}\vert<10^\circ$}
    \framebox[0.21\textwidth][c]{$10^\circ<\vert \mbox{lat}\vert<20^\circ$}
    \framebox[0.21\textwidth][c]{$20^\circ<\vert \mbox{lat}\vert<90^\circ$}\\
    \hspace{-1cm}
    \begin{turn}{90} \framebox[0.21\textwidth][c]{{\scriptsize $108^\circ<\mbox{long}<116^\circ$}} \end{turn}
    \includegraphics[width=0.21\textwidth]{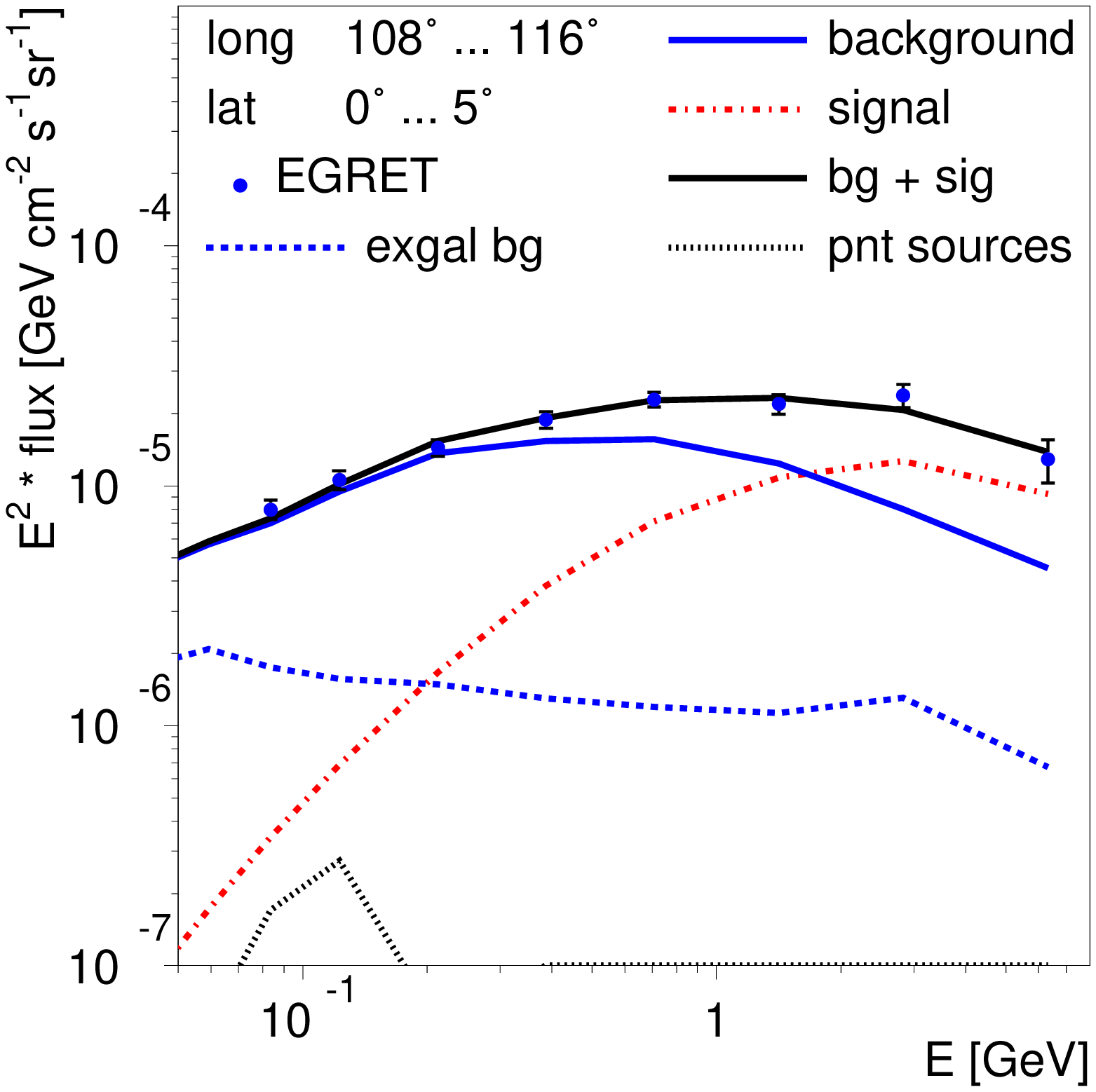}
    \includegraphics[width=0.21\textwidth]{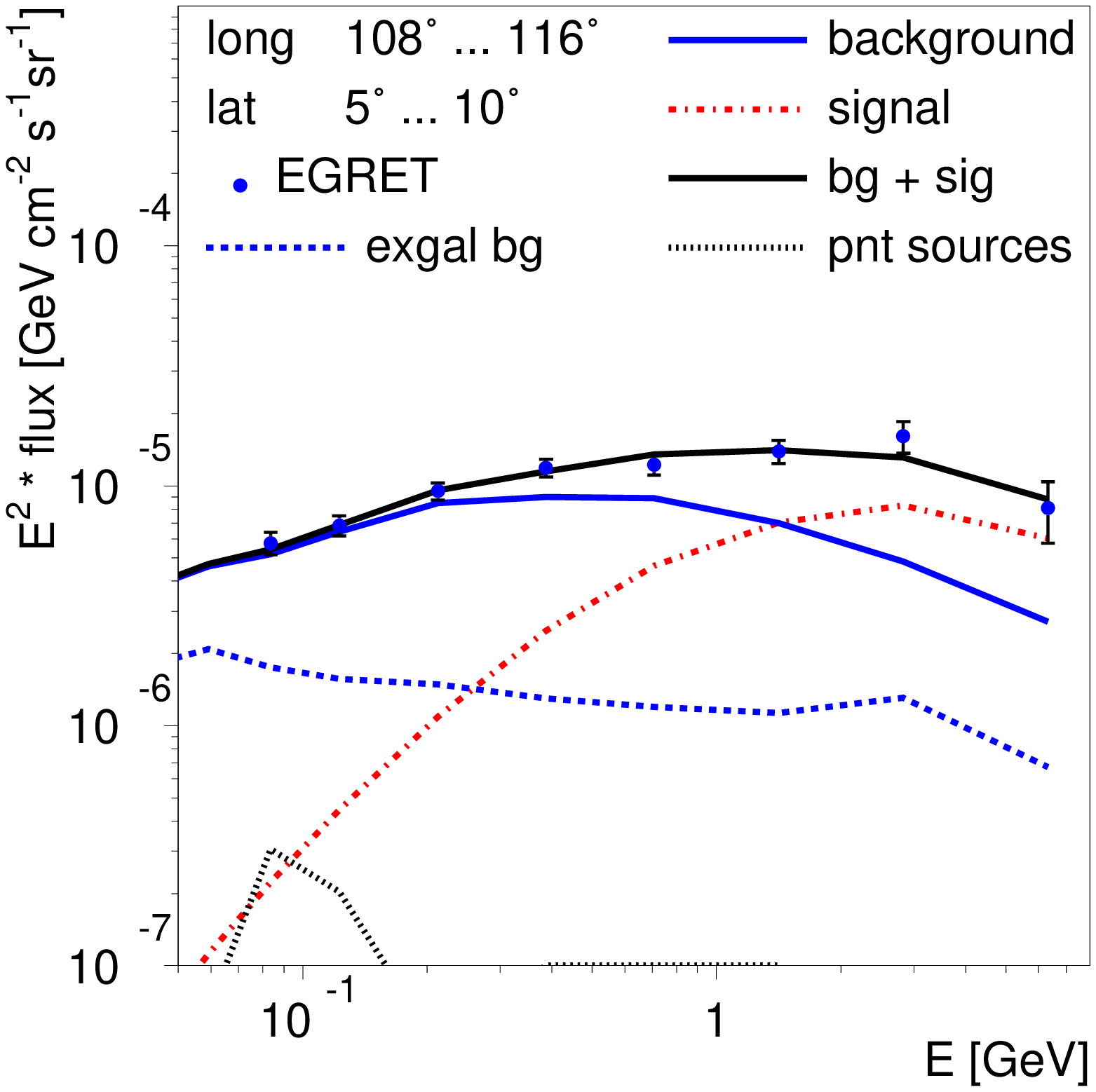}
    \includegraphics[width=0.21\textwidth]{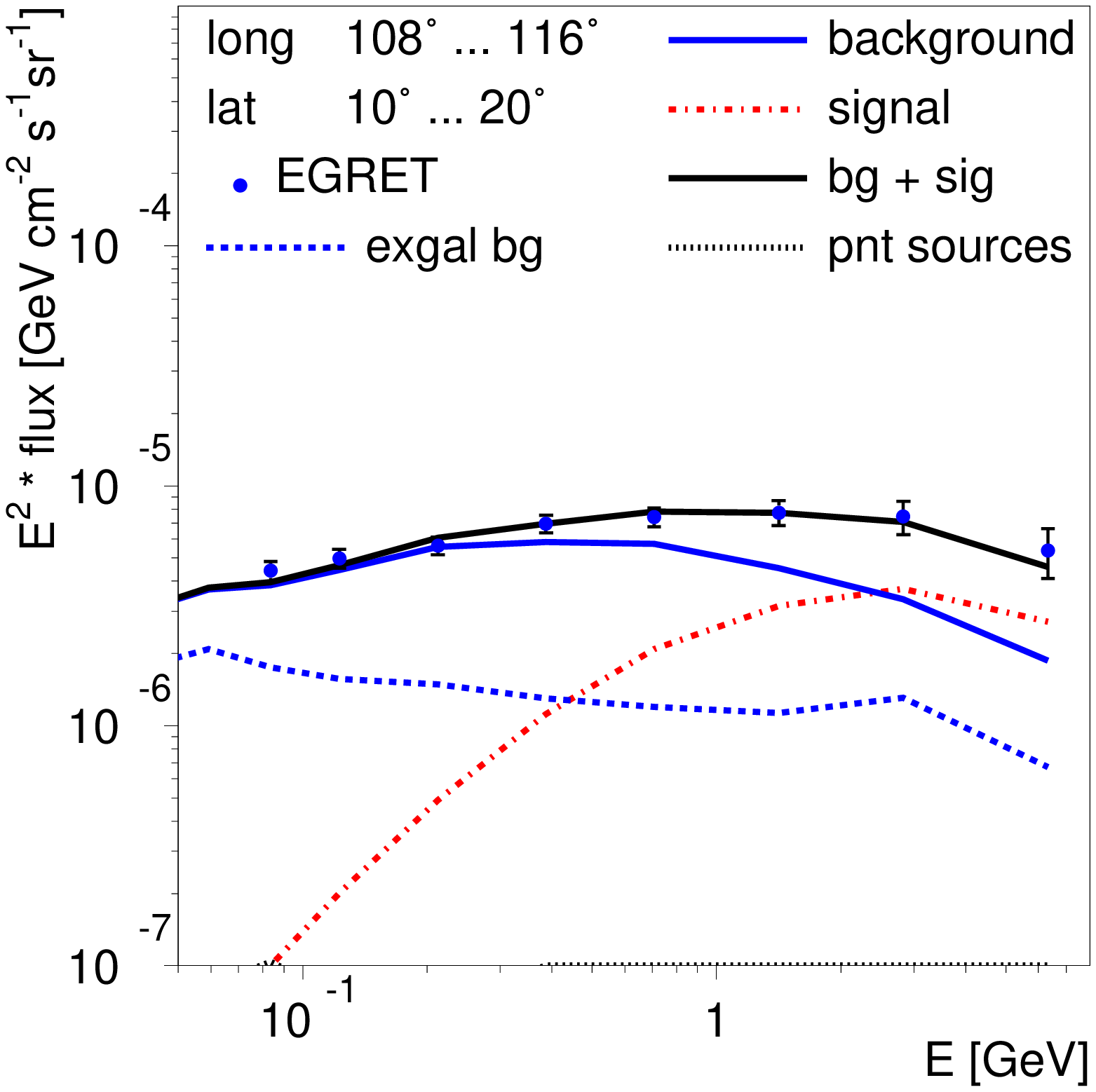}
    \includegraphics[width=0.21\textwidth]{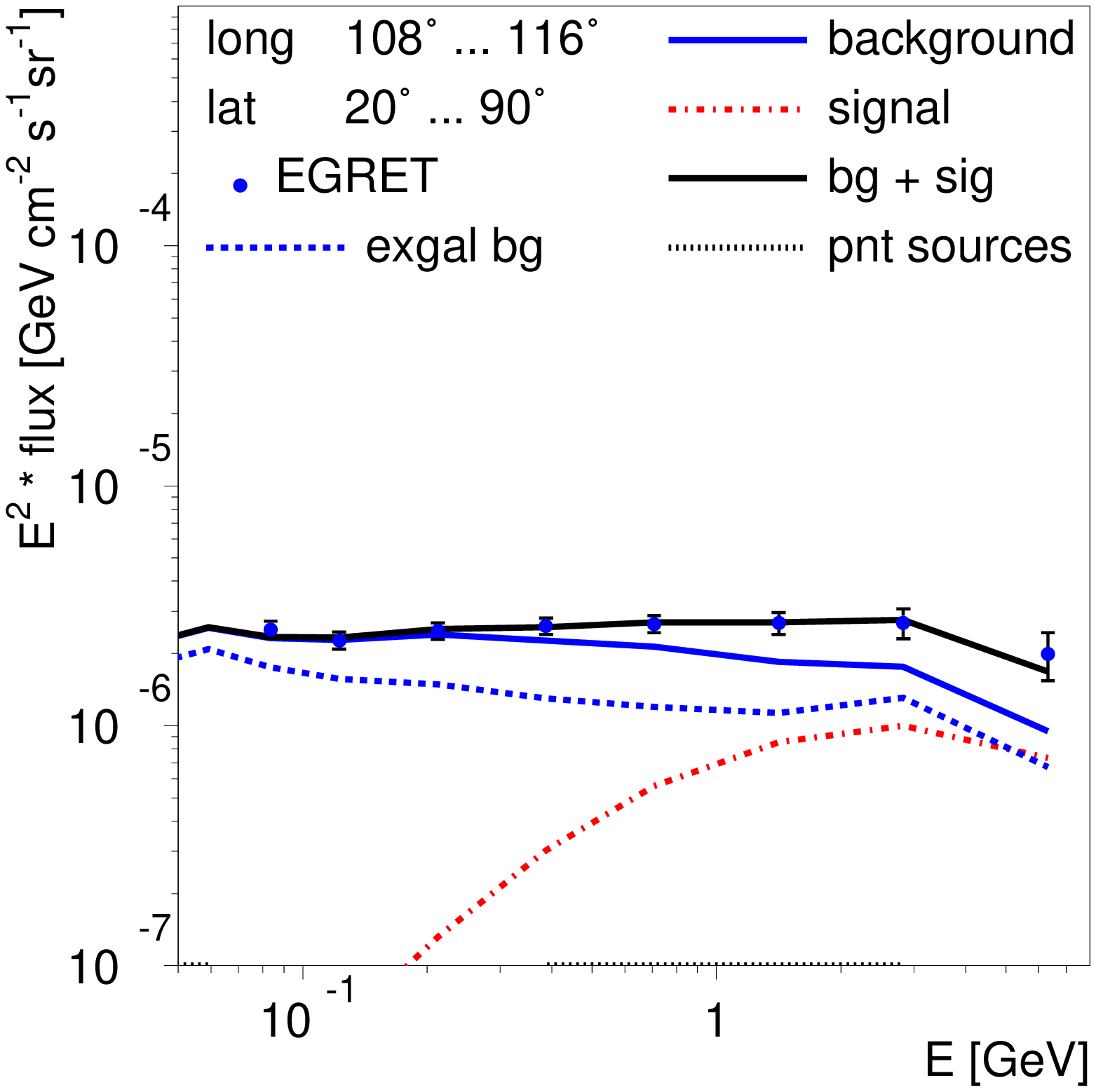}\\
    \hspace{-1cm}
    \begin{turn}{90} \framebox[0.21\textwidth][c]{{\scriptsize $116^\circ<\mbox{long}<124^\circ$}} \end{turn}
    \includegraphics[width=0.21\textwidth]{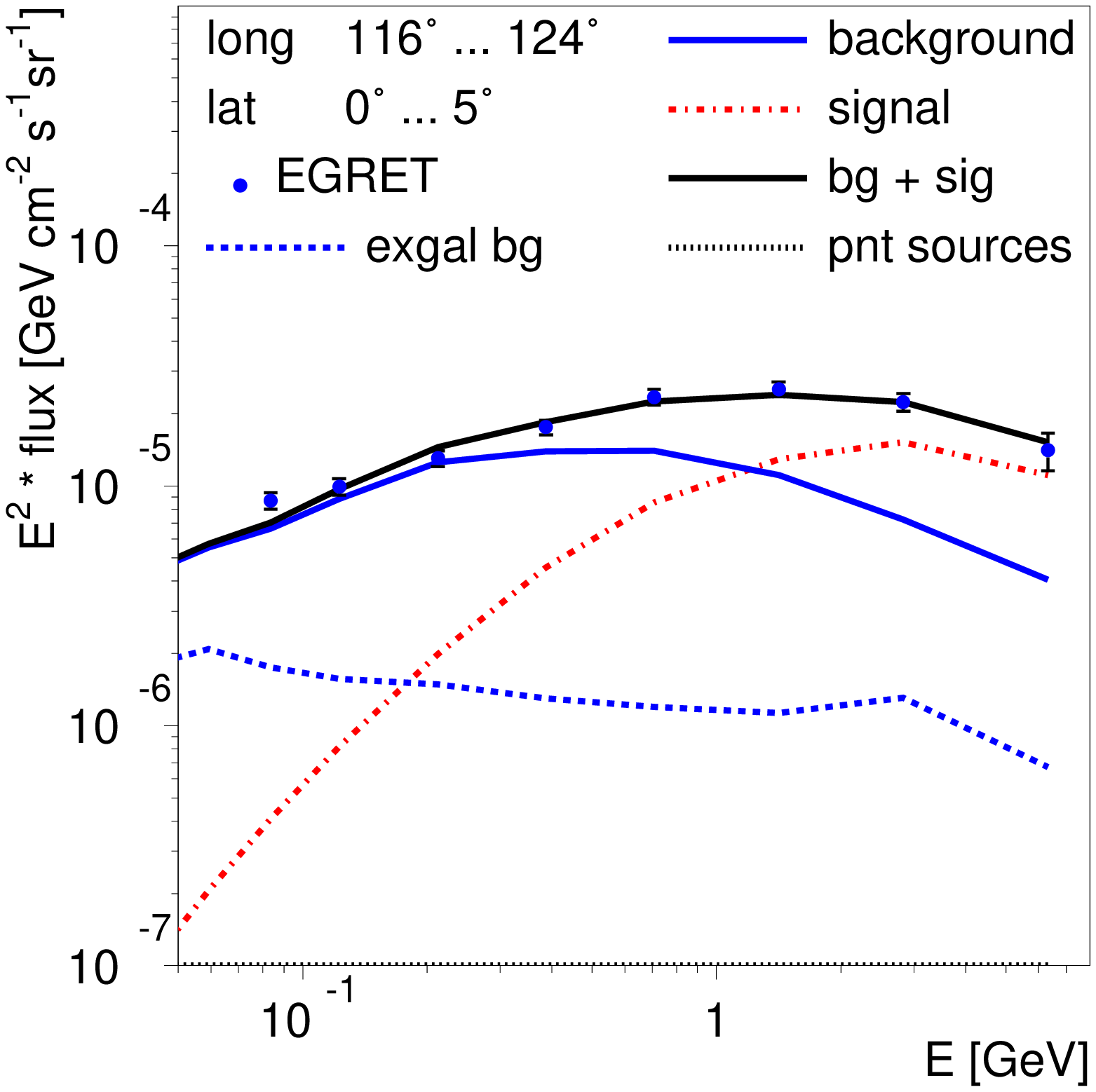}
    \includegraphics[width=0.21\textwidth]{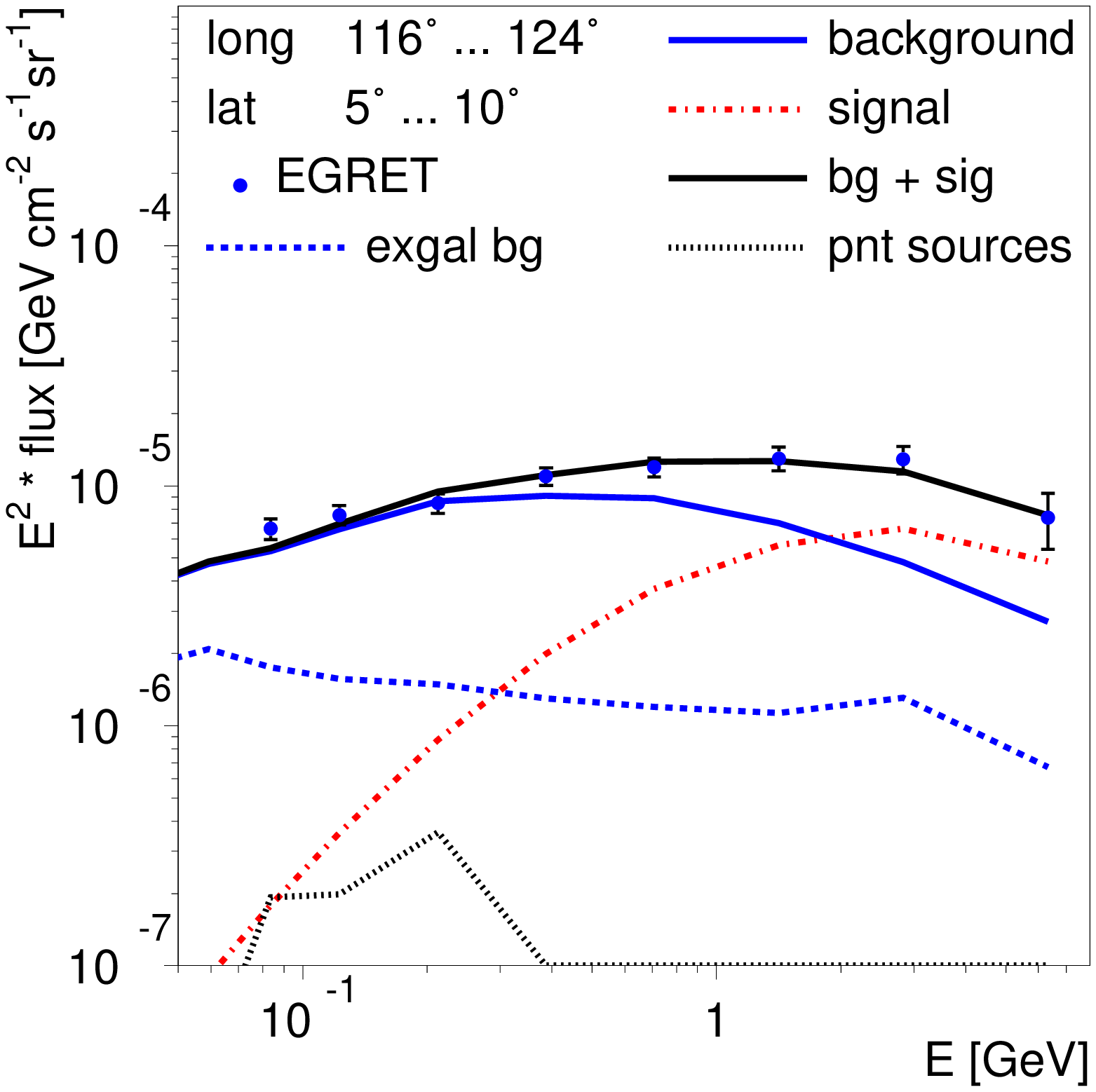}
    \includegraphics[width=0.21\textwidth]{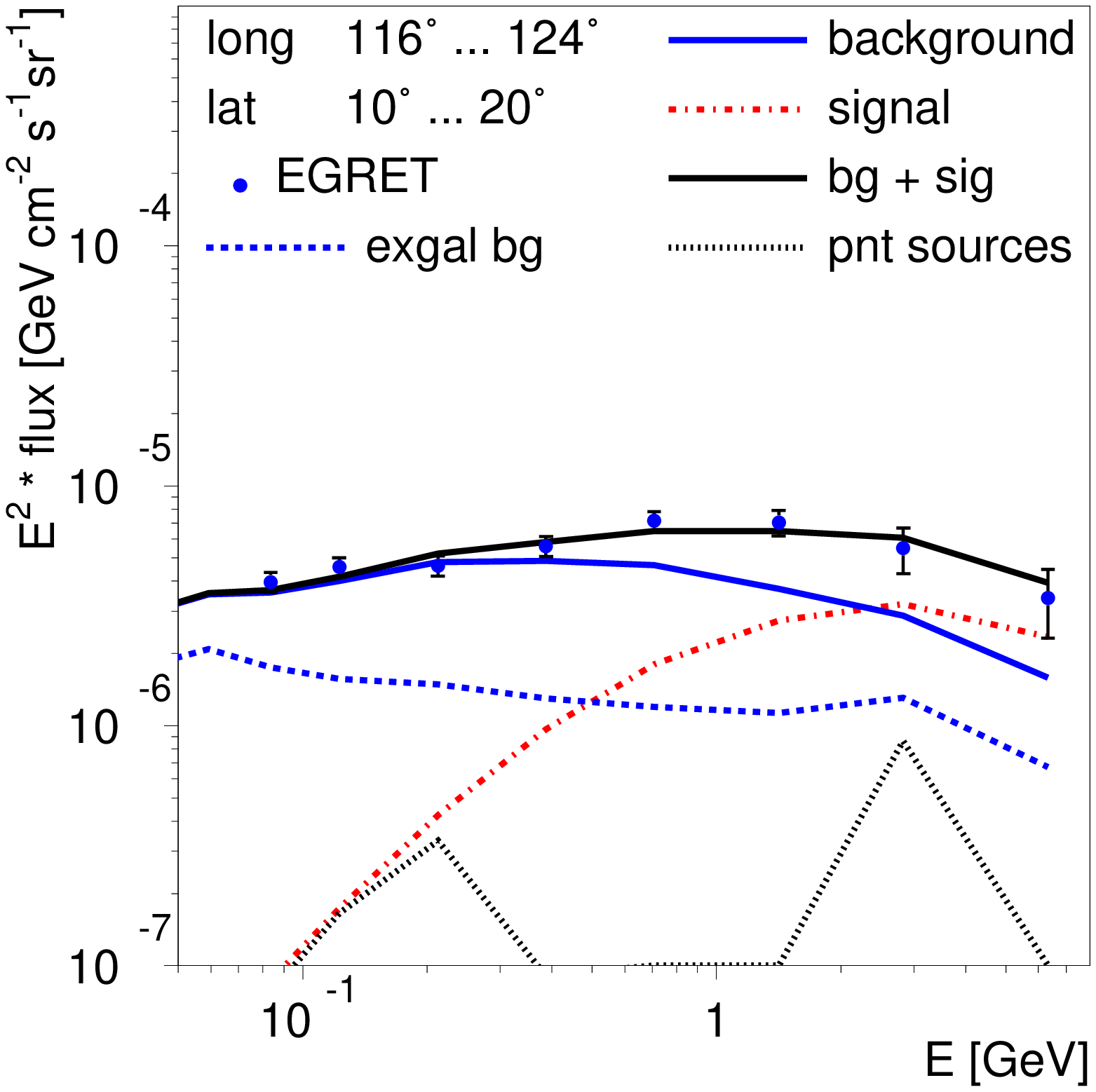}
    \includegraphics[width=0.21\textwidth]{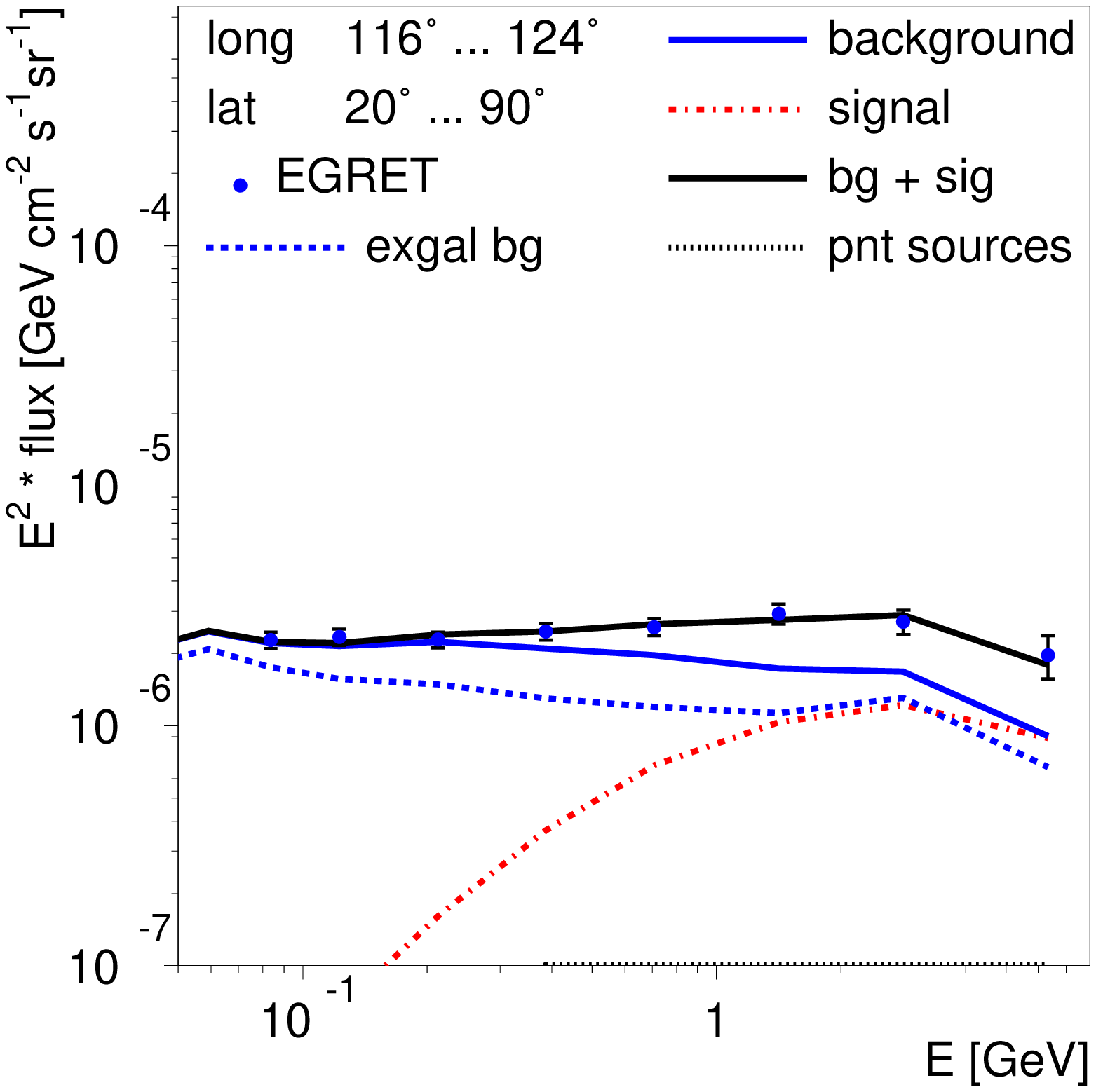}\\
    \hspace{-1cm}
    \begin{turn}{90} \framebox[0.21\textwidth][c]{{\scriptsize $124^\circ<\mbox{long}<132^\circ$}} \end{turn}
    \includegraphics[width=0.21\textwidth]{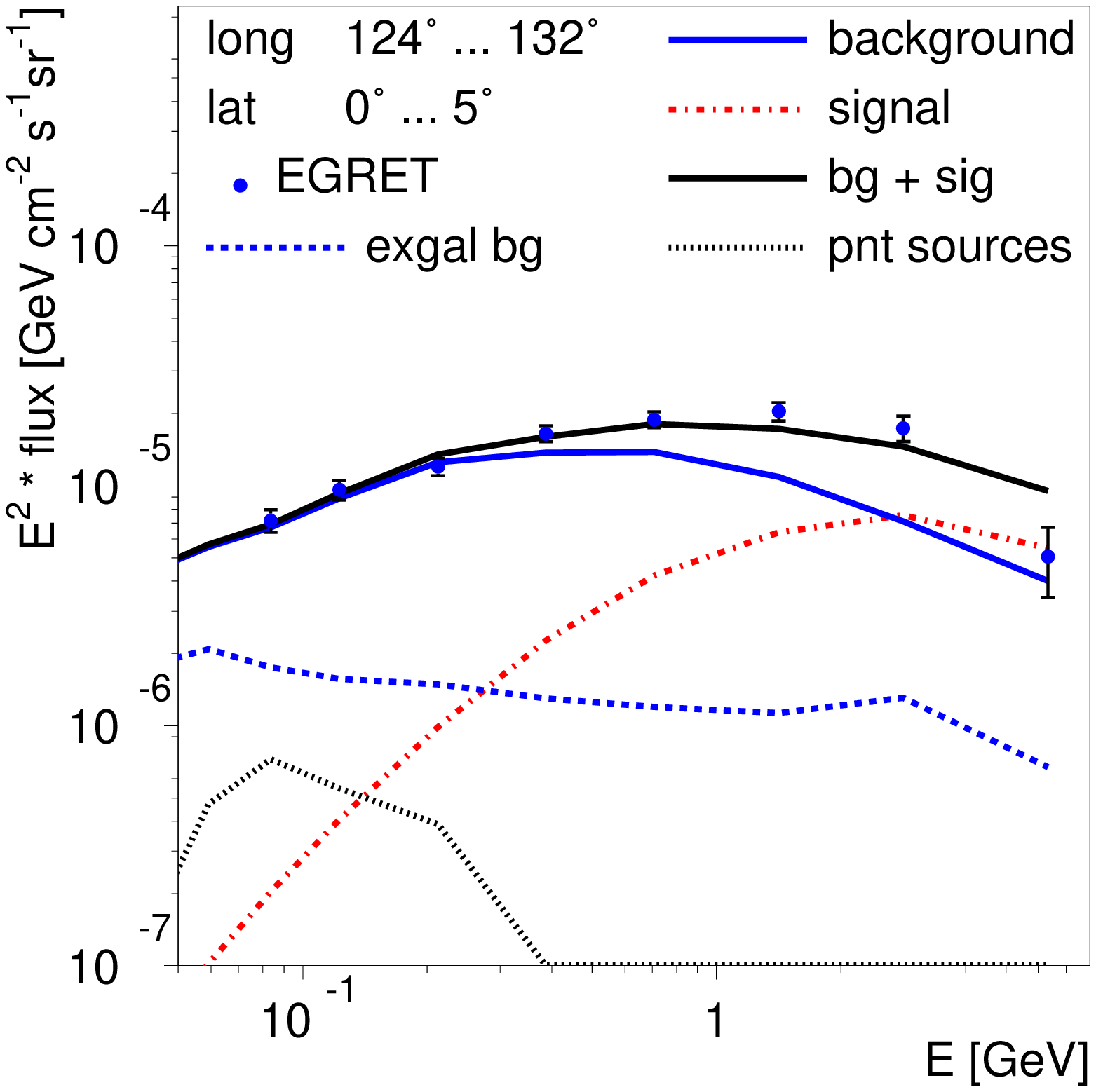}
    \includegraphics[width=0.21\textwidth]{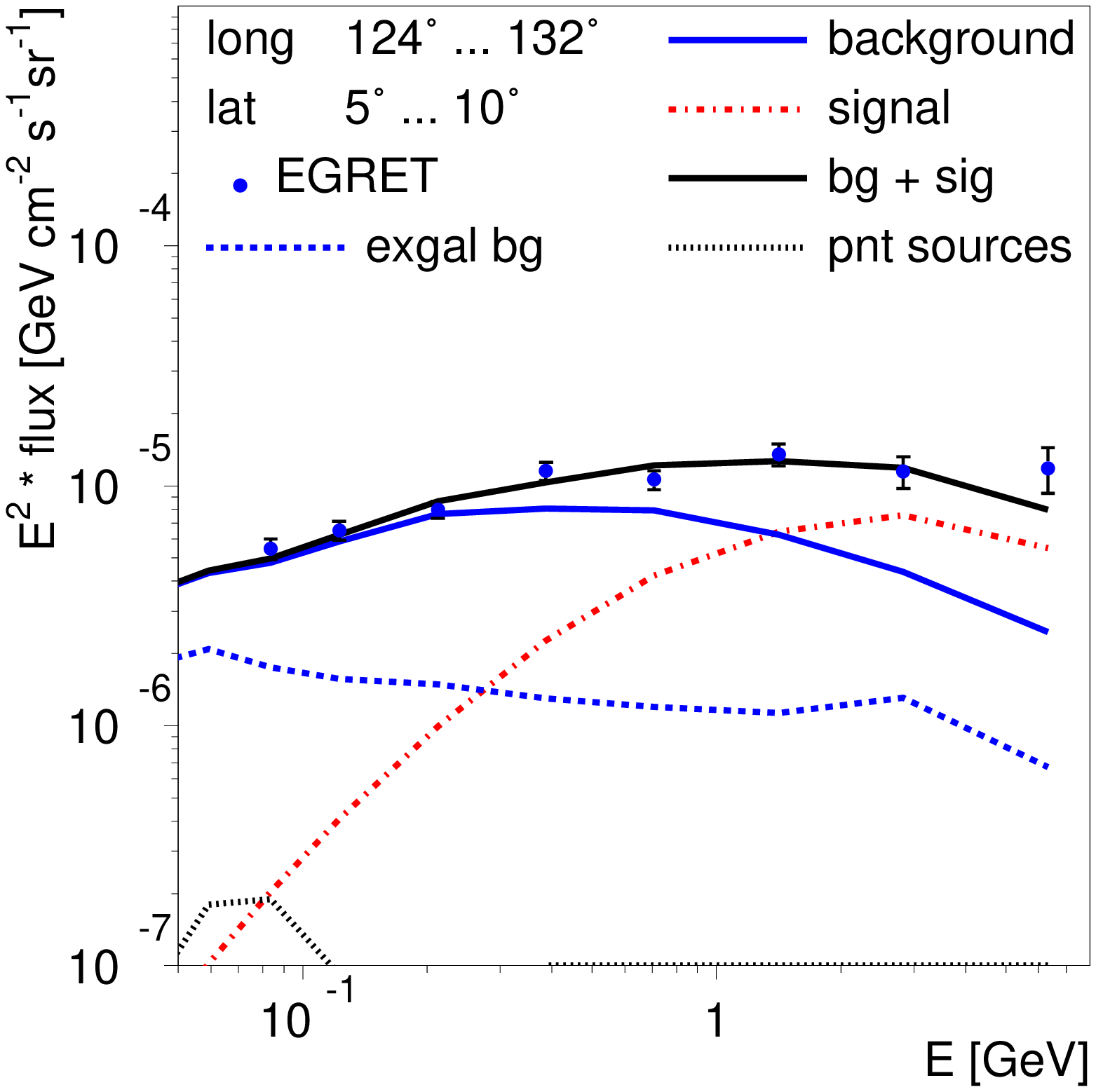}
    \includegraphics[width=0.21\textwidth]{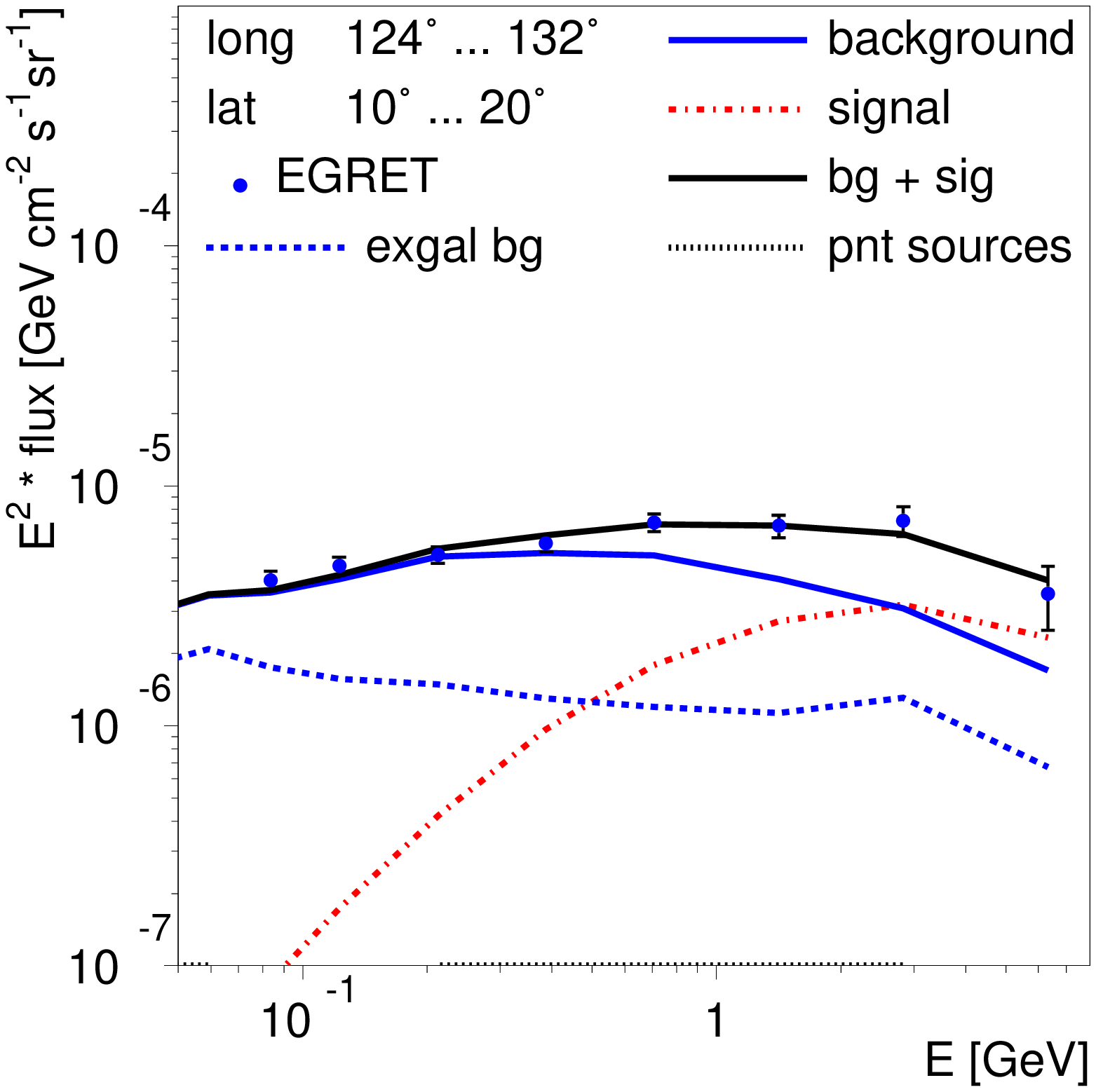}
    \includegraphics[width=0.21\textwidth]{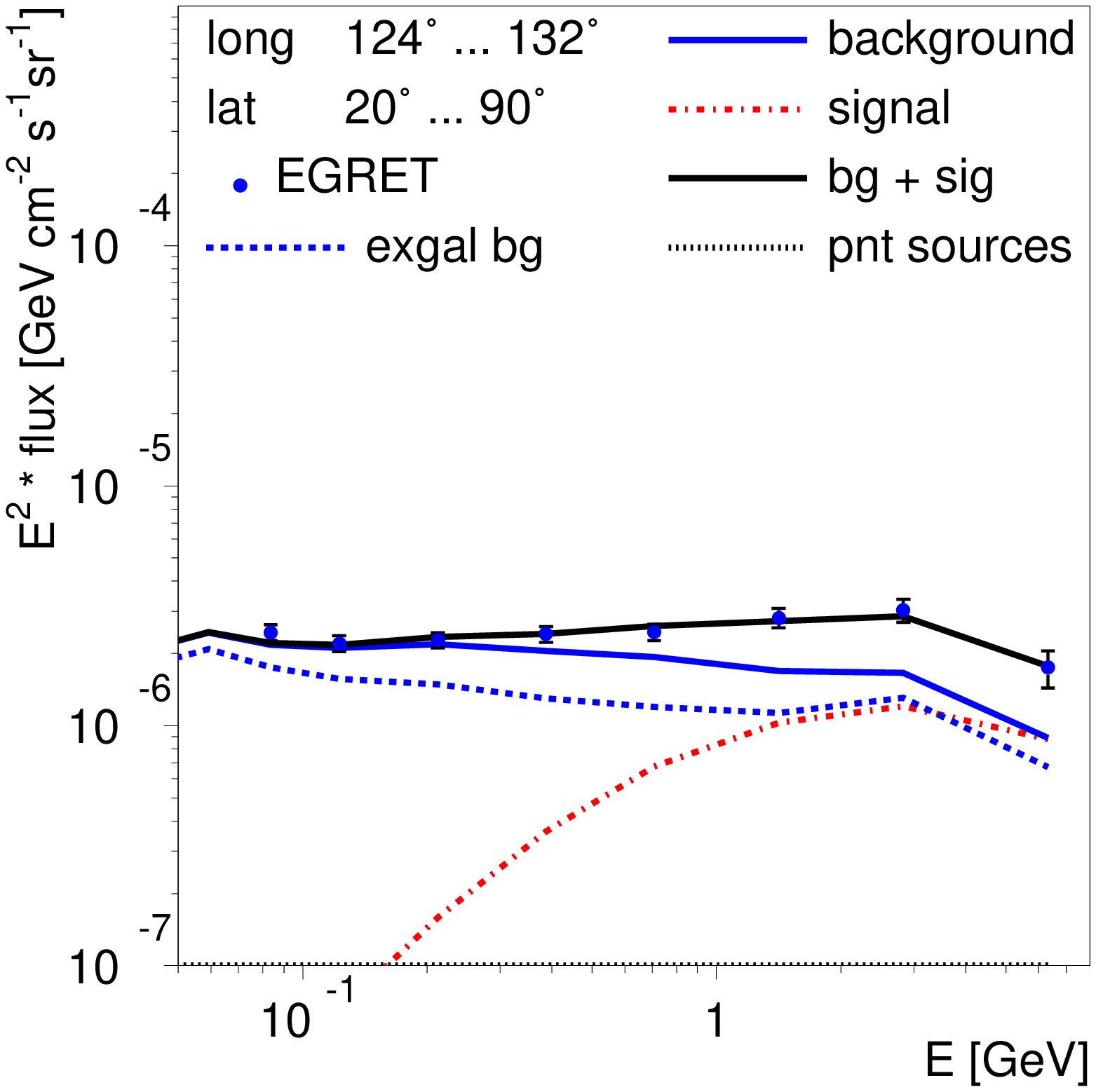}\\
    \hspace{-1cm}
    \begin{turn}{90} \framebox[0.21\textwidth][c]{{\scriptsize $132^\circ<\mbox{long}<140^\circ$}} \end{turn}
    \includegraphics[width=0.21\textwidth]{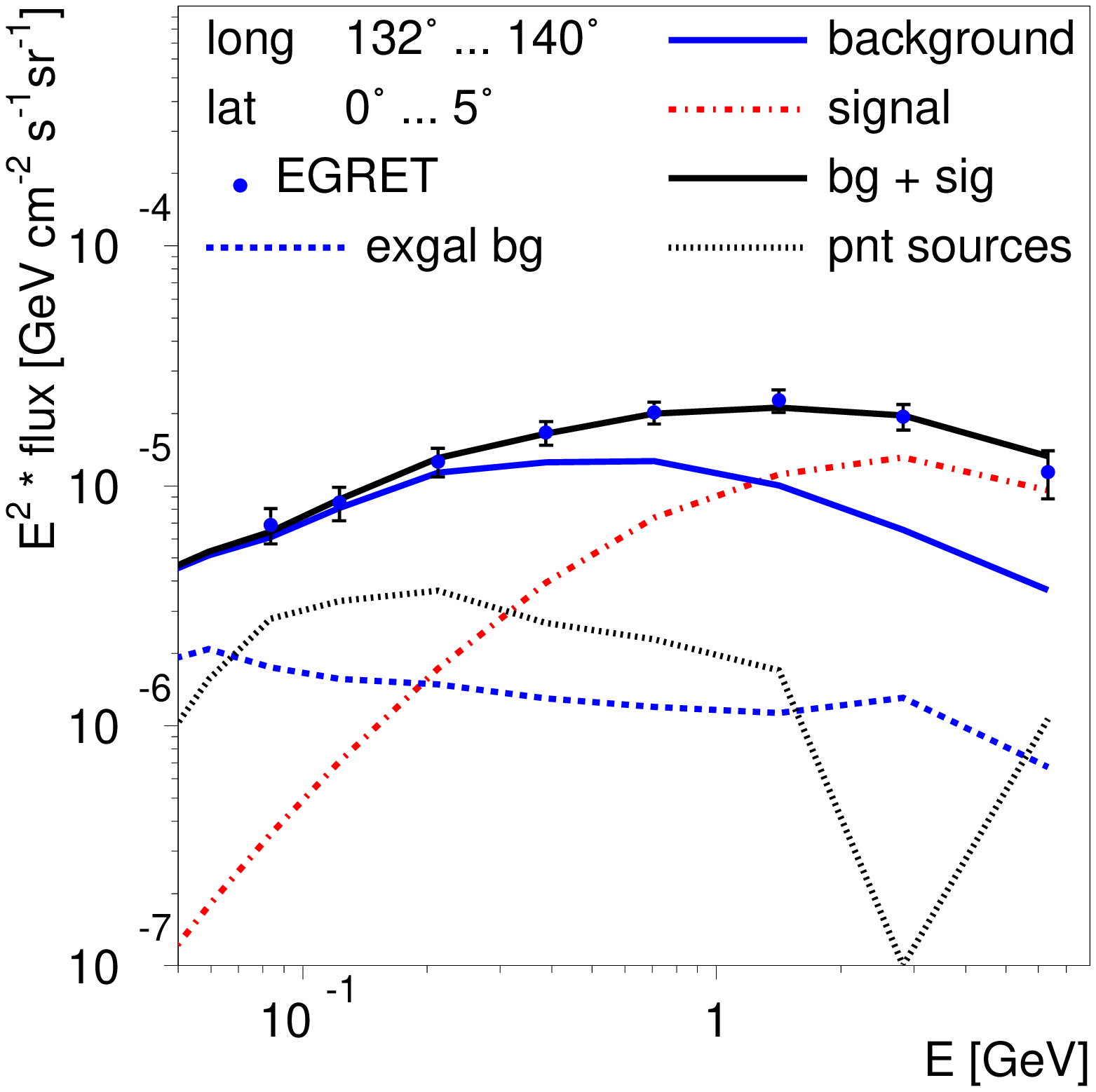}
    \includegraphics[width=0.21\textwidth]{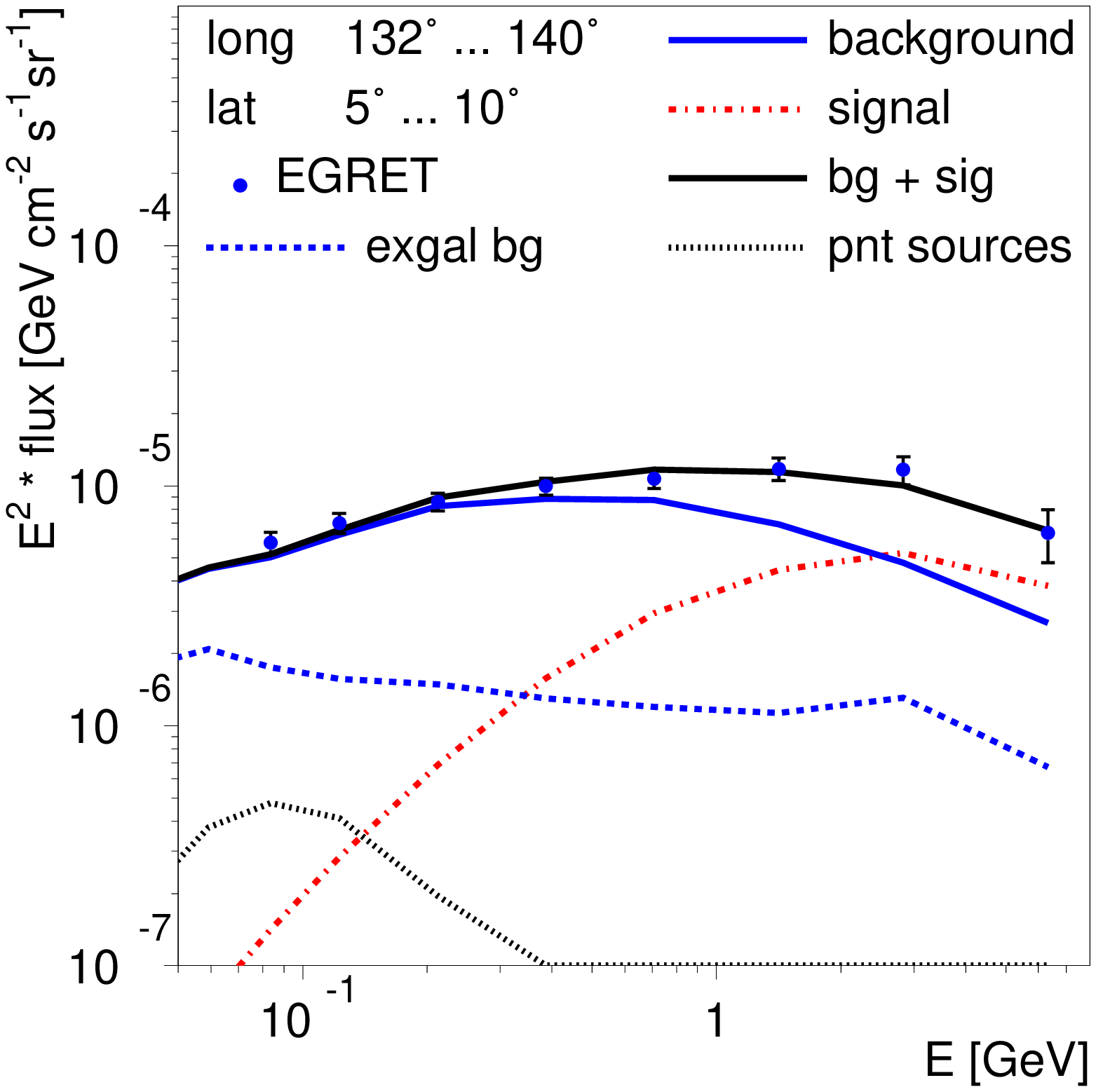}
    \includegraphics[width=0.21\textwidth]{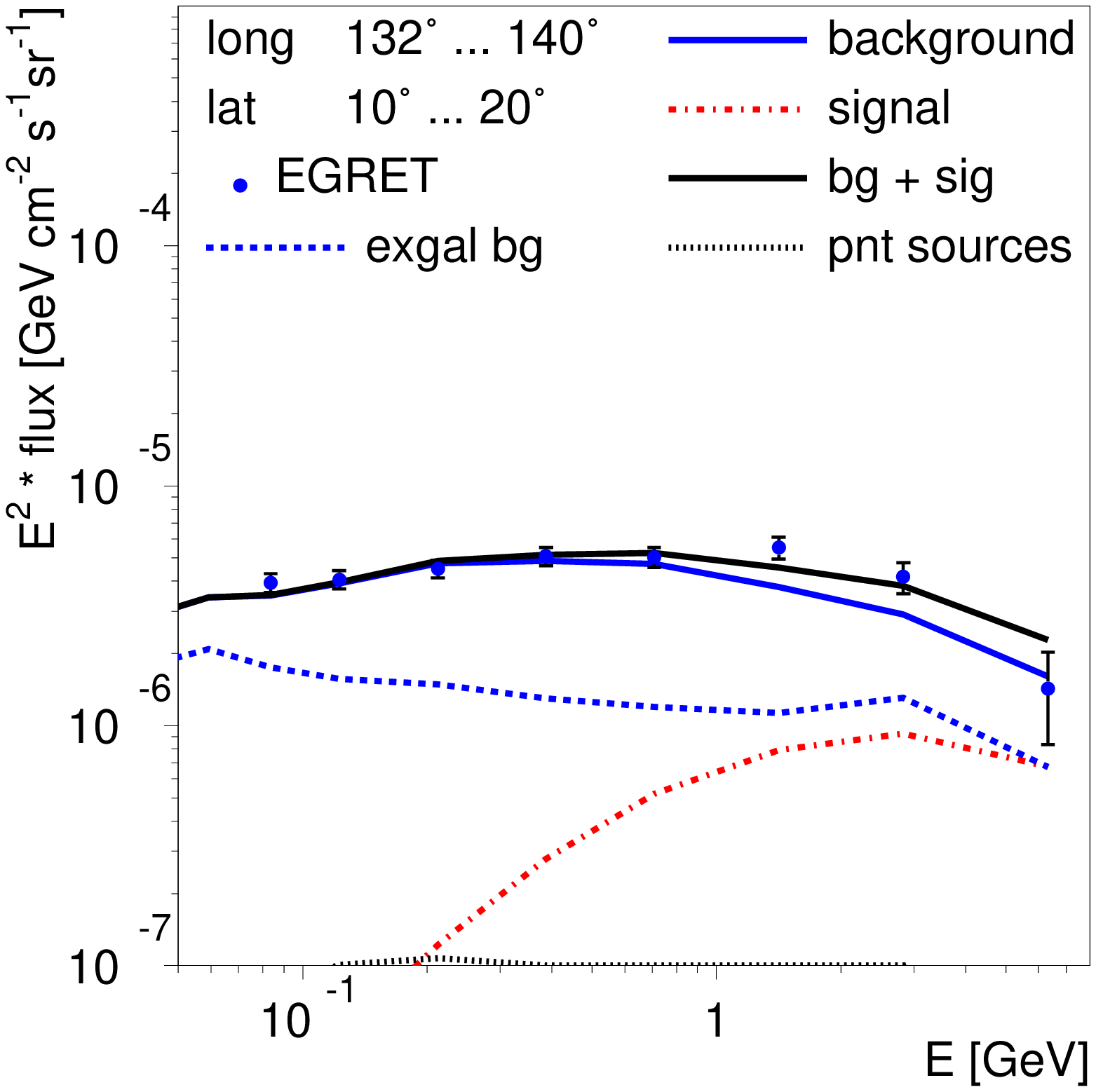}
    \includegraphics[width=0.21\textwidth]{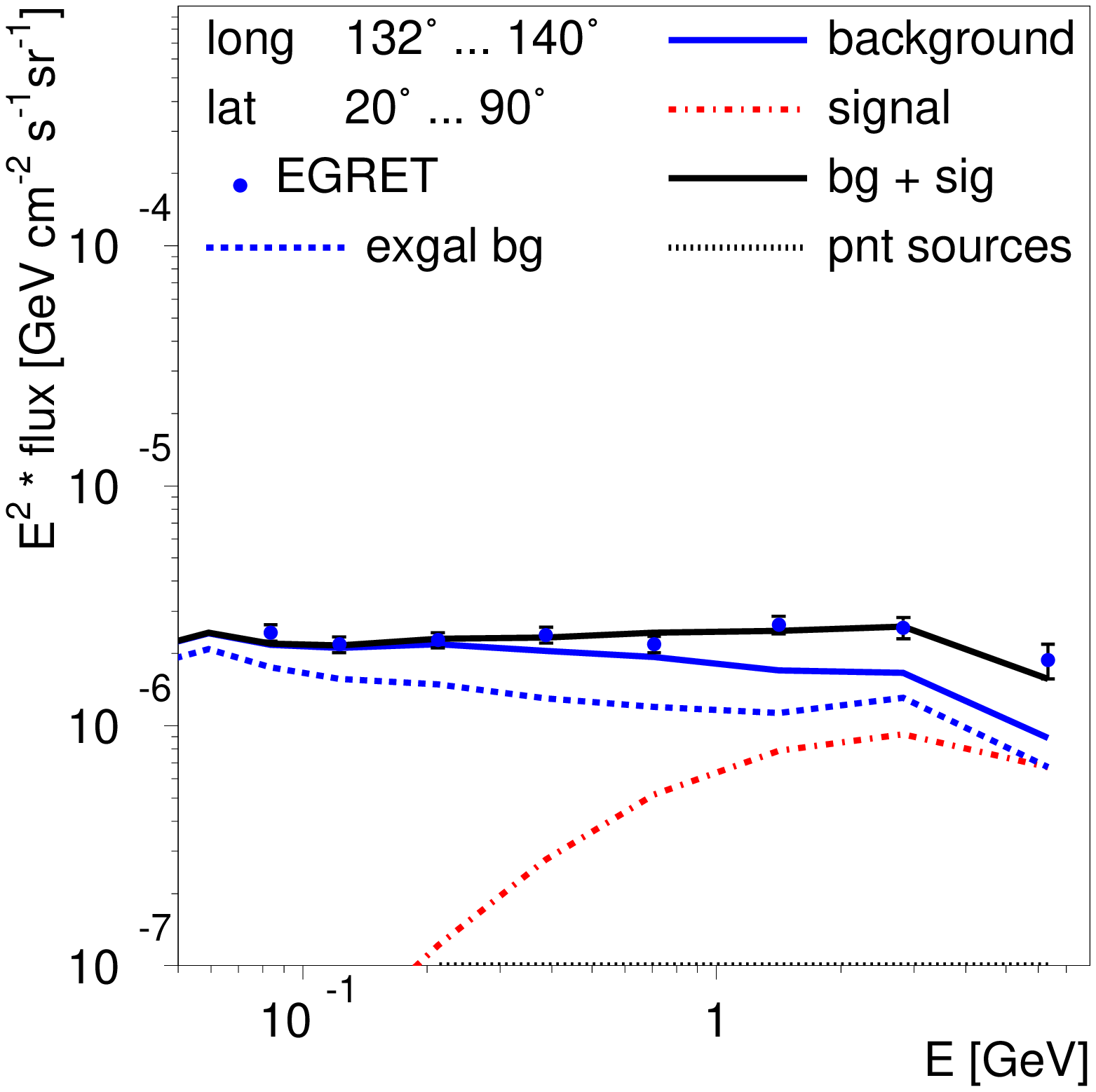}\\
    \hspace{-1cm}
    \begin{turn}{90} \framebox[0.21\textwidth][c]{{\scriptsize $140^\circ<\mbox{long}<148^\circ$}} \end{turn}
    \includegraphics[width=0.21\textwidth]{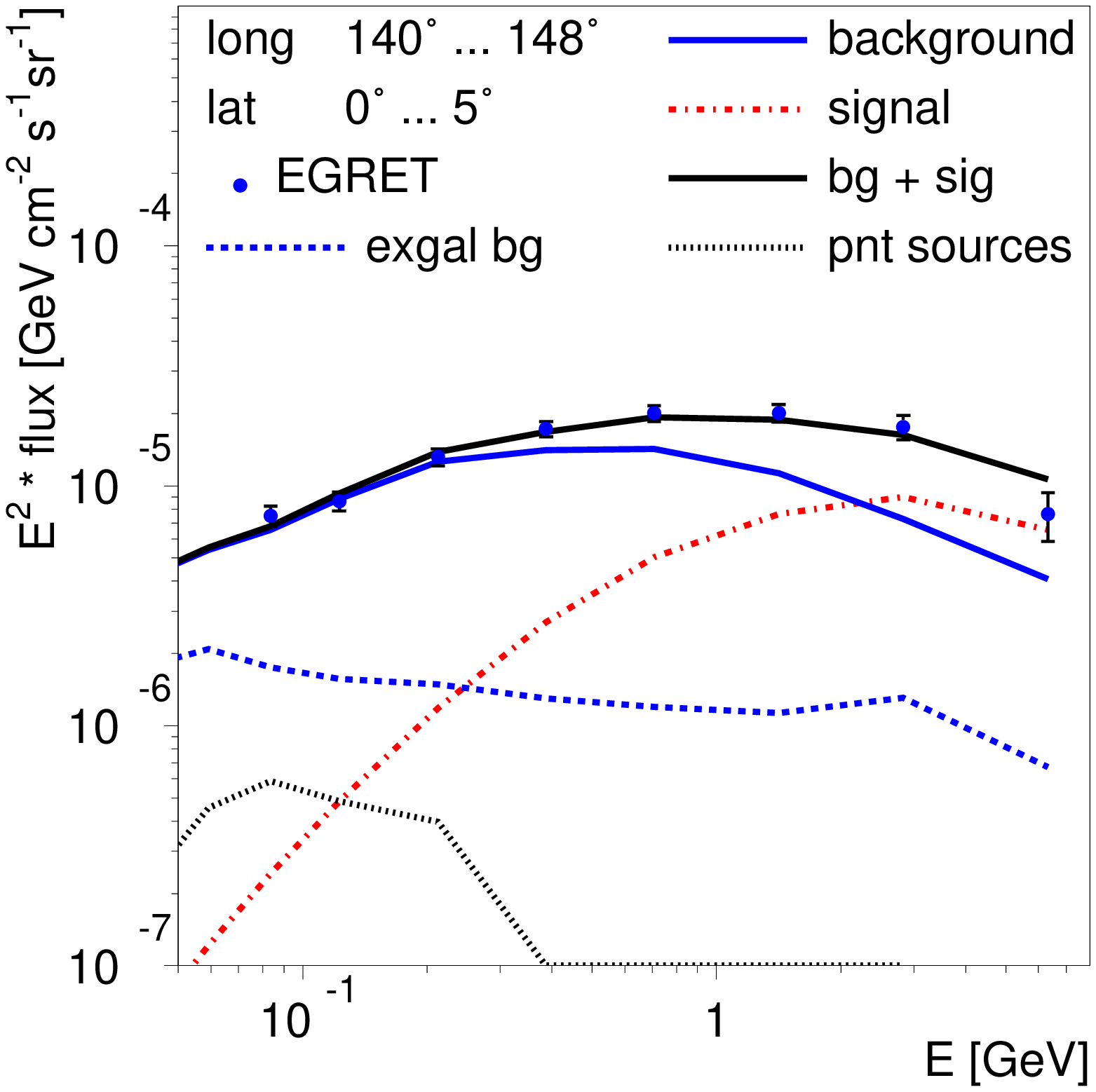}
    \includegraphics[width=0.21\textwidth]{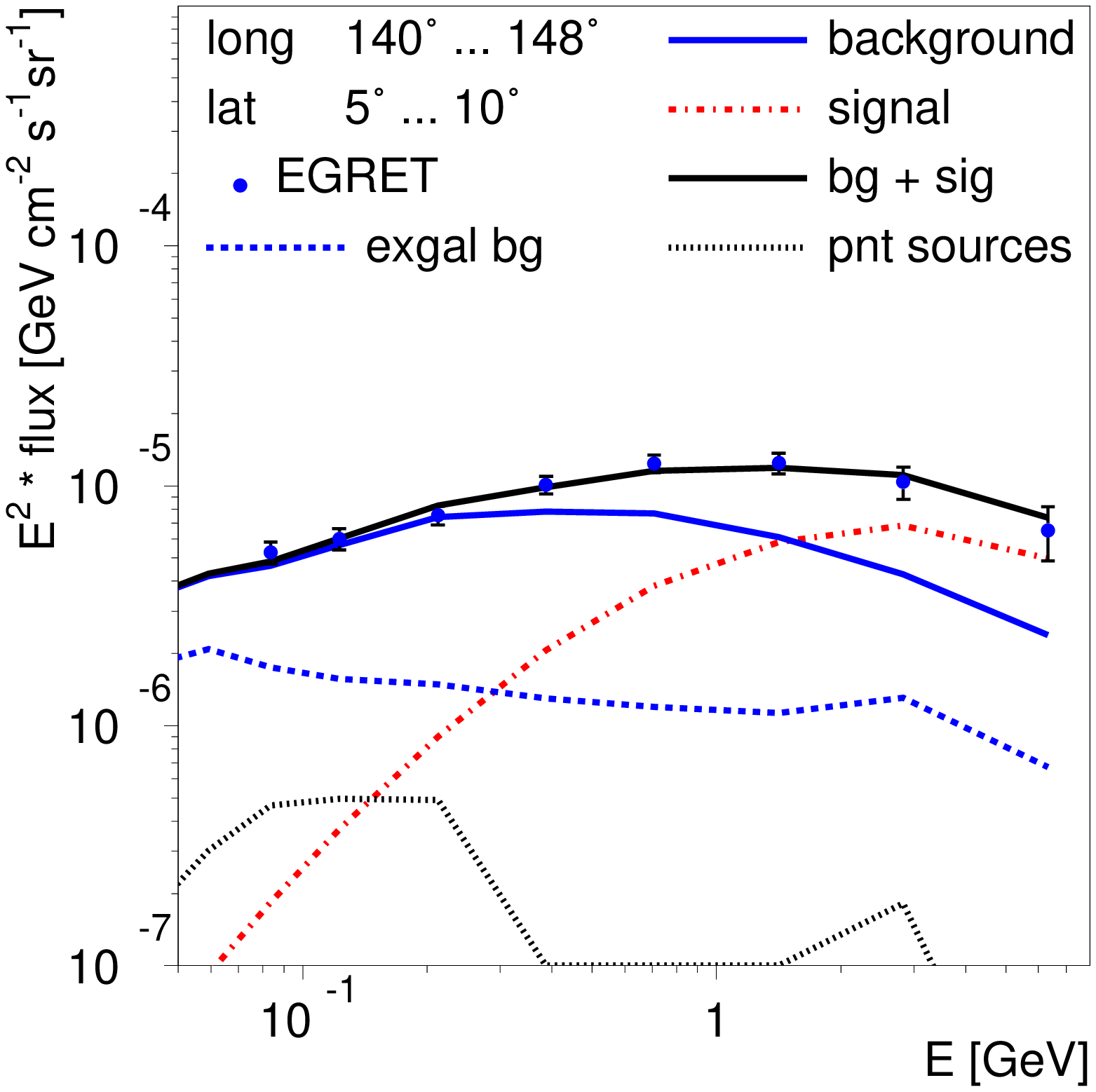}
    \includegraphics[width=0.21\textwidth]{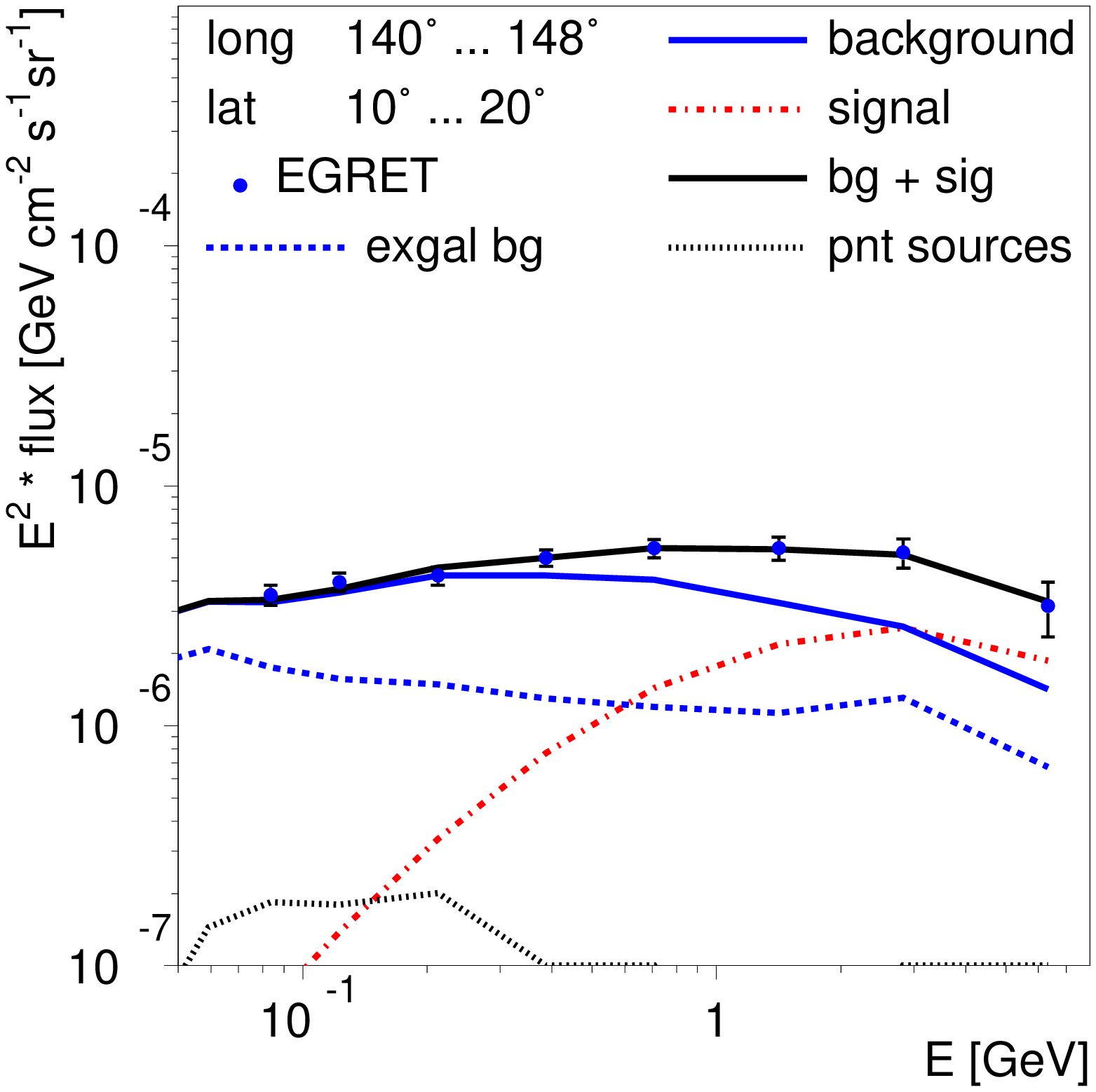}
    \includegraphics[width=0.21\textwidth]{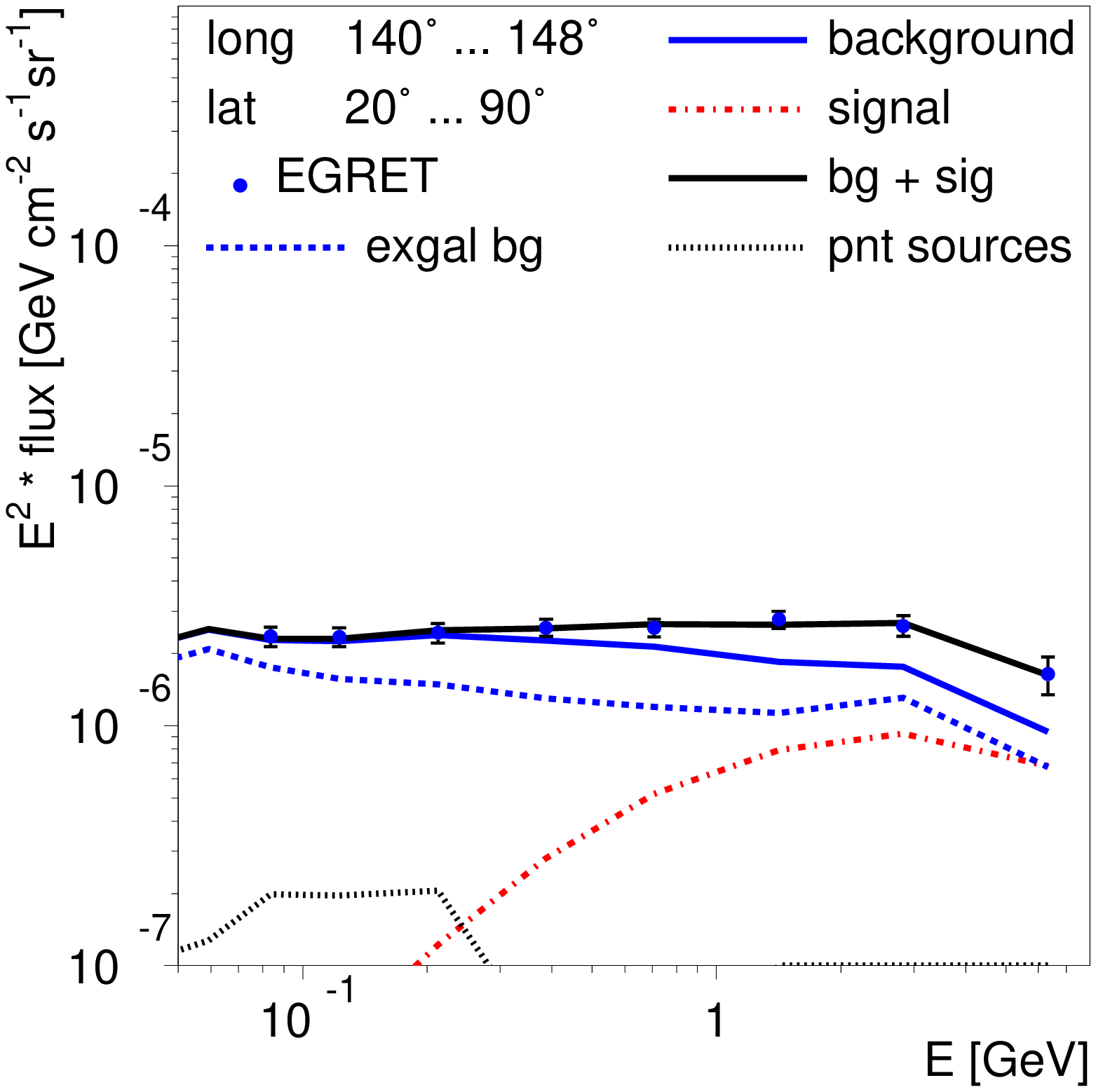}\\
    \hspace{-1cm}
    \begin{turn}{90} \framebox[0.21\textwidth][c]{{\scriptsize $148^\circ<\mbox{long}<156^\circ$}} \end{turn}
    \includegraphics[width=0.21\textwidth]{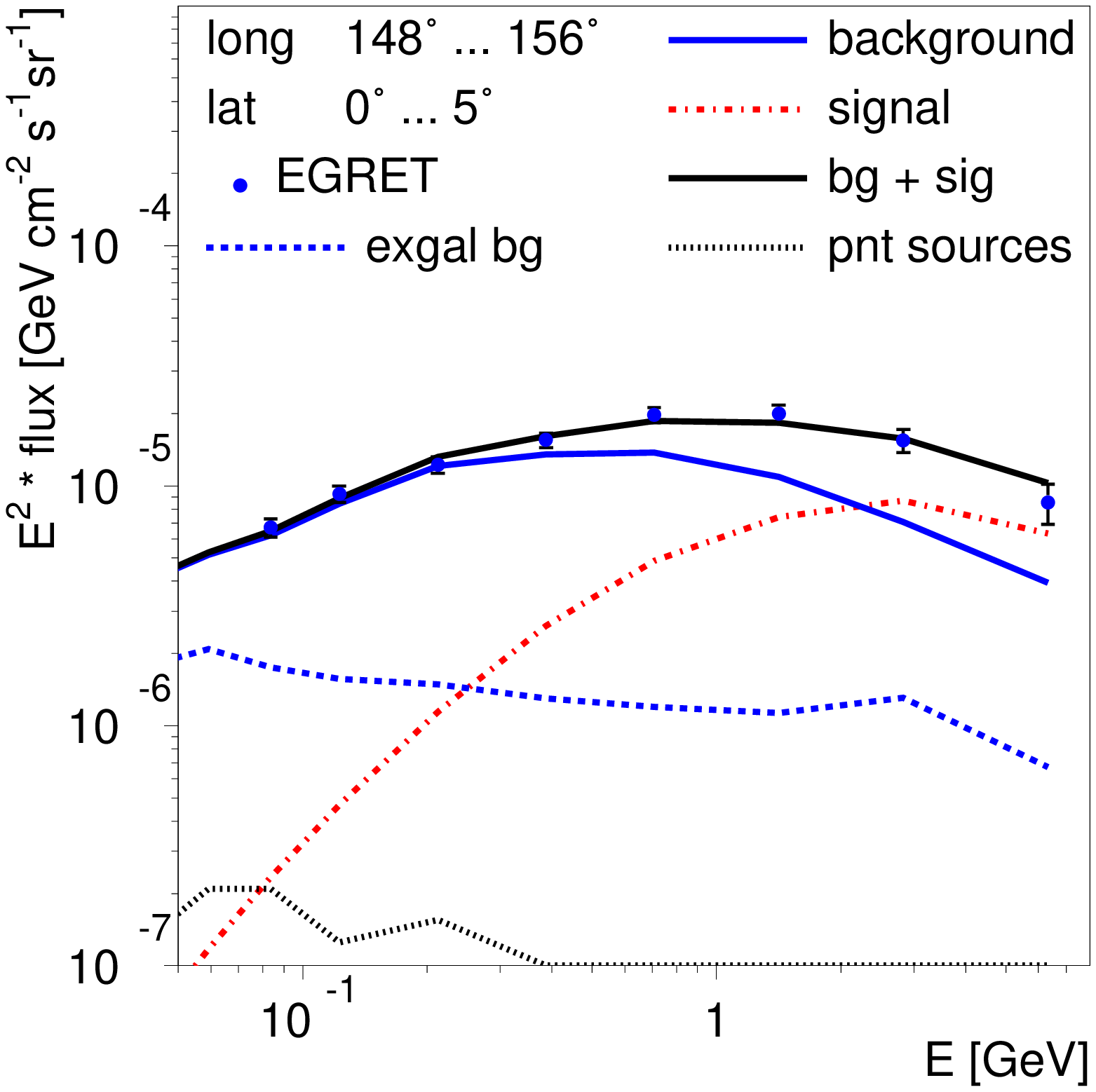}
    \includegraphics[width=0.21\textwidth]{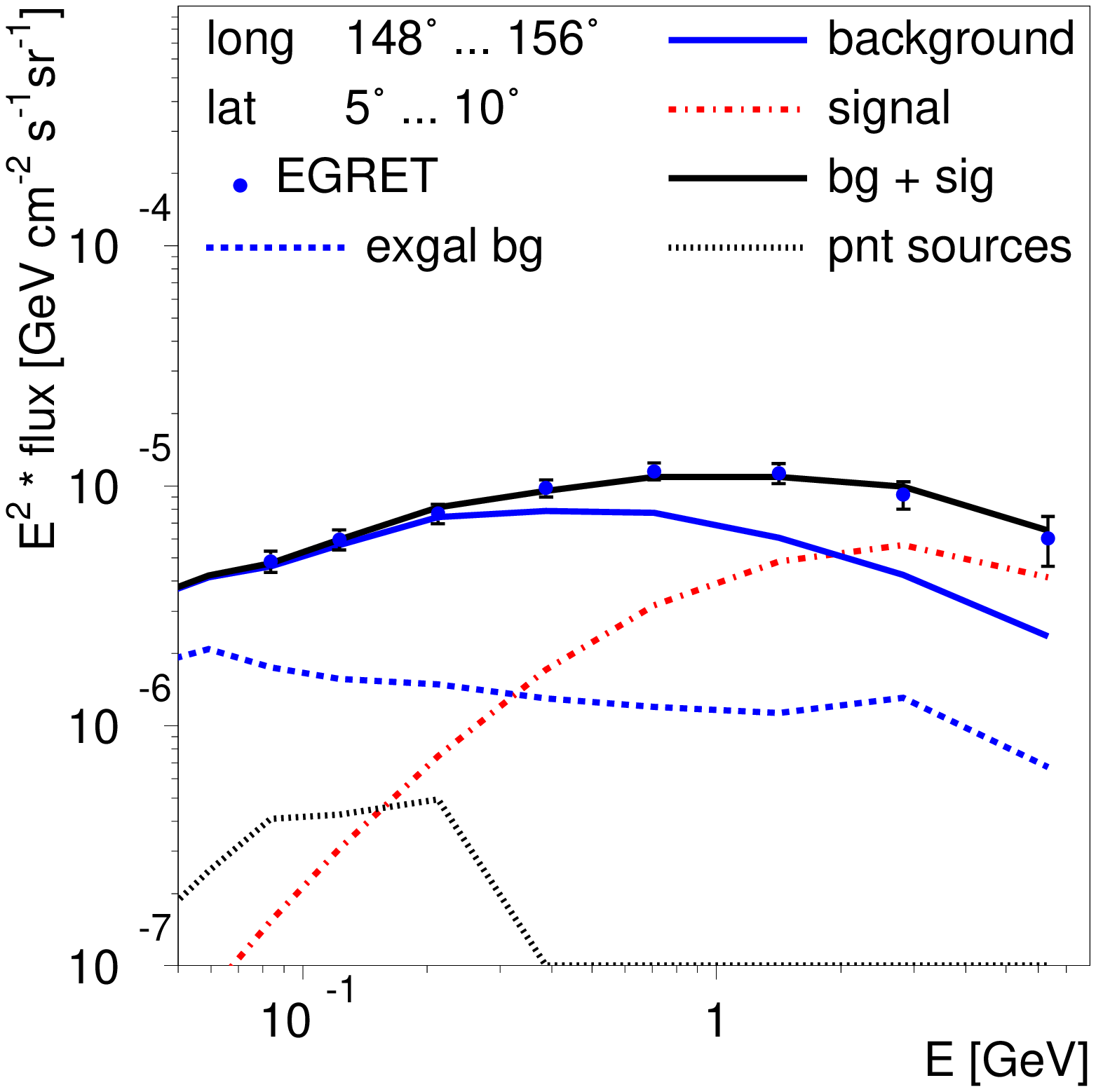}
    \includegraphics[width=0.21\textwidth]{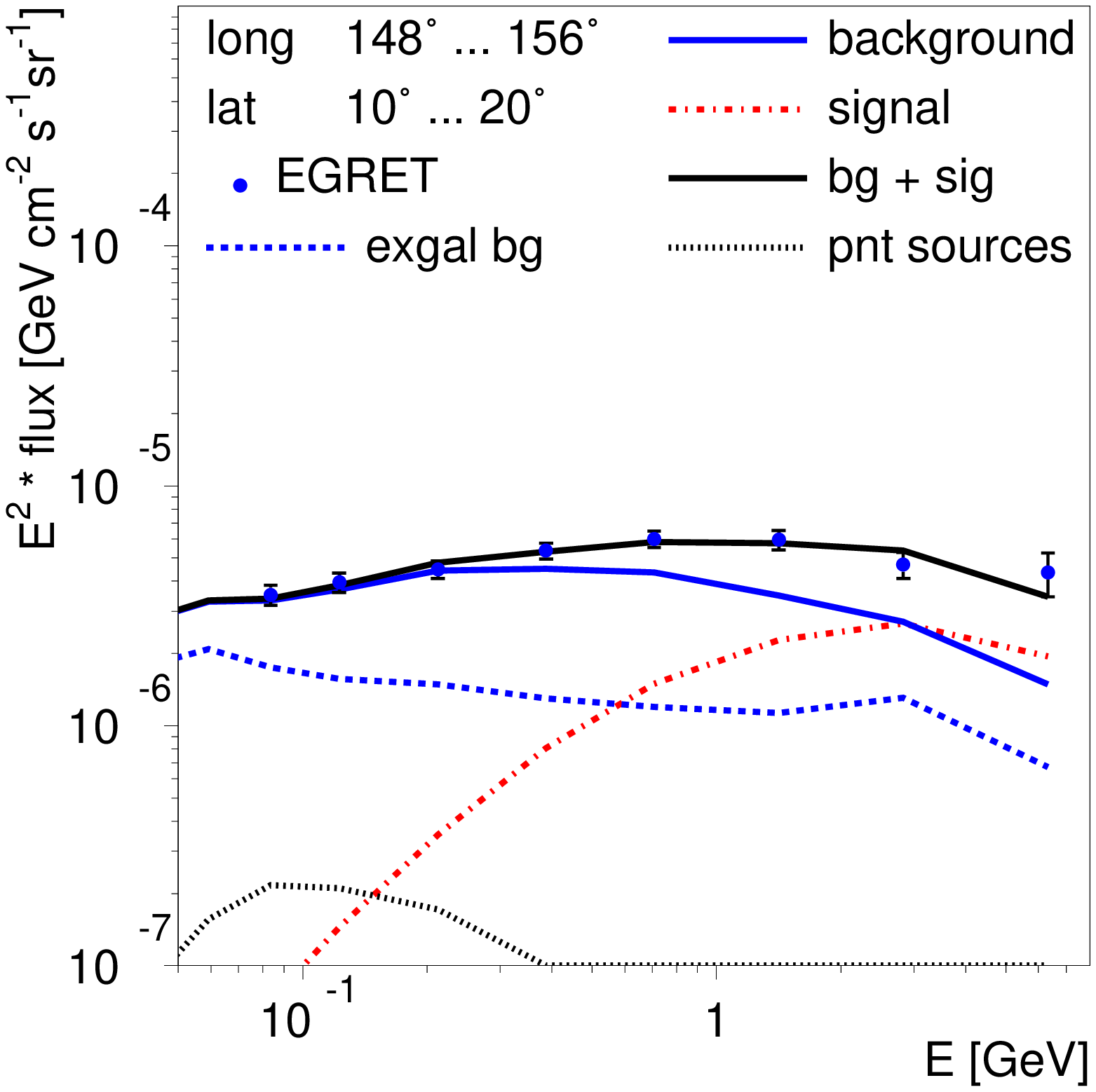}
    \includegraphics[width=0.21\textwidth]{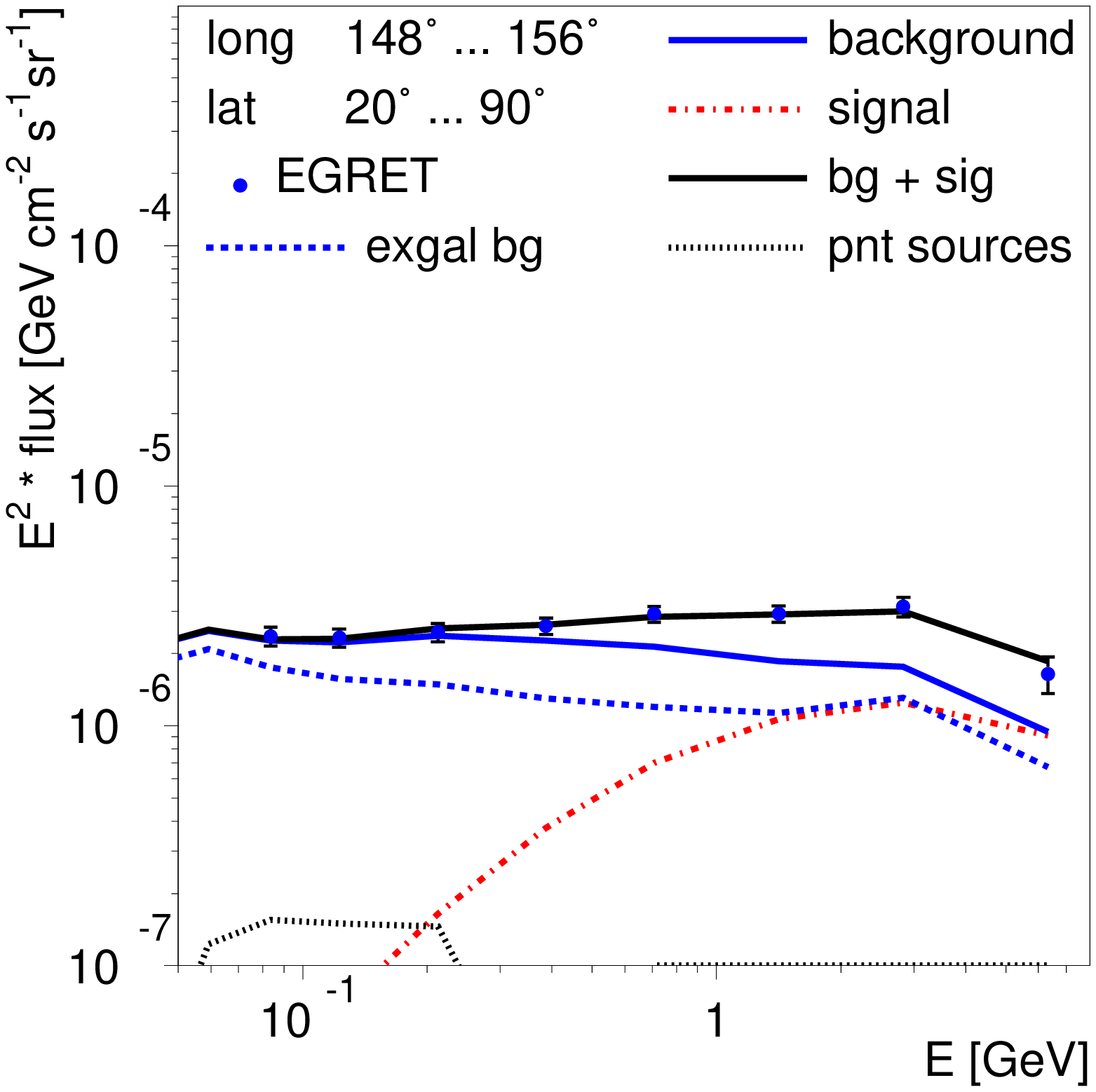}\\
  \end{center}
  \clearpage
  \begin{center}
    \framebox[0.21\textwidth][c]{$\vert \mbox{lat}\vert<5^\circ$}
    \framebox[0.21\textwidth][c]{$5^\circ<\vert \mbox{lat}\vert<10^\circ$}
    \framebox[0.21\textwidth][c]{$10^\circ<\vert \mbox{lat}\vert<20^\circ$}
    \framebox[0.21\textwidth][c]{$20^\circ<\vert \mbox{lat}\vert<90^\circ$}\\
    \hspace{-1cm}
    \begin{turn}{90} \framebox[0.21\textwidth][c]{{\scriptsize $156^\circ<\mbox{long}<164^\circ$}} \end{turn}
    \includegraphics[width=0.21\textwidth]{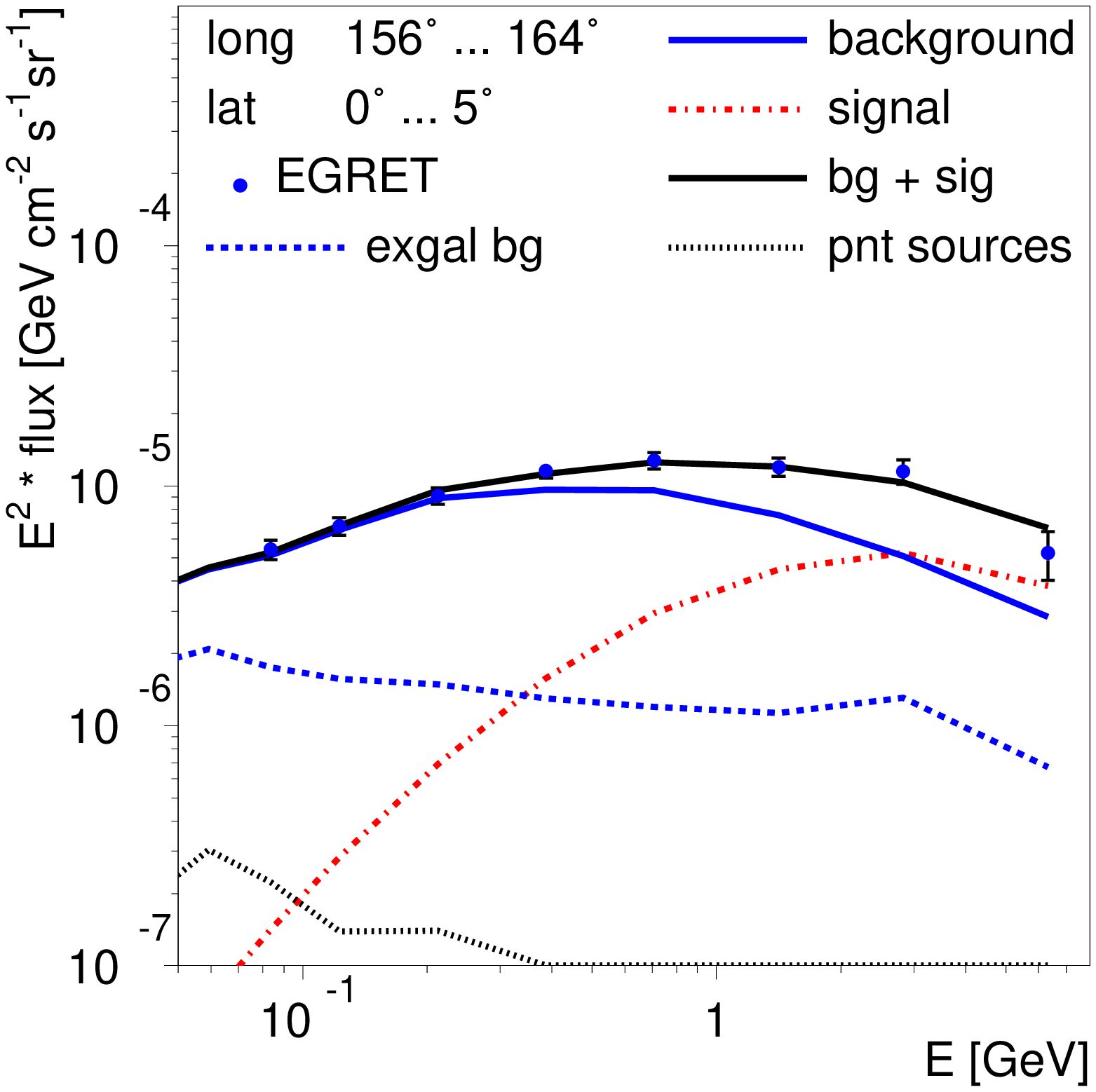}
    \includegraphics[width=0.21\textwidth]{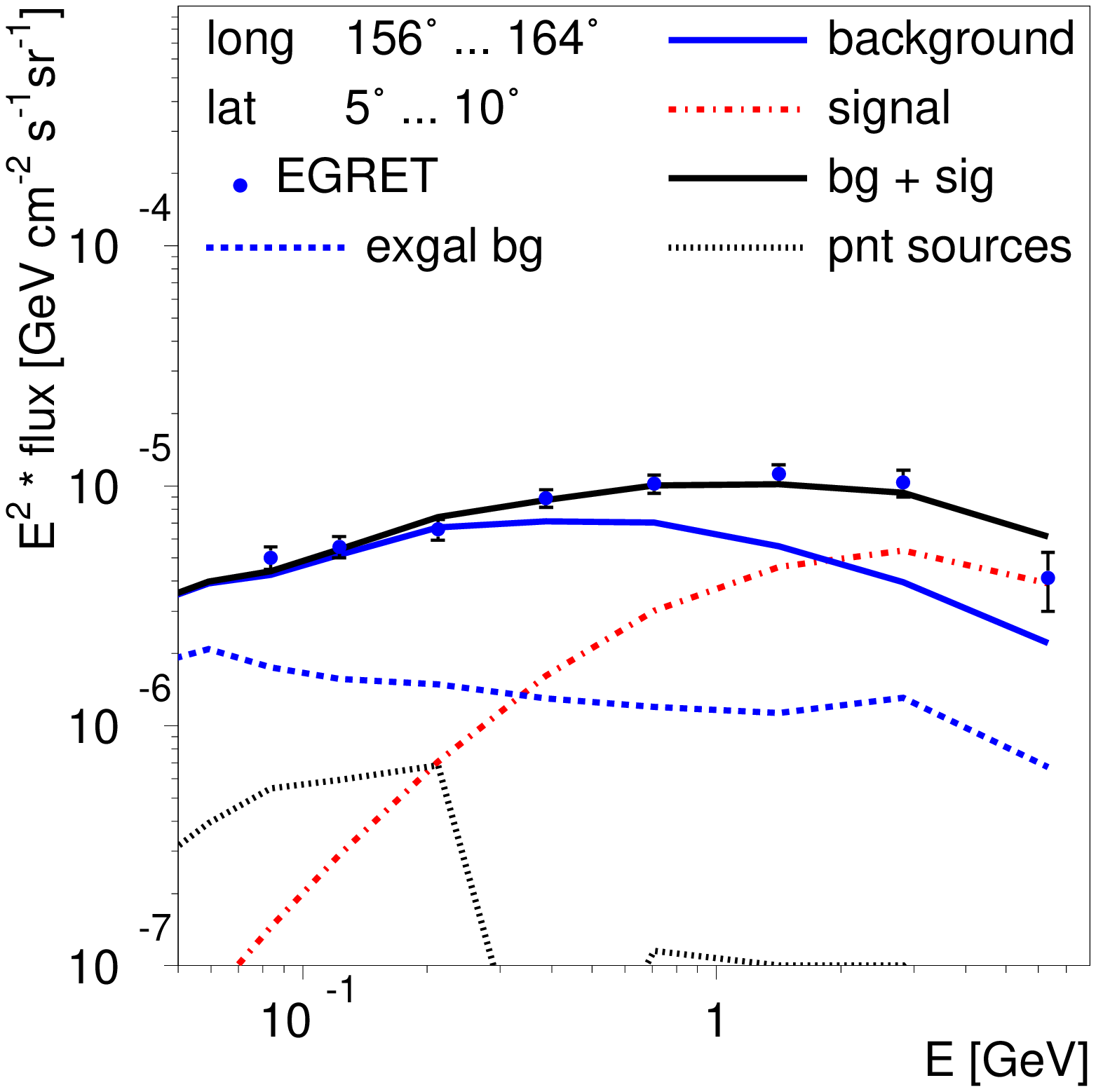}
    \includegraphics[width=0.21\textwidth]{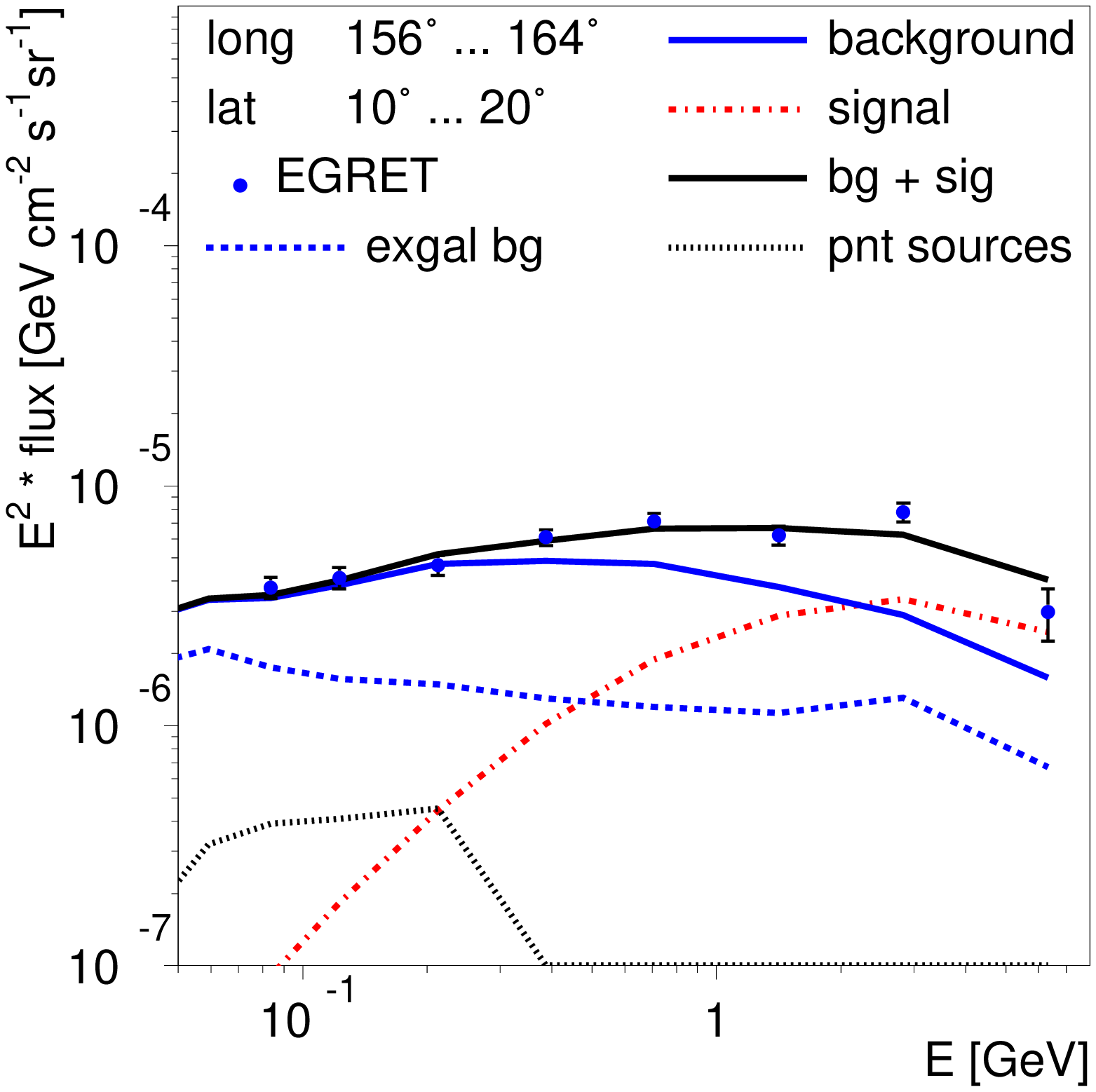}
    \includegraphics[width=0.21\textwidth]{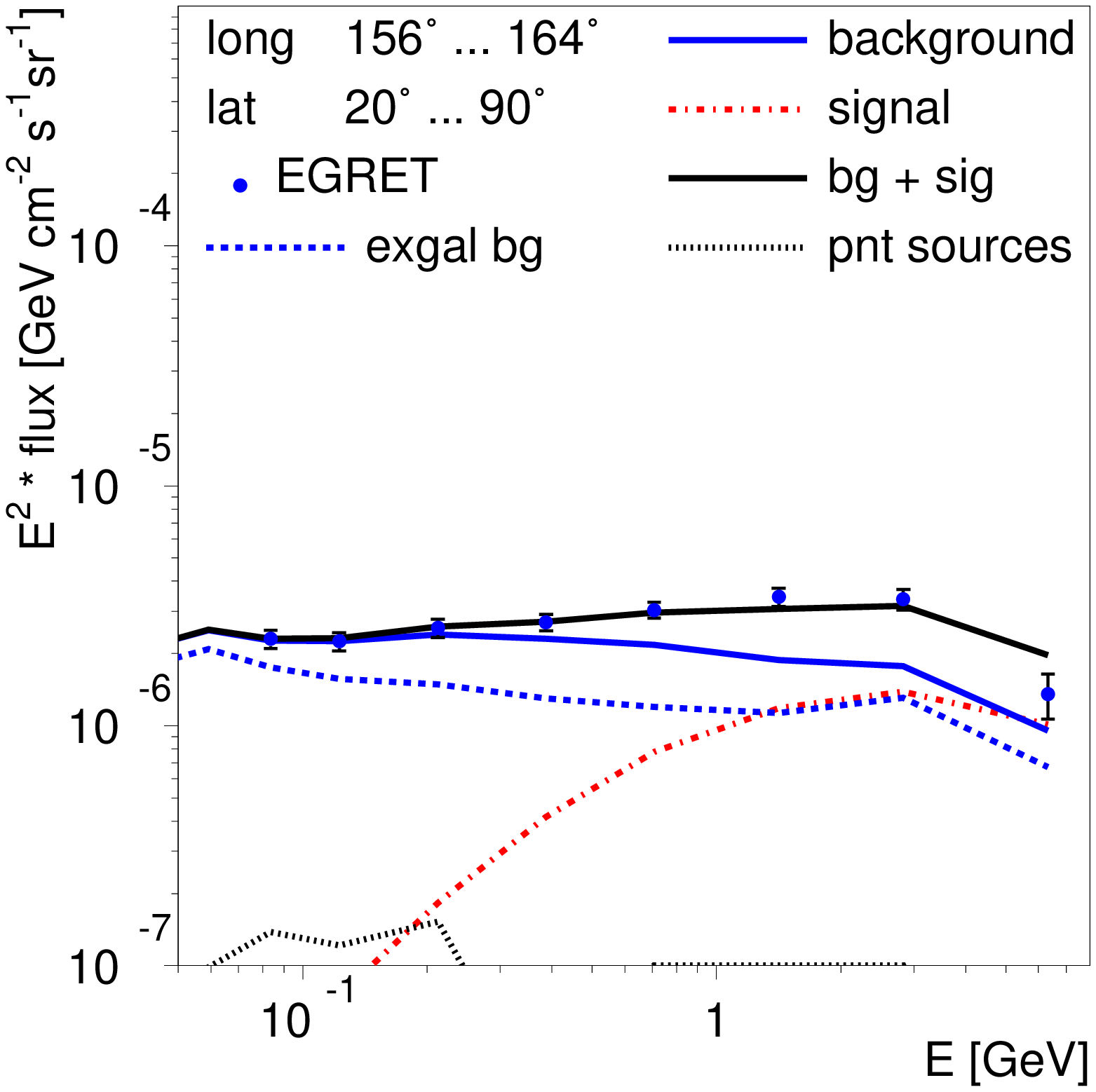}\\
    \hspace{-1cm}
    \begin{turn}{90} \framebox[0.21\textwidth][c]{{\scriptsize $164^\circ<\mbox{long}<172^\circ$}} \end{turn}
    \includegraphics[width=0.21\textwidth]{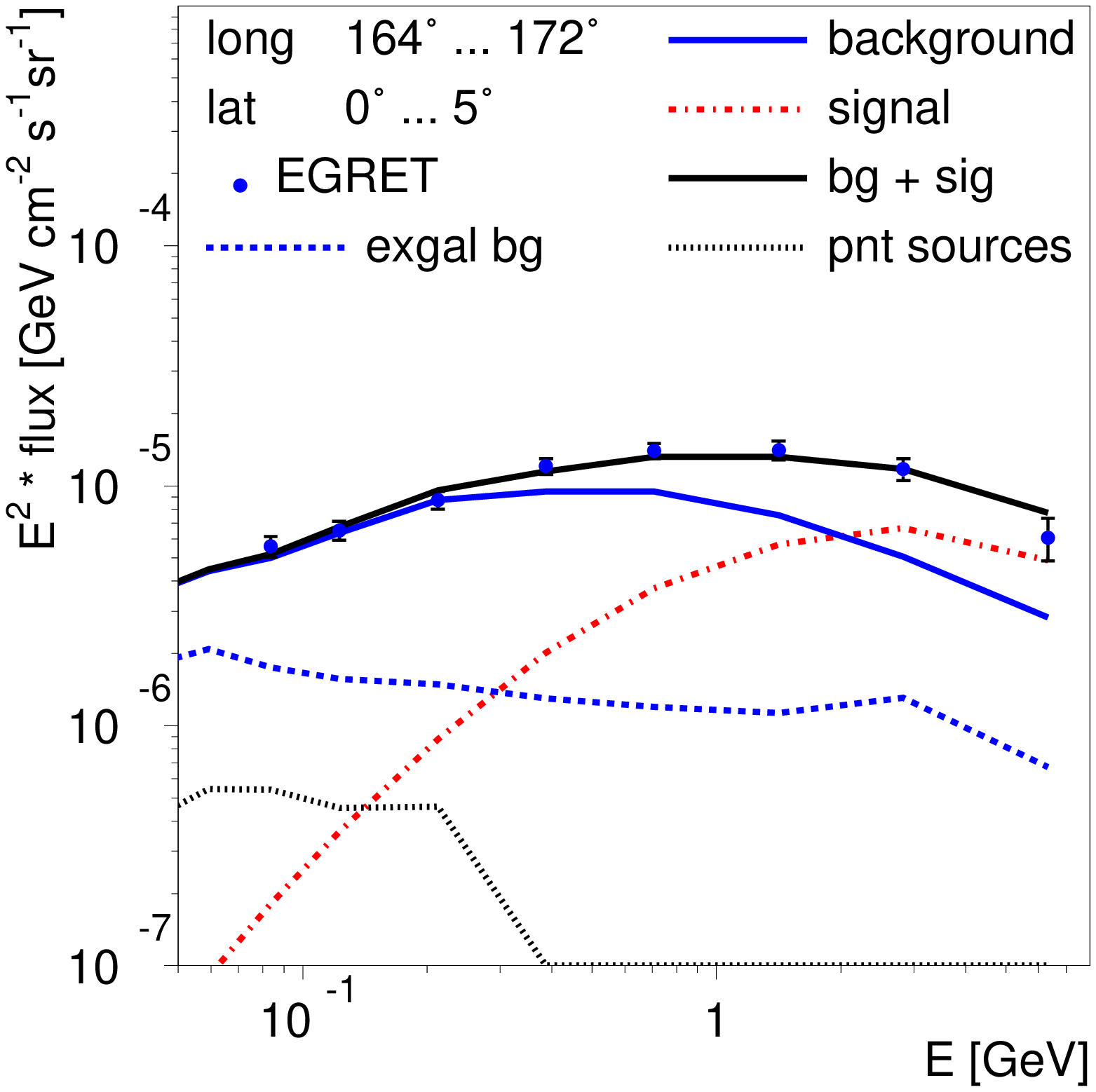}
    \includegraphics[width=0.21\textwidth]{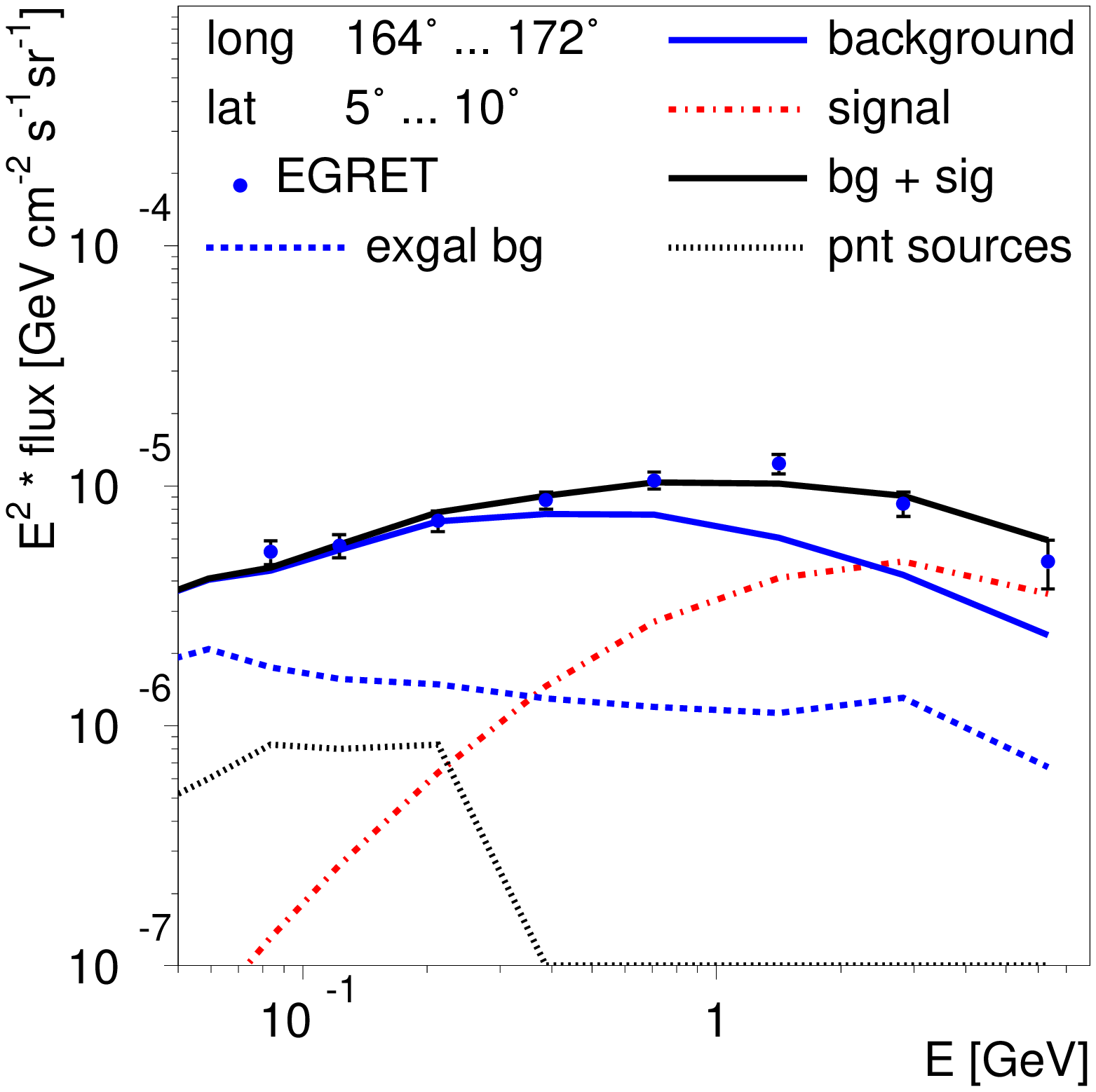}
    \includegraphics[width=0.21\textwidth]{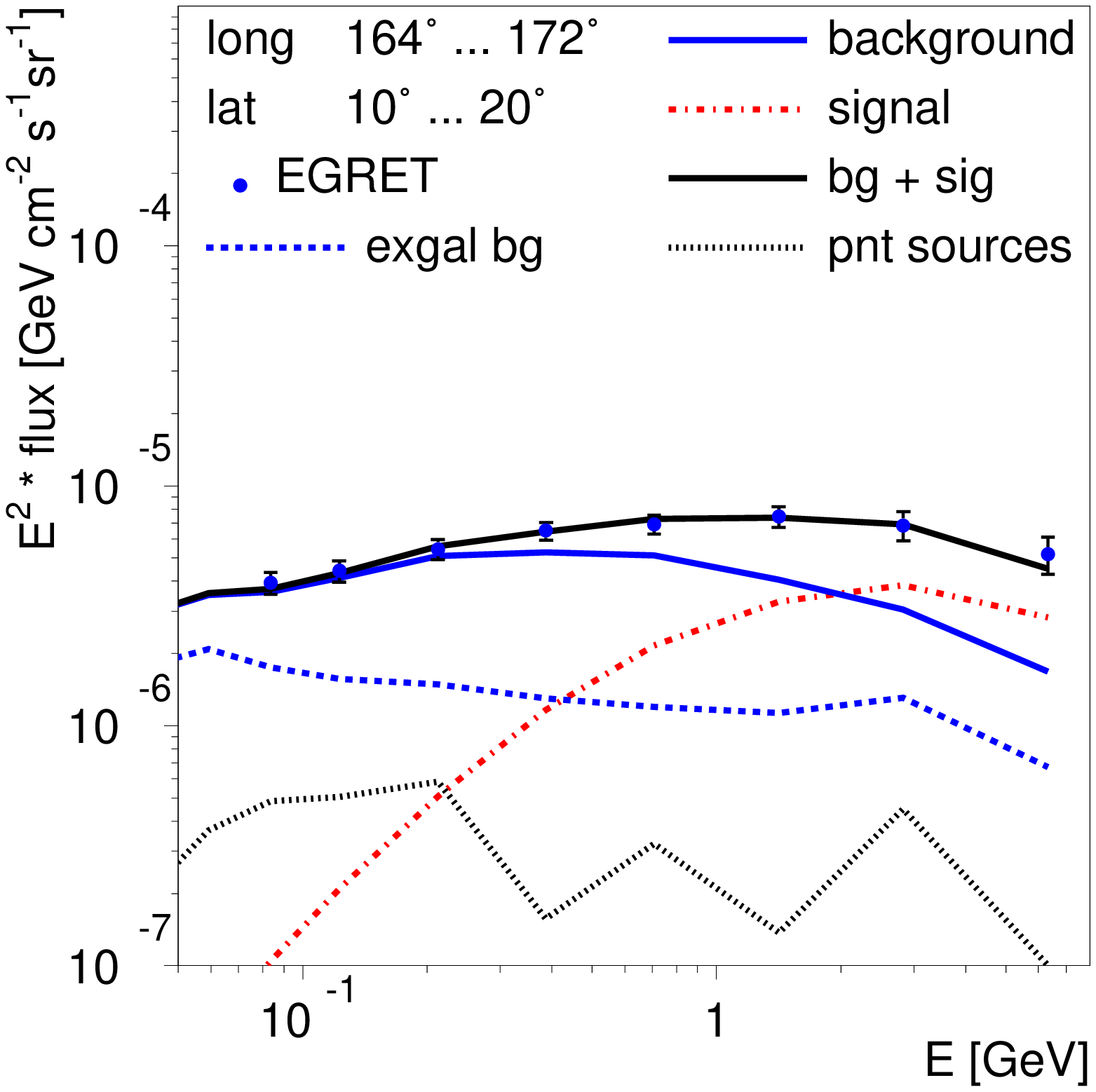}
    \includegraphics[width=0.21\textwidth]{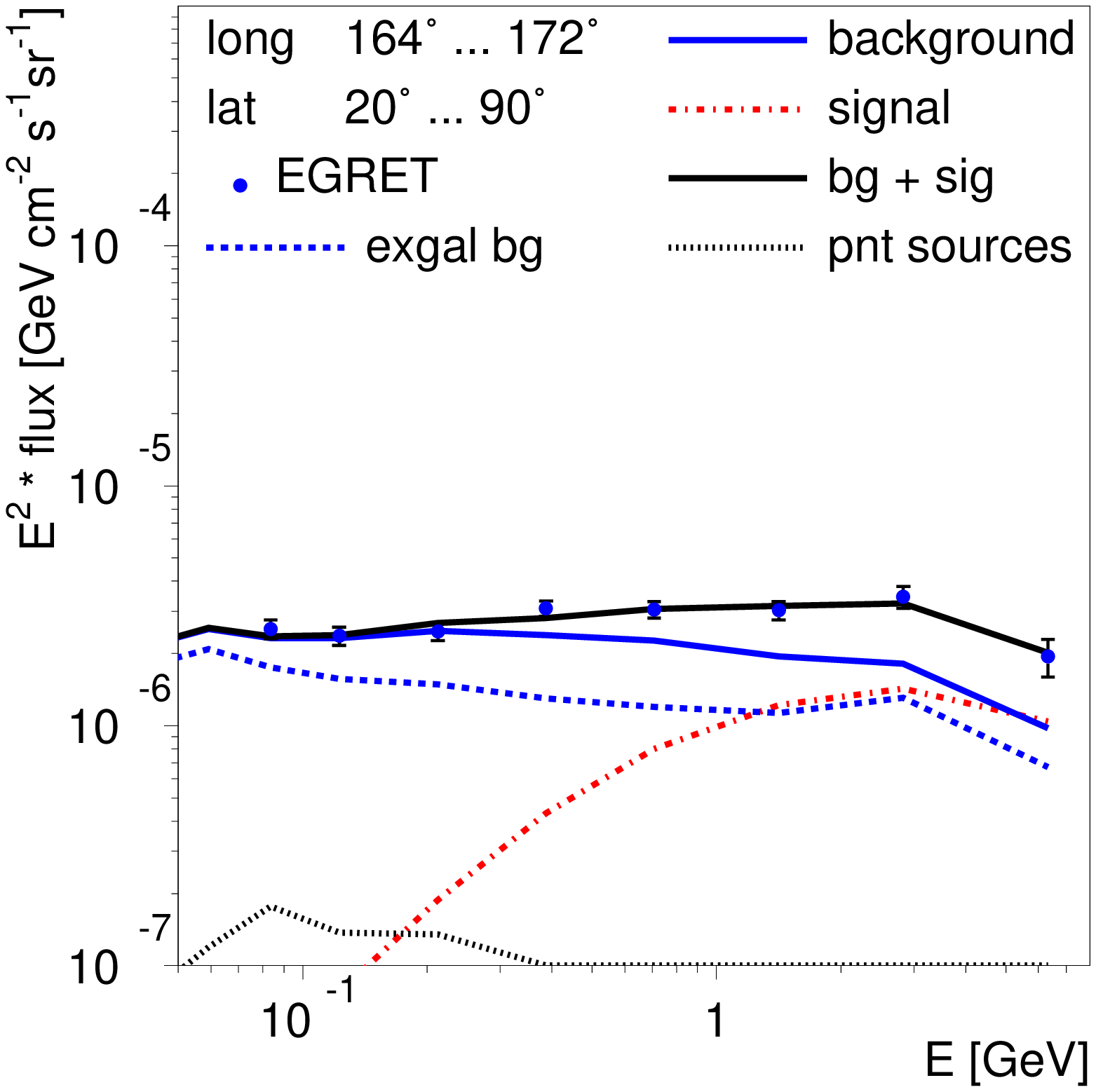}\\
    \hspace{-1cm}
    \begin{turn}{90} \framebox[0.21\textwidth][c]{{\scriptsize $172^\circ<\mbox{long}<180^\circ$}} \end{turn}
    \includegraphics[width=0.21\textwidth]{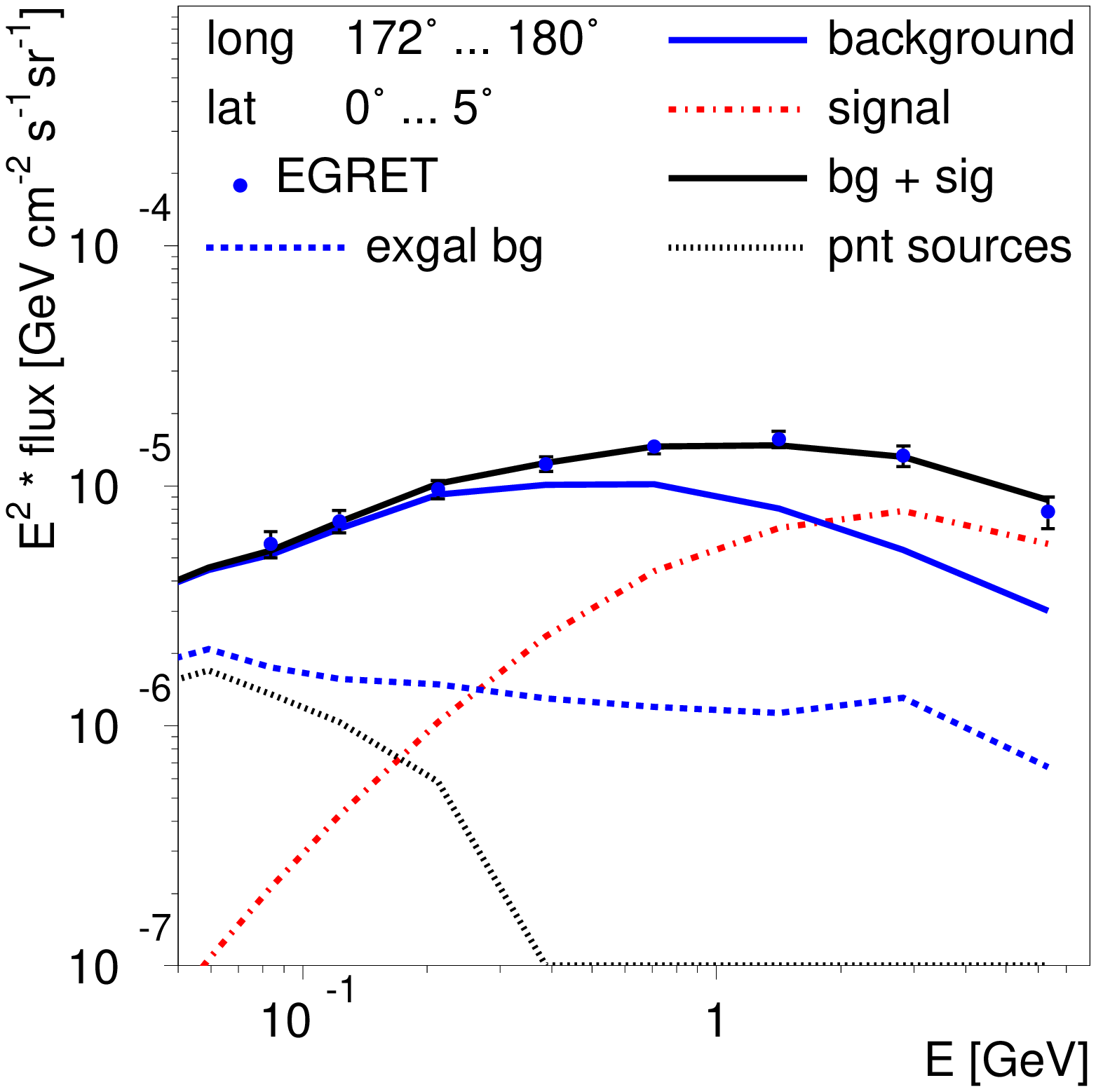}
    \includegraphics[width=0.21\textwidth]{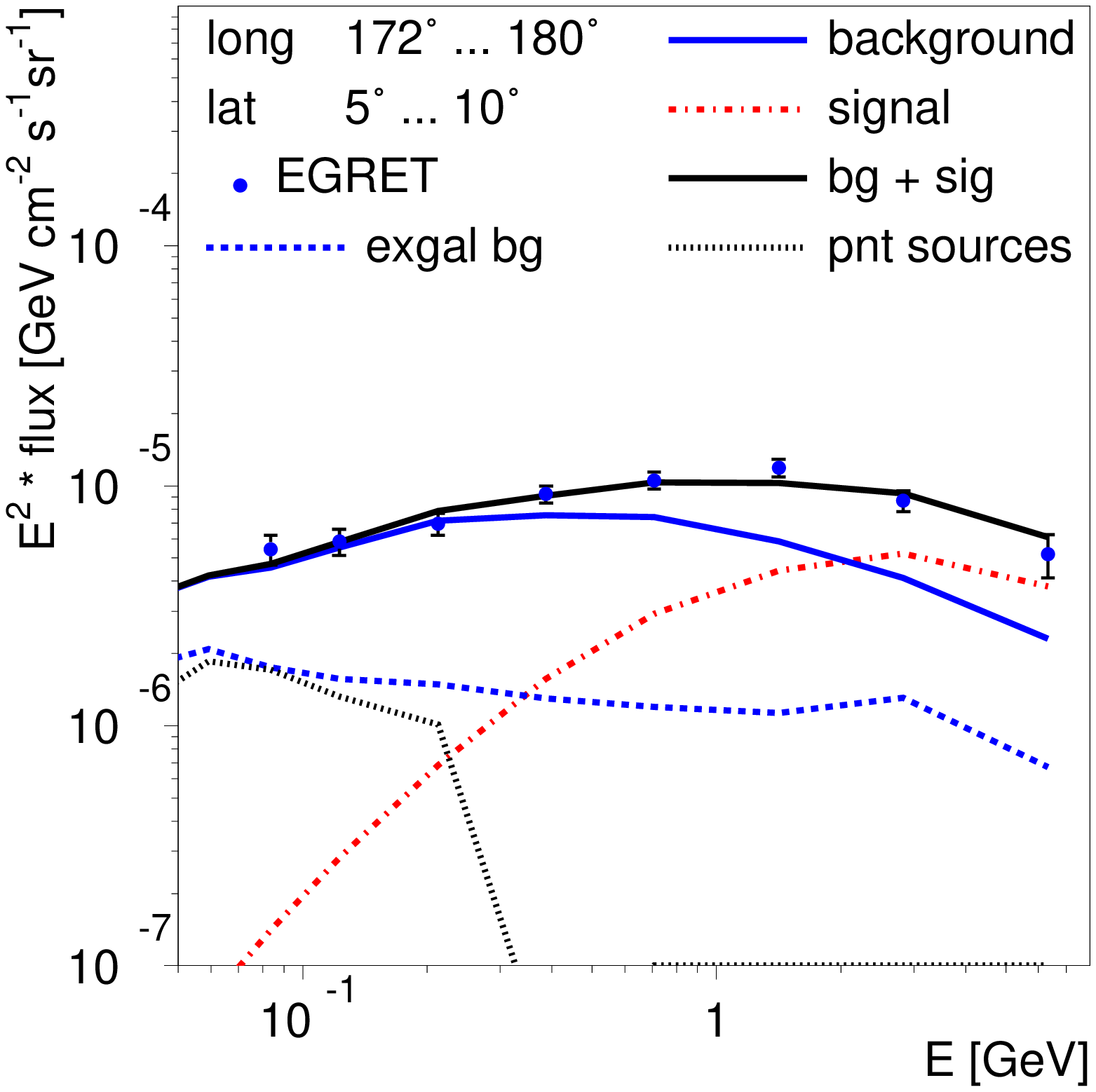}
    \includegraphics[width=0.21\textwidth]{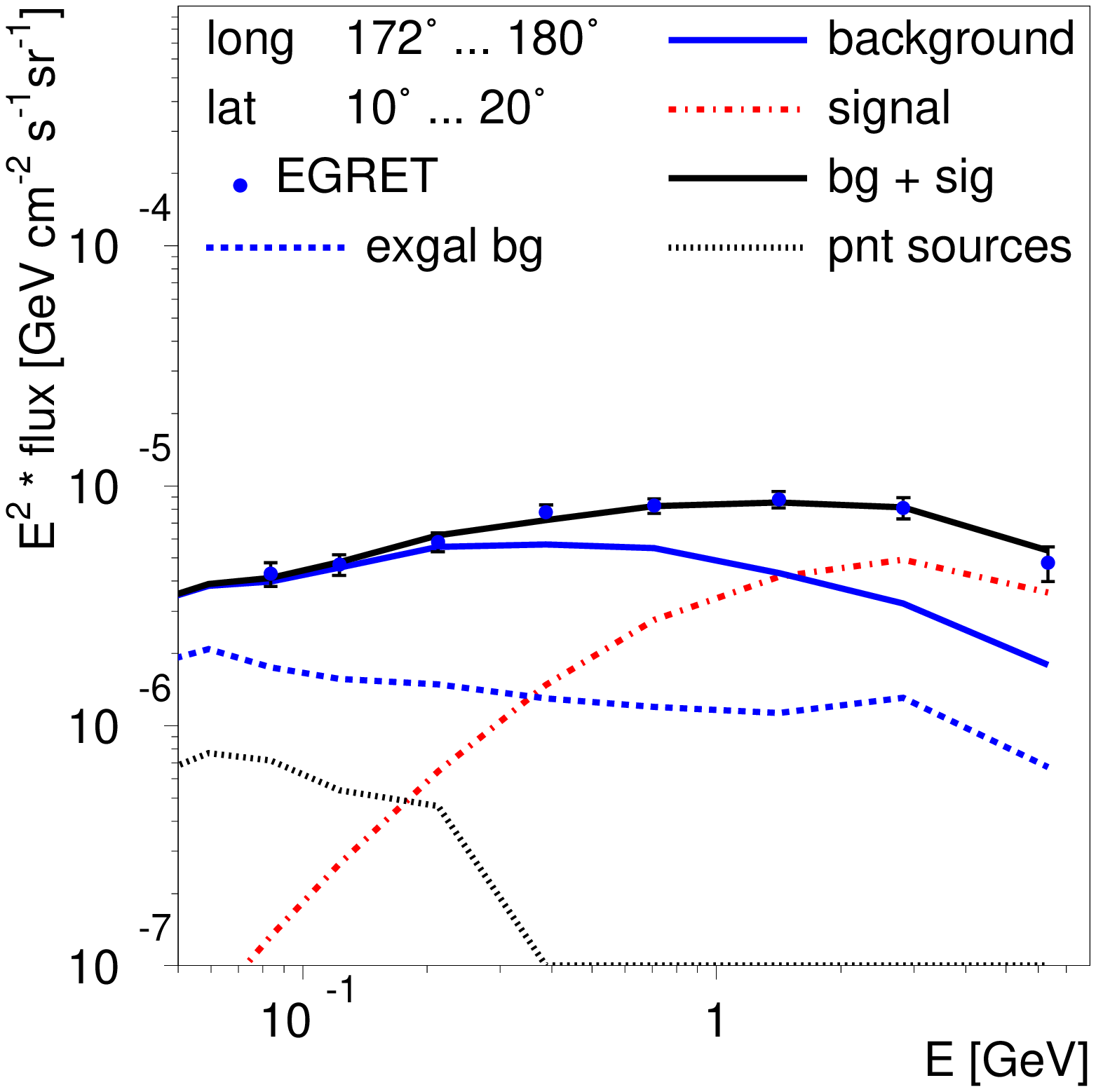}
    \includegraphics[width=0.21\textwidth]{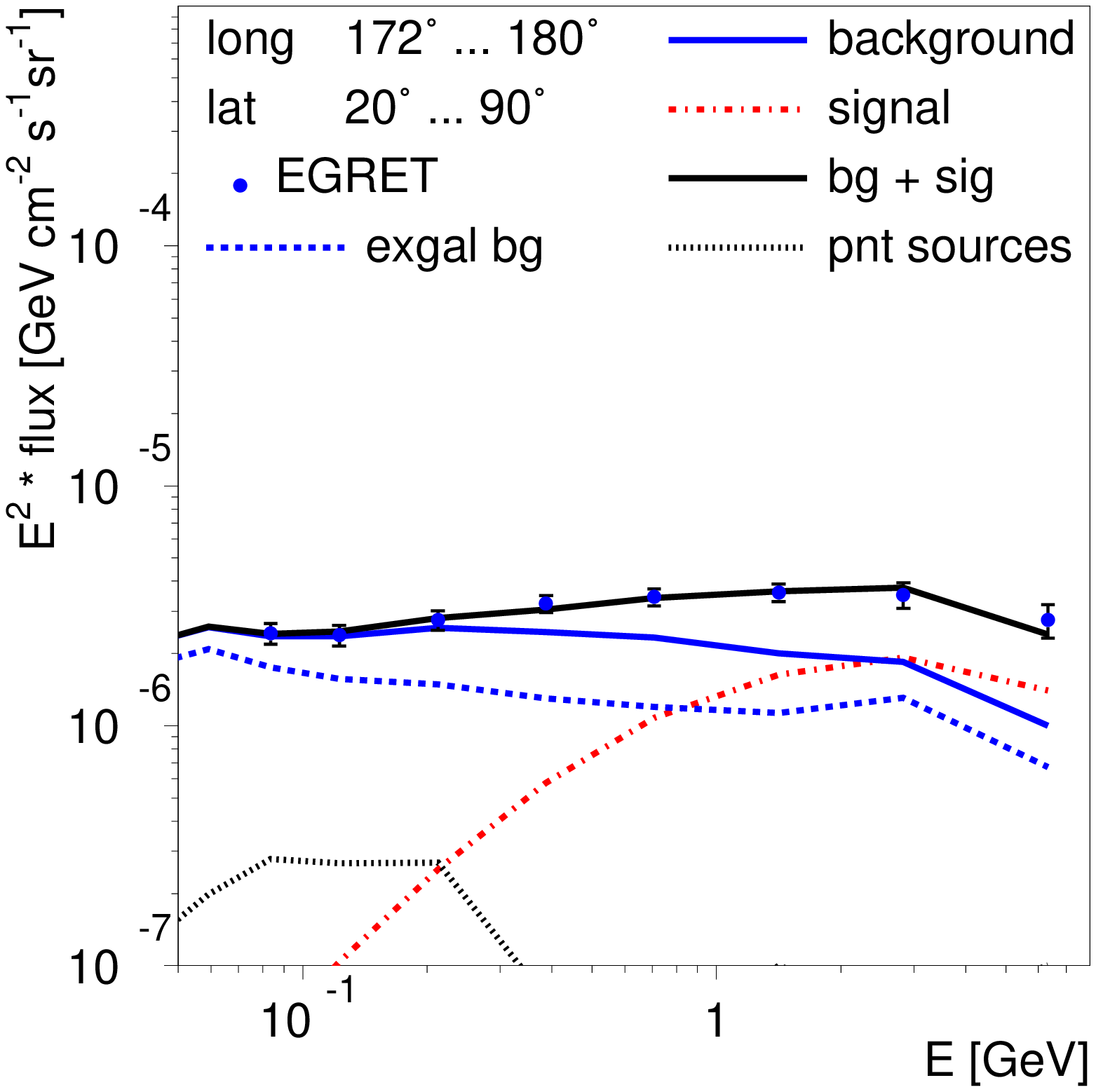}\\
  \end{center}

\end{document}